\documentclass[12pt]{article}
\usepackage{amssymb}
\usepackage{amsfonts}
\usepackage{mathrsfs}
\usepackage{amsmath}
\usepackage{slashed}
\usepackage{url}
\usepackage{bm}
\usepackage{upgreek}
\usepackage{hyperref}
\usepackage{graphicx}
\usepackage{setspace}
\usepackage[compact]{titlesec}
\usepackage[letterpaper,left=1.25in,right=1.25in,top=1in,bottom=1in,nohead]{geometry}
\usepackage{cite}
\usepackage{abstract}
\usepackage{lmodern}
\usepackage[T1]{fontenc}
\usepackage{enumerate}
\usepackage{authblk}
\usepackage{verbatim} 
\usepackage{ifthen}

\newif\iffigs
\figstrue %Figures on
\iffigs
\usepackage{tikz} 
\usepackage{epstopdf}
\usetikzlibrary{calc}
\usetikzlibrary{patterns}
\usetikzlibrary{shapes,snakes}
\usetikzlibrary{plotmarks}
\usetikzlibrary{decorations.markings}
\tikzset{->-/.style={decoration={
				markings,
				mark=at position #1 with {\arrow{>}}},postaction={decorate}}}

\fi

\usepackage{subfig}
\usepackage{amstext}
\usepackage{array} 
\usepackage{xy}
\usepackage{amsthm}

\DeclareMathOperator{\csch}{csch}
\DeclareMathOperator{\Rep}{\mathsf{Rep}}

\newcommand{\QED}{\quad\ensuremath{\square}}

\newcommand{\R}{\ensuremath{\mathbb{R}}}
\newcommand{\Q}{\ensuremath{\mathbb{Q}}}
\newcommand{\Z}{\ensuremath{\mathbb{Z}}}
\newcommand{\BH}{\ensuremath{\mathbb{H}}}

\newcommand{\C}{\ensuremath{\mathbb{C}}}

\newcommand{\Pj}{\ensuremath{\mathbb{P}}}
\newcommand{\W}{\ensuremath{\mathbb{W}}}

\newcommand{\CN}{\ensuremath{\mathcal{N}}}

\newcommand{\rJ}{\ensuremath{\mathsf{J}}}
\newcommand{\Rq}{\ensuremath{\mathsf{q}}}
\newcommand{\Rx}{\ensuremath{\mathsf{x}}}
\newcommand{\Ry}{\ensuremath{\mathsf{y}}}
\newcommand{\CI}{\ensuremath{\mathcal{I}}}

\newcommand{\CH}{\ensuremath{\mathcal{H}}}

\newcommand{\fW}{\ensuremath{\mathfrak{W}}}
\newcommand{\RA}{\ensuremath{\mathsf{A}}}
\newcommand{\RB}{\ensuremath{\mathsf{B}}}
\newcommand{\RC}{\ensuremath{\mathsf{C}}}

\newcommand{\RG}{\ensuremath{\mathsf{G}}}

\newcommand{\RR}{\ensuremath{\mathsf{R}}}

\newcommand{\RW}{\ensuremath{\mathsf{W}}}

\newcommand{\RZ}{\ensuremath{\mathsf{Z}}}
\newcommand{\ra}{\ensuremath{\mathsf{a}}}
\newcommand{\rb}{\ensuremath{\mathsf{b}}}

\newcommand{\rz}{\ensuremath{\mathsf{z}}}
\newcommand{\CO}{\ensuremath{\mathcal{O}}}
\newcommand{\SF}{\ensuremath{\mathscr{F}}}

\newcommand{\sC}{\ensuremath{\mathscr{C}}}

\newcommand{\sP}{\ensuremath{\mathscr{P}}}

\newcommand{\g}{\ensuremath{\mathfrak{g}}}
\newcommand{\h}{\ensuremath{\mathfrak{h}}}

\newcommand{\Tr}{\operatorname{\mathrm{Tr}}}

\def\^{{\wedge}}
\def\*{{\star}}
\def\bar{\overline}
\def\ul{\underline}

\def\e#1{{\rm e}^{\, #1}}
\def\wt#1{\widetilde#1}
\def\ha{\frac{1}{2}}

\def\Dsl{\,\raise.15ex\hbox{/}\mkern-13.5mu D}

\def\End{\mathop{\rm End}}

\def\Hom{\mathop{\rm Hom}}

\def\Vol{\mathop{\rm Vol}}

\def\Ker{{\mathop{\rm Ker}}}
\def\Im{\mathop{\rm Im}}
\def\Re{\mathop{\rm Re}}

\def\diag{\mathop{\rm diag}}
\def\mod{\mathop{\rm mod}}
\def\Res{{\mathop{\rm Res}}}

\def\sgn{\mathop{\rm sign}}

\def\rk{\mathop{\rm rk}}
\def\ch{{\mathop{\rm ch}}}

\def\Vol{{\mathop{\rm Vol}}}

\usepackage{hyperref}  
\usepackage{xcolor} 
\definecolor{dark-blue}{rgb}{0.15,0.15,0.4}

\hypersetup{  
    colorlinks=true,      
    linkcolor=black,         
    citecolor=black,       
    urlcolor=dark-blue          
}

\titleformat{\section}{\normalsize\bfseries}{\thesection}{1em}{}
\titleformat{\subsection}{\normalsize\bfseries}{\thesubsection}{1em}{}

\numberwithin{equation}{section}
\newcommand{\version}{\normalfont December 2018}

\begin{document}
\begin{titlepage}
\begin{flushright}
{\small\ttfamily arXiv:1812.02832}
\end{flushright}
\begin{center}
\vspace{2cm}
{\large\bfseries Anomalies and Holomorphy in Supersymmetric Chern-Simons-Matter Theories}\\
\vspace{1cm}
Nathaniel Bade\footnote{Address after Sept.\,1: Department of
  Mathematics, Northeastern University, Boston MA 02115.}\\
{\small\sl Department of Mathematics, University of British Columbia,\\
  Vancouver BC, Canada V6T 1Z2}\\
\vspace{3mm}
{\small and}\\
\vspace{3mm}
Chris Beasley\\
{\small\sl Department of Mathematics, Northeastern University, Boston
  MA 02115}
\vspace{-5mm}
\end{center}
\begin{abstract}
\baselineskip=18pt
For Chern-Simons-matter theories in three dimensions, gauge invariance
may require the Chern-Simons level $k$ to be half-integral, in which
case parity is violated.  As noted by Pasquetti for abelian theories 
with ${\CN=2}$ supersymmetry, the partition function on the ellipsoid
also admits a suitable holomorphic factorization precisely when the value
of $k$ is properly quantized.  Using known formulas for the partition
function, we investigate analytic aspects of this factorization for non-abelian
gauge groups and general matter representations.  We
verify that factorization occurs in accord with the parity anomaly for
the classical matrix groups and for the exceptional group $G_2$.  In 
an appendix, we discuss the analytic continuation of torus knot
observables in the $SU(2)$ Chern-Simons-matter theory.
\end{abstract}
\vfill\hskip 1cm\version
\end{titlepage}
\begin{onehalfspace}
\tableofcontents\noindent\hrulefill
\section{Introduction}\label{sec:int}

Remarkably, exact formulas \cite{Kapustin:2009kz} are now available
for a large class of observables in ${\CN=2}$ supersymmetric
Chern-Simons-matter theories \cite{Gaiotto:2007qi,Schwarz:2004yj}.
Many efforts -- far too numerous to mention individually -- have been made to
extract theoretical insight from those expressions, as well as to
check their physical consistency.\footnote{For a nice review of the
  background and early developments in the subject, see
  \cite{Marino:2011nm}.  See also the more recent volume
  \cite{Pestun:2016zxk} for a comprehensive review of path integral localization
  techniques for supersymmetric quantum field theories in diverse dimensions.}  The present work
falls into both categories.

Our interest lies in what is really the most basic observable, the
partition function $Z_{S^3}$ on the three-sphere.  This partition
function depends upon the choice of the following data.
\begin{enumerate}
\item A gauge group $G$.  Throughout, $G$ is a compact, connected, 
  simply-connected, and simple Lie group, eg.~${G=SU(N)}$.  We 
  consider ${G=U(1)}$ as a special case.
\item A chiral matter representation ${\Lambda \in \Rep(G)}$.  Here $\Lambda$ is a finite-dimensional, possibly
reducible, representation of $G$.  We decompose $\Lambda$ into irreducibles as 
\begin{equation}
\Lambda \,=\, \left[\lambda_1\right] \oplus \cdots \oplus \left[\lambda_n\right],
\end{equation}
where $\lambda_j$ for ${j=1,\ldots,n}$ is a highest-weight which
labels the corresponding representation.    For ${G=U(1)}$, each
${\lambda_j\in\Z}$ is the charge of the corresponding chiral matter multiplet.
\item A Chern-Simons level ${k\in\ha\Z}$.  We always assume ${k \ge
    0}$, as may be ensured by a suitable choice of orientation on $S^3$.
\item Equivariant (or `real') mass 
  parameters ${\mu\in\R^n}$, valued in the Lie algebra of the
  continuous global flavor symmetry $U(1)^n$.
\item  A Fayet-Iliopoulos (FI) parameter ${\xi\in\R}$ when ${G=U(1)}$.
\end{enumerate} 

The ${\CN=2}$ supersymmetric gauge theory may also include a
superpotential $W$, but the partition function does not depend on the
superpotential except insofar as $W$ breaks global symmetries and so
restricts the allowed values of the real mass $\mu$.
Our examples will have ${W \equiv 0}$ and $\mu$
arbitrary.\footnote{Even if the classical superpotential vanishes,
  non-perturbative effects may generate a quantum
  superpotential ${W\neq 0}$, eg.~as discussed for SQCD in
  \cite{Aharony:1997bx}.  The quantum superpotential still respects classical global symmetries and so does not alter the
  allowed values for the real mass parameter $\mu$.}
For economy of notation, we suppress
the dependence of $Z_{S^3}$ on $(G,\Lambda,k,\mu,\xi)$.  Path  
integral localization at the trivial connection on $S^3$ then produces
an effective matrix integral over the eigenvalues of
the constant mode of the adjoint scalar field $\sigma$ in the
${\CN=2}$ vector multiplet, similar to previous matrix model
expressions for bosonic Chern-Simons theory 
\cite{Beasley:2005vf,Blau:2006gh,Marino:2002fk}.  We recall the
formula for $Z_{S^3}$ momentarily.

Two properties of $S^3$ as a manifold are essential for the 
localization calculation to work so readily.  First, $S^3$ is simply-connected,
so only the contribution from the trivial connection appears in the
sum over semi-classical contributions from flat connections.  For any
other three-manifold, additional terms are included in the sum.  See
for instance
\cite{Benini:2011nc,Alday:2013lba,Gang:2009wy,Imamura:2012rq,Imamura:2013qxa} for an  
extension of the localization calculation to lens spaces, which are
finite cyclic quotients of $S^3$.  For more general Seifert manifolds,
an elegant TQFT analysis via surgery along the fiber has been performed in
\cite{Closset:2017zgf,Closset:2018ghr}.

Second, $S^3$ admits a family of metrics with at least a $U(1)$
isometry.  We shall be interested in the ellipsoid metrics induced from
the Euclidean metric on ${\R^4\simeq\C^2}$ when $S^3$ is embedded as
the subset
\begin{equation}\label{Littleb}
b^2\,|z_1|^2 + b^{-2}\,|z_2|^2 \,=\, 1\,,\qquad\qquad (z_1,z_2)\,\in\,\C^2\,.
\end{equation}
Here ${b \in \R_+}$ is a positive real parameter that labels the embedding and
hence the metric.  For ${b=1}$ the induced metric is the round
metric, but otherwise the metric is squashed.  Up to a relabeling of
the coordinates in \eqref{Littleb}, the metric is invariant under the
inversion ${b \mapsto 1/b}$.  

For all values of $b$, the ellipsoid
metric admits a ${U(1) \times U(1)}$ isometry, which acts
by separate phase rotations on $z_1$ and $z_2$.  When ${b^2 = p/q \in
  \Q}$ is rational, the left side of \eqref{Littleb} is the moment map
for a periodic $U(1)$-action on $\C^2$, and  the ellipsoid metric is
compatible with a global presentation for $S^3$ as a Seifert fibration
\begin{equation}\label{PQHOPF}
\begin{matrix}
&S^1\,\longrightarrow\,S^3\\[-0.5 ex]
&\mskip 57mu \big\downarrow\\
&\mskip 72mu \W\C\Pj^1_{p,q}
\end{matrix}\,,\qquad\qquad\qquad b^2 \,=\,p/q \,\in\,\Q\,.
\end{equation}
Here $\W\C\Pj^1_{p,q}$ is the weighted projective space, with orbifold
points at ${z_1=0}$ and ${z_2=0}$ whenever ${p,q>1}$.   In terms of the Seifert fibration, the ${U(1)\times U(1)}$ isometry acts by rotations
in the fiber and the base, preserving the pair of orbifold
points.  See Chapter $7.1$ in
\cite{Beasley:2009mb}, especially $(7.38)$ therein, for a thorough
discussion of the fibration in \eqref{PQHOPF}.  The orbifold
interpretation is implicit in the considerations of
\cite{Nishioka:2013haa}. 

With the notable exception of Appendix \ref{TorusK}, we will assume
that ${b^2\notin\Q}$ is irrational for all our calculations.
Otherwise, non-generic arithmetic behavior occurs
when $b^2$ is rational.  A complementary analysis of the
theory on the general Seifert manifold with ${b^2\in\Q}$
rational has appeared recently in \cite{Closset:2018ghr}, with which
our results about  
holomorphic factorization have some overlap.

\paragraph{The Partition Function.}

For the ellipsoid metrics on $S^3$, the partition function can
be evaluated \cite{Hama:2011ea,Imamura:2011wg} by supersymmetric localization
exactly as for the round metric, after which $Z_{S^3}$ becomes a 
function of the squashing parameter $b$ as well.
Explicitly,\footnote{Our normalization conventions for $\sigma$ in
  \eqref{ZGS} differ from those in
  \cite{Hama:2011ea,Imamura:2011wg} by ${\sigma_{\rm 
      here} = 2\pi \sigma_{\rm there}}$.} 
\begin{equation}\label{eq:GrandPartitionFunctionInt}
\begin{aligned}
&Z_{S^3} \,=\, \frac{\e{i
    \eta_0}}{|\fW|\cdot\Vol(T)}\int_{\mathfrak{h}}\!d^r\!\sigma
\;\exp{\!\left[-\frac{i\,k}{4\pi}\Tr(\sigma^2)\right]}\,\times\\
&\times\prod_{\alpha\in\Delta_+}\left[4\,\sinh\!\left(\frac{b
      \langle\alpha,\sigma\rangle}{2}\right)\sinh\!\left(\frac{
      \langle\alpha,\sigma\rangle}{2
    b}\right)\right]\cdot \prod_{j=1}^n\left[\prod_{\beta \in
\Delta_j}
s_b\!\left(\frac{\langle\beta,\sigma\rangle}{2\pi}\,+\,\mu_j\right)\right].
\end{aligned}
\end{equation}
Briefly, $|\fW|$ is the order of the Weyl group $\fW$ of $G$, and the
integral runs over a Cartan subalgebra ${\h\subset\g}$ of rank
$r$.\footnote{Eg.~for ${G=SU(N)}$, ${|\fW|=N!}$ and ${r=N-1}$.}
The argument of the exponential in the first line of
\eqref{eq:GrandPartitionFunctionInt} derives from the classical
Chern-Simons action, where `$\Tr$' is a negative-definite, invariant
form on the Lie algebra $\g$ of $G$.  For ${G=SU(N)}$, `$\Tr$' is
normalized as the trace in the fundamental $N$-dimensional
representation.  For other Lie groups, `$\Tr$' is normalized so that
the level $k$ obeys the conventional integral quantization in bosonic
Chern-Simons theory.  For more about our Lie algebra conventions, see
Appendix \ref{se:LieAlgConvention}.  The measure 
${d^r\!\sigma}$ is the $\fW$-invariant Riemannian measure on $\h$
defined using the form `$\Tr$'.  We divide ${d^r\!\sigma}$ by the
volume of a maximal torus ${T \subset G}$, where $\Vol(T)$ 
is also computed with respect to the form `$\Tr$'.  The ratio is then
independent of the normalization for `$\Tr$'.

Appearing in the second line of \eqref{eq:GrandPartitionFunctionInt}
are one-loop determinants from fluctuating, off-diagonal modes in
the vector multiplet as well as the chiral matter multiplets.  From
the vector multiplet, we find a product over the positive roots
${\alpha\in\Delta_+}$ of $\g$, and from the matter multiplet, we find a
product over the weights ${\beta\in\Delta_j}$ in each irreducible
summand $[\lambda_j]$ of the representation $\Lambda$.  Both the roots
and weights are intrinsically valued in the dual $\h^*$ of the Cartan subalgebra
$\h$.  We use ${\langle\,\cdot\,,\,\cdot\,\rangle}$ to indicate the
canonical pairing between ${\alpha,\beta\in\h^*}$ and
${\sigma\in\h}$.    

For technical convenience in Section \ref{HighRank}, we require $[\lambda_j]$ to be weight-multiplicity-free, meaning that each non-zero weight ${\beta\neq 0}$ in $\Delta_j$ has multiplicity-one.  Until that time, when we discuss the condition further, $[\lambda_j]$ is arbitrary.

The star of the show will be the double-sine function $s_b(z)$, which
appears in other guises as the quantum dilogarithm or the hyperbolic
gamma function.  We recall the definition and necessary properties of
the double-sine function in Section \ref{DoubleSine}.  Here we just
remark that $s_b(z)$ is an analytic function of its argument $z$ as
well as the parameter $b$.  Thus the 
real mass ${\mu\in\R^n}$ and the squashing parameter ${b\in\R}$ can be
continued to the complex plane, after which $Z_{S^3}$ depends
holomorphically on $(\mu,b)$.  The imaginary part of the complexified
mass parameter ${\mu_\C\in\C^n}$ is related physically to the action of
the R-symmetry \cite{Festuccia:2011ws,Hama:2010av,Hama:2011ea,Jafferis:2010un} in the
${\CN=2}$ supersymmetry algebra.  With standard conventions, ${\mu_\C
  \equiv \mu
  + \frac{i}{2} \left(b+b^{-1}\right) \RR}$, where ${\RR\in\R^n}$
describes the \mbox{R-charges} of chiral superfields.  From now on, we omit
the subscript on ${\mu_\C}$.

To ensure convergence of the matrix integral in
\eqref{eq:GrandPartitionFunctionInt}, we assume that
both $\mu$ and $k$ are given small positive imaginary parts
${+i\varepsilon}$.  The \mbox{${i\varepsilon}$-prescription} for $k$
implies that the integral over $\sigma$ converges absolutely at
infinity.  Equivalently, one can slightly tilt the integration contour
in \eqref{eq:GrandPartitionFunctionInt} away from the real slice 
${\h\subset\h_\C}$ of the Coulomb-branch.  The \mbox{${i\varepsilon}$-prescription} for
$\mu$ implies that the integrand is everywhere regular for real values
of $\sigma$.  As we review in Section \ref{DefDSine}, the double-sine
$s_b(z)$ has a pole at ${z=0}$ which otherwise collides with the real
axis in \eqref{eq:GrandPartitionFunctionInt}.

Lastly we include a phase factor $\exp{\!\left(i\,\eta_0\right)}$ in
the first line of \eqref{eq:GrandPartitionFunctionInt}.  In general
$\eta_0$ depends on the data $(G,\Lambda,\mu,b)$ which enter the one-loop
determinants, but not on the level $k$.  Even for the pure vector theory
without matter, $\eta_0$ is very delicate to determine, as the phase
of the partition function depends upon the choice of
framing.  A careful discussion of the phase $\eta_0$ for the pure vector
theory\footnote{Even for the pure vector theory, $\eta_0$ depends
  non-trivially on ${b^2=p/q}$.}
appears in Chapter $7.2$ of \cite{Beasley:2009mb} (see also 
\cite{Kallen:2011ny}), but we do not attempt to extend this analysis
of $\eta_0$ to the general ${\CN=2}$ supersymmetric Chern-Simons-matter theory.  Our results therefore apply to the integral in
\eqref{eq:GrandPartitionFunctionInt} only modulo an
undetermined overall phase, which may depend on the holomorphic
parameters $(\mu,b)$.  In practice, we often set ${\eta_0\equiv 0}$
to avoid cluttering the notation.  As will be clear in Section
\ref{OneLoop}, this convention for $\eta_0$ is not stable under renormalization
group flow, and $\eta_0$ is generically non-trivial.

When ${G=U(1)}$, the formula for $Z_{S^3}$ simplifies to
\begin{equation}\label{ZUone}
Z_{S^3} \,=\, \int_{\R} \frac{d\sigma}{2\pi}
\;\exp{\!\left[\frac{i k}{4\pi}\sigma^2 \,+\,
    i\,\xi\,\sigma\right]}\cdot \prod_{j=1}^n
s_b\!\left(\frac{\lambda_j}{2\pi}\,\sigma\,+\,\mu_j\right).
\end{equation}
Here we include the abelian FI parameter $\xi$, and
we omit the product over positive roots ${\alpha\in\Delta_+}$ in
\eqref{eq:GrandPartitionFunctionInt}.  Each ${\lambda_j\in\Z}$ is now the
charge of a corresponding chiral matter multiplet.  Like the real mass
$\mu$, the FI parameter ${\xi \equiv \xi_\C}$ is naturally
complexified, and the imaginary part of $\xi$ encodes the R-charge of the
monopole operator \cite{Willett:2011gp}.  Again, this expression for
$Z_{S^3}$ is only precise up to an overall phase.

\paragraph{The Parity Anomaly.}

In this paper we investigate two properties of
the exact expressions for $Z_{S^3}$ in
\eqref{eq:GrandPartitionFunctionInt} and \eqref{ZUone}.  
The first property is intimately related to the global parity anomaly
\cite{AlvarezGaume:1983ig,Redlich:1983kn,Redlich:1983dv} for gauge
theories in three dimensions and will be well-known to many readers,
but it provides a useful entry point to later analysis.

Let $\Dsl_A$ be the Dirac operator for complex spinors in the 
representation $\Lambda$ coupled to the gauge field $A$.
Microscopically, the global parity anomaly is the statement that the
determinant $\det(\Dsl_A)$, produced by the path integral over the
matter fermions, transforms with a minus sign 
under large, homotopically non-trivial gauge transformations.  Given
our topological
assumptions on $G$, such gauge transformations are classified by a
winding-number $w$ in ${\pi_3(G) \simeq \Z}$, and the determinant
transforms by 
\begin{equation}\label{Spect}
\det(\Dsl_A) \,\buildrel w\over\longmapsto\,
\left(-1\right)^{c_2(\Lambda)\,w}\cdot\det(\Dsl_A)\,,\qquad\qquad w
\in \pi_3(G) \simeq \Z\,.
\end{equation}
Here ${c_2(\Lambda)}$ is the quadratic Casimir of the matter representation
$\Lambda$, 
\begin{equation}
c_2(\Lambda) \,=\, \sum_{j=1}^n c_2(\lambda_j) \,\in\,\Z\,,
\end{equation}
normalized so that ${c_2({\bf N}) = 1}$ for the fundamental
representation of $SU(N)$.  More generally, the quadratic Casimir
${c_2(\g) = 2\,h_\g}$ of the adjoint representation is  
twice the dual Coxeter number $h_\g$.\footnote{Recall ${h_\g=N}$ for
  ${G=SU(N)}$.  The discussion
  ignores the gaugino in the ${\CN=2}$ vector multiplet.  Since
  ${c_2(\g)=2\,h_\g}$ is even, the gaugino does not contribute to the
  parity anomaly.}  For our conventions regarding $c_2(\Lambda)$, see
Appendix \ref{se:LieAlgConvention}.\footnote{An extensive list of values for
$c_2(\Lambda)$ for all simple Lie groups $G$ and various irreducible
representations $\Lambda$ can be found in Table 1
of \cite{McKay:1981}.}  As reviewed in \cite{Witten:1999ds}, the sign
in \eqref{Spect} is given by a spectral flow for the Dirac operator in
three dimensions and can be determined by counting fermion zero-modes
in the background of a four-dimensional instanton configuration on
${S^1 \times S^3}$.  In the latter case, the appearance of the
quadratic Casimir is familiar.

The anomaly \eqref{Spect} in the sign of $\det(\Dsl_A)$ must be
cancelled by the transformation of another term in the 
Lagrangian under the large gauge transformation.  The relevant term is
the Chern-Simons action itself,
\begin{equation}\label{CSA}
\textbf{CS}(A) \,=\, \frac{1}{4\pi}\int_{S^3} \Tr\!\left(A \^ dA +
  \frac{2}{3} A\^A\^A\right) \,\in\, \R/2\pi\Z\,.
\end{equation}
Under a large gauge transformation with winding-number $w$,
\begin{equation}\label{GtCSA}
\exp{\!\big[i\,k\,\textbf{CS}(A)\big]}\,\buildrel w\over\longmapsto\, \e{2\pi i k w}
\cdot \exp{\!\big[i\,k\,\textbf{CS}(A)\big]}\,.
\end{equation}
The product of the determinant in \eqref{Spect} with the exponential
in \eqref{GtCSA} is thus invariant when the Chern-Simons level $k$ is
equal to ${\ha c_2(\Lambda)}$ modulo 1, or
\begin{equation}\label{Quant}
k \,-\, \ha c_2(\Lambda) \,\in\,\Z\,.
\end{equation}
When $c_2(\Lambda)$ is odd, ${k\in \ha + \Z}$ must be a
half-integer and cannot vanish.  In this case, because the
Chern-Simons action is not invariant under orientation-reversal of
$S^3$, parity is broken as the price to preserve
gauge-invariance.

\begin{figure}[t]
\begin{center}
\includegraphics[scale=0.60]{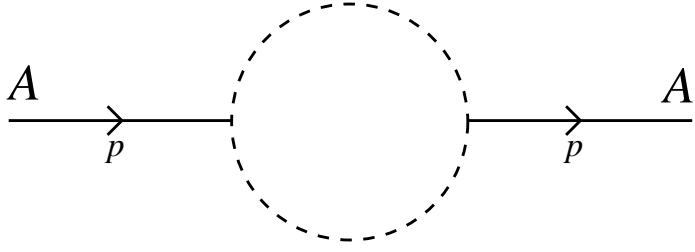}
\caption{One-loop contribution to $k_{\rm eff}$.}\label{1loop}
\end{center}
\end{figure}

The parity anomaly can also be understood perturbatively.
Consider the effective action for the gauge field $A$ which is
obtained when the ${\CN=2}$ matter multiplets are integrated-out with 
large real masses ${|\mu_j|\gg 1}$.  Through the one-loop Feynman
diagram with non-zero external momentum $p$ in Figure \ref{1loop}, the fermions
in the matter multiplets generate an effective Chern-Simons
interaction\footnote{The cubic term in the Chern-Simons action arises
  from a similar one-loop diagram with three external gauge fields at
  zero momentum.} at level 
\begin{equation}\label{KEff}
k_{\rm eff} \,=\, k \,-\, \ha \sum_{j=1}^n c_2(\lambda_j) \,\sgn(\mu_j)\,.
\end{equation}
Here $k$ is the bare Chern-Simons level in the classical Lagrangian
before the matter multiplets are integrated-out.   Quantization of
$k_{\rm eff}$ means that any corrections to $k$ can only arise at
one-loop order, and the quadratic
Casimir $c_2(\lambda_j)$ is associated to the pair of vertices
in the one-loop diagram.

Note that the shift in $k$ depends upon the sign, but not the
magnitude, of each real mass parameter $\mu_j$.\footnote{The sign of
  the real mass $\mu$ determines the sign of the fermion mass term.
  The corresponding mass term for the scalar field in the ${\CN=2}$
  chiral multiplet is always positive, and both positive and negative
  values for $\mu$ are perfectly sensible.} The overall minus
sign in \eqref{KEff} is the result of a delicate computation, first performed for
adjoint matter in \cite{Kao:1995gf}.  As a check, 
for a complex adjoint fermion with positive real mass ${\mu>0}$,
the effective Chern-Simons level is ${k_{\rm eff} = k - h_\g}$ according
to \eqref{KEff}.  This shift by $h_\g$ is twice the shift for a Majorana
fermion in the adjoint representation, determined by geometric
quantization in \cite{Witten:1999ds}.

The relation between the bare and effective Chern-Simons levels in
\eqref{KEff} refines the anomaly-cancellation
condition on $k$ in \eqref{Quant}.  As usual, 
gauge-invariance of the effective Chern-Simons action requires ${k_{\rm
  eff}\in\Z}$ to be an integer.  If ${c_2(\Lambda)\equiv
1\,\mod\,2}$ is odd, the same is true for the signed sum on the right
in \eqref{KEff}, so $k$ must be a half-integer to ensure the
integrality of $k_{\rm eff}$.

For ${G=U(1)}$, a similar version of \eqref{KEff} holds,
\begin{equation}\label{KEffAb}
k_{\rm eff} \,=\, k \,-\, \ha \sum_{j=1}^n \lambda_j^2 \, \sgn(\mu_j)\,,
\end{equation}
where the quadratic Casimir is replaced by the square of the charge
${\lambda_j\in\Z}$.  Again, $k$ is half-integral and parity broken
when the sum of charges ${\sum \lambda_j}$ is odd.

The parity anomaly is a fundamental feature of gauge theory in three
dimensions, and it is worth understanding in multiple ways.  

As a warmup, we explain in Section \ref{DoubleSine} how the formulas for $k_{\rm
  eff}$ in \eqref{KEff} and \eqref{KEffAb} can be obtained 
from the exact expression for the partition function in  \eqref{eq:GrandPartitionFunctionInt}.  Not
surprisingly, since the shift in $k$ arises by integrating-out charged
matter at one-loop, the 
shift is hidden in the asymptotic behavior of the double-sine function
$s_b(z)$ in the integrand of $Z_{S^3}$.  What is slightly 
surprising is that $s_b(z)$ has the correct asymptotic behavior to
reproduce not only the dependence of $k_{\rm eff}$ on the quadratic
Casimir $c_2(\lambda_j)$, but also the dependence on the {\sl sign} of the real mass
$\mu_j$.  Any dependence on the sign of $\mu_j$ cannot be holomorphic, so
agreement with \eqref{KEff} relies on a peculiar, though
well-known, property of the function $s_b(z)$.

\paragraph{Holomorphic Factorization.}

The parity anomaly is also related to a kind of
holomorphic/anti-holomorphic factorization for $Z_{S^3}$ in its
dependence on the real squashing parameter $b$.  Factorization was
first observed by Pasquetti 
\cite{Pasquetti:2012} in the special case ${G=U(1)}$ and
has since been studied in other examples
\cite{Chen:2013pha,Hwang:2012jh,Hwang:2015,Nieri:2013yra,Nieri:2015yia,Taki:2013opa}, typically
with gauge group ${G=SU(N)}$ or ${U(N)}$ and fundamental or
anti-fundamental matter.  See
\cite{Beem:2012mb,Benini:2013yva,Cecotti:2013mba,Fujitsuka:2013fga}
for several theoretical perspectives on the factorization.

Let us state the basic factorization conjecture, due to the authors of
\cite{Beem:2012mb}.  Introduce parameters 
\begin{equation}\label{QQtilde}
\Rq \,=\, \e{\!2\pi i b^2}\,,\qquad\qquad\qquad \wt\Rq \,=\,
\e{\!2\pi i/b^2}\,,
\end{equation}
as well as 
\begin{equation}
\Rx_j \,=\, \e{\!2\pi \mu_j b}\,,\qquad\qquad \wt\Rx_j
\,=\, \e{\!2\pi \mu_j/b}\,,\qquad\qquad j\,=\,1,\ldots,n\,.
\end{equation}
For ${b\in\R_+}$, $\Rq$ and $\wt\Rq$ define points on the unit circle in the
complex plane.  Precisely for rational ${b^2\in\Q}$, $\Rq$ and $\wt\Rq$ are
roots of unity.  Similarly for ${\mu\in\R^n}$, $\Rx$ and
$\wt\Rx$ are valued in the positive corner $(\R_+)^n$.  

More generally, $Z_{S^3}$ depends analytically on the pair $(b,\mu)$, so $b$
and $\mu$ can be allowed to take complex values.  The 
parameters $\left(\Rq,\wt\Rq\right)$ are then complementary in the
following sense. 
When $b^2$ lies in the upper half-plane, ${|\Rq|<1}$
takes values inside the unit disc, and ${|\wt\Rq|>1}$ takes values
outside the unit disc.  Conversely for ${\Im(b^2)<0}$, the roles of
$\Rq$ and $\wt\Rq$ are reversed.  Trivially, the pairs
${\Rq\!\leftrightarrow \wt\Rq}$ and ${\Rx\!\leftrightarrow\wt\Rx}$ are
swapped under the inversion ${b\mapsto 1/b}$.

Factorization amounts to the claim that the partition function can be rewritten
as a finite sum of products\footnote{Sometimes `holomorphic factorization'
  refers more literally to the situation in which a sum such as 
  \eqref{FACTOR} has only a single term.  We prefer the more expansive
notion, relevant when the partition function is the norm-square of
a holomorphic section of a higher-rank, hermitian vector bundle.}
\begin{equation}\label{FACTOR}
Z_{S^3} \,=\, \sum_{m,n\,\in\,\CI} \RG_{m n}\,
\RB^m(\Rq,\Rx)\,\wt\RB^n(\wt\Rq,\wt\Rx)\,,
\end{equation}
if ${k\in\ha\Z}$ obeys the anomaly-cancellation condition
in \eqref{Quant}.  Here ${m,n}$ take values in a finite index set $\CI$, which
depends upon the gauge theory data $(G,\Lambda,k,\mu,\xi)$.  In
general, $\RG_{m n}$ and $\RB^m$ also depend upon this data, with only
the dependence on the squashing parameter $b$ indicated
explicitly.  Thus $\RG_{m n}$ is independent of $b$, and
$\RB^m(\Rq,\Rx)$ depends on $b$ as a convergent
\mbox{$q$-hypergeometric} series in the variables $(\Rq,\Rx)$;
likewise for $\wt\RB^n(\wt\Rq,\wt\Rx)$.
Invariance of $Z_{S^3}$ under ${b\mapsto 1/b}$ implies that 
$\RG_{m n}$ is symmetric in the indices ${m,n}$.

The formula for $Z_{S^3}$ in \eqref{FACTOR} resembles the
conformal-block decomposition for the partition function of a
two-dimensional rational conformal field theory.  Following
\cite{Beem:2012mb}, we refer to the functions 
$\RB^m(\Rq,\Rx)$ as the ``blocks'' of the gauge theory.  As emphasized in
\cite{Beem:2012mb} and as we review in Section \ref{PropDSine}, the block decomposition of $Z_{S^3}$ is subtle, since
both $\RB^m(\Rq,\Rx)$ and $\wt\RB^n(\wt\Rq,\wt\Rx)$ have a natural
boundary of holomorphy on the unit circle ${|\Rq|=|\wt\Rq|=1}$.  Though
$Z_{S^3}$ is well-defined for these values of $\Rq$ and $\wt\Rq$, the
blocks themselves are only defined for values ${|\Rq|<1}$ and
${|\wt\Rq|>1}$, or vice versa.  Nonetheless, from the analytical
perspective, the block decomposition in \eqref{FACTOR} provides a very
concrete, precise conjecture about the structure of the matrix integral in
\eqref{eq:GrandPartitionFunctionInt}.

In this paper, our purpose is to verify the factorization conjecture
directly for general gauge groups $G$ and weight-multiplicity-free
representations $\Lambda$.  Along the way, we derive explicit
expressions for the blocks in low rank, eg.~when ${G=SU(2)}$, 
${SU(3)}$, ${Spin(4)}$, ${Spin(5)}$, and ${G_2}$, and for various matter
representations.

Most attempts to establish the block decomposition rely
upon the holomorphy of the integrand in
\eqref{eq:GrandPartitionFunctionInt} and the residue
theorem.\footnote{But see \cite{Closset:2018ghr} for a very elegant,
TQFT-style proof of the block decomposition for rational ${b^2\in\Q}$.}
Our proof will be no different, though we will use the
multi-dimensional residue theorem to streamline
computations.  Textbook discussions of the multi-dimensional
residue theorem appear in \cite{Griffiths:78,Tsikh:1992}, and our
workhorse version of the residue theorem is proven in
\cite{Passre:1994,Tsikh:1998}.  In Section \ref{JordanLemma} we provide an
independent, physically-motivated derivation of the latter
theorem using finite-dimensional Grassmann integration, extending the ideas in
\cite{Beasley:2003fx}.  A very similar application of
higher-dimensional residues to the evaluation of a sigma model
partition function has appeared in \cite{Gerhardus:2015sla}.   

The main technical advance in our work is not the use of the
multi-dimensional residue theorem {\sl per se}.  Rather, it is to provide a correct
justification for the use of the residue theorem at all.  As emphasized very clearly in \cite{Benini:2013yva}, and as we
review in Section \ref{RankOne}, the standard contour manipulations
used to reduce the evaluation of $Z_{S^3}$ to a residue calculation
are valid only in the special case of ``maximally chiral''
Chern-Simons-matter theories, which have
sufficiently small $|k|$ bounded in terms of $\Lambda$.\footnote{Eg.~if
${G=U(N)}$ and ${\Lambda = {\bf N}^{a_+}\!\oplus \overline{\bf
    N}{}^{a_-}}$ is the direct sum of $a_+$ copies of the fundamental and
$a_-$ copies of the anti-fundamental representation, the ``maximally
chiral'' condition is ${|k| < \ha |a_+ - a_-|}$.  This bound on $k$
is very restrictive and incompatible with the classical limit ${k
  \to \infty}$.}  Otherwise when $|k|$ is
large, the classical Gaussian factor in the first line of \eqref{eq:GrandPartitionFunctionInt}
simply obstructs the usual
step of closing the integration contour for $\sigma$ in the upper or the lower
half-plane to apply the residue theorem.

To avoid this problem, we play a familiar field theory trick:~very
massive chiral matter can be integrated both out and in.  According to the
formula in \eqref{KEff}, a Chern-Simons-matter theory
at level ${k>0}$ with representation $\Lambda$ is equivalent
to another Chern-Simons-matter theory at level ${k'=0}$ with
representation ${\Lambda\oplus\Lambda'}$.  Here $\Lambda'$ is any
representation satisfying ${\ha c_2(\Lambda') = k}$, and we eventually
take the associated real mass ${\mu'\to-\infty}$
to decouple all effects of the extra chiral matter beyond the one-loop
shift from ${k'=0}$ to $k$.  By integrating-in 
auxiliary chiral multiplets and performing a judicious swap of limits,
we reduce the general Chern-Simons-matter theory to the ``maximally
chiral'' case, for which the elementary residue calculus applies to
the matrix integral in \eqref{eq:GrandPartitionFunctionInt}.  

That said, though the idea for our trick comes directly from quantum 
field theory, we work exclusively with well-defined,
finite-dimensional integrals throughout.

\paragraph{The Plan of the Paper.}

We begin in Section \ref{DoubleSine} by reviewing the definition and
analytic properties of the double-sine function $s_b(z)$.   As a small
application, we use these results in combination with the
exact formula for $Z_{S^3}$ to rederive the one-loop shift
\eqref{KEff} of the Chern-Simons level, which in turn implies the parity
anomaly.  Our derivation is probably more useful when
run in the opposite direction, insofar as the parity anomaly provides
the raison d'etre for an otherwise unusual feature of the
double-sine function.

Next in Section \ref{RankOne}, we explain where naive attempts to
use the residue theorem to prove the factorization conjecture fail.
We illustrate the main difficulty in the elementary case ${G=U(1)}$.

In Section \ref{Residue}, we show how to salvage the residue calculus
by integrating-in massive chiral matter to reduce to level ${k'=0}$.
In this degenerate situation, convergence of the localization integral
\eqref{eq:GrandPartitionFunctionInt} over the Cartan subalgebra $\h$
is delicate.  We provide
criteria in Section \ref{SUSYBreak} under which the integral
converges when the Chern-Simons level vanishes.  In Section \ref{SusyQCD} we examine the convergence
criteria for supersymmetric QCD with gauge groups of type SU,
Sp, and SO.  We find that the integral in
\eqref{eq:GrandPartitionFunctionInt}  diverges 
precisely when supersymmetry is spontaneously-broken in the
same theory on $\R^{1,2}$.  We do not have a theoretical
explanation for this coincidence, but it seems worthy of further
investigation.

In Section \ref{JordanLemma} we prove a general version of the
multi-dimensional residue theorem for complex manifolds $M$ with
boundary.  This theorem interpolates between the Cauchy residue
formula and the celebrated Bott residue formula \cite{Bott:1967}.  The
proof is based upon the finite-dimensional supersymmetric integral in
\cite{Beasley:2003fx}, but that integral must be modified
to preserve supersymmetry when the boundary ${\partial
  M}$ is non-empty.  We deduce further consequences
when ${\partial M}$ admits a `polyhedral
decomposition,' a geometric notion introduced here implying a certain
stratification of ${\partial M}$ by CR-submanifolds.
The canonical example occurs when ${M\simeq\Sigma_1 \times \cdots \times
  \Sigma_n}$ is analytically isomorphic to a
product of Riemann surfaces, with ${\partial \Sigma_j \simeq S^1}$ for
each $j$.  When each factor ${\Sigma_j\cong\Delta}$ is a disk and the
integrand admits a suitable meromorphic structure, the result is a sum
of Jeffrey-Kirwan \cite{Jeffrey:1995} residues.  In Section
\ref{ToyEx} we illustrate combinatoric features of the multi-dimensional 
residue theorem with elementary examples.

Finally in Section \ref{HighRank}, we combine the technical
results in Sections \ref{Residue} and \ref{JordanLemma} to demonstrate
holomorphic factorization for $Z_{S^3}$.  This result relies upon the
existence of a suitable configuration of `Jordan divisors' in ${M
  \subset \h_\C \equiv \h\otimes\C}$, and much of our work in Section
\ref{HighRank} is devoted to producing these divisors.  Along the way,
in Section \ref{JordHRGG} we discuss some general features of
Jeffrey-Kirwan residues, and in Section \ref{RkTwoExs} we perform explicit computations of
the blocks $\RB^m(\Rq,\Rx)$ for the rank-two gauge groups $SU(3)$, $G_2$,
$Spin(4)$, $Spin(5)$, as well as $SU(N)$, with various matter
representations.  Some of our formulas have appeared previously in the
literature, but others are new.

The paper includes three appendices.  

In Appendix \ref{se:LieAlgConvention}, we record our Lie algebra conventions, 
including the normalization convention for the quadratic Casimir $c_2$
so that the one-loop formula in \eqref{KEff} holds universally.  

In Appendix \ref{ConvGLem}, we prove an elementary lemma about
hyperplane arrangements and convex polytopes.  This lemma is used in
the discussion in Section \ref{JordHRGG} of Jordan divisors for gauge
groups $G$ with rank larger than two.

In Appendix \ref{TorusK}, we discuss the asymptotics under analytic
continuation in $k$ for torus knot observables in $SU(2)$
Chern-Simons-matter theories.  Though not directly related to the
parity anomaly or holomorphic factorization, this material fits
broadly with our theme and is perhaps useful to include.
Surprisingly, at least for
torus knots, the chiral matter does not change the 
qualitative behavior deduced for the colored Jones polynomial in
\cite{Hikami:2007,Hikami:2010,Murakami:2004asymp}.

\paragraph{Acknowledgments.}

First and foremost, we thank Tudor Dimofte for his comments
and suggestions on a draft version of this paper.  We further
thank Cyril Closset, Thomas Dumitrescu, and Valerio Toledano Laredo for helpful
conversations.  CB additionally thanks the organizers and participants of the
2016 Simons Summer Workshop, hosted at the Simons Center for Geometry
and Physics, Stony Brook University, where some of this work was
performed.  Portions of the paper were also written while CB was
a visitor at Harvard University and at the Perimeter Institute for Theoretical Physics.  CB thanks
the members and staff of both institutions for their generous
hospitality during those visits.

This paper contains results which appear originally in the
PhD thesis \cite{BadeN:2016} of NB.

The work of CB is supported in part under National Science
Foundation Grant No.~PHY-1620637.  Any opinions, findings, and conclusions or
recommendations expressed in this material are those of the
authors and do not necessarily reflect the views of the National
Science Foundation.

\section{Parity Anomaly From Double-Sine}\label{DoubleSine}

We begin in Sections \ref{DefDSine} and \ref{PropDSine} by reviewing
the definition and properties of the double-sine function
$s_b(z)$.  The essential notion goes back over a 
century to Barnes \cite{Barnes:1901,Barnes:1904}, with more recent
revivals in \cite{Faddeev:1995nb,Kharchev:2001rs,Kurokawa:1991,Shintani:1977}.  For
additional discussion about the double-sine, see for instance the
relevant portions of 
\cite{vandeBult:2007,Bytsko:2006ut}.  A complete review appears in
\cite{Kurokawa:2003}.

In Section \ref{OneLoop}, we combine these
properties of $s_b(z)$ with the exact formula for $Z_{S^3}$ to
rederive the one-loop renormalization of the Chern-Simons level in
\eqref{KEff}.  This feature of $Z_{S^3}$ is well-known to experts and for
unitary gauge groups has been previously observed in
\cite{Aharony:2013dha,Benini:2011mf,Willett:2011gp}.

\subsection{Definition of the Double-Sine Function}\label{DefDSine}

The double-sine function admits a variety of analytic expressions
which could be used as a definition.  We will begin with the
expression which appears naturally in the localization computation
leading to \eqref{eq:GrandPartitionFunctionInt},
\begin{equation}\label{eq:DoublesineProd}
s_b(z) \,\sim\, \prod_{m,n\ge 0} \frac{\left(m+1\right) b +
  \left(n+1\right) b^{-1} + i\,z}{m\,b \,+\, n\,b^{-1} -
  i\,z}\,,\qquad\qquad b \,\in\, \R_+\,,\quad z\,\in\,\C\,.
\end{equation}
Because the double-sine arises from a one-loop determinant, $s_b(z)$ is
expressed as an infinite product over a pair of positive integers
${m,n\ge 0}$.  This infinite product does not converge for any value
of $z$ and hence is only a formal expression.  We emphasize this fact by
using the symbol `$\sim$' in \eqref{eq:DoublesineProd} rather than an equality.

The true meaning of the product formula in \eqref{eq:DoublesineProd} is that $s_b(z)$
will be defined as a meromorphic function of $z$ with zeroes and poles
at the locations below,
\begin{equation}\label{eq:PolesOfDoublesine}
\begin{aligned}
\fbox{zeroes of $s_b(z)$}:\quad z_* \,&=\, i\,(m+1)\,b\,+\,
i\,(n+1)\,b^{-1}\,,\qquad m,n\ge 0\,,\\
\fbox{poles of $s_b(z)$}:\quad z_* \,&=\, -i\,m\,b \,-\, i\,n\,b^{-1}\,.
\end{aligned}
\end{equation}
See Figure \ref{fi:zerosAndPolesofdSine} for a sketch.  For ${b>0}$ real, all
zeroes of $s_b(z)$ lie along the positive 
imaginary axis, and all poles lie along the negative
imaginary axis, including the origin.  For irrational ${b^2\notin\Q}$,
the zeroes and poles are moreover simple.  Otherwise, if ${b^2\in\Q}$ is
rational, each zero or pole at $z_*$ has the same multiplicity as the
number of positive integral pairs $(m,n)$ satisfying the relations in
\eqref{eq:PolesOfDoublesine}.  Later in Sections \ref{RankOne} and \ref{Residue}, 
we assume the generic case ${b^2\notin\Q}$ precisely to avoid 
thorny arithmetic associated to higher-order poles in $s_b(z)$.

\iffigs
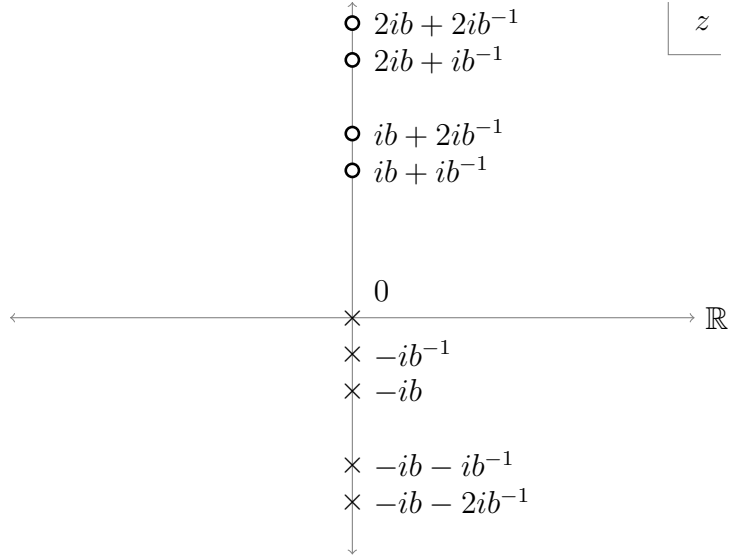
\begin{figure}
	\begin{center}
		\begin{tikzpicture}[x=.7cm,y=.7cm]
		
		\draw [<->, gray](-6.5,0)--(6.5,0) node [right, black] {$\R$};
		\draw [<->, gray](0,-4.5)--(0,6);

		\draw [thin, gray](6,6)--(6,5) -- (7,5);
		\node at (6.65,5.6) {$z$};

		%Poles
		\node at (0,0) {$\times$};		
			
		\node at (0,-.7) {$\times$};
		\node [anchor=west] at (.2,-.7) {$-ib^{-1}$};	
		
		\node at (0,-1.4) {$\times$};
		\node [anchor=west] at (.2,-1.4) {$-ib$};

		\node at (0,-2.8) {$\times$};
		\node [anchor=west] at (.2,-2.8) {$-ib-ib^{-1}$};
		\node at (0,-3.5) {$\times$};
		\node [anchor=west] at (.2,-3.5){$-ib-2ib^{-1}$};

                \node [anchor=west] at (.2,.5){$0$};
		
		%Zeros
		\fill[black] (0,2.8) circle(.15);
		\fill[white] (0,2.8) circle(.1);
		\node [anchor=west] at (.2,2.8) {$ib+ib^{-1}$};
		
		\fill[black] (0,3.5) circle(.15);
		\fill[white] (0,3.5) circle(.1);
		\node [anchor=west] at (.2,3.5) {$ib+2ib^{-1}$};
		
		\fill[black] (0,4.9) circle(.15);
		\fill[white] (0,4.9) circle(.1);
		\node [anchor=west] at (.2,4.9) {$2ib+ib^{-1}$};
		
		\fill[black] (0,5.6) circle(.15);
		\fill[white] (0,5.6) circle(.1);
		\node [anchor=west] at (.2,5.6) {$2ib+2ib^{-1}$};
		
		%	\draw [->, gray] (-3.8,3.8) --  (-4.6,4.6);
		%	\draw [->, gray] (3.8,-3.8) --  (4.6,-4.6);
		\end{tikzpicture}
		\caption{Zeroes $(\circ)$ and poles $(\times)$ of $s_b(z)$
                  for ${b\in\R_+}$.  All zeroes and poles are simple
                  when ${b^2\notin\Q}$.  The points ${z=i b}$ and ${z=i
                    b^{-1}}$ are neither zeroes nor
                  poles.}\label{fi:zerosAndPolesofdSine}
	\end{center}
\end{figure}\fi 

Once the zeroes and poles are fixed, $s_b(z)$ is determined up to
multiplication by a non-vanishing entire function of $z$.
Equivalently, $s_b(z)$ will be determined once its asymptotic behavior as
${|z|\to\infty}$ is fixed.  From either perspective, to define $s_b(z)$
honestly as a meromorphic function, the infinite product in
\eqref{eq:DoublesineProd} must be regularized.

We consider the logarithm, again a formal expression,
\begin{equation}\label{LogProdDS}
\ln s_b(z) \,\sim\, \sum_{m,n\ge 0} \ln\!\left[\left(m+1\right) b +
  \left(n+1\right) b^{-1} + i\,z\right]-\ln\!\left[m\,b \,+\, n\,b^{-1} -
  i\,z\right]\,.
\end{equation}
In logarithmic form, the divergence of the sum over $m$ and $n$ on the
right in \eqref{LogProdDS} is clear.  To obtain an
absolutely-convergent series, we differentiate the summand successively with
respect to $z$, so that 
\begin{equation}
\frac{\partial}{\partial z}\ln s_b(z) \,\sim\, \sum_{m,n\ge 0} \frac{i}{\left[\left(m+1\right) b +
  \left(n+1\right) b^{-1} + i\,z\right]} \,+\, \frac{i}{\left[m\,b \,+\, n\,b^{-1} -
  i\,z\right]}\,,
\end{equation}
and 
\begin{equation}\label{2ndDLogDS}
\begin{aligned}
\frac{\partial^2}{\partial z^2}\ln s_b(z) \,&=\, \sum_{m,n\ge 0} \frac{1}{\left[\left(m+1\right) b +
  \left(n+1\right) b^{-1} + i\,z\right]^2} \,-\, \frac{1}{\left[m\,b \,+\, n\,b^{-1} -
  i\,z\right]^2}\,,\\
&=\, \sum_{m,n\ge 0} -\frac{\big[\left(2m+1\right) b +
    \left(2n+1\right) b^{-1}\big]\big[b + b^{-1} + 2 i z\big]}{\big[m\,b \,+\, n\,b^{-1} -
  i\,z\big]^2\,\big[\left(m+1\right) b +
  \left(n+1\right) b^{-1} + i\,z\big]^2}\,.
\end{aligned}
\end{equation}
In passing to the second line of \eqref{2ndDLogDS}, we just evaluate
the difference in the first line.  Evidently, the series for the would-be
second-derivative of ${\ln s_b(z)}$ does converge absolutely and
uniformly on compact subsets of $\C$, since the
denominator in the summand grows quartically for large ${m,n\gg 1}$
and fixed $z$, while the numerator only grows linearly.

Although the formal product formula in
\eqref{eq:DoublesineProd} does not actually define $s_b(z)$, the convergent series 
in \eqref{2ndDLogDS} does define the second-derivative of ${\ln
  s_b(z)}$.  Thus 
\begin{equation}\label{LogDS}
\ln s_b(z) \,=\, C_{0,b} \,+\, C_{1,b} \, z \,+\, F_b(z)\,,
\end{equation}
where $C_{0,b}$ and $C_{1,b}$ are undetermined constants, possibly
depending upon $b$, and $F_b(z)$ is
obtained by integrating the holomorphic function in \eqref{2ndDLogDS}
twice with respect to $z$.\footnote{For instance, the integration can be
  accomplished by using the Taylor expansion of the series in
  \eqref{2ndDLogDS} about a specified point in the $z$-plane.  The choice
of the point is absorbed into the constants $C_{0,b}$ and $C_{1,b}$.}  Note that $F_b(z)$ satisfies 
\begin{equation}
F_{1/b}(z) \,=\, F_{b}(z)\,,
\end{equation}
as the series in \eqref{2ndDLogDS} is manifestly invariant under the inversion
${b\mapsto 1/b}$.  If $C_{0,b}$ and $C_{1,b}$ are themselves
invariant under ${b\mapsto 1/b}$, the double-sine obeys
\begin{equation}\label{InvertB}
s_{1/b}(z) \,=\, s_b(z)\,,
\end{equation}
ensuring that $Z_{S^3}$ in \eqref{eq:GrandPartitionFunctionInt} and \eqref{ZUone} is invariant as well.

The constants $C_{0,b}$ and $C_{1,b}$ describe potentially
$b$-dependent renormalizations 
of the partition function on $S^3$, in the sense that any choices for
these constants can be absorbed into the definitions of other
constants already appearing in our formula for $Z_{S^3}$.  From
\eqref{eq:GrandPartitionFunctionInt} and \eqref{ZUone}, the constant $C_{0,b}$ can be absorbed into the
overall normalization of $Z_{S^3}$, and the constant $C_{1,b}$ can be
absorbed into the definition of the abelian FI parameter $\xi$.  When $G$ is
non-abelian and simple, no FI term is present, and $C_{1,b}$ simply
cancels out in the product over weights of $\Lambda$ in \eqref{eq:GrandPartitionFunctionInt}.

The values of the constants $C_{0,b}$ and $C_{1,b}$ can be fixed
in various ways.  The standard approach is to use
the Barnes double-zeta function
\begin{equation}\label{def:MultipleZeta}
\zeta_2(s,x \,\big|\, \omega_1,\omega_2) \,=\,\sum_{m,n\geq 0} \left(m\,\omega_1 + n\,\omega_2 + x\right)^{-s}\,,
\end{equation}
depending on non-zero parameters ${\omega_1,\omega_2\neq 0}$.  For
${{\rm Re}(s)>0}$ sufficiently large, the sum on the right is
absolutely-convergent for all values of $x$.  Otherwise, the
definition of $\zeta_2$ is extended to other values of $s$ by analytic
continuation, with simple poles at ${s=1,2}$.  

The Barnes double-zeta function is a generalization of the classical
Hurwitz zeta function
\begin{equation}
\zeta(s,x) \,=\, \sum_{n\ge 0} \left(n + x\right)^{-s}\,.
\end{equation}
Specifically, by analogy to \eqref{def:MultipleZeta} set 
\begin{equation}
\zeta_1(s,x\,\big|\,\omega)
\,=\, \sum_{n\ge 0}
\left(n\,\omega \,+\, x\right)^{-s}\,.
\end{equation}
Then immediately,
\begin{equation}\label{ZetaOne}
\zeta_1(s,x\,\big|\,\omega) \,=\, \omega^{-s}\,\zeta(s, x/\omega)\,.
\end{equation}
For later reference, recall also the Hurwitz zeta identity 
\begin{equation}\label{HZetaId}
\zeta(0,x) \,=\, \ha \,-\, x\,.
\end{equation}

With the Barnes double-zeta function, we next introduce the Barnes 
double-gamma function\footnote{Our normalization of $\Gamma_2$ follows
  the modern convention, which differs from that of Barnes.}
\begin{equation}\label{DoubleGam}
\begin{aligned}
\Gamma_2(x \,\big|\, \omega_1,\omega_2) \,&=\,
\exp{\!\left[\frac{\partial}{\partial s}\zeta_2(s,x \,\big|\,
    \omega_1,\omega_2)\big|_{s=0}\right]}\,,\\
&\sim\, \prod_{m,n\ge 0} \frac{1}{\left(m\,\omega_1 + n\,\omega_2 + x\right)}\,.
\end{aligned}
\end{equation}
As suggested by the formal product representation in the second line
of \eqref{DoubleGam}, $\Gamma_2$ is a non-vanishing meromorphic
function of $x$ with poles at the locations 
\begin{equation}
x_* \,=\, -m\,\omega_1 - n\,\omega_2\,,\qquad\qquad m,n\ge 0\,.
\end{equation}  
If the ratio ${\omega_2/\omega_1\notin\Q}$ is irrational, each pole is
simple.  Otherwise, the poles have the multiplicities given by the product
formula.  

The definition \eqref{DoubleGam} of the double-gamma function should be
compared to the Lerch formula for the classical Euler gamma function
\begin{equation}
\Gamma(x) \,=\, \sqrt{2\pi}\,\exp{\!\left[\frac{\partial}{\partial s}\zeta(s,x)\big|_{s=0}\right]}\,,
\end{equation}
written in terms of the Hurwitz zeta.  To make the comparison precise,
let us introduce 
\begin{equation}\label{GammaOne}
\Gamma_1(x\,\big|\,\omega) \,=\, \exp{\!\left[\frac{\partial}{\partial
      s}\zeta_1(s,x\,\big|\,\omega)\big|_{s=0}\right]}\,.
\end{equation}
From \eqref{ZetaOne} and \eqref{GammaOne}, we see that $\Gamma_1$ is
related to the classical gamma function by an elementary
multiplicative factor,
\begin{equation}\label{GamOneE}
\Gamma_1(x\,\big|\,\omega) \,=\,
\frac{\omega^{-\zeta(0,x/\omega)}}{\sqrt{2\pi}}\cdot\Gamma\!\left(\frac{x}{\omega}\right)
\,=\,
\frac{\omega^{\left(x/\omega\right)}}{\sqrt{2\pi\omega}}\cdot\Gamma\!\left(\frac{x}{\omega}\right).
\end{equation}
In passing to the second equality, we apply the Hurwitz zeta identity in
\eqref{HZetaId}.

Comparing the product formulas in \eqref{eq:DoublesineProd} and
\eqref{DoubleGam}, the double-sine function $s_b(z)$ can now be
defined as the ratio of double-gamma functions 
\begin{flalign}\label{def:DSine}
\fbox{definition}\qquad\qquad
&s_b(z) \,:=\,
\frac{\Gamma_2\!\left(-i\,z\,\big|\,b,b^{-1}\right)}{\Gamma_2\!\left(i\,z \,+\, 
  Q\,\big|\,b,b^{-1}\right)}\,,\qquad\qquad Q \,\equiv\, b \,+\, b^{-1}\,.&
\end{flalign}
With this definition, $s_b(z)$ is meromorphic with zeroes and
poles at the correct locations \eqref{eq:PolesOfDoublesine} and with the correct
multiplicities.  Implicitly, the formula in \eqref{def:DSine}
specifies a choice for the renormalization constants
$C_{0,b}$ and $C_{1,b}$ in \eqref{LogDS}.  

The double-sine $s_b(z)$ is manifestly symmetric under the exchange of $b$ and
$b^{-1}$ as in \eqref{InvertB}.  Also under
inversion of $z$,
\begin{flalign}\label{InvertZ}
\fbox{inversion}\qquad\qquad
&s_b(-z) \,=\, \frac{1}{s_b(z + i\,Q)}\,,\qquad\qquad\qquad  Q \,\equiv\, b
\,+\, b^{-1}\,.&
\end{flalign}
Trivially from the definition \eqref{def:DSine}, we have the
special value
\begin{equation}
s_b\!\left(\frac{i}{2}\,Q\right) =\, 1\,.
\end{equation}

Before proceeding, let us mention one reason for the name
`double-sine'.  By comparison to \eqref{def:DSine}, 
\begin{equation}\label{Reflection}
\begin{aligned}
\Gamma_1(x\,\big|\,\omega)\cdot\Gamma_1(\omega -
  x\,\big|\,\omega) \,&=\,
\frac{1}{2\pi}\,\Gamma\!\left(\frac{x}{\omega}\right) \cdot
  \Gamma\!\left(1-\frac{x}{\omega}\right)\,,\\
&=\,\frac{1}{2\sin\!\left(\pi x/\omega\right)}\,.
\end{aligned}
\end{equation}
Here we use the relation of $\Gamma_1$ to the Euler gamma function in
\eqref{GamOneE}, along with the reflection formula 
\begin{equation}\label{RefEulGam}
\Gamma(x) \, \Gamma(1-x) \,=\, \frac{\pi}{\sin\!\left(\pi x\right)}\,.
\end{equation}
The quotient of double-gamma functions in the definition of $s_b(z)$
serves as the proper generalization of the single-gamma reflection
identity \eqref{Reflection} in which an ordinary sine enters.

\paragraph{Alternative Definitions.}  Many cousins to the double-sine function also
appear in the literature, so the reader must take care with
conventions.  Among popular variants, the non-compact quantum dilogarithm $e_b(z)$ \cite{Faddeev:2000if} 
is related to $s_b(z)$ by 
\begin{equation}
s_b(z) \,=\, \frac{\exp{\!\left[\frac{\pi
    i}{2}\!\left(z^2 - i\,Q\,z +
    \frac{1}{6}(1+Q^2)\right)\right]}}{e_b\left(z-i\,{Q}/{2}\right)}\,,\qquad\qquad
Q \,\equiv\, b + b^{-1}\,.
\end{equation}
The $G$-function $G(\omega_1,\omega_2;z)$ of Ruijsenaars \cite{Ruijsenaars:1996} is related by 
\begin{equation}
s_b(z) \,=\, G(-ib, -i/b; -i\,z - Q/2)\,.
\end{equation}
The hyperbolic gamma function $\Gamma_h(z;\omega_1,\omega_2)$ of van
de Bult \cite{vandeBult:2007} is related by 
\begin{equation}
s_b(z) \,=\, \Gamma_h(-i\,z;b,1/b)\,.
\end{equation}
Finally, the function $w_b(z)$ considered in \cite{Bytsko:2006ut}
  is related by 
\begin{equation}
s_b(z) \,=\, w_b(z-i\,Q/2)\,.
\end{equation}

\subsection{Properties of the Double-Sine Function}\label{PropDSine}

We require two further properties of $s_b(z)$.  The first property
concerns the value of the residue at each pole \eqref{eq:PolesOfDoublesine} along
the negative imaginary axis.  The second property concerns the asymptotic
behavior as ${|z|\to\infty}$.  Both properties will be essential for our
analysis of the partition function $Z_{S^3}$.

\paragraph{Quasi-periods and Residues.}

We return to the Barnes double-zeta function in
\eqref{def:MultipleZeta}.  Trivially, by rearranging terms in its defining sum,
the Barnes double-zeta function obeys 
\begin{equation}\label{ShftZ}
\begin{aligned}
\zeta_2(s,x+\omega_1\,\big|\,\omega_1,\omega_2) \,-\,
\zeta_2(s,x\,\big|\,\omega_1,\omega_2) \,&=\,
-\zeta_1(s,x\,\big|\,\omega_2)\,,\\
\zeta_2(s,x+\omega_2\,\big|\,\omega_1,\omega_2) \,-\,
\zeta_2(s,x\,\big|\,\omega_1,\omega_2) \,&=\,
-\zeta_1(s,x\,\big|\,\omega_1)\,.
\end{aligned}
\end{equation}
Taking derivatives with respect to $s$ in \eqref{ShftZ}, we find that
the double-gamma function in \eqref{DoubleGam} satisfies the
multiplicative relation
\begin{equation}
\begin{aligned}
\Gamma_2(x+\omega_1 \big|\,\omega_1,\omega_2) \,&=\,\frac{\Gamma_2(x\,\big|\,\omega_1,\omega_2)}{\Gamma_1(x\,\big|\,\omega_2)}\,,\\
\Gamma_2(x+\omega_2 \big|\,\omega_1,\omega_2)
\,&=\,\frac{\Gamma_2(x\,\big|\,\omega_1,\omega_2)}{\Gamma_1(x\,\big|\,\omega_1)}\,.
\end{aligned}
\end{equation}
The definition of $s_b(z)$ in \eqref{def:DSine}, along with the
reflection formula in \eqref{Reflection}, then implies the 
quasi-periodic transformation
\begin{flalign}\label{qPeriod}
\fbox{quasi-periods}\qquad\qquad\quad &
\begin{aligned}
s_b(z + i\,b) \,&=\, -2 i \sinh(\pi b z) \cdot s_b(z)\,,\\
s_b(z + i\,b^{-1}) \,&=\, -2 i \sinh(\pi b^{-1} z) \cdot s_b(z)\,.
\end{aligned}&
\end{flalign}

By applying the quasi-periodic transformations in succession, we
deduce
\begin{equation}\label{ShiftQ}
\begin{aligned}
s_b(z + i\, Q) \,&=\, 2 i \sinh(\pi b z) \cdot s_b(z + i b^{-1})\,,\qquad\quad Q
\,\equiv\, b + b^{-1}\,,\\
&=\, 4 \sinh(\pi b z) \sinh(\pi b^{-1} z) \cdot s_b(z)\,.\\
\end{aligned}
\end{equation}
For the sign in the first line of \eqref{ShiftQ}, note that ${\sinh(\pi
  b (z + i b^{-1})) = -\sinh(\pi b z)}$.  Combining the expression
for $s_b(z+ i Q)$ in \eqref{ShiftQ} with the inversion formula in
\eqref{InvertZ}, we obtain a product formula 
\begin{equation}\label{ProdSb}
s_b(z) \cdot s_b(-z) \,=\, \frac{1}{4 \sinh(\pi b z) \sinh(\pi b^{-1} z)}\,.
\end{equation}
The function $s_b(z)$ has a simple pole at ${z=0}$
with residue $r$ and hence a Laurent expansion
\begin{equation}
s_b(z) \,=\, \frac{r}{z} \,+\, \hbox{regular}\,,\qquad\qquad |z|^2\ll 1\,.
\end{equation}
The leading singularity at ${z=0}$ on the right in \eqref{ProdSb} then
determines 
\begin{equation}
r^2 \,=\, -\frac{1}{4\pi^2} \quad\Longrightarrow\quad r \,=\,
\pm\frac{i}{2\pi}\,.
\end{equation}
The sign of the residue apparently requires more effort to
fix, so we just quote the literature:
\begin{equation}\label{Res0}
\Res\big[s_b(z)\big]_{z=0} \,=\, \frac{i}{2\pi}\,.
\end{equation}
The quasi-periods in \eqref{qPeriod} then imply the special
values
\begin{flalign}\label{Valuessb}
\fbox{special values} \qquad&\qquad
s_b(i\,b) \,=\, b\,,\qquad\qquad s_b(i\,b^{-1}) \,=\, b^{-1}\,.&
\end{flalign}

The quasi-periodic transformations in \eqref{ShiftQ} can be iterated
to evaluate the residue of $s_b(z)$ at any other pole.  By a small
computation, for each ${m,n\ge 0}$,
\begin{equation}\label{ShiftMN}
\begin{aligned}
&\frac{s_b(z + i\,m\,b + i\,n\,b^{-1})}{s_b(z)} \,=\,
\left(-1\right)^{m n+m+n} i^{m+n}\,\times\\
&\qquad\times\prod_{u=0}^{m-1} \big[2 \sinh\!\left(\pi
    b (z + i\,u\,b)\right)\big]\cdot \prod_{v=0}^{n-1}
\big[2 \sinh\!\left(\pi b^{-1} (z +
  i\,v\,b^{-1})\right)\big].
\end{aligned}
\end{equation}
Alternatively, when the relation in \eqref{ShiftMN} is read backwards,
\begin{equation}\label{ShiftMNB}
\begin{aligned}
&\frac{s_b(z - i\,m\,b - i\,n\,b^{-1})}{s_b(z)} \,=\,
\left(-1\right)^{m n} i^{m+n}\,\times\\
&\qquad\times\prod_{u=1}^m \big[2\sinh\!\left(\pi b (z -
    i\,u\,b)\right)\big]^{-1} \cdot
\prod_{v=1}^n\big[2\sinh\!\left(\pi b^{-1} (z - 
    i\,v\,b^{-1})\right)\big]^{-1}\,.
\end{aligned}
\end{equation}
From \eqref{Res0} and \eqref{ShiftMNB}, the residue of $s_b(z)$ at
each pole on the imaginary axis is then 
\begin{flalign}\label{Residues}
\begin{aligned}
&\fbox{residues}\\
&\big[m,n\ge 0\big]
\end{aligned}\qquad\qquad&
\begin{aligned}
&\Res\big[s_b(z)\big]_{z=-i\,m\,b - i\,n\,b^{-1}}\,=\, (-1)^{m n+m+n}\cdot
\frac{i}{2\pi}\,\times\\
&\qquad\qquad\times\,\prod_{u=1}^m\left[\frac{1}{2\sin\!\left(\pi\,b^2\,u\right)}\right]
\cdot \prod_{v=1}^n\left[\frac{1}{2\sin\!\left(\pi\,b^{-2}\,v\right)}\right]\,.
\end{aligned}&
\end{flalign}
Note that the product on the right in \eqref{Residues} is finite 
for all ${m,n\ge 0}$ so long as ${b^2\notin\Q}$ is irrational, meaning
the poles of $s_b(z)$ are simple.  Otherwise, the product diverges at the
special poles of $s_b(z)$ with higher degree, due to the
vanishing of $\sin(\pi u b^2)$ or $\sin(\pi v b^{-2})$ for appropriate
${u,v}$.

\paragraph{$q$-Pochhammer Symbols.}

Both the quasi-periodicity relation in \eqref{ShiftMN} and the residue
formula in \eqref{Residues} are frequently rewritten in the language
of $q$-Pochhammer symbols.  This notation is useful for emphasizing
the algebraic as opposed to analytic properties of the double-sine
function, so we briefly recall it.

Given an integer ${N \ge 1}$, define the polynomial
\begin{equation}\label{qPoch}
\left(a;q\right)_N \,:=\, \prod_{t=0}^{N-1} \left(1 -
  a\,q^t\right) \,=\,\left(1 - a\right) \left(1 - a\,q\right)
\cdots \left(1 - a \, q^{N-1}\right).
\end{equation}
By convention ${\left(a;q\right)_0=1}$, and an important
special case will be 
\begin{equation}
\left(q;q\right)_N \,=\, \prod_{t=1}^N \left(1-q^t\right) \,=\,
\left(1-q\right) \left(1-q^2\right) \cdots \left(1-q^N\right).
\end{equation}
The finite product in \eqref{qPoch} can be extended to an infinite product 
\begin{equation}
\left(a;q\right)_\infty \,=\, \prod_{t=0}^\infty  \left(1 -
  a\,q^t\right),
\end{equation}
which converges to a holomorphic function of ${a,q\in\C}$ when ${|q|<1}$.  Thus
$\left(a;q\right)_N$ can be rewritten as the quotient
\begin{equation}
\left(a;q\right)_N \,=\, \frac{\left(a;q\right)_\infty}{\left(a
    q^N;q\right)_\infty}\,.
\end{equation}
This expression for $\left(a;q\right)_N$ also makes sense for negative
values of $N$, so we set
\begin{equation}\label{qPochN}
\begin{aligned}
\left(a;q\right)_{-N} :&= \frac{\left(a;q\right)_\infty}{\left(a
    q^{-N};q\right)_\infty} \,=\,
\frac{1}{\left(a\,q^{-N};q\right)_N} \,=\,\prod_{t=1}^N \frac{1}{\left(1 - a
    \, q^{-t}\right)}\,,\\ 
&=\, \frac{1}{\left(1-a\,q^{-1}\right)
  \left(1 - a\,q^{-2}\right) \cdots \left(1 - a\, q^{-N}\right)}\,.
\end{aligned}
\end{equation}
For example,
\begin{equation}
\begin{matrix}
\begin{aligned}
\left(a;q\right)_1 \,&=\, \left(1-a\right),\\
\left(a;q\right)_{-1} \,&=\, \frac{1}{\left(1- a\, q^{-1}\right)}\,,
\end{aligned} \qquad&\qquad
\begin{aligned}
\left(a;q\right)_2 \,&=\, \left(1-a\right) \left(1 - a \,q\right),\\
\left(a;q\right)_{-2} \,&=\, \frac{1}{\left(1 - a \,
    q^{-1}\right)\left(1 - a\, q^{-2}\right)}\,.
\end{aligned}
\end{matrix}
\end{equation}
As their essential feature, both the positive and the negative
$q$-Pochhammer symbols $(a;q)_{\pm N}$ for finite $N$ are rational
functions of $q$.

With some rearrangement of factors in \eqref{qPochN}, the negative
$q$-Pochhammer symbol can be alternatively recast
\begin{equation}\label{PosNegqPoch}
\left(a;q\right)_{-N} \,=\,\left(-\frac{1}{a}\right)^N
\cdot\frac{q^{N(N+1)/2}}{\left(q/a;q\right)_N}\,,\qquad\qquad N\,>\,0\,.
\end{equation}
Also worth remarking for its modular behavior, the infinite Pochhammer
symbol $\left(a;q\right)_\infty$ is directly related to the Dedekind
eta-function 
\begin{equation}
\eta(\tau) \,:=\, q^{1/24} \prod_{t=1}^\infty \left(1-q^t\right) \,=\, q^{1/24} \left(q;q\right)_\infty\,,\qquad\qquad
q = \e{2\pi i \tau}\,.
\end{equation}

In the language of $q$-Pochhammer symbols, the quasi-periodicity
formula \eqref{ShiftMN} becomes more succinctly
\begin{equation}\label{eq:DifferenceEqSb}
\begin{aligned}
&\frac{s_b(z + i\,m\,b + i\,n\,b^{-1})}{s_b(z)}
\,=\,\left(-1\right)^{m n} i^{m+n}\,\times\,\\
&\times\,\Big[\e{\!-\pi b m z}\,\Rq^{-m (m-1)/4}\,\big(\e{\!2\pi b
    z};\Rq\big)_{\!m}\Big] \cdot \Big[\e{\!-\pi n z/b}\,\,\wt\Rq^{-n (n-1)/4}\,\big(\e{\!2\pi z/b};\wt\Rq\big)_{\!n}\Big]\,,
\end{aligned}
\end{equation}
where we recall the identifications 
\begin{equation}\label{RecQQ}
\Rq \,=\, \e{\!2\pi i b^2}\,,\qquad\qquad\qquad \wt\Rq \,=\,
\e{\!2\pi i/b^2}\,.
\end{equation}
We omit the derivation of the formula in \eqref{eq:DifferenceEqSb}, which
follows from the identity ${2\sinh(x) = \e{x}\!\left(1-\e{-2x}\right)}$, the
definition of the $q$-Pochhammer symbol, and elementary algebraic
manipulations.  

Beyond its brevity, the $q$-Pochhammer expression for the
double-sine function ${s_b(z+i\, m\, b + i\, n\, b^{-1})}$ has
two virtues.  In the preceding 
quasi-periodicity formulas \eqref{ShiftMN} and \eqref{ShiftMNB}, the
integers $m$ and $n$ are both assumed positive, as the behavior of
$s_b(z)$ depends upon whether $z$ shifts up or down the imaginary
axis.  By contrast, the $q$-Pochhammer formula in
\eqref{eq:DifferenceEqSb} reproduces not only 
\eqref{ShiftMN} but also \eqref{ShiftMNB}, after the naive reflection
${(m,n)\mapsto (-m,-n)}$. Hence the $q$-Pochhammer expression in
\eqref{eq:DifferenceEqSb} is correct for both positive and negative values of
$m,n$.
Moreover, we already see from \eqref{eq:DifferenceEqSb} the beginning of the
desired holomorphic factorization for $Z_{S^3}$.

Reflecting $(m,n)$ and taking the limit ${z\to 0}$ in \eqref{eq:DifferenceEqSb},
we obtain a $q$-Pochhammer formula for the residues of the double-sine
function,
\begin{equation}
\begin{aligned}
&\Res\big[s_b(z)\big]_{z=-i\,m\,b - i\,n\,b^{-1}}\,=\, (-1)^{m n+m+n}\,
i^{m+n} \cdot \frac{i}{2\pi}\,\times\\
&\qquad\qquad\times\,\Big[\Rq^{-m (m+1)/4}\,\big(1;\Rq\big)_{\!-m}\Big]
\cdot \Big[\wt\Rq^{-n
  (n+1)/4}\,\big(1;\wt\Rq\big)_{\!-n}\Big]\,,\qquad m,n\ge 0\,.
\end{aligned}
\end{equation}
By the $q$-Pochhammer identity in \eqref{PosNegqPoch} with ${a=1}$,
the residue can be rewritten in the slightly more transparent fashion below,
\begin{flalign}\label{eq:DSineResidue}
\begin{aligned}
&\fbox{residues}\\
&\big[m,n\ge 0\big]
\end{aligned} \qquad\qquad&
\begin{aligned}
&\Res\big[s_b(z)\big]_{z=-i\,m\,b - i\,n\,b^{-1}}\,=\, (-1)^{m n}\,
i^{m+n} \cdot \frac{i}{2\pi}\,\times\\
&\qquad\qquad\times\left[\frac{\Rq^{m (m+1)/4}}{\big(\Rq;\Rq\big)_{\!m}}\right]
\cdot \left[\frac{\wt\Rq^{n
  (n+1)/4}}{\big(\wt\Rq;\wt\Rq\big)_{\!n}}\right].
\end{aligned}&
\end{flalign}
Expressing $\left(\Rq;\Rq\right)_{\!m}$ and
$\left(\wt\Rq;\wt\Rq\right)_{\!n}$ in terms of
products of sines, one can readily check that the result in
\eqref{eq:DSineResidue} reproduces the previous version \eqref{Residues} of
the residue formula.  According to \eqref{eq:DSineResidue}, each residue of
$s_b(z)$ is a rational function of $\Rq^{1/2}$ and $\wt\Rq^{1/2}$,
with poles at roots of unity in the complex $\Rq$-
(resp.\ $\wt\Rq$-) plane.\footnote{Recall that $\Rq$ and $\wt\Rq$
  lie on the unit circle precisely when ${b>0}$ is real.  As mentioned earlier,
  the higher-order poles in $s_b(z)$ occur for rational values of $b^2$, for
  which $\Rq$ and $\wt\Rq$ are roots of unity.}  The $q$-Pochhammer residue formula will be our 
workhorse in the analysis of holomorphic factorization.

\paragraph{Integral Representation and Asymptotics.}

To apply the residue calculus to the partition function in \eqref{eq:GrandPartitionFunctionInt}, we
require information not only about the poles and residues of $s_b(z)$ but also
about the asymptotic behavior as ${|z|\to\infty}$.  The
latter knowledge is crucial for understanding which contour
manipulations of the matrix integral are allowed.

The asymptotic behavior of $s_b(z)$ for large ${z\in\C}$ is most
easily deduced from yet another representation of the double-sine
function, this time in terms of the integral
\begin{equation}\label{logInt}
\begin{aligned}
&s_b(z) \,=\, \exp{\!\left[i \int_0^\infty \frac{du}{u}
    \left(\frac{\sin\!\left(2u\!\left(z-\frac{i}{2}Q\right)\right)}{2\sinh(b
      u)\sinh(u/b)}\,-\,\frac{z-\frac{i}{2}Q}{u}\right)\right]}\,,\\
&0 < \Im(z) < Q\,,\qquad\qquad\qquad Q \equiv b + b^{-1}\,.
\end{aligned}
\end{equation}
The integral formula for $s_b(z)$ is originally due to Ruijsenaars
\cite{Ruijsenaars:1996} and is reviewed in Chapter 2 of
\cite{vandeBult:2007}, where the reader can find a proof of
\eqref{logInt}.\footnote{Again, neither the conventions in
   \cite{vandeBult:2007} nor \cite{Ruijsenaars:1996} agree precisely
  with ours, so some translation is required to read these works.}  

Let us discuss the domain of parameters in which the integral
representation for $s_b(z)$ is valid.  The variable $u$ runs
over the positive half-line $[0,\infty)$, so the expression in
\eqref{logInt} must be integrable near both ${u=0}$ and
${u=\infty}$.  For ${u\ll 1}$, the integrand behaves
to leading-order like $u^2$ (the poles of the respective terms in
parentheses cancel), so integrability near ${u=0}$ is assured.
Otherwise, if we require the second term in the argument of
\eqref{logInt} to decay exponentially as ${u\to\infty}$, the
imaginary part of $z$ must be bounded above and below by ${0 < \Im(z)
  < Q}$.  Precisely within this strip depicted in Figure
\ref{fi:analyticStripFordSine}, $s_b(z)$ is both regular and
non-vanishing, consistent with the presentation in \eqref{logInt}.

\iffigs
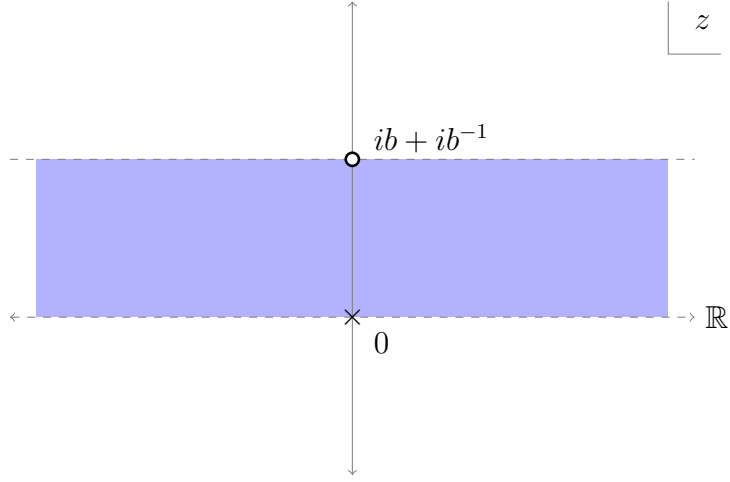
\begin{figure}[t]
	\begin{center}
		\begin{tikzpicture}[x=.7cm,y=.7cm]
		\fill[blue!30] (-6,0) rectangle (6,3);
		
		\draw [<->, gray, dashed](-6.5,0)--(6.5,0) node [right, black] {$\R$};
		\draw [<->, gray](0,-3)--(0,6);

		\draw [thin, gray](6,6)--(6,5) -- (7,5);
		\node at (6.65,5.6) {$z$};
		
		\draw [thin, gray, dashed](-6.5,3)--(6.5,3);
		\node [anchor=west] at (.2,3.4) {$ib+ib^{-1}$};
                \fill[black] (0,3) circle(.15);
		\fill[white] (0,3) circle(.1);

                \node at (0,0) {$\times$};
		\node [anchor=west] at (.2,-.5) {$0$};

		\end{tikzpicture}
		\caption{Strip ${0 < \Im z < Q \equiv b + b^{-1}}$ in which the
                  integral representation of $s_b(z)$ is
                  valid.}\label{fi:analyticStripFordSine}
	\end{center}
\end{figure}\fi 

The asymptotics of $s_b(z)$ become clearer after some 
manipulation of the integral
\begin{equation}\label{bigFz}
{\bf I}_b(z) \,=\, \int_0^\infty \frac{du}{u}
\left(\frac{\sin\!\left(2u\!\left(z-\frac{i}{2}Q\right)\right)}{2\sinh(b 
      u)\sinh(u/b)}\,-\,\frac{z-\frac{i}{2}Q}{u}\right).
\end{equation}
Because the integrand is an even function of $u$, the integration
domain can be unfolded to the real-line and ${\bf I}_b(z)$ written
more symmetrically as the principal-value integral 
\begin{equation}\label{PVFbz}
{\bf I}_b(z) \,=\, \frac{1}{2}\,\textrm{P}\!\int_\R \frac{du}{u}
\left(\frac{\e{\! 2 i u
        \left(z - \frac{i}{2}Q\right)}}{2 i \sinh(b
      u)\sinh(u/b)}\,-\,\frac{z-\frac{i}{2}Q}{u}\right).
\end{equation}
As usual, 
\begin{equation}
\textrm{P}\int_\R\,\bullet \,\,:=\, \lim_{\varepsilon\to
  0}\left[\,\int_\varepsilon^\infty\,\bullet\,\,+\,\int_{-\infty}^{-\varepsilon}\,\bullet\,\,\right].
\end{equation}

\begin{figure}[t]
\begin{center}
\includegraphics[scale=0.60]{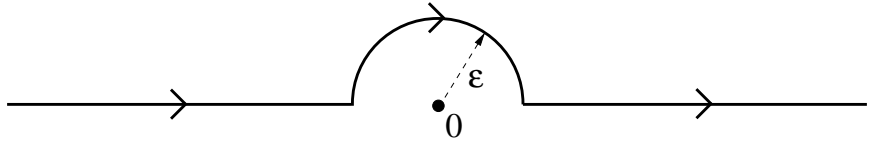}
\caption{The contour ${\R+i\varepsilon}$ in the complex $u$-plane.}\label{ContourF}
\end{center}
\end{figure}

To simplify ${\bf I}_b(z)$ still further, consider the contour denoted by
${\R+i\varepsilon}$ in Figure \ref{ContourF}.  This contour includes a
small semi-circular detour of radius $\varepsilon$ around the origin
in the upper half-plane.  If a meromorphic function $f(u)$ is
symmetrically-integrable about the origin, meaning that the Laurent
expansion of $f(u)$ has singular terms\footnote{Eg.~${f(u) = c_{-5}\,
    u^{-5} \,+\, c_{-3}\,u^{-3} \,+\, c_{-1}\,u^{-1} \,+\, c_0
    \,+\,\cdots\,}$.} with
exclusively odd degree in $u$, then 
\begin{equation}\label{ContourII}
\textrm{P}\!\int_\R\!du \, f(u) \,=\,
\int_{\R+i\varepsilon}\!du \, f(u) \,+\, i \pi\,\Res\big[f(u)\big]_{u=0}\,.
\end{equation}
Here the contribution from the small semi-circle in Figure
\ref{ContourF} has been added and subtracted to the right-side of
\eqref{ContourII} in the limit ${\varepsilon\to 0}$.
When applied to the formula for ${\bf I}_b(z)$ in \eqref{PVFbz}, we obtain
\begin{equation}\label{ContI}
{\bf I}_b(z) \,=\, -\frac{i}{4}\int_{\R+i\varepsilon} \frac{du}{u}
\cdot\frac{\e{\! 2 i u
        \left(z - \frac{i}{2}Q\right)}}{\sinh(b
      u)\sinh(u/b)} \,-\,\frac{\pi}{2}\left[\big(z-\frac{i}{2}\,Q\big)^2 +
    \frac{1}{12}\left(b^2 + b^{-2}\right)\right].
\end{equation}
By shrinking the contour at infinity, we note that
${\int_{\R+i\varepsilon} du/u^2 = 0}$ for the second term in \eqref{PVFbz}.

With the new expression \eqref{ContI} for ${\bf I}_b(z)$, 
the double-sine function can be presented more conveniently in terms
of the contour integral 
\begin{equation}\label{logCInt}
\begin{aligned}
&s_b(z) \,=\,\exp{\!\big[i\,{\bf I}_b(z)\big]}\,,\\
&\,=\,\exp{\!\left[-\frac{i \pi}{2}\big((z-\frac{i}{2}\,Q)^2 +
    \frac{1}{12}(b^2 + b^{-2})\big)+\frac{1}{4}\int_{\R+i\varepsilon} \frac{du}{u}
\frac{\e{\! 2 i u
        \left(z - \frac{i}{2}Q\right)}}{\sinh(b
      u)\sinh(u/b)}\right]}\,,
\end{aligned}
\end{equation}
again valid within the strip ${0<\Im(z)<Q}$.  We assume that
${b^2\notin\Q}$ is generic, so that the integrand in \eqref{logCInt}
has simple poles along the imaginary axis at ${u_* = 
  i\pi m b}$ or ${u_*=i\pi m/b}$ for integer ${m\neq 0}$, along with a
triple pole at ${u_*=0}$.  

The asymptotic behavior of $s_b(z)$ for
${|z|\to\infty}$ now depends upon whether the real part of $z$ is
positive or negative.  If ${\Re(z)>0}$, the integral over
${\R+i\varepsilon}$ in \eqref{logCInt} can be evaluated by closing the
contour in the upper half of the $u$-plane and applying the residue
theorem.  After a small calculation,
\begin{equation}\label{QExpsbz}
\begin{aligned}
&\int_{\R+i\varepsilon} \frac{du}{u}
\frac{\e{\! 2 i u
        \left(z - \frac{i}{2}Q\right)}}{\sinh(b
      u)\sinh(u/b)}\,=\,\\
&\qquad\qquad\qquad\sum_{m=1}^\infty \frac{4}{m}
\left[\frac{\e{\!-2\pi m 
    b z }}{1-\e{\!-2\pi i m b^2}} \,+\, \frac{\e{\!-2\pi m 
    z/b}}{1-\e{\!-2\pi i m/b^2}}\right],\qquad \Re(z)>0\,.
\end{aligned}
\end{equation}
Therefore in terms of the parameters $\Rq$ and $\wt\Rq$, the 
argument of the exponential in \eqref{logCInt} has the series expansion
\begin{equation}\label{AsympP}
\begin{aligned}
\ln s_b(z) \,&=\, -\frac{i \pi}{2}\left[(z-\frac{i}{2}\,Q)^2 +
    \frac{1}{12}(b^2 + b^{-2})\right]\,+\,\\
&\qquad\qquad\,+\, \sum_{m=1}^\infty \frac{1}{m} \left[ \frac{\e{\!-2\pi m b
    z}}{1-\Rq^{-m}} \,+\, \frac{\e{\!-2\pi m 
    z/b}}{1-\wt\Rq^{-m}}\right],\qquad\qquad \Re(z)>0\,.
\end{aligned}
\end{equation}
When the real part of $z$ is negative, the contour in \eqref{QExpsbz}
must be closed in the lower half of the $u$-plane.  From Figure \ref{ContourF},
we see that the pole at ${u_*=0}$ also contributes to the residue
computation in this case,
\begin{equation}\label{QExpsbzII}
\begin{aligned}
&\int_{\R+i\varepsilon} \frac{du}{u}
\frac{\e{\! 2 i u
        \left(z - \frac{i}{2}Q\right)}}{\sinh(b
      u)\sinh(u/b)}\,=\, 4\pi i \left[\left(z-\frac{i}{2}Q\right)^2 +
  \frac{1}{12}\left(b^2 + b^{-2}\right)\right] \,+\,\\
&\qquad\qquad\qquad\,+\,\sum_{m=1}^\infty \frac{4}{m}
\left[\frac{\e{2\pi m 
    b z }}{1-\e{2\pi i m b^2}} \,+\, \frac{\e{2\pi m 
    z/b }}{1-\e{2\pi i m/b^2}}\right],\qquad \Re(z)<0\,.
\end{aligned}
\end{equation}
The quadratic expression on the first line of \eqref{QExpsbzII} is the
contribution from the pole at ${u_*=0}$.  Hence in contrast to \eqref{AsympP},
\begin{equation}\label{AsympN}
\begin{aligned}
\ln s_b(z) \,&=\, +\frac{i \pi}{2}\left[(z-\frac{i}{2}\,Q)^2 +
    \frac{1}{12}(b^2 + b^{-2})\right]\,+\,\\
&\qquad\qquad\,+\, \sum_{m=1}^\infty \frac{1}{m} \left[ \frac{\e{2\pi m b
    z}}{1-\Rq^{m}} \,+\, \frac{\e{2\pi m 
    z/b}}{1-\wt\Rq^{m}}\right],\qquad\qquad \Re(z)<0\,.
\end{aligned}
\end{equation}

The expressions in \eqref{AsympP} and \eqref{AsympN}
are exact for $z$ in the strip ${0<\Im(z)<Q}$, so we 
deduce the asymptotic behavior
\begin{equation}\label{eq:MSineAsympta}
\begin{aligned}
&\fbox{asymptotics}\\[1 ex]
&s_b(z) \underset{|z|\to\infty}{=}
\begin{cases}
\,\,\exp{\!\Big[-\frac{i \pi}{2}\!\left((z-\frac{i}{2}\,Q)^2 +
    \frac{1}{12}(b^2 + b^{-2})\right)+\,o(1)\Big]},\quad \Re(z)>0\,,\\[1 ex]
\,\,\exp{\!\Big[+\frac{i \pi}{2}\!\left((z-\frac{i}{2}\,Q)^2 +
    \frac{1}{12}(b^2 + b^{-2})\right)+\,o(1)\Big]},\quad
\Re(z)<0\,,
\end{cases}
\end{aligned}
\end{equation}
up to exponentially-small, $o(1)$-contributions from
the respective series in \eqref{AsympP} and \eqref{AsympN}.  This
result is worthy of several remarks.
\begin{enumerate}
\item The asymptotic behavior of $s_b(z)$ as ${|z|\to\infty}$ depends
  upon the sign of the real part of $z$ and is not analytic in
  $z$.  Even for an entire function such as $\sinh(z)$, the asymptotic
  expansion as ${|z|\to\infty}$ may jump between sectors in the
  complex $z$-plane.  The non-analytic behavior in
  \eqref{eq:MSineAsympta} will ultimately be responsible for the non-analytic
  dependence on the sign of the real mass in the one-loop formula \eqref{KEff}
  for the effective Chern-Simons level $k_{\rm eff}$.
\item The definition of $s_b(z)$ in \eqref{def:DSine} amounts to a
  specific choice for the constants $C_{0,b}$ and
  $C_{1,b}$ in \eqref{LogDS}.  With some other choice, the double-sine
  function is renormalized as 
\begin{equation}\label{renDS}
s_b(z)\,\longmapsto\, \exp{\!\big[\Delta C_{0,b} + \Delta C_{1,b}\,z\big]}\cdot s_b(z)\,.
\end{equation}
Comparing to \eqref{eq:MSineAsympta}, we see that the leading, quadratic
dependence on $z$ in the argument of the exponential is determined
independently of the choice for $C_{0,b}$ and $C_{1,b}$, whereas the
subleading asymptotic behavior of $s_b(z)$ depends on the
renormalization scheme.
\item The double-sine function grows or decays very rapidly as
  ${|z|\to\infty}$, at the rate of the Gaussian 
  ${\exp{\!(i z^2)}}$.  Here we model the Gaussian, including the
  `$i$', on the effective Chern-Simons term in the matrix integral
  \eqref{ZUone}.  However, the sectors of asymptotic growth vs.\ decay
  differ sharply between the Gaussian and the double-sine, due to the
  non-analytic behavior in \eqref{eq:MSineAsympta}.  As illustrated in
  Figure \ref{fig:Conver}, the
  classical Gaussian ${\exp{\!(i z^2)}}$ 
vanishes when ${z\to\infty}$ in the first and third quadrants of the
complex plane.  By contrast, ${s_b(z)}$ vanishes when ${z\to\infty}$
in the lower half-plane.  In the complements to these regions, the
Gaussian and the double-sine respectively diverge as ${z\to\infty}$.
\end{enumerate}
\iffigs
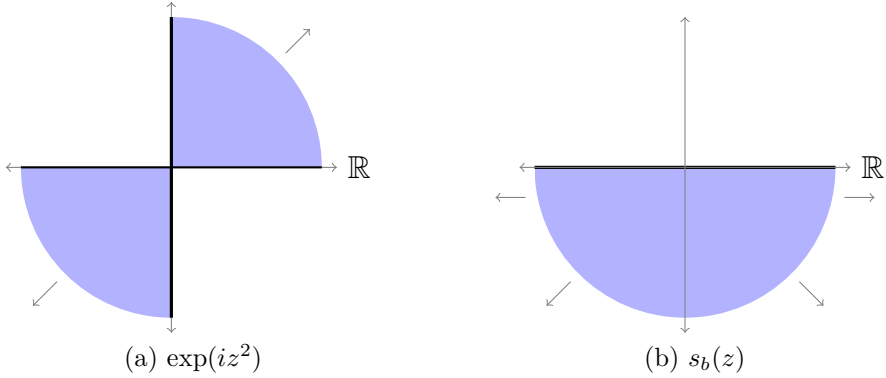
\begin{figure}[ht]
\centering
\subfloat[$\exp{\!(i z^2)}$]{
   \begin{tikzpicture}[x=.4cm,y=.4cm]
	\begin{scope}
		\clip (0,0) circle (5);
		\fill[blue!30] (5cm,5cm) rectangle (0cm,0cm);
		\fill[blue!30] (-5cm,-5cm) rectangle (0cm,0cm);
	\end{scope}

	\draw [<->, gray](-5.5,0)--(5.5,0) node [right, black] {$\R$};
	\draw [<->, gray](0,-5.5)--(0,5.5);

	\draw [thick, black](-5,0)--(5,0);
	\draw [very thick, black](0,5)--(0,-5);
	\draw [->, gray] (3.8,3.8) --  (4.6,4.6);
	\draw [->, gray] (-3.8,-3.8) --  (-4.6,-4.6);
    \end{tikzpicture}
}
\quad \quad \quad
\subfloat[$s_b(z)$]{
   \begin{tikzpicture}[x=.4cm,y=.4cm]
	\begin{scope}
		\clip (0,0) circle (5);
		\fill[blue!30] (-5,-5) rectangle (5,0);
	\end{scope}

	\draw [very thick, black](-5,0)--(5,0);

	\draw [<->, gray](-5.5,0)--(5.5,0) node [right, black] {$\R$};
	\draw [<->, gray](0,-5.5)--(0,5);

	\draw [->, gray] (5.3,-1) --  (6.3,-1);
	\draw [->, gray] (-5.3,-1) --  (-6.3,-1);
	\draw [->, gray] (-3.8,-3.8) --  (-4.6,-4.6);
	\draw [->, gray] (3.8,-3.8) --  (4.6,-4.6);
    \end{tikzpicture}
}
\caption{Asymptotic behavior of the Gaussian $\exp{\!(i z^2)}$ versus the
  double-sine $s_b(z)$.  Along rays in the shaded quadrants, the functions
decay rapidly as ${z\to\infty}$.}\label{fig:Conver}
\end{figure}
\fi

\paragraph{$\Rq$-$\wt\Rq$ Decomposition.}

The respective series expansions for ${\ln s_b(z)}$ in \eqref{AsympP}
and \eqref{AsympN} imply a factorization in the dependence on $(\Rq,\wt\Rq)$
of the sort required for the factorization of $Z_{S^3}$
in \eqref{FACTOR}.  

For later use in Sections \ref{RankOne} and \ref{Residue}, we introduce a decomposition of the double-sine $s_b(z)$ with 
argument 
\begin{equation}\label{ArgZ}
z \,=\, \mu \,+\, i\left(M + \nu\right) b \,+\, i \left(N +
  \wt\nu\right) b^{-1}\,,
\end{equation}
where 
\begin{equation}
 \mu\,,\nu\,,\wt\nu\in\,\R\,,\qquad M,N\,\in\Z\,,\qquad 0 \,\le\,
\nu,\wt\nu \,<\, 1\,.
\end{equation}
By the quasi-periodicity relation in \eqref{eq:DifferenceEqSb}, 
\begin{equation}\label{FacDSine}
\begin{aligned}
s_b(z)\,&=\, \left(-1\right)^{M N}
i^{M+N} \, \e{\!-i \pi \left(M \, {\wt\nu} + N \, \nu\right)}\,\times\\
&\times\left[\e{\!-\pi M \mu \,b}\cdot\Rq^{-\frac{1}{4} M (M -1 + 2\nu)}
  \Big(\e{\!2\pi \left(\mu b \,+\, i \, \wt\nu\right)}\,\Rq^\nu; \Rq\Big)_{\!M}\right]\times\\
&\times\left[\e{\!-\pi N \mu/b}\cdot{\wt\Rq}^{-\frac{1}{4} N (N -1 + 2\wt\nu)}
  \Big(\e{\!2\pi \left(\mu/b \,+\, i\,\nu\right)}\,\wt\Rq^{\wt\nu}; \wt\Rq\Big)_{\!N}\right]\times\\
&\times s_b\!\left(\mu \,+\, i \,\nu\, b \,+\, i \,\wt\nu\, b^{-1}\right)\,.
\end{aligned}
\end{equation}
All dependence on the integers $M$ and $N$ is absorbed by the
prefactors in the first three lines of \eqref{FacDSine}.
Clearly, $\Rq$ and $\wt\Rq$ enter each factor separately, so these
terms are consistent with the $\Rq$-$\wt\Rq$ decomposition.

Otherwise, since ${\nu,\wt\nu}$ lie in the unit interval, the series
expansions in \eqref{AsympP} and \eqref{AsympN} apply to the final
term in \eqref{FacDSine}.  For ${\mu>0}$, a brief calculation shows
\begin{equation}\label{FacDSineII}
\begin{aligned}
&s_b\!\left(\mu \,+\, i \,\nu\, b \,+\, i \,\wt\nu\, b^{-1}\right) \underset{\mu>0}{=}\,
\e{\! i \pi \left[\left(\nu-\ha\right)\left(\wt\nu-\ha\right) -
    \ha\mu^2\right]}\,\times\\
&\qquad\times\left[\e{\!\pi \mu \left(\nu-\ha\right)
  b}\cdot\Rq^{\frac{1}{4}\left(\nu^2-\nu+\frac{1}{6}\right)}\cdot
\exp{\!\left(\sum_{m=1}^\infty\frac{1}{m}\frac{\e{\!-2\pi
        m\left(\mu
          b+i\,\wt\nu\right)}\,\Rq^{-m\nu}}{1-\Rq^{-m}}\right)}\right]\times\\
&\qquad\times\left[\e{\!\pi \mu \left(\wt\nu-\ha\right)\!/b}\cdot\wt\Rq^{\frac{1}{4}\left(\wt\nu^2-\wt\nu+\frac{1}{6}\right)}\cdot
\exp{\!\left(\sum_{n=1}^\infty\frac{1}{n}\frac{\e{\!-2\pi
        n\left(\mu/b+i\,\nu\right)}\,\wt\Rq^{-n\wt\nu}}{1-\wt\Rq^{-n}}\right)}\right].
\end{aligned}
\end{equation}
We omit the similar expansion for ${\mu<0}$, which only differs by a
few signs.  Again, all dependence on
$\Rq$ and $\wt\Rq$ factorizes.  

Fix a parameter ${\zeta\in\C}$ with ${|\zeta|<1}$, and let
${0\le\nu<1}$ as above.  The preceding expansion of $s_b$ can be simplified
further via the identity
\begin{equation}\label{CircS}
f(x,\zeta) \,:=\,\sum_{m=1}^\infty \frac{1}{m} \frac{\zeta^{m}\, x^{-m \nu}}{1 -
  x^{-m}} \,=\,\Biggl\{
\begin{aligned}
&-\sum_{j=0}^\infty
\ln\!\Big[1 - \zeta\,x^{-(\nu+j)}\Big],\qquad
|x|>1\,,\\
&+\sum_{j=0}^\infty
\ln\!\Big[1 - \zeta\,x^{(1-\nu+j)}\Big],\qquad
|x|<1\,.
\end{aligned}
\end{equation}
This identity is elementary, but it illustrates that $f(\,\cdot\,,\zeta)$ has an
interesting feature as a holomorphic function of $x$.

In the first case ${|x|>1}$, we apply the Taylor expansion of
$1/(1-x^{-m})$ about ${x=\infty}$ and then reorder the
absolutely-convergent series to obtain
the sum of logarithms in \eqref{CircS}.  In the second case ${|x|<1}$,
we rewrite ${1/(1-x^{-m}) = -x^m/(1-x^m)}$ and expand similarly around
${x=0}$.  Otherwise, $f(x,\zeta)$ has poles densely distributed at all
roots of unity on the circle ${|x|=1}$, so the function is not
continuous there.  For the same reason, $f(x,\zeta)$ cannot be
continued analytically from the region ${|x|>1}$ to ${|x|<1}$, as more
or less apparent from \eqref{CircS}.  In the terminology of the classic
analysis text \cite{GoursatII:1916} (see especially Ch.\,IV, \S 87), the circle 
${|x|=1}$ is a ``natural boundary'' for $f$.  Depending upon the
magnitude of $x$, the series in \eqref{CircS} effectively represents
a {\sl pair} of distinct holomorphic functions.

Since $f(x,\zeta)$ enters the expansion
\eqref{FacDSineII}, the holomorphic functions involved in the
$\Rq$-$\wt\Rq$ decomposition of the double-sine also 
possess natural boundaries on the unit circle ${|\Rq|=1=|\wt\Rq|}$ in the
complex $\Rq$-plane, shown in Figure \ref{qHolomorphyFig}.  This circle corresponds to physical values
${b\in\R_+}$ for the squashing parameter under the identifications
${\Rq=\e{\!2\pi i b^2}}$ and ${\wt\Rq=\e{\!2 \pi i/b^2}}$.  Hence the
analytic $\Rq$-$\wt\Rq$ decomposition for $s_b$ only becomes
sensible when $b$ is continued to complex values, and the result 
depends upon whether ${|\Rq|>1}$ and ${|\wt\Rq|<1}$, or ${|\Rq|<1}$ and
${|\wt\Rq|>1}$.  We emphasize that $s_b$ itself is perfectly well-defined for
${b\in\R_+}$ (as we have assumed so far).  The existence of the natural
boundary plays an important role in \cite{Beem:2012mb}, to which we
refer the reader for a more thorough discussion of the physical
interpretation.

\iffigs
\begin{figure}[t]
	\begin{center}
		\begin{tikzpicture}[x=.7cm,y=.7cm]
		
		\fill[yellow!40] (-3.8,-3.8) rectangle (3.8,3.8);
		
		\fill[blue!40] (0,0) circle(2);
		\draw[thick, dashed] (0,0) circle(2);
		
		\node at (1.9,2.1) {$|\Rq| = 1$};
		
		\draw [<->, gray](-4,0)--(4,0) node [right, black] {$\R$};
		\draw [<->, gray](0,-4)--(0,4) node [above, black] {$i\R$};
		
		\draw [thin, gray](4,5)--(4,4) -- (6.5,4);
		\node at (5.35,4.55) {${\Rq=\e{\!2\pi i b^2}}$};
		
		\end{tikzpicture}
		\caption{The natural boundary of holomorphy for the product expansion of the
                  double-sine.  Outside the unit disk in the $\Rq$-plane, the
                  expansion of $s_b(z)$ in \eqref{FacDSineIII} converges.
                 Inside the unit disk, the expansion of $s_b(z)$ in
                  \eqref{FacDSineIV} converges.}\label{qHolomorphyFig}
	\end{center}
\end{figure}
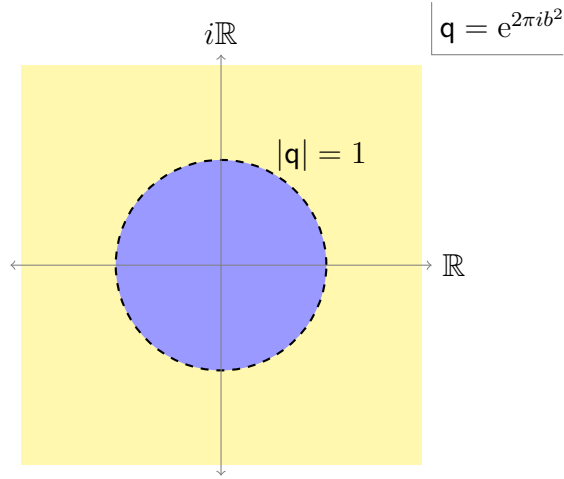\fi

For the region ${|\Rq|>1}$ and ${|\wt\Rq|<1}$, the identity 
\eqref{CircS} applied to the expansion \eqref{FacDSineII}
implies a convergent product formula for the double-sine,
\begin{equation}\label{FacDSineIII}
\begin{aligned}
&s_b\!\left(\mu \,+\, i \,\nu\, b \,+\, i \,\wt\nu\, b^{-1}\right)
\underset{\mu>0}{=} \e{\! i \pi \left[\left(\nu-\ha\right)\left(\wt\nu-\ha\right) -
    \ha\mu^2\right]} \cdot \e{\!\pi \mu \left(\nu-\ha\right)
  b + \pi \mu
  \left(\wt\nu-\ha\right) b^{-1}}\,\times\\
&\qquad\qquad\times\,\Rq^{\frac{1}{4}\left(\nu^2-\nu+\frac{1}{6}\right)}\;
\wt\Rq^{\frac{1}{4}\left(\wt\nu^2-\wt\nu+\frac{1}{6}\right)}\cdot
\prod_{j=0}^\infty\left[\frac{1 - \e{\!-2\pi
        \left(\mu b^{-1}+i\,\nu\right)}\,\wt\Rq^{(1-\wt\nu+j)}}{1-\e{\!-2\pi
        \left(\mu
          b+i\,\wt\nu\right)}\,\Rq^{-(\nu+j)}}\right].
\end{aligned}
\end{equation}
Alternatively for ${|\Rq|<1}$ and ${|\wt\Rq|>1}$, the numerator and
denominator swap,
\begin{equation}\label{FacDSineIV}
\begin{aligned}
&s_b\!\left(\mu \,+\, i \,\nu\, b \,+\, i \,\wt\nu\, b^{-1}\right)
\underset{\mu>0}{=} \e{\! i \pi \left[\left(\nu-\ha\right)\left(\wt\nu-\ha\right) -
    \ha\mu^2\right]}\cdot\e{\!\pi \mu \left(\nu-\ha\right)
  b + \pi \mu
  \left(\wt\nu-\ha\right) b^{-1}}\,\times\\
&\qquad\qquad\times\,\Rq^{\frac{1}{4}\left(\nu^2-\nu+\frac{1}{6}\right)}\;
\wt\Rq^{\frac{1}{4}\left(\wt\nu^2-\wt\nu+\frac{1}{6}\right)}\cdot
\prod_{j=0}^\infty\left[\frac{1-\e{\!-2\pi
        \left(\mu
          b+i\,\wt\nu\right)}\,\Rq^{(1-\nu+j)}}{1 - \e{\!-2\pi
        \left(\mu b^{-1}+i\,\nu\right)}\,\wt\Rq^{-(\wt\nu+j)}}\right].
\end{aligned}
\end{equation}
Both infinite products are conveniently summarized by $q$-Pochhammer
symbols \eqref{qPoch}, eg.
\begin{equation}
\prod_{j=0}^\infty\left[\frac{1 - \e{\!-2\pi
        \left(\mu/b+i\,\nu\right)}\,\wt\Rq^{(1-\wt\nu+j)}}{1-\e{\!-2\pi
        \left(\mu
          b+i\,\wt\nu\right)}\Rq^{-(\nu+j)}}\right] =\, \frac{\Big(\e{\!-2\pi
        \left(\mu/
          b+i\,\nu\right)}\,\wt\Rq^{1-\wt\nu};\wt\Rq\Big)_{\!\infty}}{\Big(\e{\!-2\pi
        \left(\mu\,
          b+i\,\wt\nu\right)}\,\Rq^{-\nu};\Rq^{-1}\Big)_{\!\infty}}\,.
\end{equation}

The product formula \eqref{FacDSineIV} for the double-sine applies
when ${\nu,\wt\nu}$ lie in the unit interval, but this result can be
immediately continued for all ${z\in\C}$ via \eqref{FacDSine}.  We
present the complete result for the $\Rq$-$\wt\Rq$ decomposition
\begin{flalign}\label{QQTSB}
\fbox{$\Rq$-$\wt\Rq$ decomposition, I} \quad & \quad
s_b\!\left(z\right) = \e{\!+i\,
  \Psi(z)}\,\SF_b^+(z; \Rq)\,\wt\SF_b^+(z; \wt\Rq)\,,\qquad \mu>0\,,&
\end{flalign}
where we set\footnote{The notation in \eqref{QQTSB} is imprecise,
  as the functions on the right depend not just on the value of
  ${z\in\C}$ but also on the way of writing the imaginary part of $z$ as the sum in
  \eqref{DecomZ}.  Nonetheless, we use this shorthand to avoid further
  cluttering the notation.}
\begin{equation}\label{DecomZ}
z \,=\, \mu \,+\, i\left(M + \nu\right) b \,+\, i \left(N +
  \wt\nu\right) b^{-1}\,.
\end{equation}
The phase $\Psi(z)$ does not depend upon the squashing parameter $b$,
\begin{equation}\label{PsiPhase}
\e{\!+i\,\Psi(z)} \,=\,
\left(-1\right)^{\left(M+\nu\right)\left(N+\wt\nu\right)}\cdot\left(-i\right)^{M
  + N + \nu + \wt\nu}\cdot\e{\!\frac{i\pi}{4}\left(1-2\mu^2\right)}\,,
\end{equation}
while $\SF_b^+(z; \Rq)$ does, both explicitly and implicitly through $\Rq$,
\begin{equation}\label{BigSF}
\begin{aligned}
\SF_b^+(z; \Rq) \,&:=\, \e{\!\pi \mu b \left(M + \nu -
      \ha\right)}\cdot\Rq^{\frac{1}{4}\left(M+\nu\right)\left(M+\nu-1\right)+\frac{1}{24}}\,\times\\
&\times\begin{cases}
\big(\e{\!-2\pi
        \left(\mu\,
          b+i\,\wt\nu\right)}\,\Rq^{1-\left(M+\nu\right)}
;\Rq\big)_{\infty}\,,\qquad\quad &|\Rq|<1\,,\\
1/\big(\e{\!-2\pi
        \left(\mu\,
          b+i\,\wt\nu\right)}\,\Rq^{-\left(M+\nu\right)}
;\Rq^{-1}\big)_{\infty}\,,\qquad\quad &|\Rq|>1\,.
\end{cases}
\end{aligned}
\end{equation}
The superscript `$+$' indicates that $\SF_b^+$ is defined only for
${\mu>0}$ positive.  Like the function $f(x,\zeta)$ 
in \eqref{CircS}, $\SF_b^+$ is an analytic function of $\Rq$ for both
${|\Rq|<1}$ and ${|\Rq|>1}$, but $\SF_b^+$ does not depend continuously on $\Rq$
across the natural boundary ${|\Rq|=1}$.  Dually,
\begin{equation}\label{BigSFTwiddle}
\begin{aligned}
\wt\SF_b^+(z;\wt\Rq) \,&:=\, \e{\!\pi \mu \left(N + \wt\nu -
      \ha\right) b^{-1}}\cdot\wt\Rq^{\frac{1}{4}\left(N+\wt\nu\right)\left(N+\wt\nu-1\right)+\frac{1}{24}}\,\times\\
&\times\begin{cases}
\big(\e{\!-2\pi
        \left(\mu
          b^{-1}+i\,\nu\right)}\,\wt\Rq^{1-\left(N+\wt\nu\right)}
;\wt\Rq\big)_{\infty}\,,\qquad\quad &|\wt\Rq|<1\,,\\
1/\big(\e{\!-2\pi
        \left(\mu
          b^{-1}+i\,\nu\right)}\,\wt\Rq^{-\left(N+\wt\nu\right)}
;\wt\Rq^{-1}\big)_{\infty}\,,\qquad\quad &|\wt\Rq|>1\,.
\end{cases}
\end{aligned}
\end{equation}

The function $\SF_b^+$ and its sibling $\wt\SF_b^+$ are very
beautiful.  As a consequence of the $q$-binomial theorem, both are
$q$-hypergeometric series.  We omit a review of the latter topic, as
it will not be important here.  Also, when evaluated at the special points 
${z=i\,b}$ and ${z=i\,b^{-1}}$ for the double-sine, both functions are directly
related to the classical eta-function.  For instance,
\begin{equation}
\SF_b^+(i\,b^{-1};\Rq)\big|_{|\Rq|<1} \,=\, \Rq^{1/24} \big(\Rq;\Rq\big)_{\!\infty} \,=\,
\eta(\tau)\,,\qquad\quad \Rq\,=\, \e{2\pi i \tau}\,,
\end{equation}
and
\begin{equation}
\SF_b^+(i\,b;\Rq)\big|_{|\Rq|>1} \,=\, 
\frac{\Rq^{1/24}}{\big(\Rq^{-1};\Rq^{-1}\big)_{\!\infty}}
\,=\, \frac{1}{\eta(\tau)}\,,\qquad\quad \Rq\,=\,\e{-2\pi i\tau}\,,
\end{equation}
including the tell-tale $\Rq^{1/24}$.

So far, ${\mu=\Re(z)}$ has been assumed to be positive.  When ${\mu<0}$
is negative, an entirely parallel $\Rq$-$\wt\Rq$ decomposition follows
from the expansion of $s_b(z)$ in \eqref{AsympN}.  In this case,
\begin{flalign}\label{QQTSBNEG}
\fbox{$\Rq$-$\wt\Rq$ decomposition, II} \quad & \quad
s_b\!\left(z\right) = \e{\!-i\,
  \Psi(z)}\,\SF_b^-(z; \Rq)\,\wt\SF_b^-(z; \wt\Rq)\,,\qquad \mu<0\,,&
\end{flalign}
where now 
\begin{equation}\label{BigFNeg}
\begin{aligned}
\SF_b^-(z; \Rq) \,:&=\, \e{\!-\pi \mu b \left(M + \nu -
      \ha\right)}\cdot\Rq^{-\frac{1}{4}\left(M+\nu\right)\left(M+\nu-1\right)-\frac{1}{24}}\,\times\\
&\times\begin{cases}
1/\big(\e{\!2\pi
        \left(\mu\,
          b+i\,\wt\nu\right)}\,\Rq^{M+\nu}
;\Rq\big)_{\!\infty}\,,\qquad\quad &|\Rq|<1\,,\\
\big(\e{\!2\pi
        \left(\mu\,
          b+i\,\wt\nu\right)}\,\Rq^{M+\nu-1};\Rq^{-1}\big)_{\!\infty}\,,\qquad\quad &|\Rq|>1\,,
\end{cases}
\end{aligned}
\end{equation}
and
\begin{equation}\label{BigFNegtwid}
\begin{aligned}
\wt\SF_b^-(z;\wt\Rq) \,:&=\, \e{\!-\pi \mu \left(N + \wt\nu -
      \ha\right) b^{-1}}\cdot\wt\Rq^{-\frac{1}{4}\left(N+\wt\nu\right)\left(N+\wt\nu-1\right)-\frac{1}{24}}\,\times\\
&\times\begin{cases}
1/\big(\e{\!2\pi
        \left(\mu
          b^{-1}+i\,\nu\right)}\,\wt\Rq^{N+\wt\nu}
;\wt\Rq\big)_{\!\infty}\,,\qquad\quad &|\wt\Rq|<1\,,\\
\big(\e{\!2\pi
        \left(\mu
          b^{-1}+i\,\nu\right)}\,\wt\Rq^{N+\wt\nu-1};\wt\Rq^{-1}\big)_{\!\infty}\,,\qquad\quad &|\wt\Rq|>1\,.
\end{cases}
\end{aligned}
\end{equation}
The distinct asymptotic behaviors in \eqref{eq:MSineAsympta} are 
captured by the innocuous flip of sign for the phase $\Psi(z)$ in
\eqref{QQTSBNEG}.  

In summary, the pair  of analytic
$\Rq$-$\wt\Rq$ decompositions  in \eqref{QQTSB} and
\eqref{QQTSBNEG} make precise a notion of 
holomorphic factorization for $s_b(z)$.

\paragraph{A Toy Model of Factorization.}

Before proceeding, let us briefly illustrate how the $\Rq$-$\wt\Rq$
decompositions in \eqref{QQTSB} and \eqref{QQTSBNEG} induce a 
factorization for associated integrals of the double-sine
$s_b(z)$.  We begin with a {\sl non}-example, in which the
factorization is obstructed, and then present the most basic example.

For the non-example, we consider the integral of the double-sine
function over the real line,
\begin{align}\label{ToyIone}
\mathbf{I}_b =\int_{\mathbb{R}}\!dz \,\,s_b(z+\mu)\,,\qquad\qquad \mu\,\in\,\BH_+\,.
\end{align}
The parameter $\mu$ here plays the role of the real mass
parameter in \eqref{eq:GrandPartitionFunctionInt}, and $\BH_+$ indicates
the upper-half of the complex plane.  We assume that
${\mu\in\BH_+}$ has a positive imaginary part so
that no poles of the integrand lie on the real axis.  Recall from
Figure \ref{fi:zerosAndPolesofdSine} that $s_b(z)$ otherwise has a
pole at ${z=0}$.  Of course, the real part of $\mu$ can be absorbed
into a shift of the integration variable $z$ in \eqref{ToyIone}.
Since $\mathbf{I}_b$ varies holomorphically with $\mu$, $\mathbf{I}_b$
must then be independent of $\mu$ for values in $\BH_+$.

\iffigs 
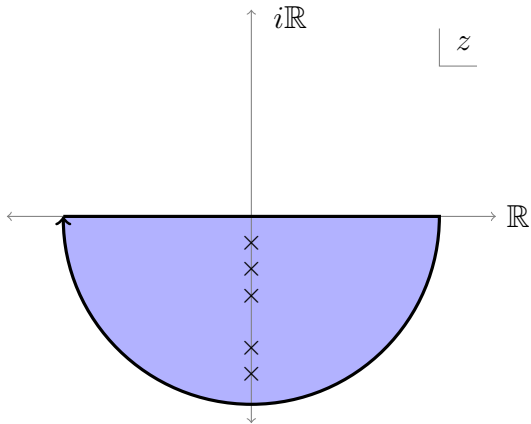
\begin{figure}[t]
	\begin{center}
	\begin{tikzpicture}[x=.5cm,y=.5cm]
	
	\begin{scope}
	\clip (0,0) circle (5);
	\fill[blue!30] (-5,-5) rectangle (5,0);
	\end{scope}
	
	\draw [<->, gray](-6.5,0)--(6.5,0) node [right, black] {$\mathbb{R}$};
	\draw [<->, gray](0,-5.5)--(0,5.5);
	\node at (.3,5.3) [right, black] {$i\mathbb{R}$};
	\draw [->,very thick, black](-5,0)--(5,0) arc(0:-180:5);

	\draw [thin, gray](5,5)--(5,4) -- (6,4);
	\node at (5.65,4.6) {$z$};
	
	%Poles	
	
	\node at (0,-.7) {\small $\times$};	
	
	\node at (0,-1.4) {\small $\times$};
	
	\node at (0,-2.1) {\small $\times$};
	\node at (0,-3.5) {\small $\times$};
	\node at (0,-4.2) {\small $\times$};	
	\end{tikzpicture}
	\end{center}
	\caption{Closed integration contour for evaluating the
          integral $\mathbf{I}_b$ of the double-sine.  Crosses
          ($\times$) indicate the schematic locations of simple poles
          for generic ${b^2\notin\Q}$ and ${\mu\in i\,\R_+}$.}\label{CntrClsFig}
\end{figure}
\fi

As indicated in Figure \ref{fig:Conver}, the double-sine function $s_b(z)$ decays rapidly as $z$ goes to infinity in the lower half-plane.  Hence $\mathbf{I}_b$ can be evaluated via the residue theorem once the real integration contour is closed as in Figure \ref{CntrClsFig}.  According to \eqref{eq:PolesOfDoublesine}, the integrand of \eqref{ToyIone} has poles at locations
\begin{align}
z_* \,=\, -i\,m\,b - i\,n\,b^{-1} -\mu\,,\qquad m,n\,\in\,\Z_{\ge 0}\,.
\end{align}
For generic ${b^2 \notin \Q}$, every pole is simple and lies in the lower half-plane.  The residue of $s_b(z)$ at each pole is given by the expression in \eqref{eq:DSineResidue}.  Hence
\begin{equation}\label{bigIsum}
\begin{aligned}
\mathbf{I}_b&= -2\pi i\sum_{m,n=0}^{\infty} \text{Res}\big[s_b(z+\mu)\big]_{z=-i\,m\,b - i\,n\,b^{-1}-\mu}
\\
&= \sum_{m,n=0}^{\infty}(-1)^{mn}\cdot
\left[i^{m}\,\frac{\Rq^{m (m+1)/4}}{\,\big(\Rq;\Rq\big)_{\!m}}\right]
\cdot \left[i^{n}\,\frac{\wt\Rq^{n
		(n+1)/4}}{\big(\wt\Rq;\wt\Rq\big)_{\!n}}\right]\,.
\end{aligned}
\end{equation}
Here the dependence on $b$ has been subsumed into a dependence on ${\Rq = \e{2\pi i b^2}}$ and ${\wt\Rq = \e{2\pi i/b^2}}$.  The double sum over integers ${m,n\ge 0}$  in the second line of \eqref{bigIsum} nearly factorizes into a product of separate $\Rq$- and $\wt\Rq$-series given by the terms in brackets, but this factorization is obstructed by the oscillating phase $(-1)^{m n}$.

To eliminate the troublesome phase, we replace the integrand in \eqref{ToyIone} by a pair of double-sine functions,
\begin{align}\label{bigJone}
\mathbf{J}_b(\mu) = \int_{\R}\! dz\,\, s_b\!\left(z+\frac{\mu}{2}\right) s_b\!\left(-z +\frac{\mu}{2}\right).
\end{align}
By fiat, we choose the signs in \eqref{bigJone} so that the integrand
is invariant under the reflection ${z \mapsto - z}$.  The dependence
on $\mu$ can no longer be eliminated by a simple shift, so
$\mathbf{J}_b(\mu)$ will now be a non-trivial holomorphic function of
$\mu$.  The coefficient of $1/2$ for $\mu$ merely serves to simplify
later formulas and could be eliminated by a rescaling of $\mu$.  As
will be clear in Section \ref{SU2GT}, the integrand mimics the
structure of the one-loop determinant for massive chiral matter in a
single copy of the fundamental representation of $SU(2)$.  The
inversion ${z\mapsto -z}$ is there interpreted as the residual Weyl
symmetry on the Coulomb branch.

To apply the residue calculus to $\mathbf{J}_b(\mu)$, we first
consider the analytic structure of the product of double-sines in
\eqref{bigJone}.  Evidently from \eqref{eq:MSineAsympta}, the leading
Gaussian terms in each factor cancel as ${|z|\to\infty}$, so that in
either half-plane
\begin{equation}\label{ProdSbz}
	s_b\!\left(z+\frac{\mu}{2}\right)\cdot
	s_b\!\left(-z+\frac{\mu}{2}\right)
	\underset{|z|\to\infty}{=}
\begin{cases}
\,\,\exp{\!\Big[-\pi \left( Q \,+\, i \,\mu\right) z \,+\,o(1)\Big]},\quad \Re(z)>0\,,\\[1 ex]
\,\,\exp{\!\Big[+\pi \left( Q \,+\, i \,\mu\right) z \,+\,o(1)\Big]},\quad
\Re(z)<0\,.
\end{cases}	
\end{equation}
Provided the combination ${Q + i\,\mu\in\R_+}$ is real and positive, as
true when $b$ is real and positive and $\mu$ is purely imaginary and
small, the result in \eqref{ProdSbz} decays exponentially along rays in both the
left and the right \mbox{half-planes}.  We assume the reality condition
${Q + i\,\mu \in \R_+}$ in order to apply the residue calculus to
$\mathbf{J}_b(\mu)$; later we 
analytically continue the result to other values of the parameters $b$ and $\mu$.

Due to the relative sign in
the arguments of the double-sines, the
product in \eqref{ProdSbz} has simple poles extending symmetrically upwards and downwards
in the imaginary direction, 
\begin{equation}
\begin{aligned}
z_* &= -i\,m\,b - i\,n\,b^{-1} -\frac{\mu}{2} \,,
\\
z_* &=  +i\,m\,b + i\,n\,b^{-1} + \frac{\mu}{2} \,,
\end{aligned}
\qquad\qquad m,n\,\in\,\Z_{\ge 0}\,.
\end{equation}
The analytic structure of the integrand in \eqref{bigJone} thus resembles that of the function ${1/\!\cosh(z)}$, and $\mathbf{J}_b(\mu)$ can be evaluated as a sum of residues by closing the integration contour in either the lower or the upper half-plane.  For convenience, we close the contour in the lower half-plane, as shown in Figure \ref{CntrClsFig2}.

\iffigs
\begin{figure}[t]
	\begin{center}
		\begin{tikzpicture}[x=.5cm,y=.5cm]
		
		\begin{scope}
		\clip (0,0) circle (5);
		\fill[blue!30] (-5,-5) rectangle (5,5);
		\end{scope}
		
		\draw [<->, gray](-6.5,0)--(6.5,0) node [right, black] {$\mathbb{R}$};
		\draw [<->, gray](0,-5.5)--(0,5.5);
		\node at (.3,5.3) [right, black] {$i\mathbb{R}$};
		\draw [->,very thick, black](-5,0)--(5,0) arc(0:-180:5);

		\draw [thin, gray](5,5)--(5,4) -- (6,4);
		\node at (5.65,4.6) {$z$};
		
		%Poles	
		
		\node at (0,-.7) {\small $\times$};	
		
		\node at (0,-1.4) {\small $\times$};
		
		\node at (0,-2.1) {\small $\times$};
		\node at (0,-3.5) {\small $\times$};
		\node at (0,-4.2) {\small $\times$};
		
		\node at (0,.7) {\small $\times$};	
		
		\node at (0,1.4) {\small $\times$};
		
		\node at (0,2.1) {\small $\times$};
		\node at (0,3.5) {\small $\times$};
		\node at (0,4.2) {\small $\times$};		
		\end{tikzpicture}
	\end{center}
	\caption{Closed integration contour for evaluating the
          integral $\mathbf{J}_b(\mu)$ of a symmetric product of
          double-sines.  Crosses ($\times$) indicate the schematic locations of simple poles for generic ${b^2\notin\Q}$ and ${\mu=i\,\varepsilon}$.}\label{CntrClsFig2}
\end{figure}
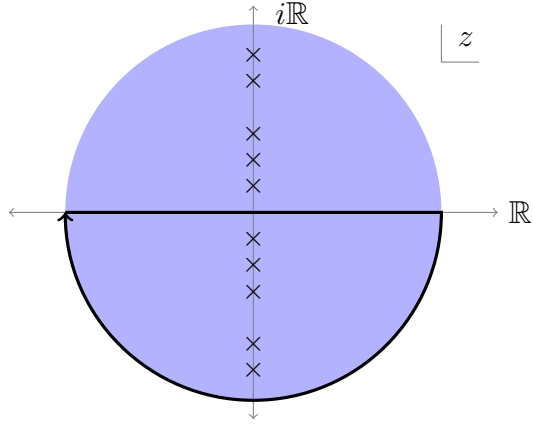
\fi

The residue calculus now implies 
\begin{equation}
\begin{aligned}\label{ResJb}
\mathbf{J}_b(\mu) 
&=-2\pi i\sum_{m,n=0}^{\infty} s_b(\mu + i\,m\,b + i\,n\,b^{-1})\cdot \text{Res}\Big[s_b\!\left(z+\frac{\mu}{2}\right)\!\Big]_{z=-i\,m\,b - i\,n\,b^{-1}-\ha\mu}
\\
&= \sum_{m,n=0}^{\infty} \left(-1\right)^{m n} s_b(\mu + i\,m\,b + i\,n\,b^{-1})\cdot\left[i^{m}\,\frac{\Rq^{m (m+1)/4}}{\,\big(\Rq;\Rq\big)_{\!m}}\right]
\cdot \left[i^{n}\,\frac{\wt\Rq^{n
		(n+1)/4}}{\big(\wt\Rq;\wt\Rq\big)_{\!n}}\right],
\end{aligned}
\end{equation}
where we again use the formula \eqref{eq:DSineResidue} for the
residues of the double-sine at each simple pole.  The $\Rq$-$\wt\Rq$
decomposition of the double-sine in \eqref{QQTSB} and \eqref{QQTSBNEG}
provides a further factorization of the term $s_b(\mu+i \,m\, b + i
\,n\, b^{-1})$ in the summand.  Upon analytic continuation, the $\Rq$-$\wt\Rq$ decomposition
depends upon the sign of the real part of $\mu$ as well as the norm of $\Rq$.
For instance, if ${\mu>0}$ and
${|\Rq|<1}$, then concretely\footnote{Recall that ${|\Rq|<1}$ implies
  the complementary bound ${|\wt\Rq|>1}$ when both $\Rq$ and $\wt\Rq$
  are expressed in terms of the squashing parameter $b^2$.}
\begin{equation}
\begin{aligned}
&s_b\!\left(\mu+i\,m\,b\,+\, i\,n\,b^{-1}\right) \,\underset{\mu>0,\,|\Rq|<1}{=}\, \left(-1\right)^{m n} \, \left(-i\right)^{m+n} \,
\e{\!\frac{i\pi}{4}\left(1-2\,\mu^2\right)}\,\times\\
&\qquad\qquad\times\,\SF^+_b\!\left(\mu + i\,m\,b +
  i\,n\,b^{-1};\Rq\right)\cdot\wt\SF_b^+\!\left(\mu+i\,m\,b+i\,n\,b^{-1};\wt\Rq\right).
\end{aligned}
\end{equation}
Upon substituting the respective definitions in \eqref{BigSF} and
\eqref{BigSFTwiddle} for $\SF_b^+$ and $\wt\SF_b^+$, we rewrite
\begin{equation}\label{sbJb}
\begin{aligned}
s_b\left(\mu+i\,m\,b + i \, n \, b^{-1}\right)
&\underset{\mu>0,\,|\Rq|<1}{=}\, \left(-1\right)^{m n} \, \left(-i\right)^{m+n} \,
\e{\!\frac{i\pi}{4}\left(1-2\,\mu^2\right)}\,\times\\
&\times\,\,\e{\!\pi \mu b \left(m-\ha\right)} \cdot \Rq^{\frac{1}{4} m
  (m-1) + \frac{1}{24}} \cdot \big(\e{\!-2\pi \mu
  b}\,\Rq^{1-m};\Rq\big)_\infty\,\times\\
&\times\,\,\e{\!\pi \mu b^{-1} \left(n-\ha\right)} \cdot
\wt\Rq^{\frac{1}{4}n(n-1) + \frac{1}{24}} \cdot
\frac{1}{\big(\e{\!-2\pi \mu b^{-1}} \wt\Rq^{-n}; \wt\Rq^{-1}\big)_\infty}\,.
\end{aligned}
\end{equation}
Because the phase $(-1)^{m n}$ enters both the residue
\eqref{ResJb} and the $\Rq$-$\wt\Rq$ decomposition \eqref{sbJb} of the
double-sine, the obstruction to holomorphic factorization of
$\mathbf{J}_b(\mu)$ vanishes.

\begin{figure}[t]
\begin{center}
\includegraphics[scale=0.50]{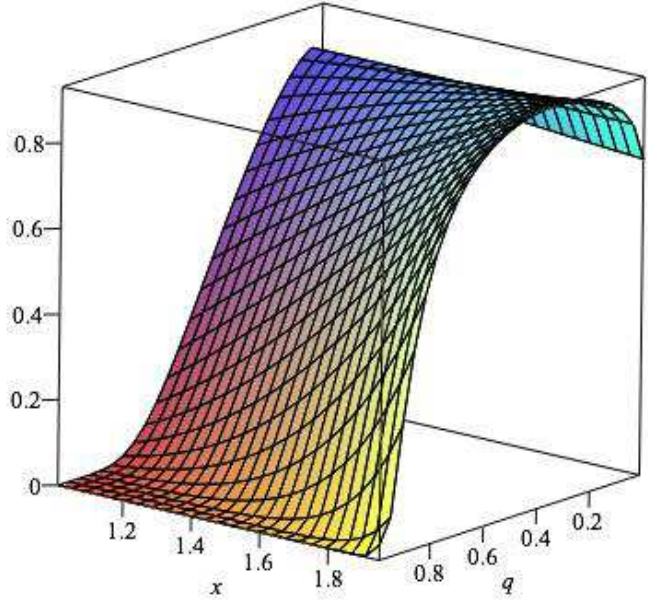}
\caption{Plot of the holomorphic block $\RB(\Rq,\Rx)$ for
  ${\Rq\in(0,1)}$ and ${\Rx\in(1,2)}$.  From its analytic expression,
  $\RB(\Rq,\Rx)$ vanishes at ${\Rq=0}$ and diverges at
  ${\Rq=1}$ for all $\Rx$, but the rapid variation of $\RB(\Rq,\Rx)$ near these
  points is not detected by the numerics.}\label{Bfunct}
\end{center}
\end{figure}

From the expressions in \eqref{ResJb} and \eqref{sbJb}, we see that
$\mathbf{J}_b(\mu)$ is given by a product 
\begin{equation}
\mathbf{J}_b(\mu) \,\underset{\mu>0,\,|\Rq|<1}{=}\, \e{\!\frac{i\pi}{4}\left(1-2\,\mu^2\right)}\cdot\RB(\Rq,\Rx)\cdot\wt\RB(\wt\Rq,\wt\Rx)\,,
\end{equation}
where we introduce the fugacity variables 
\begin{equation}
\Rx = e^{2 \pi \mu b}\qquad\hbox{ and }\qquad  \wt\Rx = e^{2 \pi \mu b^{-1}}\,.
\end{equation}
The holomorphic `blocks' $\RB(\Rq,\Rx)$ and $\wt\RB(\wt\Rq,\wt\Rx)$ in
this example then admit expansions 
\begin{equation}\label{BigBs}
\begin{aligned}
\RB(\Rq,\Rx) \,&\underset{|\Rq|<1,\,|\Rx|>1}{=}\, \sum_{m=0}^\infty\,\,\Rx^{\ha\left(m-\ha\right)} \cdot
  \Rq^{\ha m^2 + \frac{1}{24}} \cdot
  \left[\frac{\big(\Rx^{-1}\Rq^{1-m};\Rq\big)_\infty}{\big(\Rq;\Rq\big)_m}\right],\\
\wt\RB(\wt\Rq,\wt\Rx) \,&\underset{|\wt\Rq|>1,\,|\wt\Rx|>1}{=}\,
\sum_{n=0}^\infty\,\,\wt\Rx^{\ha\left(n-\ha\right)}\cdot \wt\Rq^{\ha
  n^2 + \frac{1}{24}}\cdot
\left[\frac{1}{\big(\wt\Rx^{-1}\,\wt\Rq^{-n};\wt\Rq^{-1}\big)_\infty}\cdot\frac{1}{\big(\wt
\Rq;\wt\Rq\big)_n}\right].
\end{aligned}
\end{equation}
These formulas for $\RB(\Rq,\Rx)$ and $\wt\RB(\wt\Rq,\wt\Rx)$ can be
simplified by elementary manipulations which follow from the
definition of the $q$-Pochhammer symbol in \eqref{qPoch}.  After a
small amount of algebra,
\begin{equation}\label{BigBsII}
\begin{aligned}
\RB(\Rq,\Rx) \,&\underset{|\Rq|<1,\,|\Rx|>1}{=}\, \Rx^{-1/4}\,\,\Rq^{1/24}\,
\Big(\!\left(\frac{\Rq}{\Rx}\right);\Rq\Big)_\infty\cdot
\sum_{m=0}^\infty\,\,(-1)^m \left(\frac{\Rq}{\Rx}\right)^{m/2}\left[\frac{\big(\Rx;\Rq\big)_m}{\big(\Rq;\Rq\big)_m}\right],\\
\wt\RB(\wt\Rq,\wt\Rx) \,&\underset{|\wt\Rq|>1,\,|\wt\Rx|>1}{=}\,
\wt\Rx^{-1/4}\,\,\wt\Rq^{1/24}\,\frac{1}{\big(\wt\Rx^{-1};\wt\Rq^{-1}\big)_\infty}\cdot\sum_{n=0}^\infty\,\,(-1)^n \left(\frac{\wt\Rx}{\wt\Rq}\right)^{n/2}\left[\frac{\big(\wt\Rx^{-1};\wt\Rq^{-1}\big)_n}{\big(\wt\Rq^{-1};\wt\Rq^{-1}\big)_n}\right].
\end{aligned}
\end{equation}
For fixed $\Rx$ and
$\wt\Rx$, the series in \eqref{BigBsII} are convergent for ${|\Rq|<1}$
sufficiently small and ${|\wt\Rq|>1}$ sufficiently large.  See Figure \ref{Bfunct}
for a numerical plot of $\RB(\Rq,\Rx)$ for values of ${\Rq\in(0,1)}$
and ${\Rx\in(1,2)}$.  In other regions of the parameter space,
eg.~${\mu<0}$ and ${|\Rq|>1}$, the integral $\mathbf{J}_b(\mu)$ factorizes in the
same fashion, but with different functions appearing as the
holomorphic blocks.\footnote{Due to the denominator $1/(\Rq;\Rq)_m$ in
  the series expansion \eqref{BigBsII} for $\RB(\Rq,\Rx)$, the block has a natural
  boundary of holomorphy on the unit circle ${|\Rq|=1}$, and similarly
  for $\wt\RB(\wt\Rq,\wt\Rx)$.}

The integral $\mathbf{J}_b(\mu)$ thus has exactly the structure
conjectured for the partition function $Z_{S^3}$ in 
\eqref{FACTOR}, for the trivial index set ${\CI=\{0\}}$ of one element.  All dependence on the parameter $b$ can be factored
into analytic dependence on the pairs $(\Rq,\Rx)$ and
$(\wt\Rq,\wt\Rx)$ in the respective holomorphic blocks $\RB$ and $\wt\RB$.  As the non-example involving $\mathbf{I}_b$
shows, the existence of such a factorization is a non-trivial
statement about the function $\mathbf{J}_b(\mu)$.

\subsection{One-Loop Shift in the Chern-Simons Level}\label{OneLoop}

We return to the exact formula for the Chern-Simons-matter partition
function 
\begin{equation}\label{ZGS}
\begin{aligned}
&Z_{S^3} \,=\, \frac{\e{i\eta_0}}{|\fW|\cdot\Vol(T)}\int_{\mathfrak{h}}\!d^r\!\sigma
\;\exp{\!\left[-\frac{i\,k}{4\pi}\Tr(\sigma^2)\right]}\,\times\\
&\times\prod_{\alpha\in\Delta_+}\left[4\,\sinh\!\left(\frac{b
      \langle\alpha,\sigma\rangle}{2}\right)\sinh\!\left(\frac{
      \langle\alpha,\sigma\rangle}{2
    b}\right)\right]\cdot \prod_{j=1}^n\left[\prod_{\beta \in
\Delta_j}
s_b\!\left(\frac{\langle\beta,\sigma\rangle}{2\pi}\,+\,\mu_j\right)\right].
\end{aligned}
\end{equation}
From the formula in \eqref{ZGS} and the analytic properties of the
double-sine reviewed in Section \ref{PropDSine}, we now
rederive the classic shift in the Chern-Simons level when ${\CN=2}$ chiral
multiplets in the representation 
${\Lambda = [\lambda_1]\oplus\cdots\oplus[\lambda_n]}$ are
integrated-out with large real masses ${|\mu_j|\to\infty}$, ie.\footnote{We implicitly assume
${\mu_j\to\pm\infty}$ along the real axis.  Else when ${\mu_j\in\C}$ is
complexified, $\sgn(\mu_j)$ should be replaced by the
sign of the real part $\Re(\mu_j)$.}
\begin{equation}\label{KEff2}
k_{\rm eff} \,=\, k \,-\, \ha \sum_{j=1}^n c_2(\lambda_j) \,\sgn(\mu_j)\,.
\end{equation}
For ${G=U(1)}$ with charges ${\lambda_j\in\Z}$, the formula for
$k_{\rm eff}$ is to be interpreted as 
\begin{equation}\label{KEffAbII}
k_{\rm eff} \,=\, k \,-\, \ha \sum_{j=1}^n \lambda_j^2 \, \sgn(\mu_j)\,.
\end{equation}
As we discussed in Section \ref{sec:int}, the global parity anomaly can be understood
macroscopically from the shift in \eqref{KEff2} or \eqref{KEffAbII}
and the integrality condition ${k_{\rm eff}\in\Z}$.  See for instance
\S $5.2$ in \cite{Aharony:2013dha} for a previous analysis of
$Z_{S^3}$ in the massive limit.

To ensure that the limit ${|\mu_j|\to\infty}$ is well-defined,
we must renormalize other parameters in \eqref{ZGS} beyond the level
$k$.  As we will see momentarily, $\eta_0$ must be allowed to depend
on the real masses $\mu_j$ via
\begin{equation}\label{EtaN}
\eta_0 \,=\,  \frac{\pi}{2}\,\sum_{j=1}^n\dim[\lambda_j] \left[\mu_j^2
  \,-\, i\,Q\,\mu_j
  -\frac{1}{6}\left(1 + Q^2\right)\right]\sgn(\mu_j),\qquad
Q\,\equiv\,b + b^{-1}\,.
\end{equation}
Here ${\dim[\lambda_j]}$ is the dimension of the irreducible
representation $[\lambda_j]$, and the finite, $b$-dependent terms in
\eqref{EtaN} appear by convention.  If ${G=U(1)}$, the FI
parameter $\xi$ appearing in \eqref{ZUone} must also diverge linearly
with $\mu_j$,
\begin{equation}\label{XiN}
\xi \,=\, \xi_{\rm eff} \,+\, \frac{1}{2}\,\sum_{j=1}^n \lambda_j
\left(\mu_j-\frac{i}{2}\,Q\right) \sgn(\mu_j)\,.
\end{equation}
The choice of finite part in \eqref{XiN} defines the effective FI
parameter $\xi_{\rm eff}$ in the low-energy gauge theory without
matter.  Like all one-loop matching relations, the relation
  between $\xi$ and $\xi_{\rm eff}$ depends upon the
  renormalization scheme, fixed here implicitly through definition
  \eqref{def:DSine} of the double-sine function.  

Finally, observe
  that the renormalization of the FI parameter depends upon the $U(1)$ charges
$\lambda_j$, the real masses $\mu_j$, and the metric parameter $b$.
Quantum corrections to the FI parameter in three dimensions cannot
depend upon holomorphic quantities such as couplings in the
superpotential, but dependence upon the real parameters
$(\lambda_j,\mu_j,b)$ is allowed.   Both
renormalizations in \eqref{EtaN} and \eqref{XiN} of course respect the inversion ${b\mapsto 
  1/b}$.  As explained in \cite{Closset:2012vg,Closset:2012vp}, the
term quadratic in $\mu_j$ which enters $\eta_0$ is related 
to a (possibly fractional) Chern-Simons term for the background
gauge field associated to the given real mass; our renormalization
scheme amounts to subtraction of background Chern-Simons terms.

Because the formula for $Z_{S^3}$ is exact for all
parameter values, integrating-out massive chiral matter just means
taking the limit ${\mu_j\to\pm\infty}$ in \eqref{ZGS}.  In this limit
the asymptotic expansion \eqref{eq:MSineAsympta} for the double-sine
is applicable, so 
\begin{equation}
\begin{aligned}
&s_b\!\left(\frac{\langle\beta,\sigma\rangle}{2\pi}+\mu_j\right)
\underset{\mu_j\to\pm\infty}{=} \exp{\!\left[-\left(\frac{i
      \pi}{2}\,\mu_j^2 \,+\, \frac{i}{2}\,\langle\beta,\sigma\rangle\,
    \mu_j\,+\, \frac{\pi}{2}\,Q\,\mu_j\right)\sgn(\mu_j)\right]}\times\\
&\qquad\times\,\exp{\!\left[-\left(\frac{i}{8\pi}\,\langle\beta,\sigma\rangle^2
    \,+\, \frac{Q}{4}\,\langle\beta,\sigma\rangle\,-\,\frac{i
      \pi}{12}\left(1+Q^2\right)\right)\sgn(\mu_j) + o(1)\right]}.
\end{aligned}
\end{equation}
The dependence on $\sgn(\mu_j)$ captures both asymptotic
behaviors for $s_b(z)$ in \eqref{eq:MSineAsympta}.  The one-loop contribution
to the matrix integrand from massive chiral matter in the irreducible
representation $[\lambda_j]$ becomes 
\begin{equation}\label{OneLoopmu}
\begin{aligned}
&\prod_{\beta \in
\Delta_j}
s_b\!\left(\frac{\langle\beta,\sigma\rangle}{2\pi}+\mu_j\right)
\underset{\mu_j\to\pm\infty}{=}\,\\
&\qquad\exp{\!\left[-\left(\frac{i\pi}{2}\,\mu_j^2 \,+\,
      \frac{\pi}{2}\,Q\,\mu_j \,-\,
      \frac{i \pi}{12}\left(1+Q^2\right)\right) \dim[\lambda_j]\cdot
    \sgn(\mu_j)
  \right]}\,\times\\
&\qquad\times\,\exp{\!\left[-\sum_{\beta\in\Delta_j}\left(\frac{i}{8\pi}\,\langle\beta,\sigma\rangle^2
    \,+\, \frac{i}{2}\left(\mu_j - \frac{i}{2}Q\right)\langle\beta,\sigma\rangle\right)\sgn(\mu_j)+o(1)\right]}\,.
\end{aligned}
\end{equation}
Here $\Delta_j$ denotes the finite set of weights for the
representation $[\lambda_j]$, with cardinality
${|\Delta_j|=\dim[\lambda_j]}$.  When the gauge group $G$ is simple,
the sum over weights in $[\lambda_j]$ vanishes, so 
\begin{equation}\label{SumC}
\sum_{\beta\in\Delta_j} \langle\beta,\sigma\rangle \,=\, 0\,.
\end{equation}
Also, 
\begin{equation}\label{QadC}
\sum_{\beta\in\Delta_j} \langle\beta,\sigma\rangle^2 \,=\,
-c_2(\lambda_j) \cdot \Tr\!\left(\sigma^2\right),
\end{equation}
where $c_2(\lambda_j)$ is the quadratic Casimir of the
representation.  The sum over
weights on the left in \eqref{QadC} is invariant under the action of
the Weyl group $\fW$ on ${\sigma\in\h}$, so it must be proportional to the
quadratic form ${-\Tr(\sigma^2)}$.\footnote{By convention,
  $-\Tr(\,\cdot\,)$ is positive, consistent with the manifest
  positivity on the left in \eqref{QadC}.}  See  \eqref{QadCII} in Appendix
\ref{se:LieAlgConvention} for a small proof that $c_2(\lambda_j)$ is
precisely the constant of proportionality.  

For the special case ${G=U(1)}$
with charges ${\lambda_j\in\Z}$, the limit
in \eqref{OneLoopmu} is replaced by the corresponding 
\begin{equation}\label{OneLoopmuII}
\begin{aligned}
& s_b\!\left(\frac{\lambda_j}{2\pi}\,\sigma + \mu_j\right)
\underset{\mu_j\to\pm\infty}{=}\,\\
&\qquad\exp{\!\left[-\left(\frac{i\pi}{2}\,\mu_j^2 \,+\,
      \frac{\pi}{2}\,Q\,\mu_j \,-\,
      \frac{i \pi}{12}\left(1+Q^2\right)\right)
    \sgn(\mu_j)
  \right]}\,\times\\
&\qquad\times\,\exp{\!\left[-\left(\frac{i}{8\pi}\,\lambda_j^2\,\sigma^2
    \,+\, \frac{i}{2}\,\lambda_j\,\Big(\mu_j -
    \frac{i}{2}\,Q\Big)\,\sigma\right)\sgn(\mu_j)+o(1)\right]}\,.
\end{aligned}
\end{equation}

Of the terms in the limiting expansion above for $s_b$, only the term 
proportional to $\sigma^2$ will be important.  After we substitute the
expansion for $s_b$ into the matrix integral \eqref{ZGS}, the
renormalization of $\eta_0$ in \eqref{EtaN} serves to cancel the $\sigma$-independent terms in
\eqref{OneLoopmu} and \eqref{OneLoopmuII}.  For the term which is 
linear in $\sigma$, the renormalization \eqref{XiN} of the FI 
parameter $\xi$ ensures that $Z_{S^3}$ in \eqref{ZUone} only
depends upon the effective parameter $\xi_{\rm eff}$ in the limit
${\mu_j\to\pm\infty}$.  Both renormalizations are consistent with the
underlying ambiguity \eqref{renDS} in the definition of $s_b(z)$.

Finally, the terms with quadratic dependence on $\sigma$ in
\eqref{OneLoopmu} and \eqref{OneLoopmuII} can be absorbed by a finite
shift of the bare Chern-Simons level $k$ in the original matrix
integral \eqref{ZGS}, so that 
\begin{equation}\label{ZGSren}
\begin{aligned}
\lim_{\mu_j\to\pm\infty}Z_{S^3} \,&=\, \frac{1}{|\fW|\cdot\Vol(T)}\int_{\mathfrak{h}}\!d^r\!\sigma
\;\exp{\!\left[-\frac{i\,k_{\rm eff}}{4\pi}\Tr(\sigma^2)\right]}\,\times\\
&\qquad\qquad\qquad\times\prod_{\alpha\in\Delta_+}\left[4\,\sinh\!\left(\frac{b
      \langle\alpha,\sigma\rangle}{2}\right)\sinh\!\left(\frac{
      \langle\alpha,\sigma\rangle}{2
    b}\right)\right].
\end{aligned}
\end{equation}
In this form, the limiting matrix integral describes the partition
function of pure Chern-Simons theory, with gauge group $G$
and without matter, at level $k_{\rm eff}$.  The 
coefficient of $\sigma^2$ in \eqref{OneLoopmu}, in combination with
the Lie algebra relation in \eqref{QadC}, leads precisely to the 
formula for $k_{\rm eff}$ in \eqref{KEff2}.  The abelian case
\eqref{KEffAbII} follows identically.

\paragraph{Further Remarks.}

Our derivation of effective level $k_{\rm eff}$ from the exact formula
for $Z_{S^3}$ is nice but hardly unexpected.  The finite shift in
the Chern-Simons level is famously determined at one-loop order
\cite{Redlich:1983kn,Redlich:1983dv} in perturbation theory, and the double-sine factor in 
\eqref{ZGS} also arises from a semiclassical, one-loop computation.

Yet the derivation does resolve a small paradox concerning the
analytic behavior of the double-sine function $s_b(z)$ itself.  As
claimed in \eqref{eq:MSineAsympta} on the basis of its integral
representation, $s_b(z)$ grows like a Gaussian as ${|z|\to\infty}$.
This growth means that one-loop fluctuations of
the chiral matter multiplet contribute to the matrix integral in
\eqref{ZGS} with the {\em same} asymptotic magnitude as the classical
Chern-Simons action for the vector multiplet.  The
reader might well be uneasy to hear that a quantum effect competes in
magnitude with a classical effect even far along the Coulomb-branch
${|\sigma|\gg 1}$.\footnote{By contrast, the one-loop contribution from the
vector multiplet, given by the product of hyperbolic sines in
the second line of \eqref{ZGS}, is naturally subleading for large
${|\sigma|\gg 1}$ once the integration contour for $\sigma$ is
rotated slightly away from the real axis.}  In this light, the claimed
asymptotics for $s_b(z)$ demand a physical
explanation.

Our derivation of the formula in \eqref{KEff2} reveals that the unusual asymptotics of
$s_b(z)$ in \eqref{eq:MSineAsympta} are not only correct, but are required by the parity
anomaly.  Even the peculiar asymptotic dependence on the sign of ${\Re(z)}$
is necessary to recover the corresponding physical dependence of $k_{\rm eff}$
on the sign of each real mass $\mu_j$.  For the same reason, the discrepancy
between sectors of convergence in Figure \ref{fig:Conver} must be an honest 
feature of the double-sine function, with consequences for the residue
calculus.

\section{Factorization and Obstruction in Rank-One}\label{RankOne}

As preparation to our analysis of the factorization conjecture
\eqref{FACTOR} for general gauge groups $G$ and matter representations
$\Lambda$, we revisit the abelian case ${G=U(1)}$ originally
discussed by Pasquetti \cite{Pasquetti:2012}.  Beyond reviewing the
mechanics of factorization in an elementary example, we wish to
highlight an important subtlety when the Chern-Simons level ${k}$ is
non-vanishing.  Later in Section \ref{SU2GT}, we compare to the non-abelian (but still rank one) example ${G=SU(2)}$.

\subsection{SQED at Non-Vanishing Level}\label{eq:GGU(1)}

Throughout Section \ref{eq:GGU(1)}, ${\lambda_j\in\Z}$ for
${j=1,\dots,n}$ are electric charges for chiral matter multiplets
minimally coupled to an abelian vector multiplet at Chern-Simons level
${k\ge 0}$.   Without loss, we take all charges ${\lambda_j\neq 0}$ to
be non-zero.  

The matrix integral for the partition
function on $S^3$ then reduces to an integral over ${\sigma\in\R}$,
\begin{equation}\label{ZUoneCh}
Z_{S^3} \,=\, \int_{\R} \frac{d\sigma}{2\pi}
\;\exp{\!\left[\frac{i k}{4\pi}\sigma^2 \,+\,
    i\,\xi\,\sigma\right]}\cdot \prod_{j=1}^n
s_b\!\left(\frac{\lambda_j}{2\pi}\,\sigma\,+\,\mu_j\right).
\end{equation}
To guarantee absolute-convergence, we give ${k\in\ha\Z +
  i\varepsilon}$ a small positive imaginary part.  We also assume the
real mass parameters ${\mu_j\in\BH_+}$ are non-zero, distinct, and
generic.  As indicated above, we give each $\mu_j$ a positive
imaginary part to ensure that the integrand avoids 
the pole of $s_b(z)$ at ${z=0}$ and is regular everywhere along the
real axis. Otherwise, for complex values of $\sigma$, the 
integrand is a meromorphic function with countably-many simple poles
distributed according to Figure \ref{fi:zerosAndPolesofdSine}
and with an essential singularity at infinity.

\paragraph{Chirality Bound.}

The residue calculus provides a natural tool to
evaluate the integral in \eqref{ZUoneCh}, provided we can pick a
suitable closed integration contour $\Gamma$.  The contour
$\Gamma$ must contain the real line $\R$, and the integral over 
the complement ${\Gamma - \R}$ must vanish.  The existence of $\Gamma$
is not guaranteed and depends very much on the asymptotic behavior of 
the integrand as ${|\sigma|\to\infty}$.

When ${k=0}$, the contour $\Gamma$ exists for generic choices of
$U(1)$ charges.  In this case, the asymptotic behavior of the
integrand in \eqref{ZUoneCh} is dominated entirely
by the product of double-sines arising from the chiral matter at one-loop.  
Via \eqref{eq:MSineAsympta}, as ${|\sigma|\to\infty}$ with 
fixed $\mu$, 
\begin{equation}\label{DBlSAsympta}
\prod_{j=1}^n
s_b\!\left(\frac{\lambda_j}{2\pi}\,\sigma\,+\,\mu_j\right)\underset{|\sigma|\to\infty}{=}
\begin{cases}
\,\,\exp{\!\Big[-\frac{i}{4\pi}\,\psi_2(\Lambda)\,\sigma^2 \,+\,
    O(\sigma)\Big]}\,,\qquad \Re(\sigma)>0\,,\\[1 ex]
\,\,\exp{\!\Big[+\frac{i}{4\pi}\,\psi_2(\Lambda)\,\sigma^2 \,+\,
    O(\sigma)\Big]}\,,\qquad \Re(\sigma)<0\,,
\end{cases}
\end{equation}
where $\psi_2(\Lambda)$ is the signed sum
\begin{equation}\label{PsiTwo}
\psi_2(\Lambda) \,=\, \ha\sum_{j=1}^n \lambda_j^2 \, \sgn(\lambda_j).
\end{equation}
So long as ${\psi_2(\Lambda)\neq 0}$, the contribution from the FI term in \eqref{ZUoneCh} is subleading for large $|\sigma|$ and can be ignored.  Clearly, if ${k=0}$ and 
${\psi_2(\Lambda)>0}$, the integrand decays exponentially as
${|\sigma|\to\infty}$ with ${\Im(\sigma)<0}$, so $\Gamma$ can be
closed in the lower half-plane.  Conversely for ${k=0}$ and
${\psi_2(\Lambda)<0}$, the integrand decays exponentially as
${|\sigma|\to\infty}$ with ${\Im(\sigma)>0}$, so $\Gamma$ can be
closed in the upper half-plane.  See Figure \ref{fig:Contours} for a
graphical summary of the situation.  We review the residue computation
\cite{Pasquetti:2012} with this choice of contour momentarily.

\iffigs
\begin{figure}[t]
	\centering
	\subfloat[$\psi_2(\Lambda)>0$]{
		\begin{tikzpicture}[x=.5cm,y=.5cm]
		\begin{scope}
		%		\clip (0,0) circle (5);
		\fill[blue!30] (-5.5,-5.5) rectangle (5.5,0);
		\end{scope}
		
		\draw [<->, gray](-5.5,0)--(5.5,0) node [right, black] {$\R$};
		
		\draw [black](4.3,3.5) -- (4.3,2.3) -- (5.5,2.3);
		\node at (5,3) {$\sigma$};
		
		\draw [ thick, black](-5,0)--(5,0) arc (360:180:5);
		\draw [very thick,->, black](3,0)--(3.1,0);
		\draw [very thick ,->, black](-3.21,-3.83) arc (230:229:5);
		
		\draw [<->, gray](0,-5.5)--(0,3.5);
		
		%		\draw [->, gray] (5.3,-1) --  (6.3,-1);
		%		\draw [->, gray] (-5.3,-1) --  (-6.3,-1);
		%		\draw [->, gray] (-3.8,-3.8) --  (-4.6,-4.6);
		%		\draw [->, gray] (3.8,-3.8) --  (4.6,-4.6);
		\end{tikzpicture}
	}
	\quad \quad \quad \quad
	\subfloat[$\psi_2(\Lambda)<0$]{
		\begin{tikzpicture}[x=.5cm,y=.5cm]
		\begin{scope}
		%		\clip (0,0) circle (5);
		\fill[blue!30] (-5.5,5.2) rectangle (5.5,0);
		\end{scope}
		
		\draw [<->, gray](-5.5,0)--(5.5,0) node [right, black] {$\R$};
		
		\draw [black](4.3,6.5) -- (4.3,5.3) -- (5.5,5.3);
		\node at (5,6) {$\sigma$};

		\draw [thick, black](-5,0)--(5,0) arc (0:180:5);
		\draw [very thick,->, black](3,0)--(3.1,0);
		\draw [very thick ,->, black](-3.21,3.83) arc (130:131:5);
		
		\draw [<->, gray](0,-2.5)--(0,5.5);
		
		%		\draw [->, gray] (5.3,-1) --  (6.3,-1);
		%		\draw [->, gray] (-5.3,-1) --  (-6.3,-1);
		%		\draw [->, gray] (-3.8,-3.8) --  (-4.6,-4.6);
		%		\draw [->, gray] (3.8,-3.8) --  (4.6,-4.6);
		\end{tikzpicture}
	}
	\caption{Integration contours for ${\psi_2(\Lambda)>0}$ and
          ${\psi_2(\Lambda)<0}$.  The integrand decays in the shaded
          region as ${|\sigma|\to\infty}$.}\label{fig:Contours}
\end{figure}
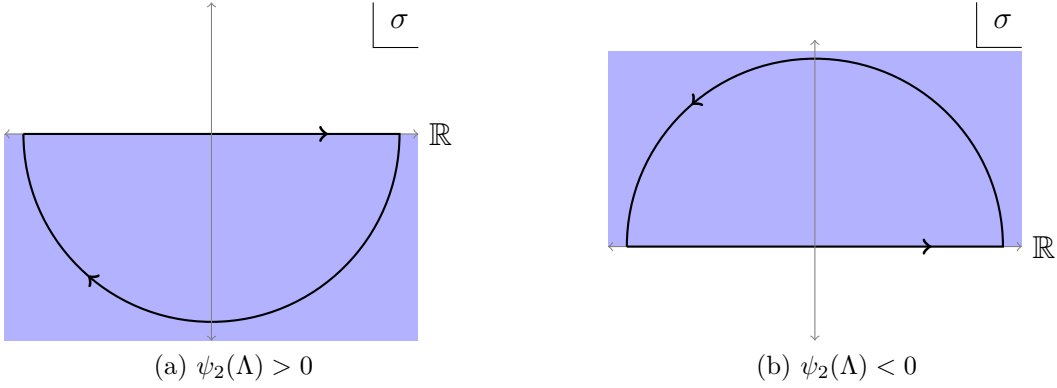
\fi

When the Chern-Simons level ${k\neq 0}$ is non-vanishing, the contour $\Gamma$ may or may not exist, as the
Gaussian term proportional to $k$ in \eqref{ZUoneCh} competes in
magnitude with the product \eqref{DBlSAsympta} of double-sines for
large $|\sigma|$.  If the Chern-Simons level is not too large, meaning
that $k$ is bounded by  
\begin{flalign}\label{ChiralK} 
\fbox{chirality bound} \qquad\qquad&\qquad\qquad
k < |\psi_2(\Lambda)|\,,&
\end{flalign}
then the contour $\Gamma$ can be chosen exactly as for the case ${k=0}$
in Figure \ref{fig:Contours}.  But if ${k > |\psi_2(\Lambda)|}$ violates the
bound in \eqref{ChiralK}, the Gaussian term dominates, and the 
integrand does not decay consistently over either the upper or the lower
half-planes.  Compare to (a) in Figure \ref{fig:Conver}.  Evidently, if the upper bound on $k$ is violated, no choice for $\Gamma$ 
exists which is amenable to the residue calculus.

Finally, in the marginal case ${k=|\psi_2(\Lambda)|}$, the subleading
$O(\sigma)$-terms omitted from the expansion in \eqref{DBlSAsympta}
become important.  With $O(\sigma)$-terms included, the one-loop matter determinant behaves as 
\begin{equation}\label{DBlSAsymptaII}
\prod_{j=1}^n
s_b\!\left(\frac{\lambda_j}{2\pi}\,\sigma\,+\,\mu_j\right)\underset{|\sigma|\to\infty}{=}
\begin{cases}
\exp{\!\Big[-\frac{i}{4\pi}\,\psi_2(\Lambda)\,\sigma^2 - \psi_1(\Lambda)\,\sigma +
    O(1)\Big]},\quad \Re(\sigma)>0\,,\\[1 ex]
\exp{\!\Big[+\frac{i}{4\pi}\,\psi_2(\Lambda)\,\sigma^2 + \psi_1(\Lambda)\,\sigma +
    O(1)\Big]},\quad \Re(\sigma)<0\,,
\end{cases}
\end{equation}
where 
\begin{equation}\label{Psi1Lamb}
\psi_1(\Lambda) \,=\, \sum_{j=1}^n\left(\ha\,Q \,+\,
  i\,\mu_j\right) \frac{|\lambda_j|}{2}\,\in\,\C\,,\qquad\qquad Q \,\equiv\, b + b^{-1}\,.
\end{equation}
Briefly, when ${\psi_2(\Lambda)=k>0}$ is positive, the contour $\Gamma$ can still be
closed in the lower half-plane, provided the FI parameter $\xi$ obeys the auxiliary bounds 
\begin{equation}\label{MargI}
\Re\!\big(\xi \,+\, i\,\psi_1(\Lambda)\big) \,<\, 0\,,\qquad\qquad
\Im\!\big(\xi \,+\, i\,\psi_1(\Lambda)\big) \,>\, 0\,.
\end{equation}
This condition ensures the decay of the integrand in the quadrant where  the Gaussian term is cancelled by the double-sine, meaning
${\Re(\sigma)>0}$ and ${\Im(\sigma)<0}$.  Conversely when
${\psi_2(\Lambda)=-k<0}$ is negative, $\Gamma$ can be 
closed in the upper half-plane if
\begin{equation}\label{MargII}
\Re\!\big(\xi \,-\, i \, \psi_1(\Lambda)\big) \,>\, 0\,,\qquad\qquad
\Im\!\big(\xi \,-\, i \, \psi_1(\Lambda)\big) \,<\, 0\,,
\end{equation}
ensuring decay of the integrand in the quadrant where ${\Re(\sigma)<0}$ and
${\Im(\sigma)>0}$.  When either bound on $\xi$ in \eqref{MargI} or \eqref{MargII} is
violated, a suitable contour $\Gamma$ does not exist.

The bound ${k < |\psi_2(\Lambda)|}$ reflects the
fact that the residue calculus is not effective for evaluating
integrals which are asymptotically Gaussian.  Clearly, the strength of
the bound depends upon the magnitude of $\psi_2(\Lambda)$, which
measures the net chirality of the matter representation $\Lambda$.  For
instance, if $\Lambda$ is a real representation of ${U(1) \simeq SO(2)}$ and so is preserved under the  charge conjugation
${\lambda_j\mapsto-\lambda_j}$, then ${\psi_2(\Lambda)=0}$.  In this case,
the residue calculus applies only when ${k=0}$, with appropriate range
of $\xi$.  Alternatively, when the bound in \eqref{ChiralK} is read backwards for fixed ${k>0}$, the residue calculus applies only when the matter spectrum is sufficiently chiral.

One might wonder whether the chirality bound ${k < |\psi_2(\Lambda)}|$
indicates an intrinsic feature of the partition function $Z_{S^3}$,
such as a failure of factorization in the semiclassical limit
${k\to\infty}$, or whether it merely describes a failure in the most
naive application of the residue calculus.  In Section
\ref{LevelZero}, we demonstrate that the second circumstance holds,
and the apparent bound on $k$ in \eqref{ChiralK} can be evaded with
more involved analytic maneuvers.

\paragraph{Residue Calculus for SQED Partition Function.}

We now review the residue calculus for $Z_{S^3}$ in the case that
the chirality bound in \eqref{ChiralK} is satisfied.  Explicit
formulas for $Z_{S^3}$ are quite complicated in general, so our focus
will be on the structure of the result and how this 
structure depends upon the gauge theory data.

According to \eqref{eq:PolesOfDoublesine}, the product of
double-sines in the integrand for $Z_{S^3}$ has poles on the union of divisors
\begin{equation}
D_j\,= \bigcup_{u,v=0}^\infty \left\{\sigma = \frac{2\pi}{\lambda_j}\!\left(-i\,u\,b \,-\,
  i\,v\,b^{-1}\,-\, \mu_j\right)\right\},\qquad j=1,\ldots,n\,.
\end{equation}
For $\mu_j$ generic, poles corresponding to different divisors 
do not coincide, ${D_j \cap D_{\ell} = \emptyset}$ for ${j\neq\ell}$,
and every pole is simple.  Also, because we
assume ${b\in\R_+}$, the poles in each $D_j$ are located
in either the upper or the lower half-plane, dictated by the sign of
the charge $\lambda_j$.\footnote{Because
  ${\mu_j\in\R+i\varepsilon}$ 
  has a small positive imaginary part, this statement remains true for
  the special pole with ${u=v=0}$, which is slightly displaced from the real axis.}
E.g.~for ${\lambda_j>0}$, all poles in $D_j$ lie in the lower half-plane.

When $Z_{S^3}$ is computed as a sum over residues, not every pole in
the integrand contributes.  As shown in Figure
\ref{fig:Contours}, the appropriate integration
contour $\Gamma$ depends upon the sign of $\psi_2(\Lambda)$ in 
\eqref{PsiTwo}.  If ${\psi_2(\Lambda)>0}$, the contour must be closed in
the lower half-plane, so only the divisors $D_j$
associated to positive charges ${\lambda_j>0}$ contribute to
the sum over residues.  Conversely if ${\psi_2(\Lambda)<0}$, only the
divisors $D_j$ for negative charges ${\lambda_j<0}$ contribute.  

We will discuss the residues themselves shortly, but let us first
abbreviate 
\begin{equation}\label{AbbRes}
\begin{aligned}
\RZ^{\,j}_{u,v} &= \Res\!\left[\exp{\!\left(\frac{i k}{4\pi}\sigma^2 \,+\,
    i\,\xi\,\sigma\right)}\cdot \prod_{\ell=1}^n
s_b\!\left(\frac{\lambda_{\ell}}{2\pi}\,\sigma\,+\,\mu_{\ell}\right)\right]_{\sigma
= \frac{2\pi}{\lambda_j}\left(-i\,u\,b -
  i\,v\,b^{-1} - \mu_j\right)}\,,\\
&u,v \ge 0\,,\qquad j\,=\,1,\ldots,n\,.
\end{aligned}
\end{equation}
Depending on the sign of $\psi_2(\Lambda)$, the residue calculus
implies 
\begin{equation}\label{ZSUMPOS}
Z_{S^3} \,=\,-i \sum_{\lambda_j>0} \sum_{u,v=0}^\infty
\RZ^{\,j}_{u,v}\,,\qquad\qquad \psi_2(\Lambda)>0\,,
\end{equation}
versus
\begin{equation}\label{ZSUMNEG}
Z_{S^3}\,=\, +i \sum_{\lambda_j<0} \sum_{u,v=0}^\infty
\RZ^{\,j}_{u,v}\,,\qquad\qquad \psi_2(\Lambda)<0\,.
\end{equation}
The minus sign in \eqref{ZSUMPOS} accounts for the orientation
of the contour in Figure \ref{fig:Contours}, and the respective sums
run over the set of $U(1)$ charges with the given sign.  Of course, when
$\psi_2(\Lambda)$ is strictly positive, at least one
charge $\lambda_j$ is necessarily positive as well, so the sum over residues
is always non-empty!  

To suppress the proliferation of signs, we restrict attention to the
case ${\psi_2(\Lambda)>0}$, meaning ${\lambda_j>0}$ for all terms in
the residue sum.

We are left to evaluate the residue $\RZ^{\,j}_{u,v}$ itself.
We have already determined the residues at poles of $s_b$
in \eqref{eq:DSineResidue}, and the $\Rq$-$\wt\Rq$ decomposition in 
\eqref{QQTSB} and \eqref{QQTSBNEG} provides a description of $s_b$ 
at regular points, as soon as we give $b$ a small imaginary part.
All that remains is careful bookkeeping.  

The essential novelty of our result for SQED concerns the arithmetic
dependence on the $U(1)$ charges.  For this reason, we
must keep track of the values of the integers ${u,v}$ in
$Z^{\,j}_{u,v}$ modulo the corresponding charge $\lambda_j$.  We decompose 
\begin{equation}\label{CONG}
\begin{aligned}
u \,&=\, M\,\lambda_j \,+\, r\,,\qquad\qquad M,N\ge 0\,,\\
v \,&=\, N\,\lambda_j \,+\, s\,,\qquad\qquad 0 \le
r,s < \lambda_j\,,
\end{aligned}
\end{equation}
meaning
\begin{equation}
u \,\equiv\, r \,\mod\, \lambda_j\,,\qquad\qquad v \,\equiv\, s \,\mod\, \lambda_j\,.
\end{equation}
After tedious but trivial algebra, which we spare the reader, we find
that each residue in \eqref{AbbRes} can be expressed as a
product\footnote{As usual for ${G=U(1)}$, ${c_2(\Lambda) = \sum \lambda_j^2}$.}
\begin{equation}\label{AbbResII}
\begin{aligned}
\RZ^{\,j}_{u,v} \,=\,
\frac{i}{\lambda_j}\cdot&\exp{\!\left[2\pi i \,\big(k + \ha
    c_2(\Lambda)\big)\, M N\right]}\,\times\\
\times\,&\exp{\!\left(i\,\Theta^{\,j}_{r,s}\right)}\cdot\RW^{\,j}_{M,r,s}(\Rq,\Rx,\Ry)\cdot{\wt\RW}^{\,j}_{N,r,s}(\wt\Rq,\wt\Rx,\wt\Ry)\,.
\end{aligned}
\end{equation}
We recall from the Introduction 
\begin{equation}
\Rq \,=\, \e{\!2\pi i b^2}\,,\qquad\qquad\qquad \wt\Rq \,=\,
\e{\!2\pi i/b^2}\,,
\end{equation}
and 
\begin{equation}
\Rx_j \,=\, \e{\!2\pi \mu_j b}\,,\qquad\qquad \wt\Rx_j
\,=\, \e{\!2\pi \mu_j/b}\,,\qquad\qquad j\,=\,1,\ldots,n\,.
\end{equation}
To account for the FI parameter in SQED, we also introduce variables
\begin{equation}
\Ry \,=\, \e{\!2\pi \xi b}\,,\qquad\qquad\qquad \wt\Ry \,=\, \e{\!2\pi \xi/b}\,.
\end{equation}
We provide expressions for $\Theta^{\,j}_{r,s}$, $\RW^{\,j}_{M,r,s}$, and
$\wt\RW^{\,j}_{N,r,s}$ below, but let us first emphasize the important
properties of \eqref{AbbResII}.

\paragraph{Properties of the Residue.}

All dependence on the squashing parameter $b$ in
  $\RZ^{\,j}_{u,v}$ is captured through the dependence on the
  variables $(\Rq,\Rx,\Ry)$ and $(\wt\Rq,\wt\Rx,\wt\Ry)$, which appear
  in the respective functions
  $\RW^{\,j}_{M,r,s}$ and $\wt\RW^{\,j}_{N,r,s}$.  The dependence
  on these variables completely factorizes.

We indicate explicitly the dependence on the integers ${M,N\ge
    0}$ as well as the congruence classes ${0 \le r,s < \lambda_j}$
  associated to $(u,v)$.  The phase $\Theta^{\,j}_{r,s}$ depends on neither $M$
  nor $N$, the function $\RW^{\,j}_{M,r,s}$ is independent of $N$, and
  the dual function $\wt\RW^{\,j}_{N,r,s}$ is independent of $M$.  The
  latter feature can be traced to the same properties of the
  formulas for $\SF_b^+$ and $\wt\SF_b^+$ in  \eqref{BigSF} and 
  \eqref{BigSFTwiddle}.

Because the SQED partition function involves a sum over 
  ${u,v\ge 0}$, we must also consider to what extent the dependence on $u$ and
  $v$ factorizes in $\RZ^{\,j}_{u,v}$.  The formula in
  \eqref{AbbResII} displays no nice factorization in terms of the
  congruence classes ${r,s}$ modulo $\lambda_j$.  But precisely
  when the anomaly-cancellation condition ${k + \ha c_2(\Lambda) \in
    \Z}$ is satisfied, so that the phase factor in the first line of
  \eqref{AbbResII} is trivial, the dependence on the integers $M$ and $N$ does
  factorize.  The combination ${k + \ha c_2(\Lambda)}$ appears in
  \eqref{AbbResII} for the same reason it appeared in Section
  \ref{OneLoop}, as the coefficient of the leading Gaussian term in
  the integrand.

The phase $\Theta^{\,j}_{r,s}$ is given in terms of the gauge
  theory data by 
\begin{equation}\label{BigTheta}
\begin{aligned}
\exp{\!\left(i\,\Theta^{\,j}_{r,s}\right)} \,&=\,
\exp{\!\left[-\frac{2\pi i}{\lambda_j^2}\left(k\,+\,\ha
      c_2(\Lambda)\right) r s\right]}\cdot\exp{\!\left[\frac{i\pi
      k}{\lambda_j^2}\mu_j^2 \,-\, \frac{2\pi
      i}{\lambda_j}\xi\,\mu_j\right]}\,\times\\
&\times\,\exp{\!\left[\sum_{\ell\neq j}\frac{i \pi}{4}\left(1 - \frac{2}{\lambda_j^2}\langle\lambda_j,\mu_\ell\rangle^2\right)\sgn\langle\lambda_j,\mu_\ell\rangle\right]},
\end{aligned}
\end{equation}
where $\langle\lambda_j,\mu_\ell\rangle$ denotes the symplectic
pairing
\begin{equation}
\langle\lambda_j,\mu_\ell\rangle \,\equiv\, \lambda_j\,\mu_\ell \,-\,
\lambda_\ell \, \mu_j\,.
\end{equation}
If ${k + \ha c_2(\Lambda) \in \Z}$, the phase
$\Theta^{\,j}_{r,s}$ still depends non-trivially on the product ${r s}$ through the
first term in \eqref{BigTheta}.  For general values of $\lambda_j$,
factorization in $(r,s)$ is spoiled by this term.

Formulas for $\RW^{\,j}_{M,r,s}(\Rq,\Rx,\Ry)$ and
  $\wt\RW^{\,j}_{N,r,s}(\wt\Rq,\wt\Rx,\wt\Ry)$, even when written in
  terms of $\SF^\pm_b(z;\Rq)$ and $\wt\SF^\pm_b(z;\wt\Rq)$ from
  Section \ref{PropDSine}, are more involved.  In detail, 
\begin{equation}\label{BigW}
\begin{aligned}
&\RW^{\,j}_{M,r,s}(\Rq,\Rx,\Ry) \,=\, \e{\!-2\pi
      i\left(k+\ha c_2(\Lambda)\right) M s/\lambda_j}\cdot i^{\left(M \lambda_j + r\right) + \sum_{\ell\neq j}\left(M +
      (r/\lambda_j)\right)\,\lambda_\ell\,
    \sgn\langle\lambda_j,\mu_\ell\rangle}\,\times\\
&\qquad\times\,\Rq^{-\frac{k}{2}\big(M + (r/\lambda_j)\big)^2} \,
\Rx_j^{-\frac{k}{\lambda_j}\big(M +
    (r/\lambda_j)\big)}\,\Ry^{\big(M +
    (r/\lambda_j)\big)}\cdot \left[\frac{\Rq^{(M \lambda_j+r)(M
        \lambda_j+r+1)/4}}{\big(\Rq;\Rq\big)_{\!\left(M\lambda_j+r\right)}}\right]\times\\
&\qquad\qquad\times
  \prod_{\ell\neq j} \SF_b^{\sgn\langle\lambda_j,\mu_\ell\rangle}\!\left(\rz_{j,\ell};\Rq\right),
\end{aligned}
\end{equation}
where 
\begin{equation}\label{rzrankone}
\rz_{j,\ell} \,=\, \frac{1}{\lambda_j}\langle\lambda_j,\mu_\ell\rangle
\,-\, i \left(M + \frac{r}{\lambda_j}\right)\lambda_\ell \, b \,-\, i
\left(N \,+\, \frac{s}{\lambda_j}\right)\lambda_\ell \, b^{-1}\,.
\end{equation}
One can check straightforwardly that $\SF_b^\pm(\rz_{j,\ell};\Rq)$
does not depend on the integer $N$ or the variables
$(\wt\Rq,\wt\Rx,\wt\Ry)$.  For instance, assuming ${|\Rq|<1}$,
\begin{equation}\label{SpecFz}
\begin{aligned}
&\SF_b^+(\rz_{j,\ell};\Rq) \,=\, \Rx_\ell^{-\ha\big[\big(M +
  (r/\lambda_j)\big)\lambda_\ell+\ha\big]}\,
\Rx_j^{\ha(\lambda_\ell/\lambda_j)\big[\big(M+(r/\lambda_j)\big)\lambda_\ell+\ha\big]}\,\times\\
&\times\Rq^{\frac{1}{4}\big[\big(M+(r/\lambda_j)\big)\lambda_\ell\big]
  \big[\big(M +
  (r/\lambda_j)\big)\lambda_\ell\,+\,1\big]+\frac{1}{24}}\cdot\Big(\e{\!2\pi
  i s (\lambda_\ell/\lambda_j)}\,\Rx_\ell^{-1}\,\Rx_j^{(\lambda_\ell/\lambda_j)}\,\Rq^{\big(M+(r/\lambda_j)\big)\lambda_\ell+1};\Rq\Big)_{\!\infty}\,.
\end{aligned}
\end{equation}

The detailed formulas in \eqref{BigW} and \eqref{SpecFz} are not
themselves important for the present work.  We provide them only to
illustrate that such formulas exist.  Also, note that $\RW^{\,j}_{M,r,s}$ can be
presented analytically as a convergent product of rational functions in $(\Rq,\Rx,\Ry)$.
Finally, $\RW^{\,j}_{M,r,s}$ does depend on the index $s$ through the
initial phase factor in \eqref{BigW} as well as the phase which enters
the final $q$-Pochhammer term in \eqref{SpecFz}.  This 
dependence further spoils factorization in the
congruence classes $(r,s)$ modulo $\lambda_j$.

For completeness (or masochism), we record the parallel formula
  for ${\wt\RW}^{\,j}_{N,r,s}$,
\begin{equation}\label{BigWtilde}
\begin{aligned}
&{\wt\RW}^{\,j}_{N,r,s}(\wt\Rq,\wt\Rx,\wt\Ry) \,=\, \e{\!-2\pi
      i\left(k+\ha c_2(\Lambda)\right) N r/\lambda_j}\cdot i^{\left(N
        \lambda_j + s\right) + \sum_{\ell\neq j}\left(N +
      (s/\lambda_j)\right)\,\lambda_\ell\,
    \sgn\langle\lambda_j,\mu_\ell\rangle}\,\times\\
&\qquad\times\,{\wt\Rq}^{-\frac{k}{2}\!\big(N + (s/\lambda_j)\big)^2} \,
{\wt\Rx}_j^{-\frac{k}{\lambda_j}\!\big(N +
    (s/\lambda_j)\big)}\,\wt\Ry^{\big(N +
    (s/\lambda_j)\big)}\cdot \left[\frac{\wt\Rq^{(N \lambda_j+s)(N
        \lambda_j+s+1)/4}}{\big(\wt\Rq;\wt\Rq\big)_{\!\left(N\lambda_j+s\right)}}\right]\times\\
&\qquad\qquad\times
  \prod_{\ell\neq j} {\wt\SF}_b^{\,\sgn\langle\lambda_j,\mu_\ell\rangle}\!\left(\rz_{j,\ell};\wt\Rq\right).
\end{aligned}
\end{equation}

\paragraph{Factorization at Higher Degree.}

From \eqref{AbbResII}, the partition
function for positive ${\psi_2(\Lambda)}$ is given by the successive sums 
\begin{equation}\label{BigZSum}
\begin{aligned}
Z_{S^3} \,=\,
\sum_{\lambda_j>0}\,\sum_{r,s=0}^{\lambda_j-1}\,\sum_{M,N=0}^\infty
&\frac{1}{\lambda_j}\exp{\!\left[2\pi i \,\big(k + \ha
    c_2(\Lambda)\big)\, M
    N\right]}\,\times\\
&\times\exp{\!\left(i\,\Theta^{\,j}_{r,s}\right)}\cdot\RW^{\,j}_{M,r,s}(\Rq,\Rx,\Ry)\cdot{\wt\RW}^{\,j}_{N,r,s}(\wt\Rq,\wt\Rx,\wt\Ry)\,,
\end{aligned}
\end{equation}
where we reduce ${u,v}$ modulo $\lambda_j$ as in \eqref{CONG}.
Precisely when ${k + \ha c_2(\Lambda)\in\Z}$, the phase in the first
line of \eqref{BigZSum} is trivial, so we can rewrite $Z_{S^3}$
in the factorized form 
\begin{equation}\label{BigZSumII}
Z_{S^3} \,=\, \sum_{\lambda_j>0}\,\sum_{r,s=0}^{\lambda_j-1} \frac{\exp{\!\left(i\,\Theta^{\,j}_{r,s}\right)}}{\lambda_j}\,\RB^{\,j}_{r,s}(\Rq,\Rx,\Ry)\,\wt\RB^{\,j}_{r,s}(\wt\Rq,\wt\Rx,\wt\Ry)\,,
\end{equation}
with blocks given by the series
\begin{equation}\label{BlockSer}
\RB^{\,j}_{r,s} \,=\, \sum_{M=0}^\infty
\,\RW^{\,j}_{M,r,s}\,,\qquad\qquad \wt\RB^{\,j}_{r,s} \,=\,
\sum_{N=0}^\infty\,{\wt\RW}^{\,j}_{N,r,s}\,.
\end{equation}
Comparing to the Factorization Conjecture in \eqref{FACTOR}, we
identify the index set $\CI$ for SQED with the set of triples
\begin{equation}
\CI \,=\, \Big\{\left(j,r,s\right)\,\Big|\, \lambda_j>0\,,\,0\le
r,s<\lambda_j\Big\}\,,
\end{equation}
and $\RG_{m n}$ is the diagonal matrix
\begin{equation}\label{BigGmn}
\RG_{m n} \,=\, \delta_{m
  n}\cdot\frac{\exp{\!\left(i\,\Theta^{\,j}_{r,s}\right)}}{\lambda_j}\,,\qquad\qquad \begin{aligned}
&m\equiv (j,r,s)\,,\\
&n \equiv (j',r',s')\,.
\end{aligned}
\end{equation}
For ${\psi_2(\Lambda)<0}$, the structure is identical, but with
contributions from negative charges.

To summarize, for each chiral multiplet whose charge $\lambda_j$ agrees
in sign with $\psi_2(\Lambda)$, we obtain chiral blocks which are
labelled by elements in the group ${\Z/\lambda_j\Z \times
  \Z/\lambda_j\Z}$.  The simplest case occurs when all chiral matter
fields have unit charge ${\lambda_j=+1}$ for ${j=1,\ldots,n}$.  In
this ``maximally-chiral'' situation, ${k \equiv n/2}$ mod $\Z$, and the
residue calculus applies if ${k < n/2}$.  Each chiral matter field
contributes exactly one chiral block ${\RB^{\,j}_{r,s=0}}$ to the
partition function, with trivial labels for congruence classes.  

For the maximally-chiral theory, the
phase $\Theta^{\,j}_{r,s}$ in \eqref{BigTheta} becomes an
uninteresting constant depending only on the real masses, and the function $\RW^{\,j}_{M,r,s}(\Rq,\Rx,\Ry)$ in \eqref{BigW} specializes to 
\begin{equation}
\begin{aligned}
\RW^{\,j}_{M,r,s=0}(\Rq,\Rx,\Ry) \,&=\, i^{M\left(1+\sum_{\ell\neq
      j}\sgn(\mu_\ell-\mu_j)\right)}\,\Ry^M\,\Rx_j^{-k M}\,\Rq^{-k
  M^2/2}\,\times\\
&\times\,\left[\frac{\Rq^{M(M+1)/4}}{\big(\Rq;\Rq\big)_M}\right]\cdot\prod_{\ell\neq j}\,\SF^{\sgn(\mu_\ell-\mu_j)}_b\!\left(\rz_{j,\ell};\Rq\right),
\end{aligned}
\end{equation}
with 
\begin{equation}
\rz_{j,\ell} \,=\, (\mu_\ell - \mu_j) \,-\, i \, M \, b \,-\, i\, N \,b^{-1}\,.
\end{equation}
For instance, if ${\mu_\ell > \mu_j}$ (meaning ${\Rx_\ell > \Rx_j}$
for ${b\in\R_+}$) and ${|\Rq|<1}$,
\begin{equation}\label{ExSbP}
\SF_b^+(\rz_{j,\ell};\Rq) \,=\,
\left(\frac{\Rx_j}{\Rx_\ell}\right)^{\ha(M+\ha)}\,\Rq^{\frac{1}{4}M(M+1)
  +\frac{1}{24}}\left(\left(\frac{\Rx_j}{\Rx_\ell}\right)\Rq^{M+1};\Rq\right)_{\!\infty}\,.
\end{equation}
See \cite{Pasquetti:2012} for a nice interpretation of
$\RB^{\,j}_{r,s=0}$ in terms of a vortex partition function.  

\subsection{Comparison to $SU(2)$ Gauge Theory}\label{SU2GT}

In marked contrast to $U(1)$, all representations of ${G=SU(2)}$ are
self-dual, either real or pseudoreal depending upon whether the spin is integral
or half-integral.  As we now explain, this fact drastically alters 
the asymptotic behavior of the matrix integrand,
with consequences for the naive residue calculus.

\paragraph{Some Preliminaries.}

According to the Lie algebra conventions in Appendix
\ref{se:LieAlgConvention}, the Cartan subalgebra ${\h\simeq\R}$ of
$SU(2)$ is generated by the anti-hermitian matrix
\begin{equation}
\hat h \,=\, 
\begin{pmatrix}
i&0\\
0&-i
\end{pmatrix}\,,
\end{equation}
with norm ${(\hat h,\hat h)=-\Tr({\hat h}^2)=2}$.  The single fundamental
weight ${\hat\omega\in\h^*}$ satisfies ${\langle\hat\omega,\hat
  h\rangle = 1}$, and the root lattice is generated by the positive
simple root 
${\hat\alpha = 2 \hat\omega}$.  

Any highest-weight $\lambda$ takes
the form 
\begin{equation}
\lambda \,=\, L\cdot\hat\omega\,,\qquad\qquad
L\,\ge\,0\,,\qquad\qquad L \in \Z\,.
\end{equation}
The representation $[\lambda]$ has dimension
$(L+1)$ and contains weights in the set
\begin{equation}
\Delta_{\lambda} =\left\{L\cdot\hat\omega,\quad
(L-2)\cdot\hat\omega,\quad\cdots,\quad -(L-2)\cdot\hat\omega,\quad
-L\cdot\hat\omega\right\}.
\end{equation}
Note that $\Delta_\lambda$ is preserved by the Weyl group
${\fW=\Z/2\Z}$, which acts by the reflection ${\hat h\mapsto -\hat h}$
and ${\hat\omega\mapsto-\hat\omega}$.  Hence $[\lambda]$ is preserved
by complex conjugation and must be real or pseudoreal.

Finally, from the general formula in \eqref{C2Vform}, the Casimir
$c_2(\lambda)$ is given in terms of the integer $L$ by the cubic polynomial
\begin{equation}\label{SU2Cas}
c_2(\lambda) \,=\, \frac{L \left(L+1\right) \left(L+2\right)}{6}\,\in\,\Z\,.
\end{equation}
For ${L=1}$, one checks ${c_2({\bf
    2})=1}$, per convention.

\paragraph{Factorization at Level Zero.}

For $SU(2)$ gauge theory at level ${k \ge 0}$ with chiral multiplets
in the representation ${\Lambda =
  [\lambda_1]\oplus\cdots\oplus[\lambda_n]}$, the 
matrix integral in
\eqref{eq:GrandPartitionFunctionInt} reduces to an integral over the
real line,
\begin{equation}\label{ZSU(2)}
Z_{S^3} \,= \int_{\R}\!\frac{d\sigma}{\pi}
\,\exp{\!\left[\frac{i k}{2\pi}\sigma^2\right]}
\sinh\!\left(b\,\sigma\right)\sinh\!\left(\sigma/b\right) \cdot
\prod_{j=1}^n \, G_{j}(\sigma)\,.
\end{equation}
Here $G_{j}(\sigma)$ is the one-loop contribution from matter
in the irreducible representation $[\lambda_j]$, of dimension $(L_j+1)$, with
real mass $\mu_j$, 
\begin{equation}\label{BigGFun}
G_{j}(\sigma) \,= \prod_{\beta\in\Delta_j}
s_b\!\left(\frac{\beta\,\sigma}{2\pi}\,+\,\mu_j\right),\qquad
\Delta_j=\left\{L_j,\,L_j-2,\,\cdots,\,2-L_j,\,-L_j\right\}.
\end{equation}
We assume ${\mu_j\in\BH_+}$ has a positive imaginary part to ensure
regularity of the integrand in 
\eqref{ZSU(2)} along the real axis.  Because the weights $\beta$ 
occur symmetrically in plus/minus pairs, ${G_j(-\sigma)=G_j(\sigma)}$
is an even function of $\sigma$.  The same is true for the integrand in
\eqref{ZSU(2)}, due to the underlying Weyl-invariance of
the matrix integral.

Let us perform a small check on the algebra
leading to \eqref{ZSU(2)} and \eqref{BigGFun}.  Observe from
\eqref{eq:MSineAsympta} that $G_j(\sigma)$ behaves asymptotically in
the massive regime ${|\mu_j|\to\infty}$ as 
\begin{equation}
\begin{aligned}
G_j(\sigma) \,&\underset{|\mu_j|\to\infty}{=}
\exp{\!\left[-\frac{i}{8\pi}\left(\sum_{\beta\in\Delta_j}\beta^2\right)\sigma^2\cdot\sgn(\mu_j)
  \,+\, O(\sigma)\right]},\\
&\underset{|\mu_j|\to\infty}{=}
\exp{\!\left[-\frac{i}{2\pi}\cdot\frac{L_j (L_j+1) (L_j+2)}{12}\,\sigma^2\cdot\sgn(\mu_j)
  \,+\, O(\sigma)\right]},
\end{aligned}
\end{equation}
where we evaluate the sum over squares explicitly in the second line.
Comparing to the polynomial formula for the $SU(2)$ Casimir in
\eqref{SU2Cas}, we see that the level $k$ effectively shifts by ${-\ha
  c_2(\lambda_j)\cdot\sgn(\mu_j)}$ when the massive multiplet is integrated-out, a
fact demonstrated for all gauge groups in Section
\ref{OneLoop}.

More relevant for the residue calculus is the behavior of the one-loop
determinant $G_j(\sigma)$ when ${|\sigma|\to\infty}$ with $\mu_j$
fixed.  For each pair of weights in the product,
the asymptotic expansion in \eqref{eq:MSineAsympta} implies
\begin{equation}
\begin{aligned}
&s_b\!\left(\frac{\beta\,\sigma}{2\pi}+\mu_j\right)\cdot
s_b\!\left(-\frac{\beta\,\sigma}{2\pi}+\mu_j\right)
\underset{|\sigma|\to\infty}{=}\\
&\qquad\exp{\!\left[-\left(\ha\,Q + i\,\mu_j\right)|\beta|\,\sigma\cdot\sgn(\Re(\sigma)) 
    + o(1)\right]}\,,\qquad Q \,\equiv\, b + b^{-1}\,.
\end{aligned}
\end{equation}
Up to an inessential rescaling of coefficients, the same expansion
appears for the toy model in \eqref{ProdSbz}.  Crucially, the leading Gaussian
terms in the respective copies of $s_b$ have cancelled.  Thus for
large $|\sigma|$, the logarithm of $G_j(\sigma)$ grows only linearly
with $\sigma$,
\begin{equation}\label{BigGFunII}
\begin{aligned}
G_j(\sigma) \,&\underset{|\sigma|\to\infty}{=}\,
\exp{\!\left[-\left(\ha\,Q + i\,\mu_j\right)\left(\ha\sum_{\beta\in\Delta_j}|\beta|\right)\sigma\cdot\sgn(\Re(\sigma))
+ O(1)\right]},\\
&\underset{|\sigma|\to\infty}{=}\,
\begin{cases}
\exp{\!\big[-\psi_1(\lambda_j)\, \sigma\,+\,O(1)\big]},\qquad\qquad
&\Re(\sigma)>0\,,\\
\exp{\!\big[+i\psi_1(\lambda_j)\,\sigma\,+\,O(1)\big]},\qquad\qquad
&\Re(\sigma)<0\,.
\end{cases}
\end{aligned}
\end{equation}
Like the analogous quantity in \eqref{Psi1Lamb}, the linear coefficient
$\psi_1(\lambda_j)$ is determined by the absolute sum over weights in
the first line of \eqref{BigGFunII},
\begin{equation}\label{Psi1LamSU2}
\psi_1(\lambda_j) \,=\, \left(\ha\,Q + i\,\mu_j\right)\times
\begin{cases}
\frac{1}{4}L_j\!\left(L_j+2\right)\,,\qquad\qquad &L_j\,\in\,2\Z\,,\\
\frac{1}{4}\left(L_j+1\right)^2,\qquad\qquad &L_j\,\in\,2\Z+1\,.
\end{cases}
\end{equation}
whose expression now depends upon whether $L_j$ is even or odd.

The complete one-loop factor in the integrand involves the
product over all $G_j(\sigma)$ for ${j=1,\ldots,n}$, 
\begin{equation}\label{ProdGj}
\prod_{j=1}^n G_j(\sigma) \underset{|\sigma|\to\infty}{=}\,
\begin{cases}
\exp{\!\big[-\psi_1(\Lambda)\, \sigma\,+\,O(1)\big]},\qquad\qquad
&\Re(\sigma)>0\,,\\
\exp{\!\big[+\psi_1(\Lambda)\,\sigma\,+\,O(1)\big]},\qquad\qquad
&\Re(\sigma)<0\,,
\end{cases}
\end{equation}
where we set ${\psi_1(\Lambda) = \psi_1(\lambda_1)
  +\cdots+\psi_1(\lambda_n)}$. So long as
${\Re(\psi_1(\Lambda))>0}$, the product in 
\eqref{ProdGj} decays exponentially when ${\sigma\to\pm\infty}$ along
the real axis.

With this discussion of $G_j(\sigma)$, let us consider the asymptotic behavior for the
$SU(2)$ integrand in \eqref{ZSU(2)} as ${|\sigma|\to\infty}$ in the complex
plane.  

If the level ${k>0}$ is non-zero, the Gaussian term
dominates the integrand along the generic ray in the complex plane,
and the naive residue calculus does not apply, for the reasons
discussed in Section \ref{eq:GGU(1)}.  Nevertheless, the integral over
$\sigma$ in \eqref{ZSU(2)} is manifestly convergent after the contour along $\R$
is rotated to ${\e{i\pi/4}\times\R}$, along which the Gaussian decays
rapidly.  In the process of rotating, the contour may well
pass through poles of the functions $G_j(\sigma)$, whose residues then
contribute to $Z_{S^3}$.  We provide an important example of this
phenomenon in Appendix \ref{TorusK}, where the residue calculus is necessary to
understand analytic features of the colored Jones polynomial.

If we wish to apply the naive residue calculus to evaluate
$Z_{S^3}$ itself, we are left to consider the degenerate case
${k=0}$, for which the classical action for the gauge field has only a
Yang-Mills term prior to localization.  In this case, there is no
guarantee that the integral over the real axis converges, and the
behavior of the integrand for ${|\sigma|\to\infty}$ depends upon a
competition between the exponentially-growing product of hyperbolic
sines in \eqref{ZSU(2)} and the exponentially-decaying product of double-sines in
\eqref{ProdGj}.  A number of authors
\cite{Aharony:2013dha,Benini:2011mf,Kapustin:2010mh,Safdi:2012re,Willett:2011gp}
have noted previously that the
matrix integral for $Z_{S^3}$ does not always converge if ${k=0}$.

Briefly, convergence of the integral
along the real axis when ${k=0}$ requires the inequality
\begin{flalign}\label{SU2Bound1}
\fbox{convergence criterion for $k=0$} \qquad\qquad &
\Re\!\big(\psi_1(\Lambda)\big) \,\ge\, Q\,,&
\end{flalign}
or by the definition in \eqref{Psi1LamSU2},
\begin{equation}\label{SU2Bound2}
\sum_{j=1}^n\left(1 - \RR_j\right)\cdot\left\{
\begin{matrix}
\frac{1}{4}L_j\!\left(L_j+2\right)\\[1ex]
\frac{1}{4}\left(L_j+1\right)^2
\end{matrix}\right\}\,\ge\, 2\,.\qquad\qquad
\begin{matrix}
 &\big[L_j\,\in\,2\Z\big]\\[1ex]
 &\big[L_j\,\in\,2\Z+1\big]
\end{matrix}
\end{equation}
Here we assume ${Q\in\R_+}$ is real, and we set ${\mu_j \equiv (\mu_j)_\R \,+\, \frac{i}{2} \, Q \,
  \RR_j}$, where ${\RR_j\in\R}$ is the
R-charge of the corresponding chiral field.  Also in
\eqref{SU2Bound2}, the summands on top and
bottom in brackets apply when $L_j$ is
respectively even or odd.

The condition in \eqref{SU2Bound2} can be read in two ways.  For a fixed
$SU(2)$ representation $\Lambda$, convergence of the matrix integral
imposes an upper-bound on the vector ${\RR\in\R^n}$ of R-charges.
Alternatively, since $\RR$ is typically bounded from below by
unitarity constraints
on dimensions of gauge-invariant chiral operators,\footnote{Recall that the dimension $D$ of any chiral
  operator in a unitary ${\CN=2}$ superconformal field
  theory is bounded from below by ${1/2}$, and for such operators
  ${D=\RR}$ is fixed by the R-charge.} convergence
of the matrix integral requires that 
the representation $\Lambda$ be sufficiently large, as measured by
the set of highest-weights.

A simple example occurs when ${\Lambda = {\bf 3}^{N_f}}$ is the sum 
of $N_f$ copies of the adjoint representation.  Anomaly-cancellation
with ${k=0}$ allows $N_f$ to be any positive integer.  Given the 
the underlying $SU(N_f)$ flavor symmetry, the R-charge $\RR$ must
be the same for all summands in $\Lambda$, so the convergence
criterion in \eqref{SU2Bound2} becomes (with ${L_j=2}$)
\begin{equation}\label{SU2Bound3}
N_f \left(1\,-\,\RR\right) \ge\,1\,.
\end{equation}
Automatically ${\RR < 1}$ by \eqref{SU2Bound3}.  Also trivially,
${N_f \ge 1}$ is a necessary condition for convergence.

Supersymmetric QCD with $N_f$ flavors of quarks, meaning
${\Lambda={\bf 2}^{2 N_f}}$, provides another example.
Anomaly-cancellation with ${k=0}$ requires the number of quarks to be 
even, so $N_f$ is an integer.  The global $SU(2 N_f)$ flavor
symmetry implies that $\RR$ is the same for each quark, so the
convergence criterion in \eqref{SU2Bound2} (with ${L_j=1}$) is
identical to \eqref{SU2Bound3}, with ${\RR < 1}$ and ${N_f \ge
  1}$.  The existence of a chiral meson operator with R-charge
${2\RR}$ provides a unitarity bound ${\RR \ge 1/4}$, and the
existence of a chiral monopole operator \cite{Aharony:1997bx} with
R-charge ${N_f (1-\RR) - 2}$ provides another unitarity bound ${N_f
  (1-\RR) \ge 5/2}$.  The meson unitary bound, in combination with
\eqref{SU2Bound3}, requires ${N_f \ge 2}$ as a necessary condition for
convergence.  The monopole unitarity bound is clearly compatible with,
but slightly stronger than, the convergence criterion.

Later in Section \ref{SUSYBreak} we perform a similar analysis for SQCD with
arbitrary gauge group of type SU, Sp, and SO.  There we find a
curious relation with the physical vacuum structure of
the same theory on $\R^{1,2}$. 

\paragraph{Residue Calculus for $SU(2)$.}

Assuming the convergence criterion at level zero is satisfied, we next
ask whether there exists a suitable integration contour
${\Gamma\subset\C}$ which allows $Z_{S^3}$ to be evaluated as a sum
over residues.  

This question again concerns the asymptotic behavior for large
${|\sigma|\gg 1}$ of the integrand.  In norm,
\begin{equation}\label{NormSU2}
\begin{aligned}
&\left|\sinh\!\left(b\,\sigma\right)\sinh\!\left(\sigma/b\right) \cdot
\prod_{j=1}^n \,
G_{j}(\sigma)\right|\underset{|\sigma|\to\infty}{=}\,\\
&\qquad\begin{cases}
\exp{\!\Big[-\big(\Re(\psi_1)-Q\big)\Re(\sigma) +
    \Im(\psi_1)\Im(\sigma)\Big]},\qquad\qquad &\Re(\sigma)>0\,,\\[1 ex]
\exp{\!\Big[+\big(\Re(\psi_1)-Q\big)\Re(\sigma) -
    \Im(\psi_1)\Im(\sigma)\Big]},\qquad\qquad &\Re(\sigma)<0\,,
\end{cases}
\end{aligned}
\end{equation}
where
\begin{equation}\label{SU2psi1}
\psi_1\equiv\psi_1(\Lambda) \,=\, \sum_{j=1}^n \left(\frac{Q}{2} + i\,\mu_j\right)\left\{
\begin{matrix}
\frac{1}{4}L_j\!\left(L_j+2\right)\\[1ex]
\frac{1}{4}\left(L_j+1\right)^2
\end{matrix}\right\}\,.
\end{equation}
As in \eqref{SU2Bound2}, the upper and lower brackets in the
expression \eqref{SU2psi1} for $\psi_1$ indicate the respective values
when $L_j$ is an even or an odd integer.  The convergence criterion \eqref{SU2Bound1} follows from
\eqref{NormSU2} with ${\sigma\in\R}$.  When we allow $\sigma$ to have
an imaginary part as well, we see from \eqref{NormSU2} that the
integrand decays consistently in either the upper or the lower
half-planes only when ${\Im(\psi_1)=0}$.  If ${\Im(\psi_1)\neq 0}$, we
are effectively stuck in the Gaussian situation shown in
Figure \ref{fig:Conver}.

Consequently, the
residue calculus for the partition function $Z_{S^3}$ is only applicable
in the special case ${k=0}$ and ${\Im(\psi_1(\Lambda))=0}$.  As
an instance of the latter condition, if the irreducible summands of ${\Lambda=[\lambda_1]\oplus\cdots\oplus[\lambda_n]}$ are identical, then 
\begin{equation}\label{CPMasses}
\Im(\psi_1(\Lambda)) \,=\,
0\qquad\underset{\lambda_1=\lambda_2=\cdots=\lambda_n}{\Longleftrightarrow}\qquad
\sum_{j=1}^n \left(\mu_j\right)_\R\,=\,0\,.
\end{equation}
On the right, $\left(\mu_j\right)_\R$ indicates the real part of
$\mu_j$, when we allow $\mu_j$ to be continued analytically to the
complex-plane.
The sum condition in \eqref{CPMasses} holds automatically in SQCD with
CP-invariant\footnote{For SQCD
  with ${n=2N_{\rm f}}$, CP-invariance requires the real masses $\left(\mu_j\right)_\R$ to appear in
  cancelling plus/minus pairs for quarks and anti-quarks.} real
masses, so it is not unreasonable to impose.  More generally, as for the
toy model in Section \ref{PropDSine}, we can analytically continue the
complexified masses $\mu_j$ to
take purely imaginary values, so that ${\Im(\psi_1(\Lambda))=0}$ holds trivially.

When ${\Im(\psi_1(\Lambda))=0}$, the partition function $Z_{S^3}$ can
be evaluated by closing the integration contour ${\Gamma\subset\C}$ in
either the lower or the upper half-plane.  By convention, we close $\Gamma$ in
the lower half-plane, after which the residue calculus proceeds much like
the $U(1)$ case in Section \ref{eq:GGU(1)}.  

Briefly, poles of the function
$G_j(\sigma)$ in \eqref{BigGFun} lie on the divisor
\begin{equation}\label{DjSU2}
D_j \,=
\bigcup_{\beta\in\Delta_j}\bigcup_{u,v=0}^\infty\left\{\sigma = \frac{2\pi}{\beta}\left(-i\,u\,b
    \,-\, i\,v\,b^{-1}-\mu_j\right)\right\},
\end{equation}
where ${\Delta_j}$ is the set of weights in \eqref{BigGFun}.  For
generic $\mu_j$ obeying the constraint in \eqref{CPMasses}, all 
poles are simple, and when the integration contour is closed in the lower
half-plane, only the poles with positive weight ${\beta>0}$ 
contribute to the sum over residues.  

Thus
\begin{equation}\label{ZSUMSU2}
Z_{S^3} \,=\,-2 i \sum_{j=1}^n \sum_{{\beta\in\Delta_j}\atop{\beta>0}} \sum_{u,v=0}^\infty
\RZ^{\,j,\beta}_{u,v}\,,
\end{equation}
where by analogy to \eqref{AbbResII} the residue is a product 
\begin{equation}\label{ResSU2}
\RZ^{\,j,\beta}_{u,v} \,=\,
\frac{i}{\beta}\cdot\exp{\!\left[i\,\Theta^{\,j,\beta}_{r,s}\right]}\cdot\RW^{\,j,\beta}_{M,r,s}(\Rq,\Rx)\cdot\wt\RW^{\,j,\beta}_{N,r,s}(\wt\Rq,\wt\Rx)\,.
\end{equation}
Similar to the abelian case, we decompose the labels ${u,v\in\Z_+}$
appearing in \eqref{DjSU2} into congruence classes modulo $\beta$,
\begin{equation}
\begin{aligned}
u \,&=\, M\,\beta \,+\, r\,,\qquad\qquad M,N\ge 0\,,\\
v \,&=\, N\,\beta \,+\, s\,,\qquad\qquad 0 \le
r,s < \beta\,,
\end{aligned}
\end{equation}
The phase $\Theta^{\,j,\beta}_{r,s}$ and the functions
$\RW^{\,j,\beta}_{M,r,s}(\Rq,\Rx)$,
$\wt\RW^{\,j,\beta}_{N,r,s}(\wt\Rq,\wt\Rx)$ are also determined in
the same manner as the corresponding expressions in Section
\ref{eq:GGU(1)}.  We state the result:
\begin{equation}\label{BigThSU2}
\begin{aligned}
&\exp{\!\left(i\,\Theta^{\,j,\beta}_{r,s}\right)} \,=\,
\exp{\!\left[2\pi i\,c_2(\Lambda)\,\frac{r s}{\beta^2}\right]}\,\times\\
&\qquad\times\,\exp{\!\left[\sum_{\ell=1}^n\sum_{\gamma\in\Delta_\ell}\frac{i
      \pi}{4}\left(1 - \frac{2}{\beta^2}\left(\beta\,\mu_\ell - \gamma\,\mu_j\right)^2\right)\sgn\!\left(\beta\,\mu_\ell - \gamma\,\mu_j\right)\right]},
\end{aligned}
\end{equation}
where ${\sgn(0)\equiv 0}$ to account for the term with
${\ell=j}$ and ${\gamma=\beta}$ in the double sum.  Also,
\begin{equation}\label{BigWSU2}
\begin{aligned}
&\RW^{\,j,\beta}_{M,r,s}(\Rq,\Rx) = \ha\,\e{\!2\pi
      i\,c_2(\Lambda) M s/\beta}\cdot
    \exp{\!\left[\frac{i\pi}{2}\!\left(\sum_{\ell=1}^n\sum_{\gamma\in\Delta_\ell}\!\left(M+\frac{r}{\beta}\right)\!\gamma
        \sgn\!\left(\beta \mu_\ell-\gamma
          \mu_j\right)+M\beta+r\right)\right]}\,\times\\
&\qquad\times\Big(\e{-2\pi i s/\beta}\,\Rq^{-(M+(r/\beta))}\,\Rx_j^{-1/\beta}\,-\,\e{2\pi i s/\beta}\,\Rq^{M+(r/\beta)}\,\Rx_j^{1/\beta}\Big)\cdot \left[\frac{\Rq^{(M \beta+r)(M
        \beta+r+1)/4}}{\big(\Rq;\Rq\big)_{\!\left(M\beta+r\right)}}\right]\times\\
&\qquad\qquad\times\prod_{\ell=1}^n\prod_{\gamma\in\Delta_\ell}
\SF_b^{\sgn(\beta \mu_\ell - \gamma \mu_j)}\!\left(\rz_{j,\ell};\Rq\right),
\end{aligned}
\end{equation}
where 
\begin{equation}
\rz_{j,\ell} \,=\, \frac{1}{\beta}(\beta \mu_\ell - \gamma \mu_j)
\,-\, i \left(M + \frac{r}{\beta}\right)\gamma \, b \,-\, i
\left(N \,+\, \frac{s}{\beta}\right)\gamma \, b^{-1}\,.
\end{equation}
Again by convention, ${\SF_b^{\sgn(0)}\equiv 1}$.  The expression for
$\wt\RW^{\,j,\beta}_{N,r,s}(\wt\Rq,\wt\Rx)$ is entirely similar, after the
exchanges ${M\leftrightarrow N}$ and ${r\leftrightarrow s}$.

What are the important features of the formulas in \eqref{ResSU2},
\eqref{BigThSU2}, and \eqref{BigWSU2}?

First, unlike \eqref{AbbResII}, no term involving the product $MN$
obstructs factorization for $\RZ^{\,j,\beta}_{u,v}$ if the level-zero
anomaly-cancellation condition ${c_2(\Lambda) \equiv 0}$ mod $2$ is
violated.  Instead, during the
calculation which leads to \eqref{ResSU2}, one 
meets only the trivial phase ${(-1)^{2 M N c_2(\Lambda)}\equiv 1}$, and
factorization {\sl per se} imposes no constraint on the value of
${c_2(\Lambda)}$.  As we discuss at the end of Section \ref{LatticeSums}, this phenomenon is not generic
and is related to the coincidence among Lie algebras that the coroot
lattice of $SU(2)$ is even, ie.~${(\hat h, \hat h)=2}$.

Second, the partition function $Z_{S^3}$ at level ${k=0}$ can be written in the
fully-factorized form 
\begin{equation}
Z_{S^3} \,=\, \sum_{j=1}^n \sum_{{\beta\in\Delta_j}\atop{\beta>0}}
\sum_{r,s=0}^{\beta-1}\,\frac{2\,\exp{\!\left(i\,\Theta^{\,j,\beta}_{r,s}\right)}}{\beta}
\, \RB^{\,j,\beta}_{r,s}(\Rq,\Rx)\,\wt\RB^{\,j,\beta}_{r,s}(\wt\Rq,\wt\Rx)\,,
\end{equation}
with blocks 
\begin{equation}\label{BlockSerSU2}
\RB^{\,j,\beta}_{r,s} \,=\, \sum_{M=0}^\infty
\,\RW^{\,j,\beta}_{M,r,s}\,,\qquad\qquad \wt\RB^{\,j,\beta}_{r,s} \,=\,
\sum_{N=0}^\infty\,{\wt\RW}^{\,j,\beta}_{N,r,s}\,.
\end{equation}
The $SU(2)$ blocks are labelled by quadruples in the set 
\begin{equation}
\CI \,=\,
\Big\{\left(j,\beta,r,s\right)\,\Big|\,\beta\in\Delta_j,\beta>0\,,\,0\le 
r,s<\beta\Big\}\,,
\end{equation}
comprising an effective $U(1)$ block for each positive weight in the
representation $[\lambda_j]$ associated to each chiral matter multiplet.  The
bilinear form $\RG_{m,n}$ in the Factorization Conjecture
\eqref{FACTOR} remains diagonal.

\section{Preliminary Analysis at Higher-Rank}\label{Residue}

In the remainder of the paper, we extend the analysis from Section
\ref{RankOne} to gauge groups of higher rank.  The present Section
\ref{Residue} is an initial grab-bag in which we broadly explain our
general strategy and then establish preliminary technical results,
some of
independent interest, about the sphere partition function $Z_{S^3}$.
A precise summary of Section \ref{Residue} follows.

In Section \ref{Chirality}, we recall by way of illustration how
the chirality bound \eqref{ChiralK} on the Chern-Simons level $k$
reappears for the unitary group ${U(N)}$.  

To evade the bound, in Section \ref{LevelZero} we reduce the analysis
for arbitrary values of $k$ to the degenerate case ${k=0}$ by integrating-in
auxiliary chiral matter, compatible with the one-loop shift in $k$
from Section \ref{OneLoop}.  As we explain, the reduction to level
${k=0}$ relies upon the existence of a suitable ``fundamental''
representation for the gauge group and so only works for simple gauge
groups of classical matrix type $\mathrm{ABCD}$, along with the
exceptional Lie group $G_2$. We also examine in detail the decoupling
of auxiliary, massive matter multiplets from the ultraviolet
partition function $Z^{\rm uv}_{S^3}$, complementing prior
observations in \cite{Aharony:2013dha}.

In Section \ref{SUSYBreak} we specialize to gauge theories with
Chern-Simons level ${k=0}$.  Convergence of the Coulomb-branch
integral \eqref{eq:GrandPartitionFunctionInt} is far from assured in
this situation and depends only upon the asymptotic, semiclassical
behavior of the integrand.  We provide sufficient criteria for general
pairs $(G,\Lambda)$ to ensure convergence and hence existence of
$Z_{S^3}$.  These criteria are phrased in terms of a Weyl-invariant
$L^1$-norm $||\,\cdot\,||_V$ on the Cartan subalgebra $\h$.  The
norm $||\,\cdot\,||_V$ is distinct from the Killing form and is
labelled by a (non-trivial) irreducible representation
${V\in\Rep(G)}$.  In the process, we explore a few geometric
properties of the norm
$||\,\cdot\,||_V$.

Finally in Section \ref{SusyQCD} we apply the convergence criteria to
supersymmetric QCD with gauge groups of type SU, Sp, and SO.  As
well-known, supersymmetry is broken and the vacuum
destabilized in SQCD on $\R^{1,2}$ when the number $N_{\rm f}$ of quark flavors
is sufficiently small relative to the rank of the gauge group
\cite{Affleck:1982as,Aharony:1997bx,Aharony:1997gp,Aharony:2011ci,Aharony:2013kma,Benini:2011mf,Karch:1997ux}.
We show that the Coulomb-branch convergence criterion
for SQCD on $S^3$ reproduces the same critical value of $N_{\rm f}$ which
determines supersymmetry-breaking on $\R^3$.  The
agreement is especially striking for gauge group $SO$, as our
Lie algebra calculation involving the norm $||\,\cdot\,||_V$ is not
even guaranteed to produce a result which is analytic in $N_{\rm f}$!  We do not have a theoretical understanding
of this coincidence, but it seems worthy of further investigation.

\subsection{Chirality Conditions for $U(N)$}\label{Chirality}

As emphasized in \cite{Benini:2013yva}, the residue calculus can be
applied to the matrix integral for $Z_{S^3}$ only when the
magnitude of the Chern-Simons level $k$ is sufficiently small.  The
precise bound on $k$ depends upon the gauge group $G$ and the matter
representation $\Lambda$, and it presents a significant
obstruction to any direct attempt to prove the Factorization
Conjecture in general.  We have already discussed the obstruction
in the abelian case ${G=U(1)}$, for which the relevant chirality bound on $k$ 
appears in \eqref{ChiralK}, but nothing is special about this case.
As a warmup, we quickly recapitulate the bound for the higher-rank,
non-abelian example ${G=U(N)}$.

For convenience, we specialize the representation $\Lambda$ to be
the direct sum of $a_+$ copies of the defining fundamental
representation ${\bf N}$ of $U(N)$ and
$a_-$ copies of the dual anti-fundamental representation
$\overline{\bf N}$,
\begin{equation}\label{FundUN}
\Lambda \,=\, {\bf N}^{a_+} \!\oplus \overline{\bf
  N}{}^{a_-}\,,
\end{equation}
for some multiplicities ${a_\pm\ge 0}$.  Cancellation of the global gauge 
anomaly requires 
\begin{equation}
k \,-\, \ha (a_+ + a_-) \,\in\,\Z\,.
\end{equation}
This quantization condition on $k$ is equivalent to 
\begin{equation}
k \,\equiv\, \psi_2(\Lambda) \,\mod\,\Z\,,
\end{equation}
where $\psi_2(\Lambda)$ is the net chirality
\begin{equation}
\qquad \psi_2(\Lambda) \,=\,
\ha \left(a_+ - a_-\right)\,.
\end{equation}
The formula for $\psi_2$ should be compared with the previous, abelian definition in \eqref{PsiTwo}.  For the
non-chiral theory with ${a_+ = a_-}$, the parity anomaly is
absent, and $k$ obeys the conventional integral quantization.

An element of the Cartan subalgebra of $U(N)$ is a diagonal
matrix
\begin{equation}
\sigma \,=\, i
\diag\!\left(\sigma_1,\,\sigma_2,\,\cdots,\,\sigma_N\right),
\end{equation} 
with ${\sigma_m\in\R}$ for ${m=1,\ldots,N}$.  In these coordinates, the
partition function becomes
\begin{equation}\label{UNPart}
\begin{aligned}
&Z_{S^3} \,=\, \frac{1}{N!}\int_{\R^N}\!\frac{d^N\!\sigma}{(2\pi)^N}
\;\exp{\!\left[\frac{i\,k}{4\pi}\!\left(\sigma_1^2 + \cdots +
      \sigma_N^2\right) \,+\, i\,\xi\left(\sigma_1 + \cdots + \sigma_N\right)\right]}\,\times\\
&\times\prod_{m < n}\left[4\,\sinh\!\left(\frac{b\left(\sigma_m -
        \sigma_n\right)}{2}\right)\sinh\!\left(\frac{\sigma_m-\sigma_n}{2b}\right)\right]\cdot
\prod_{j=1}^{a_+}G_j^{+}(\sigma) \cdot \prod_{\ell=1}^{a_-} G_\ell^{-}(\sigma).
\end{aligned}
\end{equation}
In the second line of \eqref{UNPart}, the positive
roots ${\alpha\in\Delta_+}$ of $U(N)$ are identified with the
differences ${\sigma_m  - \sigma_n}$ for pairs of indices ${m < n}$ in
the range ${m,n = 1,\ldots,N}$.  Also, each matter determinant
$G_j^+(\sigma)$ is the product over fundamental weights
\begin{equation}
G_j^+(\sigma) \,=\, \prod_{m=1}^N s_b\!\left(\frac{\sigma_m}{2\pi} +
  \mu_j\right),\qquad j\,=\,1,\ldots, a_+\,,
\end{equation}
and similarly for the anti-fundamental matter determinant
\begin{equation}
G_\ell^{-}(\sigma) \,\equiv\, G_\ell^{+}(-\sigma) \,=\, \prod_{m=1}^N s_b\!\left(-\frac{\sigma_m}{2\pi} +
  \mu_\ell\right),\qquad \ell\,=\,1,\ldots, a_-\,.
\end{equation}

The $SU(N)$ partition function can be recovered from the
$U(N)$ formula in \eqref{UNPart} by performing a further integral over the FI-parameter $\xi$ to impose the trace constraint ${\sigma_1 + \cdots +
  \sigma_N = 0}$.  Here we use the delta-function identity 
\begin{equation}
\int_\R \frac{d\xi}{2\pi} \exp{\!\big[i\,\xi\left(\sigma_1 + \cdots
      + \sigma_N\right)\big]} \,=\, \delta(\sigma_1 + \cdots
      + \sigma_N)\,.
\end{equation}

As each eigenvalue ${\sigma_m\to\pm\infty}$, the asymptotics in 
\eqref{eq:MSineAsympta} imply that the unitary matter determinant
approaches 
\begin{equation}%Correct eqn
G_j^+(\sigma) \,\underset{\sigma_m\to\pm\infty}{=}\,
\exp{\!\left[-\frac{i}{8\pi}\sum_{m=1}^N \sigma_m^2 \sgn(\Re(\sigma_m))\,+\,O(\sigma)\right]}\,,
\end{equation}
and dually under reflection of $\sigma$,
\begin{equation}
G_\ell^-(\sigma) \,\underset{\sigma_m\to\pm\infty}{=}\,
\exp{\!\left[+\frac{i}{8\pi}\sum_{m=1}^N \sigma_m^2 \sgn(\Re(\sigma_m))\,+\,O(\sigma)\right]}\,.
\end{equation}
Hence the net one-loop matter contribution in \eqref{UNPart} behaves for
large $\sigma$ as 
\begin{equation}\label{UNGs}
\prod_{j=1}^{a_+}G_j^{+}(\sigma) \cdot \prod_{\ell=1}^{a_-}
G_\ell^{-}(\sigma) \,\underset{\sigma_m\to\pm\infty}{=}\,
\exp{\!\left[-\frac{i\,\psi_2(\Lambda)}{4\pi}\sum_{m=1}^N \sigma_m^2 \sgn(\Re(\sigma_m))\,+\,O(\sigma)\right]}\,.
\end{equation}
Note that the asymptotic dependence on each element ${\sigma_m}$ factorizes in \eqref{UNGs}.  The same factorization
applies to the oscillatory Gaussian term in the first line of
\eqref{UNPart}.  Consequently, the entire integrand
${\bf I}(\sigma)$
factorizes asymptotically to leading order,
\begin{equation}
{\bf I}(\sigma) \,\underset{\sigma_m\to\pm\infty}{=}\, \prod_{m=1}^N
\exp{\!\left[\frac{i}{4\pi}
    \big(k - \psi_2(\Lambda)\cdot\sgn(\Re(\sigma_m))\big)\,\sigma_m^2 \,+\,
  O(\sigma)\right]}\,.
\end{equation}

We are now in the same situation discussed in Section \ref{eq:GGU(1)}.
When the strict chirality bound  \eqref{ChiralK} is satisfied, 
\begin{equation}\label{ChiralKUN} 
k < |\psi_2(\Lambda)|\,,
\end{equation}
the integrand ${\bf I}(\sigma)$ decays in the product
of half-planes
\begin{equation}\label{HalfP}
\BH_\pm^N \,=\, \BH_\pm \times \cdots \times \BH_\pm
  \subset \C^N\,,\qquad \pm\Im(z)>0 \hbox{ for } z \in \BH_\pm\,,
\end{equation}
where the imaginary part ${\Im(\sigma_m)}$, ${m=1,\ldots,N}$, is
respectively positive or negative.  Specifically, if 
${\psi_2(\Lambda)>0}$ the integrand decays 
for ${\sigma \in \BH_-^N}$ in the negative corner of $\C^N$, and vice versa if
${\psi_2(\Lambda)<0}$.  We depict the domain of convergence for
${N=2}$ and ${0 \le k < \psi_2(\Lambda)}$  in Figure \ref{Corner}.  Here the domain $\BH_{-}^2$ occupies the third quadrant of the plane spanned
by the imaginary parts $\Im(\sigma_1)$ and $\Im(\sigma_2)$, and 
the origin is identified with the real integration slice ${\R^2
  \subset \C^2}$ for \eqref{UNPart}.

\iffigs
\begin{figure}[t]
\centering
{
   \begin{tikzpicture}[x=.5cm,y=.5cm] % Changes size of whole picture
	\begin{scope}
		\fill[blue!30] (-5,-5) rectangle (0cm,0cm);
	\end{scope}

	\draw [<->, thick, black](-5.5,0)--(5.5,0) node [right, black] {$\text{Im}(\sigma_1)$};
	\draw [<->, thick, black](0,-5.5)--(0,5.5) node [right, black] {$\text{Im}(\sigma_2)$};
	\draw [->, black] (-3.8,-3.8) --  (-4.6,-4.6);
	\fill[black] (0,0) circle(.15);
	\node at (.9,-.7) {$\mathbb{R}^2$};
    \end{tikzpicture}
}
\caption{For ${0 \le k < \psi_2(\Lambda)}$, the domain ${\BH^2_{-} \subset
    \C^2}$ in which the matrix integrand ${\bf 
    I}(\sigma_1,\sigma_2)$ decays asymptotically along rays.}\label{Corner}
\end{figure}
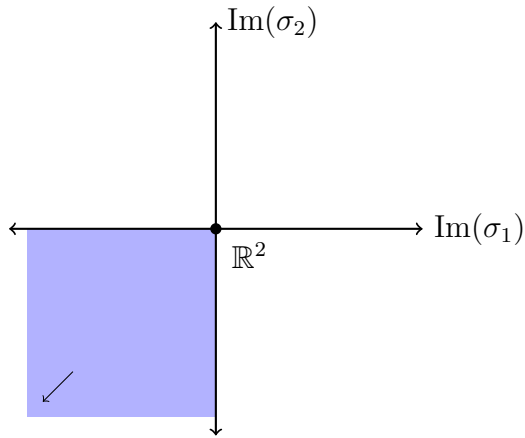
\fi

Thus when the representation $\Lambda$ is ``sufficiently'' chiral relative
to the level $k$ as in \eqref{ChiralKUN}, the residue calculus 
can be performed iteratively for each factor in $\BH^N_\pm$.  But if ${k >
  |\psi_2(\Lambda)|}$, the integrand ${\bf
  I}(\sigma)$ behaves asymptotically like a product of the Gaussians
in Figure \ref{fig:Conver}.  In that case, the residue calculus fails
to apply, at least in any obvious way.

Finally, as we discussed for abelian theories in Section \ref{eq:GGU(1)}, the
residue calculus may or may not apply in the marginal case ${k =
  |\psi_2(\Lambda)|}$, where subleading terms in the asymptotic expansion
\eqref{UNGs} become important.  Though marginal, this case is
natural and cannot be forgotten, as it occurs when ${k=0}$ and $\Lambda$ is a real 
representation of $U(N)$, with ${a_+ = a_-}$ in \eqref{FundUN}.  We
analyze the marginal case in detail for a variety of gauge groups and
matter representations in Section \ref{SUSYBreak}.

\subsection{Evading the Chirality Bound}\label{LevelZero}

Just as chiral matter can be integrated-out, so too can chiral matter
be integrated-in, through the reverse of the process in Section
\ref{OneLoop}.  In this fashion, we will evade the chirality bound.

Briefly, we start with chiral matter in a representation $\Lambda$ of
the gauge group $G$ at Chern-Simons level ${k\ge 0}$.  To
integrate-in auxiliary chiral matter in a representation $\Lambda'$ of $G$,
we consider a new theory with matter in the total representation
${\Lambda_{\rm uv} = \Lambda \oplus \Lambda'}$ and with large real masses ${|\mu'| \gg
  |\mu|}$.  Provided the new Chern-Simons level $k_{\rm uv}$ satisfies the
one-loop relation\footnote{Recall that ${c_2(\lambda_j') =
    (\lambda_j')^2}$ in the $U(1)$ case.} in \eqref{KEff2},
\begin{equation}\label{UVlevel}
k_{\rm uv} \,=\, k \,+\, \ha \sum_{j=1}^n c_2(\lambda_j') \,\sgn(\mu_j')\,,
\end{equation}
this new ultraviolet theory reduces to our original theory in
the decoupling limit ${|\mu'|\to\infty}$, with $\mu$ fixed.
Otherwise, for large but finite values of the auxiliary mass $\mu'$, the
ultraviolet theory with chiral matter in the representation
$\Lambda_{\rm uv}$ at level $k_{\rm uv}$ can be regarded as a
deformation of the original theory with matter in the representation
$\Lambda$ at level $k$.

If the sign of each auxiliary mass $\mu_j'$ is negative in
\eqref{UVlevel}, then $k_{\rm uv}$ decreases relative to the original
level $k$ (as ${c_2\ge 0}$ is always positive) after the auxiliary
matter $\Lambda'$ is introduced.  As a result, even when
the original representation $\Lambda$ is not ``sufficiently'' chiral
relative to the level $k$, the ultraviolet representation
$\Lambda_{\rm uv}$ may be sufficiently chiral relative to $k_{\rm
  uv}$, in the sense that the residue calculus applies to the
partition function $Z_{S^3}^{\rm uv}$ of the ultraviolet theory.  Whether or
not $\Lambda_{\rm uv}$ is sufficiently chiral will depend, of course,
on both the level $k$ and the choice of auxiliary
matter $\Lambda'$.

For instance, with gauge group ${G=U(N)}$ and
fundamental/anti-fundamental matter as in \eqref{FundUN}, the
integer ${k - \psi_2(\Lambda)\in\Z}$ decreases by one unit for each extra
fundamental matter multiplet with negative real
mass ${\mu_j'<0}$ which is 
added to the original theory.\footnote{If we integrate-in fundamental matter with {\sl positive} real mass ${\mu_j'>0}$,
  the difference ${k - \psi_2(\Lambda)}$ does not change.}
Alternatively, ${k + \psi_2(\Lambda)\in\Z}$ decreases by one unit
for each extra anti-fundamental matter multiplet with negative real
mass which is added.  Following this process, by integrating-in
fundamental or respectively anti-fundamental matter with negative real mass to the $U(N)$ theory, we can always arrange for the ultraviolet matter content
$\Lambda_{\rm uv}$ to be ``sufficiently-chiral'' relative to $k_{\rm uv}$,
in the sense that the chirality bound ${k_{\rm uv} \le
  |\psi_2(\Lambda_{\rm uv})|}$ in \eqref{ChiralKUN} is satisfied.
Here we include the marginal case ${k_{\rm uv} = |\psi_2(\Lambda_{\rm
    uv})|}$ within our bound, as it will play an essential technical role later.

Once the chirality bound is obeyed in the ultraviolet $U(N)$ theory, we
apply the residue calculus as in Section \ref{RankOne} to demonstrate factorization of the ultraviolet partition function 
\begin{equation}\label{uvFactor}
Z_{S^3}^{\rm uv} \,= \sum_{r, s\,\in\,\CI^{\rm uv}} \RG_{
  r  s}^{\rm uv}\,
\RB^{r}(\Rq,\Rx,\Rx')\,\wt\RB^{s}(\wt\Rq,\wt\Rx,\wt\Rx')\,.
\end{equation}
Here the ultraviolet indices ${r,s\in\CI^{\rm uv}}$ label blocks associated to
both the auxiliary and the original matter fields.  

Finally, to recover factorization for the partition function of the
original, low-energy theory, we take the decoupling limit  ${\mu'\to-\infty}$
on both sides of \eqref{uvFactor}.  In this limit, the auxiliary fugacity
variables ${\left(\Rx',\wt\Rx'\right)}$ vanish, and $\RG_{r s}^{\rm
  uv}$ degenerates on auxiliary blocks.

By factorizing the ultraviolet theory first and then flowing to
the infrared, we bypass the
technical obstruction present for ``insufficiently-chiral'' matter
content in the original $U(N)$ theory.  A schematic diagram
summarizing the argument appears in Figure \ref{Schematic}.

\iffigs
\begin{figure}[t]
\centering
	\begin{tikzpicture}[x=1.5cm,y=1.5cm]  % Overall scale of the picture

	%    Set Up Constants

	\def \tztext {$A = B+C$}			% The Equation

	\def \tzx {2.6}				% x-location of center of each box offset from origin
	\def \tzy {1.5}				% y-locataion of center of each box offset from origin
	\def \tzh {.8}				% Box height (x2)
	\def \tzw {1.6}				% Box width (x2)
	\def \tzA {.3}				% Arrow offset from side of box
	\coordinate (V) at (0,.3);			% Verticle offset of text in box from center (both above and below)

	\coordinate (A) at (-\tzx,\tzy);		% Top Left Center
	\coordinate (B) at (\tzx,\tzy);		% Top Right Center
	\coordinate (C) at (-\tzx,-\tzy);		% Bottom Left Center
	\coordinate (D) at (\tzx,-\tzy);		% Bottom Right Center

	%    Draw Boxes

	\foreach \a in {A,B,C,D}
	{
		\draw [thick, blue,rounded corners=10, draw] ($ (\a) + (\tzw,\tzh) $) -- ($ (\a) + (\tzw,-\tzh) $)-- ($ (\a) + (-\tzw,-\tzh) $) --  ($ (\a) + (-\tzw,\tzh) $) --  cycle;
    	}

	%    Box Text

	% Top Left Box Text
	\node at ($ (A) + (V) $)  {$\left(G,k_{\rm uv},\Lambda_{\rm
              uv} \simeq \Lambda\oplus\Lambda'\right)$};
	\node at ($ (A) - (V) $)  {Sufficiently-chiral};

	% Top Right Box Text
	\node at ($ (B) $)  {$\begin{aligned}&Z_{S^3}^{\rm uv}\,=\sum_{r,s\,\in\,\CI^{\rm uv}}\\
&\RG_{r s}^{\rm uv}\,\RB^{r}(\Rq,\Rx,\Rx')\,\wt\RB^{s}(\wt\Rq,\wt\Rx,\wt\Rx')\end{aligned}$};

	% Bottom Left Box Text
	\node at ($ (C) + (V) $)  {$\left(G,k,\Lambda\right)$};
	\node at ($ (C) - (V) $)  {Insufficiently-chiral};

	% Bottom Right Box Text
	\node at ($ (D) $)  {$\begin{aligned}&Z_{S^3}\,=\sum_{m,n\,\in\,\CI}\\&\quad\RG_{m n}\,\RB^m(\Rq,\Rx)\,\wt\RB^n(\wt\Rq,\wt\Rx)\end{aligned}$};

	%    Draw Arrows and Arrow Text

	\draw [->, very thick, black!30!green] ($ (C) + (0,\tzh) + (0,\tzA) $) 
		-- node [black,left] {\begin{tabular}{r} Integrate-in \\ auxiliary matter $\Lambda'$\end{tabular}} ($ (A) + (0,-\tzh) + (0,-\tzA) $);

	\draw [->, very thick, black!30!green] ($ (A) + (\tzw,0) + (\tzA,0) $) 
		-- node [above, black] {Residue} node [below, black] {calculus} ($ (B) + (-\tzw,0) + (-\tzA,0) $) ;

	\draw [->, very thick, black!30!green] ($ (B) - (0,\tzh) - (0,\tzA) $) 
		-- node [black,right] {\begin{tabular}{l} RG flow \\
                                         to infared,\\
                                         ${\Rx',\wt\Rx'\to 0}$\end{tabular}} ($ (D) + (0,\tzh) + (0,\tzA) $);

	\draw [->, very thick, black!30!green, dashed] ($ (C) + (\tzw,0) + (\tzA,0) $) 
		-- node [above, black] {Residue} node [below, black] {calculus} ($ (D) + (-\tzw,0) + (-\tzA,0) $) ;

	%    Draw Slashout and Circle

	\draw [thick, red] ($ (C)!0.5!(D) $) ellipse (.8 and .5);
	\draw [thick, red] ($ (C)!0.5!(D) +(-.7,.4) $) --++ (1.4,-.8);
	
	\end{tikzpicture}  
	
	\caption{Diagram illustrating how the chirality bound can be
          evaded to establish factorization for $Z_{S^3}$.}\label{Schematic}
\end{figure}
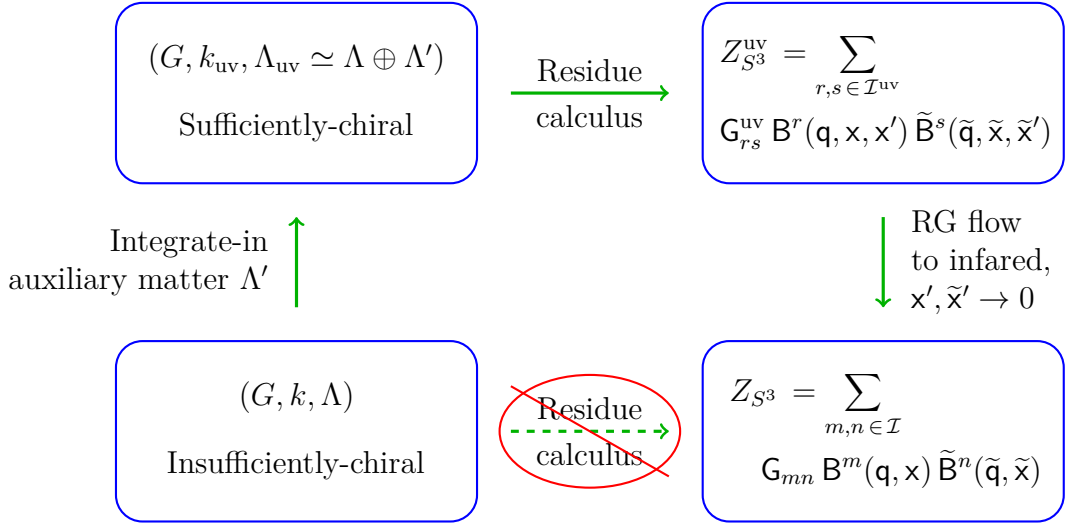
\fi

\paragraph{SQED at Non-Vanishing Level, Redux.}  Let us illustrate how the trick of integrating-in auxiliary
chiral matter allows us to evade the chirality bound in SQED at
non-zero level.

We consider a $U(1)$ vector multiplet at positive Chern-Simons level ${k >
  0}$, coupled to $N_{\rm f}$ pairs of chiral multiplets $(\Phi_j,\wt\Phi_j)$,
${j=1,\ldots,N_{\rm f}}$, each with charges $\pm 1$.
Gauge invariance implies here that ${k\in\Z}$ obeys the usual integral
quantization; throughout, we apply the $i\varepsilon$-prescription to $k$ when
necessary for convergence of various analytic expressions.  In
principle, $\Phi_j$ and $\wt\Phi_j$ can be given independent real
masses $(\mu_j,\wt\mu_j)$ for each $j$.  In practice, to reduce the
number of parameters we juggle, we assume with no essential loss that ${\mu_j =
  \wt\mu_j}$ is the maximally CP-violating real mass. 

The partition function on $S^3$ is then given by the integral
\begin{equation}\label{ZUoneK}
Z_{S^3}^{\rm SQED} \,=\, \int_{\R} \frac{d\sigma}{2\pi}
\;\exp{\!\left[\frac{i k}{4\pi}\sigma^2 \,+\,
    i\,\xi\,\sigma\right]}\cdot \prod_{j=1}^{N_{\rm f}}
s_b\!\left(\frac{\sigma}{2\pi} \,+\,\mu_j\right) s_b\!\left(-\frac{\sigma}{2\pi}\,+\,\mu_j\right).
\end{equation}
Because the matter content $\Lambda_{\rm SQED}$ transforms in a real
representation of $U(1)$, 
\begin{equation}
\Lambda_{\rm SQED} \,=\, ({\bf +1})^{N_{\rm f}}\oplus({\bf -1})^{N_{\rm f}}\,,
\end{equation}
the leading terms in the asymptotic expansions for each pair of
double-sines cancel, just as for the toy model in \eqref{ProdSbz}, and 
the asymptotic behavior of the SQED integrand is dominated by the Gaussian term
as ${|\sigma|\to\infty}$ in the complex plane.

The song remains the same:~the real integration contour for $\sigma$
in \eqref{ZUoneK} cannot be
closed in either the lower or the upper half-plane to convert the
integral into a sum over residues, suitable for
factorization.

Instead, we replace the original Chern-Simons term at level ${k>0}$ by
auxiliary matter $\Lambda'$.  Via the usual logic of renormalization, many
consistent choices for the auxiliary matter are possible,
in the sense that each choice reproduces the partition function
$Z_{S^3}^{\rm SQED}$ when auxiliary masses are large.  Here we shall
make the `marginal' choice to reduce the Chern-Simons level to ${k_{\rm uv} = 0}$
with real matter $\Lambda_{\rm uv}$, according to the notation in Figure
\ref{Schematic}.  Any choice with ${k_{\rm uv} < |\psi_2(\Lambda_{\rm uv})|}$
would allow application of the residue calculus, but some
technical conveniences occur in the marginal case where ${k_{\rm
    uv} = \psi_2(\Lambda_{\rm uv})=0}$.

To reduce the ultraviolet Chern-Simons level to zero, we integrate-in 
$k$ pairs of additional chiral multiplets
$(\Psi_\ell,\wt\Psi_\ell)$, ${\ell=1,\ldots,k}$, each with $U(1)$
charges $\pm 1$,
\begin{equation}\label{LambdaPr}
\Lambda' \,=\, ({\bf +1})^{k}\oplus({\bf -1})^{k}\,. 
\end{equation}
We again assume the real masses $(\mu'_\ell,\wt\mu'_\ell)$ for each auxiliary pair
$(\Psi_\ell,\wt\Psi_\ell)$ are equal, ${\mu'_\ell =
  \wt\mu'_\ell}$ for each $\ell$.  The auxiliary mass 
parameters $\mu'_\ell$ are otherwise distinct, generic, and large in norm, with
${|\mu'_\ell|\gg|\mu_j|}$ for all ${j=1,\ldots,N_{\rm f}}$. The net ultraviolet matter content is 
\begin{equation}
\Lambda_{\rm uv} \,=\, \Lambda_{\rm SQED} \oplus \Lambda'\,.
\end{equation}

Though $\Lambda_{\rm uv}$ also transforms in a real representation of $U(1)$, the
auxiliary masses violate parity, and an effective
Chern-Simons term can be generated when $\Lambda'$ is integrated-out.
The value of the effective Chern-Simons level depends on the signs of
the auxiliary masses.  For simplicity, we take the sign\ of each auxiliary mass
to be negative,
\begin{equation}
\Re(\mu_\ell') < 0\,,\qquad\qquad |\mu'_\ell|\gg |\mu_j|\,,\qquad\qquad
\ell=1,\ldots,k\,,\quad j=1,\ldots,N_{\rm f}\,.
\end{equation}
The matching condition in \eqref{UVlevel} then states that the
ultraviolet Chern-Simons level must be exactly zero,
\begin{equation}
k_{\rm uv} \,=\, 0\,,
\end{equation}
to reproduce SQED with $N_{\rm f}$ flavors at level $k$ at energy scales far
below the scale of the auxiliary masses $\mu'_\ell$.

The partition function for the ultraviolet theory 
is now given by an expression similar to \eqref{ZUoneK} but with no
Gaussian term for $\sigma$,
\begin{equation}\label{ZUoneUV}
\begin{aligned}
Z_{S^3}^{\rm uv} \,=\, \e{i \eta_0}\!\!\int_{\R} \frac{d\sigma}{2\pi}
\,\exp{\!\left[i\,\xi_{\rm uv}\,\sigma\right]}\cdot&\prod_{j=1}^{N_{\rm f}}
s_b\!\left(\frac{\sigma}{2\pi} \,+\,\mu_j\right)
s_b\!\left(-\frac{\sigma}{2\pi}\,+\,\mu_j\right)\times\\
&\times\prod_{\ell=1}^{k}
s_b\!\left(\frac{\sigma}{2\pi}\,+\,\mu'_\ell\right) s_b\!\left(-\frac{\sigma}{2\pi}\,+\,\mu'_\ell\right).
\end{aligned}
\end{equation}
To match the infrared SQED partition function in the limit
${\mu_\ell'\to-\infty}$ for ${\ell=1,\ldots,k}$, we must generally allow the  
phase parameter $\eta_0$ as well as the FI 
parameter $\xi_{\rm uv}$ in \eqref{ZUoneUV} to depend upon the auxiliary masses
$\mu'_\ell$.  According to the renormalization formula in
\eqref{EtaN}, $\eta_0$ depends upon the auxiliary masses via 
\begin{equation}\label{1LoopMatcheta}
\eta_0 \,=\, -\pi\sum_{\ell=1}^{k}\left[(\mu'_\ell)^2 \,-\,
  i\,Q\,\mu_\ell'\,-\,\frac{1}{6}(1+Q^2)\right]\,,\qquad Q \equiv
b \,+\, b^{-1}\,.
\end{equation}
On the other hand, according to the matching condition in \eqref{XiN},
the ultraviolet and infrared FI parameters are related by
\begin{equation}\label{SQEDXiN}
\xi_{\rm uv} \,=\, \xi \,-\, \ha \sum_{\ell=1}^k \left(\mu_\ell' \,-\, \wt\mu_\ell'\right).
\end{equation}
Since we assume the real masses for each auxiliary pair
$(\Psi_\ell,\wt\Psi_\ell)$ obey ${\mu'_\ell =
  \wt\mu'_\ell}$, the FI parameter is not renormalized in this case,
\begin{equation}\label{1LoopMatchxi}
\xi_{\rm uv} \,=\, \xi \qquad\hbox{ for }\qquad \mu'_\ell =
  \wt\mu'_\ell\,,\qquad \ell=1,\ldots,k\,.
\end{equation}
More generally, the following analysis is not
altered if we take the differences ${\mu'_\ell - \wt\mu'_\ell}$ to
remain finite in the limit ${\mu'_\ell\to-\infty}$ for
${\ell=1,\ldots,k}$, so that the FI parameter only renormalizes by a finite
amount.

We considered the properties of integrals such as \eqref{ZUoneUV} in
Section \ref{eq:GGU(1)}.  Because $\Lambda_{\rm uv}$ is a real
representation of $U(1)$, the signed sum $\psi_2(\Lambda_{\rm uv})$ 
defined in \eqref{PsiTwo} vanishes,
and the integrand $\textbf{I}(\sigma)$ of \eqref{ZUoneUV} behaves asymptotically
as\footnote{When $\sigma$ is complex, we must distinguish
  the norm 
  $|\sigma|$ from the product ${\sgn(\Re(\sigma))\cdot\sigma}$, which
  otherwise agree for real values of $\sigma$.  The
  latter quantity
  enters the asymptotic expansion \eqref{BFIsim} of the integrand
  $\textbf{I}(\sigma)$ for complex values of $\sigma$.} 
\begin{equation}\label{BFIsim}
\textbf{I}(\sigma) \,\underset{|\sigma|\to\infty}{=}\,
\exp{\!\Big[\big(i\,\xi \,-\, 
      \psi_1(\Lambda_{\rm uv})\cdot\sgn(\Re(\sigma))\big)\,\sigma
      \,+\, O(1)\Big]}\,, 
\end{equation}
where according to \eqref{Psi1Lamb},
\begin{equation}
\begin{aligned}
&\psi_1(\Lambda_{\rm uv}) \,=\, \frac{1}{2} \left(k\,+\,N_{\rm f}\right) Q
\,+\, i\,\mu_{\rm tot} \,+\, i\,\mu'_{\rm tot}\,,\\
&\mu_{\rm tot} \,=\, \sum_{j=1}^{N_{\rm f}} \mu_j\,,\qquad \mu'_{\rm
  tot} \,=\, \sum_{\ell=1}^k \mu'_\ell\,.
\end{aligned}
\end{equation}
If ${\xi\in\R}$ and ${\Re(\psi_1(\Lambda_{\rm uv}))>0}$ is positive,
$\textbf{I}(\sigma)$ decays exponentially as ${\sigma\to\pm\infty}$
along the real axis.  The ultraviolet integral in \eqref{ZUoneUV} is thus
absolutely-convergent.  In this situation, we can evaluate the
decoupling limit inside the integral to obtain the
expected identity 
\begin{equation}\label{UVtoIR}
\lim_{\stackrel{\mu'_\ell\to-\infty}{\ell=1,\ldots,k}} Z_{S^3}^{\rm uv} \,=\, \int_{\R}
\frac{d\sigma}{2\pi} \left[\lim_{\stackrel{\mu'_\ell\to-\infty}{\ell=1,\ldots,k}}\textbf{I}(\sigma)\right]
\,=\, Z_{S^3}^{\rm SQED}\,.
\end{equation}
In the second equality we apply the double-sine expansions from
Section \ref{OneLoop}, as well as the matching relations in
\eqref{1LoopMatcheta} and \eqref{1LoopMatchxi}.

Alternatively, for finite values of the auxiliary mass $\mu'$, the
ultraviolet integral in \eqref{ZUoneUV} can be computed
as a sum of residues.  Like the toy model at the end of Section
\ref{PropDSine}, we temporarily assume by analytic continuation that ${\psi_1(\Lambda_{\rm uv})\in\R_+}$ is real and
positive, eg.~$\mu_{\rm tot}$ and $\mu'_{\rm tot}$ have been rotated
to lie on the imaginary axis.  The integrand $\textbf{I}(\sigma)$ in
\eqref{BFIsim} then decays exponentially in either the lower or the
upper half-plane, depending upon the sign of the FI parameter $\xi$
(taken to be real).  For ${\xi > 0}$ the integration
contour can be closed in the upper half-plane; for ${\xi < 0}$ the
integration contour can be closed in the lower half-plane; and for
${\xi = 0}$ the contour can be closed in either the upper or the lower
half-plane.  With no essential loss, we assume ${\xi<0}$ and so close
the integration contour for \eqref{ZUoneUV} in the lower half of the
complex $\sigma$-plane.

We have already performed the requisite residue calculus for SQED in
Section \ref{eq:GGU(1)}.  After the formula in \eqref{BigZSumII} is
specialized, the ultraviolet partition function is
given by the finite sums
\begin{equation}\label{BigZBlocks}
\begin{aligned}
Z_{S^3}^{\rm uv} \,&=\, \e{i \eta_0} \sum_{j=1}^{N_{\rm f}}
\exp{\!\left(i\,\Theta^{\,j}_{\rm uv}\right)}\,\RB^{\,j}_{\rm
  uv}(\Rq,\Rx,\Rx',\Ry)\,\wt\RB^{\,j}_{\rm uv}(\wt\Rq,\wt\Rx,\wt\Rx',\wt\Ry)
\,\,+\,\\
 &\qquad\,+\,\e{i \eta_0}
 \sum_{\ell=1}^k\exp{\!\left(i\,\Theta^{\,\ell}_{\rm aux}\right)}\,\RC_{\rm
   aux}^{\,\ell}(\Rq,\Rx,\Rx',\Ry)\,\wt\RC_{\rm aux}^{\,\ell}(\wt\Rq,\wt\Rx,\wt\Rx',\wt\Ry)\,.
\end{aligned}
\end{equation}
Here we are careful to include the renormalization prefactor in \eqref{ZUoneUV},
\begin{equation}\label{RenormEt}
\e{i \eta_0} \,=\, \exp{\!\left[\frac{i \pi k}{2} \,-\, i \pi
    \sum_{\ell=1}^k (\mu'_\ell)^2\right]} \cdot \left(\Rq\,
  \wt\Rq\right)^{k/12} \cdot \prod_{\ell=1}^k \left(\Rx'_\ell\,
  \wt\Rx'_\ell\right)^{-1/2}\,,
\end{equation}
rewritten using the one-loop matching relation for $\eta_0$ in
\eqref{1LoopMatcheta}.  In this expression, $(\Rx'_\ell,\wt\Rx'_\ell)$
are fugacities associated to the mass parameters $\mu'_\ell$ for the heavy,
auxiliary chiral multiplets, while $(\Rx_j,\wt\Rx_j)$ are fugacities
associated to the mass parameters $\mu_j$ for the original, light SQED
flavors.  On the line in parameter space where $(b,\mu_j,\mu'_\ell)$ are all
real and ${b>0}$ is positive, the
assumption ${|\mu'_\ell| \gg |\mu_j|}$ with ${\mu'_\ell<0}$ negative
implies that ${|\Rx_\ell'|\ll|\Rx_j|}$ and dually
${|\wt\Rx_\ell'|\ll|\wt\Rx_j|}$.  Of course, when expressed in terms
of the variables $(\Rq,\Rx,\Rx')$ and $(\wt\Rq,\wt\Rx,\wt\Rx')$, the
renormalization prefactor \eqref{RenormEt} manifestly respects the
block decomposition for $Z_{S^3}^{\rm uv}$.

The ultraviolet partition function $Z_{S^3}^{\rm uv}$ naturally involves a sum
over two kinds of blocks.  In the first line of \eqref{BigZBlocks},
the blocks $\RB^j_{\rm uv}$ and dually $\wt\RB^j_{\rm uv}$ for ${j=1,\ldots,N_{\rm f}}$ arise from evaluating the
residue at a pole in the one-loop factor for the light SQED
multiplets, this factor given by the product of double-sines in the first line of \eqref{ZUoneUV}.  The
subscript serves to distinguish these ultraviolet blocks from the
actual blocks of the low-energy SQED theory, which will appear
shortly. By contrast, in second line of \eqref{BigZBlocks}, the 
blocks $\RC^\ell_{\rm aux}$ and $\wt\RC^\ell_{\rm aux}$ for ${\ell=1,\ldots,k}$ 
arise from poles in the one-loop factor for the
heavy auxiliary multiplets.\footnote{Note that because the integration contour is closed in the lower half of the
$\sigma$-plane, only those multiplets with positive $U(1)$-charge
contribute residues to the respective block sums.}

We verify that the block decomposition
\eqref{BigZBlocks} satisfies a pair of conditions in
the infrared limit ${\mu_\ell'\to-\infty}$, with 
${\mu_\ell'-\mu_m'}$ fixed for all pairs
${\ell,m=1,\ldots,k}$.
\begin{enumerate}
\item For the phase $\Theta_{\rm uv}^{\,j}$ and the `light' blocks $\RB^{\,j}_{\rm
    uv}$, $\wt\RB^{\,j}_{\rm uv}$ in the first line of
  \eqref{BigZBlocks}, we show that these quantities reproduce 
  the expected decomposition for the infrared SQED theory at
  level ${k}$ when ${\mu_\ell'\to-\infty}$.  This check
  is straightforward, once the renormalization prefactor
  \eqref{RenormEt} is included.
\item For the phase $\Theta^\ell_{\rm aux}$ and the `auxiliary' blocks
  $\RC^{\,\ell}_{\rm aux}$, $\wt\RC^{\,\ell}_{\rm aux}$ in the second line of
  \eqref{BigZBlocks}, we show that the contribution of these
  summands to $Z_{S^3}^{\rm uv}$ vanishes in the limit
  ${\mu_\ell'\to-\infty}$.  Naively, one expects such a decoupling of
  massive matter in the infrared.  The decoupling turns out
  to be suprisingly delicate, as the blocks $\RC^{\,\ell}_{\rm
    aux}$ and $\wt\RC^{\,\ell}_{\rm aux}$ themselves {\sl diverge} when
  ${\mu_\ell'\to-\infty}$.  The divergence of the auxiliary blocks
  $\RC^{\,\ell}_{\rm aux}$ and $\wt\RC^{\,\ell}_{\rm aux}$ in the
  limit ${\mu_\ell'\to-\infty}$ has been previously noted in \S $5.2$
  of \cite{Aharony:2013dha}, with which our analysis has some
  overlap.

  The $i\varepsilon$-prescription for $k$ in
  $\Theta^{\,\ell}_{\rm aux}$ is required to ensure convergence to
  zero for the auxiliary summands in $Z_{S^3}^{\rm uv}$.   In terms of
  the ingredients \eqref{FACTOR} for the Factorization Conjecture,
  auxiliary matter decouples as the the bilinear form $\RG_{m n}$
  degenerates in the infrared limit. 
\end{enumerate}
We now demonstrate these statements.

\paragraph{Infrared Limit of Light Blocks.}

The ingredients in the factorization of $Z_{S^3}^{\rm uv}$ can be
written much more explicitly by specializing the formulas in Section
\ref{eq:GGU(1)}.  For the light blocks in the first line of
\eqref{BigZBlocks}, the general phase in \eqref{BigTheta} becomes 
\begin{equation}\label{BigTHj}
\begin{aligned}
&\exp{\!\left(i\,\Theta^{\,j}_{\rm uv}\right)} \,=\, 
\exp{\!\left[-\frac{i \pi k}{2} \,+\, i \pi \sum_{\ell=1}^k
    \left(\mu_\ell'\right)^2\right]}\cdot\exp{\!\left(i\,\Theta^{\,j}_{\rm
      SQED}\right)}\,,
\end{aligned}
\end{equation}
where $\Theta^{\,j}_{\rm SQED}$ is the corresponding phase for the
low-energy SQED theory at level $k$,
\begin{equation}\label{SqedTHj}
\begin{aligned}
\exp{\!\left(i\,\Theta^{\,j}_{\rm SQED}\right)} &= \exp{\!\left[ i
    \pi k \, \mu_j^2 - 2 \pi i \, \xi \, \mu_j + \frac{i
      \pi}{4} \sum_{m\neq j}^{N_{\rm f}}  \left(1 - 2 \left(\mu_{m} -
        \mu_j\right)^2\right)\sgn(\mu_{m}-\mu_j)\right]}\times\\
&\times\exp{\!\left[  \frac{i \pi}{4}\sum_{m=1}^{N_{\rm f}}\left(1 - 2 \left(\mu_{m} +
        \mu_j\right)^2\right)\sgn(\mu_{m}+\mu_j)\right]}\,.
\end{aligned}
\end{equation}
All dependence on $k$ in \eqref{BigTHj} and \eqref{SqedTHj} arises not
from an ultraviolet Chern-Simons term, since ${k_{\rm uv}=0}$, but from
evaluating sums in \eqref{BigTheta}, similar to those above,
over the $2k$ auxiliary multiplets.  In the process, we use that
${\sgn(\mu'_\ell \pm \mu_j)<0}$ is negative for all indices $j$ and
$\ell$, by our assumption on the auxiliary masses $\mu'_\ell$.

By fiat, the phase factor on the right in \eqref{RenormEt}
cancels the discrepancy between ultraviolet and infrared phases in
\eqref{BigTHj}.

As for the light blocks $\RB^{\,j}_{\rm uv}$ themselves, each is a sum
of residues 
\begin{equation}\label{Bjuv}
\RB^{\,j}_{\rm uv}(\Rq,\Rx,\Rx',\Ry) \,=\, \sum_{M=0}^\infty
\RW^{\,j}_M(\Rq,\Rx,\Rx',\Ry)_{\rm uv}\,,\qquad\qquad j\,=\,1,\ldots,N_{\rm f}\,,
\end{equation}
where the residue is given by the general formula in \eqref{BigW},
specialized to the case at hand.  Explicitly,
\begin{equation}\label{BigWuv}
\begin{aligned}
&\RW^{\,j}_{M}(\Rq,\Rx,\Rx',\Ry)_{\rm uv} \,=\,  i^{M\!\big[1
      \,+\, \sum_{m\neq j}^{N_{\rm f}} \sgn(\mu_{m}-\mu_j) \,-\,
      \sum_{m=1}^{N_{\rm f}} \sgn(\mu_{m}+\mu_j)\big]}\cdot\Ry^M\cdot\left[\frac{\Rq^{M(M+1)/4}}{\left(\Rq;\Rq\right)_M}\right]\times\\
&\quad\times\,\prod_{m\neq j}^{N_{\rm f}}
\SF^{\sgn(\mu_{m}-\mu_j)}_b\big(\mu_{m}-\mu_j-
  i\,M\,b;\Rq\big)\cdot\prod_{m=1}^{N_{\rm f}}\SF^{\sgn(\mu_{m}+\mu_j)}_b\big(\mu_{m}+\mu_j+i\,M\,b;\Rq\big)\,\times\\
&\quad\times\,\prod_{\ell=1}^{k}\Big[\SF^{-}_b\big(\mu_\ell'
  - \mu_j - i \, M \, b;\Rq\big)\cdot\SF^{-}_b\big(\mu_\ell'
  + \mu_j + i \, M \, b;\Rq\big)\Big]\,.
\end{aligned}
\end{equation}
In this expression, all dependence
on the auxiliary masses $\mu'_\ell$, or equivalently on the fugacity variables
$\Rx'_\ell$, occurs in the product over $\ell$ on the last line of
\eqref{BigWuv}.  

The formula for $\SF_b^{-}$ in \eqref{BigFNeg} allows
us to make the dependence on $\Rx'$ explicit.  The result depends upon
whether $\Rq$ lies inside our outside the unit disk, due to the
natural boundary of holomorphy at ${|\Rq|=1}$ in Figure \ref{qHolomorphyFig}.  For simplicity, we
assume ${|\Rq|<1}$.  After an elementary calculation using \eqref{BigFNeg},
\begin{equation}\label{AuxProd}
\begin{aligned}
&\prod_{\ell=1}^{k}\Big[\SF^{-}_b\big(\mu_\ell'
  - \mu_j - i \, M \, b;\Rq\big)\cdot\SF^{-}_b\big(\mu_\ell'
  + \mu_j + i \, M \, b;\Rq\big)\Big] \,\underset{|\Rq|<1}{=}\,\\
&\qquad\Rq^{-\frac{1}{2} k M^2 - \frac{1}{12} k}\cdot\Rx_j^{-k M}\cdot
\prod_{\ell=1}^k\left[\frac{(\Rx_\ell')^{1/2}}{\Big(\left(\Rx_\ell'\,\Rx_j^{-1}\right)
  \Rq^{-M};\Rq\Big)_\infty \Big(\left(\Rx_\ell'\,\Rx_j\right) \Rq^M;\Rq\Big)_\infty}\right].
\end{aligned}
\end{equation}
The dependence on $k$ in \eqref{AuxProd} is such that the
ultraviolet residue $\RW^{\,j}_{M}(\Rq,\Rx,\Rx',\Ry)_{\rm uv}$ factorizes,
\begin{equation}\label{AuxProdII}
\begin{aligned}
\RW^{\,j}_{M}(\Rq,\Rx,\Rx',\Ry)_{\rm uv} \,&=\, \Rq^{-k/12} \cdot \prod_{\ell=1}^k\left[\frac{(\Rx_\ell')^{1/2}}{\Big(\left(\Rx_\ell'\,\Rx_j^{-1}\right)
  \Rq^{-M};\Rq\Big)_\infty \Big(\left(\Rx_\ell'\,\Rx_j\right)
  \Rq^M;\Rq\Big)_\infty}\right]\times\\
&\times\,\RW^{\,j}_M(\Rq,\Rx,\Ry)_{\rm SQED}\,,
\end{aligned}
\end{equation}
where $\RW^{\,j}_M(\Rq,\Rx,\Ry)_{\rm
  SQED}$ is the residue for the low-energy SQED theory
at level $k$, with no auxiliary matter.  Explicitly from \eqref{BigW},
\begin{equation}
\begin{aligned}
&\RW^{\,j}_M(\Rq,\Rx,\Ry)_{\rm SQED} \,=\, i^{M\!\big[1
      \,+\, \sum_{m\neq j}^{N_{\rm f}} \sgn(\mu_{m}-\mu_j) \,-\,
      \sum_{m=1}^{N_{\rm f}} \sgn(\mu_{m}+\mu_j)\big]}\times\\
&\qquad\times\Rq^{-\ha k M^2}\cdot\Rx_j^{-k
  M}\cdot\Ry^M\cdot\left[\frac{\Rq^{M(M+1)/4}}{\left(\Rq;\Rq\right)_M}\right]\times\\
&\qquad\times\prod_{m\neq j}^{N_{\rm f}}
\SF^{\sgn(\mu_{m}-\mu_j)}_b\!\big(\mu_{m}-\mu_j-
  i\,M\,b;\Rq\big)\cdot\prod_{m=1}^{N_{\rm f}}\SF^{\sgn(\mu_{m}+\mu_j)}_b\!\big(\mu_{m}+\mu_j+i\,M\,b;\Rq\big)\,.
\end{aligned}
\end{equation} 

Let us now consider the behavior of the block $\RB^{\,j}_{\rm uv}$ in
the infrared limit ${\mu'\to-\infty}$, for which ${\Rx'\to 0}$.  Both
$q$-Pochhammer symbols in the denominator of \eqref{AuxProdII} are
continuous at ${\Rx'=0}$ and evaluate to unity. Hence from
\eqref{AuxProdII},
\begin{equation}\label{LimitI}
\lim_{\Rx'\to 0}\left[\left(\prod_{\ell=1}^k \left(\Rx_\ell'\right)^{-1/2}\right)
\cdot \RW^{\,j}_{M}(\Rq,\Rx,\Rx',\Ry)_{\rm uv}\right]\,=\, \Rq^{-k/12}\cdot\RW^{\,j}_M(\Rq,\Rx,\Ry)_{\rm SQED} \,.
\end{equation}
The same relation applies term-by-term to the sum in
\eqref{Bjuv}, so the light ultraviolet blocks $\RB^{\,j}_{\rm uv}$ are
related to the infrared SQED blocks $\RB^{\,j}_{\rm SQED}$ by 
\begin{equation}\label{LimitII}
\lim_{\Rx'\to 0}\left[\left(\prod_{\ell=1}^k \left(\Rx_\ell'\right)^{-1/2}\right)
\cdot \RB^j(\Rq,\Rx,\Rx',\Ry)_{\rm uv}\right]\,=\, \Rq^{-k/12}\cdot\RB^j(\Rq,\Rx,\Ry)_{\rm SQED}\,.
\end{equation}
An identical relation, omitted for sake of brevity, holds for the dual blocks $\wt\RB^{\,j}_{\rm uv}$.  

The renormalization prefactor in \eqref{RenormEt} accounts precisely
for the extra factors of $\Rq$ and $\Rx'_\ell$ in the limit \eqref{LimitII}.  Recalling the phase relation in 
\eqref{BigTHj}, we thus obtain the expected infrared behavior for the first
summand in the top line of \eqref{BigZBlocks},
\begin{equation}
\begin{aligned}
&\lim_{\Rx'\to 0}\left[\e{i \eta_0} \sum_{j=1}^{N_{\rm f}}
\exp{\!\left(i\,\Theta^{\,j}_{\rm uv}\right)}\,\RB^{\,j}_{\rm
  uv}(\Rq,\Rx,\Rx',\Ry)\,\wt\RB^{\,j}_{\rm
  uv}(\wt\Rq,\wt\Rx,\wt\Rx',\wt\Ry)\right] \,=\,\\ 
&\qquad\qquad\qquad\sum_{j=1}^{N_{\rm f}}\exp{\!\left(i\,\Theta^{\,j}_{\rm
      SQED}\right)}\,
\RB^{j}(\Rq,\Rx,\Ry)_{\rm SQED}\,\wt\RB^{j}(\wt\Rq,\wt\Rx,\wt\Ry)_{\rm
  SQED}\,.
\end{aligned}
\end{equation}

\paragraph{Decoupling of Massive Matter.}

The ultraviolet partition function $Z_{S^3}^{\rm uv}$ in
\eqref{BigZBlocks} also includes contributions from blocks
$\RC^\ell_{\rm aux}(\Rq,\Rx,\Rx',\Ry)$
associated to heavy auxiliary multiplets.  To complete the proof of the
Factorization Conjecture for SQED at positive level ${k>0}$, we
must show that the sum over auxiliary blocks
$\RC^\ell_{\rm aux}(\Rq,\Rx,\Rx',\Ry)$ in the second line of 
\eqref{BigZBlocks} vanishes in the infrared limit ${\mu'\to-\infty}$
and ${\Rx'\to 0}$, consistent with naive expectations for decoupling of massive
matter.  In fact, we shall argue that each summand for
${\ell=1,\ldots,k}$ in \eqref{BigZBlocks} vanishes individually as
${\mu'\to-\infty}$.

Again, the general formulas from Section \ref{eq:GGU(1)} allow us to make the
dependence on $\mu'$ and $\Rx'$ explicit.  Briefly, for the phases,
\begin{equation}\label{AuxTHl}
\begin{aligned}
\exp{\!\left(i\,\Theta^{\,\ell}_{\rm aux}\right)} &= \exp{\!\left[-2\pi i \,
    \mu_\ell' \left(\xi - \sum_{j=1}^{N_{\rm f}} \mu_j\right) - \frac{i \pi
      k}{4} \,+\, \frac{i \pi}{2}\sum_{m=1}^k\big(\mu_m'+\mu_\ell'\big)^2\right]}\times\\
&\times\exp{\!\left[\frac{i \pi}{4}\sum_{m\neq\ell}^k\big(1 - 2\,
    (\mu_m' - \mu_\ell')^2\big)
  \sgn(\mu_m'-\mu_\ell')\right]}\,.
\end{aligned}
\end{equation}
The effective shift ${\xi \mapsto \xi - \sum \mu_j}$ in
$\Theta^\ell_{\rm aux}$ summarizes the contribution from the $N_{\rm f}$ light flavors
when ${\mu_\ell'\ll\mu_j}$ is sufficiently negative.  To extract the
limiting infrared behavior as ${\mu_\ell'\to-\infty}$ with
${\mu_m'-\mu_\ell'}$ fixed, let us expand the sum of squares
\begin{equation}
\begin{aligned}
\sum_{m=1}^k\big(\mu_m'+\mu_\ell'\big)^2 \,&=\,
\sum_{m=1}^k\big((\mu_m'-\mu_\ell') + 2 \, \mu_\ell'\big)^2\,,\\
&=\, 4 k \left(\mu_\ell'\right)^2 \,+\, 4 \mu_\ell' \sum_{m=1}^k
\left(\mu_m'-\mu_\ell'\right) \,+\, \sum_{m=1}^k
\left(\mu_m'-\mu_\ell'\right)^2\,.
\end{aligned}
\end{equation}
Thus more simply,
\begin{equation}\label{AuxTHlII}
\exp{\!\left(i\,\Theta^{\,\ell}_{\rm aux}\right)} \,=\,
\exp{\!\left[2\pi i\,k \left(\mu_\ell'\right)^2 \,+\, O(\mu_\ell')\right]}\,.
\end{equation}
After $k$ is given a small positive imaginary part ${+i\varepsilon}$, 
${0 < \varepsilon \ll 1}$, to
ensure convergence, the factor in \eqref{AuxTHlII} decays like a Gaussian as
${\mu_\ell'\to-\infty}$.  The ${i\varepsilon}$-prescription for $k$ 
previously ensured Gaussian decay of the integrand, so
it is hardly surprising to see the ${i\varepsilon}$-prescription play the same role here.

For the auxiliary blocks
themselves, each is a sum of residues,
\begin{equation}\label{AuxBlC}
\RC^\ell_{\rm aux}(\Rq,\Rx,\Rx',\Ry) \,=\, \sum_{M=0}^\infty
\RW^{\,\ell}_M(\Rq,\Rx,\Rx',\Ry)_{\rm aux}\,,
\end{equation}
where the residue is given by the product
\begin{equation}\label{ResXC}
\begin{aligned}
&\RW^{\,\ell}_M(\Rq,\Rx,\Rx',\Ry)_{\rm aux} \,=\, i^{M \big(1 + 2
    (k+N_{\rm f}) + \sum_{m\neq\ell}^k
    \sgn(\mu_m'-\mu_\ell')\big)}\cdot\Ry^M\cdot\left[\frac{\Rq^{M(M+1)/4}}{\left(\Rq;\Rq\right)_M}\right]\times\\
&\quad\times\,\prod_{m\neq \ell}^{k}
\SF^{\sgn(\mu_{m}'-\mu_\ell')}_b\big(\mu_{m}'-\mu_\ell'-
  i\,M\,b;\Rq\big)\cdot\prod_{m=1}^{k}\SF^{-}_b\big(\mu_{m}'+\mu_\ell'+i\,M\,b;\Rq\big)\,\times\\
&\quad\times\,\prod_{j=1}^{N_{\rm f}}\Big[\SF^{+}_b\big(\mu_j
  - \mu_\ell' - i \, M \, b;\Rq\big)\cdot\SF^{-}_b\big(\mu_j
  + \mu_\ell' + i \, M \, b;\Rq\big)\Big]\,.
\end{aligned}
\end{equation}
The expression for $\RW^{\,\ell}_M(\Rq,\Rx,\Rx',\Ry)_{\rm aux}$ has the same
structure as the expression for $\RW^{\,j}_M(\Rq,\Rx,\Rx',\Ry)_{\rm uv}$ in
\eqref{BigWuv}, but with a crucial difference in the dependence on the
auxiliary masses $\mu'$ and hence distinct asymptotic behavior as
${\mu'\to-\infty}$.

To bring these differences to the fore, let
us rewrite the final factors in \eqref{ResXC} in terms of the
fugacity variables $\Rx$ and $\Rx'$, where
\begin{equation}\label{ProdFsI}
\begin{aligned}
\prod_{m=1}^{k}\SF^{-}_b\big(\mu_{m}'+\mu_\ell'+i\,M\,b;\Rq\big)
\,&\underset{|\Rq|<1}{=}\, \Rq^{-\frac{k}{4}M(M-1)-\frac{k}{24}}\cdot\left(\Rx_\ell'\right)^{-\ha k (M-\ha)}\,\times\\
&\times\,\prod_{m=1}^k \left[(\Rx_m')^{-\ha
    (M-\ha)}\cdot\frac{\big(\Rx_m' \Rx_\ell';\Rq\big)_M}{\big(\Rx_m'
    \Rx_\ell';\Rq\big)_\infty}\right],
\end{aligned}
\end{equation}
and also
\begin{equation}\label{ProdFsII}
\begin{aligned}
&\prod_{j=1}^{N_{\rm f}}\Big[\SF^{+}_b\big(\mu_j
  - \mu_\ell' - i \, M \, b;\Rq\big)\cdot\SF^{-}_b\big(\mu_j
  + \mu_\ell' + i \, M \, b;\Rq\big)\Big] \,\underset{|\Rq|<1}{=}\\
&\qquad\Rq^{M N_{\rm f}/2}\cdot\left(\Rx_\ell'\right)^{N_{\rm f}/2}\cdot\prod_{j=1}^{N_{\rm f}}\left[
\Rx_j^{-M}\,\frac{\big(\Rx_\ell'\,\Rx_j;\Rq\big)_M}{\big((\Rx_\ell'\,\Rx_j^{-1})\,\Rq;\Rq\big)_M}\,\frac{\big((\Rx_\ell'\,\Rx_j^{-1})\,\Rq;\Rq\big)_\infty}{
\big(\Rx_\ell'\,\Rx_j;\Rq\big)_\infty}\right]\,.
\end{aligned}
\end{equation}
For concreteness, we assume ${|\Rq|<1}$ while evaluating the functions
$\SF^{\pm}_b$ above.  The same analysis holds for ${|\Rq|>1}$.

In the infrared, ${\Rx'_\ell\to 0}$ with fixed ratios
${\Rx'_m/\Rx'_\ell}$ for all pairs ${\ell,m=1,\ldots,k}$.  Combining the
product expansions in \eqref{ProdFsI} and \eqref{ProdFsII}, we see
that $\RW^{\,\ell}_M(\Rq,\Rx,\Rx',\Ry)_{\rm aux}$ then has the
limiting behavior
\begin{equation}\label{IRbigX}
\begin{aligned}
&\lim_{\Rx'\to 0}\left[(\Rx_\ell')^{k (M-\ha) - \ha
    N_{\rm f}}\cdot\RW^{\,\ell}_M(\Rq,\Rx,\Rx',\Ry)_{\rm aux}\right] \,=\, i^{M \big(1 + 2
    (k+N_{\rm f}) + \sum_{m\neq\ell}^k
    \sgn(\mu_m'-\mu_\ell')\big)}\,\Ry^M\,\times\\
&\qquad\times\left[\frac{\Rq^{-\frac{1}{4} (k-1) M^2+\frac{1}{4} (k+ 2 N_{f}
      + 1) M - \frac{k}{24}}}{\left(\Rq;\Rq\right)_M}\right]\cdot\prod_{m=1}^k\left(\frac{\Rx_m'}{\Rx_\ell'}\right)^{-\ha
  (M-\ha)}\cdot\,\prod_{j=1}^{N_{\rm f}}\,\Rx_j^{-M}\,\times\\
&\qquad\times\,\prod_{m\neq \ell}^{k}
\SF^{\sgn(\mu_{m}'-\mu_\ell')}_b\big(\mu_{m}'-\mu_\ell'-
  i\,M\,b;\Rq\big)\,.
\end{aligned}
\end{equation}
Equivalently, since the right side of \eqref{IRbigX} is fixed and
finite as ${\Rx'\to 0}$, 
$\RW^{\,\ell}_{M,\textrm{aux}}$ scales with $\Rx_\ell'$ in the infrared as 
\begin{equation}\label{ScalWaux}
\RW^{\,\ell}_{M,\textrm{aux}} \,=\, C_0\,(\Rx_\ell')^{-k (M-\ha) + \ha N_{\rm f}} \,+\, \cdots\,.
\end{equation}
Here $C_0$ is a constant, and terms subleading
in $\Rx_\ell'$ are indicated by the `$\ldots$' above.  Clearly for ${k>0}$ and ${M\gg 1}$ sufficiently
large, $\RW^{\,\ell}_{M,\textrm{aux}}$ diverges as ${\Rx'\to
  0}$.  Hence the auxiliary blocks $\RC^{\,\ell}_{\rm aux}$
in \eqref{AuxBlC} also diverge as ${\Rx'\to 0}$ and do not have a
well-defined infrared limit.  By an identical computation, the same divergence
afflicts the dual blocks $\wt\RC^{\,\ell}_{\rm aux}$.

We are left to address the decoupling of massive matter from
$Z_{S^3}^{\rm uv}$ in the infrared.  By definition, the fugacity variables
depend exponentially on the auxiliary masses,
\begin{equation}
\Rx'_\ell \,=\, \e{\!2\pi \mu_\ell' b}\,,\qquad\qquad \wt\Rx'_\ell = \e{\!2
  \pi \mu_\ell'/b}\,.
\end{equation}
The auxiliary residues $\RW^{\,\ell}_{M,\textrm{aux}}$ in \eqref{ScalWaux}
therefore diverge exponentially with $\mu'$ in the infrared limit
${\mu'\to-\infty}$, for all but finitely-many indices $M$.  

By contrast, 
$\exp{\!\left(i\,\Theta^{\,\ell}_{\rm aux}\right)}$ behaves to
leading-order like a Gaussian in \eqref{AuxTHlII}.  With the
renormalization prefactor from \eqref{RenormEt} included, this 
asymptotic behavior persists in the product
\begin{equation}\label{AuxThlIII}
\e{i \eta_0}\exp{\!\left(i\,\Theta^{\,\ell}_{\rm aux}\right)}\,\RW_{M,{\rm
  aux}}^{\,\ell}\,\wt\RW_{N,{\rm
   aux}}^{\,\ell} \,=\, \exp{\!\left[i
     \pi k \left(\mu_\ell'\right)^2 \,+\, O(\mu_\ell')\right]}\,,
\end{equation}
for all indices ${M,N\ge 0}$.  Via the $i\varepsilon$-prescription for $k$, the
expression in \eqref{AuxThlIII} decays rapidly when
${\mu'\to-\infty}$.  Applied term-by-term to the residue sum in
\eqref{AuxBlC}, the preceding asymptotic identity yields the
vanishing statement
\begin{equation}
\lim_{\mu'\to-\infty}\left[\e{i \eta_0}\exp{\!\left(i\,\Theta^{\,\ell}_{\rm aux}\right)}\,\RC_{\rm
  aux}^{\,\ell}(\Rq,\Rx,\Rx',\Ry)\,\wt\RC_{\rm
   aux}^{\,\ell}(\wt\Rq,\wt\Rx,\wt\Rx',\wt\Ry)\right] \,=\, 0\,.
\end{equation}

\paragraph{Reducing to Level Zero with Real Matter.}

The previous discussion shows that the schematic diagram in Figure
\ref{Schematic} commutes for SQED at arbitrary Chern-Simons level,
justifying the use of the naive residue calculus for ${k>0}$ in 
\cite{Pasquetti:2012}.  We now sketch how this argument
extends to supersymmetric Chern-Simons-matter theories with general
gauge group $G$ and matter representation $\Lambda$.  

The key step is to replace the infrared theory for the pair $(G,\Lambda)$ at level ${k>0}$ by an ultraviolet theory for a
new pair $(G,\Lambda_{\rm uv}\!=\Lambda\oplus\Lambda')$ at level
$k_{\rm uv}$, obtained by integrating-in auxiliary matter $\Lambda'$, so that two conditions are satisfied:
\begin{enumerate}
\item ${0 \,=\, k_{\rm uv} \,=\, k \,+\, \ha \sum_{j=1}^n c_2(\lambda_j')
    \,\sgn(\mu_j')}$, and 
\item ${\Lambda_{\rm uv}}$ is a real representation of $G$.
\end{enumerate}
The vanishing condition on ${k_{\rm uv}}$ ensures that the 
integrand for $Z_{S^3}^{\rm uv}$ does not contain an explicit Gaussian
term.  The reality condition on $\Lambda_{\rm uv}$ ensures that the same
integrand does not contain an implicit Gaussian term,
induced from the asymptotic behavior of the one-loop matter determinant 
\begin{equation}\label{MatDetIV}
G(\sigma)_{\rm uv} \,=\, \prod_{j=1}^n\left[\prod_{\beta \in
\Delta_j}
s_b\!\left(\frac{\langle\beta,\sigma\rangle}{2\pi}\,+\,\mu_j\right)\right],\qquad\qquad \sigma\in\h\,.
\end{equation}
Here $j$ indexes irreducible summands of $\Lambda_{\rm uv}$, and
$\Delta_j$ denotes the set of weights in each summand.  To simplify
the notation in \eqref{MatDetIV}, we do not
distinguish between the real masses $\mu$ and $\mu'$ associated to
$\Lambda$ and $\Lambda'$, respectively. 

Let us briefly discuss the implication for $G(\sigma)_{\rm uv}$ when
$\Lambda_{\rm uv}$ is real.  In that case, the set of
weights ${\cup_j\,\Delta_j}$ for $\Lambda_{\rm uv}$ is preserved under
the inversion ${\beta\mapsto-\beta}$.  The relative sign in the asymptotics
\eqref{eq:MSineAsympta} of the double-sine function $s_b(z)$ then implies a
cancellation among the leading Gaussian terms in each factor
of $G(\sigma)_{\rm uv}$.  Instead as ${|\sigma|\to\infty}$, a brief calculation shows 
\begin{equation}\label{ASYMbigG}
G(\sigma)_{\rm uv} \,\underset{|\sigma|\to\infty}{=}\,
\exp{\!\left[-\sum_{j=1}^n\left(\frac{Q}{2} + i\,\mu_j\right) ||\sigma||_{\lambda_j}
    \,+\, O(1)\right]},\qquad \Lambda_{\rm uv}^{}\,\simeq\,\Lambda_{\rm uv}^*\,.
\end{equation}
In the asymptotic formula \eqref{ASYMbigG} for $G(\sigma)_{\rm
  uv}$,  we introduce a Weyl-invariant $L^1$-norm $||\,\cdot\,||_V$ on the Cartan
subalgebra ${\h\subset\g}$ for each non-trivial representation $V$ of
the simple Lie group $G$,
\begin{flalign}\label{RepNorm}
\fbox{$L^1$-norm} \qquad\quad\qquad&
||\sigma||_V \,=\,
\ha\sum_{\beta\in\Delta_V}\big|\langle\beta,\sigma\rangle\big|\,,\qquad\qquad
\sigma\in\h\,.&
\end{flalign}
The factor of one-half in the definition \eqref{RepNorm} appears by
convention, to eliminate other factors of two later.

The properties of the norm $||\,\cdot\,||_V$ will be important in
Section \ref{SUSYBreak}, where we discuss them more fully.  For now, we emphasize that the expression in
\eqref{RepNorm} does define a norm.  Clearly ${||\sigma||_V\ge 0}$ is
positive for all ${\sigma\in\h}$.  Linearity of the dual pairing
$\langle\beta,\sigma\rangle$ means both that
${||c\,\sigma||_V=|c|\cdot||\sigma||_V}$ for    
scalars ${c\in\R}$, and the triangle inequality is obeyed,
${||\sigma_1+\sigma_2||_V \le ||\sigma_1||_V + ||\sigma_2||_V}$.
Non-degeneracy of $||\,\cdot\,||_V$, ie.~${||\sigma||_V>0}$
for ${\sigma\neq 0}$, is the remaining property to check.  Else,
${\sigma\neq 0}$ is a non-trivial element of $\h$ such that
${\langle\beta,\sigma\rangle=0}$ for all weights of $V$.
Equivalently, $\sigma$ annihilates the representation $V$.
The same statement is true for the Lie algebra ideal
${\langle\sigma\rangle\subseteq \g}$ generated by $\sigma$.  But a
simple Lie algebra has no non-trivial ideals. Thus
${\g=\langle\sigma\rangle}$ itself annihilates $V$, contrary to our
assumption that $V$ is non-trivial.\footnote{If ${G = G_1
    \times\cdots\times\,G_N}$ is semi-simple, $||\,\cdot\,||_V$ is
  non-degenerate when $V$ transforms non-trivially under each simple factor.}

Because $||\,\cdot\,||_V$ is a norm, the one-loop matter determinant
$G(\sigma)_{\rm uv}$ in \eqref{ASYMbigG} decays exponentially along
the Cartan subalgebra $\h$ whenever ${b\in\R_+}$ is positive and the imaginary
part of $\mu$, related physically to the R-charge $\RR$, is bounded
from above.  We have already encountered an
example of this behavior for gauge 
group $SU(2)$ in Section \ref{SU2GT}.  In that case, all
representations are automatically real or pseudoreal, and the
asymptotic behavior in \eqref{ASYMbigG} generalizes the analogous formula in
\eqref{BigGFunII}.  By the same token, the asymptotic features of
$G_{\rm uv}(\sigma)$ for real $\Lambda_{\rm uv}$ 
will permit us to apply the residue calculus to the ultraviolet
Chern-Simons-matter theory at level ${k_{\rm uv}=0}$, as we previously
did for $SU(2)$.

We now meet a technical question.  Given the pair $(G,\Lambda)$ and ${k>0}$,
when does an auxiliary representation $\Lambda'$ exist so that $k_{\rm
  uv}$ vanishes and the total matter content $\Lambda_{\rm uv}$ is real?

If ${G=U(1)}$, a suitable auxiliary representation $\Lambda'$ is
provided by the following construction.  By the assumption of anomaly-cancellation, 
\begin{equation}
k \,=\, \ha c_2(\Lambda) \,=\, \ha \sum_{j=1}^n \lambda_j^2\,\quad
\mod\,\Z\,,\qquad\qquad \Lambda \,=\, \bigoplus_{j=1}^n \big[\lambda_j\big]\,.
\end{equation}
Let $d$ be the integer difference
\begin{equation}
d \,=\, k \,-\, \ha c_2(\Lambda) \,\in\,\Z\,.
\end{equation}
If ${d \ge 0}$ is positive, we set
\begin{equation}\label{AuxLambda}
\Lambda' \,=\, \Lambda^* \!\oplus \big[{\bf +1}\big]^d \oplus
\big[{\bf -1}\big]^d\,,\qquad\qquad \Lambda^* =\,
\bigoplus_{j=1}^n \big[-\lambda_j\big]\,.
\end{equation}
Thus $\Lambda'$ is the direct sum of the representation dual (or conjugate) to
$\Lambda$ with $d$ chiral multiplets having $U(1)$-charges $\pm 1$.
By fiat, ${\Lambda_{\rm uv} = \Lambda\oplus\Lambda'}$ is a real
representation of $U(1)$.  Also, in the limit that the auxiliary mass
${\mu'\to-\infty}$, the one-loop matching formula \eqref{UVlevel} for $k_{\rm uv}$
implies 
\begin{equation}
k_{\rm uv} \,=\, k \,-\, \ha c_2(\Lambda') \,=\, k - \ha c_2(\Lambda)
- d \,=\, 0\,.
\end{equation}
Otherwise for ${d < 0}$, we set 
\begin{equation}
\Lambda' \,=\, \Lambda^* \!\oplus  \big[{\bf +1}\big]^{|d|} \oplus
\big[{\bf -1}\big]^{|d|}\,,
\end{equation}
so that $\Lambda_{\rm uv}$ is real.  To account for the sign of $d$,
we now give the auxiliary
$\Lambda^*$-multiplet a real mass $\mu'$, and we  
give the other auxiliary multiplets with $U(1)$-charges $\pm 1$ the opposite
real mass $-\mu'$.  In the limit ${\mu'\to-\infty}$, once again
\begin{equation}
k_{\rm uv} \,=\, k \,-\, \ha c_2(\Lambda) \,+\, |d| \,\underset{d<0}{=}\, k - \ha
c_2(\Lambda) \,-\, d \,=\, 0\,. 
\end{equation}
For ease in the following discussion, we shall assume ${d \ge 0}$.

For non-abelian $G$, the same construction of $\Lambda'$ works so long
as a fundamental representation ${\bf F}$ exists to play the role
of the unit-charge representation of $U(1)$.  Here, ${\bf F}$ must be a
representation with Casimir ${c_2({\bf F})=1}$ or ${c_2({\bf F})=2}$,
depending upon whether ${\bf F}$ is a complex representation or not.

For example, if ${G=SU(N)}$, then ${\bf F}$ is the standard
\mbox{$N$-dimensional} representation.  If ${G=Sp(2N)}$, ${\bf F}$ is the
defining \mbox{$2N$-dimensional} representation.  In both instances, 
${c_2({\bf F})=1}$ takes the minimal non-zero value with our
normalization conventions, and the preceding
choice \eqref{AuxLambda} for $\Lambda'$ immediately extends to\footnote{The fundamental representation for $Sp(2N)$ is pseudoreal, with ${{\bf F}^* \!\simeq {\bf F}}$ in
  \eqref{AuxLambdaII}.} 
\begin{equation}\label{AuxLambdaII}
\Lambda' \,=\, \Lambda^*\oplus {\bf F}^{\oplus d} \oplus
\left({\bf F}^*\right){}^{\!\oplus d}\,,\qquad\qquad G=SU(N),\,Sp(2N)\,.
\end{equation}
For $Spin(N)$, the natural guess for ${\bf F}$ is
the \mbox{$N$-dimensional} vector representation.  In this case,
though, ${c_2({\bf F})=2}$.  Since the vector representation is
already real, we set 
\begin{equation}\label{AuxLambdaIII}
\Lambda' \,=\, \Lambda^*\oplus {\bf F}^{\oplus d}\,,\qquad\qquad
G=Spin(N),\,G_2\,.
\end{equation}
The factor of two in the Casimir is compensated by summing only half
as many copies of the fundamental representation ${\bf F}$ above.  For
either ansatz in \eqref{AuxLambdaII} or \eqref{AuxLambdaIII}, one
easily checks that ${k_{\rm uv}=0}$ and ${\Lambda_{\rm
    uv}=\Lambda\oplus\Lambda'}$ is real.

What about the exceptional Lie groups?  The group $G_2$ embeds as a
subgroup of $SO(7)$ and so does possess a fundamental representation
${\bf F}$, with dimension seven.  According to Table \ref{CasTableG2}
in Appendix \ref{se:LieAlgConvention}, ${c_2({\bf F})=2}$.  Because
${\bf F}$ is manifestly real, the choice for $\Lambda'$ in
\eqref{AuxLambdaIII} also works for $G_2$.

For the exceptional Lie groups $F_4$ and $E_{6,7,8}$, a proper
fundamental representation ${\bf F}$ does not exist, in the sense that
no representation has sufficiently small Casimir.
Based upon the data in \cite{McKay:1981}, we record in Table
\ref{ExcCasmrs} the representations ${\bf F}$ which have minimal non-zero
Casimir for each of the exceptional Lie groups.  As ${c_2({\bf
    F})\ge 6}$ for $F_4$ and $E_{6,7,8}$, the construction of  
$\Lambda'$ in \eqref{AuxLambdaII} and \eqref{AuxLambdaIII} no longer
works for all values of $k$.  Eg.~for ${G=E_6}$ with ${{\bf F}={\bf
    27}}$ (a complex representation), the ansatz for $\Lambda'$ in
\eqref{AuxLambdaII} applies whenever the difference ${k - \ha
  c_2(\Lambda)}$ is divisible by ${c_2({\bf F})=6}$,
but not otherwise.  More generally, the collected Casimir values in
\cite{McKay:1981} suggest that all other choices for $\Lambda'$ also
fail to work.

\renewcommand{\arraystretch}{1.5}
\begin{table}[t]
\begin{center}
\begin{tabular}{c | c | c | c | c | c } 
& $G_2$ & $F_4$ & $E_6$ & $E_7$ & $E_8$ \\ \hline
$\dim({\bf F})$ & {\bf 7} & {\bf 26} & {\bf 27} & {\bf 56} & {\bf 248} \\ \hline
$c_2({\bf F})$ & 2 & 6 & 6 & 12 & 60 
\end{tabular}\caption{Representation with minimal non-zero value of the
  quadratic Casimir, for each of the exceptional Lie groups.  Only the
seven-dimensional representation of $G_2$ is fundamental, in the sense
that an arbitrary Chern-Simons level in $G_2$ gauge theory can be
induced by an appropriate number of massive, fundamental $G_2$
multiplets.}\label{ExcCasmrs}
\end{center}
\end{table}
\renewcommand{\arraystretch}{1.0}

Apparently, for the exceptional Lie groups other than $G_2$,
not every value for the infrared Chern-Simons level which would be
allowed by anomaly-cancellation can actually be achieved by
integrating-out massive, auxiliary matter at level zero in the
ultraviolet.  This fact prevents us from reducing the arbitrary
supersymmetric Chern-Simons-matter theory to level zero with
real matter, unless the gauge group is $SU(N)$, $Spin(N)$, $Sp(2N)$, or
$G_2$.  For the remainder of this paper we restrict to those cases,
since the conditions on ${k_{\rm uv}}$ and $\Lambda_{\rm uv}$ will be 
necessary for the application of the multi-dimensional Jordan lemma in
Section \ref{HighRank}.  

Nonetheless, we believe that the Factorization Conjecture
\eqref{FACTOR} is true in full generality for all simple Lie groups, and
the obstruction for exceptional gauge groups $F_4$ and $E_{6,7,8}$ is merely a
technical limitation of our argument.

\paragraph{Additional Remarks About Decoupling of Massive Blocks.}

In the case of SQED, we have provided a detailed analysis of
the decoupling of massive auxiliary blocks, thereby completing the
right-hand, downwards vertical arrow in Figure \ref{Schematic}.  An
identical analysis can be performed in the general non-abelian
gauge theory using the expressions for the holomorphic blocks provided
later in Section \ref{LatticeSums}.  Because the analysis is tedious and
involves no new ideas, we only briefly sketch its structure.

Provided the integral over $\h$ converges, a condition which we
examine in Section \ref{SUSYBreak}, the ultraviolet sphere partition
function $Z_{S^3}^{\rm uv}$ at level ${k_{\rm uv}=0}$ can be written
as a finite sum, schematically  
\begin{equation}\label{NonAbBL}
\begin{aligned}
&Z_{S^3}^{\rm uv} \,=\, \frac{\left(2\pi\right)^r\e{i\eta_0}}{|\fW|\cdot\Vol(T)}
\mskip-20mu\sum_{\substack{\textrm{admissible}\,\triangle_\bullet,\\
    \mathbf{r},\mathbf{s}\,\in\,(\Z/d_1\Z)\tau_1\times\cdots\times(\Z/d_r\Z)\tau_r}}\mskip-10mu
\frac{\exp{\!\left(
    i\,\Theta^{\triangle_\bullet}_{\mathbf{r},\mathbf{s},{\rm uv}}\right)}}{|\Lambda_{\rm
    W}:\Lambda_{\triangle_\bullet}|}\,\,\RB^{\triangle_\bullet}_{\mathbf{r},\mathbf{s},{\rm
  uv}}(\Rq,\Rx,\Rx')\,\wt\RB^{\triangle_\bullet}_{\mathbf{r},\mathbf{s},{\rm
uv}}(\wt\Rq,\wt\Rx,\wt\Rx')\,+\,\\
&\,+\, \frac{\left(2\pi\right)^r\e{i\eta_0}}{|\fW|\cdot\Vol(T)}
\mskip-20mu\sum_{\substack{\textrm{admissible}\,\triangle'_\bullet,\\
    \mathbf{r}',\mathbf{s}'\,\in\,(\Z/d_1\Z)\tau_1\times\cdots\times(\Z/d_r\Z)\tau_r}}\mskip-25mu
\frac{\exp{\!\left(
    i\,\Theta^{\triangle'_\bullet}_{\mathbf{r}',\mathbf{s}',{\rm aux}}\right)}}{|\Lambda_{\rm
    W}:\Lambda_{\triangle'_\bullet}|}\,\,\RC^{\triangle'_\bullet}_{\mathbf{r}',\mathbf{s}',{\rm
  aux}}(\Rq,\Rx,\Rx')\,\wt\RC^{\triangle'_\bullet}_{\mathbf{r}',\mathbf{s}',{\rm
aux}}(\wt\Rq,\wt\Rx,\wt\Rx')\,.
\end{aligned}
\end{equation}
The finite indices of summation are explained in Section \ref{LatticeSums} and
are related to the solution of a combinatorial problem involving the
weights of the representation ${\Lambda_{\rm uv} =
  \Lambda \oplus \Lambda'}$.  Otherwise,
$\RB^{\triangle_\bullet}_{\mathbf{r},\mathbf{s},{\rm uv}}$ and
$\wt\RB^{\triangle_\bullet}_{\mathbf{r},\mathbf{s},{\rm uv}}$ are blocks which arise
solely from the contributions of poles on the Coloumb branch
associated to the light matter $\Lambda$, while
$\RC^{\triangle'_\bullet}_{\mathbf{r}',\mathbf{s}',{\rm aux}}$ and 
$\wt\RC^{\triangle'_\bullet}_{\mathbf{r}',\mathbf{s}',{\rm aux}}$ involve
contributions from poles associated to the massive auxiliary matter
$\Lambda'$ and whose location on the Coulomb branch depends
upon the auxiliary real mass $\mu'$.

Explicit formulas for
$\Theta^{\triangle'_\bullet}_{\mathbf{r}',\mathbf{s}',{\rm aux}}$ and 
the non-abelian blocks can be found later in \eqref{BigNonAbPhas},
\eqref{BigWnab}, and \eqref{FormBlkSM}.  From these formulas, one
checks the following assertions.

\begin{enumerate}
\item In the decoupling limit ${\mu'\to-\infty}$ and ${\Rx'\to 0}$, the
ultraviolet blocks $\RB^{\triangle_\bullet}_{\mathbf{r},\mathbf{s},{\rm uv}}$ and
$\wt\RB^{\triangle_\bullet}_{\mathbf{r},\mathbf{s},{\rm uv}}$ reduce
by construction to infrared blocks for the pair $(G,\Lambda)$ at level
$k$.  
\item The main subtlety concerns the behavior in the limit 
${\mu'\to-\infty}$ of the
auxiliary blocks $\RC^{\triangle'_\bullet}_{\mathbf{r}',\mathbf{s}',{\rm aux}}$ and 
$\wt\RC^{\triangle'_\bullet}_{\mathbf{r}',\mathbf{s}',{\rm aux}}$.  As
in \eqref{AuxBlC} and \eqref{ScalWaux}, each block
$\RC^{\triangle'_\bullet}_{\mathbf{r}',\mathbf{s}',{\rm
    aux}}(\Rq,\Rx,\Rx')$ is an
infinite sum of terms, and each term diverges like a power of $\Rx'$ in the
limit ${\Rx'\to 0}$.  Hence the auxiliary blocks
$\RC^{\triangle'_\bullet}_{\mathbf{r}',\mathbf{s}',{\rm aux}}$ and
$\wt\RC^{\triangle'_\bullet}_{\mathbf{r}',\mathbf{s}',{\rm aux}}$
diverge in the decoupling limit.  
\item The phase
$\Theta^{\triangle'_\bullet}_{\mathbf{r}',\mathbf{s}',{\rm aux}}$
includes a quadratic dependence on $\mu'$, similar to the abelian
phase in \eqref{AuxTHlII}.  Upon
analytic continuation of parameters, the decay of the prefactor $\exp{(
    i\,\Theta^{\triangle'_\bullet}_{\mathbf{r}',\mathbf{s}',{\rm
        aux}})}$ in the second line of \eqref{NonAbBL} dominates
the termwise divergences of $\RC^{\triangle'_\bullet}_{\mathbf{r}',\mathbf{s}',{\rm aux}}$ and 
$\wt\RC^{\triangle'_\bullet}_{\mathbf{r}',\mathbf{s}',{\rm aux}}$.
The sum over auxiliary blocks in \eqref{NonAbBL} vanishes in the limit
${\mu'\to-\infty}$.
\end{enumerate} 

\subsection{Some Convergence Criteria}\label{SUSYBreak}

We now examine a fundamental issue regarding the ultraviolet partition
function $Z_{S^3}^{\rm uv}$, with ${k_{\rm uv}=0}$ and ${\Lambda_{\rm
  uv}\simeq\bigoplus_j\,[\lambda_j]}$ a real representation of $G$.  As an integral over the Cartan subalgebra ${\h\subset\g}$,
\begin{equation}\label{ZuvCmbIV}
\begin{aligned}
Z_{S^3}^{\rm uv} \,=\, \frac{\e{i
    \eta_0}}{|\fW|\cdot\Vol(T)}\int_{\mathfrak{h}}\!d^r\!\sigma&\prod_{\alpha\in\Delta_+}\left[4\,\sinh\!\left(\frac{b
      \langle\alpha,\sigma\rangle}{2}\right)\sinh\!\left(\frac{
      \langle\alpha,\sigma\rangle}{2
    b}\right)\right]\times\\
&\times\prod_{j=1}^n\left[\prod_{\beta \in
\Delta_j}
s_b\!\left(\frac{\langle\beta,\sigma\rangle}{2\pi}\,+\,\mu_j\right)\right].
\end{aligned}
\end{equation}
The product over positive roots ${\alpha\in\Delta_+}$ in the upper
line of \eqref{ZuvCmbIV} grows exponentially as $|\sigma|\to\infty$ on
$\h$, since
\begin{equation}\label{VectorL}
\begin{aligned}
\prod_{\alpha\in\Delta_+}\left[4\,\sinh\!\left(\frac{b
      \langle\alpha,\sigma\rangle}{2}\right)\sinh\!\left(\frac{
      \langle\alpha,\sigma\rangle}{2
    b}\right)\right] \,&\underset{|\sigma|\to\infty}{=}\,
\exp{\!\left[\frac{Q}{2}\sum_{\alpha\in\Delta_+}
    \big|\langle\alpha,\sigma\rangle\big| \,+\, O(1)\right]}\,,\\
&\underset{|\sigma|\to\infty}{=}\,\exp{\!\Big[\frac{Q}{2}\,||\sigma||_{\g} \,+\,
  O(1)\Big]}.
\end{aligned}
\end{equation}
In passing to the second line, we recognize the sum over positive
roots $\alpha$ as an instance of the norm in \eqref{RepNorm}, with ${V\equiv\g}$
the adjoint representation.  By comparison, we have already
observed in \eqref{ASYMbigG} that the product over weights $\beta$ of
$\Lambda_{\rm uv}$ decays exponentially,
\begin{equation}\label{ChiralL}
\prod_{j=1}^n\left[\prod_{\beta \in
\Delta_j}
s_b\!\left(\frac{\langle\beta,\sigma\rangle}{2\pi}\,+\,\mu_j\right)\right]
\,\underset{|\sigma|\to\infty}{=}\,\exp{\!\left[-\sum_{j=1}^n\left(\frac{Q}{2}
      + i\,\mu_j\right) ||\sigma||_{\lambda_j}
    \,+\, O(1)\right]}.
\end{equation}
The one-loop factors in \eqref{VectorL} and \eqref{ChiralL} for the
${\CN=2}$ supersymmetric vector and chiral multiplets thus compete in
magnitude as ${|\sigma|\to\infty}$.  

When does the ultraviolet integral in
\eqref{ZuvCmbIV} converge?

Recall that the R-charge $\RR$ determines the imaginary part of the
complexified mass ${\mu \equiv \mu_\R + \frac{i}{2}\,Q\,\RR}$.
Evidently, the decay in \eqref{ChiralL} dominates the growth in
\eqref{VectorL} provided
\begin{flalign}\label{ConvCritG}
\fbox{convergence criterion} \qquad\quad&
\sum_{j=1}^n \left(1-\RR_j\right) ||\sigma||_{\lambda_j} \,-\,
||\sigma||_{\g} \ge 0\,,\qquad \forall\sigma\in\h\,,&
\end{flalign}
with real squashing parameter ${b\in\R_+}$.  The fact that the matrix integral 
may converge only for appropriate parameter values has been noted
previously in
\cite{Aharony:2013dha,Benini:2011mf,Kapustin:2010mh,Safdi:2012re,Willett:2011gp},
where similar versions of the convergence criterion appear.

The condition in \eqref{ConvCritG} places an upper-bound on the values
of the R-charges $\RR_j$, and it also shows that the matrix integral
converges absolutely when each $\RR_j$ is sufficiently negative.
Typically, the latter
regime violates the unitarity bound for dimensions of
gauge-invariant chiral operators in the ${\CN=2}$ superconformal
algebra.  But once defined for negative values of $\RR_j$, the
partition function $Z_{S^3}^{\rm uv}$ can be defined for positive
values by analytic continuation in ${\mu\in\C^n}$.

We have already encountered a special
case of the convergence criterion for gauge group $SU(2)$ in Section \ref{SU2GT}.
For $SU(2)$, the norm ${||\,\cdot\,||_V}$ is straightforward to
evaluate explicitly as a function of the highest-weight for $V$.  As in
\eqref{Psi1LamSU2} with ${\h\simeq\R}$, we find
\begin{equation}\label{ConvCritSU2}
||\sigma||_V \,=\, \begin{cases}
&\frac{1}{4} L \left(L+2\right) |\sigma|\,,\qquad L\in 2\Z\,,\\
&\frac{1}{4} \left(L+1\right)^2 |\sigma|\,,\qquad L\in 2\Z+1\,,
\end{cases}
\qquad\quad \dim V=L+1\,.
\end{equation}
Note that the norm $||\,\cdot\,||_V$ does {\sl not} depend
analytically on the weight $L_j$ even for $SU(2)$.  The general
convergence criterion in \eqref{ConvCritG} then specializes 
to the condition for $SU(2)$ in \eqref{SU2Bound2}.

In many examples, one can argue that each ${\RR_j>0}$ must be
positive by unitarity.  For SQCD, as we consider shortly, this (weak)
statement follows from the existence of chiral meson operators.  The
convergence criterion in \eqref{ConvCritG} then implies, as a necessary
condition for convergence of the matrix integral, 
\begin{equation}\label{ConvCritGU}
\sum_{j=1}^n ||\sigma||_{\lambda_j} \,-\, ||\sigma||_{\g} \ge
0\,,\qquad\qquad\forall\sigma\in\h\,.
\end{equation}
The necessary condition in \eqref{ConvCritGU} is universal insofar as it depends
only on the pair $(G,\Lambda_{\rm uv})$ and the associated algebra
norms $||\,\cdot\,||_V$.

Our goal in the remainder of Section \ref{SUSYBreak} will be to explore
the meaning of the universal condition in \eqref{ConvCritGU}.  Ideally, one
would like to characterize all pairs $(G,\Lambda_{\rm uv})$ for which
the bound in \eqref{ConvCritGU} is true.  We have not been able to
solve this problem in complete generality, but we will give various
examples of representations which either do or do not satisfy the
bound.

\paragraph{Properties of the $L^1$-Norm.}

Let us first mention some properties of the
norm $||\,\cdot\,||_V$ on the Cartan subalgebra,
\begin{equation}\label{CoulNorm}
||\sigma||_V \,=\,
\ha\sum_{\beta\in\Delta_V}\big|\langle\beta,\sigma\rangle\big|\,,\qquad\qquad
\sigma\in\h\,,
\end{equation}
where $V$ is a non-trivial representation of the simple Lie group $G$.
By definition,
\begin{equation}
||\sigma||_{V\oplus W} \,=\, ||\sigma||_V + ||\sigma||_W\,,
\end{equation}
so we will assume $V$ to be irreducible without loss.  Also, if $V^*$
is dual to $V$, then the weights of $V^*$ are related to those of $V$
by the inversion ${\beta\mapsto-\beta}$.  Hence
\begin{equation}\label{DualNorm}
||\sigma||_{V^*} =\, ||\sigma||_V\,.
\end{equation}
Finally, the norm $||\,\cdot\,||_V$ is invariant under the action of the
Weyl group $\fW$ on ${\sigma\in\h}$.  This statement follows from the
fact that $\fW$ permutes the weights ${\beta\in\Delta_V}$ and hence
just rearranges terms in the sum.  Weyl-invariance of
$||\,\cdot\,||_V$ is a physical consequence of the residual, unbroken
gauge symmetry on the Coulomb-branch.

Because $||\,\cdot\,||_V$ is Weyl-invariant, $||\,\cdot\,||_V$ extends
to a $G$-invariant norm on the full Lie algebra $\g$.  Once we unravel
definitions as in \eqref{QadCIII}, the Lie algebra norm can be presented
in a manifestly invariant fashion as\footnote{See Appendix 
  \ref{se:LieAlgConvention} for all Lie algebra notation and conventions.} 
\begin{equation}
||x||_V \,=\, \ha\Tr_V\!\left[\sqrt{-\varphi(x)^2}\right],\qquad\qquad
  x\,\in\,\g\,,\qquad\qquad \varphi:\g\to\End(V)\,.
\end{equation}
Here $\varphi$ is the homomorphism associated to the representation
$V$.  Acting on $V$, the matrix ${-\varphi(x)^2=\varphi(x)^\dagger\varphi(x)}$ is
hermitian and so has a positive square-root, whose trace defines the
$G$-invariant norm.

By comparison, in terms of the invariant metric
\begin{equation}\label{MetricN}
\left(x,y\right) := -\Tr(x y)\,,\qquad\qquad x,y\,\in\,\g\,,
\end{equation}
normalized so that the highest root has length $\sqrt{2}$,
we have 
\begin{equation}
\ha\sqrt{\Tr_V\!\big[-\varphi(x)^2\big]} \,=\, \ha\sqrt{c_2(V)}\cdot
|x|\,,\qquad\qquad |x| \equiv (x,x)^{1/2}\,.
\end{equation}
Convexity of the square-root then implies the lower-bound
\begin{flalign}\label{LowerB}
\fbox{lower-bound}\qquad\qquad &
\ha\sqrt{c_2(V)}\cdot|x| \,\le\, ||x||_V\,,\qquad\qquad x\,\in\,\g\,.&
\end{flalign}
For ${\sigma\in\h}$, the inequality follows
equivalently from  
\begin{equation}
\sqrt{\sum_{\beta\in\Delta_V} \langle\beta,\sigma\rangle^2} \le
\sum_{\beta\in\Delta_V} \big|\langle\beta,\sigma\rangle\big|\,,
\end{equation}
where we use the identification of the Casimir $c_2(V)$ in
\eqref{QadC}.  The lower-bound in \eqref{LowerB} is not optimal,
insofar as equality need not be achieved for any ${x\in\g}$.

For the special case of the adjoint representation, the asymptotic
norm ${||\,\cdot\,||_\g}$ can be recast in a simple way.  By
Weyl-invariance, we may assume without loss that ${\sigma\in\h}$ lies
in the positive Weyl chamber $\mathcal{C}_+$, so that
${\langle\alpha,\sigma\rangle\ge 0}$ for each positive root
${\alpha\in\Delta_+}$.  Then from the definition \eqref{CoulNorm},
\begin{equation}\label{AdjNorm}
||\sigma||_\g \,=\, \sum_{\alpha\,\in\,\Delta_+}
\langle\alpha,\sigma\rangle \,=\, 2 \, \langle\rho,\sigma\rangle\,,\qquad\qquad
\sigma \,\in\,\mathcal{C}_+\,,
\end{equation}
where ${\rho\in\h^*}$ is the ubiquitous Weyl element
\begin{equation}\label{Weylvec}
\rho \,=\, \ha \sum_{\alpha\in\Delta_+}\alpha\,.
\end{equation}
The prefactor of one-half in \eqref{CoulNorm} is cancelled by the sum
over pairs $\pm\alpha$ of positive and negative roots.  Alternatively,
$\rho$ is the sum of fundamental weights
$\{\hat\omega^1,\ldots,\hat\omega^r\}$,
\begin{equation}
\rho \,=\, \hat\omega^1\,+\,\cdots\,+\,\hat\omega^r\,.
\end{equation}
So when $\sigma$ is expressed in the basis of simple coroots $\{\hat
h_1,\ldots,\hat h_r\}$ canonically dual to the fundamental weights,
ie.~${\langle\hat\omega^j,\hat h_\ell\rangle=\delta^j_\ell}$,
\begin{equation}\label{SCoRoot}
\sigma \,=\, \sigma^1 \, \hat h_1 \,+\, \cdots \,+\, \sigma^r \, \hat h_r\,,
\end{equation}
the norm of $\sigma$ becomes the sum of components
\begin{equation}\label{AdjNormII}
||\sigma||_\g \,=\, 2 \left(\sigma^1 \,+\, \cdots \,+\,
  \sigma^r\right),\qquad\qquad \sigma\,\in\,\mathcal{C}_+\,,
\end{equation}
where ${\sigma^\ell \ge 0}$ for each ${\ell=1,\ldots,r}$.  As a check,
the factor of two on the right in \eqref{AdjNormII} corresponds to the
same factor of two on the right in the bound \eqref{SU2Bound2} for
$SU(2)$.

We use the description in \eqref{AdjNorm} to
place a sharp upper-bound on $||\,\cdot\,||_\g$ in terms of the standard invariant
metric $(\,\cdot\,,\,\cdot\,)$ on $\g$ in \eqref{MetricN}.  By definition, 
\begin{equation}\label{OptUp}
||x||_\g \,\le\, \frac{|x|}{C_\g}\,,\qquad\qquad x\in\g\,,\qquad\qquad |x|\equiv
(x,x)^{1/2}\,,
\end{equation}
where $C_\g$ is the positive constant obtained by minimizing
\begin{equation}
C_\g \,=\,
\min_{\sigma\in\mathcal{C}_+}\left\{|\sigma|\,\Big|\,\langle\rho,\sigma\rangle=\ha\right\}\,.
\end{equation}
The minimum of $|\sigma|$ coincides with the minimum of the square
$(\sigma,\sigma)$, and minimizing $(\sigma,\sigma)$ on the hyperplane ${\langle\rho,\sigma\rangle=1/2}$ can be easily
accomplished with a Lagrange multiplier.  We solve the
simultaneous linear equations 
\begin{equation}
2\,(\sigma_{\rm min},\,\cdot\,) \,=\,
  \lambda\,\rho\,,\qquad\qquad\langle\rho,\sigma_{\rm min}\rangle
  \,=\, \ha\,,
\end{equation}
with Lagrange multiplier ${\lambda\in\R}$ to find  
\begin{equation}
\sigma_{\rm min} \,=\, \ha\frac{(\rho,\,\cdot\,)}{(\rho,\rho)}\,,\qquad\qquad \lambda \,=\, \frac{1}{(\rho,\rho)}\,.
\end{equation}
Note that $\sigma_{\rm min}$ lies in the positive Weyl chamber as required.
Thus
\begin{equation}
C_\g \,=\, |\sigma_{\rm min}| \,=\, \frac{1}{2\,|\rho|}\,.
\end{equation}
According to the Freudenthal-de Vries formula, proven for
instance in  Ch~13.4.2 of \cite{DiFrancesco:1997nk},
\begin{equation}
(\rho,\rho) \,=\, \frac{h_\g}{12}\,\dim \g\,,
\end{equation}
where $h_\g$ is the dual Coxeter number.  Substituting for $C_\g$ in
\eqref{OptUp} we obtain as the optimal upper-bound 
\begin{flalign}\label{UpperB}
\fbox{upper-bound} \qquad\qquad&
||x||_\g \,\le\, \sqrt{\frac{h_\g \, \dim \g}{3}}\cdot
|x|\,,\qquad\qquad x\in\g\,.&
\end{flalign}

As a small test, ${h_\g=2}$
and ${\dim\g=3}$ for $SU(2)$, so the bound becomes
\begin{equation}
||x||_{\g} \,\le\, \sqrt{2}\cdot|x|\,,\qquad\qquad G=SU(2)\,.
\end{equation}
One can check directly that the $SU(2)$ bound is correct, with
equality in this case.  As a further consistency check, the combined
inequalities in \eqref{LowerB} and \eqref{UpperB} assert, with ${V=\g}$ and
${c_2(\g)=2 h_\g}$, that ${\dim \g > 1}$, true for any simple Lie algebra.

A convenient way to picture a norm is to draw the unit sphere for it.  In Figures \ref{SU3AdjNorm} and \ref{SU3FundNorm}, we show the unit sphere for the norms $||\,\cdot\,||_\g$ and $||\,\cdot\,||_{\bf 3}$ associated to the adjoint and fundamental representations of $SU(3)$.  By the underlying Weyl symmetry, each ``sphere'' appears as a hexagon in the Cartan subalgebra, but one hexagon is rotated relative to the other.
  
According to the general description of $||\,\cdot\,||_\g$ in \eqref{AdjNorm}, the unit ball in the adjoint norm is always the convex polyhedron made from Weyl-translates of the fundamental Weyl alcove, up to scale.

\iffigs
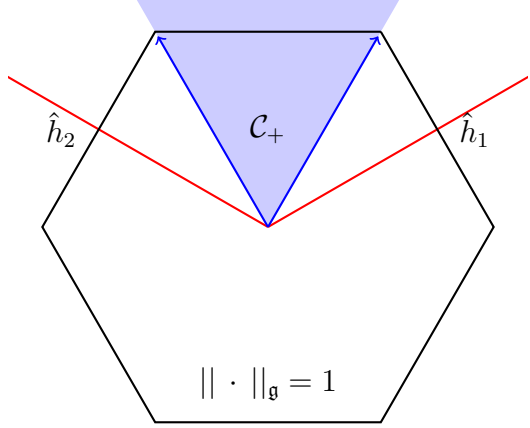
\begin{figure}[t]
\centering
	\begin{tikzpicture}[x=1.5cm,y=1.5cm] 
	\coordinate (0;0) at (0,0); 
	\foreach \c in {1,...,4}{%  
		\foreach \i in {0,...,5}{% 
			\pgfmathtruncatemacro\j{\c*\i}
			\coordinate (\c;\j) at (60*\i:\c);  
		} }
		\foreach \i in {0,2,...,10}{% 
			\pgfmathtruncatemacro\j{mod(\i+2,12)} 
			\pgfmathtruncatemacro\k{\i+1}
			\coordinate (2;\k) at ($(2;\i)!.5!(2;\j)$) ;}
		
		\foreach \i in {0,3,...,15}{% 
			\pgfmathtruncatemacro\j{mod(\i+3,18)} 
			\pgfmathtruncatemacro\k{\i+1} 
			\pgfmathtruncatemacro\l{\i+2}
			\coordinate (3;\k) at ($(3;\i)!1/3!(3;\j)$)  ;
			\coordinate (3;\l) at ($(3;\i)!2/3!(3;\j)$)  ;
		}
		
		\foreach \i in {0,4,...,20}{% 
			\pgfmathtruncatemacro\j{mod(\i+4,24)} 
			\pgfmathtruncatemacro\k{\i+1} 
			\pgfmathtruncatemacro\l{\i+2}
			\pgfmathtruncatemacro\m{\i+3} 
			\coordinate (4;\k) at ($(4;\i)!1/4!(4;\j)$)  ;
			\coordinate (4;\l) at ($(4;\i)!2/4!(4;\j)$) ;
			\coordinate (4;\m) at ($(4;\i)!3/4!(4;\j)$) ;
		}  
		
		\begin{scope}
		
		\clip ($(2;0)+(2;3)+(.3,.3)$)--($(2;6)+(2;3)+(-.3,.3)$)--($(2;6)+(2;9)+(-.3,-.3)$)--($(2;0)+(2;9)+(.3,-.3)$)--($(2;0)+(2;3)+(.3,.3)$);

		\fill[blue!20](0;0)--(3;3) -- (3;6) -- (0;0);

		\foreach \n in{1,5}{%
			\draw[thick ,red](0;0)--($2*(2;\n)$);
		}	
		
		\end{scope}

		\node [  right] at ($(2;1)+(.1,0)$) {$\hat h_1$};
		\node [  left] at ($(2;5)+(-.1,0)$) {$\hat h_2$};
		
		\draw[->,blue,thick,shorten >=2pt](0;0)--(2;2);
		\draw[->,blue,thick,shorten >=2pt](0;0)--(2;4);
		
		\draw[black, thick] (2;2) -- (2;4)--(2;6)--(2;8)--(2;10)--(2;0)--(2;2);
		
		\node at ($ .5*(2;3) $) {$\mathcal{C}_+$};

		\node at ($ .8*(2;9) $) {$||\,\cdot\,||_{\mathfrak{g}}=1$};		

		\end{tikzpicture}  
		
		\caption{The Cartan algebra $\mathfrak{h}$ of $SU(3)$, with the positive Weyl chamber $\mathcal{C}_+$ shaded in blue. The unit sphere in the asymptotic norm $||\,\cdot\,||_{\mathfrak{g}}$ is drawn in black, and rays through the simple coroots in red.}\label{SU3AdjNorm}
	\end{figure}

\begin{figure}[t]
\centering
	\begin{tikzpicture}[x=1.2cm,y=1.2cm] 

	\coordinate (0;0) at (0,0); 
	\foreach \c in {1,...,4}{%  
		\foreach \i in {0,...,5}{% 
			\pgfmathtruncatemacro\j{\c*\i}
			\coordinate (\c;\j) at (60*\i:\c);  
		} }
		\foreach \i in {0,2,...,10}{% 
			\pgfmathtruncatemacro\j{mod(\i+2,12)} 
			\pgfmathtruncatemacro\k{\i+1}
			\coordinate (2;\k) at ($(2;\i)!.5!(2;\j)$) ;}
		
		\foreach \i in {0,3,...,15}{% 
			\pgfmathtruncatemacro\j{mod(\i+3,18)} 
			\pgfmathtruncatemacro\k{\i+1} 
			\pgfmathtruncatemacro\l{\i+2}
			\coordinate (3;\k) at ($(3;\i)!1/3!(3;\j)$)  ;
			\coordinate (3;\l) at ($(3;\i)!2/3!(3;\j)$)  ;
		}
		
		\foreach \i in {0,4,...,20}{% 
			\pgfmathtruncatemacro\j{mod(\i+4,24)} 
			\pgfmathtruncatemacro\k{\i+1} 
			\pgfmathtruncatemacro\l{\i+2}
			\pgfmathtruncatemacro\m{\i+3} 
			\coordinate (4;\k) at ($(4;\i)!1/4!(4;\j)$)  ;
			\coordinate (4;\l) at ($(4;\i)!2/4!(4;\j)$) ;
			\coordinate (4;\m) at ($(4;\i)!3/4!(4;\j)$) ;
		}

		\begin{scope}
		
		\clip ($(2;0)+(2;3)+(.3,.3)$)--($(2;6)+(2;3)+(-.3,.3)$)--($(2;6)+(2;9)+(-.3,-.3)$)--($(2;0)+(2;9)+(.3,-.3)$)--($(2;0)+(2;3)+(.3,.3)$);
		
		\foreach \n in{1,5}{%
			\draw[thick ,red](0;0)--($2*(2;\n)$);
		}	
		
		\end{scope}

		\fill[blue!20](0;0)--($1.05*(3;3)$) -- ($1.05*(3;6)$) -- (0;0);

		\node [  right] at ($.75*(2;1)+(.05,-.05)$) {$\hat h_1$};
		\node [  left] at ($.75*(2;5)+(-.05,-.05)$) {$\hat h_2$};
	
		\draw[->,blue,thick,shorten >=2pt](0;0)-- ($1.16*(2;2)$);
		\draw[->,blue,thick,shorten >=2pt](0;0)--($1.16*(2;4)$);
		
		\draw[black, thick] ($1.5*(2;1)$) -- ($1.5*(2;3)$)--($1.5*(2;5)$)--($1.5*(2;7)$)--($1.5*(2;9)$)--($1.5*(2;11)$)--($1.5*(2;1)$);
		
		\node at ($ .5*(2;3) $) {$\mathcal{C}_+$};

		\node at ($ 1.1*(2;9) $) {$||\,\cdot\,||_{\bf 3}=1$};
		
		\end{tikzpicture}  
		
		\caption{The Cartan algebra $\mathfrak{h}$ of $SU(3)$, with the positive Weyl chamber $\mathcal{C}_+$ shaded in blue. The unit sphere in the asymptotic norm $||\,\cdot\,||_{\bf 3}$ is drawn in black, and rays through the simple coroots in red.}\label{SU3FundNorm}
	\end{figure}
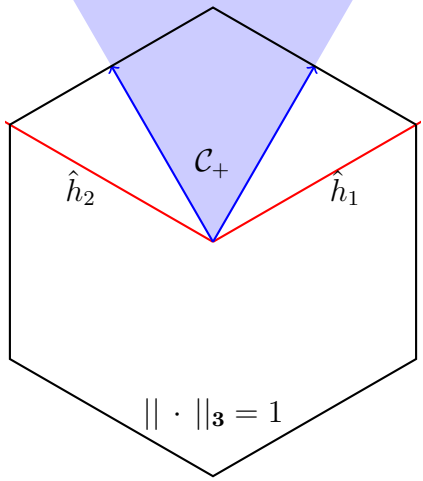
\fi

\paragraph{Numerical Condition for Convergence.}

Together, the lower- and upper-bounds in \eqref{LowerB} and
\eqref{UpperB} can used to show that all but a finite number of
representations of a fixed gauge group $G$ obey the universal necessary condition 
\eqref{ConvCritGU} for convergence of the matrix integral.  According
to those bounds, 
\begin{equation}
\sum_{j=1}^n ||\sigma||_{\lambda_j} \,-\, ||\sigma||_{\g} \,\ge\,
\sum_{j=1}^n \ha \sqrt{c_2(\lambda_j)}\cdot|\sigma| \,-\,
\sqrt{\frac{h_\g \, \dim \g}{3}}\cdot |\sigma|\,.
\end{equation}
The right side is positive if 
\begin{flalign}\label{NumBound}
\fbox{numerical condition}\qquad\qquad&
\sum_{j=1}^n \ha \sqrt{c_2(\lambda_j)} \,\ge\,\sqrt{\frac{h_\g \, \dim
    \g}{3}}\,.&
\end{flalign}
So long as $\Lambda_{\rm uv}$ has no trivial, decoupled summands, this
condition on $c_2$ is satisfied for all but a finite number of
representations.

Let us play with some numbers.  For $SU(N)$ SQCD with $N_{\rm f}$ flavors,
meaning 
\begin{equation}
\Lambda_{\rm uv} \,=\, {\bf N}^{N_{\rm f}}\!\oplus\bar{\bf
  N}{}^{N_{\rm f}}\,,
\end{equation}
we have ${c_2({\bf N}) = c_2(\overline{\bf
    N})=1}$, ${h_\g=N}$, and ${\dim\g=N^2-1}$.
After squaring both sides, the inequality in
\eqref{NumBound} is then equivalent to the condition
\begin{equation}\label{SQCDBound}
N_{\rm f}^2 \,\ge\, \frac{1}{3} N \!\left(N^2 - 1\right).
\end{equation}
Eg.~if ${N=2}$, then ${N_{\rm f} \ge 2}$ is compatible with the
necessary condition in \eqref{ConvCritGU}.

As another example, consider a general theory with gauge group $G$ and
with matter in $N_{\rm adj}$ copies of the adjoint representation,
\begin{equation}
\Lambda_{\rm uv} \,=\, \g^{N_{\rm adj}}\,.
\end{equation}
Since ${c_2(\g)=2 h_\g}$, the numerical condition in \eqref{NumBound} implies
\begin{equation}\label{AdjBound}
N^2_{\rm adj} \,\ge\, \frac{2}{3}\dim\g.
\end{equation}
On the other hand, the universal bound in \eqref{ConvCritGU} reduces
directly to the requirement 
\begin{equation} 
\left(N_{\rm adj}-1\right)\cdot||\sigma||_\g \ge 0\,,\qquad\qquad\forall\sigma\in\h\,,
\end{equation}
or more simply,
\begin{equation}
N_{\rm adj} \,\ge\, 1\,.
\end{equation}
So the numerical condition in \eqref{NumBound} cannot be sharp.

\subsection{Convergence Criteria for Supersymmetric QCD}\label{SusyQCD}  

We can determine more precise
convergence criteria than the numerical condition in \eqref{NumBound} if we examine the $L^1$-norm
$||\,\cdot\,||_V$ in detail for specific representations $V$.  We
start with the gauge group $SU(N)$ 
and take ${V\!={\bf N}}$ to be the fundamental, N-dimensional
representation.  By the observation in \eqref{DualNorm}, our results
apply equally well for the dual, anti-fundamental representation
${V\!=\bar{\bf N}}$.

The Cartan subalgebra of $SU(N)$ can be conveniently parameterized by
traceless diagonal matrices of the form 
\begin{equation}
\sigma \,=\, i
\diag\!\left(\sigma^1,\,\sigma^2-\sigma^1,\,\cdots\,,\sigma^{N-1}-\sigma^{N-2},\,-\sigma^{N-1}\right),
\end{equation}
corresponding to the basis of simple coroots in \eqref{SCoRoot}.  In
these coordinates, the positive Weyl chamber lies in the positive quadrant ${\sigma^1,\ldots,\sigma^{N-1}\ge 0}$, subject to
\begin{equation}
\mathcal{C}_+:\qquad \sigma^1 \,\ge\, \sigma^2 - \sigma^1 \,\ge\,
\cdots \,\ge\, \sigma^{N-1} - \sigma^{N-2}
\,\ge\, -\sigma^{N-1}\,.
\end{equation}
Exactly as in \eqref{AdjNormII}, the adjoint norm on the
positive Weyl chamber is given by 
\begin{equation}\label{AdjNormIII}
||\sigma||_\g \,=\, 2
\left(\sigma^1\,+\,\cdots\,+\,\sigma^{N-1}\right),\qquad\qquad
\sigma\,\in\,\mathcal{C}_+\,.
\end{equation}
The weights of the fundamental representation, when evaluated
on $\sigma$, are the diagonal entries.  Thus by definition \eqref{CoulNorm},
\begin{equation}\label{FundNorm}
\begin{aligned}
||\sigma||_{\textbf{N}} \,&=\, \ha\left[\big|\sigma^1\big| \,+\,
\sum_{j=2}^{N-1}\big|\sigma^j - \sigma^{j-1}\big| \,+\,
\big|\sigma^{N-1}\big|\right].
\end{aligned}
\end{equation}
As promised by the general observations after \eqref{RepNorm}, the sum on
the right of \eqref{FundNorm} is manifestly a Weyl-invariant norm on $\R^{N-1}$.

To put a lower-bound on the fundamental norm $||\,\cdot\,||_{\textbf{N}}$, let
$\sigma_{\rm max}$ be the maximum value in the set of real numbers $\{\sigma^1,\ldots,\sigma^{N-1}\}$, ie.
\begin{equation}
\sigma_{\rm max} \,=\, \max\big\{\sigma^1,\ldots,\sigma^{N-1}\big\}\,.
\end{equation}
For any positive reals ${\sigma^1,\ldots,\sigma^{N-1}\ge 0}$, one can
choose signs so that 
\begin{equation}\label{PosNegSUM}
\pm \sigma^1 \pm \left(\sigma^2 - \sigma^1\right) \pm \,\cdots\,
\pm\left(\sigma^{N-1} - \sigma^{N-2}\right) \pm\sigma^{N-1} \,=\, 2
\sigma_{\rm max}\,.
\end{equation}
Because each member of the set $\{\sigma^1,\ldots,\sigma^{N-1}\}$
appears exactly twice in the sum on the left, a factor of two
multiplies $\sigma_{\rm max}$ on the right in \eqref{PosNegSUM}.  The triangle inequality
then implies
\begin{equation}\label{AbsInEQ}
\sigma^1 \,+\, \sum_{j=2}^{N-1}\big|\sigma^j - \sigma^{j-1}\big| \,+\,
\sigma^{N-1} \,\ge\, 2\sigma_{\rm max}\,.
\end{equation}
See Figure \ref{TrianglePf} for an equivalent pictorial proof of
\eqref{AbsInEQ}.  Applied to the positive Weyl chamber, the inequality
in \eqref{AbsInEQ} provides the immediate bound
\begin{equation}\label{FundBound}
||\sigma||_{\textbf N} \,\ge\, \sigma_{\rm max}\,,\qquad\qquad
\sigma\,\in\,\mathcal{C}_+\,. 
\end{equation}
Moreover, the bound in \eqref{FundBound} is optimal, as equality is
achieved on the wall of the Weyl chamber where ${\sigma^1 = \cdots =
  \sigma^{N-1} = \sigma_{\rm max}}$.  

\iffigs
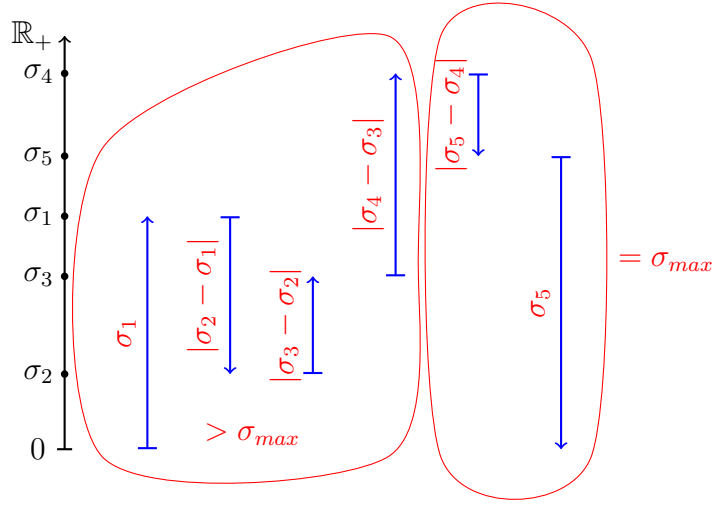
\begin{figure}[t]
\centering
	\begin{tikzpicture}[x=1cm,y=1cm]  % Overall scale of the picture

	%
	%    Set Up Constants
	%
	\coordinate (OFF) at (1.1,0);		% Offset

	% On Axis Coordinates

	\coordinate (x1) at (0,3.1);		% x1
	\coordinate (x2) at (0,1);			% x2
	\coordinate (x3) at (0,2.3);		% x3
	\coordinate (x4) at (0,5);			% x4
	\coordinate (x5) at (0,3.9);		% x5

	%
	%    Draw Axis
	%
	
	\draw [thick, black] (.1,0) -- (-.1,0) node [left] {0};
	\draw [->, thick, black] (0,0) -- (0,5.5) node [left] {$\mathbb{R}_+$};
	
	\fill [black] (x1) circle (.05) node [left] {$\sigma_1$};	
	\fill [black] (x2) circle (.05) node [left] {$\sigma_2$};
	\fill [black] (x3) circle (.05) node [left] {$\sigma_3$};
	\fill [black] (x4) circle (.05) node [left] {$\sigma_4$};
	\fill [black] (x5) circle (.05) node [left] {$\sigma_5$};

	% 
	%    Draw Intervals
	%

	% Note: for the text to have the correct oritentation, must draw downward arrows upwards and reverse | and >.

	\path[|->, blue, thick] ($ (0,0)+(OFF) $) edge  node[sloped, anchor=center, above, red] { $\sigma_1$ }  ($ (x1)+(OFF) $);
	\path[<-|, blue, thick] ($ (x2)+2*(OFF) $) edge  node[sloped, anchor=center, above, red] { $|\sigma_2-\sigma_1|$ }  ($ (x1)+2*(OFF) $);
	\path[|->, blue, thick] ($ (x2)+3*(OFF) $) edge  node[sloped, anchor=center, above, red] { $|\sigma_3-\sigma_2|$ }  ($ (x3)+3*(OFF) $);
	\path[|->, blue, thick] ($ (x3)+4*(OFF) $) edge  node[sloped, anchor=center, above, red] { $|\sigma_4-\sigma_3|$ }  ($ (x4)+4*(OFF) $);
	\path[<-|, blue, thick] ($ (x5)+5*(OFF) $) edge  node[sloped, anchor=center, above, red] { $|\sigma_5-\sigma_4|$ }  ($ (x4)+5*(OFF) $);
	\path[<-|, blue, thick] ($ (0,0)+6*(OFF) $) edge  node[sloped, anchor=center, above, red] { $\sigma_5$ }  ($ (x5)+6*(OFF) $);

	% 
	%    Draw Paths and Labels
	%

	\draw [red] plot [smooth cycle] coordinates {(.5,-.1)  (.5,4) (4.3,5.5) (4.7, 2.5) (4.3,-.1) }; % Left Circle
	\draw [red] plot [smooth cycle] coordinates {(5,-.1)  (5,5.5) (7,5.2) (7,-.1) };		         % Right Circle

	\node [red] at (2.5, .2) {$>\sigma_{max}$};
	\node [red] at (8, 2.5) {$= \sigma_{max}$};

%	\node [black] at (11, .1) {$\blacksquare$};		% QED Box
	\end{tikzpicture}  
	
	\caption{Pictorial proof of the bound for $||\,\cdot\,||_{\bf N}$.}\label{TrianglePf}
\end{figure}
\fi

Likewise, the adjoint norm
in \eqref{AdjNormIII} can be immediately bounded above by $\sigma_{\rm max}$,
\begin{equation}\label{AdjBoundII}
||\sigma||_\g \,\le\, 2 (N-1) \, \sigma_{\rm max}\,,\qquad\qquad \sigma\,\in\,\mathcal{C}_+\,.
\end{equation}

Let us apply the bounds in \eqref{FundBound} and \eqref{AdjBoundII} to
$SU(N)$ SQCD with $N_{\rm f}$ flavors, ${\Lambda_{\rm uv}={\bf
    N}^{N_{\rm f}}\!\oplus{\overline{\bf N}}{}^{N_{\rm f}}}$, each with R-charge $\RR$.
In this case, the convergence criterion \eqref{ConvCritG} becomes 
the positivity condition 
\begin{equation}\label{PosSQCD}
2 N_{\rm f} \left(1-\RR\right) ||\sigma||_{\textbf{N}} \,-\, ||\sigma||_\g
\,\ge\,0\,,\qquad\qquad \forall\sigma\,\in\,\h\,,
\end{equation}
where the factor of two multiplying $N_{\rm f}$ accounts for
quark/anti-quark pairs.  We assume ${\RR>0}$ as required by
unitarity, and note that ${\RR<1}$ is a trivial consequence of the
condition \eqref{PosSQCD}.\footnote{As a check of normalizations, the
  presence of a superpotential ${W = m_{i j} \, Q^i \wt Q^j}$, giving
  mass to each quark/anti-quark pair, implies ${\RR=1}$.  The 
  contribution from $||\,\cdot\,||_{\bf N}$ in \eqref{PosSQCD} then vanishes,
  consistent with the reduction to pure Yang-Mills theory in the
  infrared.}

By Weyl-invariance of the norms, we take $\sigma$ to lie in the positive Weyl chamber, on which 
the lower- and upper-bounds in \eqref{FundBound} and \eqref{AdjBoundII} imply 
\begin{equation}\label{SquSQCDII}
2 N_{\rm f} \left(1-\RR\right) ||\sigma||_{\textbf{N}} \,-\, ||\sigma||_\g \,\ge\, 2
\Big[N_{\rm f}\left(1-\RR\right) - (N-1)\Big] \sigma_{\rm max}\,,\qquad
\sigma\,\in\,\mathcal{C}_+\,. 
\end{equation}
The right side of \eqref{SquSQCDII} is clearly positive when
\begin{flalign}\label{SQCDCrit}
\fbox{SU(N) SQCD}\qquad\qquad\qquad&
N_{\rm f} \left(1-\RR\right)\,\ge\, N-1\,.&
\end{flalign}
Since the inequalities in \eqref{FundBound} and \eqref{AdjBoundII}
are saturated in a neighborhood of the ray
${\sigma^1=\cdots=\sigma^{N-1}>0}$, the condition in \eqref{SQCDCrit}
is necessary for convergence of the SQCD matrix
integral.  When the inequality is strict, this 
condition is also sufficient.  The marginal situation with ${N_{\rm f}
  (1-\RR)=N-1}$ requires further analysis (which we omit) to
determine convergence.

Let us make three remarks about the bound on $N_{\rm f}$ in
\eqref{SQCDCrit}.
\begin{enumerate}
\item For $SU(2)$ SQCD, the bound agrees with the result
  \eqref{SU2Bound1} of the direct analysis in Section \ref{SU2GT}.
\item As an upper-bound on the R-charge, the convergence criterion reads
\begin{equation}
\RR \,\le\,1 - \frac{N-1}{N_{\rm f}}\,.
\end{equation}
Consistency with the weak unitarity bound ${\RR>0}$ requires
\begin{equation}\label{WeakU}
N_{\rm f} \,>\, N-1\,.
\end{equation}
\item For $SU(N)$ SQCD on $\R^{1,2}$ with massless quarks,
  non-perturbative effects generate a superpotential which
  spontaneously breaks supersymmetry
  \cite{Affleck:1982as,Aharony:1997bx} when ${N_{\rm f} < 
    N-1}$, and similar quantum effects deform the classical moduli
  space of vacua when ${N_{\rm f} = N-1}$.  From this perspective, the
  appearance of the critical value for $N_{\rm f}$ on the right in
  \eqref{WeakU} is noteworthy.  

  As we demonstrate next, an identical statement is true
  for SQCD with gauge groups of type Sp and SO.  In those cases as
  well, the bound on $N_{\rm f}$ which follows from the convergence
  criterion with ${\RR>0}$ agrees with the range of $N_{\rm f}$ for
  which either a non-perturbative superpotential spontaneously breaks
  supersymmetry on $\R^{1,2}$, or the classical moduli space of vacua
  on $\R^{1,2}$ is deformed.  We do not have a
  theoretical explanation for this coincidence, but it bears further
  investigation.
 \end{enumerate}

\paragraph{Improved Criteria for Sp(2N) SQCD.}

Similar bounds determine convergence
criteria for SQCD when the gauge group $G$ is any
matrix Lie group, under which the quarks transform in the fundamental
representation ${\bf F}$ of $G$.  

We examine the cases where $G$ is $Sp(2N)$, $Spin(2N)$, and
$Spin(2N+1)$ in turn.\footnote{We work throughout with the
  simply-connected $Spin$ cover of $SO(N)$.  As usual, the geometry of
  $SO(N)$ roots and weights depends upon
whether $N$ is even or odd, so these cases must be analyzed separately.} In all these examples, the fundamental representation 
${{\bf F}\simeq{\bf F}^*}$ is self-dual, so there is no distinction
between quarks and anti-quarks.  By convention for $N_{\rm f}$ flavors, we set
\begin{equation}\label{Flavors}
\Lambda_{\rm uv} \,=\, {\bf F}^{2 N_{\rm f}}\,.
\end{equation}
Note the factor of two!  For $Sp(2N)$ at level ${k=0}$, ${N_{\rm f}}$
must be an integer to prevent global gauge anomalies.  For $Spin(2N)$
or $Spin(2N+1)$, ${N_{\rm f}}$ is allowed to be an integer or
half-integer when ${k=0}$.  As will be clear, our convention for $N_{\rm f}$ allows
us to treat all three cases uniformly, as the SQCD convergence
criterion \eqref{PosSQCD} is then given by 
\begin{equation}\label{PosSQCDF}
2 N_{\rm f} \left(1-\RR\right) ||\sigma||_{\bf F} \,-\, ||\sigma||_\g \,\ge\,
0\,,\qquad\qquad \forall\sigma\in\h\,.
\end{equation}
We assume ${0 < \RR < 1}$ as before, else for ${\RR\ge 1}$ the criterion in
\eqref{PosSQCDF} is immediately violated.

Throughout, we express the Coulomb-branch parameter $\sigma$ in
coordinates adapted to a standard set of simple coroots for $G$, in
terms of which the adjoint norm on the positive Weyl chamber assumes
the universal form in \eqref{AdjNormII}.  For the symplectic group $Sp(2N)$,
we write
\begin{equation}\label{CSASp}
\sigma \,=\, {\bf I}_{2 \times 2}\,
\diag\big(\sigma^1,\,\sigma^2-\sigma^1,\,\cdots,\,\sigma^N -
\sigma^{N-1}\big)\,,\qquad 
{\bf I}_{2 \times 2} \,=\, 
\begin{pmatrix}
i&\,0\\
0&-i
\end{pmatrix}\,.
\end{equation}
Here we abuse notation slightly, as $\sigma$ is a ${2N\times 2N}$
diagonal matrix composed of ${2\times 2}$ blocks proportional to ${\bf
  I}_{\rm 2 \times 2}$.  This presentation for $\sigma$ implicitly
embeds the Cartan subalgebra for $Sp(2N)$ inside the Cartan subalgebra
for $SU(2N)$.  The positive Weyl chamber is the subset of the positive
quadrant ${\sigma^1,\ldots,\sigma^N\ge 0}$ such that
\begin{equation}\label{PosWSp}
\mathcal{C}_+:\qquad \sigma^1 \,\ge\, \sigma^2 - \sigma^1
\,\ge\,\cdots\,\ge\,\sigma^N-\sigma^{N-1}\,\ge\, 0\,,
\end{equation}
on which 
\begin{equation}\label{AdjNormSp}
||\sigma||_\g \,=\, 2 \left(\sigma^1 \,+\,\cdots\,+\,\sigma^N\right),\qquad\qquad \sigma\,\in\,\mathcal{C}_+\,.
\end{equation}

By contrast, for the fundamental representation with dimension $2N$, the
weights are determined by the diagonal entries of $\sigma$,
and\footnote{Because the fundamental representation of $Sp(2n)$ is self-dual,
  the weights appear in plus/minus pairs, which cancel the prefactor of
  one-half in the definition \eqref{CoulNorm} of the norm
  ${||\,\cdot\,||_{\bf 2N}}$.  The same remark applies to the
  $L^1$-norms associated to the vector representations of
  $Spin(2N)$ and $Spin(2N+1)$.}
\begin{equation}\label{SpNorm}
||\sigma||_{\bf 2N} \,=\, \big|\sigma^1\big| \,+\, \sum_{j=2}^N \big|\sigma^j-\sigma^{j-1}\big|\,.
\end{equation}
The right side of \eqref{SpNorm} clearly defines a norm on $\R^N$.
Evidently, on the positive Weyl chamber in \eqref{PosWSp},
\begin{equation}
||\sigma||_{\bf 2N} \,=\, \sigma^N = \sigma_{\rm max}\,,\qquad\qquad \sigma\,\in\,\mathcal{C}_+\,,
\end{equation}
where $\sigma_{\rm max}$ is the maximum in the set
$\{\sigma^1,\ldots,\sigma^N\}$.  Note that the inequalities on the
positive Weyl chamber imply the ordering ${\sigma^N \ge \sigma^{N-1} \ge \cdots \ge
  \sigma^1}$.

Immediately for the convergence criteria in \eqref{PosSQCDF} with ${0
  < \RR < 1}$,
\begin{equation}
2 N_{\rm f} \left(1-\RR\right) ||\sigma||_{\bf 2N} \,-\, ||\sigma||_\g \,\ge\, 2
\Big[N_{\rm f}\left(1-\RR\right) - N\Big] \sigma_{\rm max},\qquad\qquad \sigma\,\in\,\mathcal{C}_+\,,
\end{equation}
with equality when ${\sigma^1=\cdots=\sigma^N}$ on the boundary of the
Weyl chamber.  Thus a necessary condition for convergence of the
matrix integral in $Sp(2N)$ SQCD is 
\begin{flalign}
\fbox{Sp(2N) SQCD}\qquad\qquad\qquad&
N_{\rm f} \left(1-\RR\right) \,\ge\, N\,.&
\end{flalign}
This condition is also sufficient when the inequality is strict.  As a
check, for ${N=1}$ the condition reduces to our previous result for
$SU(2)$ SQCD.  

Evidently, convergence of the $Sp(2N)$ matrix integral places the upper-bound
\begin{equation}
\RR \,\le\, 1 - \frac{N}{N_{\rm f}}\,.
\end{equation}
This bound is compatible with the weak unitarity constraint
${\RR>0}$ only if 
\begin{equation}\label{SpBound}
N_{\rm f} \,>\, N\,.
\end{equation}
By comparison, for $Sp(2N)$ SQCD on $\R^{1,2}$ with massless quarks,
a non-perturbative superpotential spontaneously breaks supersymmetry
when ${N_{\rm f}<N}$, and the classical moduli space of vacua is 
deformed when ${N_{\rm f}=N}$ \cite{Aharony:1997gp,Karch:1997ux}.  The
same critical value for $N_{\rm f}$ appears on the right in \eqref{SpBound}.

\paragraph{Improved Criteria for Spin(2N) SQCD.}

For gauge group $Spin(2N)$, the Cartan subalgebra can be parameterized in
terms of simple coroots via 
\begin{equation}\label{CSASp2N}
\sigma = {\bf I}_{2\times 2}
\diag\big(\sigma^1,\,\sigma^2-\sigma^1,\,\cdots,\,\sigma^{N-2}-\sigma^{N-3},\,\sigma^{N-1}-\sigma^{N-2}+\sigma^N,\,\sigma^N-\sigma^{N-1}\big),
\end{equation}
where again $\sigma$ is a ${2N\times 2N}$ block diagonal
matrix, with blocks proportional to ${\bf I}_{2 \times 2}$ in \eqref{CSASp},
embedded in the Cartan subalgebra of $SU(2N)$.  The positive Weyl
chamber lies in the portion of the positive quadrant
${\sigma^1,\ldots,\sigma^N\ge 0}$ such that
\begin{equation}\label{WeylSp2N}
\mathcal{C}_+:\quad \sigma^1 \,\ge\, \sigma^2-\sigma^1 \,\ge\, \cdots
\,\ge\,\sigma^{N-1}-\sigma^{N-2}+\sigma^N\,\ge\,\big|\sigma^N-\sigma^{N-1}\big|\,.
\end{equation}
The adjoint norm for $\sigma$ is given by the expression in
\eqref{AdjNormSp}, and the norm for the vector
representation is given by the sum of magnitudes of the diagonal entries in
\eqref{CSASp2N},
\begin{equation}\label{Norm2N}
||\sigma||_{\bf 2N} \,=\, \big|\sigma^1\big| \,+\, \sum_{j=2}^{N-2}
\big|\sigma^j-\sigma^{j-1}\big| \,+\,
\big|\sigma^{N-1}-\sigma^{N-2}+\sigma^N\big|\,+\,\big|\sigma^N-\sigma^{N-1}\big|\,.
\end{equation}
Again, one can easily see directly that this expression provides a
Weyl-invariant norm on $\R^N$.

In the positive Weyl chamber, the inequalities in \eqref{WeylSp2N}
allow us to simplify the sum of magnitudes in \eqref{Norm2N} as 
\begin{equation}
\begin{aligned}
||\sigma||_{\bf 2N} \,&=\, \sigma^{N-1} + \sigma^N \,+\,
\big|\sigma^{N-1}-\sigma^N\big|\,,\qquad\qquad \sigma\in\mathcal{C}_+\,,\\
&=\,\begin{cases}
2 \, \sigma^{N-1}\,,\quad\hbox{ if } \sigma^{N-1} \ge \sigma^N\,,\\
2 \, \sigma^N\,,\quad\quad\hbox{ if } \sigma^n \ge \sigma^{N-1}\,.
\end{cases}
\end{aligned}
\end{equation}
In fact, the inequalities in \eqref{WeylSp2N} state that either ${2
  \sigma^{N-1}}$ or ${2 \sigma^N}$ is greater than all members in the
descending chain ${\sigma^{N-2} \ge
  \sigma^{N-3} \ge \cdots \ge \sigma^1}$, so we can write 
\begin{equation}
||\sigma||_{\bf 2N} \,=\, \sigma_{\rm max}\,,\qquad\qquad \sigma\,\in\,\mathcal{C}_+\,,
\end{equation}
with 
\begin{equation}
\sigma_{\rm max} \,=\, \max\big\{\sigma^1,\ldots,\sigma^{N-2},\,2\sigma^{N-1},\,2\sigma^N\big\}\,.
\end{equation}
We also have the immediate bound on the adjoint norm
\begin{equation}
\begin{aligned}
||\sigma||_\g \,&=\, 2 \left(\sigma^1 \,+\,
  \cdots\,+\,\sigma^{N-2}\right) \,+\, 2 \sigma^{N-1} \,+\,
2\sigma^N\,,\qquad \sigma\,\in\,\mathcal{C}_+\,,\\
&\le\, 2 (N-1) \, \sigma_{\rm max}\,,
\end{aligned}
\end{equation}
with equality on the boundary where ${\sigma^1 =
  \cdots = \sigma^{N-2} = 2 \sigma^{N-1} = 2 \sigma^{N}}$.

The convergence of the matrix integral for $Spin(2N)$ SQCD in
is thence controlled by the sign of the
quantity
\begin{equation}
2 N_{\rm f} \left(1-\RR\right) ||\sigma||_{\bf 2N} \,-\, ||\sigma||_\g \,\ge\, 2 \Big[N_{\rm
  f}\left(1-\RR\right) - N + 1\Big]\sigma_{\rm max}\,,\qquad \sigma\,\in\,\mathcal{C}_+\,.
\end{equation}
Hence a necessary condition for convergence is 
\begin{flalign}\label{CritSp2N}
\fbox{Spin(2N) SQCD}\qquad\qquad\qquad&
N_{\rm f} \left(1-\RR\right) \,\ge\, N-1\,,&
\end{flalign}
which is also sufficient when the inequality is strict.  In terms of
the R-charge,
\begin{equation}
\RR \,\le\, 1 - \frac{N-1}{N_{\rm f}}\,,
\end{equation}
and positivity of ${\RR>0}$ implies
\begin{equation}\label{CritSp2NIV}
N_{\rm f} \,>\, N-1\,.
\end{equation}

\paragraph{Improved Criteria for Spin(2N+1) SQCD.}

Finally for $Spin(2N+1)$, we express $\sigma$ in a basis of simple
coroots as 
\begin{equation}
\sigma = \diag\!\big(\sigma^1\,{\bf I}_{2 \times 2},\,
\left(\sigma^2-\sigma^1\right){\bf I}_{2\times
  2},\,\cdots,\left(\sigma^{N-1}-\sigma^{N-2}\right){\bf I}_{2\times 2},\left(
2\sigma^{N}-\sigma^{N-1}\right){\bf I}_{2\times 2},\,0\big).
\end{equation}
Each entry in $\sigma$ but the last is proportional to the traceless, ${2 \times
  2}$ diagonal matrix in \eqref{CSASp}.  The final `$0$' entry is not
a ${2 \times 2}$ block.  In these coordinates, 
the positive Weyl chamber is the region in the positive quadrant
${\sigma^1,\ldots,\sigma^N\ge 0}$ where
\begin{equation}
\mathcal{C}_+:\qquad \sigma^1 \,\ge\, \sigma^2-\sigma^1
\,\ge\,\cdots\,\ge\,\sigma^{N-1}-\sigma^{N-2}\,\ge\,2\sigma^{N}-\sigma^{N-1}\,\ge\,0\,.
\end{equation}
Up to the factor of two which multiplies $\sigma^N$, this region is
the same as for $Sp(2N)$ in \eqref{PosWSp}.  The adjoint norm on the
positive chamber is given by the sum in \eqref{AdjNormSp}, and the
norm for the vector representation is the sum of
magnitudes
\begin{equation}
||\sigma||_{\bf 2N+1} \,=\, \big|\sigma^1\big| \,+\,
\sum_{j=2}^{N-1}\big|\sigma^j-\sigma^{j-1}\big| \,+\, \big|2 \sigma^N
- \sigma^{N-1}\big|\,.
\end{equation}
On the positive Weyl chamber,
\begin{equation}
||\sigma||_{\bf 2N+1} \,=\, 2 \sigma^N \,=\, \sigma_{\rm
  max}\,,\qquad\qquad \sigma\,\in\,\mathcal{C}_+\,,
\end{equation}
where 
\begin{equation}
\sigma_{\rm max} = \max\big\{\sigma^1,\ldots,\sigma^{N-1},2\sigma^N\big\}\,.
\end{equation}
Also,
\begin{equation}
||\sigma||_\g \,\le\, (2 N - 1)\,\sigma_{\rm max}\,,\qquad\qquad \sigma\,\in\,\mathcal{C}_+\,,
\end{equation}
with equality on the wall where
${\sigma^1=\cdots=\sigma^{N-1}=2\sigma^N}$.  

We thus have the bound 
\begin{equation}
2 N_{\rm f} \left(1-\RR\right) ||\sigma||_{\bf 2N+1} \,-\, ||\sigma||_\g \,\ge\, 2
\left[N_{\rm f}\left(1-\RR\right) - N + \ha\right] \sigma_{\rm max}\,,\qquad
\sigma\,\in\,\mathcal{C}_+\,,
\end{equation}
which is saturated on the wall of the Weyl chamber.
Positivity of the right-hand side implies that a necessary condition
for convergence of the matrix integral is 
\begin{flalign}\label{CritSp2NII}
\fbox{Spin(2N+1) SQCD}\qquad\qquad\qquad&
N_{\rm f} \left(1-\RR\right) \,\ge\, N-\ha\,.&
\end{flalign}
This condition is sufficient when the inequality
in \eqref{CritSp2NII} is strict.  As an upper-bound on the R-charge,
\begin{equation}
\RR \,\le\,1 - \frac{2N-1}{2N_{\rm f}}\,,
\end{equation}
and ${\RR>0}$ implies
\begin{equation}\label{CritSp2NIII}
N_{\rm f} \,>\, N - \ha\,.
\end{equation}
Recall that
$N_{\rm f}$ is allowed to be half-integral with our flavor-counting
conventions in \eqref{Flavors}, so the one-half on the right is meaningful.  

If we take the gauge group to be $Spin(N_{\rm c})$, both conditions in
\eqref{CritSp2NIV} and \eqref{CritSp2NIII} state uniformly that 
\begin{equation}\label{CritSpV}
2 N_{\rm f} \,>\, N_{\rm c} - 2\,,\qquad\qquad G = Spin(N_{\rm
  c})\,,\qquad\qquad \Lambda_{\rm uv}={\bf F}^{2 N_{\rm f}}\,.
\end{equation} 
For ${N_{\rm c}=3}$, this inequality reproduces our result for gauge group
$SU(2)$ with adjoint matter.  With the twofold counting of
flavors, the bound ${2N_{\rm f}\ge N_{\rm c}-2}$ precisely
reproduces the condition for unbroken supersymmetry
\cite{Aharony:2011ci,Aharony:2013kma,Benini:2011mf} in $Spin(N_{\rm c})$ SQCD
on $\R^{1,2}$.  For the critical value ${2 N_{\rm f} = N_{\rm c}-2}$,
the classical moduli space of vacua on $\R^{1,2}$ is deformed.

Nothing about the Lie algebra computations leading to the bound on
$N_{\rm f}$ in 
\eqref{CritSpV} guaranteed a result analytic in $N_{\rm c}$, so
the appearance of the correct critical value for
supersymmetry-breaking on $\R^{1,2}$ is doubly surprising.

\section{More About the Supersymmetric Residue Theorem}\label{JordanLemma}

Abstractly, the expression \eqref{ZuvCmbIV} for the ultraviolet partition function
$Z_{S^3}^{\rm uv}$
falls into the general class of Jordan integrals 
\begin{equation}\label{GrothZ}
\Phi_\sC \,=\, \left(\frac{1}{2\pi i}\right)^n\int_{\sC} \omega\,,\qquad\qquad \omega \,=\, \frac{g(z)\,
  dz^1\^\cdots\^dz^n}{s^1(z) \cdots s^n(z)}\,,
\end{equation}
where ${\sC\subset M}$ is a middle-dimensional, totally-real\footnote{To say that ${\sC\subset M}$ is a `totally-real' submanifold means that the
  tangent
  space at each point ${p\in\sC}$ has trivial intersection with its
  image under the complex structure tensor $\rJ$ on $M$, ie.
  ${T_p\sC\cap\left(\rJ\circ T_p\sC\right)=\{0\}}$.  This condition
  ensures ${\omega|_\sC\neq 0}$ away from zeroes of $g$.} submanifold in an
$n$-dimensional complex manifold $M$, 
with local holomorphic coordinates ${(z^1,\ldots,z^n)}$, and $\omega$ is a
meromorphic $n$-form on $M$ with poles along the union of divisors
\begin{equation}\label{PolarDs}
D_j = \Big\{z\in M \,\big|\, s^j(z)=0\Big\}\,,\qquad\qquad j\,=\,1,\ldots,n\,.
\end{equation}
Here $g$ and ${s^1,\ldots,s^n}$ are holomorphic functions on
$M$.\footnote{In our eventual analysis, we generalize slightly by
  allowing $g$ and ${s\equiv(s^1,\ldots,s^n)}$ to transform as holomorphic
  sections of complex vector bundles on $M$.}
By assumption, $\omega$ is regular on $\sC$, meaning
${\sC\cap D_j=\varnothing}$ for ${j=1,\cdots,n}$, and $\omega$ has the
appropriate asymptotic decay so that the integral converges when
$\sC$ is non-compact, as in our affine example \eqref{ZuvCmbIV} with
${\sC\equiv\h}$.  

The Jordan integral has two important features.
By holomorphy, ${d\omega=0}$ on a neighborhood of $\sC$, so
the value of $\Phi_\sC$ is not changed under small deformations of
${\sC\subset M}$.  Also, because $M$ has complex dimension $n$, the
intersection ${D_1\cap\cdots\cap D_n}$ of polar divisors is 
generically a discrete -- but possibly infinite -- set of points in
$M$.

Note that the factorization of the denominator in $\omega$ by individual
holomorphic functions ${s^1,\ldots,s^n}$, or equivalently, the
decomposition of the polar divisor of $\omega$ into a union of
precisely $n$ summands, has no intrinsic geometric
meaning and merely represents a choice in the way we present the
meromorphic form.  Any meromorphic integrand can be written as
$\omega$ in \eqref{GrothZ}, but without further assumptions (eg.~irreducibility),
the choice of the functions ${s^1,\ldots,s^n}$
in the presentation is not unique.  Later in Section \ref{HighRank},
we will say much more
about the choice of $\{s^1,\ldots,s^n\}$ for the gauge-theory
integrand in \eqref{ZuvCmbIV}.  This issue lies at the heart of our technical
results about $Z_{S^3}^{\rm uv}$.

Our goal in this subsection is to explain how the residue calculus can be applied
to the general Jordan integral in \eqref{GrothZ}.  At least in the
affine case, the multidimensional residue calculus follows from an iterative
application of the single-variable Cauchy theorem.  However, such a
naive approach quickly falls into the combinatoric swamp, and it does
not work when the complex manifold $M$ is not affine.

Instead, we analyze the Jordan integral using an effective supersymmetry 
on $M$.  The basic idea is recycled from \S $2$ of \cite{Beasley:2003fx}, in
which a more limited version of the multidimensional residue calculus is derived.  The prior
discussion pertains only to the case that $M$
is compact, without boundary, and ${\sC=\varnothing}$ is empty.  In
that case, ${\Phi_\sC=0}$ in \eqref{GrothZ}, and the residue calculus amounts
to a vanishing theorem.  

Here we wish to apply the 
residue calculus in the more interesting situation for which $M$ may
be 
non-compact, with non-empty boundary $\partial M$, and ${\Phi_\sC
  \neq 0}$.\footnote{The upper half-plane ${\BH_+}$ is the basic
one-dimensional example for $M$ to keep in mind.}   To do so requires a
non-trivial extension of ideas in \cite{Beasley:2003fx}.  This extension may be
useful for other problems and seems worth including for its own sake.\footnote{We thank
  M.\,Bertolini and R.\,Plesser for stimulating conversations about
  their ${\CN=(0,2)}$ linear sigma model examples in \cite{Bertolini:2014dna}.
  Those examples also motivate the re-analysis of the Jordan integral
  for more general complex manifolds $M$, which are non-compact with
  non-empty boundary.} 

The underlying mathematical content of the present discussion is not
new, though supersymmetry does simplify some calculations.  Our workhorse
residue theorem for the Jordan integral was originally proven in
\cite{Passre:1994,Tsikh:1998}.  See also
\cite{Griffiths:78,Tsikh:1992} for textbook references about
multidimensional residues and their applications to algebraic
geometry.  Mathematical approaches more directly related to the
effective supersymmetry on $M$ appear in \cite{Carrell:1973},
\cite{Carrell:1977}, and especially \cite{KLiu:1995}.  The ur-example
in this theory is the celebrated Bott residue formula \cite{Bott:1967}
for holomorphic vector fields.\footnote{We thank J.\,Weitsman for reminding us about the Bott
  residue formula.}

\subsection{Supersymmetric Integral on a Manifold with Boundary}

We derive the multidimensional residue theorem by considering a
supersymmetric integral on the complex manifold $M$.  Given the overlap with \cite{Beasley:2003fx}, our discussion of the background will be somewhat telegraphic.  The bosonic variables of integration are local homorphic and anti-holomorphic 
coordinates $z^j$ and ${z{}^{\bar j}\equiv \bar{z^j}}$ on $M$.  We
also introduce anti-commuting, fermionic integration variables $\theta^{\bar j}$
and $\chi^\alpha$.  The fermions $\theta^{\bar j}$ transform as
coordinates on the anti-holomorphic tangent bundle $\bar{TM}$, and
the fermions $\chi^\alpha$ transform as coordinates on a holomorphic
vector bundle $V$ of rank $r$ over $M$.  Thus ${j=1,\ldots,n}$ and ${\alpha=1,\ldots,r}$ throughout.

Along with the vector bundle $V$, we choose a holomorphic section ${s\in
  H^0_{\bar\partial}(M,V)}$.  The effective supersymmetry on $M$ then acts on the
variables $(z, \bar z, \theta, \chi)$ via 
\begin{equation}\label{HolSusy}
\begin{matrix}
\begin{aligned}
\delta z^j \,&=\, 0\,,\\
\delta \chi^\alpha \,&=\, s^\alpha\,,
\end{aligned} \qquad\qquad&\qquad\qquad
\begin{aligned}
\delta z^{\bar j} \,&=\, \theta^{\bar j}\,,\\
\delta \theta^{\bar j} \,&=\, 0\,.
\end{aligned}
\end{matrix}
\end{equation}
Since $s$ is holomorphic, ${\delta^2=0}$ by inspection.

When $M$ is compact and boundaryless, holomorphy is the
only condition which must be imposed on the section $s(z)$.  We are
interested in the opposite situation, for which $M$ is non-compact
with non-empty boundary. In this case, conditions must also be imposed on 
the behavior of $s$ at infinity as well as on the boundary
$\partial M$.  These conditions are ultimately related to the previous
assumptions about decay and regularity for the meromorphic form
$\omega$ on the real cycle $\sC$.  

First, we assume that $V$
admits a hermitian metric $h$ such that the norm-square 
${||s||^2\to\infty}$ at infinity on $M$, at a rate sufficient for absolute
convergence of all integrals which follow.  For precise analytic
statements about convergence issues for the Jordan integral, see
\cite{Passre:1994,Tsikh:1998}.  

Second, we assume that $s$ is everywhere
non-vanishing on the boundary, ${s|_{\partial M} \neq 0}$.  The
pullback of the top Chern class of $V$ to the boundary is a
topological obstruction to this requirement, so we must suppose
\begin{equation}\label{TopCV}
c_{r}(V)\big|_{\partial M} \,=\, 0 \,\in\, H^{2r}(\partial M;\R)\,,\qquad\qquad r=\rk V\,.
\end{equation}
If ${\rk V \ge \dim M}$, the condition in
\eqref{TopCV} follows by dimension-counting.  Otherwise,
the topological constraint on the boundary behavior of $V$ is
non-trivial.

Exactly as in \cite{Beasley:2003fx}, we consider a supersymmetric
integral over $M$ with the general structure
\begin{equation}\label{ZusyZ}
Z^{(0)} \,=\, \int_M g\,d\mu\,\exp{\!\left[-t\,S\right]}\,,
\end{equation}
where ${t\in\R_+}$ is a positive real parameter, ${S\equiv S(z,\bar
  z,\theta,\chi)}$ is a $\delta$-closed function which plays the role of the physical action, and ${g\,d\mu}$ is a supersymmetric
measure,
\begin{equation}\label{ZusyMeas}
g \, d\mu \,\equiv\, g(z)\, d^nz \, d^n\bar z\, d^n\theta \, d^r\chi\,.
\end{equation}
The superscript on $Z^{(0)}$ indicates that we will later have to correct this integral with a boundary term.
Globally, ${g}$
transforms as a holomorphic section of the complex line bundle
${\Omega^n_M\!\otimes\!\wedge^rV}$ over $M$,\footnote{Due to the cancellation of paired bosonic and fermionic Jacobians upon a change of coordinates, the factor ${d^n\bar z \, d^n\theta}$ in the measure ${g\, d\mu}$ transforms as a section of the trivial line bundle on $M$.}
\begin{equation}
g \,\in\, H^0_{\bar\partial}\big(M, \Omega^n_M\!\otimes\!\wedge^rV\big)\,.
\end{equation}
For our proof of the residue theorem when $M$ is non-compact, we must further assume ${|g|\to 0}$ at infinity at a rate sufficient to ensure that the integral in \eqref{ZusyZ} converges even as ${t\to 0^+}$.  This condition on $g$ depends upon the asymptotic behavior of $s$ and is related to the ``Jordan condition'' in \cite{Passre:1994,Tsikh:1998}.

For the integrand in \eqref{ZusyZ}, we take a $\delta$-trivial expression\footnote{Throught this section, we use the Einstein summation convention for tensor indices.}
\begin{equation}\label{BigWS}
S \,=\, \delta{\bf W}\,,\qquad\qquad {\bf W} = 
 h_{\bar\alpha \alpha}\,s{}^{\bar\alpha}\,\chi^\alpha\,\equiv\, (s,\chi) \,,
\end{equation}
or more explicitly,
\begin{equation}
S\,=\,  h_{\bar\alpha \alpha}\,s{}^{\bar\alpha}\,s^\alpha \,+\, h_{\bar\alpha \alpha}\,\nabla_{\bar j}s{}^{\bar\alpha}\,\theta^{\bar j}\,\chi^\alpha \,\equiv\, ||s||^2 \,+\, (\theta\cdot\nabla s,\chi)\,.
\end{equation}
For ease of notation, the hermitian metric $h$ on $V$ is subsumed into the pairing $(\,\cdot\,,\,\cdot\,)$ in the second version of the expressions for ${\bf W}$ and $S$.  Also, $\nabla$ is the
canonical \cite{Griffiths:78} covariant derivative on $V$ which is 
compatible with both the holomorphic structure and metric.  Because
${||s||^2\to\infty}$ at infinity on $M$, the 
integrand of $Z^{(0)}$ decays rapidly there, and the supersymmetric integral converges for all ${t>0}$.

The supersymmetry transformation \eqref{HolSusy} respects a
ghost-number, or grading, under which $\theta$ has charge
$+1$ and $\chi$ has charge $-1$.  With this assignment,
$\delta$ increases the ghost-number by one unit, and $S$ is
invariant.  On the other hand, the measure ${g\,d\mu}$ in
\eqref{ZusyMeas} has charge 
${n-r}$.  If this charge is non-zero, the supersymmetric integral in
\eqref{ZusyZ} vanishes identically.  Hence $Z^{(0)}$ itself is only useful
to study in the special case ${\rk V = \dim M}$, for which the ghost-number symmetry is non-anomalous.

More generally, we consider the expectation values of supersymmetric
operators
\begin{equation}\label{ExpectCO}
\langle\CO\rangle^{(0)} \,=\, \int_M  g\,d\mu\,\CO \exp{\!\left[-t\,S\right]}\,,
\end{equation}
where $\CO$ is a function of $(z, \bar z, \theta, \chi)$ which is
annihilated by $\delta$ and has ghost-number\footnote{If $\CO$ does not have the required ghost-number, the expectation value again vanishes.} 
\begin{equation}\label{GhostSel}
r \,-\, n \,=\, \rk V \,-\, \dim M\,.
\end{equation}
With this definition, ${Z^{(0)} = \langle 1 \rangle^{(0)}}$ is the expectation value of the constant function `$1$'.
If $M$ is compact and boundaryless, the expectation value depends
only upon the $\delta$-cohomology class of $\CO$, by familiar
arguments about decoupling of BRST-trivial operators.  After expanding $\CO(z, \bar z, \theta, \chi)$ in powers\footnote{Because the components of $\theta$ and
  $\chi$ are anti-commuting, the Taylor expansion of $\CO$ terminates after only a finite
  number of terms.} of the
fermionic coordinates $\theta^{\bar j}$ and $\chi^\alpha$, one obtains a more
geometric description of $\delta$-cohomology in terms of the graded complex 
\begin{equation}\label{ChainC}
\bigoplus_{(p,q)} A^{(0,q)}(M)\otimes\wedge^p
V^*,\,\qquad\qquad \delta \,\equiv\, \bar\partial \,+\, \iota_s\,.
\end{equation}
Here $A^{(0, q)}(M)$ indicates the bundle of smooth $(0,q)$ 
forms on $M$, and $V^*$
is the holomorphic bundle dual to $V$.  As usual, $\bar\partial$ is
the Dolbeault operator, and $\iota_s$ indicates the
interior product of the section $s$ with smooth sections of
${\wedge^p V^*}$.  The cohomology of this complex
\eqref{ChainC} has been studied in detail \cite{KLiu:1995} under the
name `holomorphic equivariant cohomology'.  From the geometric perspective, the following construction amounts to the study of holomorphic equivariant cohomology relative to a boundary.

When $M$ has a boundary, the usual arguments about decoupling of BRST-trivial operators do not apply.  Instead, if ${\CO=\delta{\bf U}}$ for some ${\bf U}$, the expectation value receives a boundary contribution
\begin{equation}\label{SuperStokes}
\langle\delta{\bf U}\rangle^{(0)} \,=\, \int_M g\,d\mu\, \,\delta\big({\bf U}\exp{\!\left[-t\,S\right]}\big) =\, -\frac{i}{2}\int_{\partial M} g \,d\mu\,\, {\bf U} \exp{\!\left[-t\,S\right]}\,.
\end{equation}
The second equality in \eqref{SuperStokes} is a restatement of Stokes
Theorem for the Dolbeault operator in \eqref{ChainC}, such that
${g\,d\mu}$ on the right is the induced boundary
measure.\footnote{Observe that $\delta$ acts on functions of $(z,\bar
  z,\theta,\chi)$ by the super-vector field ${\theta^{\bar j}
    \,{\partial}/{\partial z^{\bar j}} + s^\alpha
    \,{\partial}/{\partial\chi^\alpha}}$.  Only the first term in the
  vector field contributes to the Grassmann integral over $\theta$ and
  $\chi$.  The integral over the fermionic normal coordinate
  $\theta^{\bar j}$ in the vector field then produces the term
  $d^{n-1}\theta$ in the supersymmetric measure on ${\partial M}$.}
The prefactor in \eqref{SuperStokes} arises from the relation ${d^2 z
  = -\frac{i}{2}\,d\bar z\^ dz}$ between the measure and the
volume-form on the complex plane.

For a more explicit description of the boundary measure, note that the inclusion ${\partial M \subset M}$ can be modelled locally on the elementary inclusion $\R\times\C^{n-1}\subset\BH_+\times\C^{n-1}$.  Thus ${\partial M}$ is a real manifold with dimension $(2n-1)$ which admits distinguished local coordinates $(x,w,\bar w)$, where ${x\in\R}$ is a real coordinate along the boundary of the upper half-plane, and ${(w,\bar w)\in\C^{n-1}}$ are complex.  In terms of these coordinates,
\begin{equation}\label{BoundMeas}
g\,d\mu\big|_{\partial M} \,\equiv\, g(w,x) \, dx \, d^{n-1}w \, d^{n-1}\bar w \, d^{n-1}\theta \, d^r\chi\,.
\end{equation}
Globally, the local holomorphic coordinates $w$ on ${\partial M}$ are
associated to an integrable, rank-$(n-1)$ complex subbundle
${\CH\subset T_\C(\partial M)}$ of the complexified tangent bundle.
Similarly, $dx$ is the generator for a real line subbundle ${L\subset
  T^*_\R(\partial M)}$ of the (real) cotangent bundle.  Because
${\partial M}$ inherits an orientation from $M$ and ${dx\neq 0}$ is
nowhere-vanishing, ${L\simeq\R}$ is the trivial line bundle in our situation.
The description of the boundary measure in \eqref{BoundMeas} then
amounts to the global statement that $g$ restricts on ${\partial M}$
to a holomorphic section (with respect to $w$ and ${\bar w}$) of the
complex line bundle ${\wedge^{n-1}\CH^*\!\otimes\!\wedge^r
  V|_{\partial M}}$.

As a corollary to the BRST-anomaly in \eqref{SuperStokes}, the expectation value $\langle\CO\rangle^{(0)}$ for a \mbox{$\delta$-closed} operator $\CO$ generally depends upon the value of the coupling parameter $t$ and the choice of the bundle metric $h$ if ${\partial M \neq \varnothing}$.  For the dependence on ${t>0}$, 
\begin{equation}\label{ANomCO}
\begin{aligned}
\frac{d}{dt}\langle\CO\rangle^{(0)} \,&=\, -\int_M g\,d\mu\,\,S\cdot\CO \exp{\!\left[-t\,S\right]}\,,\qquad\qquad S \,=\, \delta{\bf W}\,,\\
&=\, \frac{i}{2}\int_{\partial M} g\,d\mu\,\,{\bf W}\cdot\CO \exp{\!\left[-t\,S\right]}\,,
\end{aligned}
\end{equation}
where ${\bf W}$ appears in \eqref{BigWS}.  

Let us try to cancel the variation of $\langle\CO\rangle^{(0)}$ in \eqref{ANomCO} by adding a boundary term to the expectation value.  Evidently, upon integrating the boundary integral in \eqref{ANomCO} with respect to $t$, the required boundary term is 
\begin{equation}\label{BoundTCO}
\langle\CO\rangle^{(1)} \,=\, \frac{i}{2}\int_{\partial M} g \, d\mu\, \left(\frac{{\bf W}}{S}\right)\cdot\CO\exp{\!\left[-t\,S\right]}\,,
\end{equation}
such that 
\begin{equation}\label{SumCO}
\frac{d}{dt}\langle\langle\CO\rangle\rangle \,=\, 0\,,\qquad\qquad
\langle\langle\CO\rangle\rangle \,:=\, \langle\CO\rangle^{(0)} \,+\, \langle\CO\rangle^{(1)}\,.
\end{equation}
Because the derivative vanishes for arbitrary values of ${t>0}$, the modified expectation value $\langle\langle\CO\rangle\rangle$ is now independent of $t$.

What is the meaning of the boundary term $\langle\CO\rangle^{(1)}$?
First, because we assume ${s\neq 0}$ on ${\partial M}$, division by
$S$ in \eqref{BoundTCO} is sensible.  Explicitly,
\begin{equation}\label{1overS}
\left(\frac{1}{S}\right)\!\Biggr|_{\partial M} \,=\, \frac{1}{||s||^2 + \left(\theta\cdot\nabla
    s,\chi\right)} \,=\,\frac{(-1)^{n-1} \left(\theta\cdot\nabla
    s,\chi\right)^{n-1}}{||s||^{2n}} \,+\, \cdots\,,
\end{equation}
where we expand on the right in powers of $\theta$ and $\chi$.  We write explicitly only the term which is top-degree in $\theta$ and $\chi$; the ellipses indicate terms which are lower-order in the fermions.

Second, if ${\CO=\delta{\bf U}}$ is $\delta$-trivial, then
\begin{equation}
\begin{aligned}
\langle\delta{\bf U}\rangle^{(1)} \,&=\, \frac{i}{2} \int_{\partial M} g\,d\mu \,\left(\frac{{\bf W}}{S}\right)\cdot\delta{\bf U}\exp{\!\left[-t\,S\right]}\,,\\
&=\, \frac{i}{2}\int_{\partial M} g\,d\mu\,\,\delta\!\left(\frac{{\bf W}}{S}\right)\cdot{\bf U}\exp{\!\left[-t\,S\right]}\,,\\
&=\, \frac{i}{2}\int_{\partial M} g\,d\mu\,\,{\bf U}\exp{\!\left[-t\,S\right]}\,,
\end{aligned}
\end{equation}
where we integrate-by-parts with respect to $\delta$ in the second line.  Due to the fermionic nature of ${\bf W}$, no sign appears.  In the third line, we recall that ${\delta{\bf W}=S}$.  Comparing to the anomalous boundary term in \eqref{SuperStokes}, we see that 
\begin{equation}\label{BRSTDec}
\langle\langle\delta{\bf U}\rangle\rangle \,=\, 0\,.
\end{equation}
Thus the modified expectation value $\langle\langle\,\cdot\,\rangle\rangle$ is non-anomalous, in the sense that \mbox{$\delta$-trivial} operators decouple.  This observation is of course consistent with the previous statement that $\langle\langle\CO\rangle\rangle$ is independent of $t$ when $\CO$ is $\delta$-closed.

Third, for any $\delta$-closed operator $\CO$, we have a formal relation on $M$,
\begin{equation}\label{LocTriv}
\delta\!\left(\frac{{\bf W}}{S}\cdot\CO\,\exp{\!\left[-t\,S\right]}\right)=\, \CO\exp{\!\left[-t\,S\right]}\,,\qquad\qquad s\,\neq\,0\,.
\end{equation}
This relation is formal insofar as the left side is not defined on the
vanishing locus of the section $s$.  However, at points where ${s\neq
  0}$ on $M$, we see that the operator on the right side of \eqref{LocTriv} is $\delta$-trivial.  Hence by \eqref{BRSTDec}, the non-anomalous expectation value $\langle\langle\CO\rangle\rangle$ only receives contributions from a small neighborhood of the vanishing locus of the holomorphic section $s$.

Alternatively, the localization of support for the integrand of $\langle\langle\CO\rangle\rangle$ can be seen by considering the limit ${t\to\infty}$ in the integrals over $M$ and $\partial M$.  In this case, since ${s\neq 0}$ on $\partial M$, the boundary term vanishes due to the exponential suppression in the integrand,
\begin{equation}
\lim_{t\to\infty}\langle\CO\rangle^{(1)} \,=\, 0\,,
\end{equation}
while the bulk term reduces to a sum over local contributions from components $C$ of the vanishing locus for $s$,
\begin{equation}
\lim_{t\to\infty}\langle\CO\rangle^{(0)} \,=\, \sum_{C \subset \{s=0\}} \langle\CO\rangle_C\,.
\end{equation}
Altogether,
\begin{equation}\label{ttoInf}
\lim_{t\to\infty}\langle\langle\CO\rangle\rangle \,=\, \sum_{C \subset \{s=0\}} \langle\CO\rangle_C\,.
\end{equation}

As an easy example from \cite{Beasley:2003fx}, let us suppose that
${\rk V = \dim M = n}$ and that $s$ vanishes in a \mbox{non-degenerate} fashion at an isolated point ${p\in M}$.  The non-degeneracy assumption means that the Jacobian $\det(ds)$ is non-zero at $p$,
\begin{equation}
\det(ds)(p) \,=\, \det\!\left(\frac{\partial(s^1,\ldots,s^n)}{\partial(z^1,\ldots,z^n)}\right) \,\neq\, 0\,.
\end{equation}
The local contribution $\langle\CO\rangle_p$  can then be evaluated exactly by applying the Gaussian approximation to the integral in \eqref{ExpectCO}, with the result
\begin{equation}\label{GrothRes}
\langle\CO\rangle_p \,=\,
c_n\,\frac{g(p)\,\CO(p)}{\det(ds)(p)}\,,\qquad\qquad c_n = (-1)^n\,\pi^n\,.
\end{equation}
The normalization constant $c_n$ accounts for the value of the
Gaussian integral in the bosonic directions as well as a sign from
the Grassmann integral over fermions, due to the sign which multiplies
$t$ in \eqref{ZusyZ}.

When the operator $\CO$ is evaluated at $p$, we drop all terms involving fermions and evaluate only the bosonic piece of $\CO$ to obtain a c-number.  Note that $\CO$ has a purely bosonic piece only if $\CO$ has ghost-number zero, consistent with the selection rule in \eqref{GhostSel} when ${\rk V = \dim M}$.  In the special case that ${\CO=1}$ is the identity, the right side of \eqref{GrothRes} takes precisely the form of the local Grothendieck residue at $p$.

On the other hand, so long as the integrals over $M$ and ${\partial
  M}$ converge\footnote{This assumption is always true when $M$ is a
  compact manifold with boundary.  When $M$ is non-compact, we must
  impose conditions on the asymptotic behavior of the pair $(g,s)$ at
  infinity, as mentioned previously.} absolutely in the limit ${t\to
  0^+}$, we can evaluate the non-anomalous expectation value
$\langle\langle\CO\rangle\rangle$ by setting ${t=0}$.  The bulk
contribution reduces to the integral of $\CO$, 
\begin{equation}
\lim_{t\to 0^+}\langle\CO\rangle^{(0)} \,=\, \int_M g\,d\mu\,\CO\,,
\end{equation}
which can be non-vanishing when $\CO$ has the proper ghost-number in
\eqref{GhostSel} so that fermion zero-modes can be absorbed.  In this
case, the boundary term at ${t=0}$ is generally non-zero as well,
\begin{equation}\label{BoundRes}
\lim_{t\to 0^+}\langle\CO\rangle^{(1)} \,=\, \frac{i}{2}
\int_{\partial M} g\,d\mu \left(\frac{{\bf W}}{S}\right)\cdot\CO\,,
\end{equation}
since ${\bf W}$ compensates for the change in ghost-number of the
boundary measure \eqref{BoundMeas} versus the bulk.  When interpreting the integrand of \eqref{BoundRes}, we must expand $1/S$ in powers of the fermions exactly as in \eqref{1overS}.  The resulting integrand on $\partial M$ will depend very much upon the choice of the operator $\CO$. So  without further simplification,
\begin{equation}\label{ttoZero}
\lim_{t\to 0^+}\langle\langle\CO\rangle\rangle \,=\, \int_M
g\,d\mu\,\CO \,+\,\frac{i}{2} \int_{\partial M} g\,d\mu \left(\frac{{\bf W}}{S}\right)\cdot\CO\,.
\end{equation}

Equating the two limits in \eqref{ttoInf} and \eqref{ttoZero}, we obtain a very general version of the residue theorem,
\begin{flalign}\label{ResThm}
\fbox{residue theorem}\qquad&
\int_M g\,d\mu\,\CO\,+\,\frac{i}{2} \int_{\partial M} g\,d\mu \left(\frac{{\bf W}}{S}\right)\cdot\CO\,=\, \sum_{C \subset \{s=0\}} \langle\CO\rangle_C\,.&
\end{flalign}
This theorem expresses the sum of a certain integral over the the bulk $M$ and
the boundary $\partial
M$ in terms of a sum of local contributions from each component
${C\subset M}$ of the vanishing locus for $s$.  When $M$ is compact
without boundary and ${\CO=1}$ is constant, the bulk integral on the left side of the
residue theorem is zero, due to excess fermion zero-modes, and we obtain the vanishing theorem applied
to heterotic string worldsheet instantons in \cite{Beasley:2003fx}.

\paragraph{A Special Case of the Residue Theorem.}

To produce a more familiar version of the multidimensional residue theorem, let us evaluate the boundary integral in \eqref{ResThm} in the simple case ${\CO=1}$, with $\rk V = \dim M = n$.  From the formulas for ${\bf W}$ and $1/S$ in \eqref{BigWS} and \eqref{1overS}, 
\begin{equation}
\int_{\partial M} g\,d\mu \left(\frac{{\bf W}}{S}\right) =\, (-1)^{n-1}\int_{\partial M} g\,dx\, d^{n-1}w\,d^{n-1}\bar w\,d^{n-1}\theta\,d^n\chi\,\, (s,\chi)\cdot\frac{(\theta\cdot\nabla s,\chi)^{n-1}}{||s||^{2n}}\,,
\end{equation}
since only the component of $1/S$ with top-degree in $\theta$ and $\chi$ contributes to the fermionic integral.  That integral is proportional to the product of completely anti-symmetric $\varepsilon$-tensors for $\theta$ and $\chi$, so
\begin{equation}\label{DolbRep}
\begin{aligned}
&\int_{\partial M} g\,d\mu \left(\frac{{\bf W}}{S}\right) =\\
&\qquad\int_{\partial M} g\,dx\, d^{n-1}w\,d^{n-1}\bar w\,\,\frac{\varepsilon^{\bar j_1 \cdots \bar j_{n-1}}\varepsilon^{\alpha_1 \cdots \alpha_n}\!\left(\bar s_{\alpha_n} \partial_{\bar j_1} \!\bar s_{\alpha_1} \cdots\,\partial_{\bar j_{n-1}}\!\bar s_{\alpha_{n-1}}\right)}{||s||^{2n}}\,,
\end{aligned}
\end{equation}
with ${\bar s_\alpha = h_{\bar\alpha \alpha}\,s^{\bar\alpha}}$
and the sign has been absorbed into the ordering of tensor indices.
More intrinsically, we recognize the right side of \eqref{DolbRep}
as\footnote{A factor of $(i/2)^{n-1}$ again appears when we relate the
  measure to the volume-form on $\C^{n-1}$.}
\begin{equation}
\int_{\partial M} g\,d\mu \left(\frac{{\bf W}}{S}\right) =\, \left(\frac{i}{2}\right)^{n-1} \int_{\partial M}\eta\,,
\end{equation}
where 
\begin{equation}\label{DistDolb}
\eta\,=\, (-1)^{n (n-1)/2}\,g\, dz^1\^\cdots\^dz^n\^\frac{\varepsilon^{\alpha_1 \cdots \alpha_n} \overline{\left(s_{\alpha_1}\,\partial s_{\alpha_2}\^\cdots\^\partial s_{\alpha_n}\right)}}{||s||^{2n}}\,.
\end{equation}
By construction, $\eta$ is holomorphic on the complement ${M^o=M-\{s=0\}}$ of the vanishing locus for $s$.  In the terminology of Ch $5.1$ in \cite{Griffiths:78}, $\eta$ is the ``distinguished'' Dolbeault representative of the meromorphic form $\omega$ in \eqref{GrothZ}.  The reader is invited to compare our computation of $\eta$ using supersymmetry to the fairly elaborate algebra required in \cite{Griffiths:78}.

If $s$ vanishes non-degenerately at isolated points $p$ in the interior of $M$, such that the local computation leading to \eqref{GrothRes} applies, the general residue theorem in \eqref{ResThm} specializes to the relation, including numerical factors,
\begin{equation}
\left(\frac{1}{2\pi i}\right)^n\int_{\partial M}\eta \,=\, \sum_{s(p) = 0} \Res_p\!\left(\omega\right)\,,\qquad\qquad \Res_p\!\left(\omega\right) = \frac{g(p)}{\det(ds)(p)}\,.
\end{equation}
For ${n=1}$ and ${M=\BH_+}$ the upper half-plane, this relation is the elementary Cauchy theorem.  More generally, the same relation appears in Ch $5.1$ of \cite{Griffiths:78} as the Global Residue Theorem.  

\subsection{Polyhedral Decomposition on the Boundary}

To apply the supersymmetric residue theorem to the Jordan integral $\Phi_\sC$ in \eqref{GrothZ}, we require further assumptions about the global embedding of the totally-real submanifold ${\sC \subset M}$.  Throughout, we work in the case ${\rk V = \dim M = n}$, which is implicit in the description of $\Phi_\sC$.  Relaxing the condition on the rank of $V$ could be interesting, but we do not pursue the generalization here.

First, we assume that $\sC$ is embedded in the boundary $\partial M$, as appears on the left side of \eqref{ResThm}.  Locally, ${\sC\subset M}$ is then modelled on the standard embedding of ${\R^n \subset \BH_+\times\C^{n-1}}$, where $\BH_+$ is the upper half-plane, and $\R^n$ is embedded along the real axis in each factor. 
In this local model, the normal directions ${N_{\sC/\partial M} \simeq \R^{n-1}}$ to $\sC$ in ${\partial M}$ can be given the combinatoric structure of a fan which is the union of simplicial cones ${\Sigma_1,\ldots,\Sigma_n}$, spanned by vectors ${v_1,\ldots,v_n\in\R^{n-1}}$.  By definition, each cone ${\Sigma_\alpha\subset\R^{n-1}}$ is the positive span of vectors $\{v_1,\cdots\!,\widehat v_\alpha,\cdots\!,v_n\}$,
\begin{equation}\label{ConealpH}
\Sigma_\alpha \,=\, \R_+[v_1,\cdots\!,\widehat v_\alpha,\cdots\!,v_n],\qquad\qquad \alpha=1,\ldots,n\,,
\end{equation}
where the hat means $v_\alpha$ is omitted from the generating set.
We assume $\{v_2,\cdots\!,v_n\}$ is an oriented basis for $\R^{n-1}$, with ${v_1+\cdots+v_n=0}$.  We always order the vectors in any generating set according to their subscripts, and we orient the cones so that 
\begin{equation}
\R^{n-1} \,=\, \sum_{\alpha=1}^n (-1)^{\alpha+1}\,\Sigma_\alpha\,.
\end{equation}
See Figure \ref{2dFan} for an example when ${n=3}$.

Every pair of simplicial cones intersects along a unique face 
\begin{equation}
\Sigma_{\alpha \beta} \,=\, \Sigma_\alpha\cap\Sigma_\beta \,=\, \R_+[v_1,\cdots\!,\widehat v_\alpha,\cdots\!,\widehat v_\beta,\cdots\!,v_n]\,,\qquad\qquad \alpha < \beta\,,
\end{equation}
which is itself a lower-dimensional simplicial cone, and so on.  To avoid the profusion of indices, we let ${A\equiv(\alpha_1 \cdots \alpha_p)}$ be a multi-index, with 
\begin{equation}
\Sigma_A \,=\, \R_+[v_{A^o}],\qquad\qquad A^o = \{1,\ldots,n\} - A\,.
\end{equation}
Here $A^o$ is the complement to $A$ in the index set $\{1,\ldots,n\}$.  At the bottom of the tower of incidence relations lies $\Sigma_{1 2 \cdots n} = \Sigma_1 \cap \cdots \cap \Sigma_n =\pm\sC$, identified with the origin in Figure \ref{2dFan}.  Note that the incidence relations among cones in the fan are dual to the incidence relations for the standard, $(n-1)$-dimensional simplex.

The fan also admits a boundary operator which acts in a simplicial fashion,
\begin{equation}\label{BoundOp}
\partial\Sigma_A \,=\, \sum_{\alpha \in A^o} \left(-1\right)^{(\alpha,A^o)} \Sigma_{A\cup\{\alpha\}}\,.
\end{equation}
Here $(\alpha,A^o)$ denotes the position of $\alpha$, read from the left, in the ordered set $A^o$, eg. $(4,\{1,4,7\})=2$.
As a check, for ${n=3}$, we have ${\R^2 = \Sigma_1 - \Sigma_2 + \Sigma_3}$, with boundaries
\begin{equation}
\begin{matrix}
\begin{aligned}
\partial\Sigma_1 \,&=\, -\Sigma_{12} \,+\, \Sigma_{13}\,,\\
\partial\Sigma_2 \,&=\, -\Sigma_{12} + \Sigma_{23}\,,\\
\partial\Sigma_3 \,&=\, -\Sigma_{13} + \Sigma_{23}\,,
\end{aligned} &\quad\hbox{ and }\qquad&
\partial\Sigma_{12} = \partial\Sigma_{13}=\partial\Sigma_{23}=-\Sigma_{123}\,.
\end{matrix}
\end{equation}
These relations are consistent with the orientations shown in Figure \ref{2dFan} as well as ${\partial^2=0}$.

Finally, from the analytic perspective, each simplicial cone ${\Sigma_{\alpha_1 \cdots \alpha_p}}$ admits affine coordinates modelled on $\R^p\times\C^{n-p} \subset \BH_+\times\C^{n-1}$ and thereby inherits a $\bar\partial$-operator acting in the complex directions.

\iffigs
\begin{figure}[t]
	\centering
	\begin{tikzpicture}[x=1.3cm,y=1.3cm] 
	\coordinate (0;0) at (0,0); 
	\foreach \c in {1,...,4}{%  
		\foreach \i in {0,...,5}{% 
			\pgfmathtruncatemacro\j{\c*\i}
			\coordinate (\c;\j) at (60*\i:\c);  
		} }
		\foreach \i in {0,2,...,10}{% 
			\pgfmathtruncatemacro\j{mod(\i+2,12)} 
			\pgfmathtruncatemacro\k{\i+1}
			\coordinate (2;\k) at ($(2;\i)!.5!(2;\j)$) ;}
		
		\foreach \i in {0,3,...,15}{% 
			\pgfmathtruncatemacro\j{mod(\i+3,18)} 
			\pgfmathtruncatemacro\k{\i+1} 
			\pgfmathtruncatemacro\l{\i+2}
			\coordinate (3;\k) at ($(3;\i)!1/3!(3;\j)$)  ;
			\coordinate (3;\l) at ($(3;\i)!2/3!(3;\j)$)  ;
		}
		
		\foreach \i in {0,4,...,20}{% 
			\pgfmathtruncatemacro\j{mod(\i+4,24)} 
			\pgfmathtruncatemacro\k{\i+1} 
			\pgfmathtruncatemacro\l{\i+2}
			\pgfmathtruncatemacro\m{\i+3} 
			\coordinate (4;\k) at ($(4;\i)!1/4!(4;\j)$)  ;
			\coordinate (4;\l) at ($(4;\i)!2/4!(4;\j)$) ;
			\coordinate (4;\m) at ($(4;\i)!3/4!(4;\j)$) ;
		}  
				
		\draw[->,red,thick,shorten >=20pt] (0;0)--(4;2);
		\draw[->,red,thick,shorten >=20pt] (0;0)--(4;10);	
		\draw[->,red,thick,shorten >=20pt] (0;0)--(4;18);

		\fill [red] (0,0) circle (.05);	

		\node [ ] at (0.75,-0.15) {$\Sigma_{123}=\sC$};
		
		\node [ ] at (2;11) {$\Sigma_1$};
		\node [ ] at (2;3) {$-\Sigma_2$};
		\node [ ] at (2;7) {$\Sigma_3$};
		
		\node [ ] at (2.8,1.45) {$\Sigma_{12}$};
		\node [ ] at (-2.8,1.45) {$\Sigma_{23}$};
		\node [ ] at (0.1,-3.1) {$\Sigma_{13}$};
		
		\draw [thick, draw=white, fill=white] (1.85,2.3)--(1.85,1.95)--(2.25,1.95)--(2.25,2.3);
		\draw [thick, draw=black, fill=white] (1.85,2.3)--(1.85,1.95)--(2.25,1.95);
		
		\node at (2.15,2.15){$\R^2$};		
		\end{tikzpicture} 
		\caption{The standard polyhedral fan in $\R^2$.  The origin $\Sigma_{123}$ is identified with the Jordan integration cycle $\sC$, up to orientation.}\label{2dFan}
\end{figure}
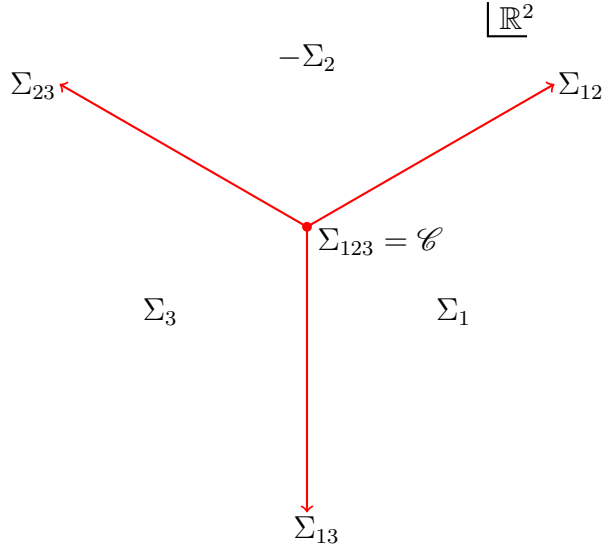
\fi

Globally, to apply the residue theorem to the Jordan integral
$\Phi_\sC$, we require that the boundary ${\partial M}$ of the complex manifold $M$ can be given
the same analytic and combinatoric structure as the fan in $\R^{n-1}$.
Thus, ${\partial M}$ decomposes as a union of connected, closed
$(2n-1)$-dimensional manifolds ${\Sigma_1,\ldots,\Sigma_n}$, each with
piecewise-smooth boundary and oriented so that $\partial M$ is the union
\begin{flalign}\label{SimpDecM}
\fbox{polyhedral decomposition}\qquad\qquad&
\partial M = \sum_{\alpha=1}^n (-1)^{\alpha+1}\,\Sigma_\alpha\,.&
\end{flalign}
The components ${\Sigma_1,\ldots,\Sigma_n}$ in the polyhedral
decomposition adjoin pairwise along connected $(2n-2)$-dimensional
submanifolds ${\Sigma_{\alpha \beta} = \Sigma_\alpha \cap \Sigma_\beta}$, and similarly in lower dimensions, with incidence relations identical to those for the simplicial cones in $\R^{n-1}$.  The totally-real Jordan integration cycle ${\sC\subset\partial M}$ is the $n$-fold intersection ${\Sigma_{12\cdots n} = \Sigma_1 \cap \cdots \cap \Sigma_n}$, up to choice of orientation.

As an analytic condition, we require each submanifold $\Sigma_{\alpha_1\cdots \alpha_p}$ for ${0<p<n}$ to admit coordinate charts
modelled on ${\R^p \times \C^{n-p}}$, compatible with the
$\bar\partial$-operator from $M$ acting in the complex directions.
This condition amounts to the geometric statement that
$\Sigma_{\alpha_1\cdots \alpha_p}$ is a CR-submanifold
\cite{Dragomir:2006} of $M$.  The CR-condition holds automatically for
the top-dimensional cones ${\Sigma_1,\ldots,\Sigma_n}$ whose oriented union is
$\partial M$.  For application to the partition function $Z^{\rm
  uv}_{S^3}$, the analytic condition will be satisfied trivially, so
we avoid a detour into CR-geometry in the present work.

For an elementary compact example of a polyhedral decomposition, let ${M\simeq \Delta^n}$ be analytically isomorphic to the polydisk, where ${\Delta = \{z\in\C\,\big|\, |z|\le1\}}$ is the closed unit disk in the complex plane.  We take $\sC$ to be the $n$-dimensional real torus ${S^1_1 \times \cdots \times S^1_n}$ in ${\partial M}$; the subscripts serve to label each $S^1$-factor.  Since 
\begin{equation}
\partial M \,=\, \sum_{\alpha=1}^n \,\Delta_1 \times \cdots \times S^1_\alpha \times \cdots \Delta_n\,,
\end{equation}
we set 
\begin{equation}\label{PolyDiskEx}
\Sigma_\alpha \,=\, (-1)^{\alpha+1}\,\Delta_1 \times \cdots \times S^1_\alpha \times \cdots \times \Delta_n\,,
\end{equation}
compatible with the orientation conventions in \eqref{SimpDecM}.  The
components ${\Sigma_1,\ldots,\Sigma_n}$ of ${\partial M}$ then satisfy
the incidence relations of the corresponding simplicial cones in
$\R^{n-1}$, and each $\Sigma_{\alpha_1\cdots \alpha_p}$ inherits the
obvious $\bar\partial$-operator from the polydisk.  With a small 
calculation,\footnote{Observe that $\partial \Sigma_1 = -\Sigma_{12} + \cdots$, $\partial\Sigma_{12} = -\Sigma_{123}+\cdots$, and so on, until $\partial\Sigma_{1 2 \cdots (n-1)} = -\Sigma_{1 2 \cdots n}$.  Alternatively, $\Sigma_1 = S^1_1 \times \Delta_2 \times \cdots \times \Delta_n$, implying $\partial\Sigma_1 = - S^1_1\times S^1_2 \times \Delta_3 \times \cdots \times \Delta_n \,+\, \cdots$.  Hence $\Sigma_{12} = S^1_1 \times S^1_2 \times \Delta_3 \times \cdots \times \Delta_n$.  Continuing inductively with care for signs, one finds $\sC = S_1^1 \times \cdots \times S_n^1 = (-1)^{n(n+1)/2+1}\,\Sigma_{12\cdots n}$.} ${\sC = (-1)^{n(n+1)/2+1}\,\Sigma_{12\cdots n}}$.  This
example underlies the residue theorems in
\cite{Griffiths:78,Passre:1994,Tsikh:1998}.  More generally, the
polydisk $\Delta^n$ can be replaced by a product of Riemann surfaces,
each with one boundary component and arbitrary genus, and the
identical decomposition works.

In addition to our assumption about the existence of a polyhedral decomposition for ${\partial M}$ into `cones' ${\Sigma_1,\ldots,\Sigma_n}$, we require the decomposition to be compatible with the geometry of the polar divisors for the meromorphic form $\omega$ in the following sense.

As in \eqref{PolarDs}, we partition the polar singularities of $\omega$ among divisors ${D_1,\ldots,D_n}$.  Let ${U_\alpha = M - D_\alpha}$ be the open set which is the complement to the divisor $D_\alpha$.  The holomorphic section $s$ vanishes precisely on the intersection ${D_1 \cap \cdots \cap D_n}$, which has support in the interior of $M$,
\begin{equation}
\{s=0\} \,=\, D_1 \cap \cdots \cap D_n \,\subset\, \hbox{int}(M)\,.
\end{equation} Hence $\{U_1,\ldots,U_n\}$ restricts to an open cover for ${\partial M}$.  Since $\omega$ is regular on ${\sC \subset \partial M}$, we have 
\begin{equation}
\sC \,\subset\, U_1 \cap \cdots \cap U_n\,.
\end{equation}
We assume that the open cover $\{U_1,\ldots,U_n\}$ is compatible with the decomposition \eqref{SimpDecM} for ${\partial M}$ in the sense that (after an appropriate choice of labels) each cone $\Sigma_\alpha$ is contained within the corresponding open set $U_\alpha$ for ${\alpha=1,\ldots,n}$,
\begin{flalign}\label{CompatC}
\fbox{compatibility} \qquad\qquad&
\Sigma_\alpha \,\subseteq\, U_\alpha \quad\Longleftrightarrow\quad D_\alpha \cap \Sigma_\alpha \,=\, \varnothing\,.&
\end{flalign}
Equivalently, the component ${s^\alpha \neq 0}$ of ${s=(s^1,\cdots,s^n)}$ is everywhere non-vanishing on the corresponding cone $\Sigma_a$.  

As will be clear in the following construction, we use the compatible cover $\{U_\alpha\}$ to produce a \v Cech representative for the supersymmetric integrand on the left side of \eqref{ResThm}.

The compatibility condition refines our previous requirement that ${s \neq 0}$ on ${\partial M}$ and is considerably more stringent.  For the polydisk example ${M=\Delta^n}$, the cones in \eqref{PolyDiskEx} are compatible with the polar divisors of $\omega$ if, for instance, each ${s^\alpha}$ for $\alpha=1,\ldots,n$ depends only on the corresponding holomorphic coordinate ${z^\alpha\in\Delta_\alpha}$, ie.~the dependence of $s$ on $z$ factorizes, and $s^\alpha$ has no zeroes on the boundary of $\Delta_\alpha$.  Later we provide several down-to-earth examples of the compatibility condition, and in Section \ref{HighRank} we analyze the meaning of the condition in detail for $Z^{\rm uv}_{S^3}$.

\subsection{Reduction to the Real Locus}

Under these assumptions, to reduce the boundary integral in the supersymmetric residue theorem to the real locus $\sC$, we apply the polyhedral decomposition of ${\partial M}$ to rewrite 
\begin{equation}\label{RealC1}
\frac{i}{2} \int_{\partial M} g\,d\mu \left(\frac{{\bf W}}{S}\right)\cdot\CO \,=\, \sum_{\alpha=1}^n\, (-1)^{\alpha+1}\cdot \frac{i}{2} \int_{\Sigma_\alpha} g\,d\mu \left(\frac{{\bf W}}{S}\right)\cdot\CO\,.
\end{equation}
Because the bulk Dolbeault operator restricts to a boundary $\bar\partial$-operator for the local holomorphic/anti-holomorphic coordinates $(w,\bar w)$ on $\partial M$, the supersymmetry $\delta$ also restricts to the boundary. 

The compatibility condition \eqref{CompatC} states that ${s^\alpha\neq 0}$ is non-vanishing on $\Sigma_\alpha$, where we have the relation
\begin{equation}
 1 \,=\ \delta\!\left(\frac{\chi^\alpha}{s^\alpha}\right) = \delta\!\left(\frac{\bf W}{S}\right).
\end{equation}
To exploit this relation termwise on the right in \eqref{RealC1}, we note trivially 
\begin{equation}\label{RealC2}
\begin{aligned}
\int_{\Sigma_\alpha} g\,d\mu\left(\frac{{\bf W}}{S}\right)\cdot\CO \,&=\, \int_{\Sigma_\alpha} g\,d\mu\left(\frac{\bf W}{S} - \frac{\chi^\alpha}{s^\alpha} + \frac{\chi^\alpha}{s^\alpha}\right)\cdot\CO\,,\\
&=\, \int_{\Sigma_\alpha} g\,d\mu \,\delta\!\left(\frac{\chi^\alpha\,{\bf W}}{s^\alpha\,S}\cdot\CO\right) \,+\, \int_{\Sigma_\alpha} g\,d\mu\left(\frac{\chi^\alpha}{s^\alpha}\right)\cdot\CO\,.
\end{aligned}
\end{equation}
In general, neither term on the right in \eqref{RealC2} vanishes.  Because $\Sigma_\alpha$ has a boundary, Stokes Theorem for the $\bar\partial$-operator implies
\begin{equation}\label{BoundOpII}
 \int_{\Sigma_\alpha} g\,d\mu \,\delta\!\left(\frac{\chi^\alpha\,{\bf W}}{s^\alpha\,S}\cdot\CO\right) \,=\, -\frac{i}{2}\int_{\partial\Sigma_\alpha} g\,d\mu\left(\frac{\chi^\alpha\,{\bf W}}{s^\alpha\,S}\right)\cdot\CO\,,
\end{equation}
where by \eqref{BoundOp},
\begin{equation}\label{BoundRel}
\partial\Sigma_\alpha \,=\, \sum_{\beta < \alpha} (-1)^\beta\,\Sigma_{\beta \alpha} \,+\, \sum_{\beta>\alpha} (-1)^{\beta-1}\,\Sigma_{\alpha \beta}\,,
\end{equation}
and the induced boundary measure on each summand is
\begin{equation}
g\,d\mu\big|_{\Sigma_{\alpha\beta}}\,\equiv\, g(x, w) \, d^2x \, d^{n-2} w \, d^{n-2} \bar w \, d^{n-2} \theta \, d^n \chi\,.
\end{equation}
Here we use the assumption that every codimension-one face ${\Sigma_{\alpha \beta}\subset\partial M}$ admits local coordinates modelled on ${\R^2 \times \C^{n-2}}$, compatible with the complex structure on $M$. 

For the second term on the right of \eqref{RealC2}, recall that $\CO$ is a $\delta$-closed operator with ghost-number zero and thus has an expansion in fermions
\begin{equation}\label{ExpandCO}
\CO \,=\, \sum_{q=0}^n \CO^{(q)}_{\bar j_1 \cdots \bar j_q;\,\beta_1\cdots\beta_q}\,\theta^{\bar j_1} \cdots \theta^{\bar j_q}\chi^{\beta_1}\cdots\chi^{\beta_q}\,,
\end{equation}
where $\CO^{(q)}$ transforms as a smooth section of the bundle $A^{(0,q)}(M)\otimes \wedge^q V^*$ appearing in the chain complex \eqref{ChainC}.  From the expression for the boundary measure in \eqref{BoundMeas}, precisely the summand $\CO^{(n-1)}$ contributes to the fermionic integral
\begin{equation}\label{COterm}
\int_{\Sigma_\alpha} g\,d\mu\left(\frac{\chi^\alpha}{s^\alpha}\right)\cdot\CO \,=\, \int_{\Sigma_\alpha}\!\!dx\,d^{n-1} w\,d^{n-1}\bar w\,\frac{g}{s^\alpha}\,\varepsilon^{\bar j_1 \cdots \bar j_{n-1}} \,\varepsilon^{\alpha\, \beta_1 \cdots \beta_{n-1}}\CO^{(n-1)}_{\bar j_1 \cdots \bar j_{n-1};\beta_1 \cdots \beta_{n-1}}\,.
\end{equation}

We now apply the formulas in \eqref{RealC2}, \eqref{BoundOpII}, and
\eqref{COterm} to simplify the left side of \eqref{RealC1}.
Performing the sum over the index $\alpha$ there, we find 
\begin{equation}\label{RealC3}
\begin{aligned}
&\frac{i}{2} \int_{\partial M} g\,d\mu \left(\frac{{\bf
      W}}{S}\right)\cdot\CO
\,=\,\Big(\frac{i}{2}\Big)^2\!\!\sum_{\alpha_1<\alpha_2}
(-1)^{\alpha_1+\alpha_2}\int_{\Sigma_{\alpha_1 \alpha_2}}\!\!g\,d\mu
\left(\frac{\chi^{\alpha_2}}{s^{\alpha_2}}-\frac{\chi^{\alpha_1}}{s^{\alpha_1}}\right)\!\frac{\bf
  W}{S}\cdot\CO\,\,+\,\\
&+\,\Big(\frac{i}{2}\Big)\sum_{\alpha=1}^n
(-1)^{\alpha+1}\int_{\Sigma_\alpha}\!\!dx\,d^{n-1} w\,d^{n-1}\bar
w\,\,\frac{g}{s^\alpha}\,\varepsilon^{\bar j_1 \cdots \bar
  j_{n-1}}\varepsilon^{\alpha\, \beta_1 \cdots
  \beta_{n-1}}\,\CO^{(n-1)}_{\bar j_1 \cdots \bar
    j_{n-1};\beta_1 \cdots \beta_{n-1}}\,.
\end{aligned}
\end{equation}
Because both $s^{\alpha_1}$ and $s^{\alpha_2}$ are non-vanishing on ${\Sigma_{\alpha_1
    \alpha_2}=\Sigma_{\alpha_1}\cap\Sigma_{\alpha_2}}$, the difference of terms in
the first line of \eqref{RealC3} is sensible.  This difference arises
from the boundary relation in \eqref{BoundRel}, after we reorder terms
in the sum over pairs of indices.

If ${n>2}$, the process continues.  We recognize that the integrand on
the right in \eqref{RealC3} can be rewritten similarly to
\eqref{RealC2} as 
\begin{equation}\label{RealC4}
\left(\frac{\chi^{\alpha_2}}{s^{\alpha_2}}-\frac{\chi^{\alpha_1}}{s^{\alpha_1}}\right)\!\frac{\bf
  W}{S}\cdot\CO \,=\, \delta\!\left(\frac{\chi^{\alpha_1} \chi^{\alpha_2}\,{\bf
      W}}{s^{\alpha_1} s^{\alpha_2}\,S}\cdot\CO\right)-\left(\frac{\chi^{\alpha_1}
    \chi^{\alpha_2}}{s^{\alpha_1} s^{\alpha_2}}\right)\cdot\CO\,.
\end{equation}
For the $\delta$-trivial term, we again apply Stokes Theorem to reduce the corresponding integral to the boundary of $\Sigma_{\alpha_1 \alpha_2}$.  For the remainder term not involving ${\bf W}$ or $S$ in \eqref{RealC4}, we do the fermionic integral directly.

Rather than carry out these steps for another special case, let us perform the general induction.  For all ${q>0}$,
\begin{equation}\label{Ind1}
\begin{aligned}
\sum_{\ell=1}^q (-1)^{\ell+1} \left(\frac{\chi^{\alpha_1} \cdots \widehat\chi^{\alpha_\ell}\cdots\chi^{\alpha_q}}{s^{\alpha_1}\cdots\widehat s^{\alpha_\ell}\cdots s^{\alpha_q}}\right)\frac{\bf W}{S}\cdot\CO\,&=\,
\delta\!\left(\frac{\chi^{\alpha_1} \cdots \chi^{\alpha_q}\, {\bf W}}{s^{\alpha_1} \cdots s^{\alpha_q}\, S}\cdot\CO\right)\,+\,\\
&+\,(-1)^{p+1} \left(\frac{\chi^{\alpha_1}\cdots\chi^{\alpha_q}}{s^{\alpha_1}\cdots s^{\alpha_q}}\right)\cdot\CO\,,
\end{aligned}
\end{equation}
where the hats indicate that $\chi^{\alpha_\ell}$ and $s^{\alpha_\ell}$ are omitted from the products on the left.  
If ${q=n}$, the $\delta$-trivial term in \eqref{Ind1} involves $(n+1)$ copies of $\chi$ -- including a copy from ${\bf W}$ -- and so vanishes, from which we obtain the algebraic relation
\begin{equation}\label{AlgInd}
\sum_{\ell=1}^n (-1)^{\ell} \left(\frac{\chi^{\alpha_1} \cdots \widehat\chi^{\alpha_\ell}\cdots\chi^{\alpha_n}}{s^{\alpha_1}\cdots\widehat s^{\alpha_\ell}\cdots s^{\alpha_n}}\right)\frac{\bf W}{S}\cdot\CO\,=\, (-1)^n \left(\frac{\chi^{\alpha_1}\cdots\chi^{\alpha_n}}{s^{\alpha_1}\cdots s^{\alpha_n}}\right)\cdot\CO\,.
\end{equation}

To induct upwards on ${q=2}$ from \eqref{RealC3}, we compute using \eqref{Ind1} that
\begin{equation}\label{Ind2}
\begin{aligned}
&\left(-\frac{i}{2}\right)^q\sum_{\#A=q} (-1)^{|A|}
\int_{\Sigma_A}\!\!g\,d\mu\!\left[\,\sum_{\alpha\in A}  (-1)^{(\alpha,A)+1}
  \left(\frac{\chi^{A-\{\alpha\}}}{s^{A-\{\alpha\}}}\right)\right]\frac{\bf
  W}{S}\cdot\CO \,=\,\\
&\qquad\,=\,\left(-\frac{i}{2}\right)^{q+1}\sum_{\#A=q} (-1)^{|A|}\int_{\partial\Sigma_A}\!\! g\,d\mu \left(\frac{\chi^A\,{\bf W}}{s^A\,S}\right)\cdot\CO\,\,+\,\\
&\qquad\qquad\,+\left(\frac{i}{2}\right)^q\!\sum_{\#A=q} (-1)^{|A|+1} \int_{\Sigma_A}\!\!g\,d\mu \left(\frac{\chi^A}{s^A}\right)\cdot\CO\,,
\end{aligned}
\end{equation}
where ${|A|=\alpha_1 + \cdots + \alpha_q}$ for a multi-index
${A=(\alpha_1 \cdots \alpha_q)}$,
\begin{equation}\label{MultiA}
|A|\,=\,\alpha_1 + \cdots + \alpha_q\,,\qquad\qquad A=(\alpha_1 \cdots \alpha_q)\,.
\end{equation}
The description of the boundary operator in \eqref{BoundOp} then implies
\begin{equation}\label{Ind3}
\begin{aligned}
&\sum_{\#A=q} (-1)^{|A|} 
\int_{\partial\Sigma_A}\!g\,d\mu\!\left(\frac{\chi^A \,{\bf
      W}}{s^A \,S}\right)\cdot\CO\,=\,\\
&\quad\,=\sum_{\#A=q} \sum_{\alpha\in A^o} (-1)^{(\alpha,A^o)+|A|}
\int_{\Sigma_{A\cup\{\alpha\}}} g\,d\mu \left(\frac{\chi^A\, {\bf
      W}}{s^A \,S}\right)\cdot\CO\,,\\
&\quad\,=\, \sum_{\#B=q+1} (-1)^{|B|} \int_{\Sigma_B}g\,d\mu
\left[\sum_{\alpha\in
    B}(-1)^{(\alpha,B)+1}\left(\frac{\chi^{B-\{\alpha\}}}{s^{B-\{\alpha\}}}\right)\right]\frac{\bf W}{S}\cdot\CO\,.
\end{aligned}
\end{equation}
To keep track of signs in the third line of \eqref{Ind3}, we use the
pigeonhole identity
\begin{equation}
(\alpha,A^o) \,=\, (\alpha,B^o\cup\{\alpha\}) \,=\,
\alpha \,-\, (\alpha,B) \,+\, 1\,,\qquad\qquad B = A \cup \{\alpha\}\,.
\end{equation}
In terms of the fermionic expansion \eqref{ExpandCO} of $\CO$, we also evaluate
\begin{equation}
\begin{aligned}
&\int_{\Sigma_A} g \, d\mu \left(\frac{\chi^A}{s^A}\right)\cdot\CO \,=\,\\
&\qquad\qquad\,=\,\int_{\Sigma_A}\!\!d^q x \, d^{n-q}w\, d^{n-q}\bar w \,\frac{g}{s^A}\, \varepsilon^{\bar j_1 \cdots \bar j_{n-q}} \varepsilon^{A\,\beta_1 \cdots \beta_{n-q}} \,\CO^{(n-q)}_{\bar j_1 \cdots \bar j_{n-q};\beta_1 \cdots \beta_{n-q}}\,.
\end{aligned}
\end{equation}
Altogether, the induction relation on $q$ in \eqref{Ind2} becomes
\begin{equation}\label{Ind4}
\begin{aligned}
&\left(-\frac{i}{2}\right)^q\sum_{\#A=q} (-1)^{|A|}
\int_{\Sigma_A}\!\!g\,d\mu\!\left[\,\sum_{\alpha\in A}  (-1)^{(\alpha,A)+1}
  \left(\frac{\chi^{A-\{\alpha\}}}{s^{A-\{\alpha\}}}\right)\right]\frac{\bf
  W}{S}\cdot\CO \,=\,\\
&=\,\left(-\frac{i}{2}\right)^{q+1}  \sum_{\#B=q+1} (-1)^{|B|} \int_{\Sigma_B}g\,d\mu\!
\left[\sum_{\alpha\in
    B}(-1)^{(\alpha,B)+1}\left(\frac{\chi^{B-\{\alpha\}}}{s^{B-\{\alpha\}}}\right)\right]\frac{\bf W}{S}\cdot\CO\,+\,\\
&+\,\left(\frac{i}{2}\right)^q \sum_{\#A=q} (-1)^{|A|+1}\int_{\Sigma_A}\!\!d^q x\, d^{n-q}w\, d^{n-q}\bar w \,\frac{g}{s^A}\, \varepsilon^{\bar j_1 \cdots \bar j_{n-q}} \varepsilon^{A\,\beta_1 \cdots \beta_{n-q}} \,\CO^{(n-q)}_{\bar j_1 \cdots \bar j_{n-m};\beta_1 \cdots \beta_{n-q}}\,.
\end{aligned}
\end{equation}

For ${q=n}$ the induction terminates, since the supersymmetric integral over $\Sigma_B$ on the right of \eqref{Ind4} vanishes in that case, and we obtain a version of the algebraic relation in \eqref{AlgInd}.  So by induction, the boundary term in the residue theorem can be evaluated as a sum of integrals over the polyhedral skeleton of ${\partial M}$,
\begin{equation}\label{BoundInt}
\begin{aligned}
&\frac{i}{2} \int_{\partial M} g\,d\mu \left(\frac{{\bf W}}{S}\right)\cdot\CO \,=\,\sum_{q=1}^n  \sum_{\#A=q} (-1)^{|A|+1}\left(\frac{i}{2}\right)^q\!\times\\
&\qquad\times\int_{\Sigma_A}\!\!d^q x\, d^{n-q}w\, d^{n-q}\bar w \,\,\frac{g}{s^A}\, \varepsilon^{\bar j_1 \cdots \bar j_{n-q}} \varepsilon^{A\,\beta_1 \cdots \beta_{n-q}} \,\CO^{(n-q)}_{\bar j_1 \cdots \bar j_{n-q};\beta_1 \cdots \beta_{n-q}}\,.
\end{aligned}
\end{equation}

To account for the bulk integral on the left side of the
supersymmetric residue theorem in \eqref{ResThm}, observe that the bulk integral
is reproduced by the right side of \eqref{BoundInt} for ${q=0}$, with empty
index set ${A=\emptyset}$, provided we identify ${M =
  \Sigma_\emptyset}$ and ${|\emptyset|=-1}$ for the sign.  Following
those conventions, we obtain a unified polyhedral reformulation of the
supersymmetric residue theorem, \begin{equation}\label{ResThmII}
\begin{aligned}
&\int_M g\,d\mu\,\CO\,+\,\frac{i}{2} \int_{\partial M} g\,d\mu \left(\frac{{\bf W}}{S}\right)\cdot\CO\\
&\qquad=\,\sum_{q=0}^n  \sum_{\#A=q}
(-1)^{|A|+1}\left(\frac{i}{2}\right)^q\,\times\\
&\qquad\qquad\times\int_{\Sigma_A}\!\!d^q x\, d^{n-q}w\, d^{n-q}\bar w
\,\,\frac{g}{s^A}\, \varepsilon^{\bar j_1 \cdots \bar j_{n-q}}
\varepsilon^{A\,\beta_1 \cdots \beta_{n-q}} \,\CO^{(n-q)}_{\bar j_1
  \cdots \bar j_{n-q};\beta_1 \cdots \beta_{n-q}}\\
&\qquad\qquad\qquad=\,\sum_{C \subset \{s=0\}} \langle\CO\rangle_C\,.
\end{aligned}
\end{equation}
The simplest situation occurs when ${\CO=1}$ is the identity, in which
case the bulk integral vanishes and only the term with ${q=n}$ in \eqref{ResThmII} is non-zero.
Explicitly from \eqref{BoundInt},
\begin{equation}\label{BoundInt2}
\begin{aligned}
\frac{i}{2} \int_{\partial M} g\,d\mu \left(\frac{{\bf W}}{S}\right)
\,&=\, (-1)^{\frac{n (n+1)}{2}+1}
\left(\frac{i}{2}\right)^n\int_{\Sigma_{1 2 \cdots
    n}}\frac{g\,d^nx}{s^1 \cdots s^n}\,.
\end{aligned}
\end{equation}
We recognize the integrand on the right as the restriction of the
meromorphic form $\omega$ in \eqref{GrothZ} to the
totally-real cycle $\sC$.  

Recall that for the polydisk example
${M=\Delta^n}$, our orientation conventions imply ${\sC = (-1)^{n
    (n+1)/2 + 1}\,\Sigma_{1 2 \cdots n}}$, which accounts neatly for
the otherwise annoying sign on the right in \eqref{BoundInt2}.  If the section $s$ vanishes
non-degenerately at isolated points in the interior of the polydisk,
and if the polar divisors ${D_1,\ldots,D_n}$ of $\omega$ are compatible with the standard polyhedral
decomposition of the boundary, we derive from \eqref{GrothRes},
\eqref{ResThmII}, and \eqref{BoundInt2} a residue theorem for the Jordan
integral\footnote{Since ${\Delta \simeq \BH_+\cup\{\infty\}}$, the Jordan lemma applies equivalently to ${\sC = \R \times \cdots \times \R}$.} over ${\sC=S^1 \times \cdots \times S^1}$,
\begin{flalign}\label{JordanLem}
\fbox{Jordan lemma}\qquad&
\frac{1}{\left(2\pi i\right)^n}\int_\sC\omega \,=\,
\sum_{s(p)=0}\Res_p(\omega)\,,\qquad\qquad p\in M=\Delta^n\,.&
\end{flalign}
This version of the supersymmetric residue theorem is stated as the
multi-dimensional Jordan lemma in \cite{Passre:1994,Tsikh:1998};
it is what we will use to study the partition function $Z^{\rm uv}_{S^3}$.

\subsection{Toy Examples in Dimensions Two and Three}\label{ToyEx}

In practice, the statement of the multi-dimensional Jordan lemma in \eqref{JordanLem} disguises considerable combinatoric subtlety related to the compatibility condition on the set of divisors $\{D_1,\ldots,D_n\}$ where the individual components of the holomorphic section ${s=(s^1,\cdots\!,s^n)}$ vanish.  We illustrate these issues with several toy examples for ${n=2}$ and ${n=3}$.  In all examples, the relevant vector bundle $V$ is trivial, so $s$ is defined by a collection of holomorphic functions.

\paragraph{Example A.}

We gradually work our way up in complexity.  For an easy start, consider the integral 
\begin{equation}\label{ExampleIone}
\begin{aligned}
\Phi(\mu,\nu) \,&=\, \frac{1}{\left(2\pi i\right)^2} \int_{\R^2}d^2 x\,\,\frac{\e{\!i\,x_2}}{\left(x_1^2 + \mu^2\right) \left((x_1+x_2)^2 + \nu^2\right)}\,,\qquad\qquad \mu,\nu\,\in\,\C\,,\\
&=\,\frac{1}{\left(2\pi i\right)^2} \int_{\R^2}d^2 x\,\,\frac{\e{\!i\,x_2}}{\left(x_1+i\,\mu\right)\left(x_1-i\,\mu\right) \left(x_1+x_2 + i\,\nu\right)\left(x_1+x_2-i\,\nu\right)}
\end{aligned}
\end{equation}
which depends upon complex parameters $\mu$ and $\nu$.  This integral converges absolutely at infinity on $\R^2$.  When $x_1$ and $x_2$ are complexified, the integrand has  poles along the four hyperplanes in $\C^2$ where 
\begin{equation}\label{HyperPs}
z_1 \,=\, \pm i\,\mu\,,\qquad\qquad z_1 + z_2 \,=\, \pm i\,\nu\,,
\end{equation}
with ${z_{1,2} \,=\, x_{1,2} \,+\, i\,y_{1,2}}$.  For ${\mu,\nu\neq 0}$ the poles are simple, and for ${\Re(\mu),\Re(\nu)\neq 0}$ the integrand is everywhere regular on $\R^2$, as we shall assume.  The dependence of $\Phi(\mu,\nu)$ on the parameters is analytic in each quadrant where ${\Re(\mu),\Re(\nu)}$ have definite signs.  When ${\mu,\nu\in\R}$ are real, $\Phi(\mu,\nu)$ must also be real and positive, as one can see by considering the inversion ${(x_1,x_2)\mapsto-(x_1,x_2)}$ in the integrand of \eqref{ExampleIone}.

Let us first evaluate $\Phi(\mu,\nu)$ directly, by applying the single-variable Cauchy theorem twice.  We fix ${x_1\in\R}$, and we evaluate the integral over $x_2$ by closing the real contour in the upper-half of the complex $z_2$-plane.  Depending on the sign of $\Re(\nu)$, we find 
\begin{equation}
\Phi(\mu,\nu) \,=\, -\frac{1}{4\pi\nu}\int_\R dx_1\,\frac{\e{\!-\nu - i\,x_1}}{(x_1^2+\mu^2)}\,,\qquad\qquad \Re(\nu)>0\,,
\end{equation}
versus
\begin{equation}
\Phi(\mu,\nu) \,=\,
\frac{1}{4\pi\nu}\int_\R dx_1\,\frac{\e{\!\nu - i\,x_1}}{(x_1^2+\mu^2)}\,,\qquad\qquad \Re(\nu)<0\,.
\end{equation}
The integral over $x_1$ is evaluated by closing the contour in the lower half of the $z_1$-plane for either case, so depending upon signs,
\begin{equation}\label{ExampleIone2}
\Phi(\mu,\nu) \,=\, \epsilon_{\mu}\,\epsilon_{\nu} \frac{\exp{\!\left(-\epsilon_\mu \,\mu - \epsilon_\nu\,\nu\right)}}{4\,\mu\,\nu}\,,\qquad \epsilon_\mu=\sgn(\Re(\mu))\,,\quad \epsilon_\nu=\sgn(\Re(\nu))\,.
\end{equation}
By comparison, the four polar hyperplanes in \eqref{HyperPs} intersect pairwise at four points in $\C^2$.  Evidently, only one of these four points makes a residue contribution to $\Phi(\mu,\nu)$, and the contributing point depends upon the signs of $\Re(\mu)$ and $\Re(\nu)$.  

Thus, the integral in \eqref{ExampleIone} can be calculated as a sum over residues, but how do we determine {\it a priori} which residues are to be included in the sum?

\iffigs
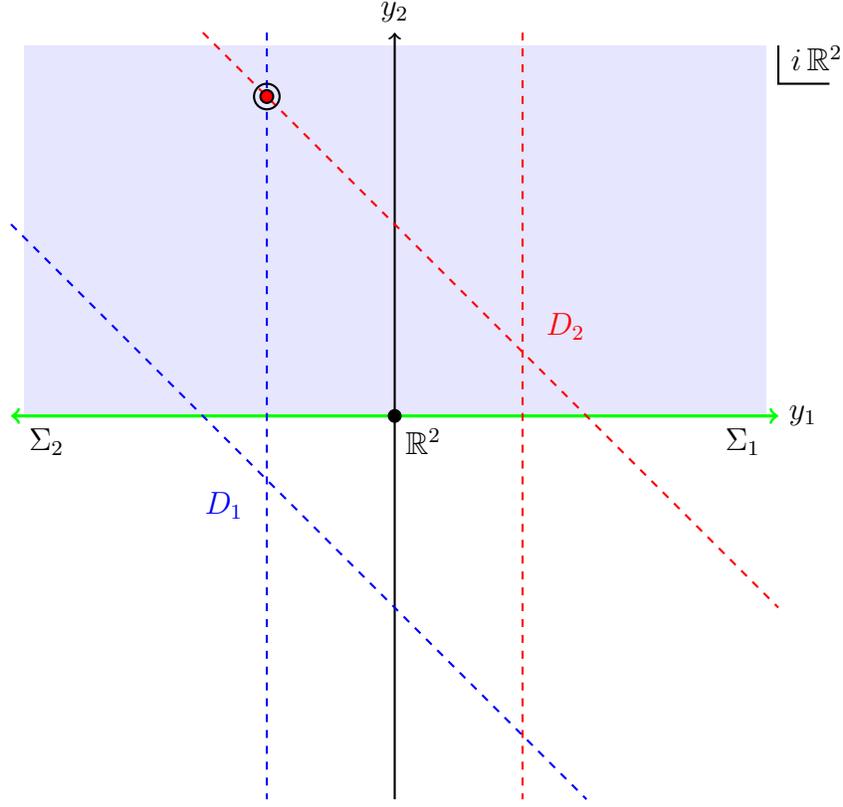
\begin{figure}
	\centering
	\begin{tikzpicture}[x=1.7cm,y=1.7cm] 
	\begin{scope}
	\fill[blue!10] (-2.9,2.9) rectangle (2.9,0);
	\end{scope}

	\draw[very thick,<->,green] (-3,0)--(3,0);
	\node  [ right ] at (3,0) {$y_1$};
	\draw[thick,->] (0,-3)--(0,3);
	\node  [ above ] at (0,3) {$y_2$};

	\draw[fill] (0,0) circle (.05);
	\node [ right ] at (0,-0.2) {$\R^2$};

	\node  [ right ] at (2.5,-0.2) {$\Sigma_1$};
	\node  [ left ] at (-2.5,-0.2) {$\Sigma_2$};	

	\draw[thick, dashed, red] (1,3)--(1,-3);
	\node [right,red] at (1.1,0.7) {$D_2$};
	\draw[thick, dashed, blue] (-1,3)--(-1,-3);
	\node [left,blue] at (-1.1,-0.7) {${D_1}$};

	\draw[thick, dashed, red] (-1.5,3)--(3,-1.5);
		
	\draw[thick, dashed, blue] (-3,1.5)--(1.5,-3);
	\draw [thick] (-1,2.5) circle (.1);	
	\draw [thick, fill=red] (-1,2.5) circle (.05);		
		
	\draw [thick, draw=black, fill=white] (3.0,2.9)--(3.0,2.6)--(3.4,2.6);
	\node  at (3.3,2.8) {$i\,\mathbb{R}^2$};
			
\end{tikzpicture} 
\caption{Polyhedral decomposition on the boundary of ${M=\C\times \BH_+}$ and divisors for $\Phi(\mu,\nu)$ in \eqref{ExampleIone}.  The real Jordan cycle ${\R^2 \subset \partial M = \C \times \R}$ is identified with the origin.  The cones ${\Sigma_{1,2}\simeq \R_+\times \R^2}$ are the half-spaces where ${y_1 \ge 0}$ and ${y_1 \le 0}$, respectively.  The four polar hyperplanes for the integrand of $\Phi(\mu,\nu)$ are indicated schematically by the dashed lines where ${y_1 = \pm\Re(\mu)}$ and ${y_1 + y_2=\pm\Re(\nu)}$.  The intersection of the hyperplanes within the real cycle $\R^2$ is not shown.  Also drawn are compatible divisors $D_1$ and $D_2$, each of which is a union of two hyperplanes.  The hyperplanes in $D_1$ do not intersect the ray corresponding to $\Sigma_1$, and similarly for $D_2$ and $\Sigma_2$.  Of the four pairwise points of intersection for the hyperplanes in $\C^2$, only one point in $M$ is contained in the intersection ${D_1\cap D_2}$ and so contributes a residue to $\Phi(\mu,\nu)$.  The special point with ${y_1\le 0}$ and ${y_2\ge 0}$ is indicated in the figure by the red dot.}\label{PolarDivsExone}
\end{figure}
\fi

The multi-dimensional Jordan lemma answers this combinatoric question if we are able to find a set of compatible divisors $D_1$ and $D_2$. Concretely, to apply the Jordan lemma in \eqref{JordanLem}, we must determine how to rewrite $\Phi(\mu,\nu)$ in the Jordan form
\begin{equation}
\Phi(\mu,\nu) \,=\, \frac{1}{\left(2\pi i\right)^2} \int_{\R^2} \omega\,,\qquad\qquad \omega\,=\, \frac{g(z)\,dz_1\^dz_2}{s^1(z) \, s^2(z)}\,.
\end{equation}
Without much thought, by comparison to \eqref{ExampleIone} we can set ${g(z) = \exp{\!\left(i\,z_2\right)}}$.  But we must consider how to choose $s^1(z)$ and $s^2(z)$ so that the quartic denominator of the integrand is reproduced,
\begin{equation}
s^1(z) \cdot s^2(z) \,=\, \left(z_1^2 + \mu^2\right) \left((z_1+z_2)^2 + \nu^2\right),
\end{equation}
and the divisors ${D_1 = \{s^1=0\}}$ and ${D_2=\{s^2=0\}}$ are compatible with a polyhedral decomposition on the boundary of a complex manifold ${M \subset \C^2}$ containing $\R^2$.

Because the integrand of $\Phi(\mu,\nu)$ decays exponentially on the half-plane where $\Im(z_2)$ is positive, we take ${M = \C \times \BH_+}$ to be the complex half-space, with Jordan cycle ${\sC = \R^2}$ embedded along the real-axis in each factor.  A polyhedral structure on the boundary of $M$ then amounts to a decomposition ${\partial M = \C \times \R = \Sigma_1 - \Sigma_2}$ for some closed $\Sigma_{1,2}$ with intersection ${\Sigma_1 \cap \Sigma_2 = \R^2}$.  A very simple decomposition is provided by the positive and negative half-spaces in ${\C \times \R \simeq \R^3}$,
\begin{equation}\label{ExCones}
\begin{aligned}
\Sigma_1 \,&=\, \quad\big\{ (x_1, y_1, x_2) \in \C\times\R\,\big|\, y_1 \,\ge\, 0\big\}\,,\\
\Sigma_2 \,&=\, -\big\{ (x_1, y_1, x_2)\in\C\times\R\,\big|\, y_1 \,\le\, 0\big\}\,,
\end{aligned}
\end{equation}
where the sign indicates that $\Sigma_2$ carries the opposite orientation to the standard ${dx_1\^dy_1\^dx_2}$.  Trivially, both $\Sigma_1$ and $\Sigma_2$ inherit a partial complex structure from $M$.

We sketch the situation in Figure \ref{PolarDivsExone}, which illustrates the crucial features of the Jordan lemma in this example.  For each choice of signs for $\Re(\mu)$ and $\Re(\nu)$, there is a {\sl unique} way to divide the four polar hyperplanes in \eqref{HyperPs} into a pair of reducible divisors $D_1$ and $D_2$ such that ${D_1 \cap \Sigma_1}$ and ${D_2 \cap \Sigma_2}$ are both empty, as required by the compatibility condition in\eqref{CompatC}.  Eg.~for ${\Re(\mu), \Re(\nu)>0}$, we set 
\begin{equation}\label{Divone}
\begin{aligned}
D_1 \,&=\, \{z_1=-i\,\mu\} \cup \{z_1+z_2=-i\,\nu\}\,,\qquad\qquad \Re(\mu),\Re(\nu)>0\,,\\
D_2 \,&=\, \{z_1=+i\,\mu\}\cup\{z_1+z_2=+i\,\nu\}\,.
\end{aligned}
\end{equation}
Equivalently, this identification of divisors corresponds to the choice of section $s$ with 
\begin{equation}\label{Sone}
s^1(z) \,=\, \left(z_1+i\,\mu\right)\left(z_1+z_2+i\,\nu\right),\qquad\quad s^2(z) \,=\, \left(z_1-i\,\mu\right) \left(z_1+z_2-i\,\nu\right).
\end{equation}
When the signs of $\Re(\mu)$ and $\Re(\nu)$ flip, we must flip the corresponding signs in \eqref{Divone} and \eqref{Sone} so that the geometric configuration of divisors in Figure \ref{PolarDivsExone} is preserved.

By contrast, see Figure \ref{IncptDivsExone} for an example of an incompatible choice for $D_1$ and $D_2$ in this example.

\iffigs
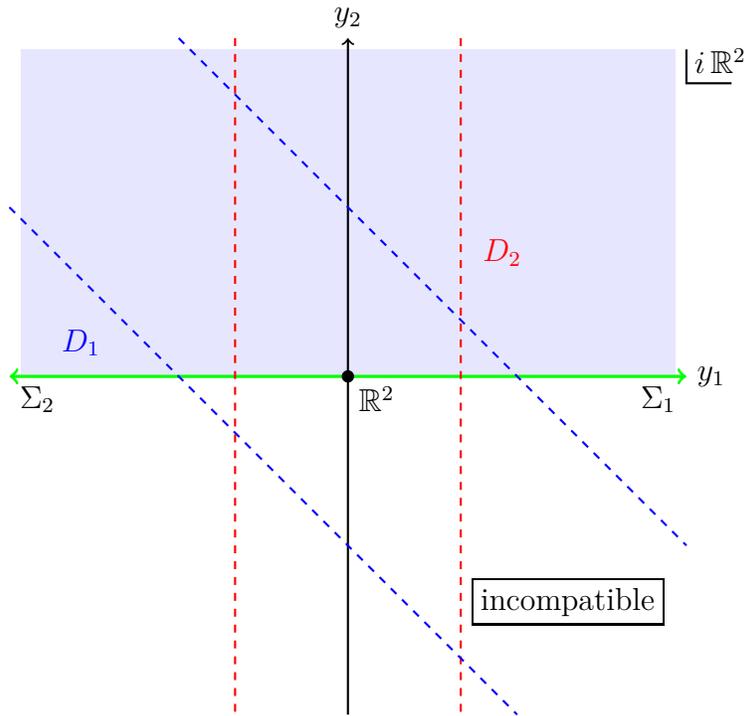
\begin{figure}
	\centering
	\begin{tikzpicture}[x=1.5cm,y=1.5cm] 
	\begin{scope}
	\fill[blue!10] (-2.9,2.9) rectangle (2.9,0);
	\end{scope}

	\draw[very thick,<->,green] (-3,0)--(3,0);
	\node  [ right ] at (3,0) {$y_1$};
	\draw[thick,->] (0,-3)--(0,3);
	\node  [ above ] at (0,3) {$y_2$};

	\draw[fill] (0,0) circle (.05);
	\node [ right ] at (0,-0.2) {$\R^2$};

	\node  [ right ] at (2.5,-0.2) {$\Sigma_1$};
	\node  [ left ] at (-2.5,-0.2) {$\Sigma_2$};	

	\draw[thick, dashed, red] (1,3)--(1,-3);
	\node [right,red] at (1.1,1.1) {$D_2$};
	\draw[thick, dashed, red] (-1,3)--(-1,-3);
	\node [left,blue] at (-2.1,0.3) {${D_1}$};

	\draw[thick, dashed, blue] (-1.5,3)--(3,-1.5);
		
	\draw[thick, dashed, blue] (-3,1.5)--(1.5,-3);

	\node[ right ] at (1,-2) {$\fbox{incompatible}$};
		
	\draw [thick, draw=black, fill=white] (3.0,2.9)--(3.0,2.6)--(3.4,2.6);
	\node  at (3.3,2.8) {$i\,\mathbb{R}^2$};
			
\end{tikzpicture} 
\caption{Incompatible set of divisors $\{D_1,\,D_2\}$ for $\Phi(\mu,\nu)$ in \eqref{ExampleIone}.  Both $D_1$ and $D_2$ have components which intersect each of $\Sigma_1$ and $\Sigma_2$, indicated by the rays in green.}\label{IncptDivsExone}
\end{figure}
\fi

The divisors $D_1$ and $D_2$ in Figure \ref{PolarDivsExone} now intersect at a {\sl unique} point in the bulk $M$, given for ${\Re(\mu),\Re(\nu)>0}$ by 
\begin{equation}
p:\quad (z_1,\,z_2) \,=\, \left(-i\,\mu,\,i\,\mu+i\,\nu\right)\,,\qquad\quad \Re(\mu),\Re(\nu)>0\,.
\end{equation}
One can readily check that the naive result for $\Phi(\mu,\nu)$ in \eqref{ExampleIone2} agrees with the local residue of $\omega$ at $p$,
\begin{equation}
\Phi(\mu,\nu) \,=\, \Res_p(\omega) \,=\, \frac{g(p)}{\det(ds)(p)}\,,
\end{equation}
 for each sign of ${\Re(\mu)}$ and ${\Re(\nu)}$.

\paragraph{Example B.}

A degenerate case of the Jordan lemma occurs when one of the divisors in a compatible set $\{D_1,\ldots,D_n\}$ is empty, corresponding to a section ${s\neq 0}$ which is everywhere non-vanishing on $M$.  In this situation, the Jordan integral on the left in \eqref{JordanLem} vanishes.

For an example of this sort,  we generalize the integral in the preceding example to depend on four parameters ${\mu_1,\mu_2,\nu_1,\nu_2\in\C}$,
\begin{equation}\label{ExampleItwo}
\begin{aligned}
&\Phi(\mu_1,\mu_2,\nu_1,\nu_2) \,=\,\\
&\qquad\frac{1}{\left(2\pi i\right)^2}  \int_{\R^2}d^2 x\,\,\frac{\e{\!i\,x_2}}{\left(x_1-i\,\mu_1\right)\left(x_1-i\,\mu_2\right) \left(x_1+x_2 - i\,\nu_1\right)\left(x_1+x_2-i\,\nu_2\right)}\,.
\end{aligned}
\end{equation}
If the real parts  ${\Re(\mu_1),\Re(\mu_2),\Re(\nu_1),\Re(\nu_2)>0}$ of these parameters are strictly positive, the polar hyperplanes are arranged schematically as in Figure \ref{EmptyD1}.  Each polar hyperplane intersects the half-space $\Sigma_1$ with ${y_1\ge 0}$ in \eqref{ExCones} but does not intersect $\Sigma_2$.  Thus a compatible set of divisors for the previous polyhedral structure on ${M = \C \times \BH_+}$ is given by 
\begin{equation}
\begin{aligned}
D_1 \,&=\, \varnothing\,,\qquad\qquad\qquad\qquad\qquad \Re(\mu_1),\Re(\mu_2),\Re(\nu_1),\Re(\nu_2)>0\,,\\
D_2 \,&=\, \big\{z_1 = i\,\mu_1\big\}\cup\big\{z_1 = i\,\mu_2\big\}\cup\big\{z_1 + z_2 = i\,\nu_1\big\}\cup\big\{z_1 + z_2 = i\,\nu_2\big\}\,.
\end{aligned}
\end{equation}
Equivalently, the denominator of $\omega$ for this example is factorized trivially as the product of 
\begin{equation}
\begin{aligned}
s^1(z) \,&=\, 1\,,\\
s^2(z) \,&=\, \left(z_1-i\,\mu_1\right)\left(z_1-i\,\mu_2\right) \left(z_1+z_2 - i\,\nu_1\right)\left(z_1+z_2-i\,\nu_2\right).
\end{aligned}
\end{equation}
The Jordan lemma immediately implies 
\begin{equation}
\Phi(\mu_1,\mu_2,\nu_1,\nu_2) \,=\, 0\,,\qquad\qquad \Re(\mu_1),\Re(\mu_2),\Re(\nu_1),\Re(\nu_2)>0\,.
\end{equation}
Note that $\Phi(\mu_1,\mu_2,\nu_1,\nu_2)$ vanishes despite the prescence of an apparent simple pole located in the interior of $M$.  

One can also check that ${\Phi(\mu_1,\mu_2,\nu_1,\nu_2) \,=\, 0}$ by iteratively applying the single-variable Cauchy theorem.  One closes the contour for $z_2$ in the upper half-plane, but one must then close the contour for $z_1$ in the lower half-plane, where the integrand is regular.

\iffigs
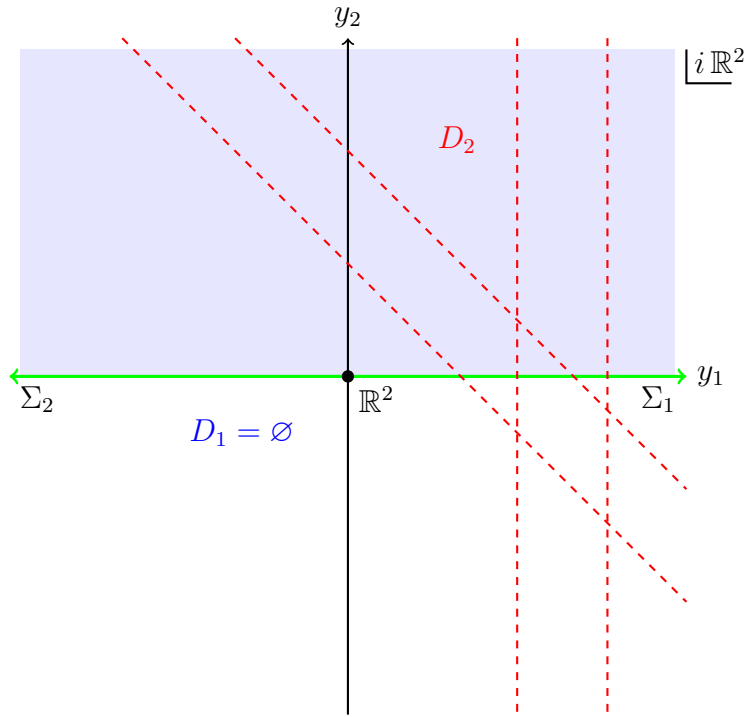
\begin{figure}
	\centering
	\begin{tikzpicture}[x=1.5cm,y=1.5cm] 
	\begin{scope}
	\fill[blue!10] (-2.9,2.9) rectangle (2.9,0);
	\end{scope}

	\draw[very thick,<->,green] (-3,0)--(3,0);
	\node  [ right ] at (3,0) {$y_1$};
	\draw[thick,->] (0,-3)--(0,3);
	\node  [ above ] at (0,3) {$y_2$};

	\draw[fill] (0,0) circle (.05);
	\node [ right ] at (0,-0.2) {$\R^2$};

	\node  [ right ] at (2.5,-0.2) {$\Sigma_1$};
	\node  [ left ] at (-2.5,-0.2) {$\Sigma_2$};	

	\draw[thick, dashed, red] (1.5,3)--(1.5,-3);
	\node [right,red] at (0.7,2.1) {$D_2$};
	\draw[thick, dashed, red] (2.3,3)--(2.3,-3);
	\node [right,blue] at (-1.5,-0.5) {${D_1}=\varnothing$};

	\draw[thick, dashed, red] (-1,3)--(3,-1);
		
	\draw[thick, dashed, red] (-2,3)--(3,-2);
		
	\draw [thick, draw=black, fill=white] (3.0,2.9)--(3.0,2.6)--(3.4,2.6);
	\node  at (3.3,2.8) {$i\,\mathbb{R}^2$};
			
\end{tikzpicture} 
\caption{A compatible set of divisors $\{D_1,D_2\}$ with $D_1$ empty.  According to the Jordan lemma, the Jordan integral vanishes, despite the presence of an apparent pole in the shaded region corresponding to the interior of $M$.}\label{EmptyD1}
\end{figure}
\fi

\paragraph{Example C.}

So far, the divisors $D_1$ and $D_2$ have been finite unions of
hyperplanes.  In our application to $Z^{\rm uv}_{S^3}$, the relevant
polar divisors will instead be unions of infinitely-many
hyperplanes, so let us give an elementary example with that feature.
We consider the Jordan integral
\begin{equation}\label{ExampleIthree}
\begin{aligned}
&\Psi(\mu_1,\mu_2,\mu_3) \,=\,\\
&\frac{1}{(2\pi
  i)^2}\int_{\R^2}d^2x\,\,\frac{1}{\cosh\!\left(\pi(x_1+\mu_1)\right)
\cosh\!\left(\pi(x_2 - x_1 + \mu_2)\right)
\cosh\!\left(\pi(x_2+\mu_3)\right)}\,,
\end{aligned}
\end{equation}
with ${\mu_1,\mu_2,\mu_3\in\C}$ such that ${-\ha<
  \Im\!\left(\mu_{1,2,3}\right) < \ha}$.  The latter condition ensures
that no pole of the integrand meets the real cycle ${\R^2 \subset \C^2}$.  

The integral over $\R^2$ converges absolutely, so $\Psi$ is an analytic function of the parameters $(\mu_1,\mu_2,\mu_3)$.  Invariance under constant shifts in the integration variables $x_1$ and $x_2$ means that $\Psi$ depends on the parameters only in the combination ${\mu_1+\mu_2-\mu_3}$.  Nonetheless, we keep all dependence on $(\mu_1,\mu_2,\mu_3)$ explicit and use the latter observation as a check of our calculation.

To ensure convergence of the infinite sum over residues in this example, we consider the more general
\begin{equation}\label{ExampleIthreeEps}
\begin{aligned}
&\Psi_\epsilon(\mu_1,\mu_2,\mu_3) \,=\,\\
&\frac{1}{(2\pi
  i)^2}\int_{\R^2}d^2x\,\,\frac{\e{\!i \epsilon_1 x_1 + i
\epsilon_2 x_2}}{\cosh\!\left(\pi(x_1+\mu_1)\right)
\cosh\!\left(\pi(x_2 - x_1 + \mu_2)\right)
\cosh\!\left(\pi(x_2+\mu_3)\right)}\,,
\end{aligned}
\end{equation}
for ${\epsilon_1, \epsilon_2 \in\R_+}$.  When ${0 < \epsilon_1 \ll \epsilon_2}$, the integrand decays in a slightly tilted version of the complex half-space ${M=\C \times \BH_+}$ that appeared previously.  See Figure \ref{CschEx} for an illustration.  After we apply the Jordan lemma to evaluate $\Psi_\epsilon(\mu_1,\mu_2,\mu_3)$, we can recover the value of the integral in \eqref{ExampleIthree} by taking the limit ${\epsilon_1,\epsilon_2\to 0^+}$.

When extended to a function on $\C^2$, the integrand in \eqref{ExampleIthreeEps} has poles along the infinite set of hyperplanes
\begin{equation}\label{HyPExThree}
\begin{aligned}
z_1 \,+\, \mu_1 \,&=\, i \left(n_1 \,+\, \ha\right)\,,\qquad n_1\,\in\,\Z\,,\\
z_2 \,-\, z_1 \,+\, \mu_2 \,&=\, i \left(n_2 \,+\, \ha\right)\,,\qquad n_2\,\in\,\Z\,,\\
z_2 \,+\, \mu_3 \,&=\, i \left(n_3 \,+\, \ha\right)\,,\qquad n_3\,\in\,\Z\,,
\end{aligned}
\end{equation}
corresponding to the three factors in the denominator in \eqref{ExampleIthreeEps}.  For the tilted version of $\Sigma_1$ and $\Sigma_2$ in \eqref{ExCones}, these hyperplanes can be assigned uniquely to the compatible pair
\begin{equation}\label{CompatDThr}
\begin{aligned}
D_1 \,&=\, \left\{z_1+\mu_1 = i \left(n_1+\ha\right) \Big|\,  n_1 < 0\right\}\\ 
&\cup \left\{ z_2 - z_1 + \mu_2 = i \left(n_2+\ha\right) \Big|\, n_2 \ge 0\right\}\\
&\cup\left\{z_2+\mu_3=i\left(n_3+\ha\right) \Big|\,n_3 \ge 0 \right\}\,,
\end{aligned}
\end{equation}
and
\begin{equation}
\begin{aligned}
D_2 \,&=\, \left\{z_1+\mu_1 = i \left(n_1+\ha\right) \Big|\,  n_1 \ge 0\right\}\\
&\cup \left\{z_2 - z_1 + \mu_2 = i \left(n_2+\ha\right) \Big|\, n_2 < 0\right\}\\
&\cup\left\{z_2+\mu_3=i\left(n_3+\ha\right) \Big|\,n_3 < 0 \right\}\,.
\end{aligned}
\end{equation}
A picture here is worth a thousand words.  See Figure \ref{CschEx} for sketches of $D_1$ and $D_2$.

\iffigs
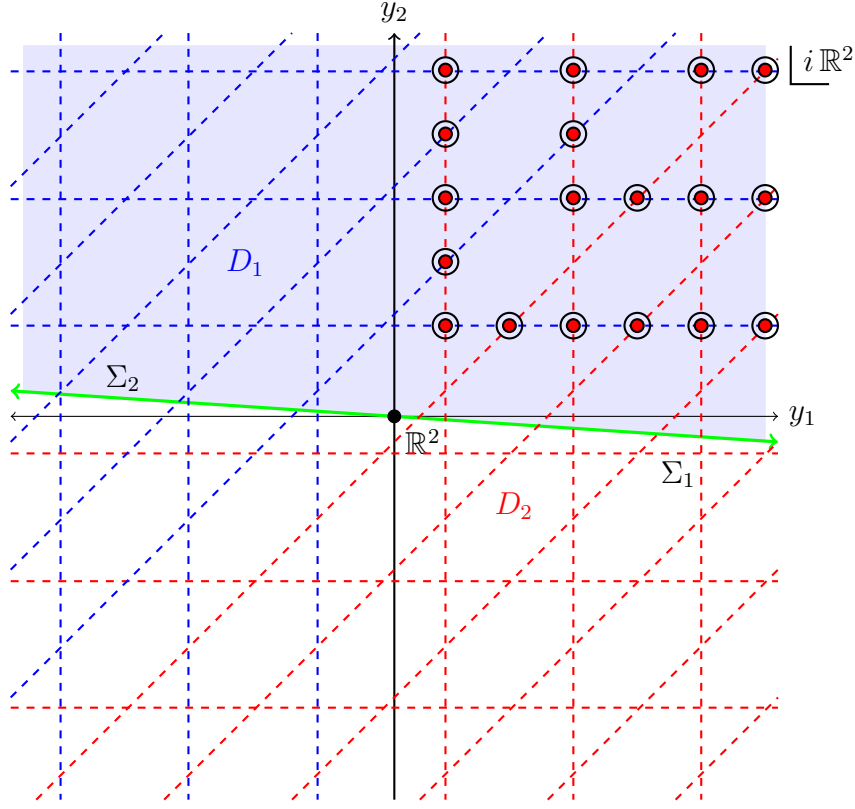
\begin{figure}
	\centering
	\begin{tikzpicture}[x=1.7cm,y=1.7cm] 
	
	\draw [fill=blue!10,blue!10] (-2.9,2.9)--(-2.9,0.2)--(2.9,-0.2)--(2.9,2.9)--(-2.9,2.9);		
	\draw[<->,black] (-3,0)--(3,0);
	\draw[very thick,<->,green] (-3,0.2)--(3,-0.2);
	\node  [ right ] at (3,0) {$y_1$};
	\draw[thick,->] (0,-3)--(0,3);
	\node  [ above ] at (0,3) {$y_2$};

	\draw[fill] (0,0) circle (.05);
	\node [ right ] at (0,-0.2) {$\R^2$};

	\node  [ right ] at (2.0,-0.45) {$\Sigma_1$};
	\node  [ left ] at (-1.9, 0.3) {$\Sigma_2$};	

	\draw[thick, dashed, blue] (-2.61,3)--(-2.61,-3);
	\draw[thick, dashed, blue] (-1.61,3)--(-1.61,-3);
	\draw[thick, dashed, blue] (-0.6,3)--(-0.6,-3);
	\draw[thick, dashed, red] (0.4,3)--(0.4,-3);
	\draw[thick, dashed, red] (1.4,3)--(1.4,-3);
	\draw[thick, dashed, red] (2.4,3)--(2.4,-3);

	\draw[thick,dashed,blue] (-3,0.71)--(3,0.71);
	\draw[thick,dashed,blue] (-3,1.7)--(3,1.7);
	\draw[thick,dashed,blue] (-3,2.7)--(3,2.7);
	\draw[thick,dashed,red] (-3,-0.29)--(3,-0.29);
	\draw[thick,dashed,red] (-3,-1.29)--(3,-1.29);
	\draw[thick,dashed,red] (-3,-2.28)--(3,-2.28);

	\draw[thick,dashed,red] (2.2,-3.0)--(3.0,-2.2);
	\draw[thick,dashed,red] (1.2,-3.0)--(3.0,-1.2);
	\draw[thick,dashed,red] (0.2,-3.0)--(3.0,-0.2);
	\draw[thick,dashed,red] (-.8,-3.0)--(3.0,0.8);
	\draw[thick,dashed,red] (-1.8,-3.0)--(3.0,1.8);
	\draw[thick,dashed,red] (-2.8,-3.0)--(3.0,2.8);

	\draw[thick,dashed,blue] (-3.0,-2.2)--(2.2,3.0);
	\draw[thick,dashed,blue] (-3.0,-1.2)--(1.2,3.0);
	\draw[thick,dashed,blue] (-3.0,-0.2)--(0.2,3.0);
	\draw[thick,dashed,blue] (-3.0,0.8)--(-0.8,3.0);
	\draw[thick,dashed,blue] (-3.0,1.8)--(-1.8,3.0);

	\draw [thick] (0.4,0.71) circle (.1);	
	\draw [thick, fill=red] (0.4,0.71) circle (.05);		
	\draw [thick] (1.4,0.71) circle (.1);	
	\draw [thick, fill=red] (1.4,0.71) circle (.05);		
	\draw [thick] (2.4,0.71) circle (.1);	
	\draw [thick, fill=red] (2.4,0.71) circle (.05);		

	\draw [thick] (0.4,1.71) circle (.1);	
	\draw [thick, fill=red] (0.4,1.71) circle (.05);		
	\draw [thick] (1.4,1.71) circle (.1);	
	\draw [thick, fill=red] (1.4,1.71) circle (.05);		
	\draw [thick] (2.4,1.71) circle (.1);	
	\draw [thick, fill=red] (2.4,1.71) circle (.05);		

	\draw [thick] (0.4,2.71) circle (.1);	
	\draw [thick, fill=red] (0.4,2.71) circle (.05);		
	\draw [thick] (1.4,2.71) circle (.1);	
	\draw [thick, fill=red] (1.4,2.71) circle (.05);		
	\draw [thick] (2.4,2.71) circle (.1);	
	\draw [thick, fill=red] (2.4,2.71) circle (.05);	

	\draw [thick] (0.9,0.71) circle (.1);	
	\draw [thick, fill=red] (0.9,0.71) circle (.05);		
	\draw [thick] (1.9,0.71) circle (.1);	
	\draw [thick, fill=red] (1.9,0.71) circle (.05);		
	\draw [thick] (2.9,0.71) circle (.1);	
	\draw [thick, fill=red] (2.9,0.71) circle (.05);		

	\draw [thick] (1.9,1.71) circle (.1);	
	\draw [thick, fill=red] (1.9,1.71) circle (.05);
	\draw [thick] (2.9,1.71) circle (.1);	
	\draw [thick, fill=red] (2.9,1.71) circle (.05);			
	\draw [thick] (2.9,2.71) circle (.1);	
	\draw [thick, fill=red] (2.9,2.71) circle (.05);

	\draw [thick] (0.4,1.21) circle (.1);	
	\draw [thick, fill=red] (0.4,1.21) circle (.05);	
	\draw [thick] (0.4,2.21) circle (.1);	
	\draw [thick, fill=red] (0.4,2.21) circle (.05);		
	\draw [thick] (1.4,2.21) circle (.1);	
	\draw [thick, fill=red] (1.4,2.21) circle (.05);		
	\node [right,red] at (0.7,-0.7) {$D_2$};
	\node [right,blue] at (-1.4,1.2) {$D_1$};

	\draw [thick, draw=black, fill=white] (3.1,2.9)--(3.1,2.6)--(3.4,2.6);
	\node  at (3.4,2.8) {$i\,\mathbb{R}^2$};
			
\end{tikzpicture} 
\caption{Polar divisors for $\Psi_\epsilon(\mu_1,\mu_2,\mu_3)$ in \eqref{ExampleIthreeEps}.  For ${0<\epsilon_1\ll\epsilon_2}$, $M$ is a slightly tilted version of the complex half-space ${\C\times\BH_+}$.  The tilt removes the ambiguity in whether to assign the horizontal hyperplanes with constant ${z_2}$ to either the divisor $D_1$ (blue) or $D_2$ (red).  Also indicated are the points in the intersection ${D_1 \cap D_2 \cap M}$ which contribute to the sum over residues.  These points lie exclusively in the positive quadrant ${y_1,y_2 > 0}$.}\label{CschEx}
\end{figure}
\fi

In this example, the factorization corresponding to the polar divisors $D_1$ and $D_2$ has nothing to do with the naive factorization of the denominator in \eqref{ExampleIthreeEps}.  Rather, we use the reflection formula \eqref{RefEulGam} for the gamma function to rewrite
\begin{equation}\label{RefEulGamII}
\frac{\pi}{\cosh\!\left(\pi x\right)} \,=\, \Gamma\!\left(\ha+i\,x\right) \Gamma\!\left(\ha-i\,x\right)\,.
\end{equation}
The meromorphic form ${\omega = g(z) \,dz_1 \^ dz_2 / s^1(z) \,s^2(z)}$ for the integrand in \eqref{ExampleIthreeEps} is then given by the Mellin-type data
\begin{equation}\label{ThreeSs}
\begin{aligned}
s^1(z) &=\, \left[\Gamma\!\left(\ha - i\left(z_1 + \mu_1\right)\right) \Gamma\!\left(\ha + i \left(z_2-z_1+\mu_2\right)\right) \Gamma\!\left(\ha+i\left(z_2+\mu_3\right)\right)\right]^{-1},\\
s^2(z) &=\, \left[\Gamma\!\left(\ha+i\left(z_1+\mu_1\right)\right) \Gamma\!\left(\ha-i\left(z_2-z_1+\mu_2\right)\right) \Gamma\!\left(\ha-i\left(z_2+\mu_3\right)\right)\right]^{-1}, 
\end{aligned}
\end{equation}
and
\begin{equation}\label{ThreeG}
g(z) \,=\, \frac{1}{\pi^3}\,\exp{\!\left(i \epsilon_1 z_1 + i \epsilon_2 z_2\right)}\,.
\end{equation}
With the definition in \eqref{ThreeSs}, $s^1(z)$ vanishes on the three families of hyperplanes which enter the divisor $D_1$ in \eqref{CompatDThr}, and similarly for $s^2(z)$.

According to the Jordan lemma, only the residues from points in the intersection ${D_1 \cap D_2 \cap M}$ contribute to $\Psi_\epsilon(\mu_1,\mu_2,\mu_3)$.  From \eqref{CompatDThr}, a brief calculation yields three infinite families of points in ${D_1 \cap D_2 \cap M}$,
\begin{equation}\label{PolesExThree}
\begin{aligned}
D_1 \cap D_2 \cap M \,&=\, \left\{\left(i\left(n_1+\ha\right)-\mu_1,\, i\left(n_1+n_2+1\right)-\mu_1-\mu_2\right) \Big|\, n_1, n_2 \ge 0\right\}\\
&\cup  \left\{\left(i\left(n_1+\ha\right)-\mu_1,\, i\left(n_3+\ha\right)-\mu_3\right) \Big|\, n_1, n_3 \ge 0\right\}\\
&\cup  \left\{\left(i\left(n_3 - n_2\right)+\mu_2-\mu_3,\, i\left(n_3+\ha\right)-\mu_3\right) \Big|\, n_2 < 0\,, n_3 \ge 0\right\}.
\end{aligned}
\end{equation}
As a check, note that all three families of points occur in the quadrant ${\Im(z_{1,2}) > 0}$ shown in Figure \ref{CschEx}.  Like the prior examples, the hyperplanes in \eqref{HyPExThree} do intersect at infinitely-many other points in $M$ beyond those in \eqref{PolesExThree}, but these extra points are not summed in the residue calculus for $\omega$.

The local residue of $\omega$ at each pole may be computed using the well-known formula for the residue of the gamma function at a negative integer,
\begin{equation}
\Res\big[\Gamma(z)\big]_{z=-n} \,=\, \frac{(-1)^n}{n!}\,,\qquad\qquad n\,\ge\,0\,.
\end{equation}
For the points labelled by the pairs of integers in
\eqref{PolesExThree}, the respective residues are 
\begin{equation}\label{ResExThree}
\begin{aligned}
\Res[\omega]_{n_1, n_2}
\,&=\,\frac{\exp{\!\left(-\ha \epsilon_1 -i \epsilon_1
      \mu_1 - \epsilon_2 - i \epsilon_2 (\mu_1 +
      \mu_2)\right)}}{\pi^2 \cosh\!\left[\pi (\mu_1+\mu_2-\mu_3)\right]}\cdot
\e{\!-\epsilon_1 n_1 - \epsilon_2 (n_1+n_2)}\,,\\
\Res[\omega]_{n_1, n_3} \,&=\,
-\frac{\exp{\!\left(-\ha\epsilon_1 - i \epsilon_1 \mu_1
    - \ha \epsilon_2 - i \epsilon_2 \mu_3\right)}}{\pi^2 \cosh\!\left[\pi
    (\mu_1+\mu_2-\mu_3)\right]}\cdot\e{\!-\epsilon_1 n_1 - \epsilon_2 n_3}\,,\\
\Res[\omega]_{n_2, n_3} \,&=\,
\frac{\exp{\!\left(i \epsilon_1 (\mu_2-\mu_3) - \ha \epsilon_2 - i
      \epsilon_2 \mu_3\right)}}{\pi^2 \cosh\!\left[\pi(\mu_1+\mu_2-\mu_3)\right]}\cdot\e{\!-\epsilon_1
  (n_3-n_2) - \epsilon_2 n_3}\,. 
\end{aligned}
\end{equation}
So by the multidimensional Jordan lemma in \eqref{JordanLem},
\begin{equation}\label{SumExThree}
\Psi_\epsilon(\mu_1,\mu_2,\mu_3) \,= \sum_{n_1, n_2 \ge 0} \Res[\omega]_{n_1, n_2} \,+ \sum_{n_1,n_3\ge
  0} \Res[\omega]_{n_1, n_3}  \,+ \sum_{n_2<0,n_3\ge
  0}\Res[\omega]_{n_2, n_3}\,.
\end{equation}
Given the dependence on $n_1$, $n_2$, and $n_3$ in \eqref{ResExThree}, the residues can be summed as power series, from which we obtain the analytic expression
\begin{equation}\label{PsiExThree}
\begin{aligned}
&\Psi_\epsilon(\mu_1,\mu_2,\mu_3)
\,=\,\frac{1}{\pi^2 \cosh\!\left[\pi (\mu_1+\mu_2-\mu_3)\right]}\,\times\\
&\times\Biggr(\frac{\e{\!-\ha \epsilon_1 -i \epsilon_1
      \mu_1 - \epsilon_2 - i \epsilon_2 (\mu_1 +
      \mu_2)}}{\left(1-\e{\!-\epsilon_1-\epsilon_2}\right)\left(1-\e{\!-\epsilon_2}\right)}\,-\,\frac{\e{\!-\ha\epsilon_1 - i \epsilon_1 \mu_1
    - \ha \epsilon_2 - i \epsilon_2
    \mu_3}}{\left(1-\e{\!-\epsilon_1}\right)\left(1-\e{\!-\epsilon_2}\right)}\,+\,
\frac{\e{\!-\epsilon_1+i \epsilon_1 (\mu_2-\mu_3) - \ha \epsilon_2 - i
      \epsilon_2
      \mu_3}}{\left(1-\e{\!-\epsilon_1}\right)\left(1-\e{\!-\epsilon_1-\epsilon_2}\right)}\Biggr)\,.
\end{aligned}
\end{equation}

The three terms on the second line of \eqref{PsiExThree} arise from the three sums in \eqref{SumExThree}.  Individually, each term diverges as ${\epsilon_1,\epsilon_2\to 0}$, but the total quantity on the right of \eqref{PsiExThree} has a well-defined limit,
\begin{equation}
\Psi(\mu_1,\mu_2,\mu_3) \,= \lim_{\epsilon_1,\epsilon_2\to 0} \Psi_\epsilon(\mu_1,\mu_2,\mu_3) \,=\, -\frac{1+4 \left(\mu_1+\mu_2-\mu_3\right)^2}{8\pi^2 \cosh\!\left[\pi (\mu_1+\mu_2-\mu_3)\right]}\,,
\end{equation}
which provides the value of the Jordan integral in \eqref{ExampleIthree}.  The existence of the limit serves as a delicate check on the residue computation.

\paragraph{Example D.}

The statement of the multi-dimensional Jordan lemma assumes the
existence of a polyhedral decomposition for ${\partial M}$ which is
compatible with the set of divisors $\{D_1,\ldots,D_n\}$.  The
existence of the decomposition is not guaranteed even when $M$ is
affine, and the situation becomes more precarious as the dimension of
$M$ grows.  We illustrate the main difficulty -- and its resolution --
with a toy example in complex dimension three, for
${M=\C^2\times\BH_+}$.  Later in Section \ref{JordHRGG}, we give a
more general and systematic discussion.

Consider the Jordan integral 
\begin{equation}\label{ExXiD}
\begin{aligned}
&\Xi(\mu_1,\mu_2,\mu_3,\mu_4) \,=\,\\
&\frac{1}{\left(2\pi i\right)^3}\int_{\R^3}\!\!d^3x\,\frac{\e{i\,x_3}}{\left(x_1-i\,\mu_1\right)\left(x_1+x_2-i\,\mu_2\right) \left(x_2+x_3-i\,\mu_3\right)\left(2\,x_1-x_2+x_3-i\,\mu_4\right)}\,.
\end{aligned}
\end{equation}
We take ${\mu_{1,2,3,4}\in\R}$ to be real and non-zero, so that the
integrand is everywhere regular on $\R^3$.  In this toy example, the
integral is only conditionally convergent, which is the price we pay to achieve a particularly simple configuration of four hyperplanes in $\C^3$ along which the integrand has a pole.  If one wishes, one can apply a version of the ${i\varepsilon}$-prescription to deform the integration contour ${\R^3 \subset \C^3}$ upwards into complex $z_3$-direction as ${||x||\to\infty}$, so that the integrand in \eqref{ExXiD} decays exponentially at infinity.

\iffigs
\begin{figure}
	\centering
	\begin{tikzpicture}[x=1.7cm,y=1.7cm] 

		\draw [fill=blue!5,blue!5] (0,2.9)--(0,0)--(2.9,0)--(2.9,2.9)--(0,2.9);
		\draw [fill=blue!10,blue!10] (-2.9,2.9)--(-2.9,-2.9)--(0,0)--(0,2.9)--(-2.9,2.9);
		\draw [fill=blue!15,blue!15] (0,0)--(-2.9,-2.9)--(2.9,-2.9)--(2.9,0)--(0,0);

		\draw[->] (-3,0.01)--(3,0.01);
		\node  [ right ] at (3,0) {$y_1$};
		\draw[->] (0,-3)--(0,3);
		\node  [ above ] at (0,3) {$y_2$};

		\draw[dashed, red] (1,-3)--(1,3.1);
		
		\draw[dashed, red] (-3,1)--(3,1);
		
		\draw[dashed, red] (-3,2)--(2,-3);
			
		\draw[dashed, red] (-2,-3)--(1.05,3.1);
		\node[ red, above] at (-0.5,0.1) {$\ell$};
		
		\draw[very thick, ->, green] (0,0.01)--(3,0.0);
		\node[ below ] at (2.7,0.0) {$\Sigma_{12}$};
		
		\draw[very thick,->,green] (0,0.01)--(0,3.01);
		\node[ left] at (0,2.7) {$\Sigma_{23}$};

		\draw[very thick,->,green] (0,0)--(-3.01,-3);
		\node[ above] at (-2.7,-2.55) {$\Sigma_{13}$};

		\node[ ] at (2,2) {\fbox{$-\Sigma_2$}};

		\node[ ] at (2,-2) {\fbox{$\Sigma_1$}};

		\node[ ] at (-2,2) {\fbox{$\Sigma_3$}};

		\draw[fill] (0,0) circle (.05);
		\node [ right ] at (0,-0.2) {$\R^3$};
		\draw [thick, draw=black, fill=white] (2.9,2.9)--(2.9,2.4)--(3.4,2.4);
		\node at (3.2,2.7) {${i\,\R^2}$};	
		\end{tikzpicture} 
		\caption{Arrangement of polar hyperplanes in ${\partial M = \C^2 \times \R}$ for $\Xi(\mu)$ in \eqref{ExXiD}.  In this figure, ${y_3=0}$, and the real directions in $\partial M$ are suppressed.  Each of the four polar hyperplanes for the integrand in  \eqref{ExXiD} is indicated by a corresponding dashed line (red).  This arrangement is appropriate for the parameter regime ${\mu_1>0}$, ${\mu_2<0}$, ${\mu_3>0}$, and ${\mu_4<0}$.  The three shaded regions describe a polyhedral decomposition of ${\partial M}$ into subsets $\Sigma_{1,2,3}$, bounded by rays $\Sigma_{12}$, $\Sigma_{23}$, and $\Sigma_{13}$ (green).  The polar hyperplane labelled by $\ell$ is incompatible with the polyhedral decomposition of ${\partial M}$, since $\ell$ passes through each of $\Sigma_{1,2,3}$.  A compatible set of divisors $\{D_1,D_2,D_3\}$ does not exist.}\label{FigExD}
	\end{figure}
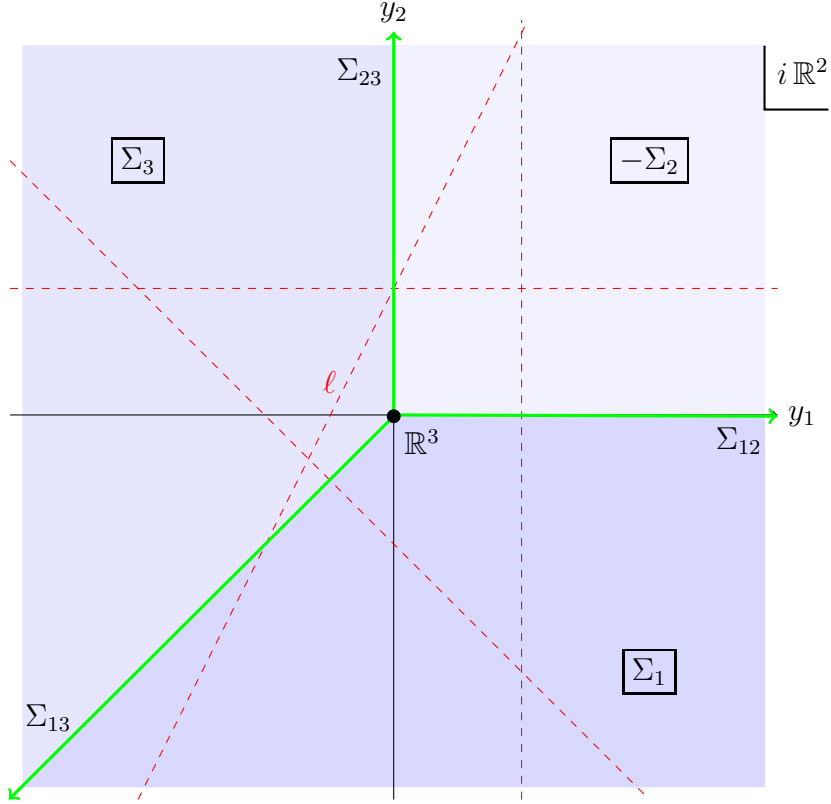
\fi

Given ${g(z) = \exp{(i\,z_3)}}$ in the numerator, we take ${M=\C^2 \times \BH_+}$.  Rather than attempt to sketch the arrangement of the four polar hyperplanes in $M$ by analogy to Figure \ref{PolarDivsExone}, which would require at least three real dimensions, we will just sketch the intersection of the polar hyperplanes with the boundary ${\partial M = \C^2 \times \R}$, where ${\Im(z_3)\equiv y_3=0}$.  See Figure \ref{FigExD}, in which the three real directions in $\partial M$ are suppressed, and the intersection of each hyperplane with ${\partial M}$ is indicated by a dashed line.  Also shown in the figure is a polyhedral decomposition ${\partial M = \Sigma_1 - \Sigma_2 + \Sigma_3}$, where 
\begin{equation}
\begin{aligned}
\Sigma_1 \,&=\, \Big\{(x_1,x_2,x_3,y_1,y_2)\in\C^2\times\R \,\Big|\,y_1-y_2\ge 0,\,y_2 \le 0\Big\}\,,\\
\Sigma_2 \,&=\, -\Big\{(x_1,x_2,x_3,y_1,y_2)\in\C^2\times\R \,\Big|\,y_1\ge 0,\,y_2 \ge 0\Big\}\,,\\
\Sigma_3 \,&=\, \Big\{(x_1,x_2,x_3,y_1,y_2)\in\C^2\times\R \,\Big|\,y_2-y_1\ge 0,\,y_1 \le 0\Big\}\,.
\end{aligned}
\end{equation}
The cones $\Sigma_{1,2,3}$ inherit the obvious (partial) complex structure from $M$.  These cones adjoin along rays ${\Sigma_{12,13,23}\simeq\BH_+\times\R^2}$, which similarly inherit a complex structure associated to the distinguished direction in the ${y_1y_2}$-plane.  Compare Figure \ref{FigExD} to the abstract polyhedral fan in Figure \ref{2dFan}.

To apply the multi-dimensional Jordan lemma directly to $\Xi(\mu_1,\mu_2,\mu_3,\mu_4)$, we must partition the four polar hyperplanes, where the denominator in \eqref{ExXiD} vanishes, among three divisors $D_{1,2,3}$ so that 
\begin{equation}
D_1\cap\Sigma_1 \,=\, D_2\cap\Sigma_2 \,=\, D_3\cap \Sigma_3 \,=\, \varnothing\,.
\end{equation}
This task is impossible.  A compatible set $\{D_1, D_2, D_3\}$ for the polyhedral decomposition in Figure \ref{FigExD} fails to exist because the hyperplane labelled in the figure by `$\ell$', where ${2 z_1 - z_2 + z_3 - i\,\mu_4=0}$, passes through all three cones $\Sigma_{1,2,3}$.

\iffigs
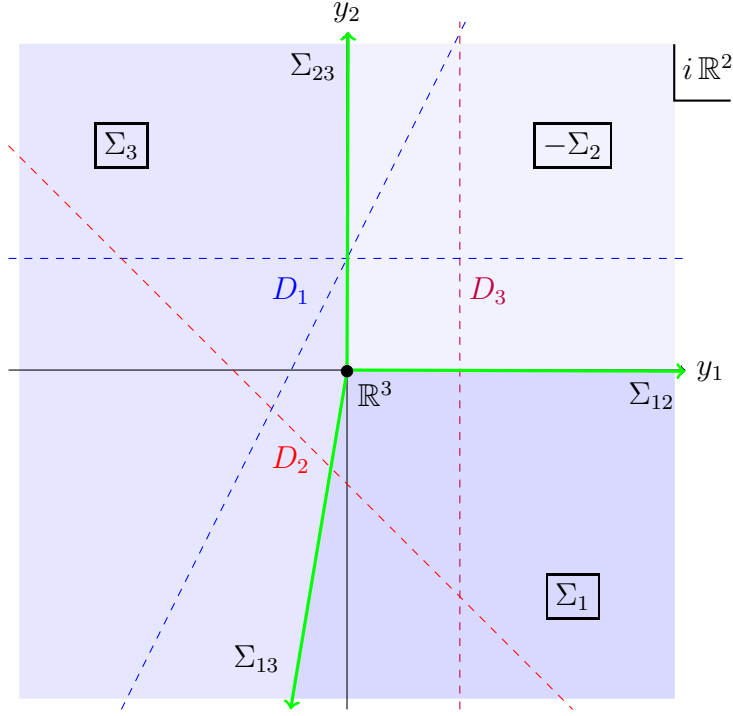
\begin{figure}
	\centering
	\begin{tikzpicture}[x=1.5cm,y=1.5cm] 

		\draw [fill=blue!5,blue!5] (0,2.9)--(0,0)--(2.9,0)--(2.9,2.9)--(0,2.9);
		\draw [fill=blue!10,blue!10] (-2.9,2.9)--(-2.9,-2.9)--(-0.5,-2.9)--(0,0)--(0,2.9)--(-2.9,2.9);
		\draw [fill=blue!15,blue!15] (0,0)--(-0.5,-2.9)--(2.9,-2.9)--(2.9,0)--(0,0);

		\draw[->] (-3,0.01)--(3,0.01);
		\node  [ right ] at (3,0) {$y_1$};
		\draw[->] (0,-3)--(0,3);
		\node  [ above ] at (0,3) {$y_2$};

		\draw[dashed, purple] (1,-3)--(1,3.1);
		\node[ above, purple] at (1.25,0.5) {$D_3$};
		
		\draw[dashed, blue] (-3,1)--(3,1);
		\node [ above, blue ] at (-0.5,0.5) {$D_1$};
		
		\draw[dashed, red] (-3,2)--(2,-3);
		\node[above,red] at (-0.5,-1) {$D_2$};
			
		\draw[dashed, blue] (-2,-3)--(1.05,3.1);
		
		\draw[very thick, ->, green] (0,0.01)--(3,0.0);
		\node[ below ] at (2.7,0.0) {$\Sigma_{12}$};
		
		\draw[very thick,->,green] (0,0.01)--(0.01,3.01);
		\node[ left ] at (0,2.7) {$\Sigma_{23}$};

		\draw[very thick,->,green] (0,0)--(-0.5,-3);
		\node[ left ] at (-0.5,-2.55) {$\Sigma_{13}$};

		\node[ ] at (2,2) {\fbox{$-\Sigma_2$}};

		\node[ ] at (2,-2) {\fbox{$\Sigma_1$}};

		\node[ ] at (-2,2) {\fbox{$\Sigma_3$}};

		\draw[fill] (0,0) circle (.05);
		\node [ right ] at (0,-0.2) {$\R^3$};
		\draw [thick, draw=black, fill=white] (2.9,2.9)--(2.9,2.4)--(3.4,2.4);
		\node at (3.2,2.7) {${i\,\R^2}$};	
		\end{tikzpicture} 
		\caption{Cones and compatible divisors for the Jordan integral in \eqref{ExXiD}.}\label{FigExDII}
	\end{figure}
\fi

At least two solutions are available.  First, we can try to deform the cones $\Sigma_{1,2,3}$ so that each hyperplane passes through at most two cones.  An example of a compatible choice is shown in Figure \ref{FigExDII} for the parameter regime ${\mu_1>0}$, ${\mu_2<0}$, ${\mu_3>0}$, and ${\mu_4<0}$.  Here
\begin{equation}
\begin{aligned}
D_1 \,&=\, \big\{z_2 + z_3 - i\,\mu_3=0\big\} \cup \big\{2\,z_1-z_2+z_3-i\,\mu_4=0\big\}\,,\\
D_2 \,&=\, \big\{z_1+z_2-i\,\mu_2=0\big\}\,,\\
D_3 \,&=\, \big\{z_1-i\,\mu_1=0\big\}\,.
\end{aligned}
\end{equation}
The intersection ${D_1 \cap D_2 \cap D_3}$ (not shown in the figure) contains two points from the two components of $D_1$, but only one of those points has ${\Im(z_3)>0}$ and so lies in $M$.  The relevant point has coordinates
\begin{equation}
p \,=\, i \left(\mu_1,\,\mu_2-\mu_1,\,\mu_3-\mu_2+\mu_1\right),
\end{equation}
so immediately by the Jordan lemma in \eqref{JordanLem},
\begin{equation}\label{SolnXio}
\Xi(\mu_1,\mu_2,\mu_3,\mu_4) \,=\, -i\,\frac{\exp{\!\left[-\mu_3+\mu_2-\mu_1\right]}}{\left(4\mu_1-2\mu_2+\mu_3-\mu_4\right)}\,,\qquad \mu_{1,3}>0\,,\quad \mu_{2,4}<0\,.
\end{equation}
As the signs of $\mu_{1,2,3,4}$ change, the assignments of divisors and cones in Figure \ref{FigExDII} generally change as well, resulting in discontinuous changes in the value of $\Xi$.

For the second solution, instead of deforming the cones $\Sigma_{1,2,3}$ which appear in the original Figure \ref{FigExD}, we modify the Jordan integrand itself.  Since the denominator in \eqref{ExXiD} is a polynomial, the integrand can be decomposed via partial fractions,
\begin{equation}\label{PartFraD}
\Xi \,=\, \Xi_1 \,+\, \Xi_2 \,+\, \Xi_3 \,+\, \Xi_4\,,
\end{equation}
where
\begin{equation}\label{PartFraDa}
\begin{aligned}
\Xi_1 \,&=\, \frac{1}{\left(2\pi i\right)^3}\int_{\R^3}\!\!d^3x\,\frac{\e{i\,x_3}\cdot(-4\lambda)}{\left(x_1+x_2-i\,\mu_2\right) \left(x_2+x_3-i\,\mu_3\right)\left(2\,x_1-x_2+x_3-i\,\mu_4\right)}\,,\\[1 ex]
\Xi_2 \,&=\, \frac{1}{\left(2\pi i\right)^3}\int_{\R^3}\!\!d^3x\,\frac{\e{i\,x_3}\cdot(2\lambda)}{\left(x_1-i\,\mu_1\right)\left(x_2+x_3-i\,\mu_3\right)\left(2\,x_1-x_2+x_3-i\,\mu_4\right)}\,,\\[1 ex]
\Xi_3 \,&=\, \frac{1}{\left(2\pi i\right)^3}\int_{\R^3}\!\!d^3x\,\frac{\e{i\,x_3}\cdot(-\lambda)}{\left(x_1-i\,\mu_1\right)\left(x_1+x_2-i\,\mu_2\right)\left(2\,x_1-x_2+x_3-i\,\mu_4\right)}\,,\\[1 ex]
\Xi_4 \,&=\, \frac{1}{\left(2\pi i\right)^3}\int_{\R^3}\!\!d^3x\,\frac{\e{i\,x_3}\cdot\lambda}{\left(x_1-i\,\mu_1\right)\left(x_1+x_2-i\,\mu_2\right) \left(x_2+x_3-i\,\mu_3\right)}\,,
\end{aligned}
\end{equation}
with 
\begin{equation}\label{PartFraDb}
\lambda \,=\, -\frac{i}{\left(4\mu_1-2\mu_2+\mu_3-\mu_4\right)}\,.
\end{equation}
We spare the reader the algebra leading to \eqref{PartFraDa} and
\eqref{PartFraDb}, as we shall eventually provide  in Section \ref{JordHRGG} a systematic discussion for a wide class of Jordan integrals.  Observe that each denominator in \eqref{PartFraDa} is cubic instead of quartic.  The integrands for $\Xi_{1,2,3,4}$ now have poles along a set of only three hyperplanes, for which a compatible polyhedral decomposition of $\partial M$ trivially exists.  See Figure \ref{PartFraFig} for the compatible polyhedral decomposition in each case.

\iffigs
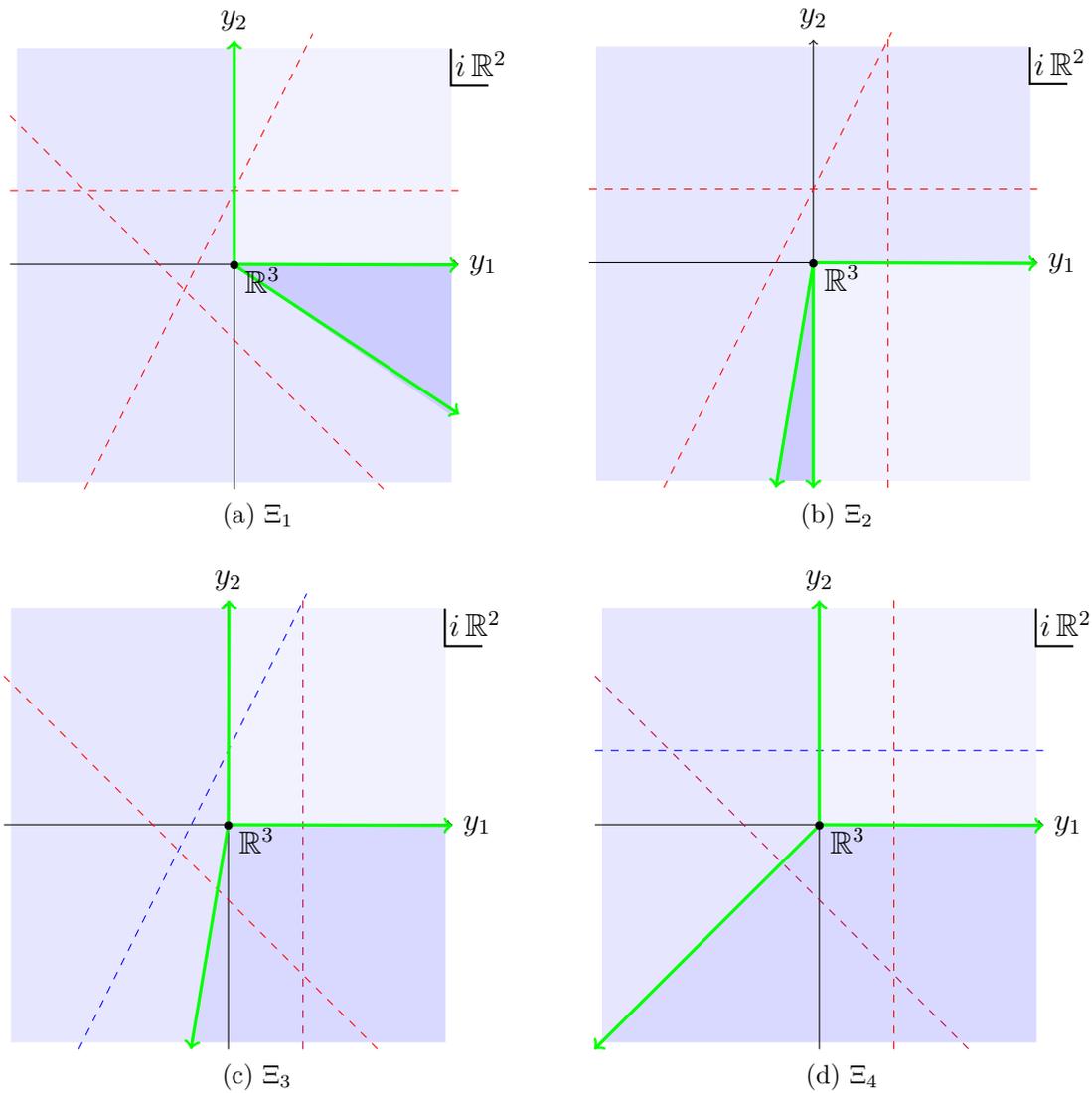
\begin{figure}
\centering
\subfloat[$\Xi_1$]{
	\begin{tikzpicture}[x=1cm,y=1cm] 

		\draw [fill=blue!5,blue!5] (0,2.9)--(0,0)--(2.9,0)--(2.9,2.9)--(0,2.9);
		\draw [fill=blue!10,blue!10] (-2.9,2.9)--(-2.9,-2.9)--(2.9,-2.9)--(2.9,-2.0)--(0,0)--(0,2.9)--(-2.9,2.9);
		\draw [fill=blue!20,blue!20] (0,0)--(2.9,-2.0)--(2.9,0)--(0,0);

		\draw[->] (-3,0.01)--(3,0.01);
		\node  [ right ] at (3,0) {$y_1$};
		\draw[->] (0,-3)--(0,3);
		\node  [ above ] at (0,3) {$y_2$};
		
		\draw[dashed, red] (-3,1)--(3,1);
		
		\draw[dashed, red] (-3,2)--(2,-3);
			
		\draw[dashed, red] (-2,-3)--(1.05,3.1);
		
		\draw[very thick, ->, green] (0,0.01)--(3,0.0);
		
		\draw[very thick,->,green] (0,0.01)--(0,3.01);

		\draw[very thick,->,green] (0,0)--(3.01,-2);

		\draw[fill] (0,0) circle (.05);
		\node [ right ] at (0,-0.2) {$\R^3$};
		\draw [thick, draw=black, fill=white] (2.9,2.9)--(2.9,2.4)--(3.4,2.4);
		\node at (3.3,2.7) {${i\,\R^2}$};	
		\end{tikzpicture}} \qquad
\subfloat[$\Xi_2$]{
\begin{tikzpicture}[x=1cm,y=1cm] 

		\draw [fill=blue!10,blue!10] (-2.9,2.9)--(-2.9,-2.9)--(-.5,-2.9)--(0,0)--(2.9,0)--(2.9,2.9)--(-2.9,2.9);
		\draw [fill=blue!20,blue!20] (0,0)--(-.5,-2.9)--(0,-2.9)--(0,0);
		\draw [fill=blue!5,blue!5] (0,0)--(0,-2.9)--(2.9,-2.9)--(2.9,0)--(0,0);
		\draw[->] (-3,0.01)--(3,0.01);
		\node  [ right ] at (3,0) {$y_1$};
		\draw[->] (0,-3)--(0,3);
		\node  [ above ] at (0,3) {$y_2$};

		\draw[dashed, red] (1,-3)--(1,3.1);
		
		\draw[dashed, red] (-3,1)--(3,1);
			
		\draw[dashed, red] (-2,-3)--(1.05,3.1);
		
		\draw[very thick, ->, green] (0,0.01)--(3,0.0);
		
		\draw[very thick,->,green] (0,0.01)--(0,-3.01);

		\draw[very thick,->,green] (0,0)--(-.5,-3);

		\draw[fill] (0,0) circle (.05);
		\node [ right ] at (0,-0.2) {$\R^3$};
		\draw [thick, draw=black, fill=white] (2.9,2.9)--(2.9,2.4)--(3.4,2.4);
		\node at (3.3,2.7) {${i\,\R^2}$};	
		\end{tikzpicture}}\\
\subfloat[$\Xi_3$]{
	\begin{tikzpicture}[x=1cm,y=1cm] 
\draw [fill=blue!5,blue!5] (0,2.9)--(0,0)--(2.9,0)--(2.9,2.9)--(0,2.9);
		\draw [fill=blue!10,blue!10] (-2.9,2.9)--(-2.9,-2.9)--(-0.5,-2.9)--(0,0)--(0,2.9)--(-2.9,2.9);
		\draw [fill=blue!15,blue!15] (0,0)--(-0.5,-2.9)--(2.9,-2.9)--(2.9,0)--(0,0);

		\draw[->] (-3,0.01)--(3,0.01);
		\node  [ right ] at (3,0) {$y_1$};
		\draw[->] (0,-3)--(0,3);
		\node  [ above ] at (0,3) {$y_2$};

		\draw[dashed, purple] (1,-3)--(1,3.1);

		\draw[dashed, red] (-3,2)--(2,-3);
		
		\draw[dashed, blue] (-2,-3)--(1.05,3.1);
		
		\draw[very thick, ->, green] (0,0.01)--(3,0.0);
		
		\draw[very thick,->,green] (0,0)--(0.01,3.01);
	
		\draw[very thick,->,green] (0,0)--(-0.5,-3);

		\draw[fill] (0,0) circle (.05);
		\node [ right ] at (0,-0.2) {$\R^3$};
		\draw [thick, draw=black, fill=white] (2.9,2.9)--(2.9,2.4)--(3.4,2.4);
		\node at (3.3,2.7) {${i\,\R^2}$};	
		\end{tikzpicture} } \qquad
\subfloat[$\Xi_4$]{
			\begin{tikzpicture}[x=1cm,y=1cm] 

		\draw [fill=blue!5,blue!5] (0,2.9)--(0,0)--(2.9,0)--(2.9,2.9)--(0,2.9);
		\draw [fill=blue!10,blue!10] (-2.9,2.9)--(-2.9,-2.9)--(0,0)--(0,2.9)--(-2.9,2.9);
		\draw [fill=blue!15,blue!15] (0,0)--(-2.9,-2.9)--(2.9,-2.9)--(2.9,0)--(0,0);

		\draw[->] (-3,0.01)--(3,0.01);
		\node  [ right ] at (3,0) {$y_1$};
		\draw[->] (0,-3)--(0,3);
		\node  [ above ] at (0,3) {$y_2$};

		\draw[dashed, red] (1,-3)--(1,3.1);
		
		\draw[dashed, blue] (-3,1)--(3,1);
		
		\draw[dashed, purple] (-3,2)--(2,-3);
			
		\draw[very thick, ->, green] (0,0.01)--(3,0.0);
		
		\draw[very thick,->,green] (0,0.01)--(0,3.01);
	
		\draw[very thick,->,green] (0,0)--(-3.01,-3);

		\draw[fill] (0,0) circle (.05);
		\node [ right ] at (0,-0.2) {$\R^3$};
		\draw [thick, draw=black, fill=white] (2.9,2.9)--(2.9,2.4)--(3.4,2.4);
		\node at (3.3,2.7) {${i\,\R^2}$};	
		\end{tikzpicture}}
\caption{Polyhedral decompositions compatible with $\Xi_{1,2,3,4}$ in \eqref{PartFraDa}.  In the top row, both ${\Xi_1=\Xi_2=0}$, since all three polar hyperplanes avoid one cone (darkest shading) in the decomposition.}\label{PartFraFig}
	\end{figure}
\fi

With no computation, we conclude from Figure \ref{PartFraFig} that ${\Xi_1 = \Xi_2 = 0}$, since in those cases there is a cone (eg.~$\Sigma_1$) which does not intersect any of the three polar hyperplanes.  Hence we can associate all three polar hyperplanes to a single divisor $D_1$, with ${D_2=D_3=\varnothing}$, so ${D_1 \cap D_2 \cap D_3 = \varnothing}$ as well.  Equivalently, the holomorphic section ${s\neq 0}$ associated to $\{D_1,D_2,D_3\}$  is everywhere non-vanishing.  Hence the sum over residues is empty.

For $\Xi_3$ and $\Xi_4$, we can assign the three polar hyperplanes compatibly to $D_1$, $D_2$, and $D_3$, which intersect in a single point.  For the parameter regime ${\mu_{1,3}>0}$ and  ${\mu_{2,4}<0}$, the point of intersection lies in $M$ for $\Xi_4$ but not for $\Xi_3$.  Thus ${\Xi_3=0}$, and by a trivial application of the residue theorem,
\begin{equation}
\Xi_4 \,=\, \lambda \cdot \exp{\!\left[-\mu_3+\mu_2-\mu_1\right]}\,,\qquad  \mu_{1,3}>0\,,\quad \mu_{2,4}<0\,.
\end{equation}
With the value for $\lambda$ in \eqref{PartFraDb}, this result agrees precisely with \eqref{SolnXio}.

Of these two strategies, either deforming the polyhedral decomposition
of ${\partial M}$ or decomposing the Jordan integrand via partial
fractions, only the latter is robust.  As both the number of polar
hyperplanes and the number of integration variables grow, finding a
``good'' polyhedral decomposition which admits a compatible set of
divisors $\{D_1,\ldots,D_n\}$ may be difficult, if not impossible, to
carry-out by hand.  By contrast, the partial-fractions decomposition
can be readily extended to a wide variety of meromorphic integrands in
arbitrary dimension, after which the multi-dimensional Jordan lemma
directly applies.   Among our goals in the next section will be to
establish a systematic procedure for the partial-fractions aka
Mittag-Leffler reduction of ${Z^{\rm uv}_{S^3}}$.

\section{Holomorphic Factorization at Higher-Rank}\label{HighRank}

We now move beyond toy models to reconsider the ultraviolet partition function
\begin{equation}\label{ZuvCmbV}
\begin{aligned}
Z_{S^3}^{\rm uv} \,=\, \frac{\e{i
    \eta_0}}{|\fW|\cdot\Vol(T)}\int_{\mathfrak{h}}\!d^r\!\sigma&\prod_{\alpha\in\Delta_+}\left[4\,\sinh\!\left(\frac{b
      \langle\alpha,\sigma\rangle}{2}\right)\sinh\!\left(\frac{
      \langle\alpha,\sigma\rangle}{2
    b}\right)\right]\times\\
&\times\prod_{j=1}^n\left[\prod_{\beta \in
\Delta_j}
s_b\!\left(\frac{\langle\beta,\sigma\rangle}{2\pi}\,+\,\mu_j\right)\right].
\end{aligned}
\end{equation}
As a reminder, we have reduced in Section \ref{LevelZero} to the case
that ${k_{\rm uv}=0}$ and $\Lambda_{\rm
  uv}\simeq\bigoplus_{j=1}^n\,[\lambda_j]$ is a real representation of
the gauge group.  This reduction is always possible for gauge groups of classical type SU-SO-Sp and for the exceptional group $G_2$.

To apply the abstract Jordan lemma in its most elementary version \eqref{JordanLem} to $Z_{S^3}^{\rm uv}$, we require the product of double-sine functions in \eqref{ZuvCmbV} to have only simple, first-order poles when $\sigma$ is continued to the complex domain ${\h_\C \equiv \h\otimes\C}$.  For this reason, we assume throughout that the continuous parameters ${b^2\notin\Q}$ and $\mu_j$ are generic.  

We must also make a stronger assumption about the matter
representation $\Lambda_{\rm uv}$.  Clearly, if any summand
$[\lambda_j]$ of $\Lambda_{\rm uv}$ has a repeated, non-zero weight
${\beta\neq 0}$, then even for generic values of $b^2$ and $\mu_j$,
the product over weights in $\Delta_j$ in the second line of
\eqref{ZuvCmbV} produces a multiple pole.  Thus, if we wish to apply
the multidimensional residue calculus in its most basic and convenient
form, we must assume that each irreducible representation
$[\lambda_j]$ is weight-multiplicity-free, ie.~all non-zero weights
occur with multiplicity-one. We do {\em not} require $\Lambda_{\rm
  uv}$ itself to be weight-multiplicity-free, since any degeneracies between irreducible summands are split by distinct mass parameters $\mu_j$.

By any measure, most irreducible representations of the general compact, simple Lie group have non-trivial weight multiplicities, so our assumption on $[\lambda_j]$ is very restrictive.  On the other hand, the low-dimensional representations which turn up in familiar examples of ${\CN=2}$ supersymmetric gauge theories tend to satisfy this condition.  Eg.~for gauge group SU, any anti-symmetric tensor power of the fundamental (or anti-fundamental) representation is weight-multiplicity-free.  For gauge groups SO and Sp, the fundamental, adjoint, and spin representations are all weight-multiplicity-free.  For gauge group $G_2$, the fundamental seven-dimensional representation and the adjoint representation are weight-multiplicity-free.  These examples are far from exhaustive.\footnote{Our examples are culled from the quasi-miniscule representations, for which the weight-multiplicity-free condition holds automatically.}  For instance, for $SU(2)$ all representations are weight-multiplicity-free.  Therefore in practice, the assumption that $[\lambda_j]$ is weight-multiplicity-free still accommodates a wide class of supersymmetric gauge theories.

Under these assumptions, we now need three pieces of geometric data to apply the elementary Jordan lemma \eqref{JordanLem} to $Z^{\rm uv}_{S^3}$:
\begin{enumerate}
\item A complex manifold $M$, with ${\dim_\C M = r}$, whose boundary contains the Cartan subalgebra ${\h \simeq \R^r}$ as a totally-real cycle.
\item A polyhedral decomposition of the boundary $\partial M = \Sigma_1 - \Sigma_2 \,+\, \cdots \,+\, (-1)^{r+1}\, \Sigma_r$ into cones, with ${\h = \pm\Sigma_{1 2 \cdots r}}$.
\item A set of Jordan divisors $\{D_1,\cdots,D_r\}$ to describe the locus in $M$ where the integrand of $Z^{\rm uv}_{S^3}$ has a simple pole.  In the construction from Section \ref{JordanLemma}, this set is equivalent to the specification of a holomorphic section $s$ of a rank-$r$ complex vector bundle over $M$.  Crucially, the Jordan divisors must be compatible with the polyhedral decomposition of the boundary in the sense of \eqref{CompatC}.
\end{enumerate}
In the present section, we explain how to obtain these data for
$Z^{\rm uv}_{S^3}$.  The most important and most subtle problem is the
determination of compatible Jordan divisors $\{D_1,\cdots,D_r\}$ for
the integrand in \eqref{ZuvCmbV}.  As Example D in Section \ref{ToyEx}
illustrates, the difficulty in finding compatible Jordan divisors
depends very much on whether the gauge group has rank ${r=2}$ or ${r >
  2}$.  We therefore analyze these cases separately in what follows.

Let us first specify the datum of the underlying complex manifold $M$.  We must select ${M \subset \h_\C \equiv \h\otimes \C}$ so that the integrand of $Z^{\rm uv}_{S^3}$, when analytically continued, decays rapidly at infinity on $M$.  In Section \ref{SUSYBreak}, we have already examined the requirement that the integrand decay exponentially at infinity along the real directions in $\h$, for which we obtain the convergence criterion \eqref{ConvCritG}.  By a similar calculation, when $\sigma$ is continued into the complex domain $\h_\C$, the norm of the integrand behaves asymptotically as 
\begin{equation}\label{AsympNorm4}
\begin{aligned}
&\left|\prod_{\alpha\in\Delta_+}\left[4\,\sinh\!\left(\frac{b
      \langle\alpha,\sigma\rangle}{2}\right)\sinh\!\left(\frac{
      \langle\alpha,\sigma\rangle}{2
    b}\right)\right]\cdot\prod_{j=1}^n\left[\prod_{\beta \in
\Delta_j}
s_b\!\left(\frac{\langle\beta,\sigma\rangle}{2\pi}\,+\,\mu_j\right)\right]\right|\\
&\,\underset{|\sigma|\to\infty}{=}\exp{\!\left[-\frac{Q}{2}\left(\sum_{j=1}^n \left(1-\RR_j\right) ||\Re(\sigma)||_{\lambda_j}-||\Re(\sigma)||_\g\right)+\sum_{j=1}^n \mu_{j,\R} \Psi_{\lambda_j}(\sigma)+O(1)\right]},
\end{aligned}
\end{equation}
where we introduce the Weyl-invariant, real-valued function
\begin{equation}\label{PsiOneG}
\Psi_V(\sigma) \,=\, \ha\sum_{\beta \in \Delta_V} \langle\beta,\Im(\sigma)\rangle \cdot \sgn\langle\beta,\Re(\sigma)\rangle\,,\qquad \Re(\sigma)\neq 0\,.
\end{equation}
We assume for convenience that ${Q\in\R_+}$ is real, and we recall our convention for the complexified `real' mass of each chiral multiplet,
\begin{equation}\label{CplxMu}
\mu \equiv \mu_\R + \frac{i}{2}\,Q\,\RR\,,
\end{equation}
where $\RR$ is the R-charge.

All asymptotic dependence on the imaginary part of ${\sigma\in\h_\C}$ occurs through $\Psi_V(\sigma)$, labelled by an irreducible representation $V$ of the gauge group $G$.  Clearly
\begin{equation}\label{PsiV1}
\Psi_V(\sigma) \,=\, \Psi_V(-\sigma) \,=\, \Psi_{V^*}(\sigma)\,,
\end{equation}
where $V^*$ is the dual of $V$.  Also, under complex-conjugation of ${\sigma\in\h_\C}$,
\begin{equation}\label{PsiV2}
 \Psi_V(\bar\sigma)\,=\,-\Psi_V(\sigma) \quad\Longrightarrow\quad \Psi_V(\sigma)=-\Psi_V(-\bar\sigma)\,.
\end{equation}
The latter relation in \eqref{PsiV2} precludes the existence of any half-space containing ${\h \subset \h_\C}$ upon which ${\Psi_V\ge 0}$ takes a definite sign, as would be necessary for the decay of the right side of \eqref{AsympNorm4} for appropriate choices of real masses $\mu_{j,\R}$.

In the special case of gauge group $SU(2)$, we have already observed this asymptotic feature of the integrand in \eqref{NormSU2}.  Similarly to the rank-one case, to ensure that the integrand of $Z^{\rm uv}_{S^3}$ is bounded in the complex domain $\h_\C$, we require the real (parts of the `real') mass parameters to satisfy
\begin{equation}\label{VanishRM}
\sum_{j=1}^n \mu_{j,\R}\,\Psi_{\lambda_j}(\sigma) \,=\, 0\,,
\end{equation}
for all values of ${\sigma\in\h_\C}$.  The $SU(2)$ version of this condition appears in \eqref{CPMasses}.  A trivial way to satisfy the vanishing condition in \eqref{VanishRM} is to set ${\mu_{j,\R}=0}$ for each ${j=1,\ldots,n}$, after which the dependence on the complex parameter $\mu_j$ in \eqref{CplxMu} is determined by analytic continuation.  For special choices of $\Lambda_{\rm uv}$, less stringent conditions on the real masses are implied by \eqref{VanishRM}.  Eg.~if ${\Lambda_{\rm uv} = \oplus_{j=1}^n \left(V_j^{} \oplus V^*_j\right)}$ with masses $(\mu_j,\wt\mu_j)$, then a relation ${\mu_{j,\R} + \wt\mu_{j,\R}=0}$ for each $j$ suffices.

If the vanishing condition in \eqref{VanishRM} as well as the earlier
convergence condition in \eqref{ConvCritG} are both satisfied, the
integrand in \eqref{ZuvCmbV} decays exponentially as
${|\sigma|\to\infty}$ along real directions and is asymptotically
constant along imaginary directions in $\h_\C$.  A toy model for this
behavior is provided by the Jordan integral \eqref{ExampleIthree}
appearing as Example C in Section \ref{ToyEx}.  In this happy situation, the complex manifold $M$ can be any complex half-space in $\h_\C$.  The orientation of ${\sC=\pm\h}$ induced from the boundary ${\partial M}$ depends upon the choice of half-space, equipped with the canonical orientation from $\h_\C$, but only the overall sign of $Z^{\rm uv}_{S^3}$ depends upon this choice.  For concreteness, we take $M$ to be the Weyl half-space
\begin{flalign}\label{WeylHalfSp}
\fbox{Weyl half-space}\qquad\qquad&
M\,=\,\BH_\rho \,:=\, \Big\{\sigma\in\h_\C \,\big|\, \big\langle\rho,\Im(\sigma)\big\rangle\ge 0\Big\}\,,&
\end{flalign}
associated to the distinguished Weyl vector ${\rho\in\h^*}$ in \eqref{Weylvec}.  

Following the analysis of Example C in Section \ref{ToyEx}, we regulate the asymptotic behavior of the integrand along complex directions in $M$ by including an $i\epsilon$-factor ${g(\sigma) = \exp{\!\left(i \epsilon \langle\rho,\sigma\rangle\right)}}$,
\begin{equation}\label{ZuvCmbVeps}
\begin{aligned}
Z_{S^3,\epsilon}^{\rm uv} \,=\, \frac{\e{i
    \eta_0}}{|\fW|\cdot\Vol(T)}&\int_{\mathfrak{h}}\!d^r\!\sigma\,\,\e{\!i\epsilon \langle\rho,\sigma\rangle}\prod_{\alpha\in\Delta_+}\left[4\,\sinh\!\left(\frac{b
      \langle\alpha,\sigma\rangle}{2}\right)\sinh\!\left(\frac{
      \langle\alpha,\sigma\rangle}{2
    b}\right)\right]\times\\
&\times\prod_{j=1}^n\left[\prod_{\beta \in
\Delta_j}
s_b\!\left(\frac{\langle\beta,\sigma\rangle}{2\pi}\,+\,\mu_j\right)\right],\qquad\qquad 0 < \epsilon \ll 1\,.
\end{aligned}
\end{equation}
The $i\epsilon$-factor breaks the underlying Weyl-invariance of the integrand, since $\rho$ itself is not invariant.  However, because the integral over $\h$ converges absolutely, the original partition function $Z^{\rm uv}_{S^3}$ can be recovered from \eqref{ZuvCmbVeps} in the limit ${\epsilon\to 0^+}$.

Throughout the following, we suppress the dependence on the $i\epsilon$-regulator.

The holomorphic vector bundle $V$ over ${M=\BH_\rho}$ is necessarily trivial, so the section ${s=(s^1,\cdots,s^r)}$ is determined by $r$ holomorphic functions on $M$.  For the remainder, we explain how to choose $s$ compatibly with the integrand of $Z^{\rm uv}_{S^3}$.

\subsection{Jordan Divisors for Rank-Two Gauge Groups}\label{JordDRk2}

We next choose a polyhedral decomposition on the boundary of $\BH_\rho$, where
\begin{equation}
\partial\BH_\rho \,=\, \big\{\sigma\in\h_\C \,\big|\, \big\langle\rho,\Im(\sigma)\big\rangle\,=\, 0\big\}\,.
\end{equation} 
When $G$ has rank-two, so that ${\dim_\C H_\rho = 2}$, the cones
${\Sigma_{1,2}\subset \partial\BH_\rho}$ can be straightforwardly
modelled on the half-spaces isomorphic to
${\R^3_+\simeq\R\times\BH_+}$ in \eqref{ExCones}.  Recall the
description of $\rho$ as a sum of fundamental weights,
\begin{equation}
\rho \,=\, \hat\omega^1\,+\,\cdots\,+\,\hat\omega^r\,,
\end{equation}
each canonically dual to the simple coroots $\{\hat h_1,\ldots,\hat h_r\}$.  Thus in rank-two, $\partial\BH_\rho$ can be decomposed into half-spaces
\begin{equation}\label{PolyRk2}
\Sigma_1 \,=\, \h \times i\,\R_+ v_+\,,\qquad\qquad \Sigma_2 \,=\, -\h \times i\,\R_+ v_-\,,
\end{equation}
associated to boundary rays
\begin{equation}
v_\pm \,=\, \pm\left(\hat h_1-\hat h_2\right).
\end{equation}
The overall minus sign in the description of $\Sigma_2$ accounts for the orientation in the decomposition ${\partial\BH_\rho = \Sigma_1 - \Sigma_2}$.  

See Figure \ref{BHrhoFig} for a sketch of $\BH_\rho$ and the cones $\Sigma_{1,2}$ when the gauge group is $SU(3)$.  This figure should be compared to Figure \ref{PolarDivsExone} in Section \ref{JordanLemma}.
\iffigs
\begin{figure}
\centering
	\begin{tikzpicture}[x=1.7cm,y=1.7cm] 
	\coordinate (0;0) at (0,0); 
	\foreach \c in {1,...,4}{%  
		\foreach \i in {0,...,5}{% 
			\pgfmathtruncatemacro\j{\c*\i}
			\coordinate (\c;\j) at (60*\i:\c);  
		} }
		\foreach \i in {0,2,...,10}{% 
			\pgfmathtruncatemacro\j{mod(\i+2,12)} 
			\pgfmathtruncatemacro\k{\i+1}
			\coordinate (2;\k) at ($(2;\i)!.5!(2;\j)$) ;}
		
		\foreach \i in {0,3,...,15}{% 
			\pgfmathtruncatemacro\j{mod(\i+3,18)} 
			\pgfmathtruncatemacro\k{\i+1} 
			\pgfmathtruncatemacro\l{\i+2}
			\coordinate (3;\k) at ($(3;\i)!1/3!(3;\j)$)  ;
			\coordinate (3;\l) at ($(3;\i)!2/3!(3;\j)$)  ;
		}
		
		\foreach \i in {0,4,...,20}{% 
			\pgfmathtruncatemacro\j{mod(\i+4,24)} 
			\pgfmathtruncatemacro\k{\i+1} 
			\pgfmathtruncatemacro\l{\i+2}
			\pgfmathtruncatemacro\m{\i+3} 
			\coordinate (4;\k) at ($(4;\i)!1/4!(4;\j)$)  ;
			\coordinate (4;\l) at ($(4;\i)!2/4!(4;\j)$) ;
			\coordinate (4;\m) at ($(4;\i)!3/4!(4;\j)$) ;
		}  
		
		\begin{scope}
		\clip (0,0) circle (2.1);

		\draw [thick, draw=black, fill=blue!10, blue!10] (4;0)--(4;4)--(4;8)--(4;12)--(4;0);;
		
		\foreach \i in {0,...,6}{% 
				\pgfmathtruncatemacro\k{\i}
				\pgfmathtruncatemacro\l{15-\i}
				\draw[thin,gray] (3;\k)--(3;\l);
				\pgfmathtruncatemacro\k{9-\i} 
				\pgfmathtruncatemacro\l{mod(12+\i,18)}   
				\draw[thin,gray] (3;\k)--(3;\l); 
				\pgfmathtruncatemacro\k{12-\i} 
				\pgfmathtruncatemacro\l{mod(15+\i,18)}   
				\draw[thin,gray] (3;\k)--(3;\l);} 
		
		\end{scope}		
		
		\draw[->,thick,shorten >=4pt,shorten <=2pt] (0;0)--(2;1);	
		
		\draw[->,thick,shorten >=4pt,shorten <=2pt] (0;0)--(2;5);
		
		\node [ above ] at ($(2;1)$) {${i\,\hat h_1}$};
		\node [ above ] at ($(2;5)$) {${i\,\hat h_2}$};
				
		\begin{scope}
		\clip (0,0) circle (2.1);

		\draw[very thick,->,green] (0,0)--(-2.1,0);
		\draw[very thick,->,green] (0,0)--(2.1,0);

		\draw[fill] (0,0) circle (.05);
		\node [ right ] at (0,-0.2) {${\h\simeq\R^2}$};
		
		\end{scope}
				
		\node[ above ] at (0,1.0) {$\BH_\rho$};

		\node [ right ] at ($1.1*(2;0)+(0,0)$) {${\big\langle\rho,\Im(\sigma)\big\rangle = 0}$};
		
		\node [ left ] at (-1.5,-0.2) {$\Sigma_2$};
		\node [ right ] at (1.5,-0.2) {$\Sigma_1$};
		
		\draw [thick, draw=black, fill=white] (1.7,2.1)--(1.7,1.7)--(2.2,1.7);
		\node at (1.95,1.9) {$i\,\mathfrak{h}$};
		
		\end{tikzpicture}  	
\caption{The complex manifold ${\BH_\rho}$ for a rank-two gauge group.  The shaded region corresponds to the interior of $\BH_\rho$, with boundary ${\langle\rho,\Im(\sigma)\rangle=0}$.  For concreteness, the coroot geometry is shown for gauge group $SU(3)$.  Each cone $\Sigma_{1,2}$ in the polyhedral decomposition of ${\partial\BH_\rho=\Sigma_1-\Sigma_2}$ is a half-space ${\R^3_+\simeq\R\times\BH_+}$, with boundary normal given by rays ${\pm i\,(\hat h_1 - \hat h_2)}$.  These rays are displayed in green.}\label{BHrhoFig}
\end{figure}
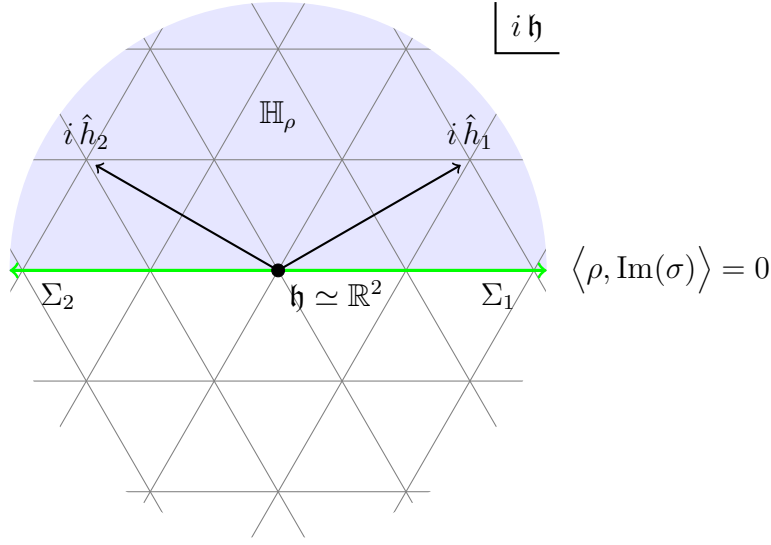\fi

The product of double-sine functions in the integrand \eqref{ZuvCmbV} of $Z^{\rm uv}_{S^3}$ has poles along the hyperplanes in $\h_\C$ where 
\begin{equation}\label{PolDivSU3}
\frac{\langle\beta,\sigma\rangle}{2\pi} \,=\, -i\,m\,b \,-\, i\,n\,b^{-1}\,-\,\mu_j\,,\qquad m,n\ge 0\,,\qquad \beta\in\Delta_j\,.
\end{equation}
Here $m$ and $n$ run over all positive integers, and $\beta$ runs over the set of weights which appear in the matter representation ${\Lambda_{\rm uv}\simeq\bigoplus_{j=1}^n [\lambda_j]}$.  In rank-two, these hyperplanes can generically be partitioned into Jordan divisors $D_{1,2}$ which are compatible with the cones $\Sigma_1$ and $\Sigma_2$ in \eqref{PolyRk2}.  But for practical computations, we must still ask how the decomposition works for a given irreducible representation of the gauge group.

As an example, in Figure \ref{FundJDFig} we indicate how the polar hyperplanes \eqref{PolDivSU3} for the fundamental and anti-fundamental representations of $SU(3)$ can be partitioned into compatible Jordan divisors $D_{1,2}$.  Recall that the weights for the fundamental representation of $SU(3)$ take values in the set $\Delta_{\bf 3}=\{\hat\omega^1,-\hat\omega^2,-\hat\omega^1+\hat\omega^2\}$.  The associated polar hyperplanes are then given by three infinite families of parallel planes, sketched as dashed lines in Figure \ref{FundJDFig}.  Evidently, after Figure \ref{FundJDFig} is overlaid on Figure \ref{BHrhoFig}, we see that the families determined by the weights $\hat\omega^1$ and $-\hat\omega^2$ in $\Delta_{\bf 3}$ comprise the divisor 
\begin{equation}\label{JDiv1}
\begin{aligned}
D_1^{(j)} \,&=\, \left\{\frac{\langle\hat\omega^1,\sigma\rangle}{2\pi}=-i\,m_1\,b\,-\,i\,n_1\,b^{-1}\,-\,\mu_j\right\}\\ &\cup \left\{\frac{\langle-\hat\omega^2,\sigma\rangle}{2\pi}=-i\,m_2\,b\,-\,i\,n_2\,b^{-1}\,-\,\mu_j\right\},
\end{aligned}
\end{equation}
and the family determined by the remaining weight ${-\hat\omega^1+\hat\omega^2}$ makes up the divisor
\begin{equation}\label{JDiv2}
D_2^{(j)} \,=\, \left\{\frac{\langle-\hat\omega^1+\hat\omega^2,\sigma\rangle}{2\pi}=-i\,m_3\,b\,-\,i\,n_3\,b^{-1}\,-\,\mu_j\right\}.
\end{equation}
Here $(m_1,n_1)$, $(m_2,n_2)$, and $(m_3,n_3)$ run over all pairs of positive integers.  For the anti-fundamental representation $\overline{\bf 3}$, the signs on all weights in $\Delta_{\bf 3}$ are reversed, leading to (b) in Figure \ref{FundJDFig}.  In general, when ${\Lambda_{\rm uv}}$ is the direct sum of multiple copies of the fundamental and anti-fundamental representations, the diagrams in Figure \ref{FundJDFig} are superimposed, with ${D_{1,2} = \cup_{j=1}^n D_{1,2}^{(j)}}$.  Compare also to Figure \ref{CschEx} in Section \ref{ToyEx}.
\iffigs
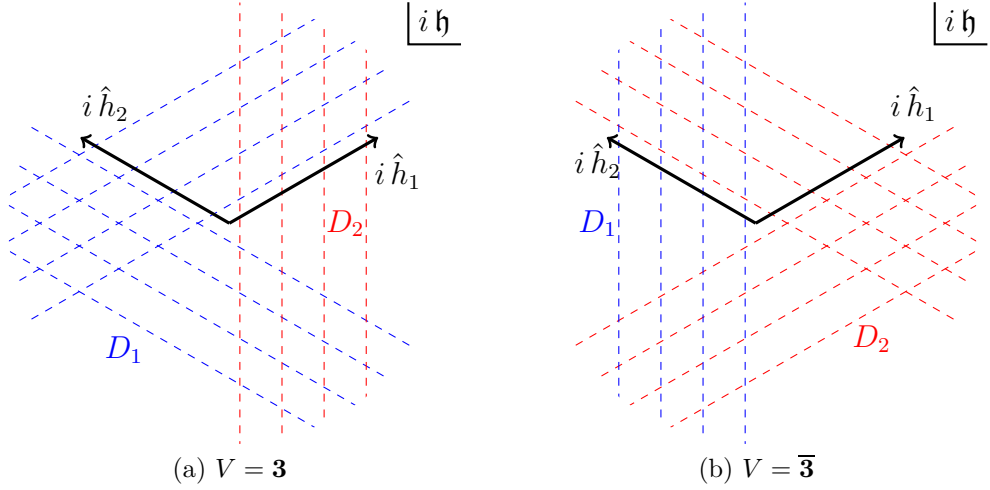
\begin{figure}
\centering
\subfloat[${V={\bf 3}}$]{
	\begin{tikzpicture}[x=1.4cm,y=1.4cm] 
	\coordinate (0;0) at (0,0); 
	\foreach \c in {1,...,4}{%  
		\foreach \i in {0,...,5}{% 
			\pgfmathtruncatemacro\j{\c*\i}
			\coordinate (\c;\j) at (60*\i:\c);  
		} }
		\foreach \i in {0,2,...,10}{% 
			\pgfmathtruncatemacro\j{mod(\i+2,12)} 
			\pgfmathtruncatemacro\k{\i+1}
			\coordinate (2;\k) at ($(2;\i)!.5!(2;\j)$) ;}
		
		\foreach \i in {0,3,...,15}{% 
			\pgfmathtruncatemacro\j{mod(\i+3,18)} 
			\pgfmathtruncatemacro\k{\i+1} 
			\pgfmathtruncatemacro\l{\i+2}
			\coordinate (3;\k) at ($(3;\i)!1/3!(3;\j)$)  ;
			\coordinate (3;\l) at ($(3;\i)!2/3!(3;\j)$)  ;
		}
		
		\foreach \i in {0,4,...,20}{% 
			\pgfmathtruncatemacro\j{mod(\i+4,24)} 
			\pgfmathtruncatemacro\k{\i+1} 
			\pgfmathtruncatemacro\l{\i+2}
			\pgfmathtruncatemacro\m{\i+3} 
			\coordinate (4;\k) at ($(4;\i)!1/4!(4;\j)$)  ;
			\coordinate (4;\l) at ($(4;\i)!2/4!(4;\j)$) ;
			\coordinate (4;\m) at ($(4;\i)!3/4!(4;\j)$) ;
		}  
		
		\begin{scope}
		
		\clip (0;0)--(2;2)--(2;4)--(0;0);
	
		\end{scope}
		
		\begin{scope}
		\clip (0,0) circle (2.1);
		
		\foreach \n in {0,1,...,3}{% 	
			\draw[dashed,blue] ($0.75*(2;1)+0.75*(2;1)+0.25*\n*(2;5)+0.1*(2;5)$) -- ($0.75*(2;7)+0.75*(2;7)+0.25*\n*(2;5)+0.1*(2;5)$);}
		
		\foreach \n in {0,1,...,3}{% 	
			\draw[dashed,blue] ($0.75*(2;5)+0.75*(2;5)-0.25*\n*(2;1)-0.1*(2;1)$) -- ($0.75*(2;11)+0.75*(2;11)-0.25*\n*(2;1)-0.1*(2;1)$);}
		
		\foreach \n in {0,1,...,3}{% 	
			\draw[dashed,red] ($0.75*(2;3)+0.75*(2;3)+0.4*(\n,0)+0.1*(1;0)$) -- ($0.75*(2;9)+0.75*(2;9)+0.4*(\n,0)+0.1*(1;0)$);}
			
		\node [blue] at (-1,-1.2) {$D_1$};
		\node [red] at (1.1,0) {$D_2$};
		\end{scope}

		\draw[->,very thick,shorten >=4pt] (0;0)--(2;1);	
		
		\draw[->,very thick,shorten >=4pt] (0;0)--(2;5);
		
		\node [ below ] at ($(2;1)+(.1,-.1)$) {$i\,\hat h_1$};
		\node [ above right ] at ($(2;5)$) {$i\,\hat h_2$};
		
		\draw [thick, draw=black, fill=white] (1.7,2.1)--(1.7,1.7)--(2.2,1.7);
		\node at (1.95,1.9) {$i\,\mathfrak{h}$};
		
		\end{tikzpicture}}\qquad
\subfloat[${V=\overline{\bf 3}}$]{  
\begin{tikzpicture}[x=1.4cm,y=1.4cm] 
\coordinate (0;0) at (0,0); 
\foreach \c in {1,...,4}{%  
	\foreach \i in {0,...,5}{% 
		\pgfmathtruncatemacro\j{\c*\i}
		\coordinate (\c;\j) at (60*\i:\c);  
	} }
	\foreach \i in {0,2,...,10}{% 
		\pgfmathtruncatemacro\j{mod(\i+2,12)} 
		\pgfmathtruncatemacro\k{\i+1}
		\coordinate (2;\k) at ($(2;\i)!.5!(2;\j)$) ;}
	
	\foreach \i in {0,3,...,15}{% 
		\pgfmathtruncatemacro\j{mod(\i+3,18)} 
		\pgfmathtruncatemacro\k{\i+1} 
		\pgfmathtruncatemacro\l{\i+2}
		\coordinate (3;\k) at ($(3;\i)!1/3!(3;\j)$)  ;
		\coordinate (3;\l) at ($(3;\i)!2/3!(3;\j)$)  ;
	}
	
	\foreach \i in {0,4,...,20}{% 
		\pgfmathtruncatemacro\j{mod(\i+4,24)} 
		\pgfmathtruncatemacro\k{\i+1} 
		\pgfmathtruncatemacro\l{\i+2}
		\pgfmathtruncatemacro\m{\i+3} 
		\coordinate (4;\k) at ($(4;\i)!1/4!(4;\j)$)  ;
		\coordinate (4;\l) at ($(4;\i)!2/4!(4;\j)$) ;
		\coordinate (4;\m) at ($(4;\i)!3/4!(4;\j)$) ;
	}  
	
	\begin{scope}
	
	\clip (0;0)--(2;2)--(2;4)--(0;0);
	
	\end{scope}
	
	\begin{scope}
	\clip (0,0) circle (2.1);
	
	\foreach \n in {0,1,...,3}{% 	
		\draw[red,dashed] ($0.75*(2;1)+0.75*(2;1)-0.25*\n*(2;5)-0.1*(2;5)$) -- ($0.75*(2;7)+0.75*(2;7)-0.25*\n*(2;5)-0.1*(2;5)$);}
	
	\foreach \n in {0,1,...,3}{% 	
		\draw[red,dashed] ($0.75*(2;5)+0.75*(2;5)+0.25*\n*(2;1)+0.1*(2;1)$) -- ($0.75*(2;11)+0.75*(2;11)+0.25*\n*(2;1)+0.1*(2;1)$);}
	
	\foreach \n in {0,1,...,3}{% 	
		\draw[blue,dashed] ($0.75*(2;3)+0.75*(2;3)-0.4*(\n,0)-0.1*(1;0)$) -- ($0.75*(2;9)+0.75*(2;9)-0.4*(\n,0)-0.1*(1;0)$);}
	
	\node [blue] at (-1.5,0) {$D_1$};
	\node [red] at (1.1,-1.1) {$D_2$};

	\end{scope}
	
	\draw[->,very thick,shorten >=4pt] (0;0)--(2;1);	
	
	\draw[->,very thick,shorten >=4pt] (0;0)--(2;5);
	
	\node [ above ] at ($(2;1)$) {$i\,\hat h_1$};
	\node [ below ] at ($(2;5)$) {$i\,\hat h_2$};
	
	\draw [thick, draw=black, fill=white] (1.7,2.1)--(1.7,1.7)--(2.2,1.7);
	\node at (1.95,1.9) {$i\,\mathfrak{h}$};
	
	\end{tikzpicture}} 
\caption{Jordan divisors $D_1$ (blue) and $D_2$ (red) associated to the fundamental and anti-fundamental representations of $SU(3)$, for the polyhedral decomposition of $\partial\BH_\rho$ in Figure \ref{BHrhoFig}.}\label{FundJDFig}
\end{figure}\fi

The assignment of compatible Jordan divisors \eqref{JDiv1} and \eqref{JDiv2} for the fundamental and anti-fundamental representations of $SU(3)$ immediately generalizes to all irreducible representations $V$ of any rank-two group $G$.  If ${\beta\in\Delta_V}$ is a weight of $V$, expand
\begin{equation}
\beta \,=\, w_1\,\hat\omega^1 \,+\, w_2\,\hat\omega^2\,.
\end{equation}
As throughout, we assume the real mass parameter $\mu_j$ has a small positive imaginary part.  The polar hyperplanes \eqref{PolDivSU3} associated to the weight $\beta$ then intersect the cone $\Sigma_1$ in \eqref{PolyRk2} precisely when 
\begin{equation}\label{ConeIneq}
\langle\beta,v_+\rangle < 0 \quad\Longleftrightarrow\quad w_2 \,>\, w_1\,.
\end{equation}
In this case, the family of polar hyperplanes for the weight $\beta$ belongs to the Jordan divisor $D_2$.  

Conversely,  if ${w_1 > w_2}$, the associated family of polar hyperplanes belongs to $D_1$.

Finally, in the special case ${w_1=w_2}$, when $\beta$ is
proportional to the Weyl vector $\rho$ itself, the polar hyperplanes
in \eqref{PolDivSU3} are parallel to the boundary ${\partial\BH_\rho}$
and intersect neither $\Sigma_1$ nor $\Sigma_2$.  The same degeneracy
occurs for a subset of the polar hyperplanes shown in Figure
\ref{CschEx}.  As in that toy example, we break the degeneracy between
$\Sigma_1$ and $\Sigma_2$ by tilting ${\BH_\rho\subset\h_\C}$ slightly in the direction of ${-i \hat h_2}$ to
\begin{equation}
\BH_{\wt\rho} \,=\, \big\{\sigma\in\h_\C \,\big|\, \big\langle\wt\rho,\Im(\sigma)\big\rangle\,\ge\, 0\big\}\,,\qquad \wt\rho \,=\, \rho - \epsilon\,\hat\omega^2\,,\qquad 0 \ll \epsilon < 1\,.
\end{equation}
See Figure \ref{BHwtrhoFig} for a sketch of $\BH_{\wt\rho}$.  Polar hyperplanes with ${w_1=w_2>0}$ then intersect $\Sigma_1$ and so belong to the Jordan divisor $D_2$, and conversely for ${w_1=w_2<0}$.

\iffigs
\begin{figure}[t!]
\centering
	\begin{tikzpicture}[x=1.7cm,y=1.7cm] 
	\coordinate (0;0) at (0,0); 
	\foreach \c in {1,...,4}{%  
		\foreach \i in {0,...,5}{% 
			\pgfmathtruncatemacro\j{\c*\i}
			\coordinate (\c;\j) at (60*\i:\c);  
		} }
		\foreach \i in {0,2,...,10}{% 
			\pgfmathtruncatemacro\j{mod(\i+2,12)} 
			\pgfmathtruncatemacro\k{\i+1}
			\coordinate (2;\k) at ($(2;\i)!.5!(2;\j)$) ;}
		
		\foreach \i in {0,3,...,15}{% 
			\pgfmathtruncatemacro\j{mod(\i+3,18)} 
			\pgfmathtruncatemacro\k{\i+1} 
			\pgfmathtruncatemacro\l{\i+2}
			\coordinate (3;\k) at ($(3;\i)!1/3!(3;\j)$)  ;
			\coordinate (3;\l) at ($(3;\i)!2/3!(3;\j)$)  ;
		}
		
		\foreach \i in {0,4,...,20}{% 
			\pgfmathtruncatemacro\j{mod(\i+4,24)} 
			\pgfmathtruncatemacro\k{\i+1} 
			\pgfmathtruncatemacro\l{\i+2}
			\pgfmathtruncatemacro\m{\i+3} 
			\coordinate (4;\k) at ($(4;\i)!1/4!(4;\j)$)  ;
			\coordinate (4;\l) at ($(4;\i)!2/4!(4;\j)$) ;
			\coordinate (4;\m) at ($(4;\i)!3/4!(4;\j)$) ;
		}  
		
		\begin{scope}
		\clip (0,0) circle (2.1);
		
		\draw [thick, draw=black, fill=blue!10, blue!10] (-2.1,0.2)--(2.1,-0.2)--(2.1,2.1)--(-2.1,2.1)--(-2.1,0.2);
		
		\foreach \i in {0,...,6}{% 
				\pgfmathtruncatemacro\k{\i}
				\pgfmathtruncatemacro\l{15-\i}
				\draw[thin,gray] (3;\k)--(3;\l);
				\pgfmathtruncatemacro\k{9-\i} 
				\pgfmathtruncatemacro\l{mod(12+\i,18)}   
				\draw[thin,gray] (3;\k)--(3;\l); 
				\pgfmathtruncatemacro\k{12-\i} 
				\pgfmathtruncatemacro\l{mod(15+\i,18)}   
				\draw[thin,gray] (3;\k)--(3;\l);} 
		
		\end{scope}		
		
		\draw[->,thick,shorten >=4pt,shorten <=2pt] (0;0)--(2;1);	
		
		\draw[->,thick,shorten >=4pt,shorten <=2pt] (0;0)--(2;5);
		
		\node [ above ] at ($(2;1)$) {${i\,\hat h_1}$};
		\node [ above ] at ($(2;5)$) {${i\,\hat h_2}$};
				
		\begin{scope}
		\clip (0,0) circle (2.1);

		\draw[very thick,->,green] (0,0)--(-2.1,0.2);
		\draw[very thick,->,green] (0,0)--(2.1,-0.2);

		\draw[fill] (0,0) circle (.05);
		\node [ right ] at (-0.2,-0.2) {${\h\simeq\R^2}$};
		
		\end{scope}
				
		\node[ above ] at (0,1.0) {$\BH_{\wt\rho}$};

		\node [ right ] at ($1.1*(2;0)+(0,0)$) {${\big\langle\wt\rho,\Im(\sigma)\big\rangle = 0}$};
		
		\node [ left ] at (-1.5,-0.1) {$\Sigma_2$};
		\node [ right ] at (1.5,-0.4) {$\Sigma_1$};
		
		\draw [thick, draw=black, fill=white] (1.7,2.1)--(1.7,1.7)--(2.2,1.7);
		\node at (1.95,1.9) {$i\,\mathfrak{h}$};
		
		\end{tikzpicture}  	
\caption{Tilted version ${\BH_{\wt\rho}}$ of the positive Weyl halfspace.  The tilt resolves the degeneracy between $\Sigma_1$ and $\Sigma_2$ for polar hyperplanes with weight proportional to the Weyl vector $\rho$.}\label{BHwtrhoFig}
\end{figure}
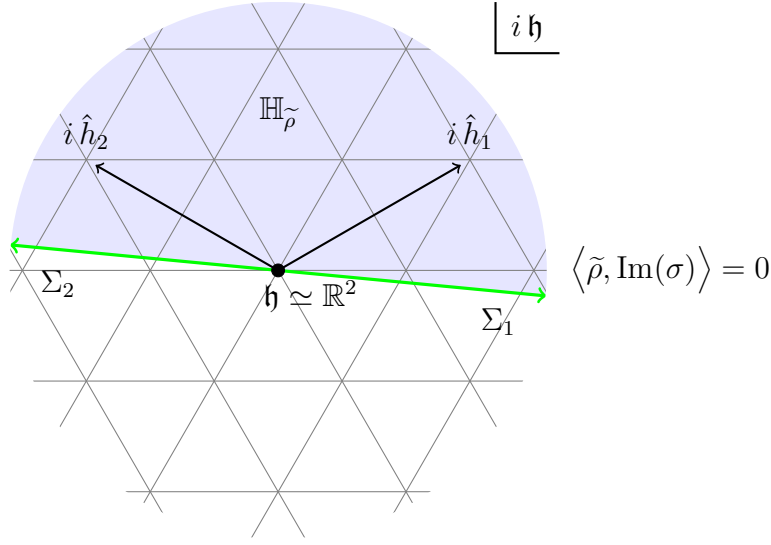\fi

These observations are summarized in Table \ref{Rk2JrdDiv}.  Altogether,
\begin{equation}\label{BigJD1}
D_1 \,=\, \bigcup_{j=1}^n \bigcup_{w_1>w_2}\,\bigcup_{w_1=w_2<0}\,\bigcup_{m,n\ge 0} \left\{\frac{\langle\beta,\sigma\rangle}{2\pi} \,=\, -i\,m\,b \,-\, i\,n\,b^{-1}\,-\,\mu_j\right\}\,,
\end{equation}
and
\begin{equation}\label{BigJD2}
D_2 \,=\,   \bigcup_{j=1}^n \bigcup_{w_1<w_2}\,\bigcup_{w_1=w_2>0}\,\bigcup_{m,n\ge 0} \left\{\frac{\langle\beta,\sigma\rangle}{2\pi} \,=\, -i\,m\,b \,-\, i\,n\,b^{-1}\,-\,\mu_j\right\}\,.
\end{equation}
The general description of the Jordan divisors in rank-two specializes to the preceding formulas \eqref{JDiv1} and \eqref{JDiv2} for the fundamental representation of $SU(3)$.
\renewcommand{\arraystretch}{1.5}
\begin{table}[t!]
\begin{center}
\begin{tabular}{ c | c }
Weight ${\beta = w_1\,\hat\omega^1\,+\,w_2\,\hat\omega^2}$ & Jordan divisor \\ \hline
${w_1\,>\,w_2}$ & $D_1$ \\
${w_1\,<\,w_2}$ & $D_2$ \\
${w_1\,=\,w_2\,<\,0}$ & $D_1$\\
${w_1\,=\,w_2\,>\,0}$ & $D_2$
\end{tabular}
\caption{Assignment of polar hyperplanes \eqref{PolDivSU3} for a given non-zero weight ${\beta\in\Delta_V}$ to Jordan divisors.  The assignments for weights with ${w_1=w_2}$ depends upon the choice of tilt in Figure \ref{BHwtrhoFig}.}\label{Rk2JrdDiv}
\end{center}
\end{table}
\renewcommand{\arraystretch}{1.0}

Via the multi-dimensional Jordan lemma, $Z^{\rm uv}_{S^3}$ can be evaluated as a residue sum over points in the intersection ${D_1 \cap D_2 \cap \BH_\rho}$,
\begin{equation}\label{ResSumRk2}
Z^{\rm uv}_{S^3} \,=\, (2\pi i)^2\mskip -2mu\sum_{p \in D_1 \cap D_2 \cap \BH_\rho} \Res_p(\omega)\,,
\end{equation}
where $\omega$ is the meromorphic form defined by the integrand in \eqref{ZuvCmbV}.  Given any hyperplanes ${H_1\subset D_1}$ and ${H_2\subset D_2}$ which are components of the respective Jordan divisors, the intersection is either transverse or empty, in the degenerate case that $H_1$ and $H_2$ are both parallel to the boundary ${\partial\BH_\rho}$.  If non-empty, ${H_1 \cap H_2}$ is a unique point in ${\h_\C\simeq\C^2}$.  The only question for determining the range of summation in \eqref{ResSumRk2} is whether that point in ${H_1 \cap H_2}$ (now assumed non-empty) lies in the positive Weyl halfspace $\BH_\rho$ or not.  

Recall that points in $\BH_\rho$ are those for which ${\langle\rho,\Im(\sigma)\rangle\ge 0}$.\footnote{By the compatibility condition on Jordan divisors, ${H_1 \cap H_2}$ never lies in the boundary ${\partial\BH_\rho}$.}  By our assumption that each $\mu_j$ has a small positive imaginary part, the right-hand sides of both \eqref{BigJD1} and \eqref{BigJD2} have strictly-negative imaginary part for all ${m,n\ge 0}$.  Determining whether ${H_1\cap H_2}$ lies in $\BH_\rho$ is then an elementary problem of linear algebra.  We omit details and state the result.  Suppose that $H_1$ is given by a weight ${\beta\in\h^*}$ in \eqref{BigJD1}, and $H_2$ is given by another weight ${\gamma\in\h^*}$ in \eqref{BigJD2}.   By assumption that ${H_1\cap H_2}$ is non-empty, $\gamma$ is not parallel to $\beta$, so the pair $(\beta,\gamma)$ provides a basis for $\h^*$.  The point of intersection ${H_1 \cap H_2}$ then lies in the positive Weyl half-space $\BH_\rho$ exactly when 
\begin{equation}\label{SignCritHr}
H_1 \cap H_2 \,\in\,\BH_\rho \quad\Longleftrightarrow\quad \sgn(\beta\^\gamma) < 0\,,
\end{equation}
where $\sgn(\beta\^\gamma)$ denotes the orientation of the basis $(\beta,\gamma)$ compared to the fixed basis $(\hat\omega^1,\hat\omega^2)$.  Note that this criterion is independent of the integers ${m,n\ge 0}$ and the mass parameter $\mu_j$, which additionally label each hyperplane component in the Jordan divisors $D_1$ and $D_2$.  Eg.~for $\Delta_{\bf 3}=\{\hat\omega^1,-\hat\omega^2,-\hat\omega^1+\hat\omega^2\}$, $\sgn(\hat\omega^1\^(-\hat\omega^1+\hat\omega^2))=+1$ and $\sgn(-\hat\omega^2\^(-\hat\omega^1+\hat\omega^2))=-1$.  Hence only the intersection of the hyperplanes with weights ${\beta=-\hat\omega^2}$ (extending in the direction of ${i\,\hat h_1}$) and ${\gamma=-\hat\omega^1+\hat\omega^2}$ lie in the positive Weyl half-space $\BH_\rho$, as clear in Figure \ref{FundJDFig}(a).

\renewcommand{\arraystretch}{1.5}
\begin{table}
\begin{center}
\begin{tabular}{| c | c | c |}
\multicolumn{3}{c}{${SU(3)}$,\quad${\Lambda_{\rm uv}={\bf 3}\oplus\bar{\bf 3}}$} \\
	\hline
	$D_1$ & $D_2$&${H_1\cap H_2}$ in $\mathbb{H}_\rho$\\
	\hline
	$\hat\omega^1$&$-\hat\omega^1$
	& $\left(-\hat\omega^2,-\hat\omega^1\right)$
	\\	
	\cline{1-2}
	 $-\hat\omega^2$ & $\hat\omega^2$
	&  $\left(-\hat\omega^2, -\hat\omega^1+\hat\omega^2\right)$ 
	\\
	\cline{1-2}
	$\hat\omega^1-\hat\omega^2$ & $-\hat\omega^1+\hat\omega^2$ 
	& $\left(\hat\omega^1-\hat\omega^2,-\hat\omega^1\right)$
	 \\ 
	\hline
\end{tabular}
\begin{tabular}{| c | c | c |}
	\multicolumn{3}{c}{${SU(3)}$,\quad${\Lambda_{\rm uv}=\mathbf{8}}$} \\
	\hline	
	$D_1$ & $D_2$&${H_1\cap H_2}$ in $\mathbb{H}_\rho$\\	
	\hline
	${-\hat\omega^1-\hat\omega^2}$&${\hat\omega^1+\hat\omega^2}$
	& $\left(-\hat\omega^1-\hat\omega^2,-\hat\omega^1+2\hat\omega^2\right)$ 
	\\
	\cline{1-2}
	${\hat\omega^1-2\hat\omega^2}$ & ${-\hat\omega^1\,+\,2\hat\omega^2}$ 
	&$\left(-\hat\omega^1-\hat\omega^2,-2\hat\omega^1+\hat\omega^2\right)$ 
	\\
	\cline{1-2}
	${2\hat\omega^1-\hat\omega^2}$ & ${-2\hat\omega^1+\hat\omega^2}$ 
	&$\left(\hat\omega^1-2\hat\omega^2,-2\hat\omega^1+\hat\omega^2\right)$ 
	\\
	\hline
\end{tabular}
\begin{tabular}{| c | c | c |}
		\multicolumn{3}{c}{${G_2}$,\quad${\Lambda_{\rm uv}={\bf 7}}$} \\
		\hline	
		$D_1$ & $D_2$&${H_1\cap H_2}$ in $\mathbb{H}_\rho$\\	
		\hline
		$\hat\omega^1$&$-\hat\omega^1$
		& $\left(\hat\omega^1-\hat\omega^2,-\hat\omega^1\right)$
		\\
		\cline{1-2}	
		${\hat\omega^1-\hat\omega^2}$ & ${-\hat\omega^1+\hat\omega^2}$
		&
		$\left(\hat\omega^1-\hat\omega^2,-2\hat\omega^1+\hat\omega^2\right)$ 
		\\
		\cline{1-2}
		${2\hat\omega^1-\hat\omega^2}$ &
                ${-2\hat\omega^1+\hat\omega^2}$ 
                & $\left(2\hat\omega^1-\hat\omega^2,-\hat\omega^1\right)$  
		\\	
		\hline
\end{tabular}\caption{Partitions of non-zero weights of $\Lambda_{\rm uv}$ into Jordan divisors $D_1$ and $D_2$, for select rank-two groups and representations.  In the third column appear the pairs of weights for which the associated hyperplanes ${H_1\subset D_1}$ and ${H_2\subset D_2}$ intersect in $\BH_\rho$ and hence contribute to the residue sum.}\label{WtH1H2Tab}
\end{center}
\end{table}
\renewcommand{\arraystretch}{1.0}

In Table \ref{WtH1H2Tab} we provide further examples of the analysis for the rank-two groups $SU(3)$ and $G_2$ and select low-dimensional representations $\Lambda_{\rm uv}$.  On the left side of each table, we display the partition of non-zero weights into Jordan divisors $D_1$ and $D_2$ according to the criteria in Table \ref{Rk2JrdDiv}.  On the right side, we list the pairs of weights for polar hyperplanes ${H_1\subset D_1}$ and ${H_2 \subset D_2}$ such that ${H_1\cap H_2 \subset \BH_\rho}$ enters the residue sum, via the criterion in \eqref{SignCritHr}.

Later in Section \ref{RkTwoExs}, we evaluate the residue formula \eqref{ResSumRk2} for $Z^{\rm uv}_{S^3}$ explicitly in various rank-two cases.

\subsection{Jordan Divisors for Higher-Rank Gauge Groups}\label{JordHRGG}

For gauge groups of rank three and higher, we run into the problem illustrated by Example D in Section \ref{ToyEx}.  Namely, we are generally unable to subdivide the infinite set of polar hyperplanes for the integrand in \eqref{ZuvCmbV} into a finite set of Jordan divisors $\{D_1,\ldots,D_r\}$ which are compatible with some polyhedral decomposition of $\partial\BH_\rho$.  A version of the elementary assignment for ${r=2}$ in Table \ref{Rk2JrdDiv} does not seem to exist in higher rank.  This technical problem is not specific to the supersymmetric partition function $Z^{\rm uv}_{S^3}$ but is present, for purely dimensional reasons, for the general multivariate Mellin-Barnes integral considered in \cite{Passre:1994,Tsikh:1998}.

To circumvent the problem for ${r\ge 3}$, we apply a partial-fractions decomposition to the integrand.  This decomposition simplifies the underlying configuration of poles so that a compatible set of Jordan divisors exists for each summand in the decomposition.  In principle, the partial-fractions decomposition works in exactly the same way as the elementary example in \eqref{PartFraD}, \eqref{PartFraDa}, and \eqref{PartFraDb}.  Here we extend that example systematically.

The Coulomb-branch integrand in \eqref{ZuvCmbV} has poles along an infinite union of hyperplanes.  Rather than start with the infinite case, we first consider a meromorphic form $\omega$ which has simple poles along only a finite arrangement of ${N\ge r}$ hyperplanes in a complex vector space ${\h_\C\simeq\C^r}$ of dimension $r$,
\begin{equation}\label{MeromHRGG}
\omega \,=\, \frac{g(z) \, dz_1\^\cdots\^dz_r}{f_1(z) \cdots f_N(z)}\,,\qquad\qquad z\in\h_\C\,.
\end{equation}
Each $f_j(z)$ for ${j=1,\ldots,N}$ is an affine linear function,
\begin{equation}\label{AffLinfj}
f_j(z) \,=\, \langle \ra_j , z\rangle \,+\, \rb_j\,,\qquad\qquad \ra_j \in \h_\C^*\,,\qquad\qquad \rb_j\in\C\,,
\end{equation}
and $g(z)$ is any entire function with sufficiently rapid decay at
infinity on a complex half-space $\BH_\rho =
\{\langle\rho,\Im(z)\rangle\ge0\}$, which is parametrized by a
non-vanishing real covector ${\rho\in\h^*}$.  In the application to
gauge theory, $\rho$ is the Weyl vector, but we leave $\rho$ arbitrary
here.  For ${N<\infty}$ and ${g(z)=\exp{[i\langle\rho,z\rangle]}}$, the residue calculus for such a meromorphic form has been analyzed in detail by Jeffrey and Kirwan \cite{Jeffrey:1995}.  We wish to reproduce a few of their results using the technology in Section \ref{JordanLemma}.

We impose several conditions on the affine linear functions in the denominator of $\omega$ and hence on the geometry of the polar hyperplane arrangement.
\begin{enumerate}
\item The functions $f_j(z)$ for ${j=1,\ldots,N}$ are non-constant, with ${\ra_j\neq 0}$, and distinct.
\item Each covector ${\ra_j\in\h^*}$ is real.  The hyperplane ${H_j=\{f_j(z)=0\}}$ respects the direct-sum decomposition ${\h_\C\simeq\h\oplus i\h}$ into real and imaginary parts.  Under the projection ${\Im:\h_\C\to i\h}$ onto the imaginary part, $\Im(H_j)$ is a real codimension-one hyperplane in $i\h$, as shown in sketches such as Figure \ref{FundJDFig}. 
\item Each affine constant $\rb_j$ satisfies ${\Im(\rb_j)\neq 0}$.  No hyperplane passes through the origin, and $\omega$ is regular on the real subspace ${\h\subset\h_\C}$.  After multiplying $f_j(z)$ by a sign, which can be absorbed into the definition of $g(z)$, we assume without loss that ${\Im(\rb_j)>0}$ is positive.
\end{enumerate}
These conditions are satisfied by any finite subset of polar hyperplanes for the gauge theory integrand in \eqref{ZuvCmbV}.  The first condition follows from the assumption that $(b,\mu_j)$ are generic and each $[\lambda_j]$ is weight-multiplicity-free.  The second condition amounts to the reality of the weights ${\beta\in\Delta_j}$.  The third condition follows from the identical assumption on $\mu_j$ as well as the location \eqref{eq:PolesOfDoublesine} of poles  for $s_b(z)$.

Under these conditions, the integral of $\omega$ over ${\h\subset\h_\C}$ depends analytically on the parameters $(\ra_j,\rb_j)$.  Without loss, we take $(\ra_j,\rb_j)$ to be generic, subject to the preceding.\footnote{For the Coulomb-branch integrand in \eqref{ZuvCmbV}, the generiticity assumption on $\ra_j$ is violated.  As shown in Figure \ref{FundJDFig}, the polar hyperplanes appear in infinite families of (non-generic) parallel planes in that case.  By analyticity, however, one is free to make small deformations of each plane, and the underlying residue calculus is unchanged.}  By generiticity, any $r$ hyperplanes intersect at a unique point, and no more than $r$ hyperplanes meet at any given point.  Equivalently, at least ${(N-r)}$ members of the set $\{f_1(z),\,\cdots,f_N(z)\}$ are non-vanishing for all ${z\in\h_\C}$.

\paragraph{(a) Initial case ${N=r}$.}
We begin with the easiest case\footnote{For ${N<r}$ the Jordan integral vanishes, as for instance follows from the vanishing argument to be discussed subsequently for the case ${N=r}$ and ${\delta_{\ul e}=0}$.} when ${N=r}$.  For this value of $N$, a partition of the polar hyperplanes associated to the functions $\{f_1,\ldots,f_N\}$ into Jordan divisors $\{D_1,\ldots,D_r\}$, compatible with some polyedral decomposition of $\partial\BH_\rho$, always exists.  However the naive, one-to-one assignment ${D_1=\{f_1(z)=0\}}$, $\,\ldots\,$, ${D_r=\{f_r(z)=0\}}$, modulo permutation of indices, is not necessarily correct.  This subtlety is illustrated by Figure \ref{PartFraFig}(a-b), for which two Jordan divisors are empty!

To demonstrate existence of compatible Jordan divisors, introduce the imaginary subspace of the boundary 
\begin{equation}
\begin{aligned}
\Im(\partial\BH_\rho) \,&=\, i\h \cap \partial\BH_\rho\,\subset\h_\C\,,\\
&=\, \big\{z\in\h_\C \,\big|\, \Re(z)=0,\,\langle\rho,\Im(z)\rangle=0\big\} \,\simeq\,\R^{r-1}\,.
\end{aligned}
\end{equation}
The subspace $\Im(\partial\BH_\rho)$ is depicted schematically in Figure \ref{FigExD} for the case ${r=3}$.  Again generically, the imaginary part $\Im(H_j)$ of each hyperplane ${H_j=\{f_j(z)=0\}}$ intersects $\Im(\partial\BH_\rho)$ in real codimension-one.  In Figure \ref{FigExD}, those intersections are shown as dashed lines (red).  For the remainder, we work exclusively in $\Im(\partial\BH_\rho)$ and so do not distinguish between the complex hyperplane ${H_j\subset\h_\C}$ and the real hyperplane ${\Im(H_j)\cap\Im(\partial\BH_\rho) \subset \Im(\partial\BH_\rho)}$.

By generiticity, any $(r-1)$ hyperplanes in the arrangement $\{H_1,\,\cdots,H_r\}$ intersect at a unique point in $\Im(\partial\BH_\rho)$, from which we obtain vectors
\begin{equation}\label{Polytvert}
\begin{aligned}
e_1 &= \Im(H_2) \cap \Im(H_3) \cap \cdots \cap \Im(H_r) \cap \Im(\partial\BH_\rho)\,,\\
e_2 &= \Im(H_1) \cap \Im(H_3) \cap \cdots \cap \Im(H_r) \cap \Im(\partial\BH_\rho)\,,\\
&\,\,\,\vdots\\
e_r &= \Im(H_1) \cap \Im(H_2) \cap \cdots \cap \Im(H_{r-1}) \cap \Im(\partial\BH_\rho)\,.
\end{aligned}
\end{equation}
The boundary vectors ${e_1,\ldots,e_r\in \Im(\partial\BH_\rho)}$ are distinct and non-zero. 

\iffigs
\begin{figure}
\centering
\subfloat[${\delta_{\ul e}=0}$]{
	\begin{tikzpicture}[x=1cm,y=1cm] 

		\draw [fill=red!10,red!10] (-2,1)--(-.667,-.333)--(0,1)--(-2,1);
		\draw[->] (-3,0)--(3,0);
		\draw[->] (0,-3)--(0,3);
		
		\draw[dashed, red] (-3,1)--(3,1);
		
		\draw[dashed, red] (-3,2)--(2,-3);
			
		\draw[dashed, red] (-2,-3)--(1.05,3.1);

		\draw[fill] (0,0) circle (.07);
		\node [ right ] at (0,-0.2) {${\h}$};

		\draw[fill,red] (-2,1) circle (0.05);
		\node[ above ] at (-2,1) {$e_1$};

		\draw[fill,red] (-.667,-.333) circle (0.05);
		\node[ below ] at (-.6,-.4) {$e_2$};

		\draw[fill,red] (0,1) circle (0.05);
		\node[ right ] at (0,1.1) {$e_3$};			

		\node[ above ] at (-.8,0.2) {$\sP_{\ul e}$};	

		\draw [thick, draw=black, fill=white] (2.2,2.9)--(2.2,2.4)--(3.4,2.4);
		\node at (3.1,2.7) {${\Im(\partial\BH_\rho)}$};	
		\end{tikzpicture}} \qquad
\subfloat[${\delta_{\ul e}=1}$]{
		\begin{tikzpicture}[x=1cm,y=1cm] 

		\draw [fill=red!10,red!10] (-2,1)--(1,-2)--(1,1)--(-2,1);

		\draw[->] (-3,0)--(3,0);
		\draw[->] (0,-3)--(0,3);

		\draw[dashed, red] (1,-3)--(1,3);
		
		\draw[dashed, red] (-3,1)--(3,1);
		
		\draw[dashed, red] (-3,2)--(2,-3);

		\draw[fill] (0,0) circle (.07);
		\node [ right ] at (0,-0.2) {${\h}$};

		\draw[fill,red] (-2,1) circle (0.05);
		\node[ above ] at (-1.9,0.9) {$e_1$};

		\draw[fill,red] (1,-2) circle (0.05);
		\node[ left ] at (1,-2) {$e_2$};

		\draw[fill,red] (1,1) circle (0.05);
		\node[ right ] at (1,1.1) {$e_3$};		

		\node[ above ] at (-.8,0.2) {$\sP_{\ul e}$};	
	
		\draw [thick, draw=black, fill=white] (2.2,2.9)--(2.2,2.4)--(3.4,2.4);
		\node at (3.1,2.7) {${\Im(\partial\BH_\rho)}$};	
		\end{tikzpicture}}
\caption{Examples of convex polytopes $\sP_{\ul e}$ for ${r=3}$.  Under projection to the imaginary subspace $\Im(\partial\BH_\rho)$ of the boundary, the origin is identified with the real cycle ${\h\subset\h_\C}$.  The hyperplane arrangements in (a) and (b) cannot be deformed into each other while preserving regularity of the meromorphic form $\omega$ along $\h$.}\label{delsPFig}
	\end{figure}
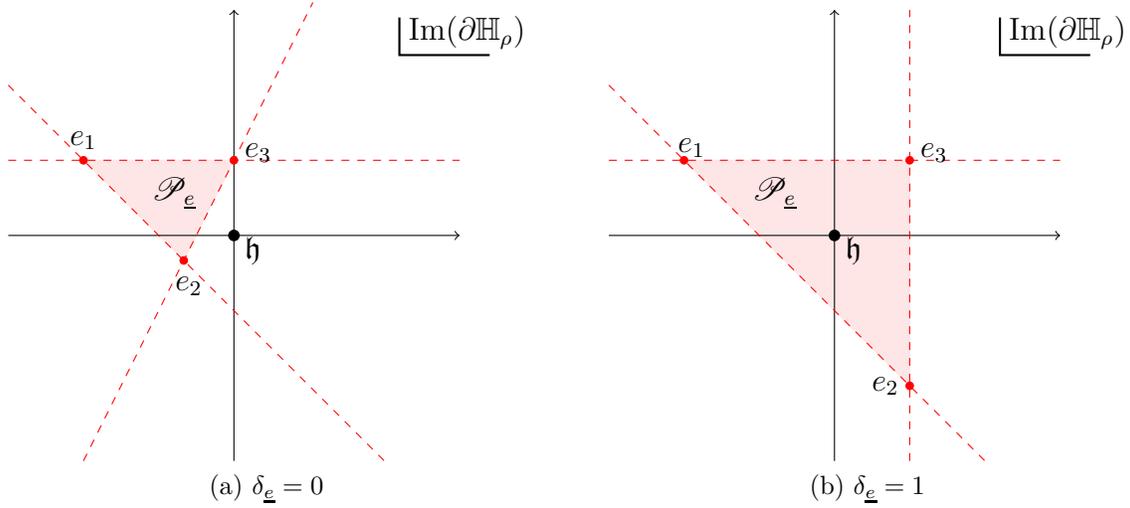
\fi

The polyhedral decomposition on $\partial\BH_\rho$ depends upon an simple analytic invariant.  To describe the invariant, let ${\sP_{\ul e}\subset \Im(\partial\BH_\rho)}$ be the convex polytope with vertices $\{e_1,\,\cdots,e_r\}$, ie.
\begin{equation}
\sP_{\ul e} \,=\, \left\{\sum_{j=1}^r t_j\,e_j \,\Big|\, t_1,\ldots,t_r \in \R_+\,, \sum_{j=1}^r t_j = 1\right\}\,.
\end{equation}
Geometrically, $\sP_{\ul e}$ is the compact region in $\Im(\partial\BH_\rho)$ bounded by the hyperplanes ${\Im(H_1),\,\ldots,\Im(H_r)}$.  Of course, only for ${N=r}$ does the arrangement determine a single bounded region!  

Depending upon the values of the parameters $(\ra_j,\rb_j)$, the polytope $\sP_{\ul e}$ may or may not contain the origin in $\Im(\partial\BH_\rho)$.  See 
Figure \ref{delsPFig} for an illustration.  We distinguish the two situations with a numerical quantity
\begin{equation}\label{deltaP}
\delta_{\ul e} \,=\, \begin{cases}
&1,\qquad\qquad \{0\} \in \sP_{\ul e}\,,\\
&0,\qquad\qquad \hbox{otherwise}\,.
\end{cases}
\end{equation}
Regularity of $\omega$ along the real cycle ${\h\subset\h_\C}$ means that no hyperplane $\Im(H_j)$ passes through the origin in $\Im(\partial\BH_\rho)$.  As a result, $\delta_{\ul e}$ does not change under any continuous variation of the parameters ${\ra_j\in\h^*}$ and ${\rb_j\in\C}$ which preserves regularity of $\omega|_\h$.  In this sense, $\delta_{\ul e}$ is an ``analytic'' invariant of the hyperplane arrangement.

The invariant $\delta_{\ul e}$ can be described dually in terms of the covectors ${\ra_1,\ldots,\ra_r\in\h^*}$ and the element ${\rho\in\h^*}$ which defines the half-space $\BH_\rho$.  The dual description is more useful in practice, since the covectors $\ra_j$ enter directly in the definition of the meromorphic form $\omega$ and correspond to weights ${\beta\in\Delta_V}$ in gauge theory, whereas the vertices ${e_1,\ldots,e_r\in\Im(\partial\BH_\rho)}$ can only be found after solving a linear system.  

Let $\R_+[\ra_1,\ldots,\ra_r]$ be the cone in ${\h^*\simeq\R^r}$ generated by the set of covectors.  Then $\delta_{\ul e}$ is either $1$ or $0$ precisely when ${\pm\rho}$ either does or does not lie within the cone $\R_+[\ra_1,\ldots,\ra_r]$,
\begin{equation}\label{deltaPII}
\delta_{\ul e}  \,=\, \delta_{\ul\ra,\pm\rho} \,=\, \begin{cases}
&1,\qquad\qquad \pm\rho \in \R_+[\ra_1,\ldots,\ra_r\big]\,,\\
&0,\qquad\qquad \hbox{otherwise}\,.
\end{cases}
\end{equation}
In the first line of \eqref{deltaPII}, one of $\rho$ or $-\rho$ lies within the cone; in the second line, neither $\rho$ nor $-\rho$ lies within the cone.  The preceding definition \eqref{deltaP} of $\delta_{\ul e}$ involves only the geometry of hyperplanes in the boundary ${\Im(\partial\BH_\rho)}$, which is invariant under a reversal in the sign of $\rho$, so the dual characterization must have the same symmetry.  Using \eqref{deltaPII}, we write ${\delta_{\ul e}\equiv \delta_{\ul\ra,\pm\rho}}$ at times to emphasize the dual dependence.

The proof of equivalence between the definitions in \eqref{deltaP} and \eqref{deltaPII} is a small exercise in linear algebra.  Because the proof is unenlightening, we relegate it to Appendix \ref{ConvGLem}.

We now construct a polyhedral decomposition 
\begin{equation}\label{PolyBHrho}
\partial\BH_\rho \,=\, \Sigma_1 \,-\, \Sigma_2 \,+\, \cdots \,+\, (-1)^{r+1} \, \Sigma_r\,,
\end{equation}
which will ultimately be compatible with a set of Jordan divisors
${D_1,\ldots,D_r}$ for the meromorphic form in \eqref{MeromHRGG}.
With respect to the decomposition ${\h_\C = \h \oplus i\h}$, we make
the ansatz that each $(2r-1)$-dimensional cone $\Sigma_j$ is a product 
\begin{equation}\label{SomeSigs}
\Sigma_j \,=\, \h \times i \R_+\big[v_1,\,\cdots,\hat{v}_j,\,\cdots,v_r\big]\,,\qquad\qquad j\,=\,1,\ldots,r\,,
\end{equation}
for some non-zero vectors ${v_1,\ldots,v_r\in\Im(\partial\BH_\rho)}$.  Compare to the identical polyhedral cones in \eqref{ConealpH}, where the same notation is used.

The construction depends upon the value of the invariant $\delta_{\ul e}$ in \eqref{deltaP}.

If ${\delta_{\ul e}=0}$, meaning that the convex polytope $\sP_{\ul e}$ does not contain the origin in ${\Im(\partial\BH_\rho)\simeq\R^{r-1}}$, then by definition there exists a ray extending from the origin which does not intersect any hyperplane ${\Im(H_1),\ldots,\Im(H_r)}$ forming a wall of $\sP_{\ul e}$.  This ray is characterized by an open condition,  so we can find a polyhedral cone $\R_+[v_2,\,\cdots,\,v_r] \subset \Im(\partial\BH_\rho)$ of dimension $(r-1)$ which contains the given ray and which also does not intersect the hyperplanes ${\Im(H_1),\ldots,\Im(H_r)}$.  We assume the generators $\{v_2,\ldots,v_r\}$ are oriented compatibly with ${\Im(\partial\BH_\rho)}$.  Now let ${v_1\in\Im(\partial\BH_\rho)}$ be any generic (non-zero) vector which is not contained within the preceding cone, eg.~${v_1 = -v_2 - v_3 - \cdots - v_r}$.  See Figure \ref{delsZeroFig} for an illustration when ${r=3}$.

\iffigs
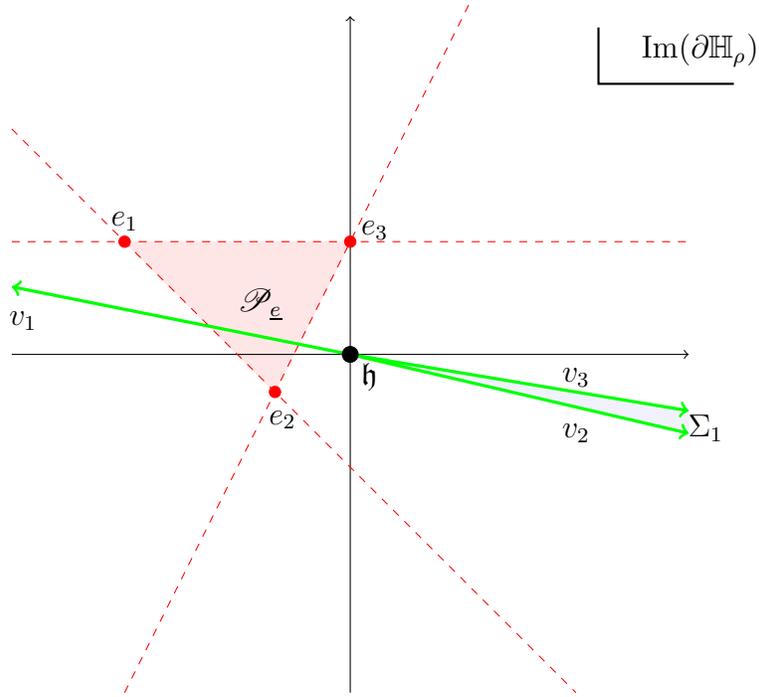
\begin{figure}
\centering
	\begin{tikzpicture}[x=1.5cm,y=1.5cm] 
		\draw [fill=red!10,red!10] (-2,1)--(-.667,-.333)--(0,1)--(-2,1);
		\draw [fill=blue!5,blue!5] (0,0)--(3,-0.7)--(3,-0.5)--(0,0);
		\draw[->] (-3,0)--(3,0);
		\draw[->] (0,-3)--(0,3);
		
		\draw[dashed, red] (-3,1)--(3,1);
		
		\draw[dashed, red] (-3,2)--(2,-3);
			
		\draw[dashed, red] (-2,-3)--(1.05,3.1);

		\draw[very thick, green, ->] (0,0)--(3,-0.5);
		\node [] at (2,-0.2) {$v_3$};

		\draw[very thick, green, ->] (0,0)--(3,-0.7);
		\node [ ] at (2,-0.7) {$v_2$};

		\draw[very thick, green, ->] (0,0)--(-3,0.6);
		\node [ ] at (-2.9,0.3) {$v_1$};

		\draw[fill] (0,0) circle (.07);
		\node [ right ] at (0,-0.2) {${\h}$};

		\draw[fill,red] (-2,1) circle (0.05);
		\node[ above ] at (-2,1) {$e_1$};

		\draw[fill,red] (-.667,-.333) circle (0.05);
		\node[ below ] at (-.6,-.4) {$e_2$};

		\draw[fill,red] (0,1) circle (0.05);
		\node[ right ] at (0,1.1) {$e_3$};			

		\node[ above ] at (-.8,0.2) {$\sP_{\ul e}$};	
		
		\node[ right ] at (2.9,-0.65) {$\Sigma_1$};
		
		\draw [thick, draw=black, fill=white] (2.2,2.9)--(2.2,2.4)--(3.4,2.4);
		\node at (3.1,2.7) {${\Im(\partial\BH_\rho)}$};	
		\end{tikzpicture}
\caption{Vectors in a compatible polyhedral decomposition of ${\partial\BH_\rho}$ when ${\delta_{\ul e}=0}$.  The vectors ${v_2,v_3}$ (green) span a small cone which does not intersect any of the three polar hyperplanes (dashed red) forming the walls of the polytope $\sP_{\ul e}$.  The remaining vector $v_1$ lies outside the cone spanned by $\{v_2,v_3\}$.  Since no polar hyperplane intersects $\Sigma_1=\h \times i \R_+[v_2,v_3]$, all can be compatibly assigned to the Jordan divisor $D_1$, with the other Jordan divisors ${D_2,D_3}$ empty.  The Jordan integral ${\Phi_\sC=0}$ for ${\sC=\pm\h}$ therefore vanishes.}\label{delsZeroFig}
\end{figure}
\fi

Given the set $\{v_1,\ldots,v_r\}$, define $\Sigma_j$ via \eqref{SomeSigs} to obtain the polyhedral decomposition.  By convention, each $\Sigma_j$ is oriented according to the signs in \eqref{PolyBHrho}; the orientation data is omitted from \eqref{SomeSigs}.  As for compatibility \eqref{CompatC}, by the preceding choice of $\{v_2,\ldots,v_r\}$, the polar hyperplanes ${H_1,\ldots,H_r}$ do not intersect the distinguished cone $\Sigma_1 = \h \times \R_+[v_2,\,\cdots,\,v_r]$.  Hence a compatible set of Jordan divisors is provided by the trivial partition
\begin{equation}
\begin{aligned}
D_1 \,&=\, \big\{f_1(z)=0\big\} \cup \cdots \cup \big\{f_r(z)=0\big\}\,,\\
D_2 \,&=\, \cdots \,=\, D_r \,=\, \varnothing\,,
\end{aligned}
\end{equation}
under which all polar hyperplanes are assigned to $D_1$, and the other Jordan divisors are empty.  Equivalently, in terms of the section ${s\equiv(s^1,\,\cdots,s^r)}$ from Section \ref{JordanLemma},
\begin{equation}
s^1(z) \,=\, f_1(z) \cdots f_r(z)\,,\qquad\qquad s^2(z) \,=\, \cdots \,=\, s^r(z) \,=\, 1\,.
\end{equation} 
Thus if ${\delta_{\ul e}=0}$ the integral of $\omega$ over $\h$ vanishes,
\begin{equation}
\frac{1}{\left(2\pi i\right)^r}\int_{\h}\omega \,=\, 0\,.
\end{equation}
This discussion generalizes the vanishing of $\Xi_1$ and $\Xi_2$ in Figure \ref{PartFraFig}.

If ${\delta_{\ul e}=1}$, meaning that the origin lies inside the convex polytope $\sP_{\ul e}$, then by definition there exist positive constants ${t_1,\ldots,t_r>0}$, ${t_1 + \cdots + t_r=1}$, so that 
\begin{equation}\label{ZerosP}
t_1\,e_1 \,+\, \cdots \,+\, t_r\,e_r \,=\, 0\,.
\end{equation}
Let ${v_1,\ldots,v_r}$ be vectors in $\Im(\partial\BH_\rho)$ which generate a system of rays dual to the faces of $\sP_{\ul e}$.  That is, each ray $\R_+[v_j]$ for each ${j=1,\ldots,r}$ passes transversely through the face of $\sP_{\ul e}$ formed by the corresponding hyperplane $\Im(H_j)$ and does not intersect the other hyperplanes $\Im(H_\ell)$ for ${j\neq\ell}$.  See Figure \ref{delsOneFig} for an illustration of the dual rays when ${r=3}$.  In terms of the pairs $(\ra_j,\rb_j)$ which appear in \eqref{AffLinfj}, the duality condition on $v_j$ amounts to the system of inequalities
\begin{equation}\label{ConeIneqII}
\langle\ra_j,\,v_j\rangle < 0\,,\qquad\qquad \langle\ra_\ell,\,v_j\rangle > 0\,,\qquad \ell\neq j\,.
\end{equation}
This condition relies on the convention ${\Im(\rb_j)>0}$ for all $j$.  Compare also to the similar condition \eqref{ConeIneq} which appears in rank-two.  

\iffigs
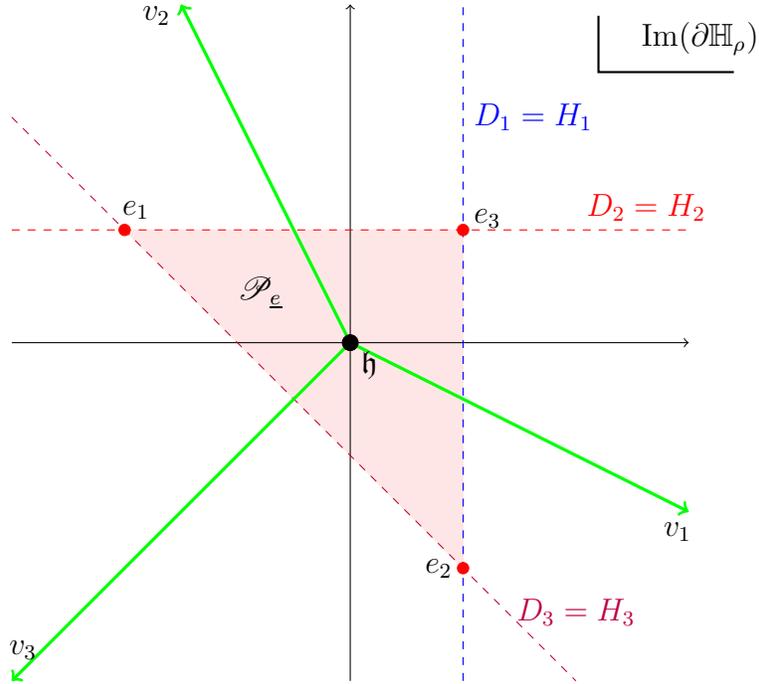
\begin{figure}
\centering
\begin{tikzpicture}[x=1.5cm,y=1.5cm] 

		\draw [fill=red!10,red!10] (-2,1)--(1,-2)--(1,1)--(-2,1);

		\draw[->] (-3,0)--(3,0);
		\draw[->] (0,-3)--(0,3);

		\draw[dashed, blue] (1,-3)--(1,3);
		\node [blue,right] at (1,2) {$D_1=H_1$};
		
		\draw[dashed, red] (-3,1)--(3,1);
		\node[red,right] at (2,1.2) {$D_2=H_2$};
		
		\draw[dashed,purple] (-3,2)--(2,-3);
		\node[above,purple] at (2,-2.6) {$D_3=H_3$};

		\draw[very thick, green, ->] (0,0)--(-3,-3);
		\node [ above ] at (-2.9,-2.9) {$v_3$};

		\draw[very thick, green, ->] (0,0)--(-1.5,3);
		\node [ left ] at (-1.5,2.9) {$v_2$};

		\draw[very thick, green, ->] (0,0)--(3,-1.5);
		\node [ below ] at (2.9,-1.5) {$v_1$};

		\draw[fill] (0,0) circle (.07);
		\node [ right ] at (0,-0.2) {${\h}$};

		\draw[fill,red] (-2,1) circle (0.05);
		\node[ above ] at (-1.9,1) {$e_1$};

		\draw[fill,red] (1,-2) circle (0.05);
		\node[ left ] at (1,-2) {$e_2$};

		\draw[fill,red] (1,1) circle (0.05);
		\node[ right ] at (1,1.1) {$e_3$};		

		\node[ above ] at (-.8,0.2) {$\sP_{\ul e}$};	
	
		\draw [thick, draw=black, fill=white] (2.2,2.9)--(2.2,2.4)--(3.4,2.4);
		\node at (3.1,2.7) {${\Im(\partial\BH_\rho)}$};	
\end{tikzpicture}
\caption{Vectors in a compatible polyhedral decomposition of ${\partial\BH_\rho}$ when ${\delta_{\ul e}=1}$.  The rays (green) generated by ${v_1,v_2,v_3}$ are dual to the faces of the polytope $\sP_{\ul e}$ delimited by the hyperplanes ${H_1,H_2,H_3}$  (dashed lines).  Each polar hyperplane ${H_1,H_2,H_3}$ is assigned in a one-to-one fashion to a Jordan divisor ${D_1,D_2,D_3}$.  The cone ${\Sigma_1 = \h \times \R_+[v_2,v_3]}$ does not intersect the divisor $D_1$, and similarly in a cyclic fashion for the other cones, so the assignment of Jordan divisors is compatible with the decomposition of $\partial\BH_\rho$.  For clarity in the figure, we have not shaded the cones ${\Sigma_1,\Sigma_2,\Sigma_3}$, unlike the examples in Section \ref{JordanLemma}.  The Jordan integral ${\Phi_\sC=\Res_p(\omega)}$ is given by the local residue at ${p=H_1 \cap H_2 \cap H_3 \cap \BH_\rho}$ (if non-empty).}\label{delsOneFig}
\end{figure}
\fi

Though duality guarantees the existence of vectors $\{v_1,\ldots,v_r\}$ which obey the bounds in \eqref{ConeIneqII}, a concrete choice can be made in terms of the vertices of $\sP_{\ul e}$ with the assignment
\begin{equation}\label{DualRays}
v_j \,=\, -e_j\,,\qquad\qquad j=1,\ldots,r\,.
\end{equation}
For by the definitions in \eqref{AffLinfj} and \eqref{Polytvert},
\begin{equation}\label{ConeIneqIII}
\Im(f_\ell)(e_j)\,=\, 0 \qquad\Longrightarrow\qquad \langle\ra_\ell,\,e_j\rangle \,=\, -\Im(\rb_j) < 0\,,\qquad \ell\neq j\,.
\end{equation}
Also, from the identity \eqref{ZerosP} and the preceding \eqref{ConeIneqIII},
\begin{equation}
\langle\ra_j,\,e_j\rangle \,=\, -\sum_{\ell\neq j} \frac{t_\ell}{t_j}\,\langle\ra_j,\,e_\ell\rangle \,=\, \sum_{\ell\neq j} \frac{t_\ell}{t_j}\,\Im(\rb_\ell) > 0\,.
\end{equation}
The sign flip in \eqref{DualRays} ensures that ${v_j=-e_j}$ obeys the inequalities in \eqref{ConeIneqII}.

We use the vectors $\{v_1,\ldots,v_r\}$ to define cones $\Sigma_j$ according to \eqref{SomeSigs}.  

\ul{Claim}:~the given polyhedral decomposition of ${\partial\BH_\rho}$ is compatible with the naive, one-to-one assignment of Jordan divisors
\begin{equation}\label{JDdelone}
D_1 = H_1 = \{f_1(z)=0\}\,,\quad\ldots\,,\quad D_r = H_r = \{f_r(z)=0\}\,.
\end{equation}
Equivalently, in terms of the holomorphic section ${s\equiv (s^1,\cdots,s^r)}$ of the rank-$r$ trivial bundle over $\BH_\rho$,
\begin{equation}
s^1(z) \,=\, f_1(z)\,,\quad\ldots\,,\quad s^r(z) \,=\, f_r(z)\,.
\end{equation}
The claim follows just by unraveling the definition of the generating set $\{v_1,\ldots,v_r\}$.  Recall that compatibility means that the intersection $\Sigma_j \cap D_j = \varnothing$ is empty for each ${j=1,\ldots,r}$.  Because the imaginary directions in $\Sigma_j$ are generated by rays in the direction of the vectors ${v_\ell}$ for ${\ell\neq j}$, we require ${\R_+[v_\ell] \cap \Im(D_j) = \varnothing}$ for all ${\ell\neq j}$.  Given the assignments in \eqref{JDdelone} and the definition of the affine linear function $f_j(z)$ in \eqref{AffLinfj}, with $\Im(\rb_j)>0$ by convention, the latter condition can be restated as the inequality
\begin{equation}
\R_+[v_\ell] \cap \Im(D_j) = \varnothing \quad\Longleftrightarrow\quad \langle\ra_j,\,v_\ell\rangle > 0\,,\quad \ell\neq j\,.
\end{equation}
This inequality already appears in \eqref{ConeIneqII} as part of the
definition for the set of rays dual to $\sP_{\ul e}$.\quad$\square$  

The compatibility condition can also be verified by inspection for the example in Figure \ref{delsOneFig}.  Note that for clarity, we have not shaded the cones ${\Sigma_1,\Sigma_2,\Sigma_3}$ in Figure \ref{delsOneFig}, as we did for the examples in Figure \ref{PartFraFig}(c-d).

Because the polyhedral decomposition of $\partial\BH_\rho$ is compatible with the one-to-one assignment of Jordan divisors in \eqref{JDdelone}, the Jordan lemma now implies that the integral of $\omega$ is given by the residue at the special point 
\begin{equation}
p \in  H_1 \cap \cdots \cap H_r \cap \BH_\rho\,,
\end{equation}
when such a point exists.  By generiticity, the hyperplanes ${H_1,\ldots,H_r}$ always intersect at a unique point $p$ in $\h_\C$, so the only question is whether the point $p$ lies in $\BH_\rho$ or $-\BH_\rho$.  A simple criterion is provided by the convexity lemma proven in Appendix \ref{ConvGLem}.  Briefly, if ${\delta_{\ul e}=1}$ as we assume, then ${p\in\BH_\rho}$ precisely when the covector ${-\rho\in\h^*}$ lies in the positive cone $\R_+[\ra_1,\ldots,\ra_r]$.

To summarize our results for the case ${N=r}$, we refine the definition of the analytic invariant $\delta_{\ul\ra,\pm\rho}$ in \eqref{deltaPII} by setting 
\begin{equation}
\delta_{\ul\ra,-\rho} \,=\, 
\begin{cases}
&1\,\qquad -\!\rho\,\in\,\R_+[\ra_1,\ldots,\ra_r]\,,\\
&0\,\qquad \hbox{otherwise}\,.
\end{cases}
\end{equation}
Then
\begin{equation}\label{JorNeqR}
\begin{aligned}
\frac{1}{\left(2\pi i\right)^r} \int_{\h} \omega \,&=\, \delta_{\ul\ra,-\rho}\cdot \left(-1\right)^{|\rho|}\,\Res_p(\omega)\,,\qquad p\in H_1 \cap \cdots \cap H_r\,,\\
&=\, \delta_{\ul\ra,-\rho}\cdot \left(-1\right)^{|\rho|}\frac{g(p)}{\det(\ra)}\,.
\end{aligned}
\end{equation}
The nontrivial content of the residue theorem \eqref{JorNeqR} is the prefactor $\delta_{\ul\ra,-\rho}$, which states that the integral vanishes unless $-\rho$ lies inside the positive cone $\R_+[\ra_1,\ldots,\ra_r]$. 

For completeness, we evaluate the local residue at $p$ explicity in the second line.  Here $\ra$ is the ${r\times r}$ matrix whose rows are given by the covectors ${\ra_j\in\h^*}$, expressed in some choice of basis.  More invariantly, ${\det(\ra) = \ra_1\^\cdots\^\ra_r/dz_1\^\cdots\^dz_r|_\h}$ is a ratio of volume-forms on $\h$.  
Finally, the overall sign $\left(-1\right)^{|\rho|}$ depends upon the
element ${\rho\in\h^*}$ which specifies the half-space $\BH_\rho$
along which $g(z)$ decays.  The boundary $\partial\BH_\rho$ inherits a
canonical orientation from ${\BH_\rho\subset\h_\C}$, and this
orientation determines the orientation of the real cycle ${\sC=\pm\h}$
which enters the Jordan lemma.  As indicated, the orientation of $\sC$
may or may not agree with the original orientation of $\h$.  The
discrepancy is measured by $|\rho|$, to be determined next.

Both to fix the sign and to provide an elementary example, suppose that $g(z)$ in \eqref{MeromHRGG} is given by the exponential
\begin{equation}
g(z) \,=\, \exp{\!\left(i\,\langle\rho,z\rangle\right)} \,=\, \exp{\!\left(i \rho_1 z_1 \,+\, \cdots \,+\, i \rho_r z_r\right)}\,,
\end{equation}
where we choose dual (oriented) bases $\{\hat\omega^1,\ldots,\hat\omega^r\}$ and
$\{\hat h_1,\ldots,\hat h_r\}$ for $\h^*$ and $\h$ in which to express
$\rho$ and $z$.\footnote{The notation agrees with that in Appendix \ref{se:LieAlgConvention} for
  the positive simple weights and coroots when ${\h\subset\g}$ is the Cartan subalgebra.  The reader should take care not to confuse the weights $\hat\omega^j$ with the meromorphic form $\omega$.}  We choose the sign in the exponent so that $g(z)$ decays at infinity on $\BH_\rho$, where ${\Im\langle\rho,z\rangle\ge 0}$.   The ansatz for $g(z)$ and the volume-form ${dz_1\^\cdots\^dz_r|_\h}$ are preserved by $SL(\h)$.  So without loss, we assume that ${\ra_1,\ldots,\ra_r}$ in \eqref{AffLinfj} are proportional to the basis for $\h^*$,
\begin{equation}\label{rabasis}
\ra_1 \,=\, a_1\,\hat\omega^1\,,\qquad \ra_2\,=\, a_2\,\hat\omega^2\,,\qquad\cdots\,,\qquad \ra_r \,=\, a_r\,\hat\omega^r\,,
\end{equation}
for some non-vanishing coefficients ${a_1,\ldots,a_r \in \R}$. 

With these assumptions, the Jordan integral factorizes,
\begin{equation}\label{ItIntJorNeqR}
\frac{1}{\left(2\pi i\right)^r}\int_\h \omega \,=\, \frac{1}{\left(2\pi i\right)^r} \int_{\h} d^r\!z\,\frac{\e{i \rho_1 z_1}}{\left(a_1 z_1\,+\,\rb_1\right)}\times\cdots\times\frac{\e{i \rho_r z_r}}{\left(a_r z_r\,+\,\rb_r\right)}\,.
\end{equation}
Each factor can be evaluated by closing the contour in the upper or lower half-plane, depending upon the signs of ${\rho_1,\ldots,\rho_r}$.  Since we assume ${\Im(\rb_j)>0}$ for all $j$, whether or not the contour for $z_j$ surrounds the pole depends upon the sign of $a_j$.  Specifically, the Jordan integral vanishes unless the sign of each $a_j$ is opposite to the sign of $\rho_j$,
\begin{equation}\label{Signsrhoa}
\sgn(\rho_j)=-\sgn(a_j)\,,\qquad\qquad j=1,\ldots,r\,.
\end{equation}
Given \eqref{rabasis}, the sign condition is the same as the geometric
criterion that $-\rho$ lies in the positive cone ${\R_+[\ra_1,\ldots,\ra_r]}$.

Comparing the right side of \eqref{JorNeqR} to \eqref{ItIntJorNeqR}, 
we also determine the orientation-induced sign.  For each component
${\rho_1,\ldots,\rho_r}$ which is negative, the corresponding contour
integral acquires an extra sign when closed in the lower half-plane,
so 
\begin{equation}
\begin{aligned}
|\rho| \,&=\, \#\big\{\sgn(\rho_j)<0,\,j=1,\ldots,r\big\}\,,\\
&=\, \#\big\{\sgn(a_j)>0,\,j=1,\ldots,r\big\}\,.
\end{aligned}
\end{equation}
In the second line, we apply the condition in \eqref{Signsrhoa}.  To describe the sign invariantly, we observe that  for the basis in \eqref{rabasis},
\begin{equation}\label{signrho}
(-1)^{|\rho|} \,=\, (-1)^r \cdot \sgn(\det(\ra)) \,=\, (-1)^r \cdot \sgn(\det(df_p))\,.
\end{equation}
The sign of $\det(\ra)$ is invariant under the dual action by $SL(\h)$, so the latter expression for $(-1)^{|\rho|}$ is true independent of the choice of basis.  Geometrically, ${\sgn(\det(\ra))=\pm 1}$ depending upon whether or not the orientation induced by ${df_1 \^ \cdots \^ df_r|_\h}$ at $p$ agrees with the fixed orientation of $\h$, indicated on the right in \eqref{signrho}.

The formula for the Jordan integral in \eqref{JorNeqR} can then be rewritten more simply as 
\begin{equation}\label{JorNeqRII}
\frac{1}{\left(2\pi i\right)^r} \int_{\h} \omega \,=\, \delta_{\ul\ra,-\rho}\cdot \left(-1\right)^r \frac{g(p)}{|\!\det(\ra)|}\,,\qquad\qquad p\in H_1 \cap \cdots \cap H_r\,.
\end{equation}
The appearance of the absolute-value of $\det(\ra)$ can also be
understood as follows.  Clearly, the assignment of indices to the covectors $\{\ra_1,\ldots,\ra_r\}$ is arbitrary, as these indices just label the functions which appear in the denominator \eqref{MeromHRGG} of $\omega$.  Any formula for the Jordan integral must be invariant under permutations of $\{\ra_1,\ldots,\ra_r\}$.  However, the sign of $\det(\ra)$ is alternating under permutations.  Bose-symmetry thus requires the absolute-value to make its appearance.

\paragraph{(b) Induction for ${N>r}$.}
For ${r < N < \infty}$, we proceed by induction on $N$.  Given the meromorphic form $\omega$ in\eqref{MeromHRGG}, we attempt to decompose the denominator as a finite sum
\begin{equation}
\frac{1}{f_1(z) \cdots f_N(z)} \,=\, \sum_{j=1}^N\,\frac{c_j}{f_1(z) \cdots \widehat{f_j}(z) \cdots f_N(z)}\,,
\end{equation}
for some constants ${c_1,\ldots,c_N\in\C}$.  The notation indicates that $\widehat{f_j}(z)$ is omitted from the product in the denominator on the right, effectively reducing the value of $N$ by one.  The constants must obey 
the partial-fractions equation
\begin{equation}
c_1\,f_1(z) \,+\, \cdots \,+\, c_N\,f_N(z) \,=\, 1\,,
\end{equation}
for all values of ${z\in\h_\C}$.  

Because each function ${f_j(z) = \langle\ra_j,z\rangle \,+\, \rb_j}$ is affine linear, the partial-fractions equation is possible to solve.  Explicitly, in terms of the pairs $(\ra_j,\rb_j)$, we require 
\begin{equation}\label{PartFrA}
\sum_{j=1}^N c_j \, \ra_j \,=\, 0\,,\\
\end{equation}
and
\begin{equation}\label{PartFrB}
\sum_{j=1}^N c_j \, \rb_j \,=\, 1\,.
\end{equation}
Because ${N>r}$, the set $\{\ra_1,\ldots,\ra_N\}$ is an overcomplete basis for ${\h^*\simeq\R^r}$, and an $(N-r)$-dimensional family of solutions exists for the vanishing equation in \eqref{PartFrA}.  By generiticity, ${\sum_j c_j\,\rb_j \neq 0}$ for a solution to \eqref{PartFrA}, and by a complex scaling we ensure the normalization condition in \eqref{PartFrB}.  The moduli space of solutions for the partial-fractions problem is thus a copy of $\C\Pj^{N-r-1}$.

Inducting downwards to the case ${N=r}$, we obtain a complete decomposition
\begin{equation}\label{CcPartFr}
\frac{1}{f_1(z) \cdots f_N(z)} \,=\, \sum_{\#J=N-r}\,\frac{c_J}{f_{J^o}(z)}\,,\qquad\qquad z\,\in\h_\C \simeq\C^r\,,
\end{equation}
where ${J=(j_1\cdots j_{N-r})}$ is a multi-index, ${J^o=\{1,\ldots,N\}-J = (\ell_1 \cdots \ell_r)}$ is the complementary index set, and ${f_{J^o}(z)=f_{\ell_1}(z) \cdots f_{\ell_r}(z)}$ is the product over functions labelled by $J^o$.  By the previous discussion, the values of the constants ${c_J\in\C}$ are not unique, but a particular choice can be made geometrically as follows.  

For each value of the multi-index $J$, let ${p_J\in\h_\C}$ be the point of intersection for the complementary hyperplanes,
\begin{equation}\label{DefPJ}
p_J \,=\, H_{\ell_1} \cap \cdots \cap H_{\ell_r}\,,\qquad\qquad J^o = (\ell_1 \cdots \ell_r)\,.
\end{equation}
Equivalently, $p_J$ is the point where 
\begin{equation}
f_{\ell_1}(p_J) \,=\, \cdots \,=\, f_{\ell_r}(p_J) \,=\, 0\,,\qquad J^o = (\ell_1 \cdots \ell_r)\,.
\end{equation}
Trivially, for any pair of multi-indices $(J,K)$ with ${\#J=\#K=N-r}$, 
\begin{equation}\label{DelJK}
f_K(p_J) \,=\, f_{k_1}(p_J) \cdots f_{k_{N-r}}(p_J) \,=\, 0\quad\hbox{unless}\quad J=K\,.
\end{equation}
The constants $c_J$ in \eqref{CcPartFr} satisfy 
\begin{equation}\label{PartFrC}
\sum_{\#K=N-r} c_K\, f_K(z) \,=\, 1\,.
\end{equation}
When we evaluate the left-side of \eqref{PartFrC} at the point $p_J$ and use \eqref{DelJK}, we obtain
\begin{equation}
c_J f_J(p_J) \,=\, 1 \quad\Longrightarrow\quad c_J \,=\, \frac{1}{f_J(p_J)}\,.
\end{equation}
Thus, with the definition of the intersection point ${p_J\in\h_\C}$ in \eqref{DefPJ}, the partial-fractions decomposition in \eqref{CcPartFr} can be written canonically as 
\begin{equation}\label{CcPartFrII}
\frac{1}{f_1(z) \cdots f_N(z)} \,=\, \sum_{\#J=N-r}\,\frac{1}{f_J(p_J) \cdot f_{J^o}(z)}\,.
\end{equation}
By fiat, the right side of \eqref{CcPartFrII} has an identical structure of poles and residues as the left.

Applied to the Jordan integral of $\omega$ in \eqref{MeromHRGG}, we obtain
\begin{equation}\label{JorNeqRIII}
\begin{aligned}
\frac{1}{\left(2\pi i\right)^r}\int_\h \omega \,&=\,  \frac{1}{\left(2\pi i\right)^r}\sum_{\#J=N-r} \int_\h \frac{g(z)\, dz_1\^\cdots\^dz_r}{f_J(p_J)\cdot f_{J^o}(z)}\,,\\
&=\,(-1)^r \sum_{\#J=N-r} \delta_{\ra_{J^o},-\rho}\cdot\frac{g(p_J)}{f_J(p_J) \left|\det(\ra_{J^o})\right|}\,.
\end{aligned}
\end{equation}
In the second line, we use the previous residue formula in \eqref{JorNeqRII} for ${N=r}$, where ${\ra_{J^o}}$ is the ${r\times r}$ matrix spanned by rows ${\ra_{\ell1},\ldots,\ra_{\ell_r}}$, ${J^o=(\ell_1 \cdots \ell_r)}$.
On the other hand, up to a sign, the ratio on the right in \eqref{JorNeqRIII} is the local residue of $\omega$ at the point ${p_J}$, where $\omega$ has a simple pole.  So more geometrically,
\begin{equation}\label{JorNeqRIV}
\frac{1}{\left(2\pi i\right)^r}\int_\h \omega \,=\, (-1)^r \sum_{\#J=N-r} \delta_{\ra_{J^o},-\rho} \cdot \sgn(\det(df_{p_J}))\cdot\Res_{p_J}(\omega)\,.
\end{equation}
Following the notation in \eqref{signrho}, ${\sgn(\det(df_{p_J}))=\pm 1}$ depending upon whether the orientation at $p_J$ induced by ${df_{\ell_1}\^\cdots\^df_{\ell_r}|_\h}$ agrees with the fixed orientation on $\h$.  By convention, we always order the indices ${J^o=(\ell_1\cdots\ell_r)}$ from least to greatest, and we use the same ordering to define the sign of the local residue of $\omega$ at $p_J$.  With this convention, the dependence of sign on the order of multi-indices in $J^o$ cancels.\footnote{The overall sign $(-1)^r$ in \eqref{JorNeqRIV} is an unfortunate consequence of our convention that ${\Im(\rb_j)>0}$ be positive as opposed to negative.  As the latter convention is motivated by analytic features of the double-sine function $s_b(z)$, whose definition is fixed, we have decided to accept the inelegant sign on the right in \eqref{JorNeqRIV}.}

The crucial feature of either \eqref{JorNeqRIII} or \eqref{JorNeqRIV} is again the prefactor ${\delta_{\ra_{J^o},-\rho}}$.  This prefactor states that a given pole $p_J$ of $\omega$ contributes to the residue sum exactly when $-\rho$ lies in the positive cone ${\R_+[\ra_{\ell_1},\ldots,\ra_{\ell_r}]\subset\h^*}$ spanned by the differentials ${df_{\ell_1}\equiv\ra_{\ell_1},\cdots,df_{\ell_r}\equiv\ra_{\ell_r}}$ of the linear functions which vanish at $p_J$.  If $\omega$ factorizes with respect to a given set of linear coordinates on $\h$, as in \eqref{ItIntJorNeqR} but allowing more linear factors in each denominator, then the criterion provided by ${\delta_{\ra_{J^o},-\rho}}$ is clearly correct.  However, for ${N>r}$ not every multivariate denominator can be factorized by a suitable choice of linear coordinates, so the residue formulas in \eqref{JorNeqRIII} and \eqref{JorNeqRIV} carry non-trivial geometric content.

\paragraph{(c) Mittag-Leffler expansion.}

The integrand for $Z^{\rm uv}_{S^3}$ in \eqref{ZuvCmbV} has poles
along an infinite -- as opposed to finite -- union of hyperplanes in ${\h_\C\simeq\C^r}$.  This case can be treated using the Mittag-Leffler expansion (see eg.\,Theorem $4$ in Ch.\,5 of \cite{Ahlfors:1979}) in place of the finite partial-fractions decomposition.  

Briefly, let ${w_n \in \C}$ be a sequence of points in the complex plane so that ${|w_n|\to\infty}$ (the sequence has no accumulation point), and let ${r_n\in\C}$ be any other sequence of complex numbers.  The Mittag-Leffler theorem states that there exists a meromorphic function $F(w)$ with a simple pole at each $w_n$ having specified residue $r_n$.\footnote{When $F(w)$ has poles of higher-order, the Mittag-Leffler expansion extends in the natural way.  We make the assumption about simple poles only for notational convenience.}  Each such function has a convergent expansion
\begin{equation}\label{MittagLefE}
F(w) \,=\, \sum_{n=1}^\infty \left[\frac{r_n}{w-w_n} \,+\, q_n(w)\right] \,+\, h(w)\,,
\end{equation}
where $\{q_n(w)\}_{n=1,\ldots,\infty}$ is a sequence of polynomials depending upon $(w_n,r_n)$, necessary for uniform convergence on compact sets, and $h(w)$ is any entire function, the choice of which obviously does not alter the poles and residues of $F$.  For our application, the double-sine function $s_b(w)$ for ${b^2\notin\Q}$ is such a meromorphic function and so in particular has a Mittag-Leffler expansion.

Let ${\ra_1,\ldots,\ra_N\in\h^*}$ be a finite set of weights, and introduce meromorphic functions ${F_1(w),\ldots,F_N(w)}$ with expansions as in \eqref{MittagLefE}.  By analogy to the meromorphic form with polynomial denominator in \eqref{MeromHRGG}, we consider 
\begin{equation}
\omega \,= \left[F_1\big(\langle\ra_1,z\rangle\big) \cdots F_N\big(\langle\ra_N,z\rangle\big)\right] g(z) \, dz_1 \^ \cdots \^ dz_r\,,\qquad\qquad z\in\h_\C\,,
\end{equation}
where $g(z)$ is an entire function decaying rapidly at infinity on $\BH_\rho$ as before.  Explicitly, the product of ${F_1,\ldots,F_N}$ has an expansion
\begin{equation}\label{MitLefII}
\begin{aligned}
&F_1\big(\langle\ra_1,z\rangle\big) \cdots F_N\big(\langle\ra_N,z\rangle\big) =\!\!\!\!\!\sum_{n_1,\ldots,n_N=1}^\infty \left(\left[\frac{r_{n_1}}{\langle\ra_1,z\rangle-w_{n_1}} \,+\, q_{n_1}\big(\langle\ra_1,z\rangle\big)\right] \,+\, h_1\big(\langle\ra_1,z\rangle\big)\right)\\
&\qquad\qquad\qquad\qquad\qquad\times \cdots \times\left(\left[\frac{r_{n_N}}{\langle\ra_N,z\rangle-w_{n_N}} \,+\, q_{n_N}\big(\langle\ra_N,z\rangle\big)\right] \,+\, h_N\big(\langle\ra_N,z\rangle\big)\right).
\end{aligned}
\end{equation}
Collecting terms, the product can be rewritten succinctly
\begin{equation}
F_1\big(\langle\ra_1,z\rangle\big) \cdots F_N\big(\langle\ra_N,z\rangle\big) =\!\!\sum_{n_1,\ldots,n_N=1}^\infty \frac{H_{n_1 \cdots n_N}(z)}{\left(\langle\ra_1,z\rangle-w_{n_1}\right) \cdots \left(\langle\ra_N,z\rangle-w_{n_N}\right)},
\end{equation}
where each $H_{n_1 \cdots n_N}(z)$ is an entire function, made from the polynomials $\{q_{n_1},\ldots,q_{n_N}\}$ and the entire functions $\{h_1,\ldots,h_N\}$ which appear in \eqref{MitLefII}.

The meromorphic form $\omega$ thus admits its own expansion 
\begin{equation}\label{Mittagom}
\omega \,=\, \sum_{n_1,\ldots,n_N=1}^\infty \frac{H_{n_1\cdots n_N}(z)\,g(z)\,dz_1\^\cdots\^dz_r}{\left(\langle\ra_1,z\rangle-w_{n_1}\right) \cdots \left(\langle\ra_N,z\rangle-w_{n_N}\right)}\,.
\end{equation}
Each term in the series has the same structure considered previously for the inductive Case\,(b).  Provided the new numerator ${H_{n_1 \cdots n_N}(z)\,g(z)}$ decays sufficiently rapidly for all ${n_1,\ldots,n_N}$, we can apply the residue formula in \eqref{JorNeqRIV} term-by-term, 
\begin{equation}\label{JorNeqRV}
\frac{1}{\left(2\pi i\right)^r}\int_\h \omega \,=\, (-1)^r \!\!\sum_{n_1,\ldots,n_N=1}^\infty \sum_{\#J=N-r} \delta_{\ra_{J^o},-\rho} \cdot \sgn(\det(df_{p_{J;\ul{n}}}))\cdot\Res_{p_{J;\ul{n}}}(\omega)\,,
\end{equation}
where ${p_{J;\ul{n}}\equiv p_{J;n_1,\ldots,n_N}}$ is any point at which $r$ polar hyperplanes intersect,
\begin{equation}
\langle\ra_{\ell_1}, p_{J;\ul{n}}\rangle - w_{n_{\ell_1}} \,=\, \cdots \,=\, \langle\ra_{\ell_r}, p_{J;\ul{n}}\rangle - w_{n_{\ell_r}} \,=\, 0\,,\qquad J^o = (\ell_1 \cdots \ell_r)\,.
\end{equation}

To determine whether such a point contributes to the residue sum, we use the same local criterion from Case\,(b).  Multiplying each of ${\ra_{\ell_1},\ldots,\ra_{\ell_r}}$ by $\pm 1$ as necessary, we arrange by convention that ${\Im(w_{n_{\ell_1}}),\ldots,\Im(w_{n_{\ell_r}})<0}$ for the term in question in the expansion \eqref{Mittagom}.  With that choice of signs, the residue contributes when $-\rho$ lies inside the positive cone $\R_+[\ra_{\ell_1},\ldots,\ra_{\ell_r}]$.  

For residue computations involving $s_b(z)$, the sign of ${\Im(w_n)<0}$ is negative for all $n$, so a term-by-term ``renormalization'' of the signs of ${\ra_1,\ldots,\ra_N}$ is unnecessary to apply the admissibility criterion ${-\rho\in\R_+[\ra_1,\ldots,\ra_{\ell_r}]}$.

\paragraph{Example:~Supersymmetric QCD.}

Let us illustrate the admissibility criterion for the residues which contribute to the partition function $Z^{\rm uv}_{S^3}$ in the example of $SU(N)$ SQCD with ${N_{\rm f}=1}$.  We work in the basis of fundamental weights $\{\hat\omega^1,\ldots,\hat\omega^{N-1}\}$ from Appendix \ref{se:LieAlgConvention}, so that the Weyl vector is the sum ${\rho = \hat\omega^1 + \cdots + \hat\omega^{N-1}}$, and ${\h\simeq\R^{N-1}}$ is oriented by ${\hat\omega^1\^\cdots\^\hat\omega^{N-1}}$.  

The matter representation ${\Lambda_{\rm uv}={\bf N}\oplus\overline{\bf N}}$ has $2N$ distinct weights 
\begin{equation}\label{SUNwts}
\Delta_{{\bf N}\oplus\overline{\bf N}} \,=\, \left\{\pm\hat\omega^1,\,\pm\left(\hat\omega^1-\hat\omega^2\right),\,\cdots,\,\pm\left(\hat\omega^{N-2}-\hat\omega^{N-1}\right),\,\pm\hat\omega^{N-1}\right\}\,.
\end{equation}
To determine which residues contribute to the integral \eqref{ZuvCmbV}
for $SU(N)$ SQCD with one flavor, we seek to classify all
$(N-1)$-tuples $(\ra_1,\ldots,\ra_{N-1})$ in $\Delta_{{\bf
    N}\oplus\overline{\bf N}}$ such that 

\medskip\noindent
\begin{tabular}{l l}
\fbox{admissibility} & 
{1.\quad$(\ra_1,\ldots,\ra_{N-1})$ is a positively-oriented basis for
  $\h$, and}\\[1 ex]
&{2.\quad$-\rho$ lies in the positive cone $\R_+[\ra_1,\ldots,\ra_{N-1}]$.}
\end{tabular}

\medskip\noindent
The orientation condition is cosmetic and ensures that the sign of $\det(f_{p_{J;\ul{n}}})$ in \eqref{JorNeqRV} is positive for every residue.

For each $(N-1)$-tuple $(\ra_1,\ldots,\ra_{N-1})$ that obeys these conditions, one sums over an infinite set of residues at the points ${\sigma_\bullet\in\h_\C}$ where 
\begin{equation}\label{QarukHyp}
\frac{\langle\ra_j,\sigma_\bullet\rangle}{2\pi} \,=\, -i \, u_j \, b \,-\, i \, v_j \, b^{-1} - \mu_{{\bf N},\overline{\bf N}}\,,\qquad u_j, v_j\ge 0\,,\quad j=1,\ldots,N-1\,.
\end{equation}
Here we allow distinct real masses $\mu_{\bf N}$ and $\mu_{\overline{\bf N}}$ for quarks and anti-quarks.  Whether $\mu_{\bf N}$ or $\mu_{\overline{\bf N}}$ appears in the hyperplane equation \eqref{QarukHyp} is determined by whether the weight $\ra_j$ comes from the fundamental or anti-fundamental representation of $SU(N)$.  To avoid even more cumbersome notation, this distinction is indicated only schematically in \eqref{QarukHyp}.

For instance, to make contact with the previous analysis in Section \ref{JordDRk2}, for $SU(3)$
\begin{equation}\label{SU3wts}
\Delta_{{\bf 3}\oplus\overline{\bf 3}} \,=\, \left\{\pm\hat\omega^1,\,\pm\left(\hat\omega^1-\hat\omega^2\right),\,\pm\hat\omega^2\right\}.
\end{equation}
From this set of weights, precisely three admissible doubles ${\triangle\equiv(\ra_1,\ra_2)}$ exist,
\begin{equation}\label{SU3admd}
\begin{aligned}
\triangle_1 \,&=\, \left(-\hat\omega^1+\hat\omega^2,\,-\hat\omega^2\right),\\
\triangle_2 \,&=\, \left(-\hat\omega^1,\,-\hat\omega^2\right),\\
\triangle_3 \,&=\, \left(-\hat\omega^1,\,\hat\omega^1-\hat\omega^2\right),\\
\end{aligned}
\end{equation}
such that $\triangle$ is a positive basis for $\h$, and ${-\rho=-\hat\omega^1-\hat\omega^2}$ lies in the cone $\R_+[\triangle]$.  The identical list of doubles, derived using reasoning specific to two dimensions, appears at the top in Table \ref{WtH1H2Tab}.  

For $SU(3)$ SQCD with ${N_{\rm f}>1}$ flavors, the weights in \eqref{SU3wts} are decorated with flavor indices, but the same admissibility criterion holds.  So one obtains the same list of admissible doubles in \eqref{SU3admd}, such that $\ra_1$ and $\ra_2$ are decorated with arbitrary flavor indices.  For each choice of $\Delta_1$, $\Delta_2$, or $\Delta_3$ and values of flavor indices, one performs an infinite sum over residues at the points in \eqref{QarukHyp}.

More generally, for $SU(N)$ SQCD with ${N_{\rm f}=1}$, the set
${\Delta_{{\bf N}\oplus\overline{\bf N}}}$ contains $N$ pairs of
weights.  Modulo signs, any subset of $(N-1)$ weights from the set in
\eqref{SUNwts} provides a basis for ${\h\simeq \R^{N-1}}$.  There are
$N$ such subsets; fix one.  Including the choice of signs on each basis element, we divide $\h$ into $2^{N-1}$ chambers, precisely one of which contains $-\rho$ and hence is admissible.  Thus we obtain $N$ admissible $(N-1)$-tuples ${\triangle\equiv(\ra_1,\ldots,\ra_{N-1})}$.  With a little thought, they are given by 
\begin{equation}\label{SUNadmd}
\begin{aligned}
\triangle_1 \,&=\left(-1\right)^{N-1}\cdot\left(-\hat\omega^1+\hat\omega^2,\,-\hat\omega^2+\hat\omega^3,\,-\hat\omega^3+\hat\omega^4,\,\cdots,\,-\hat\omega^{N-2}+\hat\omega^{N-1},\,-\hat\omega^{N-1}\right),\\
\triangle_2 \,&=\left(-1\right)^{N-1}\cdot\left(-\hat\omega^1,\,-\hat\omega^2+\hat\omega^3,\,-\hat\omega^3+\hat\omega^4,\,\cdots,-\hat\omega^{N-2}+\hat\omega^{N-1},-\hat\omega^{N-1}\right),\\
\triangle_3 \,&=\left(-1\right)^{N-1}\cdot\left(-\hat\omega^1,\,\hat\omega^1-\hat\omega^2,\,-\hat\omega^3+\hat\omega^4,\,\cdots,\,-\hat\omega^{N-2}+\hat\omega^{N-1},\,-\hat\omega^{N-1}\right),\\
&\,\,\vdots\\
\triangle_a \,&=\left(-1\right)^{N-1}\cdot\left(-\hat\omega^1,\,\hat\omega^1-\hat\omega^2,\,\cdots,\,\hat\omega^{a-2}-\hat\omega^{a-1},\,-\hat\omega^{a}+\hat\omega^{a+1},\,\cdots,\,-\hat\omega^{N-1}\right),\\
&\,\,\vdots\\
\triangle_{N-1} \,&= \left(-1\right)^{N-1}\cdot\left(-\hat\omega^1,\,\hat\omega^1-\hat\omega^2,\,\cdots,\,\hat\omega^{N-3}-\hat\omega^{N-2},\,-\hat\omega^{N-1}\right),\\
\triangle_N \,&=\left(-1\right)^{N-1}\cdot\left(-\hat\omega^1,\,\hat\omega^1-\hat\omega^2,\,\cdots,\,\hat\omega^{N-3}-\hat\omega^{N-2},\,\hat\omega^{N-2}-\hat\omega^{N-1}\right).
\end{aligned}
\end{equation}
The sign $(-1)^{N-1}$ indicates whether the basis is positively- or negatively-oriented with respect to $\h$.  Each tuple specifies a countable set of residues which must be summed to calculate $Z^{\rm uv}_{S^3}$.  

Similarly, the admissible tuples for $SU(N)$ SQCD with ${N_{\rm f}>1}$ are given by the same tuples of weights in \eqref{SUNadmd}, decorated by a choice of flavor for each weight.

As the SQCD example illustrates, the admissibility criterion is easy
to apply in practice.  We use it to perform some explicit residue computations in Section \ref{RkTwoExs}.

\subsection{Local Residues, Lattice Sums, and Holomorphic Blocks}\label{LatticeSums}

We now apply the results in Sections \ref{JordDRk2} and \ref{JordHRGG}
to evaluate the local residues which contribute to holomorphic blocks
for the partition function $Z_{S^3}^{\rm uv}$.  Like the rank-one
examples in Section \ref{RankOne}, the main goals are to check and
determine the general structure of the block-decomposition
\begin{equation}
Z_{S^3}^{\rm uv} \,=\, \sum_{m,n\,\in\,\CI} \RG_{m n}\,
\RB^m(\Rq,\Rx)\,\wt\RB^n(\wt\Rq,\wt\Rx)\,.
\end{equation}

\paragraph{Admissible Poles of the Gauge Theory Integrand.}

We start with a description of the poles in ${\h_\C\simeq\C^r}$ whose
residues contribute to $Z_{S^3}^{\rm uv}$.  

Fix an admissible $r$-tuple ${\triangle_\bullet = \{\beta_1,\ldots,\beta_r\}}$
of weights which appear in the total matter representation
$\Lambda_{\rm uv}$.  Later in Section \ref{RkTwoExs}, we classify
admissible $r$-tuples for some easy examples of $\Lambda_{\rm uv}$.  Because $\triangle_\bullet$ is a basis for $\h^*$, the weights in $\triangle_\bullet$ generate a finite-index sublattice of the weight lattice $\Lambda_{\rm W}$,
\begin{equation}\label{Lamtrabul}
\Lambda_{\triangle_\bullet} \,:=\, \Z[\beta_1,\ldots,\beta_r] \,\subseteq\,\Lambda_{\rm W}\,.
\end{equation}
As reviewed in Appendix \ref{se:LieAlgConvention}, the coroot lattice ${\Lambda^{\rm R}\simeq\Lambda_{\rm W}^*\subset\h}$
is canonically dual over $\Z$ to the weight lattice $\Lambda_{\rm W}$.  We introduce
similarly the lattice dual to $\Lambda_{\triangle_\bullet}$, 
\begin{equation}\label{DualAdS}
\Lambda^{\triangle_\bullet} \,:=\, \big\{t\in\h\,\big|\,\langle\beta,t\rangle\in\Z\,,\forall\beta\in\Lambda_{\triangle_\bullet}\big\}\subset\h\,.
\end{equation}
Trivially, the coroot lattice ${\Lambda^{\rm R} \subseteq
  \Lambda^{\triangle_\bullet}}$ is a sublattice of finite-index in
$\Lambda^{\triangle_\bullet}$, and the quotient ${\Lambda^{\triangle_\bullet}/\Lambda^{\rm R}}$ is a finite abelian group
\begin{equation}\label{CharGp}
\Gamma_{\triangle_\bullet} \,:=\,
\Lambda^{\triangle_\bullet}/\Lambda^{\rm R} \,\simeq\,\Lambda_{\rm
  W}/\Lambda_{\triangle_\bullet}\,.
\end{equation}
The latter isomorphism follows by duality.  See Figure \ref{AdmLattices} for an example of the lattices
$\Lambda_{\triangle_\bullet}$ and $\Lambda^{\triangle_\bullet}$ for the group $SU(3)$ and a suitable
choice for the admissible set $\triangle_\bullet$.
\iffigs
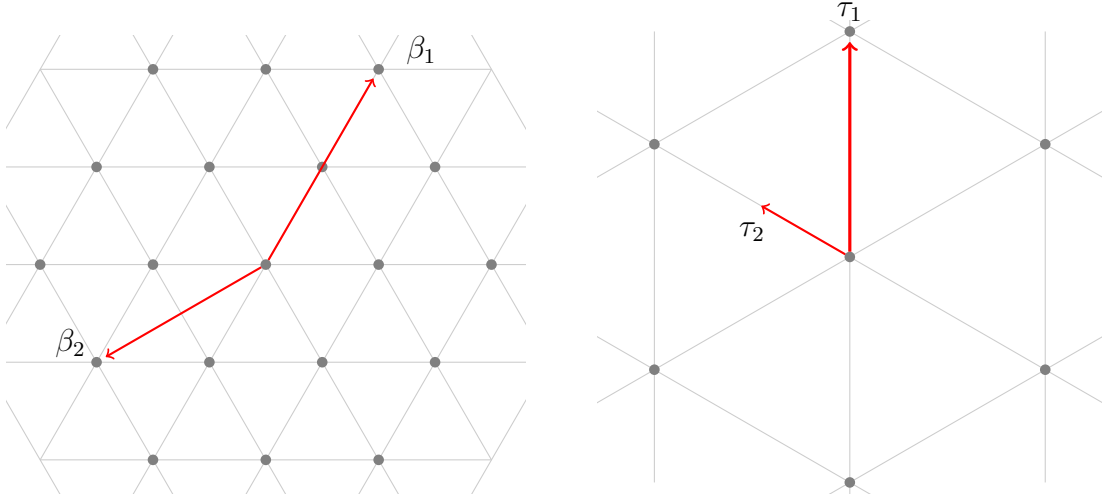
\begin{figure}[t!]
	\centering
	\begin{tikzpicture}[x=1.5cm,y=1.5cm] 
	
	\coordinate (0;0) at (0,0); 
	\foreach \c in {1,...,4}{%  
		\foreach \i in {0,...,5}{% 
			\pgfmathtruncatemacro\j{\c*\i}
			\coordinate (\c;\j) at (60*\i:\c);  
		} }
		\foreach \i in {0,2,...,10}{% 
			\pgfmathtruncatemacro\j{mod(\i+2,12)} 
			\pgfmathtruncatemacro\k{\i+1}
			\coordinate (2;\k) at ($(2;\i)!.5!(2;\j)$) ;}
		
		\foreach \i in {0,3,...,15}{% 
			\pgfmathtruncatemacro\j{mod(\i+3,18)} 
			\pgfmathtruncatemacro\k{\i+1} 
			\pgfmathtruncatemacro\l{\i+2}
			\coordinate (3;\k) at ($(3;\i)!1/3!(3;\j)$)  ;
			\coordinate (3;\l) at ($(3;\i)!2/3!(3;\j)$)  ;
		}
		
		\foreach \i in {0,4,...,20}{% 
			\pgfmathtruncatemacro\j{mod(\i+4,24)} 
			\pgfmathtruncatemacro\k{\i+1} 
			\pgfmathtruncatemacro\l{\i+2}
			\pgfmathtruncatemacro\m{\i+3} 
			\coordinate (4;\k) at ($(4;\i)!1/4!(4;\j)$)  ;
			\coordinate (4;\l) at ($(4;\i)!2/4!(4;\j)$) ;
			\coordinate (4;\m) at ($(4;\i)!3/4!(4;\j)$) ;
		}

		\begin{scope}
		
		\clip ($(2;0)+(2;3)+(.3,.3)$)--($(2;6)+(2;3)+(-.3,.3)$)--($(2;6)+(2;9)+(-.3,-.3)$)--($(2;0)+(2;9)+(.3,-.3)$)--($(2;0)+(2;3)+(.3,.3)$);
		
		\foreach \i in {0,...,6}{% 
			\pgfmathtruncatemacro\k{\i}
			\pgfmathtruncatemacro\l{15-\i}
			\draw[thin,gray!40] (3;\k)--(3;\l);
			\pgfmathtruncatemacro\k{9-\i} 
			\pgfmathtruncatemacro\l{mod(12+\i,18)}   
			\draw[thin,gray!40] (3;\k)--(3;\l); 
			\pgfmathtruncatemacro\k{12-\i} 
			\pgfmathtruncatemacro\l{mod(15+\i,18)}   
			\draw[thin,gray!40] (3;\k)--(3;\l);} 
		
		\end{scope}

		\fill [gray] (0;0) circle (2pt);
		\foreach \c in {1,...,2}{%
			\pgfmathtruncatemacro\k{\c*6-1}    
			\foreach \i in {0,...,\k}{% 
				\fill [gray] (\c;\i) circle (2pt);}}  
		
		\draw[->,red,thick,shorten >=4pt,shorten <=2pt](0;0)--(2;7);
		\draw[->,red,thick,shorten >=4pt,shorten <=2pt](0;0)--(2;2);
		
		\node [ above right] at ($(2;2)+(.15,-.05)$) {$\beta_1$};
		\node [ above left] at ($(2;7)+(0,-.05)$) {$\beta_2$};
		
		\end{tikzpicture} 
		\qquad
		\begin{tikzpicture}[x=1.5cm,y=1.5cm] 
	
		\coordinate (0;0) at (0,0); 
		\foreach \c in {1,...,4}{%  
			\foreach \i in {0,...,5}{% 
				\pgfmathtruncatemacro\j{\c*\i}
				\coordinate (\c;\j) at (60*\i+30:\c);  
			} }
			\foreach \i in {0,2,...,10}{% 
				\pgfmathtruncatemacro\j{mod(\i+2,12)} 
				\pgfmathtruncatemacro\k{\i+1}
				\coordinate (2;\k) at ($(2;\i)!.5!(2;\j)$) ;}
			
			\foreach \i in {0,3,...,15}{% 
				\pgfmathtruncatemacro\j{mod(\i+3,18)} 
				\pgfmathtruncatemacro\k{\i+1} 
				\pgfmathtruncatemacro\l{\i+2}
				\coordinate (3;\k) at ($(3;\i)!1/3!(3;\j)$)  ;
				\coordinate (3;\l) at ($(3;\i)!2/3!(3;\j)$)  ;
			}
			
			\foreach \i in {0,4,...,20}{% 
				\pgfmathtruncatemacro\j{mod(\i+4,24)} 
				\pgfmathtruncatemacro\k{\i+1} 
				\pgfmathtruncatemacro\l{\i+2}
				\pgfmathtruncatemacro\m{\i+3} 
				\coordinate (4;\k) at ($(4;\i)!1/4!(4;\j)$)  ;
				\coordinate (4;\l) at ($(4;\i)!2/4!(4;\j)$) ;
				\coordinate (4;\m) at ($(4;\i)!3/4!(4;\j)$) ;
			}

			\begin{scope}
			
			\clip ($(2;2)+(2;5)+(-.5,.1)$)--($(2;8)+(2;5)+(-.5,-.1)$) -- ($(2;8)+(2;11)+(.5,-.1)$)-- ($(2;2)+(2;11)+(.5,.1)$);
			
			\foreach \i in {1,3,5}{% 
				\pgfmathtruncatemacro\k{\i}
				\pgfmathtruncatemacro\l{15-\i}
				\draw[thin,gray!40] (3;\k)--(3;\l);
				\pgfmathtruncatemacro\k{9-\i} 
				\pgfmathtruncatemacro\l{mod(12+\i,18)}   
				\draw[thin,gray!40] (3;\k)--(3;\l); 
				\pgfmathtruncatemacro\k{12-\i} 
				\pgfmathtruncatemacro\l{mod(15+\i,18)}   
				\draw[thin,gray!40] (3;\k)--(3;\l);} 
			
			\end{scope}
			
			\fill [gray] (0;0) circle (2pt);
			\foreach \c in {2,...,2}{%
				\pgfmathtruncatemacro\k{\c*6-1}    
				\foreach \i in {0,2,...,\k}{% 
					\fill [gray] (\c;\i) circle (2pt);}}  
			
			\draw[->,red,very thick,shorten >=4pt,shorten <=2pt](0;0)--(2;2);
			\draw[->,red,thick,shorten >=4pt,shorten <=2pt](0;0)--(1;2);
			
			\node [ above ] at ($(2;2)+(0,0)$) {$\tau_1$};
			\node [ below ] at ($(1;2)+(0,-.1)$) {$\tau_2$};
			
			\end{tikzpicture}   
			
	\caption{Examples of the dual lattices
          ${\Lambda_{\triangle_\bullet}=\Z[\beta_1,\beta_2]\subset\h^*}$ and
          ${\Lambda^{\triangle_\bullet}=\Z[\tau_1,\tau_2]\subset\h}$ for the
          admissible double ${\triangle_\bullet=\{\beta_1,\beta_2\}}$.  The
          grey dots
          on the left and right indicate the weight
          $\Lambda_{\rm W}$ and
          coroot $\Lambda^{\rm R}$ lattices of $SU(3)$, respectively.  Clearly
          ${\Lambda_{\triangle_\bullet}\subseteq\Lambda_{\rm W}}$ and
          ${\Lambda^{\rm R}\subseteq\Lambda^{\triangle_\bullet}}$.}\label{AdmLattices}
\end{figure}
\fi

Let
$\{\tau_1,\ldots,\tau_r\}$ be integral generators of
$\Lambda^{\triangle_\bullet}$ which are dual to the 
weights in the admissible set $\triangle_\bullet$,
\begin{equation}\label{TauGens}
\langle\beta_m,\tau_n\rangle \,=\, \delta_{m n}\,,\qquad m,n=1,\ldots,r\,.
\end{equation}
For each generator, there exists a least positive integer ${d_m>0}$ so that 
\begin{equation}\label{TauDegs}
d_m\,\tau_m\,\in\,\Lambda^{\rm R}\,,\qquad\qquad d_m\,\in\,\Z\,,\qquad\qquad m=1,\ldots,r\,.
\end{equation}
In terms of the generators $\{\tau_1,\ldots,\tau_r\}$, any element
${t\in\Lambda^{\triangle_\bullet}}$ can be expressed as a sum 
\begin{equation}\label{Cosett}
\begin{aligned}
t \,&=\, \zeta_1 \tau_1 \,+\, \cdots \,+\, \zeta_r \tau_r \,+\, h\,,\\
\left(\zeta_1,\ldots,\zeta_r\right)&\in\Z/d_1\Z
\times\cdots\times\Z/d_r\Z\,,\qquad h \in
\Z\big[d_1\tau_1,\cdots,d_r\tau_r\big]\subset\Lambda^{\rm R}\,.
\end{aligned}
\end{equation}
Note that $\Gamma_{\triangle_\bullet}$ in \eqref{CharGp} is usually a
strict quotient of ${\Z/d_1\Z
\times\cdots\times\Z/d_r\Z}$, due to non-trivial relations between the generators
$\{\tau_1,\ldots,\tau_r\}$ modulo $\Lambda^{\rm R}$.

This formalism provides the correct structure to classify admissible
poles of the integrand in \eqref{ZuvCmbV}.  Such a polar point
${\sigma_\bullet\in\h_\C}$ satisfies the linear equations
\begin{equation}\label{LinPole}
\begin{aligned}
\frac{\langle\beta_1^{[j_1]},\sigma_\bullet\rangle}{2\pi} \,&=\,
-i\,u_1\, b - i\,v_1\,b^{-1}\,-\,\mu_{j_1}\,,\qquad\qquad u_1,v_1\ge 0\,,\\
&\hspace{.5em}\vdots\\
\frac{\langle\beta_r^{[j_r]},\sigma_\bullet\rangle}{2\pi} \,&=\,
-i\,u_r\, b - i\,v_r\,b^{-1}-\mu_{j_r}\,,\qquad\qquad u_r,v_r\ge 0\,,
\end{aligned}
\end{equation}
for positive integers $u_m,v_m \in \Z_{\ge 0}$, $m=1,\ldots,r$, and mass
parameters ${\mu_{j_1},\ldots,\mu_{j_r}}$.  Here
${j_1,\ldots,j_r=1,\ldots,n}$ are flavor indices for the total matter
representation $\Lambda_{\rm uv}=\oplus_{j=1}^n[\lambda_j]$.  We
also attach (not-necessarily distinct) flavor indices to the admissible weights
$\beta_1^{[j_1]},\ldots,\beta_r^{[j_r]}\in\triangle_\bullet$ to indicate from
which summand the weight originates.  

In terms of the dual basis $\{\tau_1,\ldots,\tau_r\}$ for
${\Lambda^{\triangle_\bullet}\subset\h}$, the linear equations
\eqref{LinPole} are solved by 
\begin{equation}
\sigma_\bullet \,=\, -2\pi
i\sum_{m=1}^r\left(u_m\,b \,+\,
  v_m\,b^{-1}\,-\, i\,\mu_{j_m}\right)\tau_m\,.
\end{equation}
Equivalently, since ${\sum u_m\tau_m}$ and ${\sum v_m\tau_m}$ are
elements of the lattice $\Lambda^{\triangle_\bullet}$ refining $\Lambda^{\rm
  R}$, these elements can be decomposed
via \eqref{Cosett} as 
\begin{equation}\label{TauGensII}
\begin{aligned}
\sum_{m=1}^r u_m\,\tau_m \,&=\, \sum_{m=1}^r \zeta_m\,\tau_m \,+\,
{\bf M}\,,\qquad\qquad {\bf M},{\bf N}\in\Z_{\ge
  0}\big[d_1\tau_1,\cdots,d_r\tau_r\big]\subset\Lambda^{\rm R}\,,\\
\sum_{m=1}^r v_m\,\tau_m \,&=\, \sum_{m=1}^r \eta_m\,\tau_m \,+\, {\bf
  N}\,,\qquad\qquad \zeta,\eta\in\Z/d_1\Z
\times\cdots\times\Z/d_r\Z\,.
\end{aligned}
\end{equation}
By analogy to the notation \eqref{CONG} in rank-one,
we introduce the shorthand for the characteristic elements on the
right in \eqref{TauGensII},
\begin{equation}
\mathbf{r} \,=\, \sum_{m=1}^r \zeta_m\,\tau_m\,,\qquad\qquad
\mathbf{s} \,=\, \sum_{m=1}^r \eta_m\,\tau_m\,.
\end{equation}
In this notation,
\begin{equation}\label{Bultsgm}
\sigma_\bullet \,=\, -2\pi
i\left(\mathbf{r}\,b \,+\,  \mathbf{s}\,b^{-1} \,+\,{\bf M}\,b \,+\, {\bf
    N}\,b^{-1}\right) -\, 2\pi\sum_{m=1}^r \mu_{j_m}\tau_m\,.
\end{equation}
The sum over admissible poles for $\triangle_\bullet$ thereby reduces
to a finite double-sum over characteristics
$\mathbf{r},\mathbf{s}\in (\Z/d_1\Z)\tau_1
\times\cdots\times(\Z/d_r\Z)\tau_r$ and an infinite double-sum over
coroots 
$\mathbf{M},\mathbf{N}\in\Z_{\ge 0}[d_1\tau_1,\cdots,d_r\tau_r]$.

The decomposition in \eqref{Bultsgm} displays the integral structure in the solution $\sigma_\bullet$.
Automatically for any weight ${\beta\in\Lambda_{\rm W}}$, we have integral pairings $\langle\beta,{\bf
    M}\rangle,\langle\beta,{\bf N}\rangle\in\Z$, while
${\langle\beta,\mathbf{r}\rangle,\langle\beta,\mathbf{s}\rangle\in\Q}$
are only rational in general.  But for the special weights ${\beta\in\triangle_\bullet}$ in the
admissible set, the pairings $\langle\beta,\mathbf{r}\rangle,\langle\beta,\mathbf{s}\rangle\in\Z$
are integral by \eqref{TauGens} and \eqref{TauGensII}.

\paragraph{General Formula for Local Residues.}

We now evaluate the local residue of the gauge theory integrand at the
point ${\sigma_\bullet\in\h_\C}$ in \eqref{Bultsgm}.  By way of
notation, introduce
\begin{equation}\label{BigZresuv}
\begin{aligned}
\RZ^{{\triangle_\bullet}}_{u,v} &= \Res\!\left[
		\e{\!-\frac{i\,k}{4\pi}\Tr(\sigma^2)}\cdot
		\prod_{\alpha\in\Delta_+}\left[4\,\sinh\!\left(\frac{b
			\langle\alpha,\sigma\rangle}{2}\right)\sinh\!\left(\frac{
			\langle\alpha,\sigma\rangle}{2
			b}\right)\right]\times
              \right.
		\\
		&
		\hspace{12em}
                \left.
		\times
		\prod_{j=1}^{n}
		\prod_{\beta \in
			\Delta_j}
		s_b\!\left(\frac{\langle\beta,\sigma\rangle}{2\pi}\,+\,\mu_j\right)
		\right]_{\sigma = \sigma_\bullet}\,.
\end{aligned}
\end{equation}
The residue is labelled by the admissible set ${\triangle_\bullet}$ of weights and the
positive integers ${u\equiv (u_1,\ldots,u_r)}$ and ${v\equiv
  (v_1,\ldots,v_r)}$.  Though the ultraviolet partition function \eqref{ZuvCmbV} does
not include a microscopic Chern-Simons term, such a term is generally
induced after heavy matter multiplets are integrated-out, according to
Section \ref{LevelZero}.  To account 
for the latter effect, we include an explicit Gaussian when
we evaluate the residue $\RZ^{\triangle_\bullet}_{u,v}$ in \eqref{BigZresuv}.

Recall the definitions of the block variables
\begin{equation}\label{BlckVarsIV}
\begin{matrix}
\begin{aligned}
\Rq \,&=\, \e{\!2\pi i b^2}\,,\\
\Rx_j \,&=\, \e{\!2\pi \mu_j b}\,,
\end{aligned} \qquad\quad&\quad\qquad 
\begin{aligned}
\wt\Rq \,&=\,\e{\!2\pi i/b^2}\,,\\
\wt\Rx_j \,&=\, \e{\!2\pi \mu_j/b}\,,\qquad\qquad j\,=\,1,\ldots,n\,.
\end{aligned}
\end{matrix}
\end{equation}
The entire factors on the first line of \eqref{BigZresuv} can be
immediately expressed in these variables as follows.

\smallskip\noindent{\sl Chern-Simons term.}\quad The Lie algebra norm of $\sigma_\bullet$ in \eqref{Bultsgm} is given by 
\begin{equation}
\begin{aligned}
\frac{\left(\sigma_\bullet,\sigma_\bullet\right)}{4\pi^2} \,&=\, -b^2\,
|\mathbf{r}+{\bf M}|^2 \,-\, b^{-2}\, |\mathbf{s}+{\bf N}|^2\,+\,\\ 
&+\,2 i b
\left(\mathbf{r}+{\bf M}, \sum_{m=1}^r \mu_{j_m} \tau_m\right) \,+\, 2 i
b^{-1} \left(\mathbf{s} + {\bf N}, \sum_{m=1}^r\mu_{j_m} \tau_m\right)+\,\\
&+\, \big|\sum_{m=1}^r\mu_{j_m}\tau_m\big|^2 \,-\, 2 \left(\mathbf{r}+{\bf M},\mathbf{s}+{\bf N}\right),
\end{aligned}
\end{equation}
where we use the shorthand ${|\,\cdot\,|^2 \equiv -\Tr(\,\cdot\,,\cdot\,)}$.
Hence in terms of the block variables in \eqref{BlckVarsIV},
\begin{equation}\label{CSBlckVars}
\begin{aligned}
\exp{\!\left[-\frac{i\,k}{4\pi}\Tr(\sigma_\bullet^2)\right]}
\,&=\,\e{\!-2\pi i k\,\left(\mathbf{r},\mathbf{s}\right)}\cdot\e{\!-2\pi i k\,\left(\mathbf{M},\mathbf{N}\right)} 
		\cdot \exp{\!\left[i \pi k\,\big|\!\sum_{m=1}^r \mu_{j_m} \tau_m\big|^2\right]}\,\times\\
&\times\e{-2\pi i k\,(\mathbf{s},\mathbf{M})}\,\Rq^{\!-\frac{k}{2} |\mathbf{r}+\mathbf{M}|^2}
		\prod_{m=1}^r\Rx_{j_m}^{-k\left(\mathbf{r}+\mathbf{M},\,\tau_m\right)}\,\times\\
&\times\e{-2\pi i k\,(\mathbf{r},\mathbf{N})}\,\wt\Rq^{\!-\frac{k}{2}|\mathbf{s}+\mathbf{N}|^2}
		\prod_{m=1}^r\wt\Rx_{j_m}^{-k\left(\mathbf{s}+\mathbf{N},\,\tau_m\right)}\,.
\end{aligned}
\end{equation}
Note that the dependence on $(\Rq,\Rx)$ and $(\wt\Rq,\wt\Rx)$
factorizes in the last two lines of \eqref{CSBlckVars}.  However for
${k\neq 0}$, the dependence on the coroots $\mathbf{M}$ and
$\mathbf{N}$ generally does not, due to the phase in the first line.

\smallskip\noindent{\sl Vector determinant.}\quad For the one-loop contribution from the vector multiplet,
\begin{equation}\label{GnVectLp}
\begin{aligned}
&\prod_{\alpha\in\Delta_+}\left[4\,\sinh\!\left(\frac{b\langle\alpha,\sigma_\bullet\rangle}{2}\right)\sinh\!\left(\frac{
			\langle\alpha,\sigma_\bullet\rangle}{2
			b}\right)\right]=\,\\
&=\,\prod_{\alpha\in\Delta_+}\Biggr[(-1)^{(\alpha,\mathbf{N})}\cdot 2\sinh\!\left({\pi\left\langle\alpha,i\left(\mathbf{r}+\mathbf{M}\right)
      b^2\,+\,\sum_{m=1}^r\mu_{j_m}\tau_m\,b\,+\,i\,\mathbf{s}\right\rangle}\right)\times\\
&\hspace{3em}\times(-1)^{(\alpha,\mathbf{M})}\cdot 2\sinh\!\left({\pi
			\left\langle\alpha,i
                          \left(\mathbf{s}+\mathbf{N}\right)
                          b^{-2}\,+\,\sum_{m=1}^r\mu_{j_m}\tau_m\,b^{-1}+i\,\mathbf{r}\right\rangle}\right)\Biggr]\,.
\end{aligned}
\end{equation}
To express the right side of \eqref{GnVectLp} additively in terms of the block
variables, we use the Weyl denominator formula (see Lemma 24.3 in \cite{Fulton}
for a proof),
\begin{equation}\label{WeylDen}
\prod_{\alpha\in\Delta_+}\left[2\,\sinh\!\left(\frac{\alpha}{2}\right)\right]\,=\,
\sum_{w\in \fW} \left(-1\right)^w \e{w(\rho)}\,,
\end{equation}
where $(-1)^w$ is positive or negative as the action by the Weyl transformation
${w\in\fW}$ preserves or reverses the orientation of
$\h$.  Recall also that ${\rho=\ha\sum_{\alpha\in\Delta_+}\alpha}$ is the Weyl vector.  By \eqref{WeylDen},
\begin{equation}\label{VectBlckVars}
\begin{aligned}
&\prod_{\alpha\in\Delta_+}\left[4\,\sinh\!\left(\frac{b\langle\alpha,\sigma_\bullet\rangle}{2}\right)\sinh\!\left(\frac{
			\langle\alpha,\sigma_\bullet\rangle}{2
			b}\right)\right]=\,\\
&\hspace{3em}=\,(-1)^{2(\rho,\mathbf{M})}\left[\sum_{w\in
    \fW}\left(-1\right)^w \e{2 \pi i \langle
    w(\rho),\mathbf{s}\rangle}\,\Rq^{\left\langle
      w(\rho),\mathbf{r}+\mathbf{M}\right\rangle} \prod_{m=1}^{r}\Rx_{j_m}^{\left\langle w(\rho), \tau_m\right\rangle}\right]\times\\
&\hspace{3em}\times\,(-1)^{2(\rho, \mathbf{N})}\left[\sum_{w\in
    \fW}\left(-1\right)^w \e{2 \pi i \langle w(\rho),\mathbf{r}\rangle}\,
  \wt\Rq^{\left\langle w(\rho), \mathbf{s}+\mathbf{N}\right\rangle} \prod_{m=1}^{r}\wt\Rx_{j_m}^{\left\langle w(\rho),\tau_m\right\rangle}\right].
\end{aligned}
\end{equation}
In this expression, all dependence on the variables $(\Rq,\Rx)$ vs
$(\wt\Rq,\wt\Rx)$ and the coroots $\mathbf{M}$ vs $\mathbf{N}$ factorizes.

\smallskip\noindent{\sl Matter determinant.}\quad For the residue of the one-loop matter determinant, we factor
the determinant into a regular term and a singular term at $\sigma_\bullet$,
\begin{equation}\label{MatDetRes}
\begin{aligned}
&\Res\!\left[\prod_{j=1}^{n}\prod_{\beta
  \in\Delta_j}s_b\!\left(\frac{\langle\beta,\sigma\rangle}{2\pi}\,+\,\mu_j\right) 
\right]_{\sigma=\sigma_\bullet}\,=\,\\
&\quad\left[\prod_{\ell=1}^n\prod_{\substack{\beta \in
				\Delta_\ell,\\
                                \beta\not\in{\triangle_\bullet}}}s_b\!\left(\frac{\langle\beta,\sigma_\bullet\rangle}{2\pi}\,+\,\mu_\ell\right)\right]\times\Res\!\left[\prod_{j'=1}^n\prod_{\substack{\beta \in
				\Delta_{j'},\\
                                \beta\in{\triangle_\bullet}}}s_b\!\left(\frac{\langle\beta,\sigma\rangle}{2\pi}\,+\,\mu_{j'}\right)\right]_{\sigma=\sigma_\bullet}\,.
\end{aligned}
\end{equation}
We further factorize the regular piece using the functions $\SF^\pm_b$ and
$\wt\SF^\pm_b$ introduced in \eqref{BigSF}, \eqref{BigSFTwiddle},
\eqref{BigFNeg}, and \eqref{BigFNegtwid}.  By analogy to the rank-one variable
$\rz_{j,\ell}$ in \eqref{rzrankone}, let 
\begin{equation}
\begin{aligned}
\rz_{\beta,\ell}^{\triangle_\bullet} \,&=\, \frac{\langle\beta,\sigma_\bullet\rangle}{2\pi}\,+\,\mu_\ell\,,\\
&=\,\mu_\ell \,-\, \langle \beta,
\sum_{m=1}^r\mu_{j_m}\tau_m\rangle \,-\, i\,\langle
\beta, \mathbf{r}+\mathbf{M}\rangle\,b - i\,\langle
\beta, \mathbf{s}+\mathbf{N}\rangle\,b^{-1}\,,
\end{aligned}
\end{equation}
which is the argument of the double-sine $s_b(z)$ in the first product on the
right of \eqref{MatDetRes}.  The single-variable $\Rq$-$\wt\Rq$
decompositions in \eqref{QQTSB} and \eqref{QQTSBNEG} then imply
\begin{equation}\label{MatBlckVarsI}
\begin{aligned}
&\prod_{\ell=1}^n\prod_{\substack{\beta \in
				\Delta_\ell,\\
                                \beta\not\in{\triangle_\bullet}}}s_b\!\left(\frac{\langle\beta,\sigma_\bullet\rangle}{2\pi}\,+\,\mu_\ell\right)
                            =\,\\ 
&\prod_{\ell=1}^{n}\prod_{\substack{\beta \in
				\Delta_\ell,\\
                                \beta\not\in{\triangle_\bullet}}}
\e{\!i \sgn\left(\Re(\rz_{\beta,\ell}^{\triangle_\bullet})\right)\,
  \Psi(\rz_{\beta,\ell}^{\triangle_\bullet})}\,
		\SF_b^{\sgn\left(\Re(\rz_{\beta,\ell}^{\triangle_\bullet})\right)}\!\left(\rz_{\beta,\ell}^{\triangle_\bullet};\Rq\right)\,
		\wt\SF_{b}^{\;\sgn\left(\Re(\rz_{\beta,\ell}^{\triangle_\bullet})\right)}\!\left(\rz_{\beta,\ell}^{\triangle_\bullet};\wt\Rq\right)\,,
\end{aligned}
\end{equation}
where $\Psi$ is the phase in \eqref{PsiPhase}.  This phase is independent of the squashing
parameter $b$.  Explicitly,
\begin{equation}\label{PsiPhaseMN}
\e{\!i\,
  \Psi(\rz_{\beta,\ell}^{\triangle_\bullet})}\,=\,\left(-1\right)^{\langle\beta,\mathbf{r}+\mathbf{M}\rangle
  \langle\beta,\mathbf{s}+\mathbf{N}\rangle}\cdot
i^{\langle\beta,\mathbf{M}+\mathbf{N}+\mathbf{r}+\mathbf{s}\rangle}\cdot\e{\!\frac{i\pi}{4}\left[1-2(\mu_\ell
  - \langle\beta,\sum_{m=1}^r\mu_{j_m}\tau_m\rangle)^2\right]}\,.
\end{equation}
Since ${\langle\beta,\mathbf{M}\rangle,\langle\beta,\mathbf{N}\rangle\in\Z}$
are integers, $\SF_b^\pm(\rz_{\beta,\ell}^{\triangle_\bullet};\Rq)$ is
independent of both the coroot $\mathbf{N}$ and the variables
$(\wt\Rq,\wt\Rx)$.  Dually
$\wt\SF_b^\pm(\rz_{\beta,\ell}^{\triangle_\bullet};\wt\Rq)$ is 
independent of $\mathbf{M}$ and the variables $(\Rq,\Rx)$.  The
product in \eqref{MatBlckVarsI} therefore factorizes up to the
quadratic dependence on $(\mathbf{M},\mathbf{N})$ in the phase
\eqref{PsiPhaseMN}.

For the residue on the right in \eqref{MatDetRes}, 
quasi-periodicity \eqref{eq:DifferenceEqSb} of the double-sine
function yields
\begin{equation}\label{MatDetResII}
\begin{aligned}
&\Res\!\left[\prod_{j'=1}^n\prod_{\substack{\beta \in
				\Delta_{j'},\\
                                \beta\in{\triangle_\bullet}}}s_b\!\left(\frac{\langle\beta,\sigma\rangle}{2\pi}\,+\,\mu_{j'}\right)\right]_{\sigma=\sigma_\bullet}\,=\,\\
&\qquad\prod_{\beta\in\triangle_\bullet}\Biggr[\left(-1\right)^{\langle\beta,\mathbf{r}+\mathbf{M} \rangle\langle\beta,\mathbf{s}+\mathbf{N}\rangle}\cdot
		i^{\langle\beta,\mathbf{M}+\mathbf{N}+\mathbf{r}+\mathbf{s}\rangle}\,\times\\
&\qquad\qquad\times\,\left(\frac{\Rq^{\langle\beta,
      \,\mathbf{r}+\mathbf{M}\rangle\left(\langle\beta,\,\mathbf{r}+\mathbf{M}
        \rangle+1\right)/4}}{\big(\Rq;\Rq\big)_{\!\langle\beta,
      \,\mathbf{r}+\mathbf{M}\rangle}}\right)\cdot
\left(\frac{\wt\Rq^{\langle\beta, \,\mathbf{s}+\mathbf{N}
      \rangle\left(\langle\beta,
        \,\mathbf{s}+\mathbf{N}\rangle+1\right)/4}}{\big(\wt\Rq;\wt\Rq\big)_{\!\langle\beta,
      \,\mathbf{s}+\mathbf{N} \rangle}}\right)\Biggr]\times\\
&\qquad\qquad\qquad\times\,\Res\!\left[
		\prod_{\beta\in\triangle_\bullet}s_b\!\left(
		\frac{\langle\beta,\sigma\rangle}{2\pi}
		\right)
		\right]_{\sigma = 0}\,.
\end{aligned}
\end{equation}
Note that
$\langle\beta,\mathbf{r}\rangle,\langle\beta,\mathbf{s}\rangle\in\Z$
are integral here since ${\beta\in\triangle_\bullet}$ lies in the set of
admissible weights.

Finally, recall from \eqref{Res0} the normalization
\begin{equation}
\Res\big[s_b(z)\big]_{z=0} \,=\, \frac{i}{2\pi}\,.
\end{equation}
Hence
\begin{equation}
\Res\!\left[\prod_{\beta\in\triangle_\bullet}s_b\!\left(\frac{\langle\beta,\sigma\rangle}{2\pi}\right)\right]_{\sigma
  = 0} \,=\, \frac{i^r}{\det(\triangle_\bullet)},
\end{equation}
where $\det(\triangle_\bullet)$ is the determinant of the matrix
spanned by the admissible weights ${\beta\in\triangle_\bullet}$, expressed in the
basis of fundamental weights $\{\hat\omega^1,\ldots,\hat\omega^r\}$.  With
our orientation conventions, ${\det(\triangle_\bullet)}>0$ is always a
positive integer and is equal to the index of the admissible lattice
$\Lambda_{\triangle_\bullet}$, defined in \eqref{Lamtrabul}, as a
sublattice of the weight lattice $\Lambda_{\rm
  W}$,
\begin{equation}
\det(\triangle_\bullet) \,=\, |\Lambda_{\rm
  W}:\Lambda_{\triangle_\bullet}| \,\in\,\Z\,.
\end{equation}
Thus the residue in \eqref{MatDetResII} is given altogether by 
\begin{equation}\label{MatDetResIII}
\begin{aligned}
&\Res\!\left[\prod_{j'=1}^n\prod_{\substack{\beta \in
				\Delta_{j'},\\
                                \beta\in{\triangle_\bullet}}}s_b\!\left(\frac{\langle\beta,\sigma\rangle}{2\pi}\,+\,\mu_{j'}\right)\right]_{\sigma=\sigma_\bullet}\,=\,\\
&\qquad \frac{i^r}{|\Lambda_{\rm W}:\Lambda_{\triangle_\bullet}|}\prod_{\beta\in\triangle_\bullet}\Biggr[\left(-1\right)^{\langle\beta,\mathbf{r}+\mathbf{M} \rangle\langle\beta,\mathbf{s}+\mathbf{N}\rangle}\cdot
		i^{\langle\beta,\mathbf{M}+\mathbf{N}+\mathbf{r}+\mathbf{s}\rangle}\,\times\\
&\qquad\qquad\times\,\left(\frac{\Rq^{\langle\beta,
      \,\mathbf{r}+\mathbf{M}\rangle\left(\langle\beta,\,\mathbf{r}+\mathbf{M}
        \rangle+1\right)/4}}{\big(\Rq;\Rq\big)_{\!\langle\beta,
      \,\mathbf{r}+\mathbf{M}\rangle}}\right)\cdot
\left(\frac{\wt\Rq^{\langle\beta, \,\mathbf{s}+\mathbf{N}
      \rangle\left(\langle\beta,
        \,\mathbf{s}+\mathbf{N}\rangle+1\right)/4}}{\big(\wt\Rq;\wt\Rq\big)_{\!\langle\beta,
      \,\mathbf{s}+\mathbf{N} \rangle}}\right)\Biggr]\,.
\end{aligned}
\end{equation}
Again, the quadratic dependence of the phase on $\mathbf{M}$ and
$\mathbf{N}$ in the second line of \eqref{MatDetResIII} obstructs a
complete factorization of the residue.

For reference, the complete matter residue \eqref{MatDetRes}
becomes
\begin{equation}\label{MatDetResIV}
\begin{aligned}
&\Res\!\left[\prod_{j=1}^{n}\prod_{\beta
  \in\Delta_j}s_b\!\left(\frac{\langle\beta,\sigma\rangle}{2\pi}\,+\,\mu_j\right) 
\right]_{\sigma=\sigma_\bullet}\,=\,\\
&\qquad \frac{i^r}{|\Lambda_{\rm W}:\Lambda_{\triangle_\bullet}|}\prod_{\beta\in\triangle_\bullet}\Biggr[\left(-1\right)^{\langle\beta,\mathbf{r}+\mathbf{M} \rangle\langle\beta,\mathbf{s}+\mathbf{N}\rangle}\cdot
		i^{\langle\beta,\mathbf{M}+\mathbf{N}+\mathbf{r}+\mathbf{s}\rangle}\,\times\\
&\qquad\qquad\times\,\left(\frac{\Rq^{\langle\beta,
      \,\mathbf{r}+\mathbf{M}\rangle\left(\langle\beta,\,\mathbf{r}+\mathbf{M}
        \rangle+1\right)/4}}{\big(\Rq;\Rq\big)_{\!\langle\beta,
      \,\mathbf{r}+\mathbf{M}\rangle}}\right)\cdot
\left(\frac{\wt\Rq^{\langle\beta, \,\mathbf{s}+\mathbf{N}
      \rangle\left(\langle\beta,
        \,\mathbf{s}+\mathbf{N}\rangle+1\right)/4}}{\big(\wt\Rq;\wt\Rq\big)_{\!\langle\beta,
      \,\mathbf{s}+\mathbf{N} \rangle}}\right)\Biggr]\,\times\\
&\times \prod_{\ell=1}^{n}\prod_{\substack{\beta \in
				\Delta_\ell,\\
                                \beta\not\in{\triangle_\bullet}}}
\e{\!i \sgn\left(\Re(\rz_{\beta,\ell}^{\triangle_\bullet})\right)\,
  \Psi(\rz_{\beta,\ell}^{\triangle_\bullet})}\,
		\SF_b^{\sgn\left(\Re(\rz_{\beta,\ell}^{\triangle_\bullet})\right)}\!\left(\rz_{\beta,\ell}^{\triangle_\bullet};\Rq\right)\,
		\wt\SF_{b}^{\;\sgn\left(\Re(\rz_{\beta,\ell}^{\triangle_\bullet})\right)}\!\left(\rz_{\beta,\ell}^{\triangle_\bullet};\wt\Rq\right)\,.
\end{aligned}
\end{equation}
The total matter residue includes a phase
\begin{equation}
\prod_{j=1}^n\prod_{\beta\in\Delta_j}
\left(-1\right)^{\langle\beta,\mathbf{M}\rangle\langle\beta,\mathbf{N}\rangle}
\,=\,
\prod_{j=1}^n\left(-1\right)^{c_2(\lambda_j)\cdot\left(\mathbf{M},\mathbf{N}\right)
}\,, 
\end{equation}
induced from the explicit term on the second line in
\eqref{MatDetResIV} as well as the phase 
$\Psi(\rz^{\triangle_\bullet}_{\beta,\ell})$ in \eqref{PsiPhaseMN}.
Here $\left(\mathbf{M},\mathbf{N}\right)$ is the inner-product of
coroots in the invariant Lie algebra metric, and we use the identity
\eqref{QadCII} in Appendix \ref{se:LieAlgConvention} to sum over
weights.  As will be clear, this phase encodes the parity anomaly from
Section \ref{OneLoop} in each holomorphic block.

\paragraph{General Formula for Holomorphic Blocks.}

The local residue formulas in \eqref{CSBlckVars},
\eqref{VectBlckVars}, and \eqref{MatDetResIV} can be combined to determine
the general holomorphic block for supersymmetric Chern-Simons-matter
theories on $S^3$.  

By analogy to the abelian expression in \eqref{AbbResII}, we
write the local residue in \eqref{BigZresuv} as 
\begin{equation}\label{BigZresuvII}
\begin{aligned}
\RZ^{\triangle_\bullet}_{u,v} \,&=\, \frac{i^r}{|\Lambda_{\rm
    W}:\Lambda_{\triangle_\bullet}|}\exp{\!\left[-2\pi i \left( k +
      \ha c_2(\Lambda)\right)(\mathbf{M},\mathbf{N})\right]}\,\times\\
&\qquad\times\exp{\!\left(
    i\,\Theta^{\triangle_\bullet}_{\mathbf{r},\mathbf{s}}\right)}\cdot
\RW^{\triangle_\bullet}_{\mathbf{M},\mathbf{r},\mathbf{s}}(\Rq,\Rx)\cdot
\wt\RW^{\triangle_\bullet}_{\mathbf{N},\mathbf{r},\mathbf{s}}(\wt\Rq,\wt\Rx)\,,
\end{aligned}
\end{equation}
where the admissible $r$-tuple $\triangle_\bullet$ and the positive
integers ${u,v\in\Z^r_{\ge 0}}$ determine coroots
${\mathbf{M},\mathbf{N}\in\Z_{\ge 0}[d_1\tau_1,\cdots,d_r\tau_r]}$ and
characteristics
${\mathbf{r},\mathbf{s}\in (\Z/d_1\Z)\tau_1 \times\cdots\times (\Z/d_r\Z)\tau_r}$ via
\eqref{TauGensII}.  For the phase factor in the second line of
\eqref{BigZresuvII}, we find
\begin{equation}\label{BigNonAbPhas}
\begin{aligned}
&\exp{\!\left(i\,\Theta^{\triangle_\bullet}_{\mathbf{r},\mathbf{s}}\right)}
=\, \exp{\!\left[-2\pi i\left(k+\ha c_2(\Lambda)\right) (\mathbf{r},\mathbf{s})\right]}\cdot\exp{\!\left[i
    \pi k\,\big|\!\sum_{m=1}^r\mu_{j_m}\tau_m\big|^2\right]}\times\\
&\times\exp{\!\left[\frac{i\pi}{4}\sum_{\ell=1}^n\sum_{\substack{\beta \in
				\Delta_\ell,\\
                                \beta\not\in{\triangle_\bullet}}}\left(1-2\left(\mu_\ell
  - \langle\beta,\sum_{m=1}^r\mu_{j_m}\tau_m\rangle\right)^2\right)\sgn\!\left(\mu_\ell
  - \langle\beta,\sum_{m=1}^r\mu_{j_m}\tau_m\rangle\right)\right]}\,.
\end{aligned}
\end{equation}
Compare to the abelian phase in \eqref{BigTheta}.  Similarly for the
block summand,  
\begin{equation}\label{BigWnab}
\begin{aligned}
&\RW^{\triangle_\bullet}_{\mathbf{M},\mathbf{r},\mathbf{s}}(\Rq,\Rx)
\,=\,\e{-2 \pi i \left(k +
    \ha c_2(\Lambda)\right)\!(\mathbf{s},\,\mathbf{M})}\,\times\\
&\quad\times\,i^{\sum_{\beta\in\triangle_\bullet}\langle\beta,\,\mathbf{r}+\mathbf{M}\rangle+
                  \sum_{\ell=1}^n\sum_{\substack{\beta \in
				\Delta_\ell,\\
                                \beta\not\in{\triangle_\bullet}}} \langle\beta,\,\mathbf{r}+\mathbf{M}\rangle
			\cdot\sgn\left(\Re(\rz_{\beta,\ell}^{\triangle_\bullet})\right)}\,\times\\
&\quad\times\,\Rq^{-\!\frac{k}{2}|\mathbf{r}+\mathbf{M}|^2}\cdot\prod_{m=1}^r\Rx_{j_m}^{-k(\mathbf{r}+\mathbf{M},\,\tau_m)}\cdot\prod_{\beta\in\triangle_\bullet}\left[\frac{\Rq^{\langle\beta, \mathbf{r}+\mathbf{M} \rangle \left(\langle\beta, \mathbf{r}+\mathbf{M} \rangle+1\right)/4}}{\left(\Rq;\Rq\right)_{\!\langle\beta, \mathbf{r}+\mathbf{M} \rangle}}\right]\times\\
&\quad\times\,(-1)^{2(\rho, \mathbf{M})}
		\left[\sum_{w\in \fW}(-1)^w\, \e{2\pi i \langle w(\rho),\mathbf{s}\rangle}\,\Rq^{\langle w(\rho),\,\mathbf{r}+\mathbf{M} \rangle}\prod_{m=1}^{r}\Rx_{j_m}^{\langle w(\rho), \tau_m\rangle }\right]\times\\
&\quad\times\,\prod_{\ell=1}^{n}\prod_{\substack{\beta \in
				\Delta_\ell,\\
                                \beta\not\in{\triangle_\bullet}}} \SF_b^{\sgn\left(\Re(\rz_{\beta,\ell}^{\triangle_\bullet})\right)}\!\left(\rz_{\beta,\ell}^{\triangle_\bullet};\Rq\right).
\end{aligned}
\end{equation}
Compare again to the abelian analogue $\RW^{\,j}_{M,r,s}$ in
\eqref{BigW}.\footnote{Because the Fayet-Iliopoulos parameter is
  absent in the non-abelian theory with simple gauge group, the formal variable $\Ry$ does not
  appear in $\RW^{\triangle_\bullet}_{\mathbf{M},\mathbf{r},\mathbf{s}}$.}  The expression
for $\RW^{\triangle_\bullet}_{\mathbf{M},\mathbf{r},\mathbf{s}}$ has
the same structure, but now includes a
contribution from the vector multiplet in the fourth line of
\eqref{BigWnab} through the Weyl sum over
${w\in\fW}$.   Crucially for factorization,
$\RW^{\triangle_\bullet}_{\mathbf{M},\mathbf{r},\mathbf{s}}$ is
independent of the coroot $\mathbf{N}$ as well as the variables
$(\wt\Rq,\wt\Rx)$.  We omit an entirely similar expression for the
dual summand
$\wt\RW^{\triangle_\bullet}_{\mathbf{N},\mathbf{r},\mathbf{s}}$, which
is likewise independent of $\mathbf{M}$ and $(\Rq,\Rx)$.

The sphere partition function is the residue sum\footnote{We omit the
  superscript from $Z_{S^3}^{\rm uv}$ in \eqref{FactoZS3}, as our formulas now include the
  effect of the Chern-Simons term in \eqref{CSBlckVars}.}
\begin{equation}\label{FactoZS3}
Z_{S^3} \,=\, \frac{\left(-2\pi i\right)^r\e{i\eta_0}}{|\fW|\cdot\Vol(T)}
\sum_{\textrm{admissible}\,\triangle_\bullet}\sum_{u,v\in\Z^r_{\ge 0}}\RZ^{\triangle_\bullet}_{u,v}\,.
\end{equation}
Here we include the overall normalization from \eqref{ZuvCmbV} as well as a
factor $(-2\pi i)^r$ from the residue formula in \eqref{JorNeqRV}.
Substituting for $\RZ^{\triangle_\bullet}_{u,v}$ in
\eqref{BigZresuvII}, we obtain
\begin{equation}\label{FactoZS3II}
\begin{aligned}
&Z_{S^3} \,=\,\frac{\left(2\pi\right)^r\e{i\eta_0}}{|\fW|\cdot\Vol(T)}
\sum_{\substack{\textrm{admissible}\,\triangle_\bullet,\\ \mathbf{r},\mathbf{s}\,\in\,(\Z/d_1\Z)\tau_1\times\cdots\times(\Z/d_r\Z)\tau_r}}
\frac{\exp{\!\left(
    i\,\Theta^{\triangle_\bullet}_{\mathbf{r},\mathbf{s}}\right)}}{|\Lambda_{\rm
    W}:\Lambda_{\triangle_\bullet}|}\,\,\times\\
&\times\sum_{\mathbf{M},\mathbf{N}\in\Z_{\ge 0}[d_1\tau_1,\cdots,d_r\tau_r]}\exp{\!\left[-2\pi i \left( k +
      \ha c_2(\Lambda)\right)\cdot(\mathbf{M},\mathbf{N})\right]}\,
\RW^{\triangle_\bullet}_{\mathbf{M},\mathbf{r},\mathbf{s}}(\Rq,\Rx)\,
\wt\RW^{\triangle_\bullet}_{\mathbf{N},\mathbf{r},\mathbf{s}}(\wt\Rq,\wt\Rx)\,.
\end{aligned}
\end{equation}
If for all positive coroots ${\mathbf{M},\mathbf{N}\in\Lambda^{\rm
    R}_+}$, the integrality condition below is obeyed,
\begin{equation}\label{FactoCond}
\left( k + \ha c_2(\Lambda)\right)\cdot(\mathbf{M},\mathbf{N}) \,\in\,\Z\,,
\end{equation}
then the phase in the second line of \eqref{FactoZS3II} is unity, and the sum
over $\mathbf{M}$ and $\mathbf{N}$ factorizes.  In that case we obtain the block
decomposition
\begin{equation}\label{FactoZS3III}
Z_{S^3} \,=\, \frac{\left(2\pi\right)^r\e{i\eta_0}}{|\fW|\cdot\Vol(T)}
\sum_{\substack{\textrm{admissible}\,\triangle_\bullet,\\
    \mathbf{r},\mathbf{s}\,\in\,(\Z/d_1\Z)\tau_1\times\cdots\times(\Z/d_r\Z)\tau_r}}
\frac{\exp{\!\left(
    i\,\Theta^{\triangle_\bullet}_{\mathbf{r},\mathbf{s}}\right)}}{|\Lambda_{\rm
    W}:\Lambda_{\triangle_\bullet}|}\,\,\RB^{\triangle_\bullet}_{\mathbf{r},\mathbf{s}}(\Rq,\Rx)\,\wt\RB^{\triangle_\bullet}_{\mathbf{r},\mathbf{s}}(\wt\Rq,\wt\Rx)\,,
\end{equation}
where each block is given by a formal sum
\begin{equation}\label{FormBlkSM}
\begin{aligned}
\RB^{\triangle_\bullet}_{\mathbf{r},\mathbf{s}}(\Rq,\Rx)\,&=\,
\sum_{\mathbf{M}\in\Z_{\ge 0}[d_1\tau_1,\cdots,d_r\tau_r]} \RW^{\triangle_\bullet}_{\mathbf{M},\mathbf{r},\mathbf{s}}(\Rq,\Rx)\,,\\
\wt\RB^{\triangle_\bullet}_{\mathbf{r},\mathbf{s}}(\wt\Rq,\wt\Rx)
\,&=\, \sum_{\mathbf{N}\in\Z_{\ge 0}[d_1\tau_1,\cdots,d_r\tau_r]} \wt\RW^{\triangle_\bullet}_{\mathbf{N},\mathbf{r},\mathbf{s}}(\wt\Rq,\wt\Rx)\,.
\end{aligned}
\end{equation}

We conclude with some structural remarks about the non-abelian
factorization formula in \eqref{FactoZS3III}.

First, with the Lie algebra conventions in Appendix
\ref{se:LieAlgConvention}, the invariant metric
$(\,\cdot\,,\,\cdot\,)$ on the coroot lattice $\Lambda^{\rm R}$ is 
integral, meaning ${(\mathbf{M},\mathbf{N})\in\Z}$.  The
anomaly-cancellation condition ${k-\ha
  c_2(\Lambda) \in \Z}$ therefore implies the integrality condition
in \eqref{FactoCond} and is sufficient for holomorphic
factorization \eqref{FactoZS3III} of $Z_{S^3}$.  

However, the anomaly-cancellation
condition is not necessary for holomorphic factorization.  If the coroot lattice
$\Lambda^{\rm R}$ happens to be even, so that
${(\mathbf{M},\mathbf{N})\in 2\Z}$ in \eqref{FactoCond}, then
$Z_{S^3}$ factorizes as in \eqref{FactoZS3III} whether or not ${k\in\ha\Z}$
is properly quantized.  We have already encountered an elementary
instance of the discrepancy between anomaly-cancellation and
holomorphic factorization for gauge group $SU(2)$ in Section
\ref{SU2GT}.  See the discussion following \eqref{BigWSU2}; trivially
for $SU(2)$, the coroot lattice is even.  

For Lie groups of higher
rank, examples with $\Lambda^{\rm R}$ an even lattice are easy to find.  In 
rank-two, the coroot lattices for $Spin(4)$ and ${Spin(5)\simeq
  Sp(4)}$ are even, demonstrated directly in Appendix \ref{se:LieAlgConvention}, so
$Z_{S^3}$ factors regardless of anomaly-cancellation in
those examples.  More generally, the coroot lattice of the symplectic
group $Sp(2N)$ is
even for all values of $N$ (proof omitted), and holomorphic factorization holds for arbitrary
${k\in\ha\Z}$ in $Sp(2N)$ Chern-Simons-matter theories.  

A short
exercise shows that these examples, all based upon the symplectic
group, provide the exhaustive list of the
compact, simple, and simply-connected Lie groups (types ABCDEFG) such that
$\Lambda^{\rm R}$ is an even lattice.  For other, non-symplectic Lie groups,
holomorphic factorization of $Z_{S^3}$ is true if and only if
${k\in\ha\Z}$ obeys the anomaly-cancellation condition.

Second, comparing the result in \eqref{FactoZS3III} to the 
Factorization Conjecture in \eqref{FACTOR}, we see that the holomorphic
blocks are labelled by an index set $\CI$ consisting of triples
\begin{equation}\label{BigCI}
\CI \,=
\big\{\left(\triangle_\bullet,\mathbf{r},\mathbf{s}\right)\,\big|\,\triangle_\bullet\,\hbox{admissible
$r$-tuple},\,
\mathbf{r},\mathbf{s}\in (\Z/d_1\Z)\tau_1\times\cdots\times(\Z/d_r\Z)\tau_r \big\}\,.
\end{equation}
Recall that $\triangle_\bullet$ is an $r$-tuple subset of weights
in the matter representation $\Lambda$, such that $\triangle_\bullet$ satisfies the
admissibility conditions following \eqref{SUNwts}, and
$\{\tau_1,\ldots,\tau_r\}$ is dual to $\triangle_\bullet$.  Roughly
speaking, the elements
${\mathbf{r},\mathbf{s}}$ in the finite group
$(\Z/d_1\Z)\tau_1\times\cdots\times(\Z/d_r\Z)\tau_r $ play the role of 
theta-characteristics for the blocks.  Of particular note, the index
set $\CI$ is necessarily 
finite, consistent with the Factorization Conjecture, and depends only upon
the pair $(G,\Lambda)$.  In Section \ref{RkTwoExs} we classify admissible
doubles $\triangle_\bullet$ and hence determine $\CI$ for a variety of rank-two examples.

Finally, we observe that the bilinear form $\RG_{m n}$ on blocks in
\eqref{FACTOR} is diagonal when expressed in the basis labelled by 
triples $(\triangle_\bullet,\mathbf{r},\mathbf{s})$,
\begin{equation}\label{BigGmnnab}
\RG_{m n} \,=\, \delta_{m n}\cdot \frac{\exp{\!\left(i\,\Theta^{\triangle_\bullet}_{\mathbf{r},\mathbf{s}}\right)}}{|\Lambda_{\rm
    W}:\Lambda_{\triangle_\bullet}|}\,,\qquad\qquad
\begin{aligned}
m&\equiv\left(\triangle_\bullet,\mathbf{r},\mathbf{s}\right),\\
n&\equiv\left(\triangle'_\bullet,\mathbf{r}',\mathbf{s}'\right).
\end{aligned}
\end{equation}
For brevity, we omit the overall normalization constants in
\eqref{FactoZS3III} from the definition of $\RG_{m n}$.  The formula
for $\RG_{m n}$ in \eqref{BigGmnnab} is a natural generalization of
the previous abelian formula in \eqref{BigGmn}.

\subsection{Some Examples in Rank-Two}\label{RkTwoExs}

We finally present some explicit examples of the block decomposition for
various choices of the gauge group $G$ and the matter representation
$\Lambda$.  Our goal is to illustrate how the previous structural
results in Sections \ref{JordDRk2}, \ref{JordHRGG}, and
\ref{LatticeSums}, including the abstract factorization formula
for $Z_{S^3}$, appear in practice.

According to the factorization formula in \eqref{FactoZS3III},
understanding the classification of holomorphic blocks associated to a
pair $(G,\Lambda)$ amounts to solving a combinatoric problem in
discrete geometry.
\begin{quotation}\noindent
{\bf Problem}:~Determine all admissible $r$-tuples $\Delta_\bullet$ which are
subsets of the set of weights in $\Lambda$, and compute the degrees
$(d_1,\ldots,d_r)$ for the generators
$\{\tau_1,\ldots,\tau_r\}$ dual to each $\triangle_\bullet$.
\end{quotation}
The computation of degrees is straightforward, so
the fundamental
problem is the classification of admissible $r$-tuples
$\triangle_\bullet$ for
$(G,\Lambda)$.

Let us begin with a general remark about the classification problem.

Suppose that ${\Lambda = \bigoplus_{j=1}^n [\lambda_j]}$ is the direct sum
of weight-multiplicity-free irreducible representations $[\lambda_j]$
of $G$, and let
${\Delta_\Lambda = \bigsqcup_{j=1}^n \Delta_j}$ be the set of (non-zero)
weights in $\Lambda$.  We allow for the possibility that some summands
in $\Lambda$ are isomorphic, in which case repeated weights in
$\Delta_\Lambda$ are distinguished by a flavor index.  The set
${\Delta_\Lambda = \Delta_\Lambda^+ \cup \Delta_\Lambda^-}$ decomposes
as a union of positive and negative weights.  We assume that
$\Lambda$ is a real representation of $G$ (as true previously for
$\Lambda_{\rm uv}$), in which case ${\Delta_\Lambda^- =
  -\Delta_\Lambda^+}$ in $\h^*$.  All these assumptions hold, for instance, for
the $SU(N)$ SQCD example discussed at the end of Section \ref{JordHRGG}.

In this situation, if ${\{\beta_1,\ldots,\beta_r\}\subseteq\Delta_\Lambda^+}$ is any basis for
$\h^*$ composed of elements in the positive component
$\Delta_\Lambda^+$, then each of the $2^r$ subsets
$\{\pm\beta_1,\ldots,\pm\beta_r\} \subseteq \Delta_\Lambda$ (with
signs assigned independently to each element) is also a basis for
$\h^*$.  Of these $r$-tuple subsets, precisely one is admissible, in
the sense that $-\rho$ lies in the positive cone for the given choice of signs.  This
statement follows by the same observation made previously for $SU(N)$
SQCD.  For all choices of signs, the cones over
$\{\pm\beta_1,\ldots,\pm\beta_r\}$ divide ${\h\simeq\R^r}$ into $2^r$
chambers, and ${-\rho>0}$ is positive on a unique chamber.

Thus, for every positively-oriented basis
${\{\beta_1,\ldots,\beta_r\}\subset\Delta_\Lambda^+}$ contained within
the positive component of $\Delta_\Lambda$, there is a unique
admissible $r$-tuple $\Delta_\bullet$, for some
choice of signs on the generators ${\beta_1,\ldots,\beta_r}$.  The
classification of admissible $r$-tuples for $(G,\Lambda)$ is then
equivalent to the classification of bases for $\h^*$ (up to
permutation of generators) which are
contained as subsets of $\Delta_\Lambda^+$.  We
applied this correspondence to classify admissible $(N-1)$-tuples for
$SU(N)$ SQCD with ${N_{\rm f}=1}$ in \eqref{SUNadmd}.  

The classification is particularly easy when the gauge group $G$ has rank-two, for
which the specialized analysis in Section \ref{JordHRGG} can be
applied algorithmically.  In the remainder, we present several
easy examples of holomorphic factorization for $Z_{S^3}$ in rank-two.
We conclude with a discussion of $SU(N)$ SQCD.

\paragraph{Blocks for $SU(3)$ gauge theory with adjoint matter.}

We consider $SU(3)$ gauge theory coupled to ${N_{\rm adj}}$ copies of
the adjoint representation,
\begin{equation}
\Lambda \,=\, \mathbf{8}^{N_{\rm adj}}\,,\qquad\qquad
G\,=\, SU(3)\,.
\end{equation}
When ${N_{\rm adj}=1}$, the set ${\Delta_\Lambda^+=\Delta_+}$ is just
the set of three positive roots, any two of which provide a basis.
Hence there are three admissible doubles
${\triangle_{1,2,3}}$.  These doubles are already
listed in Table \ref{WtH1H2Tab}.\footnote{Since the Weyl vector $\rho$
  is itself a positive root, a degeneracy in the polar hyperplanes
  occurs and is resolved according to Figure \ref{BHwtrhoFig}.}  We reproduce them below in the basis
of fundamental weights,
\begin{equation}\label{AdDblSU3v0}
\begin{aligned}
\triangle_1 \,&=\, \big\{-2\hat\omega^1+\hat\omega^2,\,\hat \omega^1-2\hat
\omega^2\big\}\,,\qquad\qquad \big[N_{\rm adj}=1\big]
\\
\triangle_2 \,&=\, \big\{-\hat\omega^1+2\hat \omega^2,\,-\hat\omega^1-\hat\omega^2\big\}\,,
\\
\triangle_3 \,&=\, \big\{-2\hat\omega^1+\hat\omega^2,\,-\hat\omega^1-\hat\omega^2\big\}\,.
\end{aligned}
\end{equation}
For ${N_{\rm adj}>1}$, essentially the same classification holds, but
each weight now carries a flavor index to label the chiral multiplet from
which it arises,
\begin{equation}\label{AdDblSU3}
\begin{aligned}
\triangle_{1;j_1 j_2} \,&=\, \big\{\left(-2\hat\omega^1+\hat\omega^2\right){}_{\!j_1},\left(\hat \omega^1-2\hat
\omega^2\right){}_{\!j_2}\big\}\,,\qquad\qquad j_1,j_2=1,\ldots,N_{\rm adj}\,,\\
\triangle_{2;j_1 j_2} \,&=\, \big\{\left(-\hat\omega^1+2\hat \omega^2\right){}_{\!j_1},\left(-\hat\omega^1-\hat\omega^2\right){}_{\!j_2}\big\}\,,
\\
\triangle_{3;j_1 j_2} \,&=\, \big\{\left(-2\hat\omega^1+\hat\omega^2\right){}_{\!j_1},\left(-\hat\omega^1-\hat\omega^2\right){}_{\!j_2}\big\}\,.
\end{aligned}
\end{equation}
Altogether, there are ${3 N_{\rm adj}^2}$ admissible doubles for
$SU(3)$ gauge theory with adjoint matter.

For each of the admissible doubles $\triangle_\bullet$ in \eqref{AdDblSU3}, the index of
the sublattice $\Lambda_{\triangle_\bullet}$ in $\Lambda_{\rm W}$ is
given by 
\begin{equation}
|\Lambda_{\rm W}:\Lambda_{\triangle_\bullet}| \,=\,
\det(\triangle_\bullet) \,=\, 3\,.
\end{equation}
Hence with no calculation whatsoever,
\begin{equation}
\Gamma_{\triangle_\bullet} \,\simeq\, \Z/3\Z\,.
\end{equation}
In terms of coroots dual to the simple weights, the refinement
${\Lambda^{\triangle_{1;j_1 j_2}}=\Z[\tau_1,\tau_2]}$ associated to the
admissible double 
$\triangle_{1;j_1 j_2}$ in \eqref{AdDblSU3} (see also Figure
\ref{AdmLattices}) is generated by 
\begin{equation}
\triangle_{1;j_1 j_2}:\quad\tau_1\,=\, -\frac{2}{3}\hat h_1 - \frac{1}{3} \hat h_2\,,\qquad\qquad
\tau_2\,=\,-\frac{1}{3}\hat h_1 - \frac{2}{3} \hat h_2\,,
\end{equation}
with degrees
\begin{equation}
d_1 \,=\, d_2 \,=\, 3\,.
\end{equation}
Following \eqref{Bultsgm}, the set of poles in
$\h$ which contribute to this first block occur at locations
\begin{equation}
\begin{aligned}
\triangle_{1;j_1 j_2}:\quad&\sigma_\bullet = -2\pi i\left(\mathbf{r}\,b + \mathbf{s}\,b^{-1} +
  \mathbf{M}\,b + \mathbf{N}\,b^{-1}\right) -\, 2\pi \mu_{j_1}\tau_1
\,-\, 2\pi\mu_{j_2}\,\tau_2\,,\\
&\mathbf{M},\mathbf{N}\in\big\{3m\,\tau_1+3n\,\tau_2\,\big|\,
m,n\in\Z_{\ge 0}\}\subset \Lambda^{\rm R}\,,\\
&\mathbf{r},\mathbf{s}\in(\Z/3\Z)\tau_1 \times (\Z/3\Z)\tau_2\,.
\end{aligned}
\end{equation}
For the other doubles in \eqref{AdDblSU3}, a
similar description applies with the generators
\begin{equation}
\triangle_{2;j_1 j_2}:\quad \tau_1\,=\,-\frac{1}{3}\hat h_1 +
\frac{1}{3}\hat h_2\,,\qquad \tau_2 \,=\, -\frac{2}{3}\hat h_1 -
\frac{1}{3}\hat h_2\,,
\end{equation}
and
\begin{equation}
\triangle_{3;j_1 j_2}:\quad \tau_1\,=\,-\frac{1}{3}\hat h_1 +
\frac{1}{3}\hat h_2\,,\qquad \tau_2 \,=\, -\frac{1}{3}\hat h_1 -
\frac{2}{3}\hat h_2\,.
\end{equation}

Concretely, the block decomposition for $SU(3)$ gauge theory with
adjoint matter takes the form
\begin{equation}
Z_{S^3} \,=\, \frac{C_{SU(3)}}{3}\,
\sum_{a=1}^3 \sum_{j_1,j_2=1}^{N_{\rm adj}} \sum_{\mathbf{r},\mathbf{s}\in(\Z/3\Z)\tau_1\times(\Z/3\Z)\tau_2}
\exp{\!\left(
    i\,\Theta^{\triangle_{a;j_1
        j_2}}_{\mathbf{r},\mathbf{s}}\right)}\,\RB^{\triangle_{a;j_1
    j_2}}_{\mathbf{r},\mathbf{s}}(\Rq,\Rx)\,\wt\RB^{\triangle_{a;j_1 j_2}}_{\mathbf{r},\mathbf{s}}(\wt\Rq,\wt\Rx)\,,
\end{equation}
where $C_{SU(3)}$ is the group prefactor in \eqref{FactoZS3III}, which we
leave implicit.  Evidently, the
partition function is a sum over ${|\CI|=3^5
  N_{\rm adj}^2}$ blocks.

\paragraph{Blocks for $G_2$ gauge theory with fundamental matter.}
Our techniques work equally well for exceptional gauge groups such as
$G_2$.  We consider ${N_{\rm f}\ge 1}$ chiral multiplets which each transform
in a copy of the fundamental, seven-dimensional representation,
\begin{equation}
\Lambda\,=\,\mathbf{7}^{N_{\rm f}}\,,\qquad\qquad G = G_2\,.
\end{equation}
The fundamental representation is real with six non-zero weights.  For
${N_{\rm f}=1}$, the positive component $\Delta_\Lambda^+$ contains
three elements, any two of which span, so there are again three
basic admissible doubles $\triangle_{1,2,3}$.  See Table
\ref{WtH1H2Tab}.  Allowing for flavor indices, 
\begin{equation}\label{AdDblG2}
\begin{aligned}
\triangle_{1;j_1 j_2} \,&=\, \big\{\left(-\hat\omega^1\right){}_{\!j_1},\left(\hat \omega^1-\hat
\omega^2\right){}_{\!j_2}\big\}\,,\qquad\qquad j_1,j_2=1,\ldots,N_{\rm f}\,,\\
\triangle_{2;j_1 j_2} \,&=\, \big\{\left(-2\hat\omega^1+\hat\omega^2\right){}_{\!j_1},\left(\hat\omega^1-\hat\omega^2\right){}_{\!j_2}\big\}\,,
\\
\triangle_{3;j_1 j_2} \,&=\, \big\{\left(-\hat\omega^1\right){}_{\!j_1},\left(2\hat\omega^1-\hat\omega^2\right){}_{\!j_2}\big\}\,.
\end{aligned}
\end{equation}
The reader is invited to check that
$-\rho=-\hat\omega^1-\hat\omega^2$ lies in each positive cone
$\R_+[\triangle_{1,2,3}]$.  In total, $G_2$ gauge theory with $N_{\rm
  f}$ fundamental flavors admits ${3 N_{\rm f}^2}$ admissible doubles.

For each admissible double $\triangle_\bullet$ in \eqref{AdDblG2}, the
index of $\Lambda_{\triangle_\bullet}$ in the weight lattice
$\Lambda_{\rm W}$ is unity,
\begin{equation}
|\Lambda_{\rm W}:\Lambda_{\triangle_\bullet}| \,=\,
\det(\triangle_\bullet) \,=\, 1\,.
\end{equation}
The finite group $\Gamma_{\triangle_\bullet}$ is therefore
trivial in this example,
\begin{equation}
\Gamma_{\triangle_\bullet} \,\simeq\, \{1\}\,.
\end{equation}
The dual lattice ${\Lambda^{\triangle_{\bullet}}\simeq\Lambda^{\rm
    R}}$ is identical to the coroot lattice and is explicitly generated by 
\begin{equation}
\begin{aligned}
&\triangle_{1;j_1 j_2}:\quad \tau_1\,=\,-\hat h_1-\hat h_2\,,\qquad
\tau_2\,=\,-\hat h_2\,,\\
&\triangle_{2;j_1 j_2}:\quad \tau_1\,=\, -\hat h_1-\hat h_2\,,\qquad
\tau_2 \,=\, -\hat h_1-2\hat h_2\,,\\
&\triangle_{3;j_1 j_2}:\quad \tau_1\,=\, -\hat h_1-2\hat h_1\,,\qquad \tau_2\,=\,-\hat h_2\,,
\end{aligned}
\end{equation}
with degrees ${d_1=d_2=1}$ in all cases.

Since the group of characteristics for ${\mathbf{r},\mathbf{s}}$ is trivial, $Z_{S^3}$ decomposes
via \eqref{FactoZS3III} as a sum over ${|\CI|=3 N_{\rm f}^2}$ holomorphic blocks,
\begin{equation}
Z_{S^3} \,=\, C_{G_2} \sum_{a=1}^3 \sum_{j_1,j_2=1}^{N_{\rm f}}\exp{\!\left(
    i\,\Theta^{\triangle_{a;j_1
        j_2}}\right)}\,\RB^{\triangle_{a;j_1
    j_2}}(\Rq,\Rx)\,\wt\RB^{\triangle_{a;j_1 j_2}}(\wt\Rq,\wt\Rx)\,.
\end{equation}

\paragraph{Blocks for $Spin(4)$ gauge theory with vector matter.}  The
Lie algebra for $Spin(4)$ is not simple, but we include this example
nonetheless.  We consider $N_{\rm f}$ chiral multiplets which
transform in the fundamental, vector representation of $Spin(4)$,
\begin{equation}
\Lambda \,=\, \mathbf{4}^{N_{\rm f}}\,,\qquad\qquad G\,=\,Spin(4)\,.
\end{equation}
The fundamental representation has four non-zero weights, so
$\Delta_\Lambda^+$ contains only two elements.  Modulo flavor indices,
only one admissible double exists,
\begin{equation}
\triangle_{j_1 j_2}
\,=\,\big\{\left(-\hat\omega^1+\hat\omega^2\right){}_{\!j_1}\,,\left(-\hat\omega^1-\hat\omega^2\right){}_{\!j_2}\big\}\,,\qquad\qquad
j_1,j_2=1,\ldots,N_{\rm f}\,.
\end{equation}
This example, like the preceding example for $SU(3)$, is degenerate in
the sense that the Weyl vector $\rho$ lies as a weight in the
fundamental representation.  Equivalently, $-\rho$ lies on the
boundary of the positive cone $\R_+[\triangle_{j_1 j_2}]$ generated by
the elements in $\triangle_{j_1 j_2}$.  We make the choice for Jordan
divisors in Table \ref{Rk2JrdDiv} to break the degeneracy.

On the other hand,
\begin{equation}
|\Lambda_{\rm W}:\Lambda_{\triangle_{j_1 j_2}}| \,=\,
  \det(\triangle_{j_1 j_2}) \,=\, 2\,,
\end{equation}
so the group $\Gamma_{\triangle_{j_1 j_2}}$ is non-trivial,
\begin{equation}
\Gamma_{\triangle_{j_1 j_2}} \,\simeq\, \Z/2\Z\,.
\end{equation}
In terms of simple coroots, the dual lattice $\Lambda^{\triangle_{j_1
    j_2}}=\Z[\tau_1,\tau_2]$ is generated by 
\begin{equation}
\tau_1 \,=\, -\ha \hat h_1 + \ha \hat h_2\,,\qquad\qquad \tau_2\,=\,
-\ha \hat h_1 - \ha \hat h_2\,,
\end{equation}
with degrees
\begin{equation}
d_1 \,=\, d_2 \,=\, 2\,.
\end{equation}
Following \eqref{Bultsgm}, the set of poles in
$\h$ which contribute to the holomorphic blocks occur at locations
\begin{equation}\label{Spin4Pole}
\begin{aligned}
\triangle_{j_1 j_2}:\quad&\sigma_\bullet = -2\pi i\left(\mathbf{r}\,b + \mathbf{s}\,b^{-1} +
  \mathbf{M}\,b + \mathbf{N}\,b^{-1}\right) -\, 2\pi \mu_{j_1}\tau_1
\,-\, 2\pi\mu_{j_2}\,\tau_2\,,\\
&\mathbf{M},\mathbf{N}\in\big\{2m\,\tau_1+2n\,\tau_2\,\big|\,
m,n\in\Z_{\ge 0}\}\subset \Lambda^{\rm R}\,,\\
&\mathbf{r},\mathbf{s}\in(\Z/2\Z)\tau_1 \times (\Z/2\Z)\tau_2\,.
\end{aligned}
\end{equation}

By contrast with the preceding $G_2$ example, the group of characteristics for
$\mathbf{r},\mathbf{s}$ is non-trivial while the set of admissible
doubles is trivial.  The partition function decomposes as a sum over
${|\CI|=4^2 N_{\rm f}^2}$ blocks,
\begin{equation}
Z_{S^3} \,=\, \frac{C_{Spin(4)}}{2}\sum_{j_1,j_2=1}^{N_{\rm f}}
\sum_{\mathbf{r},\mathbf{s}\in (\Z/2\Z)\tau_1\times(\Z/2\Z)\tau_2} \exp{\!\left(
    i\,\Theta^{\triangle_{j_1
        j_2}}_{\mathbf{r},\mathbf{s}}\right)}\,\RB^{\triangle_{j_1
    j_2}}_{\mathbf{r},\mathbf{s}}(\Rq,\Rx)\,\wt\RB^{\triangle_{j_1 j_2}}_{\mathbf{r},\mathbf{s}}(\wt\Rq,\wt\Rx)\,.
\end{equation}

\paragraph{Blocks for $Spin(5)$ gauge theory with vector and spinor
  matter.}

By virtue of the rank-two identification ${Spin(5)\simeq Sp(4)}$, both
vector and spinor representations count as weight-multiplicity-free
fundamental representations.  We analyze the block decomposition in
either case.

We first consider purely vector matter,
\begin{equation}
\Lambda \,=\, \mathbf{5}^{\rm N_f}\,,\qquad\qquad G = Spin(5)\,.
\end{equation}
The vector representation has four non-zero weights (the fifth weight vanishes),
\begin{equation}
\Delta_\Lambda \,=\, \big\{\hat{\omega}^1,-\hat{\omega}^1+2\hat{\omega}^2\,,\hat{\omega}^1-2\hat{\omega}^2,-\hat{\omega}^1 \big\}\,.\qquad\qquad \big[N_{\rm f}=1\big]
\end{equation}
Modulo flavor indices, there is precisely one admissible double in
this case,
\begin{equation}
\triangle_{j_1 j_2} \,=\,
\big\{\left(-\hat\omega^1\right){}_{\!j_1}\,,\left(\hat\omega^1-2\hat\omega^2\right){}_{\!j_2}\big\}\,,\qquad\qquad
j_1,j_2=1,\ldots,N_{\rm f}\,.
\end{equation}
Like the example for $Spin(4)$,
\begin{equation}
|\Lambda_{\rm W}:\Lambda_{\triangle_{j_1 j_2}}| \,=\,
\det(\triangle_{j_1 j_2}) \,=\, 2\,,
\end{equation}
and
\begin{equation}
\Gamma_{\triangle_{j_1 j_2}}\,\simeq\,\Z/2\Z\,.
\end{equation}
Generators for the dual lattice ${\Lambda^{\triangle_{j_1
      j_2}}=\Z[\tau_1,\tau_2]}$ are
\begin{equation}
\tau_1 \,=\, -\hat h_1 - \ha\hat h_2\,, \qquad\qquad \tau_2 \,=\, -\ha
\hat h_2\,,
\end{equation} 
with degrees 
\begin{equation}
d_1 \,=\, d_2 \,=\, 2\,.
\end{equation}
Therefore the set of poles in $\h$ which contribute to the holomorphic
blocks occur at locations
\begin{equation}
\begin{aligned}
\triangle_{j_1 j_2}:\quad&\sigma_\bullet = -2\pi i\left(\mathbf{r}\,b + \mathbf{s}\,b^{-1} +
  \mathbf{M}\,b + \mathbf{N}\,b^{-1}\right) -\, 2\pi \mu_{j_1}\tau_1
\,-\, 2\pi\mu_{j_2}\,\tau_2\,,\\
&\mathbf{M},\mathbf{N}\in\big\{2m\,\tau_1+2n\,\tau_2\,\big|\,
m,n\in\Z_{\ge 0}\}\subset \Lambda^{\rm R}\,,\\
&\mathbf{r},\mathbf{s}\in(\Z/2\Z)\tau_1 \times (\Z/2\Z)\tau_2\,,
\end{aligned}
\end{equation}
identical in form to those \eqref{Spin4Pole} of $Spin(4)$.

Finally, the partition function decomposes as a sum over ${|\CI|=4^2
  N_{\rm f}^2}$ blocks,
\begin{equation}
Z_{S^3} \,=\, \frac{C_{Spin(5)}}{2}\sum_{j_1,j_2=1}^{N_{\rm f}}
\sum_{\mathbf{r},\mathbf{s}\in (\Z/2\Z)\tau_1\times(\Z/2\Z)\tau_2} \exp{\!\left(
    i\,\Theta^{\triangle_{j_1
        j_2}}_{\mathbf{r},\mathbf{s}}\right)}\,\RB^{\triangle_{j_1
    j_2}}_{\mathbf{r},\mathbf{s}}(\Rq,\Rx)\,\wt\RB^{\triangle_{j_1 j_2}}_{\mathbf{r},\mathbf{s}}(\wt\Rq,\wt\Rx)\,.
\end{equation}

By contrast, consider $Spin(5)$ gauge theory with spinor matter,
\begin{equation}
\Lambda = \mathbf{4}^{N_{\rm sp}}\,,\qquad\qquad G\,=\,Spin(5)\,.
\end{equation}
The spinor representation has weights
\begin{equation}
\Delta_\Lambda \,=\,  \big\{\hat{\omega}^2,\,\,-\hat{\omega}^2 ,\,\,
\hat{\omega}^1-\hat{\omega}^2,\,\,-\hat{\omega}^1+\hat{\omega}^2\big\}\,.\qquad\qquad
\big[N_{\rm sp}=1\big]
\end{equation}
Trivially, there is one admissible double modulo flavor indices,
\begin{equation}
\triangle_{j_1 j_2} \,=\,
\big\{\left(-\hat\omega^1+\hat\omega^2\right){}_{\!j_1}\,,\left(-\hat\omega^2\right){}_{\!j_2}\big\}\,,\qquad\qquad
j_1,j_2=1,\ldots,N_{\rm sp}\,.
\end{equation}
Unlike the case for vector matter,
\begin{equation}
|\Lambda_{\rm W}:\Lambda_{\triangle_{j_1 j_2}}| \,=\,
\det(\triangle_{j_1 j_2}) \,=\, 1\,,
\end{equation}
so
\begin{equation}
\Gamma_{\triangle_{j_1 j_2}} \,\simeq\, \{1\}\,.
\end{equation}
The lattice ${\Lambda^{\triangle_{j_1 j_2}}=\Z[\tau_1,\tau_2]}$ is
generated by 
\begin{equation}
\tau_1 \,=\, -\hat h_1\,,\qquad\qquad \tau_2 \,=\, -\hat h_1 - \hat h_2\,,
\end{equation} 
each with degree ${d_1=d_2=1}$.  The group of characteristics for
$\mathbf{r},\mathbf{s}$ is therefore trivial.
Hence the $Spin(5)$ partition function with spinor matter decomposes
as a sum over only ${|\CI|=N_{\rm sp}^2}$ blocks,
\begin{equation}
Z_{S^3} \,=\, C_{Spin(5)} \sum_{j_1,j_2=1}^{N_{\rm sp}}
 \exp{\!\left(
    i\,\Theta^{\triangle_{j_1
        j_2}}\right)}\,\RB^{\triangle_{j_1
    j_2}}(\Rq,\Rx)\,\wt\RB^{\triangle_{j_1 j_2}}(\wt\Rq,\wt\Rx)\,.
\end{equation}

\paragraph{Blocks for $SU(N)$ SQCD.}

We have already classified admissible tuples for $SU(N)$ SQCD in
\eqref{SUNadmd}.  We now use this classification to determine the
structure of the SQCD block decomposition.

Including flavor indices ${j_1,\cdots,j_{N-1}=1,\ldots,N_{\rm f}}$,
the admissible tuples are given up to orientation by 
\begin{equation}\label{SUNadmdII}
\begin{aligned}
\triangle_{1;j_1 \cdots j_{N-1}}
&=\!\left([-\hat\omega^1+\hat\omega^2]_{j_1}, [-\hat\omega^2+\hat\omega^3]_{j_2},
  [-\hat\omega^3+\hat\omega^4]_{j_3},\cdots,
  [-\hat\omega^{N-2}+\hat\omega^{N-1}]_{j_{N-2}}, [-\hat\omega^{N-1}]_{j_{N-1}}\right),\\
\triangle_{2;j_1 \cdots j_{N-1}} &=\!\left([-\hat\omega^1]_{j_1},
  [-\hat\omega^2+\hat\omega^3]_{j_2}, [-\hat\omega^3+\hat\omega^4]_{j_3},\cdots,[-\hat\omega^{N-2}+\hat\omega^{N-1}]_{j_{N-2}},[-\hat\omega^{N-1}]_{j_{N-1}}\right),\\
\triangle_{3;j_1 \cdots j_{N-1}} &=\!\left([-\hat\omega^1]_{j_1},
  [\hat\omega^1-\hat\omega^2]_{j_2}, [-\hat\omega^3+\hat\omega^4]_{j_3},\cdots,[-\hat\omega^{N-2}+\hat\omega^{N-1}]_{j_{N-2}},[-\hat\omega^{N-1}]_{j_{N-1}}\right),\\
&\,\,\vdots\\
\triangle_{a;j_1 \cdots j_{N-1}} &=\!\left([-\hat\omega^1]_{j_1},
  [\hat\omega^1-\hat\omega^2]_{j_2},\cdots,[\hat\omega^{a-2}-\hat\omega^{a-1}]_{j_{a-1}},[-\hat\omega^{a}+\hat\omega^{a+1}]_{j_a},
  \cdots,[-\hat\omega^{N-1}]_{j_{N-1}}\right),\\
&\,\,\vdots\\
\triangle_{N-1;j_1 \cdots j_{N-1}} &=\!\left([-\hat\omega^1]_{j_1},
  [\hat\omega^1-\hat\omega^2]_{j_2},\cdots,[\hat\omega^{N-3}-\hat\omega^{N-2}]_{j_{N-2}},[-\hat\omega^{N-1}]_{j_{N-1}}\right),\\
\triangle_{N;j_1 \cdots j_{N-1}} &=\!\left([-\hat\omega^1]_{j_1},[\hat\omega^1-\hat\omega^2]_{j_2},\cdots,[\hat\omega^{N-3}-\hat\omega^{N-2}]_{j_{N-2}},[\hat\omega^{N-2}-\hat\omega^{N-1}]_{j_{N-1}}\right).
\end{aligned}
\end{equation}
In total, there are ${N N_{\rm f}^{N-1}}$ admissible tuples.

Each tuple has index
\begin{equation}
|\Lambda_{\rm W}:\Lambda_{\triangle_{\bullet}}| \,=\,
\det(\triangle_\bullet) \,=\, 1\,,
\end{equation}
so
\begin{equation}
\Gamma_{\triangle_\bullet} \,=\, \{1\}\,.
\end{equation}
For eg.~the first tuple $\triangle_{1;j_1\cdots j_{N-1}}$, the dual lattice
${\Lambda^{\triangle_{1;j_1 \cdots
      j_{N-1}}}=\Z[\tau_1,\cdots,\tau_{N-1}]}$ has generators
\begin{equation}
\begin{aligned}
\tau_1 \,&=\, -\hat h_1\,,\\
\tau_2 \,&=\, -\hat h_1\,-\,\hat h_2\,,\\
\tau_3 \,&=\, -\hat h_1\,-\,\hat h_2\,-\,\hat h_3\,,\\
&\,\,\,\vdots\,\\
\tau_{N-2} \,&=\, -\hat h_1\,-\,\hat h_2\,-\,\hat h_3 \,-\, \cdots
\,-\, \hat h_{N-2}\,,\\
\tau_{N-1} \,&=\,-\hat h_1\,-\,\hat h_2\,-\,\hat h_3 \,-\, \cdots
\,-\, \hat h_{N-2}\,-\, \hat h_{N-1}\,,
\end{aligned}
\end{equation}
each with degrees
\begin{equation}
d_1 \,=\, d_2 \,=\, \cdots \,=\, d_{N-1} \,=\, 1\,.
\end{equation}
More generally, because $\Lambda^{\triangle_\bullet}\simeq\Lambda^{\rm
    R}$ for each admissible tuple, 
$\tau_1,\ldots,\tau_{N-1}$ have unit degree in every case.
Consequently the group of characteristics is trivial for all
$\triangle_{a;j_1 \cdots j_{N-1}}$, ${a=1,\ldots,N}$.

Thus the SQCD partition function decomposes as a sum over
${|\CI|=N N_{\rm f}^{N-1}}$ blocks,
\begin{equation}
Z_{S^3} \,=\, C_{SU(N)} \sum_{a=1}^N \sum_{j_1,\,\cdots,j_{N-1}=1}^{N_{\rm f}}
 \exp{\!\left(
    i\,\Theta^{\triangle_{a;j_1\cdots
        j_{N-1}}}\right)}\,\RB^{\triangle_{a;j_1\cdots
    j_{N-1}}}(\Rq,\Rx)\,\wt\RB^{\triangle_{a;j_1 \cdots j_{N-1}}}(\wt\Rq,\wt\Rx)\,.
\end{equation}
\appendix
\section{Lie Algebra Conventions}\label{se:LieAlgConvention}

We record our conventions for Lie groups and Lie algebras.  A
basic reference is \cite{Fulton}.

The Lie group $G$ is compact, connected, simply-connected,
and simple.  We fix a maximal torus ${T \subset G}$.  By assumption
${T \simeq U(1)^r}$, where ${r=\rk(G)}$ is the rank of $G$.  The
associated Cartan subalgebra is ${\h\subset\g}$, and the Weyl group
${\fW}$ acts on $\h$ by outer automorphisms induced from conjugation in $G$.

The set of roots is denoted by $\Delta$.  Individual roots
  ${\alpha\in\h^*}$ are valued in the dual of the Cartan subalgebra.
  The complexification ${\g_\C = \g \otimes \C}$ admits a rootspace
  decomposition which diagonalizes the adjoint action of $\h$,
\begin{equation}
\g_\C \,=\, \h_\C \oplus \bigoplus_{\alpha\in\Delta}
\g_\alpha\,.
\end{equation}
For any elements ${h\in\h}$ and ${x\in\g_\alpha}$, 
\begin{equation}\label{Rootsp}
\left[h,x\right] \,=\, i\,\langle\alpha,h\rangle\,x\,.
\end{equation}
Here ${\langle\,\cdot\,\,,\,\cdot\,\rangle}$ indicates the canonical
pairing between $\h^*$ and $\h$.  The factor of `$i$' in
\eqref{Rootsp} is consistent with the convention that elements of $\g$
be represented by anti-hermitian matrices.  The lattice
${\Lambda_{\rm R}\subset\h^*}$ generated by $\Delta$ is the root
lattice of $G$.

Each rootspace $\g_\alpha$ is one-dimensional.  We select
  generators ${e_\alpha\in\g_\alpha}$ so that the triple below,
\begin{equation}
e_\alpha\in\g_\alpha\,,\qquad
e_{-\alpha}\,=\,\bar{e_\alpha}\in\g_{-\alpha}\,,\qquad h_\alpha\,=\,
    -i\,[e_\alpha, e_{-\alpha}]\in\h\,,
\end{equation}
satisfies the canonical $\mathfrak{sl}_2(\C)$ algebra
\begin{equation}\label{Sl2C}
[h_\alpha, e_\alpha] \,=\, 2 i\,e_\alpha\,,\qquad [h_\alpha,
e_{-\alpha}] \,=\, -2 i\,e_{-\alpha}\,,\qquad [e_\alpha, e_{-\alpha}] =
i\,h_\alpha\,.
\end{equation}
Concretely, $\{e_\alpha, e_{-\alpha}, h_\alpha\}$ correspond under 
algebra isomorphism to the respective ${2\times 2}$ matrices
\begin{equation}
e = \begin{pmatrix}0&1\\0&0\end{pmatrix},\qquad f = \begin{pmatrix}0&0\\1&0\end{pmatrix}\,,\qquad h = \begin{pmatrix}i&0\\0&-i\end{pmatrix}\,.
\end{equation}
With this convention, the choice of ${e_\alpha\in\g_\alpha}$ is fixed up to a
phase.  The coroot ${h_\alpha\in\h}$ is then uniquely determined by \eqref{Sl2C}
and satisfies ${\langle\alpha,h_\alpha\rangle = 2}$.  

The elements $\{h_\alpha\}$ for ${\alpha\in\Delta}$ generate the
coroot lattice ${\Lambda^{\rm R}\subset\h}$. 
Because $G$ is simply-connected, $\Lambda^{\rm R}$ is isomorphic to
the lattice of homomorphisms from $U(1)$ to $T$,
\begin{equation}\label{Coroot}
\Lambda^{\rm R} \simeq \Hom(U(1),T)\,.
\end{equation}

We next choose a decomposition of the roots into
  positive and negative subsets, 
\begin{equation}\label{Ordering}
\Delta \,=\, \Delta_+\!\cup \Delta_-\,,\qquad\qquad \Delta_- = -\Delta_+\,.
\end{equation}
Each ${\alpha\in\Delta_+}$ can be written uniquely as a positive integral
combination of simple roots ${\hat\alpha^1,\ldots,\hat\alpha^r
  \in \Delta_+}$, which provide a basis for $\h^*$.  The set of
positive roots contains a distinguished highest root ${\vartheta\in\Delta_+}$,
determined by the condition ${[e_\alpha,e_{\vartheta}]=0}$ for all
${\alpha\in\Delta_+}$.

The root decomposition \eqref{Ordering} also determines a positive Weyl
chamber ${\mathcal{C}_+ \subset \h}$.  By definition, $\mathcal{C}_+$
consists of those ${h\in\h}$ for which
${\langle\hat\alpha^j,h\rangle \ge 0}$ for 
all ${j=1,\ldots,r}$.  The Weyl group $\fW$ acts transitively by
permutations on the set of Weyl chambers, and each chamber is a convex
polyhedral cone with dimension ${r = \rk(G)}$.

The Lie algebra $\g$ is equipped with an
  invariant, {\sl 
    negative-definite} quadratic form `$\Tr$' which defines a metric
\begin{equation}\label{Met}
\left(x,y\right) := -\Tr(x y)\,,\qquad\qquad x,y\,\in\,\g\,.
\end{equation}
Because $\g$ is simple, any invariant metric is unique up to normalization.
The form `$\Tr$' is normalized so that the highest root $\vartheta$ has
length $\sqrt{2}$,
\begin{equation}
(\vartheta,\vartheta) \,=\, 2\,.
\end{equation}
For a simple Lie algebra, roots have at most
two possible lengths, either ``long'' or ``short''.  The highest root
$\vartheta$ is always long.  Short roots have ${(\alpha,\alpha)=2/n}$ for
${n=1,2,3}$.  The algebra $\g$ is simply-laced when all roots have the 
same length (ie.~${n=1}$).  Through the isomorphism ${\g\simeq\g^*}$ induced by the
metric, coroots and roots are related by ${h_\alpha =
  2\alpha/(\alpha,\alpha)}$, and ${(h_\alpha,h_\alpha) =
  4/(\alpha,\alpha)}$.

Under the exponential map, the weight lattice ${\Lambda_{\rm
    W}\subset\h^*}$ is identified with the character lattice
\begin{equation}
\Lambda_{\rm W} \simeq\,\Hom(T,U(1))\,.
\end{equation}
By comparison to \eqref{Coroot}, $\Lambda_{\rm W}$ is canonically dual
to $\Lambda^{\rm R}$ over $\Z$.  Consequently, 
the weight lattice is generated by fundamental
weights ${\hat\omega^1,\ldots,\hat\omega^r\in\h^*}$ dual to the
coroots ${\hat h_1 \equiv h_{\hat\alpha^1},\ldots,\hat h_r\equiv h_{\hat\alpha^r}}$
associated to the positive simple roots, ie.
\begin{equation}\label{FWeig}
\langle\hat\omega^j,\,\hat h_\ell\rangle \,=\, \delta^j_\ell\,,\qquad\qquad j,\ell=1\,,\ldots,r\,.
\end{equation}
The root lattice ${\Lambda_{\rm R} \subseteq \Lambda_{\rm W}}$ is a
sublattice of finite index in the weight lattice,
\begin{equation}
\left[\Lambda_{\rm W}:\Lambda_{\rm R}\right] = |\mathfrak{z}_G|\,,
\end{equation}
where the index is given by the order of the center $\mathfrak{z}_G$ of $G$.

\paragraph{Representations and Casimirs.}

Let $V$ be an irreducible representation of $G$
with highest weight ${\lambda\in\h^*}$.  Since $G$ is compact, $V$ has
finite dimension automatically.  Implicitly, $V$ is equipped with a
Lie algebra homomorphism 
${\varphi:\g\to\End(V)}$.  When the action of $\h$ on $V$ is
diagonalized, the complexification ${V_\C=V\otimes\C}$ splits into
weight spaces  
\begin{equation}
V_\C \,=\, \bigoplus_{\beta \in \Delta_V} V_\beta\,,
\end{equation}
where ${\Delta_V}$ is the set of weights for
$V$.  Each weight space $V_\beta$ is
one-dimensional (allowing for repeated weights in $\Delta_V$), so
${|\Delta_V|=\dim V}$.  The Weyl group $\fW$ 
permutes the weights in $\Delta_V$.  For any elements
${h\in\h}$ and ${v\in V_\beta}$, 
\begin{equation}\label{Weight}
\varphi(h)\cdot v \,=\, i \, \langle\beta,h\rangle\,v\,.
\end{equation}
The highest weight ${\lambda\in\Delta_V}$ is the unique weight for
which ${\langle\lambda,h_\alpha\rangle\ge 0}$ for all
${\alpha\in\Delta_+}$.  A generator ${v_\lambda\in V_\lambda}$ is a
highest-weight vector, which is annihilated by the action of each
raising-operator $e_\alpha$ for ${\alpha\in\Delta_+}$,
\begin{equation}\label{Annpsi}
\varphi(e_\alpha)\cdot v_\lambda \,=\, 0\,,\qquad\qquad \alpha\,\in\,\Delta_+\,.
\end{equation}

Casimir operators are central elements in the universal
enveloping algebra $\mathcal{U}(\g)$. Let $\{t_a\}$ for
${a=1,\ldots,\dim\g}$ be a basis of $\g$, and let ${g_{a b} = (t_a,
  t_b)}$ be the matrix of inner-products with respect to this basis.
Any simple Lie algebra admits the quadratic Casimir operator 
\begin{equation}
\CO_2 \,=\, \sum_{a,b=1}^{\dim\g}\,g^{a b}\,t_a t_b \,,\qquad\qquad g^{a
  b} = (g_{a b})^{-1}\,.
\end{equation}
We define the quadratic Casimir of the representation
$V$ by the trace
\begin{equation}\label{CtwoV}
c_2(V) \,=\, -\frac{\Tr_{V}\!\big[\varphi(\CO_2)\big]}{\dim\g}\,.
\end{equation}
Because we divide by the dimension of $\g$ in the normalization of
$c_2(V)$, the Casimir $c_2(V)$ satisfies
\begin{equation}\label{CtwoVII}
-\Tr_V\!\big[\varphi(t_a)\,\varphi(t_b)\big] \,=\,
c_2(V)\cdot g_{a b}\,.
\end{equation}
The negative sign in \eqref{CtwoVII} agrees with the convention in
\eqref{Met} and ensures that ${c_2(V)\ge 0}$ is positive.
Since $V$ is specified by the highest weight $\lambda$, we frequently
write ${c_2(\lambda) \equiv c_2(V)}$ in the body of the paper.  By the
Schur Lemma, $\CO_2$ acts on $V$ as a scalar multiple of the identity,
\begin{equation}\label{Schur}
\varphi(\CO_2) \,=\, -\psi(V)\cdot{\bf 1}_{V}\,,\qquad\qquad \psi(V)\,\in\,\R\,,
\end{equation}
whence\footnote{Some authors distinguish $\psi(V)$ as the Casimir and
  $c_2(V)$ as the index of the representation $V$.  Because $\psi(V)$
  plays no role for us, we just refer to $c_2(V)$ as the Casimir.}
\begin{equation}
c_2(V) \,=\, \psi(V)\cdot\frac{\dim V}{\dim\g}\,.
\end{equation}

For a direct sum of
representations,
\begin{equation}
c_2(V\oplus W) \,=\, c_2(V) + c_2(W)\,,
\end{equation}
and for the tensor product,
\begin{equation}\label{TensorP}
c_2(V\otimes W) \,=\, c_2(V) \dim W \,+\, c_2(W) \dim V\,.
\end{equation}
Also, if ${V^*}$ is the representation dual to $V$,
\begin{equation}\label{DualV}
c_2(V^*) = c_2(V)\,.
\end{equation}

To evaluate $c_2(V)$, a useful basis for $\g$ is given by the simple coroots
${\hat h_1\,,\ldots\,,\hat h_r}$, along with
the raising/lowering pairs ${(e_\alpha, e_{-\alpha})}$ for
${\alpha\in\Delta_+}$.  The $\mathfrak{sl}_2(\C)$ algebra in
\eqref{Sl2C} and the invariance of the metric on $\g$ together imply
\begin{equation}\label{RootPR}
\left(e_{-\alpha}, e_\alpha\right) =\, \frac{i}{2}
\left([h_\alpha,e_{-\alpha}],e_\alpha\right) = \frac{i}{2}
\left(h_\alpha,[e_{-\alpha},e_\alpha]\right) = \ha \left(h_\alpha,
  h_\alpha\right) =\, 2/(\alpha,\alpha)\,.
\end{equation}
We set ${\RA_{j\ell} = (\hat h_j,\hat h_\ell)}$ for
${j,\ell=1,\ldots,r}$.  All other inner-products beyond
$\left(e_{-\alpha}, e_\alpha\right)$ and $(\hat h_j,\hat
  h_\ell)$ vanish.  Thus
\begin{equation}\label{QCasII}
\CO_2 \,=\, \sum_{j,\ell=1}^r {\RA}^{j \ell}\,\hat h_{j}
\hat h_{\ell} \,+\, \sum_{\alpha\in\Delta_+}
\frac{(\alpha,\alpha)}{2}\left(e_{-\alpha}\,e_\alpha \,+\, e_\alpha\,
  e_{-\alpha}\right).
\end{equation}
The constant $\psi(V)$ in \eqref{Schur} can be evaluated by acting
with $\varphi(\CO_2)$ on the highest-weight vector $v_\lambda$,
\begin{equation}\label{BigPsiV}
\begin{aligned}
\varphi(\CO_2)\cdot v_\lambda \,=\,
-\left(\lambda,\lambda+2\rho\right)\cdot v_\lambda
\quad\Longleftrightarrow\quad \psi(V) \,=\, \left(\lambda,\lambda+2\rho\right),
\end{aligned}
\end{equation}
where ${\rho\in\h^*}$ is the Weyl element,
\begin{equation}
\rho \,=\, \ha \sum_{\alpha\in\Delta_+}\alpha\,.
\end{equation}
The formula in \eqref{BigPsiV} follows from the expression for $\CO_2$
in \eqref{QCasII}, along with the defining conditions in \eqref{Sl2C},
\eqref{Weight}, and \eqref{Annpsi}.  Thus
\begin{equation}\label{C2Vform}
c_2(V) \,=\, \left(\lambda,\lambda+2\rho\right)\cdot\frac{\dim V}{\dim\g}\,.
\end{equation}
Via the Weyl character formula, the dimension of $V$
can also be expressed algebraically in terms of the highest weight
$\lambda$,
\begin{equation}\label{WeylDim}
\dim V \,=\, \prod_{\alpha\in\Delta_+} \frac{\left(\lambda+\rho,\alpha\right)}{\left(\rho,\alpha\right)}\,.
\end{equation}

As a universal example, for the adjoint representation ${V=\g}$, the highest
weight ${\lambda=\vartheta}$ is the highest root.  So with our
conventions,
\begin{equation}
c_2(\g) \,=\, \left(\vartheta,\vartheta+2\rho\right) =\, 2\,h_\g\,.
\end{equation}
Here $h_\g$ is the dual Coxeter number of $\g$, 
\begin{equation}
(\vartheta,\rho) \,=\, h_\g - 1\,.
\end{equation}

The normalization for $c_2(V)$ has two consequences.  First,
after the definitions are unraveled, the relation in \eqref{CtwoVII}
means that the normalized metric on $\g$ is given by 
\begin{equation}\label{Normg}
(x,y) = -\frac{1}{c_2(V)}\Tr_V\!\big[\varphi(x)
\varphi(y)\big]\,,\qquad\qquad x,y\in\g\,.
\end{equation} 
If ${c_2(V)=1}$, the normalized metric is simply the trace in $V$.
The fundamental and anti-fundamental representations
of $SU(N)$ provide the basic examples,
\begin{equation}
c_2({\bf N}) \,=\, c_2(\bar{\bf N}) \,=\, 1\,.
\end{equation}

Second, as appears in \eqref{QadC} in Section \ref{OneLoop}, we have the identity
\begin{equation}\label{QadCII}
\sum_{\beta\in\Delta_V} \langle\beta,x\rangle^2 \,=\, c_2(V)\cdot
(x,x),\qquad\qquad x\,\in\,\h\,. 
\end{equation}
To prove this identity, endow $V$ with a $G$-invariant metric and 
pick unit generators ${v_\beta\in V_\beta}$ for each weight space,
\begin{equation}
(v_\beta,v_\beta) \,=\, 1\,,\qquad\qquad \beta\in \Delta_V\,.
\end{equation}
The set $\{v_\beta\}$ provides a normalized eigenbasis for ${x\in\h}$ acting
on $V$, so via \eqref{Weight}
\begin{equation}\label{QadCIII}
\sum_{\beta\in\Delta_V} \langle\beta,x\rangle^2 \,=\,
-\sum_{\beta\in\Delta_V} \left(v_\beta,\,\varphi(x)^2\cdot
  v_\beta\right) \,=\, -\Tr_V\!\big[\varphi(x)^2\big]\,.
\end{equation}
The required identity \eqref{QadCII} follows immediately from the
relation in \eqref{Normg}, with ${x=y}$.

\paragraph{Conventions for $SU(3)$.}

We make these conventions explicit for ${G=SU(3)}$.
In this case, the positive simple coroots which span ${\h\simeq\R^2}$ can be
taken to be 
\begin{align}\label{CoRSU3}
\hat{h}_1 &=\begin{pmatrix}
i&0&0\\
0&-i&0\\
0&0&0
\end{pmatrix},&
\hat{h}_2 &=\begin{pmatrix}
0&0&0\\
0&i&0\\
0&0&-i
\end{pmatrix}.
\end{align}
For any diagonal matrix in $\h$ of the form ${i \diag(x_1,x_2,x_3)}$ with
${x_1+x_2+x_3=0}$, the positive simple roots
$\hat\alpha^1$ and $\hat\alpha^2$ are given by the respective
differences ${x_1 - x_2}$ and ${x_2 - x_3}$.  Hence 
\begin{equation}\label{SimpelR}
\begin{matrix}
\begin{aligned}
\langle\hat\alpha^1,\hat h_1\rangle \,&=\, 2\,,\\
 \langle\hat\alpha^2,\hat h_1\rangle \,&=\, -1\,,
\end{aligned} \qquad&\qquad
\begin{aligned}
\langle\hat\alpha^1,\hat h_2\rangle \,&=\, -1\,\\
\langle\hat\alpha^2,\hat h_2\rangle \,&=\, 2\,.
\end{aligned}
\end{matrix}
\end{equation}
The highest root $\vartheta$ evaluates the difference ${x_1 - x_3}$,
so 
\begin{equation}
\vartheta \,=\, \hat\alpha^1 + \hat\alpha^2\,,
\end{equation}
and the set of positive roots is given by 
\begin{equation}
\Delta_+ =
  \{\hat\alpha^1,\hat\alpha^2,\hat\alpha^1+\hat\alpha^2\}\,.
\end{equation}

\iffigs
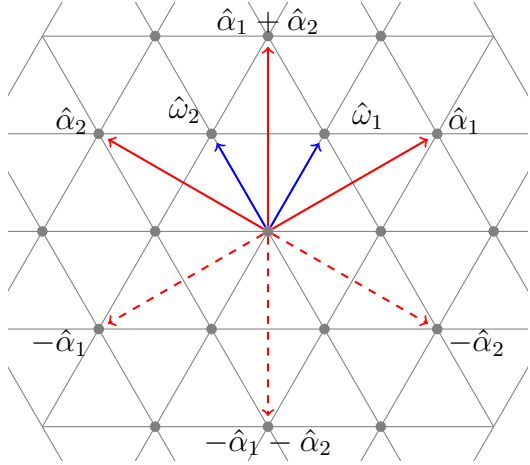
\begin{figure}[t]
	\centering
	\begin{tikzpicture}[x=1.5cm,y=1.5cm] 

	\coordinate (0;0) at (0,0); 
	\foreach \c in {1,...,4}{%  
		\foreach \i in {0,...,5}{% 
			\pgfmathtruncatemacro\j{\c*\i}
			\coordinate (\c;\j) at (60*\i:\c);  
		} }
		\foreach \i in {0,2,...,10}{% 
			\pgfmathtruncatemacro\j{mod(\i+2,12)} 
			\pgfmathtruncatemacro\k{\i+1}
			\coordinate (2;\k) at ($(2;\i)!.5!(2;\j)$) ;}
		
		\foreach \i in {0,3,...,15}{% 
			\pgfmathtruncatemacro\j{mod(\i+3,18)} 
			\pgfmathtruncatemacro\k{\i+1} 
			\pgfmathtruncatemacro\l{\i+2}
			\coordinate (3;\k) at ($(3;\i)!1/3!(3;\j)$)  ;
			\coordinate (3;\l) at ($(3;\i)!2/3!(3;\j)$)  ;
		}
		
		\foreach \i in {0,4,...,20}{% 
			\pgfmathtruncatemacro\j{mod(\i+4,24)} 
			\pgfmathtruncatemacro\k{\i+1} 
			\pgfmathtruncatemacro\l{\i+2}
			\pgfmathtruncatemacro\m{\i+3} 
			\coordinate (4;\k) at ($(4;\i)!1/4!(4;\j)$)  ;
			\coordinate (4;\l) at ($(4;\i)!2/4!(4;\j)$) ;
			\coordinate (4;\m) at ($(4;\i)!3/4!(4;\j)$) ;
		}

		\begin{scope}
		%\clip (0,0) circle (2.5);
		
		\clip ($(2;0)+(2;3)+(.3,.3)$)--($(2;6)+(2;3)+(-.3,.3)$)--($(2;6)+(2;9)+(-.3,-.3)$)--($(2;0)+(2;9)+(.3,-.3)$)--($(2;0)+(2;3)+(.3,.3)$);
		
		\foreach \i in {0,...,6}{% 
			\pgfmathtruncatemacro\k{\i}
			\pgfmathtruncatemacro\l{15-\i}
			\draw[thin,gray] (3;\k)--(3;\l);
			\pgfmathtruncatemacro\k{9-\i} 
			\pgfmathtruncatemacro\l{mod(12+\i,18)}   
			\draw[thin,gray] (3;\k)--(3;\l); 
			\pgfmathtruncatemacro\k{12-\i} 
			\pgfmathtruncatemacro\l{mod(15+\i,18)}   
			\draw[thin,gray] (3;\k)--(3;\l);} 
		
		\end{scope}

		\fill [gray] (0;0) circle (2pt);
		\foreach \c in {1,...,2}{%
			\pgfmathtruncatemacro\k{\c*6-1}    
			\foreach \i in {0,...,\k}{% 
				\fill [gray] (\c;\i) circle (2pt);}}

		\foreach \n in{1,3,5}{%
			\draw[->,red,thick,shorten >=4pt,shorten <=2pt](0;0)--(2;\n);
		}
		
		\foreach \n in{7,9,11}{%
			\draw[->,dashed, red,thick,shorten >=4pt,shorten <=2pt](0;0)--(2;\n);
		}		
		
		\node [ above right] at ($(2;1)+(0,-.1)$) {$\hat \alpha_1$};
		\node [ above left] at ($(2;5)+(0,-.1)$) {$\hat \alpha_2$};
		\node [ above] at ($(2;3)+(0,-.08)$) {$\hat \alpha_1+\hat \alpha_2$};
		
		\node [ below left] at ($(2;7)+(0,.1)$) {$-\hat \alpha_1$};
		\node [ below right] at ($(2;11)+(0,.1)$) {$-\hat \alpha_2$};
		\node [ below] at ($(2;9)+(0,.08)$) {$-\hat \alpha_1-\hat \alpha_2$};

		\draw[->,blue,thick,shorten >=4pt,shorten <=2pt](0;0)--(1;1);
		\draw[->,blue,thick,shorten >=4pt,shorten <=2pt](0;0)--(1;2);
		
		\node [ above right] at ($(1;1)+(.15,-.05)$) {$\hat\omega_1$};
		\node [ above left] at (1;2) {$\hat \omega_2$};
		
		\end{tikzpicture}  
		
		\caption{The weight lattice of $SU(3)$. The positive simple roots are $\hat \alpha_{1,2}$ and the fundamental weights are $\hat\omega_{1,2}$.}\label{SU3diag}
	\end{figure}
	\fi

The fundamental weights $\hat\omega^1$ and $\hat\omega^2$ are
determined as the duals \eqref{FWeig} to the positive simple coroots
$\hat h_1$ and $\hat h_2$.  Thus from \eqref{SimpelR}, 
\begin{equation}\label{WtsSU3}
\hat\omega^1 \,=\, \frac{2}{3}\hat\alpha^1 \,+\,
\frac{1}{3}\hat\alpha^2\,,\qquad\qquad \hat\omega^2 \,=\,
\frac{1}{3}\hat\alpha^1 \,+\, \frac{2}{3}\hat\alpha^2\,.
\end{equation}
Any highest weight $\lambda$ is a positive integral combination of fundamental weights,
\begin{equation}\label{MaxLamSU}
\lambda \,=\, m \, \hat\omega^1 \,+\,
n\,\hat\omega^2\,,\qquad\qquad m,n\ge 0\,.
\end{equation}
We let $V_{m,n}$ be the associated irreducible
representation.  In the intrinsic labelling by the weight lattice, the
fundamental, anti-fundamental, and adjoint representations of
$SU(3)$ are  respectively
\begin{equation}
V_{1,0} \,\equiv\, {\bf 3}\,,\qquad\quad V_{0,1} \,\equiv\,\bar{\bf
  3}\,,\qquad\quad V_{1,1} \,\equiv\,{\bf 8}\,.
\end{equation}

The properly normalized metric on the Lie algebra of $SU(3)$ is given by the 
trace in the fundamental representation $V_{1,0}$,
\begin{equation}\label{MetSU3}
\left(x,y\right) \,=\, -\Tr_{V_{1,0}}\!\left[x\,y\right]\,,\qquad\qquad x,y\,\in\,\g\,.
\end{equation}
We denote the matrix of inner-products with
respect to the coroot basis in \eqref{CoRSU3} by 
\begin{align}
\RA_{j \ell} \,=\, (\hat{h}_j,\hat{h}_\ell) \,=\,
\begin{pmatrix}
2&-1
\\
-1&2
\end{pmatrix},\qquad\qquad j,\ell=1,2\,.
\end{align}
Tautologically, the matrix of inner-products for the dual fundamental weights is
the inverse of $\RA$,
\begin{equation}\label{QuadSU3}
\left(\RA^{-1}\right){}^{\!j \ell} \,=\, (\hat\omega^j,\hat\omega^\ell)
\,= \frac{1}{3}\begin{pmatrix}
2&1
\\
1&2
\end{pmatrix}.
\end{equation}
From the relation between roots and weights in \eqref{WtsSU3} and the
presentation of the inner-product in \eqref{QuadSU3}, one can directly
check that all roots of $SU(3)$ have length $\sqrt{2}$, including the
highest root $\vartheta$.  This statement confirms the claim in \eqref{MetSU3}.

For completeness, we evaluate the Casimir invariant $c_2(V_{m,n})$ as
a function of the positive integers ${m,n\in\Z_{\ge 0}}$ in \eqref{MaxLamSU}.  In terms of
the fundamental weights, the Weyl element is given by 
\begin{equation}
\rho \,=\, \ha \sum_{\alpha\in\Delta_+}\alpha \,=\, \hat\omega^1 + \hat\omega^2\,.
\end{equation}
By a small calculation using the inner-product in \eqref{QuadSU3},
\begin{equation}
\psi\!\left(V_{m,n}\right) = \left(\lambda,\lambda+2\rho\right) \,=\,
\frac{2}{3}\left(m^2 + m n + n^2\right) +\, 2 m \,+\, 2 n\,.
\end{equation}
Similarly from the Weyl dimension formula,
\begin{equation}
\dim V_{m,n} \,=\,  \prod_{\alpha\in\Delta_+}
\frac{\left(\lambda+\rho,\alpha\right)}{\left(\rho,\alpha\right)}
\,=\, \ha \left(m+1\right) \left(n+1\right) \left(m+n+2\right).
\end{equation}
Thus, dividing by ${\dim\g = 8}$,
\begin{equation}
c_2(V_{m,n}) \,=\, \frac{1}{8}\,\psi(V_{m,n}) \dim V_{m,n}\,.
\end{equation}
As a small check, note that the formula for $c_2(V_{m,n})$ is
symmetric in the integers $(m,n)$.  This symmetry follows abstractly from
the isomorphism ${V_{m,n}^* \simeq V_{n,m}}$ and the identity
${c_2(V_{m,n}^*) = c_2(V_{m,n})}$.\footnote{The $SU(3)$ identity ${V_{m,n}^*
    \simeq V_{n,m}}$ is induced from the Dynkin involution in type A
  and is not true for the group $G_2$, as we consider next.} 

For instance with ${m=2}$ and ${n=1}$,
\begin{equation}
\begin{aligned}
\psi(V_{2,1}) \,&=\, \frac{32}{3}\\
\dim V_{2,1} \,&=\, 15
\end{aligned}\,\,\Biggr\}\quad\Longrightarrow\quad c_2(V_{2,1}) \,=\, 20\,.
\end{equation}
These numerics illustrate that ${\psi(V)\in\Q}$ is only rational 
whereas ${c_2(V)\in\Z}$ is integral, as follows
inductively from the tensor product relation in \eqref{TensorP}.

\paragraph{Conventions for $G_2$.}

For a non-simply-laced example, we consider the exceptional Lie group
${G_2}$, also with rank two.

Let  $(x^1,\ldots,x^7)$ be
Euclidean coordinates on $\R^7$, and introduce the three-form
\begin{equation}\label{BigPhi}
\Phi \,=\, \theta^{1 2 3} + \theta^{1 4 5} + \theta^{167} +
\theta^{246} - \theta^{257} - \theta^{347} - \theta^{356} \,\in\,\wedge^3\R^7\,.
\end{equation}
Here we use the shorthand ${\theta^{i j k} \equiv dx^i\^dx^j\^dx^k}$.
The Lie group $G_2$ can be defined most elegantly \cite{Bryant:1987} as
the subgroup of ${GL(7,\R)}$ which preserves $\Phi$ under the linear action
on $\R^7$.
Because $\Phi$ is selected to lie in an open orbit of $GL(7,\R)$, the
stabilizer of $\Phi$ has dimension 
\begin{equation}
\dim G_2 \,=\, \dim GL(7,\R) - \dim \wedge^3\R^7 \,=\, 14\,.
\end{equation}
Thus, elements of $G_2$ can be presented concretely as 
invertible ${7\times 7}$ matrices.  A tiny bit of further work shows
that $G_2$ actually sits as a subgroup of $SO(7)$.

A maximal torus ${T\simeq U(1)^2}$ can be exhibited as
a pair of commuting rotations in $\R^7$ which preserve the three-form
$\Phi$.  These rotations will lie in distinct $SO(2)$ subgroups of the
ambient $SO(7)$, with generators 
\begin{equation}\label{H1G2}
\hat{h}_1 \,=\,
\begin{pmatrix}
0&0&0&0\\
0&{\bf J}_{2\times 2}&0&0\\
0&0&-2{\bf J}_{2 \times 2}&0\\
0&0&0&{\bf J}_{2 \times 2}
\end{pmatrix},\qquad\qquad 
{\bf J}_{2 \times 2} \,=\, 
\begin{pmatrix}
0&-1\\
1&0
\end{pmatrix},
\end{equation}
and
\begin{equation}\label{H2G2}
\hat{h}_2 \,=\, 
\begin{pmatrix}
0&0&0&0\\
0&{\bf 0}_{2 \times 2}&0&0\\
0&0&{\bf J}_{2\times 2}&0\\
0&0&0&-{\bf J}_{2 \times 2}
\end{pmatrix}.
\end{equation}
For clarity, both generators are written in terms of a 1-2-2-2 block
decomposition of the ${7\times 7}$ matrix.
Geometrically, $\hat{h}_2$ generates a rotation which fixes
$x^{1,2,3}$ and otherwise acts with equal magnitude and opposite
direction in the $x^{4,5}$- and $x^{6,7}$-planes.  The generator
$\hat{h}_1$ is a linear combination of $-2\hat{h}_2$ with a similar
rotation in the $x^{2,3}$- and $x^{6,7}$-planes.  Clearly
${[\hat{h}_1,\hat{h}_2]=0}$, and we leave the reader
to check that both generators preserve $\Phi$ in \eqref{BigPhi} and hence
are elements of the Lie algebra for $G_2$.

\iffigs
\begin{figure}
	\centering
	\begin{tikzpicture}[x=1.5cm,y=1.5cm] 
	\coordinate (0;0) at (0,0); 
	\foreach \c in {1,...,4}{%  
		\foreach \i in {0,...,5}{% 
			\pgfmathtruncatemacro\j{\c*\i}
			\coordinate (\c;\j) at (60*\i:\c);  
		} }
		\foreach \i in {0,2,...,10}{% 
			\pgfmathtruncatemacro\j{mod(\i+2,12)} 
			\pgfmathtruncatemacro\k{\i+1}
			\coordinate (2;\k) at ($(2;\i)!.5!(2;\j)$) ;}
		
		\foreach \i in {0,3,...,15}{% 
			\pgfmathtruncatemacro\j{mod(\i+3,18)} 
			\pgfmathtruncatemacro\k{\i+1} 
			\pgfmathtruncatemacro\l{\i+2}
			\coordinate (3;\k) at ($(3;\i)!1/3!(3;\j)$)  ;
			\coordinate (3;\l) at ($(3;\i)!2/3!(3;\j)$)  ;
		}
		
		\foreach \i in {0,4,...,20}{% 
			\pgfmathtruncatemacro\j{mod(\i+4,24)} 
			\pgfmathtruncatemacro\k{\i+1} 
			\pgfmathtruncatemacro\l{\i+2}
			\pgfmathtruncatemacro\m{\i+3} 
			\coordinate (4;\k) at ($(4;\i)!1/4!(4;\j)$)  ;
			\coordinate (4;\l) at ($(4;\i)!2/4!(4;\j)$) ;
			\coordinate (4;\m) at ($(4;\i)!3/4!(4;\j)$) ;
		}

		\begin{scope}
		%	\clip (0,0) circle (2.5);
		\clip ($(2;0)+(.3,0)$)--($(2;2)+(.3,.3)$)--($(2;4)+(-.3,.3)$)--($(2;6)+(-.3,0)$)--($(2;8)+(-.3,-.3)$)--($(2;10)+(.3,-.3)$)--($(2;0)+(.3,0)$);
		
		\foreach \i in {0,...,6}{% 
			\pgfmathtruncatemacro\k{\i}
			\pgfmathtruncatemacro\l{15-\i}
			\draw[thin,gray] (3;\k)--(3;\l);
			\pgfmathtruncatemacro\k{9-\i} 
			\pgfmathtruncatemacro\l{mod(12+\i,18)}   
			\draw[thin,gray] (3;\k)--(3;\l); 
			\pgfmathtruncatemacro\k{12-\i} 
			\pgfmathtruncatemacro\l{mod(15+\i,18)}   
			\draw[thin,gray] (3;\k)--(3;\l);} 
		
		\end{scope}
		
		\fill [gray] (0;0) circle (2pt);
		\foreach \c in {1,...,2}{%
			\pgfmathtruncatemacro\k{\c*6-1}    
			\foreach \i in {0,...,\k}{% 
				\fill [gray] (\c;\i) circle (2pt);}}  

		\foreach \n in{1,3,5}{%
			\draw[->,red,thick,shorten >=4pt,shorten <=2pt](0;0)--(2;\n);
		}
		
		\foreach \n in{7,9,11}{%
			\draw[->,dashed, red,thick,shorten >=4pt,shorten <=2pt](0;0)--(2;\n);
		}
		
		\foreach \n in{0,1,2}{%
			\draw[->,red,thick,shorten >=4pt,shorten <=2pt](0;0)--(1;\n);
		}
		
		\foreach \n in{3,4,5}{%
			\draw[->,dashed, red,thick,shorten >=4pt,shorten <=2pt](0;0)--(1;\n);
		}

		\node [ above right] at ($(1;0)+(.08,-.07)$) {$\hat \alpha_1$};
		\node [ above right] at ($(2;5)+(.02,-.08)$) {$\hat \alpha_2$};
		
		\draw[->,blue,thick,shorten >=4pt,shorten <=2pt](0;0)--(1;1);
		\draw[->,blue,thick,shorten >=4pt,shorten <=2pt](0;0)--(2;3);
		\node [ above right] at ($(1;1)+(.15,-.05)$) {$\hat \omega_1$};
		\node [ above left] at ($(2;3)+(-.1,-.05)$) {$\hat \omega_2$};
		
		\end{tikzpicture}  
		
		\caption{The weight lattice of $G_2$. The positive
                  simple roots are $\hat \alpha_{1,2}$ and the
                  fundamental weights are $\hat\omega_{1,2}$.}\label{fi:G2lattice}
	\end{figure}
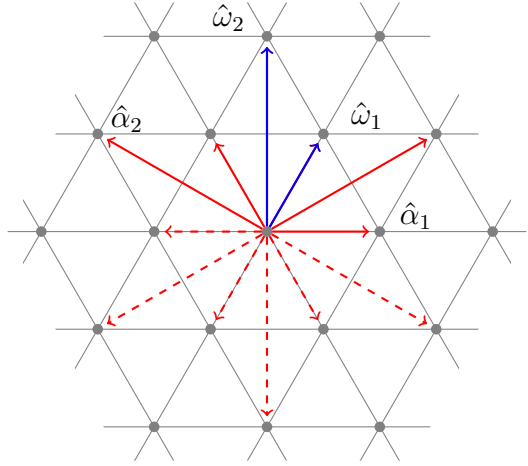
	\fi

With malice aforethought, we have selected $\hat{h}_1$ and $\hat{h}_2$
to be the coroots associated to the positive simple roots $\hat\alpha^1$ and
$\hat\alpha^2$ of $G_2$.  Thus $\hat\alpha^{1,2}$ pair with
$\hat{h}_{1,2}$ according to the Cartan matrix 
\begin{equation}\label{CartanG2}
\begin{matrix}
\begin{aligned}
\langle\hat\alpha^1,\hat{h}_1\rangle \,&=\, 2\,,\\
\langle\hat\alpha^2,\hat{h}_1\rangle \,&=\,-3\,,\\
\end{aligned} \qquad&\qquad
\begin{aligned}
\langle\hat\alpha^1,\hat{h}_2\rangle \,&=\, -1\,,\\
\langle\hat\alpha^2,\hat{h}_2\rangle \,&=\,2\,.\\
\end{aligned}
\end{matrix}
\end{equation}
With this labelling of roots, $\hat\alpha^1$ will be a short root, and
$\hat\alpha^2$ will be a long root.  The other four positive roots of
$G_2$ are the sums
\begin{equation}\label{RootsG2}
\begin{matrix}
\begin{aligned}
\alpha^3 \,&=\, \hat\alpha^1 + \hat\alpha^2\,,\\
\alpha^5 \,&=\, 3\,\hat\alpha^1 + \hat\alpha^2\,,
\end{aligned} \qquad&\qquad 
\begin{aligned}
\alpha^4 \,&=\, 2\,\hat\alpha^1 + \hat\alpha^2\,,\\
\alpha^6 \,&=\, 3\,\hat\alpha^1 + 2\,\hat\alpha^2\,,
\end{aligned}
\end{matrix}
\end{equation}
so 
\begin{equation}\label{PosRotsG2}
\Delta_+ \,=\,
\left\{\hat\alpha^1,\hat\alpha^2,\alpha^3,\alpha^4,\alpha^5,\alpha^6\right\}\,.
\end{equation}

As for $SU(3)$, the fundamental weights $\hat\omega^{1,2}$ are
determined in terms of the simple roots $\hat\alpha^{1,2}$ by the duality
relation in \eqref{FWeig}.  Comparing to \eqref{CartanG2},
\begin{equation}
\hat\omega^1 \,=\, 2\,\hat\alpha^1 \,+\,
\hat\alpha^2\,,\qquad\qquad \hat\omega^2 \,=\, 3\,\hat\alpha^1
\,+\, 2 \,\hat\alpha^2\,.
\end{equation}
Unlike for $SU(3)$, the fundamental weights are integral
combinations of the simple roots.  The root and weight lattices for
$G_2$ therefore coincide, ${\Lambda_{\rm
    R} = \Lambda_{\rm W}}$, and the center of $G_2$ is trivial.

We again parametrize each highest weight $\lambda$ as 
\begin{equation}
\lambda \,=\, m \, \hat\omega^1 \,+\, n \,
\hat\omega^2\,,\qquad m,n\ge 0\,,
\end{equation}
with associated irreducible representation $V_{m,n}$.  As
we will check later, ${V_{1,0}}$ is the 
representation of dimension seven implicit in the definition of ${G_2
  \subset GL(7,\R)}$.  Also from \eqref{RootsG2},
${\hat\omega^2=\alpha^6=\vartheta}$ is the highest root, so ${V_{0,1}
  \simeq \g}$ is the adjoint.  In the notation from particle physics,
\begin{equation}\label{StdRepsG2}
V_{1,0} \,\equiv\, {\bf 7}\,,\qquad\qquad V_{0,1}\,\equiv\,{\bf 14}\,.
\end{equation}
Both $V_{1,0}$ and $V_{0,1}$ are real representations, hence
self-dual.

As we shall check directly, the properly-normalized metric on the Lie algebra of
$G_2$ is given by {\em half} the trace in the seven-dimensional
representation $V_{1,0}$,
\begin{equation}\label{MetG2}
\left(x,y\right) \,=\,
-\ha\Tr_{V_{1,0}}\!\left[x\,y\right]\,,\qquad\qquad x,y\,\in\,\g\,.
\end{equation}
For the simple coroots $\hat{h}_1$ and $\hat{h}_2$ in
\eqref{H1G2} and \eqref{H2G2}, the matrix $\RA$ of inner-products is
straightforward to evaluate.  We find 
\begin{align}
\RA_{j \ell} \,=\, (\hat{h}_j,\hat{h}_\ell) \,=\,
\begin{pmatrix}
6&-3
\\
-3&2
\end{pmatrix},\qquad\qquad j,\ell=1,2\,.
\end{align}
The inner-product matrix for the fundamental weights is
then the inverse
\begin{equation}\label{QuadG3}
\left(\RA^{-1}\right){}^{\!j \ell} \,=\, (\hat\omega^j,\hat\omega^\ell)
\,= \frac{1}{3}\begin{pmatrix}
2&3
\\
3&6
\end{pmatrix}.
\end{equation}
As required, ${\vartheta=\hat\omega^2}$ has length $\sqrt{2}$, and
${\hat\omega^1}$ is a short root with length $\sqrt{2/3}$.

These data suffice to compute the Casimir $c_2(V_{m,n})$ as a function
of ${m,n}$.  First, the Weyl vector for $G_2$ is 
\begin{equation}
\rho \,=\, \ha\sum_{\alpha\in\Delta_+}\alpha \,=\, \hat\omega^1 \,+\,
\hat\omega^2\,.
\end{equation}
A quick calculation using the inner-product in \eqref{QuadG3} yields
\begin{equation}\label{PsiG2}
\psi(V_{m,n}) \,=\, \left(\lambda,\lambda+2\rho\right) \,=\,
\frac{2}{3} m^2 \,+\, 2 n^2 \,+\, 2 m n \,+\, \frac{10}{3} m \,+\, 6 n\,.
\end{equation}
The application of the Weyl dimension formula is only marginally more involved,
\begin{equation}\label{DimG2}
\begin{aligned}
\dim V_{m,n} \,&= \prod_{\alpha\in\Delta_+}
\frac{\left(\lambda+\rho,\alpha\right)}{\left(\rho,\alpha\right)} =\,
\frac{1}{120}\left(m+1\right)\left(n+1\right)\left(m+n+2\right)\times\\
&\qquad\qquad\qquad\qquad\times\left(m+2n+3\right)\left(m+3n+4\right)\left(2m+3n+5\right).
\end{aligned}
\end{equation}
The six linear factors on the right in \eqref{DimG2} arise from the
six positive roots of $G_2$.  One can
also easily check that ${\dim V_{1,0}=7}$ and ${\dim V_{0,1}=14}$
according to the dimension formula.  These results for $\psi(V_{m,n})$
and ${\dim V_{m,n}}$ have appeared previously in
\cite{Macfarlane:2001}, among other places.  

Together, the expressions
\eqref{PsiG2} and \eqref{DimG2} can be used to evaluate the
quadratic Casimir for any irreducible representation of $G_2$,
\begin{equation}
c_2(V_{m,n}) = \frac{1}{14}\,\psi(V_{m,n}) \dim V_{m,n}\,.
\end{equation}
Note that ${c_2(V_{1,0})=2}$ for the defining seven-dimensional
representation, consistent with the normalization of the trace in
\eqref{MetG2}.  We record the dimension and quadratic
Casimir for all representations of $G_2$ with ${m,n=0,\ldots,5}$ in
Tables \ref{RepTableG2} and \ref{CasTableG2}.  As suggested by Table
\ref{CasTableG2}, $c_2(V_{m,n})$ is even for all $G_2$ representations.
\renewcommand{\arraystretch}{1.5}
\begin{table}[t]
\begin{center}
\begin{tabular}{c | c | c | c | c | c | c} 
$\dim V_{m,n}$ & 0 & 1 & 2 & 3 & 4 & 5 \\ \hline
0 & 1 & 7 & 27 & 77 & 182 & 378 \\ \hline
1 & 14 & 64 & 189 & 448 & 924 & 1728 \\ \hline
2 & 77 & 286 & 729 & 1547 & 2926 & 5103 \\ \hline
3 & 273 & 896 & 2079 & 4096 & 7293 & 12096 \\ \hline
4 & 748 & 2261 & 4914 & 9177 & 15625 & 24948 \\ \hline
5 & 1729 & 4928 & 10206 & 18304 & 30107 & 46656 \\ \hline
\end{tabular}\caption{Dimensions of irreducible $G_2$
  representations.  The label $m$ runs horizontally, and $n$ runs
  vertically.  Note that $G_2$ admits a pair of irreducible
  representations $V_{3,0}$ and $V_{0,2}$ with dimension $77$.}\label{RepTableG2}
\end{center}
\end{table}
\begin{table}[t]
\begin{center}
\begin{tabular}{c | c | c | c | c | c | c} 
$c_2(V_{m,n})$ & 0 & 1 & 2 & 3 & 4 & 5 \\ \hline
0 & 0 & 2 & 18 & 88 & 312 & 900 \\ \hline
1 & 8 & 64 & 288 & 960 & 2640 & 6336 \\ \hline
2 & 110 & 572 & 1944 & 5304 & 12540 & 26730 \\ \hline
3 & 702 & 2944 & 8514 & 20480 & 43758 & 85824 \\ \hline
4 & 2992 & 10982 & 28548 & 62928 & 125000 & 230472 \\ \hline
5 & 9880 & 33088 & 79704 & 164736 & 309672 & 544320 \\ \hline
\end{tabular}\caption{Casimirs of irreducible $G_2$
  representations.  The label $m$ runs horizontally, and $n$ runs
  vertically.  The two representations with dimension $77$ have
  distinct Casimirs, ${c_2(V_{3,0})\neq c_2(V_{0,2})}$.}\label{CasTableG2}
\end{center}
\end{table}
\renewcommand{\arraystretch}{1.0}

\paragraph{Conventions for $Spin(4)$.}
Since $\mathfrak{so}(4)\simeq\mathfrak{su}(2)\oplus\mathfrak{su}(2)$
is reducible, conventions for this rank-two example follow from those
for $SU(2)$ in Section \ref{SU2GT}.

\paragraph{Conventions for $Spin(5)$.}

The positive simple coroots which span ${\h\simeq\R^2}$
can be identified with the ${5\times 5}$ anti-symmetric matrices 
\begin{equation}\label{BasisSpin(5)}
\widehat{h}_1 = \left(
\begin{matrix}
\mathbf{J}_{2\times 2} &\mathbf{0}_{2\times 2} & 0\\
\mathbf{0}_{2\times 2}& -\mathbf{J}_{2\times 2} & \vdots\\
0&\ldots&0
\end{matrix}\right)
\,,
\qquad\qquad \widehat{h}_2 = \left(
\begin{matrix}
\mathbf{0}_{2\times 2}&\mathbf{0}_{2\times 2} & 0\\
\mathbf{0}_{2\times 2}& 2\mathbf{J}_{2\times 2} & \vdots\\
0&\ldots&0
\end{matrix}\right)\,,
\end{equation}
where $\mathbf{J}_{2\times 2}$ appears in \eqref{H1G2}.  The pairing
of the positive simple roots $\hat{\alpha}^{1,2}$ with these
generators is determined by the Cartan matrix of ${Spin(5)}$ to be
\begin{equation}
\begin{aligned}
\langle \hat{\alpha}^1, \widehat{h}_1\rangle \,&=\, 2\,,
&\qquad\qquad&&
\langle \hat{\alpha}^1, \widehat{h}_2\rangle \,&=\, -2\,, \\
\langle \hat{\alpha}^2, \widehat{h}_1\rangle \,&=\, -1\,,
&&&
\langle \hat{\alpha}^2, \widehat{h}_2\rangle \,&=\, 2\,.
\end{aligned}
\end{equation}
Hence the fundamental weights $\widehat{\omega}^{1,2}$ which are
canonically dual to $\hat h_{1,2}$ are given by 
\begin{equation}
\begin{aligned}
\widehat{\omega}^{1} \,=\, \hat{\alpha}^1\,+\,\hat{\alpha}^2\,,
\qquad\qquad
\widehat{\omega}^{2} \,=\, \ha\hat{\alpha}^1\,+\,\hat{\alpha}^2\,.
\end{aligned}
\end{equation}
See Figure \ref{Spin5diag} for a diagram of the weight and root
lattices for $Spin(5)$.

\iffigs
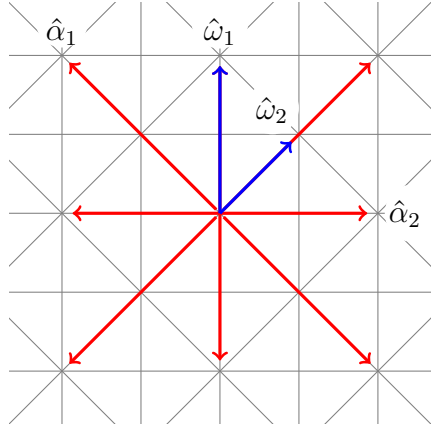
\begin{figure}[t]
	\centering
	\begin{tikzpicture}[x=.7cm,y=.7cm] 

	\coordinate (0;0) at (0,0);

	\begin{scope}
	
	\clip (-4,-4)--(-4,4)--(4,4)--(4,-4)--(-4,-4);
	
	\foreach \i in {-3,-1.5,0,1.5,3}{% 
		\draw[thin,gray] (6,\i)--(-6,\i);
		\draw[thin,gray] (\i,6)--(\i,-6);	
	} 
	
	\foreach \i in {-3,-1.5,0,1.5,3}{% 
		\draw[thin,gray] ($(\i,\i) +(6,-6) $)--($(\i,\i) +(-6,6) $);	
		\draw[thin,gray] ($(\i,-\i) +(6,6) $)--($(\i,-\i) +(-6,-6) $);	
	} 		
	\end{scope}

	\draw[->,red,very thick,shorten >=4pt,shorten <=2pt](0,0)--(3,3);	
	\draw[->,red,very thick,shorten >=4pt,shorten <=2pt](0,0)--(-3,3);	
	\draw[->,red,very thick,shorten >=4pt,shorten <=2pt](0,0)--(-3,-3);	
	\draw[->,red,very thick,shorten >=4pt,shorten <=2pt](0,0)--(3,-3);
	
	\draw[->,red,very thick,shorten >=4pt](0,0)--(0,3);	
	\draw[->,red,very thick,shorten >=4pt](0,0)--(0,-3);
	\draw[->,red,very thick,shorten >=4pt](0,0)--(3,0);
	\draw[->,red,very thick,shorten >=4pt](0,0)--(-3,0);	
	
	\fill[white] (-3,3.65) circle(.5);
	\node [ above] at (-3,3) {$\hat \alpha_1$};
	\fill[white] (3.68,0) circle(.55);
	\node [ right] at (3,0) {$\hat \alpha_2$};
	
	\draw[->,blue,very thick,shorten >=4pt=](0,0)--(1.5,1.5);
	\draw[->,blue,very thick,shorten >=4pt](0,0)--(0,3);
	
	\fill[white] (0,3.68) circle(.55);
	\node [ above ] at (0,3) {$\hat \omega_1$};
	\fill[white] (1,2) circle(.5);
	\node [ above left] at (1.5,1.5) {$\hat\omega_2$};
	
	\end{tikzpicture}  
	
	\caption{The weight lattice of $Spin(5)$. The positive simple roots are $\hat \alpha_{1,2}$ and the fundamental weights are $\hat\omega_{1,2}$.}\label{Spin5diag}
\end{figure}
\fi

We parametrize each highest weight $\lambda$ as 
\begin{equation}
\lambda \,=\, m \, \hat\omega^1 \,+\, n \,
\hat\omega^2\,,\qquad\qquad m,n\ge 0\,,
\end{equation}
with associated irreducible representation $V_{m,n}$.  In this
notation, $V_{1,0}$ is the vector representation, and $V_{0,1}$ is the
spinor representation,
\begin{equation}
V_{1,0}\,\equiv\,\mathbf{5}\,,\qquad\qquad V_{0,1}\,\equiv\,\mathbf{4}\,.
\end{equation}
Since ${Spin(5)\simeq Sp(4)}$, both the vector and the spinor
representations have valid claims to be the `fundamental' representation.

As we shall verify, the properly-normalized metric on the Lie
algebra is {\em half} the trace in the vector representation,
\begin{align}\label{Spin5Kil}
(x,y) = -\ha\Tr_{V_{1,0}}[x\,y]\,,\qquad\qquad x,y\,\in\,\g\,.
\end{align}
With respect to the dual bases of simple coroots and fundamental
weights, the metric is represented by the matrix of inner-products
\begin{align}\label{KillingFormSpin5}
\RA_{j \ell} \,=\, (\hat{h}_j,\hat{h}_\ell) \,=\,
\begin{pmatrix}
2&-2
\\
-2&4
\end{pmatrix},\qquad\qquad
\left(\RA^{-1}\right){}^{\!j \ell} \,=\, (\hat\omega^j,\hat\omega^\ell)
\,= \frac{1}{4}\begin{pmatrix}
4&2
\\
2&2
\end{pmatrix}.
\end{align}

For the irreducible representations $V_{m,n}$, the Weyl dimension formula states
\begin{align}\label{DimSpin(5)}
\dim\,V_{m,n} = \prod_{\alpha\in \Delta_+}\frac{(\lambda+\rho,\alpha)}{(\rho,\alpha)} = \frac16(m+1)(n+1)(2m+n+3)(m+n+2)\,.
\end{align}
Similarly, we can use the expression for the Killing form in \eqref{KillingFormSpin5} to evaluate
\begin{align}\label{PhiSpin(5)}
\psi(V_{m,n}) = (\lambda,\lambda+2\rho) = m^2+mn+\frac12n^2+3m+2n\,.
\end{align}
Using the expressions in \eqref{DimSpin(5)} and \eqref{PhiSpin(5)}, we
calculate the quadratic Casimir for any irreducible representation  via
\begin{align}
c_2(V_{m,n}) = \frac{1}{10}\,\psi(V_{m,n})\,\dim\,V_{m,n}\,.
\end{align}
Values of the Casimir for $Spin(5)$ representations with low dimension are
collected in Table \ref{CasTableSpin5}.  Since ${c_2(V_{1,0})=2}$, the
normalization in \eqref{Spin5Kil} follows from the general formula in \eqref{Normg}.

\renewcommand{\arraystretch}{1.5}
\begin{table}[t]
	\begin{center}
		\begin{tabular}{c | c | c | c | c | c | c} 
			$\dim V_{m,n}$ & 0 & 1 & 2 & 3 & 4 & 5 \\ \hline
			0 & 1 & 5   & 14  & 30  & 55  & 91    \\ \hline
			1 & 4 & 16  & 40  & 80  & 140 & 224   \\ \hline
			2 & 10& 35  & 81  & 154 & 260 & 405   \\ \hline
			3 & 20& 64  & 140 & 256 & 420 & 640   \\ \hline
			4 & 35& 105 & 220 & 390 & 625 & 935   \\ \hline
			5 & 56& 160 & 324 & 560 & 880 & 1296  \\ \hline
		\end{tabular}\caption{Dimensions of irreducible $Spin(5)$
		representations.  The label $m$ runs horizontally, and $n$ runs
		vertically.}\label{RepTableSpin5}
\end{center}
\end{table}
\begin{table}[t]
	\begin{center}
		\begin{tabular}{c | c | c | c | c | c | c} 
			$c_2(V_{m,n})$ & 0 & 1 & 2 & 3 & 4 & 5 \\ \hline
			0 & 0   & 2   & 14   & 54   & 154  & 364   \\ \hline
			1 & 1   & 12  & 58   & 188  & 483  & 1064  \\ \hline
			2 & 6   & 42  & 162  & 462  & 1092 & 2268  \\ \hline
			3 & 21  & 112 & 371  & 960  & 2121 & 4192  \\ \hline
			4 & 56  & 252 & 748  & 1794 & 3750 & 7106  \\ \hline
			5 & 126 & 504 & 1377 & 3108 & 6204 & 11340 \\ \hline
		\end{tabular}\caption{Casimirs of irreducible $Spin(5)$
		representations.  The label $m$ runs horizontally, and $n$ runs
		vertically.  The Casimir of the vector
                representation $V_{1,0}$ is equal to 2, while the
                Casimir of the spin representation $V_{0,1}$ is equal
                to 1.}\label{CasTableSpin5}
\end{center}
\end{table}
\renewcommand{\arraystretch}{1.0}

Finally, for use in Section \ref{RkTwoExs}, let us record the complete
set of weights in the fundamental and spin representations, 
\begin{equation}
\begin{aligned}
\Delta_{V_{1,0}} \,&=\, \big\{\hat{\omega}^1,\,\, -\hat{\omega}^1+2\hat{\omega}^2,\,\,0,\,\,\hat{\omega}^1-2\hat{\omega}^2,\,\,-\hat{\omega}^1	\big\}\,,\\
\Delta_{V_{0,1}} \,&=\, \big\{\hat{\omega}^2,\,\,-\hat{\omega}^2 ,\,\, \hat{\omega}^1-\hat{\omega}^2,\,\,-\hat{\omega}^1+\hat{\omega}^2\big\}\,.
\end{aligned}
\end{equation}
Though the vanishing weight in $\Delta_{V_{1,0}}$ is not relevant for the example in
Section \ref{RkTwoExs}, we include it to make the
dimension-counting clear.  Note also that
${-\Delta_{V_{0,1}}=\Delta_{V_{0,1}}}$, consistent with pseudoreality
of the spin representation for $Spin(5)$.

\section{Convexity Lemma}\label{ConvGLem}

In this appendix, we prove the equivalence between the two characterizations of the invariant ${\delta_{\ul e}\equiv \delta_{\ra,\pm\rho}}$ in \eqref{deltaP} and \eqref{deltaPII}.  The equivalence boils down to a statement in convex geometry.  For clarity, we adopt a slightly simpler notation here than used in Section \ref{JordHRGG}.

Let $\{H_1,\ldots,H_n\}$ be an arrangement of hyperplanes in general position\footnote{Ie.~all intersections have the expected dimension.} in a real vector space ${V \simeq \R^n}$.  Each hyperplane is described as the vanishing locus for an affine linear function 
\begin{equation}\label{HyperSys}
\begin{aligned}
H_1:\quad &\langle a_1,x\rangle \,+\, b_1 \,=\, 0\,,\qquad x\,\in\,V\\
&\qquad\qquad\vdots\\
H_n:\quad &\langle a_n,x\rangle \,+\, b_n \,=\, 0\,,
\end{aligned}
\end{equation}
where each ${a_1,\ldots,a_n\in V^*}$ and ${b_1,\cdots,b_n\in\R}$ are
non-zero and generic.  Multiplying each function by $\pm 1$ as
necessary, we assume without loss that ${b_1,\ldots,b_n > 0}$ are
strictly-positive.  If we wished, we could also normalize each
${b_1,\ldots,b_n}$ to unity by scaling ${a_1,\ldots,a_n}$, but we will leave the values arbitrary in the following.

Introduce a non-zero covector ${\rho\in V^*}$ used to define the half-space ${\BH_\rho \subset V}$,
\begin{equation}
\BH_\rho \,=\, \big\{x\in V \,\big|\, \langle\rho,x\rangle \ge 0\big\},
\end{equation}
with boundary ${\partial\BH_\rho = \Ker(\rho) \simeq \R^{n-1}}$.  Let ${h_1,\ldots,h_n}$ be the boundary intersections 
\begin{equation}
h_1 \,=\, H_1 \cap \partial\BH_\rho\,,\qquad\cdots\,,\qquad h_n \,=\, H_n \cap \partial\BH_\rho\,,
\end{equation}
each of which is now a hyperplane in $\partial\BH_\rho$.

Finally, let $\sP$ be the polytope in $\partial\BH_\rho$ which is bounded by ${h_1,\ldots,h_n}$.  The polytope $\sP$ is the convex hull of vertices
\begin{equation}
\begin{aligned}
e_1 \,&=\, \widehat{h_1} \cap h_2 \cap \cdots \cap h_n \,\in\,\partial\BH_\rho\,,\\
e_2 \,&=\, h_1 \cap \widehat{h_2} \cap \cdots \cap h_n \,\in\,\partial\BH_\rho\,,\\
&\qquad\qquad\qquad\vdots\\
e_n \,&=\, h_1 \cap h_2 \cap \cdots \cap \widehat{h_n} \,\in\,\partial\BH_\rho\,,
\end{aligned}
\end{equation}
where the hat means that the indicated hyperplane is omitted from the list.  By assumption, ${e_1,\ldots,e_n}$ are distinct points in general position in $\partial\BH_\rho$.

Equivalence of \eqref{deltaP} and \eqref{deltaPII} amounts to the geometric statement:
\begin{quote}
($\star$) The origin ${\{0\}}$ lies inside ${\sP\subset\partial\BH_\rho}$ if and only if $\rho$ or $-\rho$ lies inside the positive cone ${\R_+[a_1,\ldots,a_n]\subset V^*}$ generated by the covectors ${a_1,\ldots,a_n}$.
\end{quote}
Since only $\partial\BH_\rho$ and not $\BH_\rho$ plays a role on the left-hand side of ($\star$), the equivalence must be symmetric under sign reversal of $\rho$.  By contrast, once we fix ${b_1,\ldots,b_n>0}$, the sign of each ${a_1,\ldots,a_n\in V^*}$ carries geometric meaning, as enters the right-hand side of ($\star$).

The proof of ($\star$) follows by a short calculation in either direction.

\medskip\noindent{($\Longrightarrow$):}\quad Suppose that ${\{0\}\in\partial\BH_\rho}$ lies in the polytope $\sP$.  As $\sP$ is the convex hull of $\{e_1,\ldots,e_n\}$, there exist positive parameters ${t_1,\ldots,t_n>0}$, ${t_1+\cdots+t_n=1}$, so that 
\begin{equation}\label{ConvHL}
t_1\,e_1 \,+\, \cdots \,+\, t_n\,e_n \,=\, 0\,.
\end{equation}
Contracting with $a_1$ yields
\begin{equation}\label{ConveqI}
t_1\,\langle a_1,e_1\rangle \,+\, t_2\,\langle a_1,e_2\rangle \,+\, \cdots \,+\, t_n \, \langle a_1,e_n\rangle \,=\, 0\,.
\end{equation}
The points other than $e_1$, ie.~${e_2,\ldots,e_n}$, lie in the hyperplane ${h_1 = H_1 \cap \partial\BH_\rho}$ and so satisfy
\begin{equation}
\langle a_1, e_2 \rangle \,=\, \cdots \,=\, \langle a_1, e_n\rangle \,=\, - b_1\,.
\end{equation}
More generally,
\begin{equation}\label{AEoffd}
\langle a_j, e_\ell\rangle \,=\, -b_j\,,\qquad j\neq\ell\,,\qquad j,\ell=1,\ldots,n\,.
\end{equation}
Substituting into \eqref{ConveqI}, we solve for $\langle a_1, e_1\rangle$ as 
\begin{equation}
\langle a_1, e_1\rangle \,=\, \frac{b_1}{t_1}\left(t_2\,+\,\cdots\,+\,t_n\right) \,=\, b_1\left(\frac{1-t_1}{t_1}\right)\,.
\end{equation}
Contracting \eqref{ConvHL} with ${a_2,\ldots,a_n}$ yields similarly
\begin{equation}\label{AEdiag}
\langle a_j, e_j\rangle \,=\, b_j \left(\frac{1-t_j}{t_j}\right),\qquad j=1,\ldots,n\,.
\end{equation}

We wish to show that ${\pm\rho\in\R_+[a_1,\ldots,a_n]}$, which is equivalent to the algebraic relation
\begin{equation}\label{ConvAR}
w_1\,a_1 \,+\, \cdots \,+\, w_n\,a_n \,=\, \pm\rho\,,
\end{equation}
for some positive parameters ${w_1,\ldots,w_n>0}$ and choice of sign on the right.  
Since $\{a_1,\ldots,a_n\}$ is a basis for $V^*$ and $\{e_1,\ldots,e_n\}$ span the hyperplane ${\Ker(\rho)\subset V}$, the algebraic relation \eqref{ConvAR} is implied by the linear equations
\begin{equation}\label{wLinSyst}
\langle w_1\,a_1 \,+\, \cdots \,+\, w_n\,a_n,\,e_j\rangle \,=\, 0\,,\qquad j=1,\ldots,n\,,
\end{equation}
for ${w_1,\ldots,w_n>0}$.  Using \eqref{AEoffd} and \eqref{AEdiag}, the linear system \eqref{wLinSyst} can be written explicitly in terms of the positive parameters $(b,t)$ as 
\begin{equation}\label{KerA}
{\bf A}\cdot w \,=\, 0\,,
\end{equation}
with ${n\times n}$ coefficient matrix
\begin{equation}
{\bf A} \,=\, 
\begin{bmatrix}
b_1 \left(\frac{1-t_1}{t_1}\right) & -b_2 & -b_3 & \quad\cdots & -b_n \\
-b_1 & b_2 \left(\frac{1-t_2}{t_2}\right) & -b_3 & \quad\cdots & -b_n \\
\vdots & \vdots & \vdots & \quad\vdots & \vdots \\
-b_1 & -b_2 & -b_3 & \quad\cdots & b_n \left(\frac{1-t_n}{t_n}\right)
\end{bmatrix}
\end{equation}
and 
\begin{equation}\label{wvector}
w \,=\, 
\begin{bmatrix}
w_1\\
\vdots\\
w_n
\end{bmatrix}
\end{equation}

Because ${e_1,\ldots,e_n}$ are not linearly independent in $V$ but instead lie on the hyperplane ${\Ker(\rho)}$ (and are otherwise generic), the matrix ${\bf A}$ has a kernel with dimension one.  The only question is whether this kernel is generated by a vector with positive entries ${w_1,\ldots,w_n>0}$, in which case we are done.  But by inspection, a positive solution to \eqref{KerA} is given by 
\begin{equation}\label{SolvW}
w_j \,=\, t_j\, b_1 \cdots \widehat{b}_j \cdots b_n > 0\,,\qquad\qquad j=1,\ldots,n\,.
\end{equation}
Again, the hat indicates that $b_j$ is omitted from the product, and we use the relation ${t_1 + \cdots + t_n =1}$ in verifying the solution.\QED

\medskip\noindent{($\Longleftarrow$):}\quad  We run the preceding calculation in reverse.  Suppose ${\pm\rho\in\R_+[a_1,\ldots,a_n]}$, or 
\begin{equation}\label{ConvARII}
w_1\,a_1 \,+\, \cdots \,+\, w_n\,a_n \,=\, \pm\rho\,,\qquad\qquad w_1,\ldots,w_n>0\,,
\end{equation}
for some choice of sign on the right.  Since ${e_1,\ldots,e_n\in\Ker(\rho)}$ and the relation in \eqref{AEoffd} still holds, contraction of \eqref{ConvARII} yields
\begin{equation}\label{DiagWs}
\langle a_j, e_j \rangle \,=\, \frac{b_1 w_1 \,+\, \cdots \,+\, \widehat{b_j w_j} \,+\, \cdots \,+\, b_n w_n}{w_j}\,,\qquad\qquad j=1,\ldots,n\,.
\end{equation}
where the $j$th term is omitted from the sum.

We wish to show that 
\begin{equation}
t_1 \,e_1 \,+\, \cdots \,+\, t_n \,e_n \,=\, 0\,,
\end{equation}
for some positive parameters ${t_1,\ldots,t_n>0}$ with ${t_1+\cdots+t_n=1}$.
As $\{a_1,\ldots,a_n\}$ is a basis for $V^*$, it suffices to show that each covector annihilates the left-hand side, ie.
\begin{equation}
t_1 \,\langle a_j, e_1\rangle \,+\, \cdots \,+\, t_n \,\langle a_j ,e_n\rangle \,=\, 0\,,\qquad j=1,\ldots,n\,.
\end{equation}
Via \eqref{AEoffd} and \eqref{DiagWs}, we obtain the linear system
\begin{equation}
{\bf B}\cdot t \,=\, 0\,,
\end{equation}
where 
\begin{equation}
{\bf B} = \begin{bmatrix}
\frac{b_2 w_2 \,+\, b_3 w_3 \,+\, \cdots \,+\, b_n w_n}{w_1} & -b_1 & -b_1 & \quad\cdots\quad & -b_1 \\
-b_2 & \frac{b_1 w_1 \,+\, b_3 w_3 \,+\, \cdots \,+\, b_n w_n}{w_2} & -b_2 & \quad\cdots\quad & -b_2 \\
\vdots & \vdots & \vdots & \quad\vdots\quad & \vdots\\
-b_n & -b_n & -b_n &\quad\cdots\quad & \frac{b_1 w_1 \,+\, b_2 w_2 \,+\, \cdots \,+\, b_{n-1} w_{n-1}}{w_n}
\end{bmatrix}
\end{equation}
and
\begin{equation}
t \,=\, \begin{bmatrix}
t_1\\
\vdots\\
t_n
\end{bmatrix}.
\end{equation}
Again, because the points ${e_1,\ldots,e_n}$ lie in a hyperplane, the ${n\times n}$ matrix ${\bf B}$ has a one-dimensional kernel.  By inspection, the kernel is generated by the vector with entries
\begin{equation}\label{SolvT}
t_j \,=\, \frac{b_j w_j}{b_1 w_1 \,+\, \cdots \,+\, b_n w_n}>0\,,\qquad\qquad j=1,\ldots\,n.
\end{equation}
The denominator ensures ${t_1+\cdots+t_n=1}$.\QED

The explicit relations \eqref{SolvW} and \eqref{SolvT} between $w$ and $t$ also show that as any wall of the polytope $\sP$ approaches ${\{0\}\in\partial\BH_\rho}$, a wall of the cone $\R_+[a_1,\ldots,a_n]$ approaches the covector ${\pm\rho\in V^*}$, for some choice of sign.

The ambiguity in sign on $\rho$ can be removed with an extra geometric condition.  Let ${x_\bullet\in H_1 \cap \cdots \cap H_n \in V}$ be the unique point of intersection for the hyperplanes in the arrangement.  By our assumption of general position, $x_\bullet$ exists and lies in either the half-space $\BH_\rho$ or $-\BH_\rho$.  We refine the geometric equivalence ($\star$) using $x_\bullet$:
\begin{quote}
($\star\star$) The origin ${\{0\}}$ lies in ${\sP\subset\partial\BH_\rho}$ and $x_\bullet$ lies in ${\BH_\rho\subset V}$ if and only if $-\rho$ lies in the positive cone ${\R_+[a_1,\ldots,a_n]\subset V^*}$.
\end{quote}
\noindent{Proof:}\quad The only issue here is to keep track of signs.  By \eqref{HyperSys} the intersection point ${x_\bullet\in V}$ solves the inhomogeneous system 
\begin{equation}
\langle a_1, x_\bullet\rangle \,=\, -b_1\,,\qquad\cdots\,,\qquad \langle a_n, x_\bullet\rangle \,=\, -b_n\,.
\end{equation}
Expand $\rho$ in the basis $\{a_1,\ldots,a_n\}$ for $V^*$ as 
\begin{equation}
\rho \,=\, w_1\,a_1 \,+\, \cdots \,+\, w_n\,a_n\,,
\end{equation}
for some coefficients ${w_1,\ldots,w_n\in\R}$.  Then 
\begin{equation}
\begin{aligned}
\langle\rho, x_\bullet\rangle \,&=\, w_1 \langle a_1, x_\bullet\rangle \,+\, \cdots \,+\, w_n \langle a_n, x_\bullet\rangle\,,\\
&=\, -\left(w_1 \, b_1 \,+\, \cdots \,+\, w_n \, b_n\right).
\end{aligned}
\end{equation}
Recall ${b_1,\ldots,b_n>0}$ by assumption.  Hence if $\pm\rho\in\R_+[a_1,\ldots,a_n]$, then ${x_\bullet \in \mp\BH_\rho}$.  The claim $(\star\star)$ now follows from the preceding lemma $(\star)$.\QED

\section{Asymptotic Behavior of Torus Knot Observables}\label{TorusK}

In this appendix, we specialize to ${\CN=2}$ supersymmetric
Chern-Simons-matter theory with gauge group $SU(2)$ and matter in the
irreducible representation of dimension ${L+1}$.  Equivalently, the
matter has spin-$L/2$.  Following the notation
from Section \ref{SU2GT},
\begin{equation}
\Lambda \,=\, [\mathbf{L+1}]\,,\qquad\qquad G\,=\, SU(2)\,.
\end{equation}
According to the formula for $c_2$ in \eqref{SU2Cas}, anomaly-cancellation
\eqref{Quant} requires the level $k$ to be half-integral when
${L\equiv 1 \mod 4}$ and integral otherwise.

The ${\CN=2}$ supersymmetric Chern-Simons theory possesses
\cite{Kapustin:2009kz} a class of
half-BPS Wilson loop operators which wrap torus knots
${\mathcal{K}_{p,q}\subset S^3}$, appearing as the generic fiber of
the orbifold bundle in \eqref{PQHOPF}.  In particular, the squashing
parameter of the metric on $S^3$ takes the distinguished rational
value 
\begin{equation}
b^2 \,=\, \frac{p}{q}\,\in\,\Q\,.
\end{equation}
Throughout, ${p,q\in\Z}$ are
relatively-prime integers with ${p>q>1}$.  We
also allow the degenerate case ${p=q=1}$, for which
${\mathcal{K}_{1,1}\equiv\bigcirc}$ is the unknot.  As another
example, $\mathcal{K}_{3,2}$ is the trefoil knot.  See Chapter $7.1$
in \cite{Beasley:2009mb} for more about the geometry of torus knots in $S^3$.

Let $\mathsf{W}_{\mathbf{n}}\!\left(\mathcal{K}_{p,q}\right)$ be the
supersymmetric Wilson loop operator which wraps a $(p,q)$-torus knot
and is decorated with the $n$-dimensional irreducible representation of $SU(2)$.
When $\mathsf{W}_{\mathbf{n}}\!\left(\mathcal{K}_{p,q}\right)$ is
inserted into the path integral, the localization formula
\eqref{ZSU(2)} for the partition function $Z_{S^3}$ naturally 
generalizes to
\begin{equation}\label{ZTorusK}
\begin{aligned}
&Z_{S^3}\big(k,\mu;\mathcal{K}_{p,q},n,L\big)\,=\, \frac{1}{2 \pi i} \frac{1}{\sqrt{pq}} \, \exp{\!\left[-{{i
\pi}\over{2 k}}\left(\frac{p}{q} + \frac{q}{p}
 + pq \,
(n^2-1)\right)\right]} \,\times\,\\  
&\qquad\times \, \int_\R \! dz \;
\ch_{\mathbf{n}}\!\left(\zeta\,\frac{z}{2}\right)
\sinh{\!\left(\zeta\,\frac{z}{2 p}\right)}
\sinh{\!\left(\zeta\,\frac{z}{2 q}\right)} 
\exp{\!\left[-\frac{k}{8\pi}
\left(\frac{z^2}{p q}\right)\right]}\,\times\,\\
&\qquad\times\prod_{\substack{\beta\in\Delta\,,\\ \Delta=\{L, L-2,\, \cdots,
2-L, -L\}}}
s_{b=\sqrt{p/q}}\!\left(\zeta\,\frac{\beta\,z}{4 \pi
  \sqrt{pq}}\,+\,\mu\right)\,.
\end{aligned}
\end{equation}
Throughout, ${\zeta\equiv\e{\!i\pi/4}}$ is an eighth-root of unity, and $\ch_{\mathbf{n}}$ is
the character for the $n$-dimensional representation of $SU(2)$,
\begin{equation}\label{SU2Char}
\ch_{\mathbf{n}}(z) \,=\, \frac{\sinh(n\,z)}{\sinh(z)} \,=\,
\e{\!(n-1) z} \,+\, \e{\!(n-3) z} \,+\, \cdots \,+\, \e{\!-(n-3) z}
\,+\, \e{\!-(n-1) z}\,.
\end{equation}
The Wilson loop operator
$\mathsf{W}_{\mathbf{n}}\!\left(\mathcal{K}_{p,q}\right)$ is itself
represented on the Coulomb branch by the $SU(2)$ character
$\ch_{\mathbf{n}}$ in the integrand of \eqref{ZTorusK}.  The other
factors in the integrand follow from \eqref{ZSU(2)} after the
substitution ${\sigma = \zeta \, z/2\sqrt{pq}}$ and a contour rotation
to the real axis.  By analyticity, the value of the integral does not change under the contour rotation.  The $(p,q)$-dependence in the overall phase is
taken from $(7.73)$ in \cite{Beasley:2009mb}, which describes the pure
Chern-Simons gauge theory without matter.  Note that
$k_{\textrm{here}}$ is identified with $(k+2)_{\textrm{there}}$ in
\cite{Beasley:2009mb}, as ${\CN=2}$ supersymmetry prevents the
renormalization of the Chern-Simons level.

By convention when ${L=0}$ in \eqref{ZTorusK}, we factor off the 
contribution $s_{b=\sqrt{p/q}}(\mu)$ from the
decoupled, free ${\CN=2}$ chiral multiplet with mass $\mu$,
\begin{equation}\label{ZTorusKLZero}
\begin{aligned}
&Z_{S^3}\big(k;\mathcal{K}_{p,q},n\big)\Big|_{L=0}\,:=\, \frac{1}{2 \pi i} \frac{1}{\sqrt{pq}} \, \exp{\!\left[-{{i
\pi}\over{2 k}}\left(\frac{p}{q} + \frac{q}{p}
 + pq \,
(n^2-1)\right)\right]} \,\times\,\\  
&\qquad\times \, \int_\R \! dz \;
\ch_{\mathbf{n}}\!\left(\zeta\,\frac{z}{2}\right)
\sinh{\!\left(\zeta\,\frac{z}{2 p}\right)}
\sinh{\!\left(\zeta\,\frac{z}{2 q}\right)} 
\exp{\!\left[-\frac{k}{8\pi}
\left(\frac{z^2}{p q}\right)\right]}\,.
\end{aligned}
\end{equation}
Famously, in the special case \eqref{ZTorusKLZero} of pure Chern-Simons gauge
theory, the $SU(2)$ 
Wilson loop path integral reproduces the $n$-colored Jones polynomial
in terms of the ratio
\begin{equation}\label{nJones}
J_n\!\left({\mathcal{K}_{p,q}}\right) =\,
\frac{Z_{S^3}\big(k;\mathcal{K}_{p,q},n\big)}{Z_{S^3}\big(k;\mathcal{K}_{1,1},n\big)}\Biggr|_{L=0}\,,\qquad\qquad
\mathsf{q} \,=\, \e{\!-2\pi i/k}\,,
\end{equation}
where $\mathsf{q}$ is the Laurent argument of $J_n\!\left({\mathcal{K}_{p,q}}\right)$.
Eg.~for ${n=2}$, the original, uncolored Jones
polynomial is 
\begin{equation}\label{JTorusK}
J_2\!\left(\mathcal{K}_{p,q}\right) =\,
\frac{\mathsf{q}^{\ha(p-1)(q-1)}}{1-\mathsf{q}^2}\left[1 \,+\,
  \mathsf{q}^{p+q} \,-\, \mathsf{q}^{p+1} \,-\, \mathsf{q}^{q+1}\right],
\end{equation}
which can be obtained directly from \eqref{ZTorusKLZero} by
evaluating a sum of Gaussian integrals.  As a small check, note that
the bracketed polynomial factor in \eqref{JTorusK} vanishes if ${\mathsf{q}=\pm 1}$ and
${\gcd(p,q)=1}$, so the factor is divisible by $(1-\mathsf{q}^2)$, and the right-hand side of \eqref{JTorusK} is indeed a
Laurent polynomial in $\mathsf{q}$.

Both the partition function $Z_{S^3}(k;\mathcal{K}_{p,q},n)|_{L=0}$ and the colored Jones polynomial are known to have exotic, discontinuous
asymptotic behavior in the limit ${k,n\to\infty}$ with fixed
\begin{equation}
\gamma \,=\, \frac{n}{k} \,\in\,\C\,.
\end{equation}
Implicitly, $k$ has been complexified, as sensible in
expressions such as \eqref{ZTorusKLZero} and \eqref{JTorusK}.  By contrast, 
  ${n\ge 1}$ is a positive integer throughout.  For
$(p,q)$-torus knots, the asymptotic behavior of $J_n\!\left(\mathcal{K}_{p,q}\right)$
as a function of ${\gamma\in\C}$ was originally determined by Hikami and
Murakami \cite{Hikami:2007,Hikami:2010,Murakami:2004asymp} from the finite-dimensional, Coulomb-branch 
integral in \eqref{ZTorusKLZero}.  Later, Witten \cite{Witten:2010cx}
explained how the same result could be obtained more generally from the infinite-dimensional Chern-Simons path
integral.

In this appendix, we explore the asymptotic behavior of the
supersymmetric partition function
$Z_{S^3}\big(k,\mu;\mathcal{K}_{p,q},n,L\big)$ when ${L\neq 0}$, including
the non-trivial contribution from the charged chiral multiplet in the
third line of \eqref{ZTorusK}.  Perhaps unsurprisingly, in the presence of
matter the asymptotic behavior of $Z_{S^3}$ depends non-trivially
upon both $\gamma$ and the mass $\mu$.  We find qualitatively similar
analytic phenomena to those discussed for pure Chern-Simons gauge theory in
\cite{Hikami:2007,Hikami:2010,Murakami:2004asymp,Witten:2010cx}, but the presence of
supersymmetric matter changes the detailed, quantitative behavior of
$Z_{S^3}$ as ${n\to\infty}$.

This appendix has two parts.  In Appendix \ref{Recollect}, we offer a
streamlined derivation of the essential results in
\cite{Hikami:2007,Hikami:2010,Murakami:2004asymp} concerning the asymptotic
behavior of the colored Jones polynomial.  In Appendix \ref{Coupling}, we
generalize the asymptotic analysis to the supersymmetric $SU(2)$
Chern-Simons theory with spin-$L/2$ matter.

\subsection{Recollections About the Colored Jones Polynomial}\label{Recollect}

We first review the asymptotic analysis for the Wilson loop partition function in pure
Chern-Simons theory, without matter.  This analysis was performed rigorously in \cite{Hikami:2007,Hikami:2010,Murakami:2004asymp}, including the expansion to all orders in $1/n$.  Our goals are more modest.  We sketch a simplified analysis, valid to leading-order as ${n\to\infty}$, which suffices to exhibit the more interesting semiclassical features of the partition function.

Rewrite the contour integral
\eqref{ZTorusKLZero} as 
\begin{equation}\label{ZTorusKII}
Z_{S^3}\big(k;\mathcal{K}_{p,q},n\big)\Big|_{L=0} \,=\, \frac{1}{2\pi i}\frac{1}{\sqrt{pq}}\,
\Phi\cdot\ha\int_{\R}dz \Big[F(z)\,\e{f_+(z)}\,-\, F(z)\,\e{f_-(z)}\Big],
\end{equation}
where 
\begin{equation}\label{BigFx}
\begin{aligned}
\Phi \,&=\, 
\exp{\!\left[-\frac{i\pi}{2k}\left(\frac{p}{q}+\frac{q}{p} +pq\,(n^2-1)\right)\right]}\,,\\
F(z) \,&=\, 
\frac{\sinh\!\left(\zeta z/2p\right)\sinh\!\left(\zeta z/2q\right)}
{\sinh(\zeta z/2)}\,,\qquad\qquad \zeta\,\equiv\,\e{\!i\pi/4}\,,\\
f_\pm(z)&=
-\frac{k}{8\pi}\left(\frac{z^2}{pq}\right) \pm\, \ha\zeta n z\,.
\end{aligned}
\end{equation}
In passing from \eqref{ZTorusKLZero} to \eqref{ZTorusKII}, we use the
explicit description of the $SU(2)$ character in \eqref{SU2Char} as a
ratio of hyperbolic sines.  We then decompose $\sinh(n z)$ as a sum of exponentials.

Under the sign reversal ${z\mapsto-z}$, corresponding to the residual 
Weyl symmetry on the Coulomb branch, ${F(z)\mapsto -F(z)}$ is odd,
and the functions ${f_\pm(z)\mapsto f_\mp(z)}$ exchange.  Via this symmetry,
\begin{equation}\label{ZTorusKIII}
Z_{S^3}\big(\gamma;\mathcal{K}_{p,q},n\big)\Big|_{L=0} \,=\,\frac1{2\pi
  i}\frac{1}{\sqrt{pq}}\,\Phi\cdot\int_{\R}dz\;
F(z)\,\e{f_+(z)}\,.
\end{equation}
Here we consider $Z_{S^3}$ as a function of the integer $n$ and the complex scaling variable ${\gamma=n/k\in\C}$.  Expressed in these variables,
\begin{equation}\label{AsymPhi}
\Phi \,=\,\exp{\!\left[
-\frac{i\pi}{2} p q \gamma n\right]} \cdot \exp{\!\left[-\frac{i\pi}{2}\left(\frac{p}{q} + \frac{q}{p} - p q\right)\frac{\gamma}{n}\right]}\,,
\end{equation}
and
\begin{equation}\label{fplus}
f_\pm(z)\,=\,-\frac{n}{8\pi pq \gamma}z^2 \,\pm\, \frac{1}{2}\zeta n z\,,
\end{equation}
while $F(z)$ is independent of $n$ and $\gamma$.  

A reflection in the real part of $\gamma$ induces complex conjugation on the partition function (up to an overall phase),
\begin{equation}\label{ccZS3}
\bar{Z_{S^3}\!\left(\gamma;\mathcal{K}_{p,q},n\right)}\big|_{L=0} \,=\, i\,Z_{S^3}\!\left(-\bar\gamma;\mathcal{K}_{p,q},n\right)\!\big|_{L=0}\,,
\end{equation}
as can be seen by changing variables in \eqref{ZTorusKIII}.  For the remainder, we use this relation to fix ${\Re(\gamma)\ge 0}$ without loss.  When the real part of $\gamma$ is strictly-positive,
the integrand of \eqref{ZTorusKIII} decays like a Gaussian as
${z\to\pm\infty}$, and the integral over $\R$ converges absolutely.
More generally, the sectors of convergence and hence the defining contour rotate with the phase of $\gamma$.  See Figure
\ref{fi:asymptBehaviorOfWLPartitionFunc} for a
sketch of the sectors of convergence when ${\gamma\in\R_+}$ is real
and positive.

\iffigs
\begin{figure}
	\begin{center}
		\begin{tikzpicture}[x=.6cm,y=.6cm]
		\begin{scope}
		\clip (0,0) circle (5);
		\fill[blue!20] (6,-6) -- (0,0) -- (6,6)  --(6,-6) ;
		\fill[blue!20] (-6,6) -- (0,0) -- (-6,-6)  --(-6,6) ;
		\end{scope}
		
		\draw [red](-5.5,0)--(5.5,0) node [right, red] {${\R}$};
		\draw [gray](0,-4.5)--(0,4);
		\node at (4,4) {$\zeta\cdot\R$};
		
		\draw [->, thick, red](-5,0)--(-2,0);
		\draw [->, thick, red](-2,0)--(2.5,0);
		
		\draw [thick, red](2.5,0)--(5,0);
		
		\draw [ gray, dashed](-4,-4)--(4,4);
		\draw [ gray, dashed](-4,4)--(4,-4);
		
		\node at (1,1) {$\times$};
		\node at (3,3) {$\times$};
		
		\node at (-1,-1) {$\times$};
		\node at (-3,-3) {$\times$};
		
		\node at (2.3,1.5) {$z_*$};
		
		\fill[black] (2,2) circle(.15);

		\end{tikzpicture}
		\caption{Sectors of convergence,
                  stationary-phase point ${z_*=2\pi \zeta pq \gamma}$,
                  and polar points $(\times)$ for the Wilson loop partition
                  function in \eqref{ZTorusKIII} with real, positive
                  ${\gamma\in\R_+}$.  The shaded regions are those for which ${|\exp{\!\left[
			f_+(x)\right]|} \to 0}$ as ${|z|\to\infty}$.
                  The function $F(z)$ in \eqref{BigFx} has simple poles along the line
                  ${\zeta\cdot\R}$, located at the points $\{2\pi
                  \zeta \cdot \Z\}-\{2\pi \zeta p\cdot \Z\} - \{2\pi
                  \zeta q \cdot \Z\}$ when ${p>q>1}$ (a nontrivial
                  torus knot).  At the special value ${\gamma=1/pq}$, the
                  critical point collides with the pole of $F(z)$ nearest to the
                  origin.}\label{fi:asymptBehaviorOfWLPartitionFunc}
	\end{center}
\end{figure}
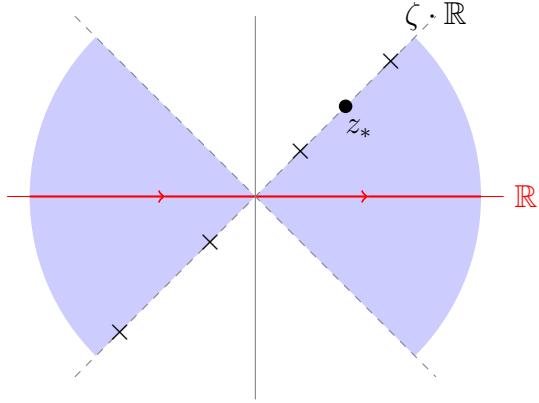\fi

Because $f_+(z)$ in \eqref{fplus} scales linearly with $n$, the method of steepest-descents determines the asymptotic behavior of $Z_{S^3}(\gamma;\mathcal{K}_{p,q},n)|_{L=0}$ in the limit ${n\to\infty}$.  For a nice review of the method of steepest-descents, with attention to various subtleties, see \S $6.6$ in \cite{Bender:1978}.  Moreover, $f_+(z)$ is quadratic and so has a unique critical point in the complex $z$-plane, at
\begin{equation}\label{CritX}
z_* \,=\, 2\pi \zeta p q \gamma\,.
\end{equation}
Naively, one might wonder how the function on the left of
\eqref{ZTorusKIII} can display any interesting asymptotic dependence
on $\gamma$, given that a steepest-descent contour must pass through a
single critical point whose location varies linearly with
$\gamma$.  The answer is provided by the
seemingly-innocuous prefactor $F(z)$ in \eqref{BigFx}.

For the unknot, with ${p=q=1}$, the prefactor $F(z)$ is entire and
truly innocuous.  Otherwise, for nontrivial torus knots with relatively-prime
${p>q>1}$, $F(z)$ has an infinite series of simple poles which are
located along the tilted line ${\zeta\cdot\R}$ at the points
\begin{equation}\label{PolFx}
\big\{\text{Poles of $F(z)$}\big\} \,=\, \left[2\pi \zeta \cdot \Z\right]\,-\,\left[2\pi
\zeta \cdot p\Z\right] \,-\, \left[2\pi \zeta \cdot q\Z\right]\,,\qquad\qquad p>q>1\,.
\end{equation}
The three terms on the right in \eqref{PolFx} correspond to the zeroes
of the three hyperbolic sines in \eqref{BigFx}.  As hopefully
clear by context, `$-$' indicates that the given subsets of \mbox{$p$-tuple}
and \mbox{$q$-tuple} points are to be removed from the set
${2\pi\zeta\cdot\Z}$.  Equivalently, the poles of $F(z)$ occur at
points ${2\pi \zeta j}$ for integer $j$ not divisible by $p$ or $q$, ie.~${p,q\nmid j}$.  For instance, if ${p=3}$ and
${q=2}$, the poles occur at points
$\{\pm2\pi\zeta,\,\pm2\pi\zeta\cdot5,\,\pm2\pi\zeta\cdot7\,,\pm2\pi\zeta\cdot11,\,\ldots\}$.
For all ${p>q>1}$, the prefactor $F(z)$ is regular at the origin and
has its closest nearby poles at $\pm2\pi\zeta$.

To apply the method of steepest-descents, we deform the real
integration contour to pass through the critical point $z_*$, such that a small neighborhood of the contour near $z_*$ is a path of steepest-descent.  As
visible in Figure \ref{fi:asymptBehaviorOfWLPartitionFunc}, the
contour may cross finitely-many poles of
$F(z)$ during the deformation.  If so, by the Cauchy theorem these
poles contribute residues which must be included, along with the naive
Gaussian contribution from $z_*$, in the leading
approximation to $Z_{S^3}(\gamma;\mathcal{K}_{p,q},n)|_{L=0}$ as ${n\to\infty}$.

Whether or not the integration contour passes through a pole of $F(z)$ during its deformation away from the real axis
depends upon the value of ${\gamma\in\C}$.  This dependence
underlies the unusual asymptotic behaviour of 
$Z_{S^3}(\gamma;\mathcal{K}_{p,q},n)|_{L=0}$ when the torus knot
$\mathcal{K}_{p,q}$ is nontrivial.  Moreover, if supersymmetric matter is coupled to the gauge theory, poles of the
double-sine functions in the third line of \eqref{ZTorusK} will
also play a role in the contour analysis.

\paragraph{A Family of Integration Contours.}  For the following, we take ${p>q>1}$ unless otherwise noted.

To evaluate $Z_{S^3}(\gamma;\mathcal{K}_{p,q},n)|_{L=0}$ as ${n\to\infty}$, we must specify a smooth family of contours $\mathcal{C}_\gamma$, each of which passes through the critical point $z_*$ determined by $\gamma$ in \eqref{CritX} and along which the integral  converges.  A canonical choice would be to take $\mathcal{C}_\gamma$ to be the straight-line steepest-descent path through $z_*$.  This choice is made in \cite{Hikami:2007,Hikami:2010,Murakami:2004asymp}, but it requires some tedious geometry to analyze precisely.  We instead adopt a more heuristic, graphical approach to defining $\mathcal{C}_\gamma$ as a homotopy from the initial real contour in Figure \ref{fi:asymptBehaviorOfWLPartitionFunc}.

The contour will depend on the complex parameter $\gamma
  \equiv |\gamma|\,\e{\!-i\varphi}$ through both the norm and
phase.  Suppose first that the norm ${|\gamma|\neq 0}$ is fixed, and
the phase $\varphi$ is continuously increased or decreased through the
range ${-\pi/2 \le \varphi \le \pi/2}$.\footnote{Recall that
  ${\Re(\gamma)\ge 0}$ is positive throughout.} There are
two qualitatively distinct cases, depicted in Figure \ref{fi:analyContiOfGammaAsympt}.

\iffigs
\begin{figure}
	\centering
	\subfloat[${\Im(\gamma)<0}$.]{\label{fi:anContGammaImLEQ0}
		\begin{tikzpicture}[x=.45cm,y=.45cm]
		\begin{scope}
		\clip (0,0) circle (5);
		\fill[blue!20] (3,-6) -- (0,0) -- (6,3) -- (6,-6) -- (3,-6);
		\fill[blue!20] (-3,6) -- (0,0) -- (-6,-3) -- (-6,6) -- (-3,6);
		\end{scope}
		
		\draw [ gray, dashed](-2.5,5)--(2.5,-5);
		\draw [ gray, dashed](5,2.5)--(-5,-2.5);
		
		\draw [->, black] (3.5,3.5) arc(45:20:3.5);
		\draw [->, black] (-3.5,-3.5) arc(225:200:3.5);
		
		\draw [red](-5.5,0)--(5.5,0) node [right, red]
                {${\R}$};
                 \draw [<->, black](0,-4.5)--(0,4);
                	\node [right] at (4,4) {$\zeta\cdot\R$};
		
		\draw [->, thick, red](-5,0)--(-2,0);
		\draw [->, thick, red](-2,0)--(2.5,0);
		\draw [thick, red](2.5,0)--(5,0);
		
		\draw [ black, dotted](-4,-4)--(4,4);
		
		\node at (1,1) {$\times$};
		\node at (3,3) {$\times$};
		
		\node at (-1,-1) {$\times$};
		\node at (-3,-3) {$\times$};

		\node at (3,.5) {$z_*$};
		\fill[black] (2.4,1.2) circle(.15);		
		\end{tikzpicture}
	}
	\qquad\qquad
	\subfloat[${\Im(\gamma)>0}$.]{\label{fi:anContGammaImGEQ0}
		\begin{tikzpicture}[x=.45cm,y=.45cm]
		\begin{scope}
		\clip (0,0) circle (5);
		\fill[blue!20] (6,-3) -- (0,0) -- (3,6) -- (6,6) -- (6,-3);
		\fill[blue!20] (-6,3) -- (0,0) -- (-3,-6) -- (-6,-6) -- (-6,3);
		\end{scope}
		
		\draw [ gray, dashed](5,-2.5)--(-5,2.5);
		\draw [ gray, dashed](2.5,5)--(-2.5,-5);
		
		\draw [->, black] (3.95,3.95) arc(45:70:3.95);
		\draw [->, black] (-3.95,-3.95) arc(225:250:3.95);
		
		\draw [red](-5.5,0)--(5.5,0) node [right, red]
                {${\R}$};
                \draw [<->, gray](0,-4.5)--(0,4);
                \node [right] at (4,4) {$\zeta\cdot\R$};
		
		\draw [->, thick, red](-5,0)--(-2,0);
		\draw [->, thick, red](-2,0)--(2.5,0);
		\draw [thick, red](2.5,0)--(5,0);
		
		\draw [ black, dotted](-4,-4)--(4,4);
		
		\node at (1,1) {$\times$};
		\node at (3,3) {$\times$};
		
		\node at (-1,-1) {$\times$};
		\node at (-3,-3) {$\times$};
		
		\node at (.8,3) {$z_*$};
		\fill[black] (1.2,2.4) circle(.15);		
		\end{tikzpicture}
	}
	\caption{Change in sectors of convergence (shaded) and location of the 
          critical point $z_*$ with the phase of ${\gamma\in\C}$, as $\gamma$
          rotates off the positive real axis.  The locations of the poles
          ($\times$) of $F(z)$ do not depend on $\gamma$.}\label{fi:analyContiOfGammaAsympt}
\end{figure}
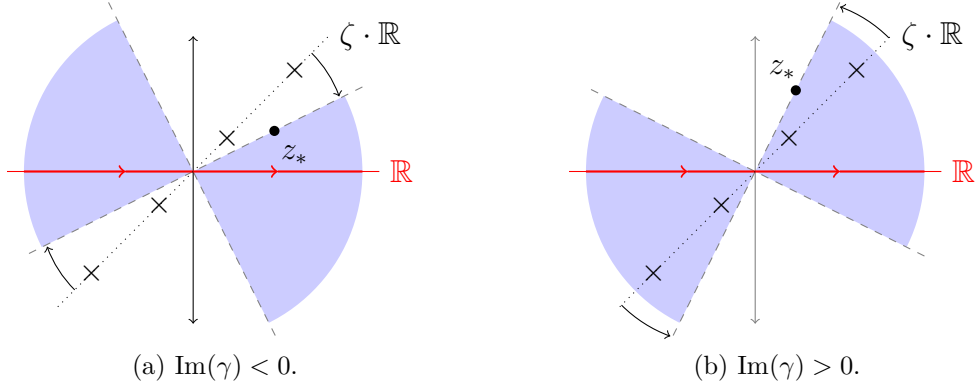\fi

\begin{enumerate}
\item[(a.)] If ${0<\varphi\le\pi/2}$, meaning ${\Im(\gamma)<0}$, the sectors of
  convergence rotate clockwise by angle $\varphi$, as in Figure \ref{fi:anContGammaImLEQ0}. In the process,
  the critical point $z_*$ also rotates clockwise off the $\zeta$-line, away from the
  poles of $F(z)$.  The initial real contour is deformed to a contour $\mathcal{C}_\gamma$
  through $z_*$ {\em without} passing through any pole of $F(z)$.  See Figure \ref{fig:imGammaLt0} for a picture of the typical $\mathcal{C}_\gamma$ in this case.  The naive Gaussian approximation about $z_*$ in the limit
  ${n\to\infty}$ is applicable.
\item[(b.)] If ${-\pi/2\le\varphi\le 0}$, meaning ${\Im(\gamma)\ge 0}$, the sectors
  of convergence rotate counterclockwise as in Figure \ref{fi:anContGammaImGEQ0}.
 In the process, the critical point $z_*$ also rotates counterclockwise
  off the $\zeta$-line of poles.  The typical contour $\mathcal{C}_\gamma$ through $z_*$ is shown in Figures \ref{fi:anContGammaImGEQ0Contour} and \ref{fi:anContGammaSmallContour}.  Evidently, whether or not the initial real contour crosses a pole of $F(z)$ in the deformation to $\mathcal{C}_\gamma$ depends upon the norm $|\gamma|$.
\end{enumerate}
We separate (b.) into three subcases depending on the norm $|\gamma|$.

\iffigs
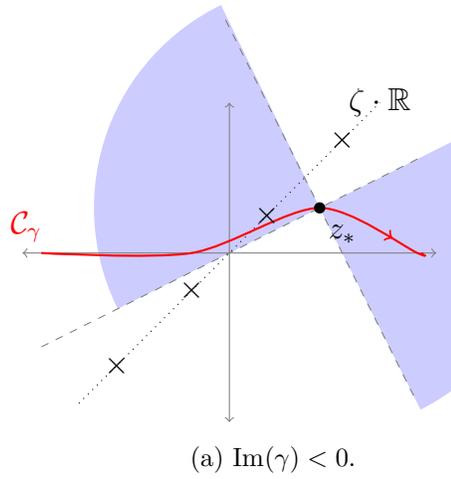
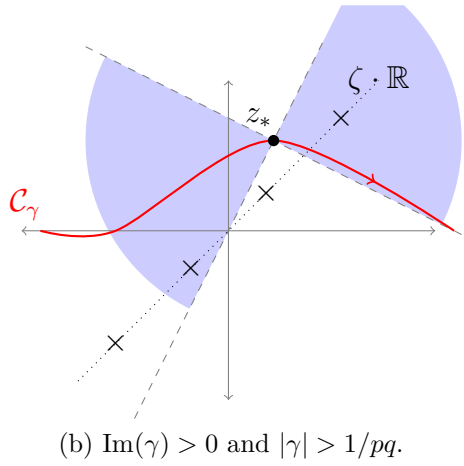
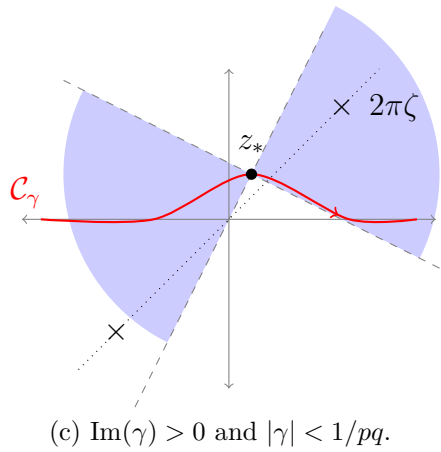
\begin{figure}
	\centering
	\subfloat[${\Im(\gamma)<0}$.]{\label{fi:anContGammaImLEQ0Contour}
		\begin{tikzpicture}[x=.5cm,y=.5cm]
		\begin{scope}
		\clip (2.4,1.2) circle (6);
		\fill[blue!20] (5.4,-4.8) -- (2.4,1.2) -- (6,3) -- (6,-6) -- (3,-6);
		\fill[blue!20] (-3,6)--(-1,8.2) -- (2.4,1.2) -- (-6,-3) -- (-6,6) -- (-3,6);
		\end{scope}
		
		\draw [ gray, dashed](-.1,6.2)--(4.9,-3.8);
		\draw [ gray, dashed](5,2.5)--(-5,-2.5);
		
		\draw [<->, gray](-5.5,0)--(5.5,0);
		\draw [<->, gray](0,-4.5)--(0,4);
                \node at (4,4) {$\zeta\cdot\R$};
		
		\draw [ black, dotted](-4,-4)--(4,4);
		
		\draw [thick,red, ->-=.9] plot [smooth] coordinates {(-5,0) (-1,0) (2.4,1.2) (5,0) (5.1,0)};
		\node [red] at (-5.4, .7) {$\mathcal{C}_\gamma$};
		
		\node at (1,1) {$\times$};
		\node at (3,3) {$\times$};
		
		\node at (-1,-1) {$\times$};
		\node at (-3,-3) {$\times$};
		
		\node at (3,.5) {$z_*$};
		\fill[black] (2.4,1.2) circle(.15);		
		\end{tikzpicture}\label{fig:imGammaLt0}
	}\\
        	\subfloat[${\Im(\gamma)>0}$ and ${|\gamma|>1/pq}$.]{\label{fi:anContGammaImGEQ0Contour}
		\begin{tikzpicture}[x=.5cm,y=.5cm]
		\begin{scope}
		\clip (1.2,2.4) circle (5);
		\fill[blue!20] (7.2,-.6) -- (1.2,2.4) -- (3,6) -- (6,6) -- (7.2,-.6);
		\fill[blue!20] (-4.8,5.4) -- (1.2,2.4) -- (-3,-6) -- (-6,-6) -- (-6,3);
		\end{scope}
		
		\draw [ gray, dashed](6.2,-.1)--(-3.8,4.9);
		\draw [ gray, dashed](2.5,5)--(-2.5,-5);
		
		\draw [<->, gray](-5.5,0)--(5.5,0);
                \draw [<->, gray](0,-4.5)--(0,4);
		\node at (4,4) {$\zeta\cdot\R$};
		
		\draw [thick,red,->-=.8] plot [smooth] coordinates {(-5,0) (-3,0) (1.2,2.4) (6,0)};
		\node [red] at (-5.4, .7) {$\mathcal{C}_\gamma$};
	
		\draw [ black, dotted](-4,-4)--(4,4);
		
		\node at (1,1) {$\times$};
		\node at (3,3) {$\times$};
		
		\node at (-1,-1) {$\times$};
		\node at (-3,-3) {$\times$};
		
		\node at (.8,3) {$z_*$};
		\fill[black] (1.2,2.4) circle(.15);		
		\end{tikzpicture}\label{fig:imGammaGrt0}
	}\\
	\subfloat[${\Im(\gamma)>0}$ and ${|\gamma|<1/pq}$.]{\label{fi:anContGammaSmallContour}
		\begin{tikzpicture}[x=.5cm,y=.5cm]
		\begin{scope}
		\clip (.6,1.2) circle (5);
		\fill[blue!20] (6.6,-1.8) -- (.6,1.2) -- (3,6) -- (6,6) -- (6.6,-1.8);
		\fill[blue!20] (-4.2,3.6) -- (.6,1.2) -- (-3,-6) -- (-6,-6) -- (-6,3);
		\end{scope}
		
		\draw [ gray, dashed](5.6,-1.3)--(-4.4,3.7);
		\draw [ gray, dashed](2.5,5)--(-2.5,-5);
		
		\draw [<->, gray](-5.5,0)--(5.5,0);
		\draw [<->, gray](0,-4.5)--(0,4);
		
		\draw [thick,red,->-=.8] plot [smooth] coordinates {(-5,0) (-2,0) (.6,1.2) (3.2,0) (5,0)};
		\node [red] at (-5.4, .7) {$\mathcal{C}_\gamma$};
		
		\draw [ black, dotted](-4,-4)--(4,4);
		
		\node at (3,3) {$\times$};
		\node at (4.4,3) {$2\pi\zeta$};
		\node at (-3,-3) {$\times$};
				
		\node at (.6,2) {$z_*$};
		\fill[black] (.6,1.2) circle(.15);		
		\end{tikzpicture}\label{fig:Gamma1onpq}
	}
	\caption{Typical integration contours ${\mathcal{C}_\gamma}$ for
          various values of ${\gamma\in\C}$.  For ${\Im(\gamma)<0}$ or
          ${|\gamma|<1/pq}$, the initial real contour
          is deformed to $\mathcal{C}_\gamma$ without crossing a pole ($\times$)
          on the $\zeta$-line.  Otherwise, the difference ${\mathcal{C}_\gamma - \R}$ is a closed contour encircling a finite number of poles, whose
          residues contribute to the torus knot observable in \eqref{ZTorusKIII}.}\label{fi:anContGammaContour}
	\end{figure}\fi

\begin{enumerate}
\item[(b1.)] If ${\Im(\gamma)\ge 0}$ and ${|\gamma|<1/pq}$, the critical point ${z_*=2\pi \zeta
    p q \gamma}$ lies closer
  to the origin than the smallest poles of $F(z)$ at the points
  $\pm2\pi\zeta$.  The real contour can be deformed to
  a contour $\mathcal{C}_\gamma$ through $z_*$
  {\em without} passing through any pole of $F(z)$.  This situation is
  shown in Figure \ref{fig:Gamma1onpq}.  The naive Gaussian
  approximation in the limit ${n\to\infty}$ is applicable.
\item[(b2.)] If ${\Im(\gamma)\ge 0}$, ${|\gamma|\ge 1/pq}$, and the
  critical point $z_*$ is not
  coincident with a pole of $F(z)$, the deformation from
  the real contour to $\mathcal{C}_\gamma$ crosses finitely-many
  poles, whose residues contribute in addition to the naive Gaussian
  integral about $z_*$.  See Figure \ref{fig:imGammaGrt0}.
\item[(b3.)] If ${\gamma\in\R}$, ${|\gamma|\ge 1/pq}$, and $z_*$
  is {\em coincident} with a pole of $F(z)$, an analysis which
  incorporates subleading effects in ${n\gg 1}$ is required.  We illustrate
  with the most relevant example, when ${\gamma=1/pq}$ and $z_*$ is coincident with the minimal pole at $2\pi\zeta$.
\end{enumerate}
We perform the analysis for cases (a)-(b1), (b2), and (b3) in turn.

\paragraph{Case I: ${\Im(\gamma)<0}$ or ${|\gamma|<1/pq}$.}  

This case is elementary.  The initial real contour is deformed to pass
through the critical point $z_*$ without passing 
through any pole of $F(z)$.  The standard Gaussian approximation for
the resulting integral along the new contour $\mathcal{C}_\gamma$ applies.

Trivially for the critical point in \eqref{CritX},
\begin{equation}
f_+(z_*) \,=\,\, \frac{i}{2}\pi p q \gamma n\,,\qquad\qquad f''_+(z_*)
\,=\,  -\frac{n}{4\pi p q \gamma}\,\neq\,0\,,
\end{equation}
and
\begin{equation}
F(z_*) \,=\, i\,\frac{\sin(\pi q \gamma) \sin(\pi p \gamma)}{\sin(\pi p
  q \gamma)}\,.
\end{equation}
The assumptions on $\gamma$ ensure that the denominator of $F(z_*)$ is non-zero.
Hence in the semiclassical regime ${k,n\to\infty}$ with ${\gamma=n/k}$
fixed, the torus knot observable \eqref{ZTorusKIII}  behaves as 
\begin{equation}\label{eq:stationaryPhaseTerm}
\begin{aligned}
Z_{S^3}\big(\gamma;\mathcal{K}_{p,q},n\big)\Big|_{L=0} \,&\underset{n\to\infty}{=}\, \frac{1}{2\pi
  i} \frac{1}{\sqrt{pq}}\exp{\!\left[-\frac{i}{2}\pi p q \gamma
    n\right]}\times\sqrt{\frac{2\pi}{-f_+''(z_*)}}\,F(z_*)\,\e{f_+(z_*)}\,,\\
&\underset{n\to\infty}{=}\, \sqrt{\frac{2\gamma}{n}}\cdot\frac{\sin(\pi q \gamma) \sin(\pi p
  \gamma)}{\sin(\pi p q \gamma)} \,+\, O\!\left(n^{-3/2}\right)\,.
\end{aligned}
\end{equation}
Of particular note, the torus knot observable remains finite and
vanishes like $n^{-1/2}$ in the semiclassical limit. 

For the unknot with ${p=q=1}$, the Gaussian analysis works
for all values of ${\gamma\in\C}$, since $F(z)$ is entire.  The formula in \eqref{eq:stationaryPhaseTerm} specializes to 
\begin{equation}\label{AsymUnknt}
Z_{S^3}\big(\gamma;\bigcirc,n\big)\Big|_{L=0}\,\underset{n\to\infty}{=}\, \sqrt{\frac{2
    \gamma}{n}}\,\sin(\pi\gamma) \,+\, O\!\left(n^{-3/2}\right)\,, 
\end{equation}
which agrees with the asymptotic behavior of the exact result
\begin{align}
Z_{S^3}\big(k;\bigcirc,n\big)\Big|_{L=0}\,=\, \sqrt{\frac{2}k}\,\sin\!\left(\frac{\pi n}{k}\right).
\end{align}

For the $n$-colored Jones polynomial \eqref{nJones} of the
general $(p,q)$-torus knot, the ratio of the asymptotic formulas in
\eqref{eq:stationaryPhaseTerm} and \eqref{AsymUnknt} implies
\begin{equation}
\begin{aligned}
 J_n\!\left(\mathcal{K}_{p,q}\right) \,&\underset{k,n\to\infty}{=}\,
 \frac{\sin(\pi p\gamma)\sin(\pi q\gamma)}{\sin(\pi \gamma)\sin(\pi
   pq\gamma)}\,,\\
&\,\,\,\,=\,\frac{1}{\Delta_{\mathcal{K}_{p,q}}(\mathsf{t})}\,,
 \qquad\qquad \mathsf{t} \,=\, \e{2\pi i\gamma}\,,
\end{aligned}
\end{equation}
where $\Delta_{\mathcal{K}_{p,q}}(\mathsf{t})$ is the Alexander-Conway polynomial
of the torus knot,
\begin{equation}
\Delta_{\mathcal{K}_{p,q}}(\mathsf{t}) \,=\,
  \frac{\left(\mathsf{t}^{pq/2} \,-\,
  \mathsf{t}^{-pq/2}\right)\left(\mathsf{t}^{1/2} \,-\,
  \mathsf{t}^{-1/2}\right)}{\left(\mathsf{t}^{p/2} \,-\,
  \mathsf{t}^{p/2}\right)\left(\mathsf{t}^{q/2} \,-\,
  \mathsf{t}^{-q/2}\right)}\,.
\end{equation}
The asymptotic relation between the $n$-colored Jones polynomial and
the Alexander polynomial is a special case of the
Melvin-Morton-Rozansky conjecture, proven in
\cite{BarNatan:1996,Garoufalidis:2005} and true for all knots.

\paragraph{Case II: ${\Im(\gamma)\ge 0}$ and ${|\gamma|\ge1/pq}$, with
  $\gamma$ generic.}

In this case the deformation from the real contour to $\mathcal{C}_\gamma$ passes through poles of
the prefactor $F(z)$ in \eqref{ZTorusKIII}; see Figure
\ref{fig:imGammaGrt0}.  By Cauchy's theorem, the Wilson loop 
partition function is equal to the contour integral along $\mathcal{C}_\gamma$
plus a finite sum of residues.  Schematically,
\begin{equation}
Z_{\R} \,=\, Z_{\mathcal{C}_\gamma} \,+\, Z_{\textrm{res}}\,.
\end{equation}
In the limit ${n\to\infty}$, the contour integral $Z_{\mathcal{C}_\gamma}$ is given by the
finite expression in \eqref{eq:stationaryPhaseTerm}, and the residue term
$Z_{\textrm{res}}$ can be evaluated exactly for all $n$.

For concreteness, assume $\gamma$ lies in an open neighborhood of
${1\in\C}$, meaning $z_*$ lies in an open neighborhood of ${2\pi p q
  \zeta}$.  As will be clear shortly, the assumption on the particular
value of $\gamma$ becomes inessential in the limit
${n\to\infty}$.  By \eqref{PolFx}, $Z_{\textrm{res}}$ is a sum of
residues at the points ${z=2\pi \zeta j}$ for ${j=1,\ldots,pq}$ such
that ${p,q\nmid 
  j}$.  Explicitly for the contour integral in \eqref{ZTorusKIII},
\begin{equation}
Z_{\textrm{res}} \,=\,
\frac{1}{\sqrt{pq}}\Phi\cdot\sum^{pq-1}_{\substack{j = 1\\ p,q\nmid
    j}} \Res\!\left[F(z)\,\e{f_+(z)}\right]_{z = 2\pi \zeta j}\,,
\end{equation}
and by a brief calculation,
\begin{equation}\label{eq:residuesOfPolesFromVec}
\Res\!\left[F(z)\,\e{f_+(z)}\right]_{z = 2\pi \zeta j} \,=\, \frac{i}{\pi}\sin\!\left(\frac{\pi  j}{p}\right)\sin\!\left(\frac{\pi  j}{q}\right)\exp{\!\left[
i\pi n j \left(1-\frac{j}{2pq\gamma}\right)\right]}\,.
\end{equation}
Substituting the expression \eqref{AsymPhi} for $\Phi$,
\begin{equation}\label{ZresII}
\begin{aligned}
Z_{\textrm{res}} \,&=\, \frac{i}{\pi\sqrt{pq}} \exp{\!\left[
-\frac{i\pi}{2}pq \gamma n\,+\,O\!\left(\frac{1}{n}\right)
\right]}\,\times\\
&\times\sum^{pq-1}_{\substack{j = 1\\ p,q\nmid
    j}}\sin\!\left(\frac{\pi  j}{p}\right)\sin\!\left(\frac{\pi  j}{q}\right)\exp{\!\left[
i\pi n j \left(1-\frac{j}{2pq\gamma}\right)\right]}\,.
\end{aligned}
\end{equation}

When ${\gamma\in\R_+}$ is real, $Z_{\textrm{res}}$ remains finite and
oscillates with $n$.

Else if ${\Im(\gamma)>0}$ is positive, the exponential in the first
line of \eqref{ZresII} grows as ${n\to\infty}$, whereas the
exponential in the second line decays as ${n\to\infty}$.  In this
limit, the dominant term in the sum over $j$ is the term with ${j=1}$,
from the pole at $2\pi\zeta$ which is nearest the origin, and all
terms with ${j>1}$ are exponentially suppressed relative to the
leading term.  Hence
\begin{equation}\label{AsymZres}
Z_{\textrm{res}} \,\underset{n\to\infty}{=}\, \frac{i}{\pi\sqrt{pq}}
\sin\!\left(\frac{\pi}{p}\right)\sin\!\left(\frac{\pi}{q}\right)\exp{\!\left[-\frac{i}{2}\pi
    p q \gamma n \,+\, i \pi n \,-\, \frac{i \pi n}{2 p q
      \gamma}\,+\, O\!\left(\frac{1}{n}\right)\right]}.
\end{equation}
Because only the pole at $2\pi\zeta$ contributes as ${n\to\infty}$, 
the initial assumption on the particular value of 
$\gamma$, which was used to set the upper limit on the range of
summation over $j$ in \eqref{ZresII}, is irrelevant.  

When ${\Im(\gamma)>0}$ and ${|\gamma|>1/pq}$, the exponential in
\eqref{AsymZres} diverges as ${n\to\infty}$.  As
$Z_{\mathcal{C}_\gamma}$ remains finite in the limit, we obtain the
same asymptotic behavior for the full Wilson loop partition function
\begin{equation}\label{ExpZS3}
\begin{aligned}
&Z_{S^3}\big(\gamma;\mathcal{K}_{p,q},n\big)\big|_{L=0}
\,\underset{n\to\infty}{=}\,\\
&\qquad\frac{i}{\pi\sqrt{pq}}
\sin\!\left(\frac{\pi}{p}\right)\sin\!\left(\frac{\pi}{q}\right)\exp{\!\left[-\frac{i}{2}\pi
    p q \gamma n \,+\, i \pi n \,-\, \frac{i \pi n}{2 p q
      \gamma}\,+\, O\!\left(\frac{1}{n}\right)\right]},
\end{aligned}
\end{equation}
or for the free energy,
\begin{equation}\label{FreeZ}
\lim_{n\to\infty}\frac{\ln{Z_{S^3}\big(\gamma;\mathcal{K}_{p,q},n\big)\big|_{L=0}}}{n}
\,=\, i\pi\left(1-\frac{p q \gamma}{2} - \frac{1}{2 p q
    \gamma}\right).
\end{equation}

Finally, on the semicircle where ${\Im(\gamma)>0}$ and
${|\gamma|=1/pq}$, the exponential in \eqref{AsymZres} is bounded and oscillates with
$n$, similar to the case with ${\gamma\in\R_+}$ real.

We summarize the asymptotic behavior of the semiclassical Wilson loop
observable as a function of ${\gamma\in\C}$ in Figure
\ref{fi:asymptBehaviorOfZwrtGamma}.  The results in Cases I and II
agree with the more detailed analysis by Murakami in
\cite{Murakami:2004asymp}.  

Note that the unknot partition function in
\eqref{AsymUnknt} is finite as ${n\to\infty}$ for all values of
${\gamma\in\C}$.  Hence the exponential divergence \eqref{ExpZS3} of the
$(p,q)$-torus knot partition function for ${\Im(\gamma)>0}$
and ${|\gamma|>1/pq}$, together with the definition \eqref{nJones} of the colored
Jones polynomial, implies that $J_n(\mathcal{K}_{p,q})$ also diverges exponentially
as ${n\to\infty}$ for these values of $\gamma$.

\iffigs
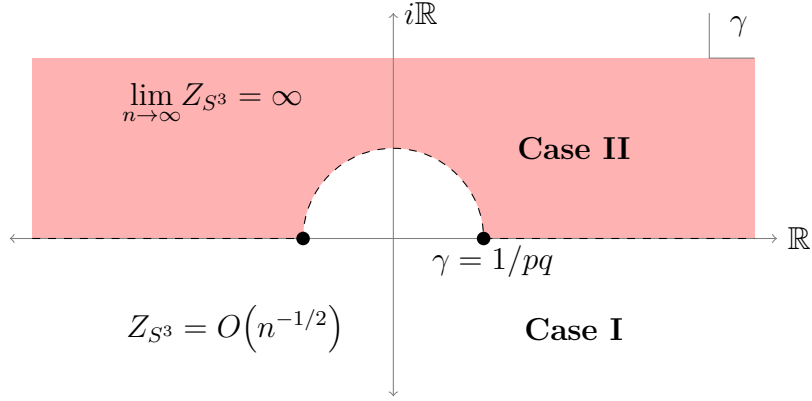
\begin{figure}
	\begin{center}
		\begin{tikzpicture}[x=.6cm,y=.6cm]

		\fill[red!30] (-8,0)--(-8,4)--(8,4)--(8,0)--(-8,0);
		
		\fill[white] (0,0) circle(2);
		
		\draw [<->, gray](-8.5,0)--(8.5,0) node [right, black] {$\R$};
		\draw [<->, gray](0,-3.5)--(0,5) node [right, black] {$i\R$};
		
		\draw [black, dashed](-8,0)--(-2,0)  arc(180:0:2) -- (8,0);		
		
		\draw [thin, gray](7,5)--(7,4) -- (8,4);
		\node at (7.65,4.7) {$\gamma$};
		
		\fill[black] (2,0) circle(.15);
		\node at (2.2,-0.5) {${\gamma=1/pq}$};
                \fill[black] (-2,0) circle(.15);
		
		\node at (4,-2) {\textbf{Case I}};
		\node at (4,2) {\textbf{Case II}};
		
		\node at (-4,3) {${\underset{n\to\infty}{\lim}Z_{S^3} = \infty}$};
		
		\node at (-3.5,-2) {${Z_{S^3}=O\!\left(n^{-1/2}\right)}$};

		%	\draw [->, gray] (-3.8,3.8) --  (-4.6,4.6);
		%	\draw [->, gray] (3.8,-3.8) --  (4.6,-4.6);
		\end{tikzpicture}
		\caption{Asymptotic behavior of the Wilson loop
                  partition function
                  $Z_{S^3}(\gamma;\mathcal{K}_{p,q},n)$ as
                  ${n\to\infty}$, ${p>q>1}$.  For ${\Im(\gamma)<0}$ or
                  ${|\gamma|<1/pq}$, the partition function vanishes
                  like $n^{-1/2}$.
                  For ${\Im(\gamma)>0}$ and ${|\gamma|>1/pq}$, the
                  partition function diverges exponentially with $n$
                  at the rate in \eqref{FreeZ}.  For generic values of
                  $\gamma$ on the boundary between these regions, the
                  value of $Z_{S^3}$ oscillates with $n$.  At the
                  special points ${\gamma=\pm 1/pq}$, the partition
                  function has a finite, non-zero limit.}\label{fi:asymptBehaviorOfZwrtGamma}
	\end{center}
\end{figure}\fi

\paragraph{Case III: ${\gamma\in\frac{1}{pq}\Z - \frac{1}{p}\Z -
    \frac{1}{q}\Z}$ is non-generic.}

We finally consider the special case in which the critical point $z_*$
coincides with a pole of the prefactor $F(z)$ in \eqref{BigFx}.  We
focus on the most relevant example in which $z_*$ coincides with the
minimal pole at $2\pi\zeta$, corresponding to ${\gamma=1/pq}$.

To carry out the semiclassical analysis in this more delicate situation, we
rewrite the contour integral in \eqref{ZTorusKIII} for ${\gamma=1/pq}$
as 
\begin{equation}\label{ZTorusKdel}
Z_{S^3}\big(\gamma;\mathcal{K}_{p,q},n\big)\Big|_{L=0} \,=\,\frac1{2\pi
  i}\frac{1}{\sqrt{pq}}\,\Phi\cdot\int_{\R}dz\,\exp{\!\left[G(z)\right]}\,,
\end{equation}
where now
\begin{equation}
G(z) = -\frac{n}{8\pi}z^2 \,+\, \ha \zeta n z
\,+\,\ln\sinh\!\left(\frac{\zeta z}{2p}\right) +
\ln\sinh\!\left(\frac{\zeta z}{2q}\right) \,-\, \ln\sinh\!\left(\frac{\zeta z}{2}\right).
\end{equation}
The additional terms in the definition of $G(z)$ relative to $f_+(z)$
in \eqref{fplus} arise from the logarithm of the prefactor $F(z)$.
Also, for ${\gamma=1/pq}$,
\begin{equation}
\Phi \,=\, \exp{\!\left[-\frac{i\pi}{2} n \,+\, O\!\left(\frac{1}{n}\right)\right]}\,.
\end{equation}

Subleading corrections will shift the critical point of $G(z)$
slightly away from the naive critical point at $2\pi\zeta$, where
$G(z)$ diverges.  We compute this correction perturbatively in the
expansion parameter ${1/n\ll 1}$.  The exact expression for the derivative is given by 
\begin{equation}
G'(z) \,=\, -\frac{n}{4\pi}z \,+\, \ha \zeta n  \,+\,\frac{\zeta}{2}\left[\frac{1}{p}\coth\!\left(\frac{\zeta z}{2p}\right) \,+\, \frac{1}{q}\coth\!\left(\frac{\zeta z}{2q}\right) \,-\, \coth\!\left(\frac{\zeta z}{2}\right)\right].
\end{equation}
We now solve ${G'(z_*)=0}$ perturbatively with the ansatz
\begin{equation}
z_* \,=\, 2\pi \zeta \left(1-\frac{\delta}{n}\right),\qquad\qquad
\zeta\equiv \e{\!i\pi/4}\,,
\end{equation}
for a rescaled variable $\delta$.  In terms of the new variable, the critical point equation ${G'(z_*)=0}$ becomes
\begin{equation}\label{CritPtG}
0 \,=\, \delta \,+\,
  \frac{1}{p}\coth\!\left[\frac{i\pi}{p}\!\left(1-\frac{\delta}{n}\right)\right]
  +
  \frac{1}{q}\coth\!\left[\frac{i\pi}{q}\!\left(1-\frac{\delta}{n}\right)\right]-\coth\!\left[i\pi\!\left(1-\frac{\delta}{n}\right)\right].
\end{equation}
We expand each hyperbolic cotangent in Taylor series to obtain at leading-order
\begin{equation}\label{CritPtGP}
\left(1 \,+\, \frac{\mathsf{A}}{n}\right)\delta \,+\, \mathsf{B} \,-\,
\frac{i\,n}{\pi\,\delta} \,+\, O\!\left(n^{-2}\right)\,=\, 0\,.
\end{equation}
Here $\mathsf{A}$ and $\mathsf{B}$ are finite constants depending upon the coprime pair $(p,q)$,
\begin{equation}
\begin{aligned}
\mathsf{A} \,&=\, \frac{i\pi}{p^2} \csch^2\!\left(\frac{i\pi}{p}\right) \,+\,
\frac{i\pi}{q^2}\csch^2\!\left(\frac{i\pi}{q}\right) \,+\,
\frac{i\pi}{3}\,,\\
\mathsf{B} \,&=\, \frac{1}{p}\coth\!\left(\frac{i\pi}{p}\right) \,+\, \frac{1}{q}\coth\!\left(\frac{i\pi}{q}\right),
\end{aligned}
\end{equation}
and the singular term proportional to $n/\delta$ in \eqref{CritPtGP} arises from the pole
in the last term of \eqref{CritPtG}.  The perturbative equation for
$\delta$ has solutions
\begin{equation}
\delta_\pm \,=\, \frac{-\mathsf{B} \,\pm\,\sqrt{\mathsf{B}^2 \,+\, 4
    \left(1+\frac{\mathsf{A}}{n}\right)\left(\frac{i\,
        n}{\pi}\right)}}{2\left(1\,+\,\frac{\mathsf{A}}{n}\right)} \,=\,
\pm\zeta\sqrt{\frac{n}{\pi}} \,+\, O(1)\,,
\end{equation}
independent of $\mathsf{A}$ and $\mathsf{B}$ to
leading-order.

The initial real contour in \eqref{ZTorusKdel} can be deformed to pass
through the critical point described by $\delta_+$ without crossing
the pole at $2\pi\zeta$.  After this contour deformation, the Wilson loop partition function is
evaluated semiclassically via the naive Gaussian
approximation around 
\begin{equation}\label{Pertzstar}
z_* \,=\, 2\pi\zeta\left(1-\zeta\sqrt\frac{1}{\pi n}\right).
\end{equation}
By comparison to the Gaussian formula in
\eqref{eq:stationaryPhaseTerm}, the small, $O(n^{-1/2})$-correction to
$z_*$ is only necessary to include when we evaluate the polar term in
$F(z_*)$; the perturbative correction to $z_*$ can otherwise be
ignored in \eqref{eq:stationaryPhaseTerm} to leading-order.
Substituting the perturbative result for $z_*$ into the first line of
\eqref{eq:stationaryPhaseTerm}, we obtain the non-vanishing limit
\begin{equation}\label{eq:stationaryPhaseTermII}
\lim_{n\to\infty} Z_{S^3}\big(\gamma=\frac{1}{pq};\mathcal{K}_{p,q},n\big)\Big|_{L=0}
\,=\,\zeta^{-1}
\sqrt{\frac{2}{\pi p q}}\sin\!\left(\frac{\pi}{p}\right) \sin\!\left(\frac{\pi}{q}
  \right)\,.
\end{equation}
At the critical value ${\gamma=1/pq}$, the Wilson loop partition
function neither vanishes nor diverges as ${n\to\infty}$.

For the unknot, note that
\begin{equation}\label{eq:Zunknot}
Z_{S^3}\big(\gamma=\frac{1}{pq};\bigcirc,n\big)\Big|_{L=0}\,=\,
\sqrt{\frac{2}{p q n}} \sin\!\left(\frac{\pi}{p q}\right).
\end{equation}
When evaluated at the special value ${\gamma=1/pq}$, the colored Jones
polynomial of the $(p,q)$-torus knot thus grows asymptotically with
$n$ as\footnote{The asymptotic formula \eqref{specialJKpq} for
  $J_n(\mathcal{K}_{p,q})$ at ${\gamma=1/pq}$ agrees with Theorem 3.1 in \cite{Hikami:2007} up to the replacement
  of $\sqrt{\pi}$ with $\sqrt{2}$ in the denominator.  This small discrepancy seems to be a typo.}
\begin{equation}\label{specialJKpq}
J_n\big(\mathcal{K}_{p,q}\big) \,\underset{n\to\infty}{=}\, \zeta^{-1}
\frac{\sin\!\left(\frac{\pi}{p}\right)
  \sin\!\left(\frac{\pi}{q}\right)}{
  \sin\!\left(\frac{\pi}{p q}\right)}\,\sqrt{\frac{n}{\pi}} \,+\,
O(1)\,,\qquad \gamma=1/pq\,.
\end{equation}
Unlike the exponential growth for values of $\gamma$ in the shaded
region of Figure \ref{fi:asymptBehaviorOfZwrtGamma},
$J_n(\mathcal{K}_{p,q})$ has only a power-law divergence with $n$ at the
point ${\gamma=1/pq}$.

\subsection{Coupling to Supersymmetric Matter}\label{Coupling}

A similar asymptotic analysis works for the Wilson loop partition
function in \eqref{ZTorusK} when the $SU(2)$ gauge theory is coupled
to matter.  

Again via the Weyl symmetry on the Coulomb-branch, the
partition function with matter in the irreducible $SU(2)$ representation of
dimension $[L+1]$ takes the form 
\begin{equation}\label{ZmatTorusII}
Z_{S^3}\big(\gamma,\mu;\mathcal{K}_{p,q},n,L\big) \,=\,\frac1{2\pi
  i}\frac{1}{\sqrt{pq}}\,\Phi\cdot\int_{\R}dz\;
H(z)\,\e{f_+(z)}\,,
\end{equation}
where $\Phi$ and $f_+(z)$ are the same functions in \eqref{BigFx}, and
the new prefactor $H(z)$ now includes a one-loop contribution from the matter
as well as the vector multiplet,
\begin{equation}\label{BigHfix}
H(z) \,=\, \frac{\sinh\!\left(\zeta z/2p\right)\sinh\!\left(\zeta z/2q\right)}
{\sinh(\zeta z/2)}\,\times\mskip-50mu\prod_{\substack{\beta\in\Delta\,,\\ \Delta=\{L, L-2,\, \cdots,
2-L, -L\}}}\mskip-50mu
s_{b=\sqrt{p/q}}\!\left(\zeta\,\frac{\beta\,z}{4 \pi
  \sqrt{pq}}\,+\,\mu\right).
\end{equation}
Again, ${\zeta \equiv \exp{\!(i\pi/4)}}$ is an eighth-root of unity.
In the limit ${n\to\infty}$ with\footnote{The
  semiclassical limit of ``spinning-matter'' with ${n,L\to\infty}$
  jointly might also
  be interesting to explore, but we do not analyze that limit here.}  ${L\in \Z}$ and ${\gamma,\mu\in\C}$ fixed, a
steepest-descent contour for the integrand in \eqref{ZmatTorusII} passes through the
same critical point $z_*$ of $f_+(z)$ in \eqref{CritX}.  

However, as we
have already observed, the semiclassical behavior of $Z_{S^3}$ as a
function of both ${\gamma,\mu\in\C}$ depends very much on the analytic
structure of the prefactor $H(z)$, which has poles
at the locations in \eqref{PolFx} as well as at points determined
by poles of the double-sine factors
in \eqref{BigHfix},
\begin{equation}\label{PolHx}
\begin{aligned}
&\big\{\text{Poles of $H(z)$}\big\} \,=\, \left[2\pi \zeta \cdot \Z\right]\,-\, \left[2\pi
\zeta \cdot p\Z\right] \,-\, \left[2\pi \zeta \cdot q\Z\right]\,\,+\,\\
&+\,\sum_{\substack{\beta\neq 0\in\Delta\,,\\ \Delta=\{L, L-2,\, \cdots,
2-L, -L\}}} \mskip-20mu\left[\frac{4\pi\zeta}{\beta}\cdot\big(p \Z_{\le 0} \,+\, q
\Z_{\le 0} \,+\, i \,\mu \sqrt{pq}\big)\right].
\end{aligned}
\end{equation}
The poles in $H(z)$ appearing in the second line of \eqref{PolHx} generally
occur with non-trivial multiplicity, due both to the sum over weights $\beta$
as well as the dual sum over negative integers 
multiplying $p$ and $q$.  Significantly, $H(z)$ always has simple
poles at the special points $\pm4\pi i\zeta \mu \sqrt{pq}/L$, which for
small mass ${0<|\mu|\ll 1}$ are the poles of $H(z)$ lying nearest the
origin.  See Figure \ref{PolesofHz} for a schematic diagram of the
poles of $H(z)$.

\iffigs
\begin{figure}[h]
	\begin{center}
		\begin{tikzpicture}[x=.6cm,y=.6cm]
		\begin{scope}
		\clip (0,0) circle (5);
		\fill[blue!20] (6,-6) -- (0,0) -- (6,6)  --(6,-6) ;
		\fill[blue!20] (-6,6) -- (0,0) -- (-6,-6)  --(-6,6) ;
		\end{scope}
		
		\draw [red](-5.5,0)--(5.5,0) node [right, red] {${\R}$};
		\draw [gray](0,-4.5)--(0,4);
		\node at (4,4) {$\zeta\cdot\R$};
		
		\draw [->, thick, red](-5,0)--(-2,0);
		\draw [->, thick, red](-2,0)--(2.5,0);
		
		\draw [thick, red](2.5,0)--(5,0);
		
		\draw [ gray, dashed](-4,-4)--(4,4);
		\draw [ gray, dashed](-4,4)--(4,-4);
		
		\node at (1,1) {$\times$};
		\node at (3,3) {$\times$};
		
		\node at (-1,-1) {$\times$};
		\node at (-3,-3) {$\times$};
		
		\node at (2.3,1.5) {$z_*$};

                \fill[black] (1.5,1) circle(.15);
		\fill[white] (1.5,1) circle(.1);
			
		\fill[black] (3.5,3) circle(.15);
		\fill[white] (3.5,3) circle(.1);
		
		\fill[black] (-1.5,-1) circle(.15);
		\fill[white] (-1.5,-1) circle(.1);
		
		\fill[black] (-2.5,-2) circle(.15);
		\fill[white] (-2.5,-2) circle(.1);
		
		\fill[black] (-3,-2.5) circle(.15);
		\fill[white] (-3,-2.5) circle(.1);
		
		\fill[black] (-3.5,-3) circle(.15);
		\fill[white] (-3.5,-3) circle(.1);
		
		\fill[black] (2,2) circle(.15);

		\end{tikzpicture}
		\caption{Schematic distribution of poles of the
                  prefactor $H(z)$.   Poles from the vector multiplet
                  are indicated by $(\times)$, and the poles from the
                  matter multiplet are indicated by $(\circ)$.  The
                  critical point of $f_+(z)$ lies at $z_*$.  Compare
                  to Figure \ref{fi:asymptBehaviorOfWLPartitionFunc}.}\label{PolesofHz}
	\end{center}
\end{figure}
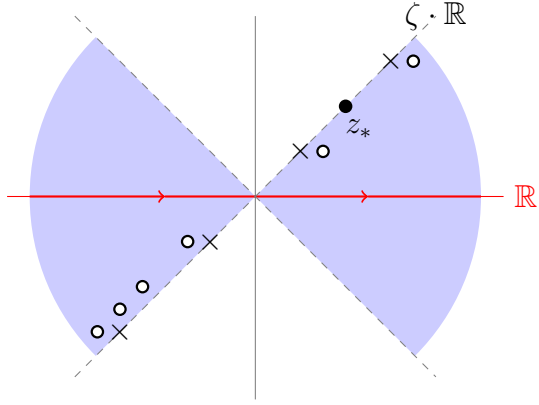\fi

The situation is particularly simple when ${\mu\in i\R}$ is
pure imaginary, corresponding physically to the theory with
vanishing real mass.  As evident from \eqref{PolHx}, all poles of
$H(z)$ then lie along the line ${\zeta\cdot\R}$, just like the
prefactor $F(z)$ in \eqref{BigFx} which controls the analytic behavior
of the colored Jones polynomial.  In this case, the qualitative
features of the discussion in Section \ref{Recollect} remain unchanged when
${\mu\in i\R}$ is imaginary, though quantitative details do depend on
the value of $\mu$.

In the remainder, we specialize our analysis to the case ${\mu\in
  i\R}$ is imaginary.  For non-zero real masses and general
${\mu\in\C}$, the refined stationary-phase techniques used in
\cite{Hikami:2007,Hikami:2010,Murakami:2004asymp} should be equally
applicable, but we do not consider the generalization here.

\paragraph{Case I: ${\Im(\gamma)<0}$ or ${|\gamma| < \min\!\big\{1/pq,\,
    2|\mu|/L\sqrt{pq}\big\}}$.}

When $\Im(\gamma)<0$ or $|\gamma|$ is smaller than both ${1/pq}$ and
$2{|\mu|/L\sqrt{pq}}$, the naive stationary-phase analysis of
\eqref{ZmatTorusII} works, with no
contributions from poles of $H(z)$.
The single critical point of $f_+(z)$ occurs at ${z_*=2\pi\zeta p q \gamma}$ as before, with 
\begin{equation}
H(z_*) \,=\,  i\,\frac{\sin(\pi q \gamma) \sin(\pi p \gamma)}{\sin(\pi p
  q \gamma)}\,\times\mskip-50mu\prod_{\substack{\beta\in\Delta\,,\\ \Delta=\{L, L-2,\, \cdots,
2-L, -L\}}}\mskip-50mu
s_{b=\sqrt{p/q}}\!\left(\frac{i}{2}\beta\gamma\sqrt{pq}\,+\,\mu\right).
\end{equation}
The conditions ${|\gamma|<1/pq}$ and ${|\gamma|< 2|\mu|/L\sqrt{pq}}$
ensure that $H(z_*)$ is non-singular.  As in \eqref{eq:stationaryPhaseTerm},
\begin{equation}\label{ZmatStatPhas}
\begin{aligned}
&Z_{S^3}\big(\gamma,\mu;\mathcal{K}_{p,q},n,L\big)
\,\underset{n\to\infty}{=}\,\\
&\quad\sqrt{\frac{2\gamma}{n}}\cdot\frac{\sin(\pi q \gamma) \sin(\pi p
  \gamma)}{\sin(\pi p q \gamma)}\,\times\mskip-50mu\prod_{\substack{\beta\in\Delta\,,\\ \Delta=\{L, L-2,\, \cdots,
2-L, -L\}}}\mskip-50mu
s_{b=\sqrt{p/q}}\!\left(\frac{i}{2}\beta\gamma\sqrt{pq}\,+\,\mu\right)
\,+\, O\!\left(n^{-3/2}\right).
\end{aligned}
\end{equation}
Just like the theory without matter, $Z_{S^3}$ vanishes asymptotically
as ${n^{-1/2}}$.  Only the coefficient of ${n^{-1/2}}$ depends upon
  the mass $\mu$, through one-loop corrections encoded by the
  double-sine functions.

\paragraph{Case II: ${\Im(\gamma)\ge 0}$ and ${|\gamma| > \min\!\big\{1/pq,\,
    2|\mu|/L\sqrt{pq}\big\}}$, with $\gamma$ generic.}

In this case the deformation from the real contour in \eqref{ZmatTorusII} to the
stationary-phase contour $\mathcal{C}_\gamma$ passes through at
least one pole of $H(z)$, so that again
\begin{equation}
Z_{\R} \,=\, Z_{\mathcal{C}_\gamma} \,+\, Z_{\textrm{res}}\,.
\end{equation}
The situation now depends upon whether poles from the vector
multiplet or the matter multiplet contribute to the residue sum
$Z_{\textrm{res}}$.

If ${1/pq < |\gamma| <  2|\mu|/L\sqrt{pq}}$, which is only possible
when the mass $\mu$ is sufficiently large, poles from the matter
multiplet do not contribute to $Z_{\textrm{res}}$.  All contributions
to the residue sum arise from the vector multiplet.  These
contributions were considered previously in Case II of Appendix \ref{Recollect}.
The dominant contribution to $Z_{\textrm{res}}$ as
${n\to\infty}$ arises from the pole in $H(z)$ at $2\pi\zeta$.  By the
same computation which leads to \eqref{AsymZres},
\begin{equation}\label{AsymZresMat}
\begin{aligned}
Z_{\textrm{res}} \,&\underset{n\to\infty}{=}\, \frac{i}{\pi\sqrt{pq}}
\sin\!\left(\frac{\pi}{p}\right)\sin\!\left(\frac{\pi}{q}\right)\,\times\mskip-50mu\prod_{\substack{\beta\in\Delta\,,\\ \Delta=\{L, L-2,\, \cdots,
2-L, -L\}}}\mskip-50mu
s_{b=\sqrt{p/q}}\!\left(\frac{i\,\beta}{2
  \sqrt{pq}}\,+\,\mu\right)\times\,\\
&\qquad\times\,\exp{\!\left[-\frac{i}{2}\pi
    p q \gamma n \,+\, i \pi n \,-\, \frac{i \pi n}{2 p q
      \gamma}\,+\, O\!\left(\frac{1}{n}\right)\right]}.
\end{aligned}
\end{equation}
Our assumptions on $\gamma$ imply that the exponential diverges with
$n$.  Since $Z_{\mathcal{C}_\gamma}$ vanishes as ${n\to\infty}$, the
free energy behaves identically to the pure vector theory,
\begin{equation}\label{FreeZMat}
\lim_{n\to\infty}\frac{\ln{Z_{S^3}\big(\gamma,\mu;\mathcal{K}_{p,q},n,L\big)}}{n}
\,=\, i\pi\left(1-\frac{p q \gamma}{2} - \frac{1}{2 p q
    \gamma}\right),\qquad 1/pq < |\gamma| <  2|\mu|/L\sqrt{pq}\,.
\end{equation}

By contrast, when ${2|\mu|/L\sqrt{pq} < |\gamma| < 1/pq}$, poles from
the matter multiplet but not the vector multiplet contribute to
$Z_{\textrm{res}}$.  These poles occur at locations labelled by a
triple $(a,b,\beta)$,
\begin{equation}\label{zabs}
\begin{aligned}
z_{a,b} \,&=\, \frac{4\pi\zeta}{\beta}\left(-p\,a - q\,b \,+\, i \mu
  \sqrt{pq}\right),\qquad a,b\,\in\,\Z_{\ge 0}\,,\\
\beta \,&\in\, \big\{L,\,L-2,\,\cdots,\,2-L,\,-L\big\}\,.
\end{aligned}
\end{equation}
All dependence on $n$ occurs through the prefactor $\Phi$ and the
function $f_+(z)$ in \eqref{ZmatTorusII}.  The residue at
$z_{a,b}$ for given $\beta$ scales with $n$ as 
\begin{equation}\label{AsymZresMatII}
\begin{aligned}
&\frac{\ln Z_{\textrm{res}}}{n}
\,\underset{n\to\infty}{=}\,\\
&-i\pi\left[\frac{2}{\beta^2 p q
    \gamma}\left(p\,a+q\,b-i\mu\sqrt{pq}\right)^2\,+\,\frac{2}{\beta}\left(p\,a+q\,b-i\mu\sqrt{pq}\right)
+\, \frac{p q \gamma}{2} \,+\, O\!\left(\frac{1}{n}\right)\right].
\end{aligned}
\end{equation}
This result should be compared to the corresponding scaling in the
second line of \eqref{AsymZresMat}.  Inessential prefactors are
captured by the $O(1/n)$-terms in \eqref{AsymZresMatII}.  

Since by assumption ${\mu\in i\R}$ is
imaginary, the scaling of the norm $|Z_{\textrm{res}}|$ is much
simpler, with 
\begin{equation}\label{AsymZresMatIII}
\frac{\ln|Z_{\textrm{res}}|}{n}
\,\underset{n\to\infty}{=}\, \frac{\pi p q}{2} \Im(\gamma) \left[1
  \,-\, \frac{4}{\beta^2 p^2 q^2 |\gamma|^2}\left(p\,a + q \, b - i
    \mu \sqrt{pq}\right)^2\,+\, O\!\left(\frac{1}{n}\right)\right].
\end{equation}
For fixed ${\gamma\in\C}$ with ${\Im(\gamma)>0}$,
the pole which makes the dominant contribution
to $Z_{\textrm{res}}$ is that for which the pair ${a,b\in\Z_{\ge 0}}$ minimizes the quantity $\left(p\,a + q\,b -
i\mu\sqrt{pq}\right)^2$ and for which
${\beta^2=L^2}$ is maximum.  Evidently the dominant pole depends upon
${\mu\in i\R}$ and the knot $\mathcal{K}_{p,q}$ through the first condition, but when either ${\Im(\mu)>0}$ or
${|\mu|\ll p,q}$ is small, the dominant pole is always the distinguished pole
of $H(z)$ lying nearest to the origin, with ${a=b=0}$ and
${|\beta|=L}$ in \eqref{zabs}.  As for signs, the dominant pole lies
in the upper half-plane.  Whether ${\beta=\pm L}$ for the dominant
pole is determined by the sign of $\Im(\mu)$.  For ${\Im(\mu)>0}$,
${\beta=-L}$, and vice versa.

In summary, if ${2|\mu|/L\sqrt{pq} < |\gamma| < 1/pq}$ and ${\mu\in
  i\R}$ with either 
${\Im(\mu)>0}$ or ${|\mu|\ll p,q}$, the free energy diverges
exponentially with $n$ at the rate in 
\eqref{AsymZresMatII},
\begin{equation}\label{FreeZMatII}
\lim_{n\to\infty}\frac{\ln{Z_{S^3}\big(\gamma,\mu;\mathcal{K}_{p,q},n,L\big)}}{n}
\,=\, i\pi\left( \frac{2 |\mu| 
    \sqrt{pq}}{L}  \,-\, \frac{p q \gamma}{2} \,-\, \frac{2}{L^2}\frac{|\mu|^2}{\gamma}\right).
\end{equation}
For other values of ${\mu\in i\R}$ with ${\Im(\mu)<0}$, the free
energy still diverges 
exponentially with $n$, but the rate of divergence has a complicated,
non-analytic dependence on $\mu$, determined by minimizing 
$\left(p\,a + q\,b - i\mu\sqrt{pq}\right)^2$ over ${a,b\in\Z_{\ge 0}}$.

Finally, when ${\Im(\gamma) \ge 0}$ and $|\gamma|$ is greater than
${2|\mu|/L\sqrt{pq}}$ and $1/pq$, poles from both the vector
and the matter multiplets contribute to $Z_{\textrm{res}}$.  Depending
upon the value of the mass ${\mu\in i\R}$, the leading asymptotic divergence in
$Z_{S^3}$ is given by one or the other of the prior results \eqref{FreeZMat} and
\eqref{FreeZMatII}.  For ${|\mu|>L/2\sqrt{pq}}$, the vector
contribution is dominant,
\begin{equation}\label{FreeZMatIII}
\lim_{n\to\infty}\frac{\ln{Z_{S^3}\big(\gamma,\mu;\mathcal{K}_{p,q},n,L\big)}}{n}
\,\stackrel{|\mu|>L/2\sqrt{pq}}{=}\, i\pi\left(1-\frac{p q \gamma}{2} - \frac{1}{2 p q
    \gamma}\right).
\end{equation}
While for ${|\mu|<L/2\sqrt{pq}}$, the matter contribution is dominant, 
\begin{equation}\label{FreeZMatIV}
\lim_{n\to\infty}\frac{\ln{Z_{S^3}\big(\gamma,\mu;\mathcal{K}_{p,q},n,L\big)}}{n}
\,\stackrel{|\mu|<L/2\sqrt{pq}}{=}\, i\pi\left(\frac{2 |\mu| 
    \sqrt{pq}}{L} \,-\,\frac{p q \gamma}{2} \,-\, \frac{2}{L^2}\frac{|\mu|^2}{\gamma}\right).
\end{equation}
Understanding the scaling results in
\eqref{FreeZMat}, \eqref{FreeZMatII}, \eqref{FreeZMatIII}, and
\eqref{FreeZMatIV} more simply and directly by a semiclassical analysis in the
Chern-Simons-matter theory could be very interesting.

\paragraph{Case III:  $\gamma$ is non-generic.}

We lastly consider the special case in which the critical point
${z_*=2\pi\zeta p q \gamma}$ of $f_+(z)$ is coincident with one of the
poles \eqref{PolHx} in the prefactor $H(z)$ in \eqref{ZmatTorusII}.
When ${\gamma \in \frac{1}{pq}\Z - \frac{1}{p}\Z - \frac{1}{q}\Z}$
corresponds to a pole arising from the vector multiplet, the
asymptotic analysis proceeeds exactly as in Appendix \ref{Recollect},
with the result that $Z_{S^3}$ scales independently of $n$ as
${n\to\infty}$.  

Otherwise, we are left to consider the case that
$\gamma$ corresponds to a pole arising from the matter multiplet, so that
\begin{equation}
\begin{aligned}
\gamma \,&=\, \frac{2}{\beta}\!\left(-\frac{a}{q}\,-\,\frac{b}{p}\,+\,
  i\frac{\mu}{\sqrt{pq}}\right),\quad a,b\in\Z_{\ge 0}\,,\\
\beta \,&\in\, \big\{L,\,L-2,\,\cdots,\,2-L,\,-L\big\}\,.
\end{aligned}
\end{equation}
For concreteness, we focus on the case ${\beta=-L}$ and ${a=b=0}$, or
\begin{equation}\label{SpecGam}
\gamma \,=\, -\frac{2i\mu}{L \sqrt{pq}}\,,
\end{equation}
which describes the pole nearest to the origin and lying in the
upper half-plane when ${\Im(\mu)>0}$.

We rewrite the original contour integral \eqref{ZmatTorusII} for the
special value of $\gamma$ in \eqref{SpecGam} as 
\begin{equation}\label{ZmatTorusIII}
Z_{S^3}\big(\gamma=-\frac{2i\mu}{L \sqrt{pq}},\mu;\mathcal{K}_{p,q},n,L\big) \,=\,\frac1{2\pi
  i}\frac{1}{\sqrt{pq}}\,\Phi\cdot\int_{\R}dz\;
\e{I(z)}\,,
\end{equation}
where 
\begin{equation}\label{BigIz}
\begin{aligned}
I(z) \,&=\, -\frac{i n L}{16\pi \mu \sqrt{pq}}\,z^2 \,+\, \ha n \zeta
z \,+\,\ln\sinh\!\left(\frac{\zeta z}{2p}\right) +
\ln\sinh\!\left(\frac{\zeta z}{2q}\right) -
\ln\sinh\!\left(\frac{\zeta z}{2}\right)+\,\\
&+\,\sum_{\substack{\beta\in\Delta\,,\\ \Delta=\{L, L-2,\, \cdots,
2-L, -L\}}}\mskip-50mu
\ln s_{b=\sqrt{p/q}}\!\left(\zeta\,\frac{\beta\,z}{4 \pi
  \sqrt{pq}}\,+\,\mu\right),
\end{aligned} 
\end{equation}
and
\begin{equation}
\Phi \,=\, \exp{\!\left[-\frac{\pi \mu \sqrt{pq}}{L}\, n \,+\, O\!\left(\frac{1}{n}\right)\right]}\,.
\end{equation}
Subleading corrections shift the critical point of
$I(z)$ away from the naive value ${z_{*,{\rm naive}}=4\pi\mu\sqrt{pq}/\zeta
  L}$, where the double-sine summand for ${\beta=-L}$ in the second line of
\eqref{BigIz} diverges.  Following the procedure in Appendix
\ref{Recollect}, we compute the correction to the critical
point of $I(z)$ perturbatively in ${1/n\ll 1}$.  

To start, the
derivative $I'(z)$ is given exactly by 
\begin{equation}\label{BigIzD}
\begin{aligned}
I'(z) \,&=\, -\frac{i n L}{8\pi \mu \sqrt{pq}}\,z \,+\, \ha n \zeta
\,+\,\frac{\zeta}{2}\left[\frac{1}{p}\coth\!\left(\frac{\zeta z}{2p}\right) \,+\, \frac{1}{q}\coth\!\left(\frac{\zeta z}{2q}\right) \,-\, \coth\!\left(\frac{\zeta z}{2}\right)\right]+\,\\
&+\,\frac{\zeta}{4\pi\sqrt{pq}}\mskip-30mu\sum_{\substack{\beta\in\Delta\,,\\ \Delta=\{L, L-2,\, \cdots,
2-L, -L\}}}
\frac{\beta\,s_{b=\sqrt{p/q}}'\!\left(\zeta\,\frac{\beta\,z}{4 \pi
  \sqrt{pq}}\,+\,\mu\right)}{s_{b=\sqrt{p/q}}\!\left(\zeta\,\frac{\beta\,z}{4 \pi
  \sqrt{pq}}\,+\,\mu\right)}.
\end{aligned} 
\end{equation} 
We are not aware of a convenient analytic expression for the
derivative $s'_b$ of the double-sine function, but such an expression
will not be necessary to determine the leading asymptotic behavior of
$Z_{S^3}$ as ${n\to\infty}$.  With the ansatz
\begin{equation}
z_* \,=\, \frac{4\pi \mu \sqrt{pq}}{\zeta\,L}\left(1 \,-\,
  \frac{\delta}{n}\right)\,,\qquad\qquad \zeta \equiv \e{i\pi/4}\,,
\end{equation}
we next solve the critical point equation ${I'(z_*)=0}$ perturbatively for
$\delta$.  In terms of the new variable $\delta$, the critical point
equation \eqref{BigIzD} becomes
\begin{equation}\label{BigIzDII}
\begin{aligned}
&0 \,=\, \delta \,+\,
\frac{1}{p}\coth\!\left[\frac{2\pi\mu}{L}\sqrt{\frac{q}{p}}\left(1-\frac{\delta}{n}\right)\right] 
+
\frac{1}{q}\coth\!\left[\frac{2\pi\mu}{L}\sqrt{\frac{p}{q}}\left(1-\frac{\delta}{n}\right)\right]\\
&\qquad-\coth\!\left[\frac{2\pi\mu\sqrt{pq}}{L}\left(1-\frac{\delta}{n}\right)\right]+\,\\
&\qquad\qquad+\,\frac{1}{2\pi\sqrt{pq}}\sum_{j=0}^L \left(2j-L\right)\cdot
\frac{s_{b=\sqrt{p/q}}'\Big(\left(\frac{2j}{L} \,+\, \left(1-\frac{2j}{L}\right)\frac{\delta}{n}\right)\mu\Big)}{s_{b=\sqrt{p/q}}\Big(\left(\frac{2j}{L} \,+\, \left(1-\frac{2j}{L}\right)\frac{\delta}{n}\right)\mu\Big)}\,=\,0\,,
\end{aligned}
\end{equation}
where we have changed the index of summation via the substitution
${\beta=2j-L}$ in the final line.  Compare to the previous, simpler
critical point equation in  \eqref{CritPtG}.  Because $s_b(z)$ has a simple pole
at ${z=0}$, the summand in \eqref{BigIzDII} for ${j=0}$ is singular at
${\delta=0}$.  Otherwise, for generic values of $\mu$, all other terms
in \eqref{BigIzDII} are regular at ${\delta=0}$.  

In the perturbative expansion for large ${n\gg 1}$, the critical point
equation \eqref{BigIzDII} takes the schematic form 
\begin{equation}\label{BigIzDIII}
\left(1 \,+\, \frac{\mathsf{A}}{n}\right)\delta \,+\, \mathsf{B} \,+\,
\frac{L}{2\pi\mu\sqrt{pq}}\cdot\frac{n}{\delta} \,+\, O\!\left(n^{-2}\right)\,=\, 0\,.
\end{equation}
Like the analogous equation in \eqref{CritPtGP}, $\mathsf{A}$ and
$\mathsf{B}$ are constants independent of $\delta$ and $n$, but
depending on the parameters $(p,q,\mu,L)$.  These constants arise from
the Taylor expansions of the hyperbolic cotangents and double-sine functions in
\eqref{BigIzDII}.  To leading-order in $n$, the values of $\mathsf{A}$
and $\mathsf{B}$ are inessential, so we do not evaluate the
constants here.  

The singular term in \eqref{BigIzDIII},
proportional to $1/\delta$, arises from the aforementioned simple pole
of $s_b(z)$ at ${z=0}$.  As a result of this term, the pertubative
equation for $\delta$ has the pair of solutions
\begin{equation}
\delta_\pm \,=\, \frac{-\mathsf{B} \,\pm\,\sqrt{\mathsf{B}^2 \,-\, 4
    \left(1+\frac{\mathsf{A}}{n}\right)\left(\frac{L\,n}{2\pi\mu\sqrt{pq}}\right)}}{2\left(1\,+\,\frac{\mathsf{A}}{n}\right)} \,=\,
\pm i \sqrt{\frac{L\,n}{2\pi\mu\sqrt{pq}}} \,+\, O(1)\,.
\end{equation}
For ${\Im(\mu)>0}$ the critical point described by $\delta_+$ is
distinguished, since the original real contour in \eqref{ZmatTorusIII}
can be deformed to pass through the $\delta_+$-critical point without passing
over the pole in $H(z)$ at ${4\pi\mu\sqrt{pq}/\zeta L}$.  After the contour
deformation, $Z_{S^3}$ can then be 
evaluated by the naive stationary-phase approximation at the corrected
critical point 
\begin{equation}\label{CorZCaseIII}
z_* \,=\, \frac{4\pi \mu \sqrt{pq}}{\zeta\,L} \,-\,\zeta
\sqrt{\frac{8\pi\mu\sqrt{pq}}{L\,n}} \,+\, O\!\left(\frac{1}{n}\right).
\end{equation}
When we apply the stationary-phase formula in \eqref{ZmatStatPhas},
the subleading $O\big(1/\sqrt{n}\big)$-correction in
\eqref{CorZCaseIII} is only
necessary to resolve the singularity in the prefactor $H(z_*)$.  For
non-singular terms, the correction can be neglected at leading-order.
We thus find the non-vanishing limit
\begin{equation}\label{eq:ZmatstationaryPhaseTermIII}
\begin{aligned}
&\lim_{n\to\infty}
Z_{S^3}\!\left(\gamma=-\frac{2i\mu}{L\sqrt{pq}},\mu;\mathcal{K}_{p,q},n,L\right) 
\,=\,\\
&\qquad\frac{\zeta}{L}
\sqrt{\frac{2}{\pi}}\cdot\frac{\sinh\!\left(\frac{2\pi\mu}{L}\sqrt{\frac{q}{p}}\right)
  \sinh\!\left(\frac{2\pi\mu}{L}\sqrt{\frac{p}{q}}\right)}{\sinh\!\left(\frac{2\pi\mu}{L}\sqrt{pq}\right)}\times\,\mskip-50mu\prod_{\substack{\beta\in\Delta\,,\\ \Delta=\{L, L-2,\, \cdots,
2-L\}}}\mskip-45mu
s_{b=\sqrt{p/q}}\!\left(\mu\left(1+\frac{\beta}{L}\right)\right)
\end{aligned}
\end{equation}
Observe that the term for ${\beta=-L}$ is omitted in the product
over weights, so the argument of each double-sine is non-zero for
${\mu\neq 0}$, and the limit is non-singular.

The formula \eqref{eq:ZmatstationaryPhaseTermIII} for
${\lim_{n\to\infty} Z_{S^3}}$ in the theory with supersymmetric matter
superficially resembles the prior result
\eqref{eq:stationaryPhaseTermII} in the pure gauge theory, insofar as
both imply a finite, non-vanishing limit.  However,
these formulas have an important qualitative difference.  

The limit
formula \eqref{eq:stationaryPhaseTermII} for $Z_{S^3}$ in pure gauge
theory applies only when the torus knot $\mathcal{K}_{p,q}$ is {\em
  non-trivial}.  For the unknot, $Z_{S^3}$ in pure gauge theory
vanishes as $1/\sqrt{n}$ according to \eqref{eq:Zunknot}.  By
contrast, in the theory with supersymmetric matter, the result for
${\lim_{n\to\infty} Z_{S^3}}$ in
\eqref{eq:ZmatstationaryPhaseTermIII} applies to all torus knots,
including the unknot with ${p=q=1}$.  Concretely, the one-loop matter
determinant creates a singularity in the prefactor $H(z)$ which is present for all torus
knots, whereas the corresponding prefactor $F(z)$ for the pure gauge
theory -- recall \eqref{BigFx} -- is non-singular for the unknot.  
The Jones polynomial $J_n(\mathcal{K}_{p,q})$ then
has a power-law divergence with $n$ at the special point
${\gamma=1/pq}$, but the analogous ratio of supersymmetric partition
functions in the $SU(2)$ gauge theory with spin-$L/2$ matter remains finite as ${n\to\infty}$,
\begin{equation}
\begin{aligned}
&\lim_{n\to\infty}\frac{Z_{S^3}\!\left(\gamma=-\frac{2i\mu}{L\sqrt{pq}},\mu;\mathcal{K}_{p,q},n,L\right)}{Z_{S^3}\!\left(\gamma=-\frac{2i\mu}{L},\mu;\mathcal{K}_{1,1},n,L\right)}
\,=\,\\
&\qquad\qquad\frac{\sinh\!\left(\frac{2\pi\mu}{L}\sqrt{\frac{q}{p}}\right)
  \sinh\!\left(\frac{2\pi\mu}{L}\sqrt{\frac{p}{q}}\right)}{\sinh\!\left(\frac{2\pi\mu}{L}\sqrt{pq}\right)
  \sinh\!\left(\frac{2\pi\mu}{L}\right)}\,\times\mskip-45mu\prod_{\substack{\beta\in\Delta\,,\\ \Delta=\{L, L-2,\, \cdots,
2-L\}}}\mskip-5mu \left[\frac{s_{b=\sqrt{p/q}}\!\left(\mu\left(1+\frac{\beta}{L}\right)\right)}{s_{b=1}\!\left(\mu\left(1+\frac{\beta}{L}\right)\right)}\right].
\end{aligned}
\end{equation}

\end{onehalfspace}

\end{document}